\documentclass[11pt,hidelinks]{report}
\usepackage{latexsym}
\usepackage{amssymb}
\usepackage{amsmath}
\usepackage{amscd}
\usepackage{amsthm}
\usepackage{hyperref}
\usepackage{slashed}
\usepackage{cite}

\usepackage[papersize={8.5 in, 11 in}, nohead, includeheadfoot, left=1.5 in, right = 1 in, vmargin= 1 in]{geometry}

\allowdisplaybreaks
\usepackage{amsmath}

\theoremstyle{definition}

\usepackage{setspace}
\theoremstyle{remark}

\numberwithin{equation}{section}

\newcommand{\e}{\emph}

\global\long\def\bR{\mathbb{R}}%

\def\thetitle{THE STRONGLY COUPLED $E_8\times E_8$ HETEROTIC STRING:\\ GEOMETRY $\&$ PHENOMENOLOGY}
\def\theauthor{Sebastian Dumitru}   
\def\theadvisor{Burt A. Ovrut}

\def\theyear{2022}

\usepackage{tocloft}
\usepackage{color}
\usepackage{array}

\usepackage[compat=1.1.0]{tikz-feynman}

\usepackage{graphicx}
\usepackage{booktabs} 

\usepackage{subcaption}

\usepackage{float}

\usepackage{caption}

\usepackage{breqn}

\usepackage{amsmath}

\newcommand{\Z}{\ensuremath{\mathbb{Z}}}

\newcommand{\R}{\ensuremath{{\mathbb{R}}}}
\newcommand{\C}{\ensuremath{{\mathbb{C}}}}
\newcommand{\Rep}[1]{\ensuremath{\boldsymbol{\underline{#1}}}}

\newcommand{\Kcone}{\ensuremath{\mathcal{K}}}

\newcommand{\Ocal}{\ensuremath{{\cal O}}}
\newcommand{\Vvis}{\ensuremath{V^{(1)}}}
\newcommand{\Vhid}{\ensuremath{V^{(2)}}}
\newcommand{\Rhat}{\ensuremath{{\widehat R}}}
\newcommand{\ah}{\hat{\alpha}_{\text{GUT}}}

\global\long\def\repd#1{\boldsymbol{#1}}%
\global\long\def\repb#1{\overline{\boldsymbol{#1}}}%
\global\long\def\dd{\text{d}}%
\global\long\def\ii{\text{i}}%
\global\long\def\ee{\text{e}}%
\global\long\def\SO#1{{SO}(#1)}%
\global\long\def\Uni#1{{U}(#1)}%
\global\long\def\SU#1{{SU}(#1)}%
\global\long\def\Ex#1{{E}_{#1}}%
\global\long\def\ex#1{\mathfrak{e}_{#1}}%
\global\long\def\kernel{\operatorname{ker}}%
\global\long\def\eqspace{\mathrel{\phantom{{=}}{}}}%
\global\long\def\op#1{\operatorname{#1}}%

\global\long\def\id{\operatorname{id}}%

\DeclareMathOperator{\tr}{tr}
\DeclareMathOperator{\Span}{span}
\DeclareMathOperator{\rank}{rank}
\DeclareMathOperator{\re}{Re}

\newcommand{\Tr}{\mbox{Tr$\;$}}

\newcommand{\pmns}{{\mbox{\tiny PMNS}}}

\newcommand{\trix}[1]{\left(\begin{array}{#1}}
\newcommand{\notrix}{\end{array}\right)}
\newcommand{\comment}[1]{}
\newcommand{\rulesep}{\unskip\ \vrule\ }

\def\beq{\begin{equation}}
\def\eeq{\end{equation}}
\def\bea{\begin{eqnarray}}
\def\eea{\end{eqnarray}}

	\DeclareMathOperator{\imag}{Im}

\newcommand{\madgraph}{\textsc{MG5\_aMC@NLO}}

\def\tr{\text{Tr}\,}

\begin{document}

\pagenumbering{roman}
\large\newlength{\oldparskip}\setlength\oldparskip{\parskip}\parskip=.15in
\thispagestyle{empty}
\begin{center}


\thetitle

\theauthor

A DISSERTATION 

in 

Physics and Astronomy

Presented to the Faculties of the University of Pennsylvania

in 

Partial Fulfillment of the Requirements for the

Degree of Doctor of Philosophy


\theyear
\end{center}

\vspace{1.5cm} 

\noindent Supervisor of Dissertation

\bigskip
\bigskip

\noindent Burt A. Ovrut, Professor of Physics\\

\normalsize\parskip=\oldparskip

\newpage
\doublespacing

\chapter*{Acknowledgments}
\addcontentsline{toc}{chapter}{Acknowledgements}%

The work in this thesis has been a long time in the making and there are many people
who have helped and supported me over the last five years.

Firstly, I want to thank my advisor, Burt Ovrut, for the research opportunities he provided me with and for his constant presence and guidance along my studies.

I am also indebted to my
collaborators, Anthony Ashmore, and Austin Purves, for their significant contributions to our research, as well as for the many enlightening conversations we had. Furthermore, the study of the R-parity violating decays of superparticles was partly inspired from the collaboration with Evelyn Thomson and her research group.

I would be remiss not to thank (in alphabetical order) my good friends Alex Breitweiser and Ian Graham, who made my time
in the US really fun and memorable, and who put up with me as their roommate for three years.

Finally, my greatest thanks go to my partner Alecsandra, for her love and support, and for being a continuous source of excitement.

My work supported in part by the research grant
DOE No. DESC0007901.


\newpage

\begin{center}
  ABSTRACT\\
\thetitle\\
\vspace{.5in}
  \theauthor\\
  \theadvisor
\end{center}

\noindent

Working within the context of the strongly coupled $E_8\times E_8$ heterotic string theory, we analyze the $B-L$ MSSM, a realistic supersymmetric extension of the Standard Model, from both a low-energy phenomenology
and high-energy string perspective. From a formal point of view, we examine different constructions of string vacua, which satisfy a series of
theoretical and phenomenological constraints. Such vacua configurations are anomaly-free,
preserve $N = 1$ SUSY at the compactification scale and yield the correct value for the
$SO(10)$ unification scale and gauge coupling in the visible particle sector. 
Furthermore, we analyze a possible SUSY-breaking mechanism for the type of 
vacua we construct, via gaugino condensation in the hidden sector. In an attempt to connect these findings to experiment, 
we study the prospects of detecting the $B-L$ MSSM at the LHC in the near future. 
Within this specific context, we show that R-parity violating decays of supersymmetric particles could be amenable to direct detection at the ATLAS and CMS detectors. Detection of
these processes would not only be an explicit indication of "Beyond the Standard Model" physics,
but would also hint strongly at the existence of $N=1$ SUSY with spontaneously broken R-parity.
\addcontentsline{toc}{chapter}{Abstract}%

\vspace*{\fill}


\clearpage

\tableofcontents


\clearpage
\addcontentsline{toc}{chapter}{List of Tables}
\listoftables


\clearpage
\addcontentsline{toc}{chapter}{List of Figures}
\listoffigures

\newpage
\pagenumbering{arabic}
\include{introdept}
\include{back}
\include{finitedept}
\include{infinitedept}

\chapter{Introduction}

\section{String Theory - Going Beyond the Standard Model}

The Standard Model (SM) of high-energy physics provides a remarkably
successful description of presently known phenomena. To date, experiments advancing toward the TeV range have failed to find any additional structure
beyond the Standard Model. This is puzzling, because, despite its success, we know that the Standard Model of particles must still be a work in progress.

 One of the unsatisfactory characteristics of the SM is that it has many free
parameters, that have no underlying explanation. There are three
gauge coupling constants $g_1$, $g_2$, $g_3$, the QCD $\theta$ parameter, the masses of the quarks and leptons, the CP-violating phase, in addition to Higgs and Yukawa couplings,
all of which are external to the model, rather than predicted by it. Moreover, the relatively recent discovery of the non-zero neutrino masses offers strong hints of "Beyond the Standard Model Physics".

Another compelling motivation for "Beyond the SM Physics" is that the gauge coupling constants seem to unify around the $10^{16}$GeV scale, commonly referred to as the Grand Unification scale $M_U$. Indeed, there is
strong evidence for such unification, coming from the evolution of the three SM gauge
couplings to high energies with the renormalization group equations. If that is the case, there must be a larger unification group at high energies, which contains
the SM group $SU(3)\times SU(2)_L\times U(1)_Y$ and admits complex representations to accommodate a chiral fermion spectrum.

In contrast, this scale is much larger than the Electroweak breaking scale $\sim 100$GeV. 
The fact that the ratio $M_U/M_{EW}$ is so large provides another hint to the character of
physics beyond the Standard Model, because of the well-known "hierarchy problem", related to the mass of the Higgs boson. The 2012 discovery of the Higgs boson found its mass near 125GeV.
However, the Higgs mass receives enormous quantum
corrections from the virtual effects of every particle or other phenomenon that couples, directly or
indirectly, to the Higgs field. A question can immediately be raised as to why the Higgs boson is so much lighter than the Planck mass or the unification scale. Although this problem could
simply have a fine-tunning explanation, a more elegant solution is provided by the existence of supersymmetry, which protects the mass of the Higgs boson from receiving divergent loop corrections.

Although it has never been observed, supersymmetry plays a central role within the most accepted theories of nature. 
From a phenomenological perspective, supersymmetric models constitute a rich playground to construct particle physics
models, with supersymmetry broken at low energy scales, potentially attainable by the LHC in the near future. There are, however, several other motivations to consider supersymmetry, mostly coming from string theory, which will be discussed next. For example, spacetime configurations with 
preserved supersymmetry are simpler to study from a theoretical point of view; furthermore, they are automatically stable and do not contain tachyons. 

A particular feature of supersymmetry is that it predicts a super-partner for each of the known particles of the Standard Model, identical in mass and any other quantum numbers except for spin.
We do not observe, however, such particles in nature. For this reason, if supersymmetry is indeed a manifest theory of our universe, it must have been broken above the Electroweak scale. As a result of this supersymmetry breaking, all these superparticles must have attained a very large mass. Two questions instantly arise: 1. by which mechanism can supersymmetry be broken? and 2. 
where are all the massive superparticles? The answer to the first question is more complex. We will not answer it here, but we point out that the second chapter of this thesis shows how this problem can be solved in the context of string theory. The second question has two answers, depending on if the lightest supersymmetric particle (the LSP) is stable or not. The LSP can be stable only
if it is protected by a symmetry called "R-parity". Therefore, in models with conserved R-parity, the LSP is a possible dark matter candidate and must be neutral to avoid a disallowed density of charged relics. For this reason, R-parity severely narrows the SUSY phenomenological landscape. Models with broken R-parity make it possible to assume that the LSP will simply decay into Standard Model particles. Such R-parity violating (RPV) decays are potentially amenable for detection at the LHC, thus opening a large arena of particle physics phenomenology. The study of such RPV decays is the main topic of the third chapter of this thesis.

So far, we have pointed out many reasons to believe the Standard Model is not the final theory of nature, before even addressing what is probably the biggest of them all: gravity. Gravity cannot be quantized in the same way that all the other forces do. Furthermore, if one is to have a truly unified
description of the interactions in our universe, it is natural to consider gravity on the same
footing with the interactions of the SM.

Though many of these problems have been individually addressed, the only consistent framework that can potentially solve all of these issues is that of string theory. The premise of string theory is that, at the fundamental level, matter does not consist
of point-particles, but rather of tiny loops of string, which already implies that string theory cannot be a QFT. However, the string length is
usually supposed to be close to the Planck length $\sim 10^{-33}$cm, and hence,
at lower energies, string theory resembles an effective QFT theory. The extended nature of a string becomes only
relevant at the string scale, where it smears out the location of the interaction, which helps to
avoid ultraviolet divergences. From this simple premise, all theories of general relativity, electromagnetism, and Yang-Mills emerge naturally. The different particle species are interpreted as different
vibrational modes of the strings. Among this vibrational spectrum, we find a massless mode that exactly has the
properties of a spin-2 graviton. String
theories thus potentially constitute the unified theory of all interactions including the SM
and gravitation. 

 It turns out that when trying to quantize these vibrational modes of the string, we can obtain a vacuum state free of tachyons only if we include the supersymmetric spectrum of string vibrational excitations. That is, the only
physical string theories we know how to solve are, in fact, superstring theories. Furthermore, the superstring quantization is possible only in 10-dimensional spacetime. This interesting fact can be consistent with our observation of a seemingly four-dimensional universe, as long as six dimensions are
compact. Finally, there isn't just one possible consistent superstring theory, but five of them. Their different characteristics are shown in Table \ref{tab:string_theories}. Throughout this thesis, we will explore only one such string theory model, namely the $E_8\times E_8$ heterotic string, and show how to obtain a
low-energy particle content and effective action close to the SM of particle physics.

Finally, we point out that string theory is not necessarily a theory of loops of strings only; it can contain extended objects with different numbers of spatial dimensions, such as membranes. The strings appear to play the role
of the fundamental object in the small coupling perturbative limit only, in which all other higher dimensional states are heavy and decouple from the dynamics; at finite coupling all branes
are expected to be on equal footing, leading to a complicated structure for the theory, for which no full theoretical description is known.
We explore the rich structure of such theories in the next section and demonstrate their importance in reproducing low-energy physics observations.

\begin{table}[t]
\begin{center}
\begin{tabular}{|c||c|c|c|c|}
\hline
 & String Type & SUSY & Gauge Group & Massless Bosonic Spectrum
\\
\hline
\hline
IIA& \begin{tabular}{@{}c@{}} oriented \\  closed \end{tabular}&\begin{tabular}{@{}c@{}} $N=2$ \\   non-chiral \end{tabular}& $U(1)$ & $g_{IJ}, \>B_{IJ}, \>\phi, \>C_I, \>C_{IJK}$\\
\hline
IIB& \begin{tabular}{@{}c@{}} oriented \\  closed \end{tabular}& \begin{tabular}{@{}c@{}} $N=2$ \\   chiral \end{tabular} & none  & $g_{IJ}, \>B_{IJ}, \>\phi, \>C_0, \>C_{IJ},\>C_{IJKL}$\\
\hline
I&\begin{tabular}{@{}c@{}} non-oriented \\  open, closed \end{tabular}& \begin{tabular}{@{}c@{}} $N=1$ \\  chiral \end{tabular}& $SO(32)$ & $g_{IJ}, \>\phi, \>A_I^a, \>C_{IJ}$\\
\hline
\begin{tabular}{@{}c@{}}$SO(32)$ \\  heterotic \end{tabular}
&oriented closed&\begin{tabular}{@{}c@{}} $N=1$ \\  chiral \end{tabular}& $SO(32)$ & $g_{IJ},\>B_{IJ}, \>\phi, \>A_I^a$\\
\hline
\begin{tabular}{@{}c@{}}$E_8\times E_8$ \\  heterotic \end{tabular} &\begin{tabular}{@{}c@{}}oriented \\  closed \end{tabular}& \begin{tabular}{@{}c@{}} $N=1$ \\   chiral \end{tabular}& $E_8\times E_8$ & $g_{IJ},\>B_{IJ}, \>\phi, \>A_I^a$\\
\hline
\end{tabular}
\end{center}
\caption{Superstring theory types in 10D.}
\label{tab:string_theories}
\end{table}

\section{String Dualities and M-theory}

Before presenting the theoretical context of the strongly coupled heterotic string, we need to introduce a theory that is not a theory of strings, called M-theory.
M-theory is a theory in physics believed to unify all consistent versions of superstring theory, shown in Table \ref{tab:string_theories}. Although a complete formulation of M-theory is not yet known, such a formulation should describe two- and five-dimensional objects called M$p$-branes and should be approximated by eleven-dimensional supergravity at low energies~\cite{Witten:1996mz}. Its massless fields are the
11D metric $G_{\hat I\hat J}$, the 3-index antisymmetric tensor $C_{\hat I\hat J\hat K}$ and the 11D gravitino $\psi_{\hat I\alpha}$, with 11D indices. We denote the eleven-dimensional
coordinates of the eleven-dimensional space $\mathcal{M}_{11}$ by $x^0,\dots,x^9,x^{11}$ and the corresponding (hatted) indices by $\hat I, \hat J, \hat K, \dots = 0, . . . , 9, 11$. 

The bosonic part of the 11D supergravity action is
\begin{equation}
S_{SG}=\frac{1}{2\kappa_{11}^2} \left[\int_{\mathcal{M}_{11}}\sqrt{-g}
\left(-R-\tfrac{1}{2}| G|^2\right)-\frac{1}{6}\int_{\mathcal{M}_{11}} C\wedge G\wedge G
\right]\ ,
\label{eq:supergravity11}
\end{equation}
where the four-form $G$ is the field strength of the three-form potential $C_{IJK}$
\begin{equation}
G=dC\ .
\end{equation}
This is a theory of pure supersymmetric gravity, with no conventional gauge symmetries or matter content. 
The expression \eqref{eq:supergravity11} can also be written explicitly in terms of the 11D tensor components, such that
\begin{equation}
S_{SG}=\frac{1}{2\kappa_{11}^2}\int_{\mathcal{M}_{11}}\sqrt{-g} \left[
-R-\tfrac{1}{24}G_{\hat I\hat J\hat K \hat L}G^{\hat I\hat J\hat K\hat L}-\frac{\sqrt{2}}{1728}\epsilon^{I_ {\hat 1}\dots I_{\hat{11}}}C_{I_{\hat 1}I_{\hat2}I_{\hat3}}G_{I_{\hat 4}\dots I_{\hat 7}}
G_{I_{\hat 8}\dots I_{\hat {11}}}
\right]\ ,
\end{equation}
where $G_{\hat I\hat J \hat K\hat L}=24\partial_{[\hat I}C_{\hat I\hat J\hat K]}$. 

It is perhaps the simplest to explain the duality between M-theory and type IIA theory at strong coupling. 
Type IIA string theory contains, apart from the field content presented in the previous section (of which the bosonic components were shown in Table \ref{tab:string_theories}), non-perturbative states called D$p$-branes with even $p$. These states are excited states of the vacuum, with tension and mass approximated by~\cite{Ibanez:2012zz}
\begin{equation}
\begin{split}
T_p&\sim \frac{{\alpha^{\prime}}^{-(p+1)/2}}{g_s}\ ,\\
M_p&\sim {\alpha^{\prime}}^{-1/2}g_s^{-1/(p+1)}\ .
\end{split}
\end{equation}

In the strong coupling limit $g_s\rightarrow \infty$, the D0-branes of the IIA theory become the dominant light modes of the theory. Let us consider bound states formed of stacks of $k$ D$0$ branes. The mass of each of those bound states is proportional to $k/g_s$ and goes to zero in the limit $g_s\rightarrow \infty$. Therefore, in the limit of strong coupling, this system resembles a tower of KK modes, suggesting that the strongly coupled type IIA theory is related to a higher-dimensional 11D theory with a compact dimension of type $S^1$; this theory is called M-theory and the strongly coupled type IIA string theory is its $S^1$ decompactification limit. In this limit, the 11-dimensional space of M-theory has the structure 
$\mathcal{M}_{11}=M_{10}\times S^1$, where $M_{10}$ is the 10-dimensional space in which the type IIA theory is embedded. 
We will use unhatted indices $I, J, K, \dots  = 0, \dots , 9$ to label the ten-dimensional coordinates. When
we later further compactify the theory on a Calabi-Yau three-fold, we will use indices $A, B, C, \dots =
4\dots 9$ for the Calabi-Yau coordinates, and indices $\mu, \nu \dots= 0, \dots, 3$ for the coordinates of the
remaining, uncompactified, four-dimensional space. Holomorphic and antiholomorphic coordinates
on the Calabi-Yau space will be labeled by $a, b, c, \dots$ and $\bar a, \bar b, \bar c\dots$ .

To make the duality more clear, note that the field content of the low energy action of M-theory, that is the 11D supergravity action, reduces to the type IIA string theory $N=2$ non-chiral matter content given in Table \ref{tab:string_theories}, when compactified on $S^1$:
\begin{alignat*}{3}
g_{\hat I\hat J}\quad \longrightarrow\quad &g_{IJ}  \qquad && \text{graviton}\\
			                &g_{I,11} \equiv C_I\qquad &&\text{RR 1-form}\\
				       &g_{11,11} \equiv \phi \qquad &&\text{dilaton}\\
C_{\hat  I\hat  J \hat K}\quad \longrightarrow\quad &C_{IJK}  \qquad && \text{RR 3-form}\\
			                &C_{IJ,11} \equiv B_{IJ}\qquad &&\text{NSNS 2-form}\\
\psi_{\hat I\alpha}\quad \longrightarrow\quad &\psi_{I\alpha},\>\psi_{I\dot\alpha}  \qquad && \text{gravitinos}\\
			                &\psi_{11\alpha},\>\psi_{11\dot\alpha} \equiv \lambda_{\alpha}, \lambda_{\dot \alpha}\qquad &&\text{dilatinos}\\		
\end{alignat*}

This correspondence between the fields of type IIA string theory and the M-theory fields implies the 
existence of an 11D lift of the non-perturbative type IIA $p$-branes to the excited states of M-theory, called M$p$-branes. 
We have already shown that the type IIA D0 branes are lifted to a tower of KK modes, which are more precisely, the KK momenta of 11D 
supergravitons. Similarly, the IIA string is lifted to an extended M2 brane wrapped on a $S^1$ cycle:
\begin{figure}[H]
  \centering
  \includegraphics[width=0.6\textwidth]{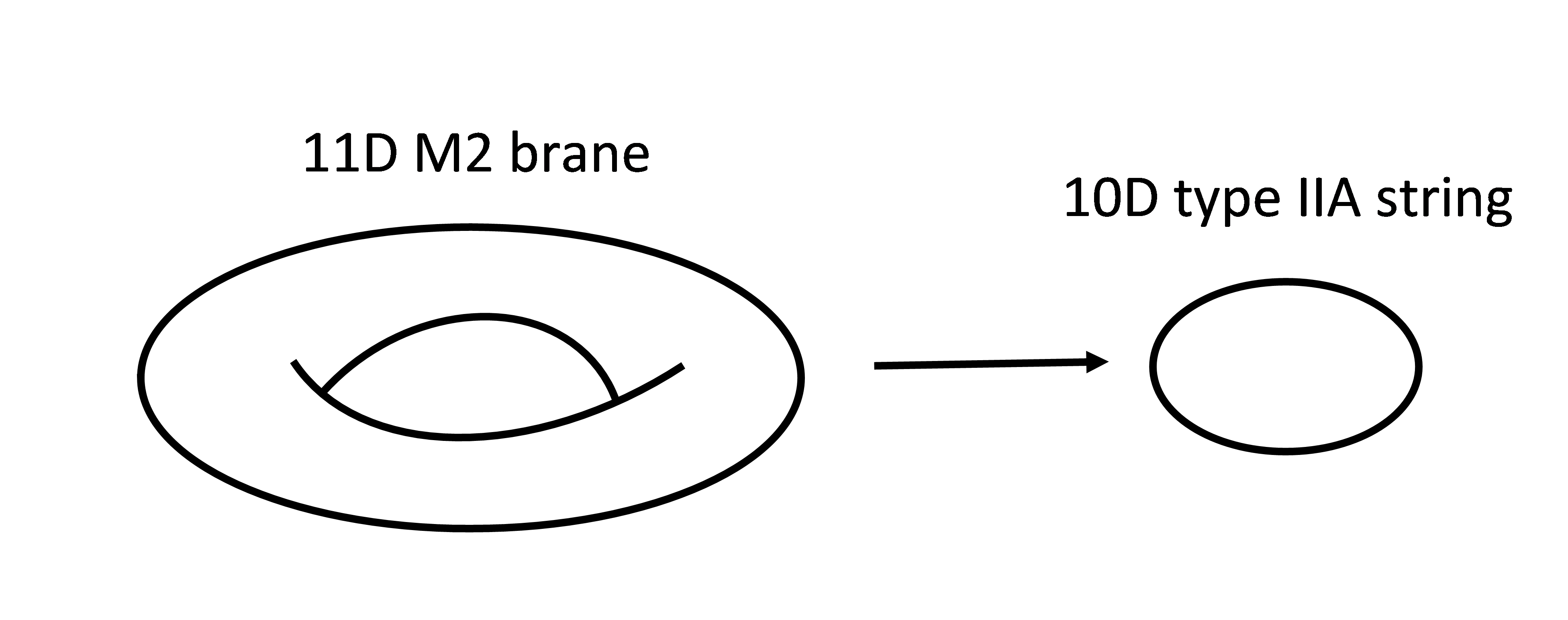} 
  \label{fig:mbrane_string}
\end{figure}

\section{Hořava–Witten theory}

 In the previous section, we showed that M-theory compactified on $S^1$ results in type IIA theory. There is, however, one more possible compactification of M-theory to 10 dimensions, on the quotient $S^1/\Z_2$. The resultant theory is the strongly coupled $E_8\times E_8$ heterotic string, also known as Hořava–Witten theory\cite{Horava:1995qa,Horava:1996ma}.

The
orbifold $I=S^1/\Z_2$ is chosen in the $x^{11}$ direction, so we assume $x^{11} \in [-\pi \rho, \pi\rho]$, with the endpoints
identified as $x^{11} \sim x^{11} + 2\pi\rho$. The $Z_2$ symmetry acts as $x^{11}\sim -x^{11}$. There exist two
ten-dimensional hyperplanes, $M^{10}_i$ with $i=1,2$, locally specified by the conditions $x^{11} = 0$ and $x^{11}=\pi\rho$, which are fixed under the action of the $\Z_2$ symmetry. This is called the “upstairs” picture, in which the
eleventh coordinate is considered as the full circle with singular points at the fixed hyperplanes. Most often, however, we will use the
“downstairs” picture, in which the orbifold is considered as an interval $x^{11} \in [0, \pi \rho]$, with the fixed
hyperplanes forming boundaries to the eleven-dimensional space (see Figure  \ref{fig:Vacum_structure2}). 

\begin{figure}[t]
  \centering
  \includegraphics[width=0.6\textwidth]{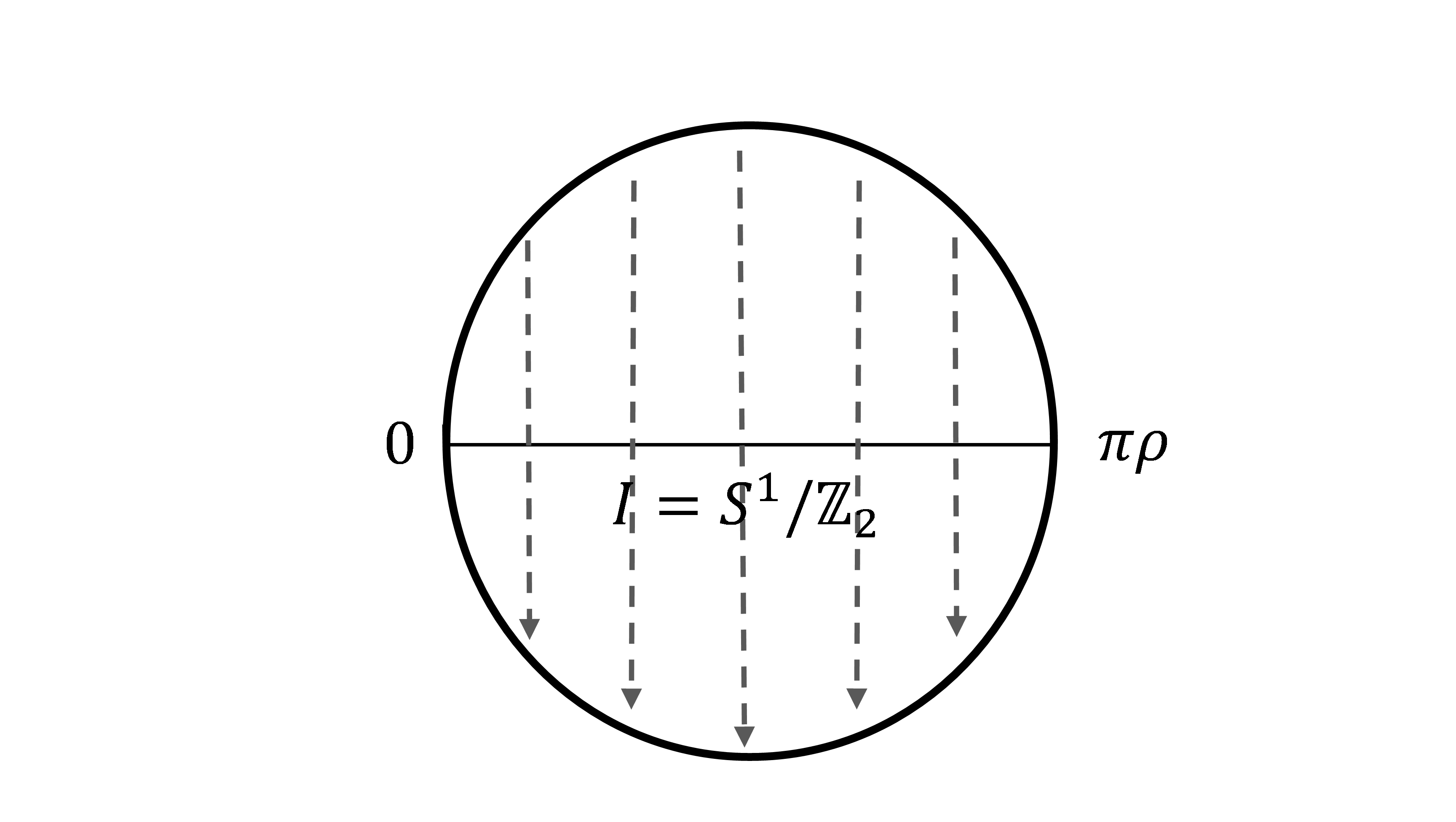}
  \label{fig:Orbififold}
\caption{The $I=S^1/\Z_2$ orbifold interval across the 11th dimension, with two fixed points at $0$ and $\pi \rho$, invariant under the $\Z_2$ action.}
\end{figure}

As explained above, to lowest order, the effective action for M-theory on $S^1/\Z_2$ is that of 11D supergravity. The bosonic part of this action was given in eq. \ref{eq:supergravity11}. However, the components of the supergravity multiplet $(g_{\hat I\hat J},
\psi_{\hat I\alpha}, C_{\hat I\hat J\hat K})$ are restricted by the orbifold $Z_2$ symmetry, which acts on the $x^{11}$ coordinate and 
the 3-form field $C_3\equiv C_{\hat I\hat J\hat K}$, as
\begin{equation}
\begin{split}
\theta:x^{11}\quad &\longrightarrow \quad-x^{11}\ ,\\
\theta: C_3\quad &\longrightarrow \quad -C_3\ .
\end{split}
\end{equation}
The sign flip of the 3-form $C_3$ is required to keep the 11D Chern-Simons coupling $C\wedge G\wedge G$ invariant. As consequence, the 11-dimensional bosonic fields are either even or odd under $\Z_2$. Specifically:
\begin{itemize}
\item even: $g_{IJ}, \> g_{11,11}, \> C_{11, IJ}$\ ,
\item odd: $g_{11, J},\> g_{I,11}, \>C_{IJK}$\ ,
\end{itemize}
where, as a reminder, the indices $ I,J,K=0,\dots,9$ run over the 10 dimensional space $M_{10}$. For the components of the 11-dimensional 
gravitino, the condition is
\begin{equation}
\psi_I(x^{11})=\Gamma_{11}\psi_{I}(-x^{11})\ ,\qquad \psi_{11}(x^{11})=-\Gamma_{11}\psi_{11}(-x^{11})\ .
\end{equation}

In the limit of large separation between the 10-dimensional hyperplanes $M^{1,2}_{10}$, and for slowly varying fields across the orbifold dimension, one can ask what is the 10-dimensional effective theory. This corresponds to projecting the $\Z_2$ invariant massless spectrum of M-theory into 10D:
\begin{alignat*}{3}
g_{\hat I\hat J}\quad \longrightarrow\quad &g_{IJ}  \qquad && \text{graviton}\\
				       &g_{11,11} \equiv \phi \qquad &&\text{dilaton}\\
C_{\hat  I\hat  J \hat K}\quad \longrightarrow\quad 
			                &C_{IJ,11} \equiv B_{IJ}\qquad &&\text{NSNS 2-form}\\
\psi_{\hat I\alpha}\quad \longrightarrow\quad &\psi_{I\alpha} \qquad && \text{gravitino}\\
			                &\psi_{11\dot\alpha} \equiv \lambda_{\dot \alpha}\qquad &&\text{dilatino}\\		
\end{alignat*}
We recover the 10D $N=1$ gravity supermultiplet, implying that the action of $\Z_2$ breaks half of the
supersymmetries. 

The theory is chiral from a ten-dimensional perspective and has a gravitational anomaly, localized on the fixed
hyperplanes and therefore, it appears to be inconsistent. The key to this problem is that as one approaches the fixed points
$x^{11}=0,\pi R$ on the orbifold, the dynamics of M-theory on $S^1/\Z_2$ cannot be accurately described by 
supergravity alone. The action $\theta: x^{11}\rightarrow -x^{11}$ relates points infinitesimally close together, 
which leads to effects beyond the supergravity approximation. It can be shown that this effect leads to 
two 10-dimensional $N=1$ $E_8$ gauge supermultiplets, one on each of the two fixed 10-dimensional hyperplanes of the orbifold, $M^1_{10}$, $M_{10}^2$. 
The structure of this 11-dimensional vacuum configuration is shown in Figure \ref{fig:Vacum_structure2}.

\begin{figure}[t]
  \centering
  \includegraphics[width=0.76\textwidth]{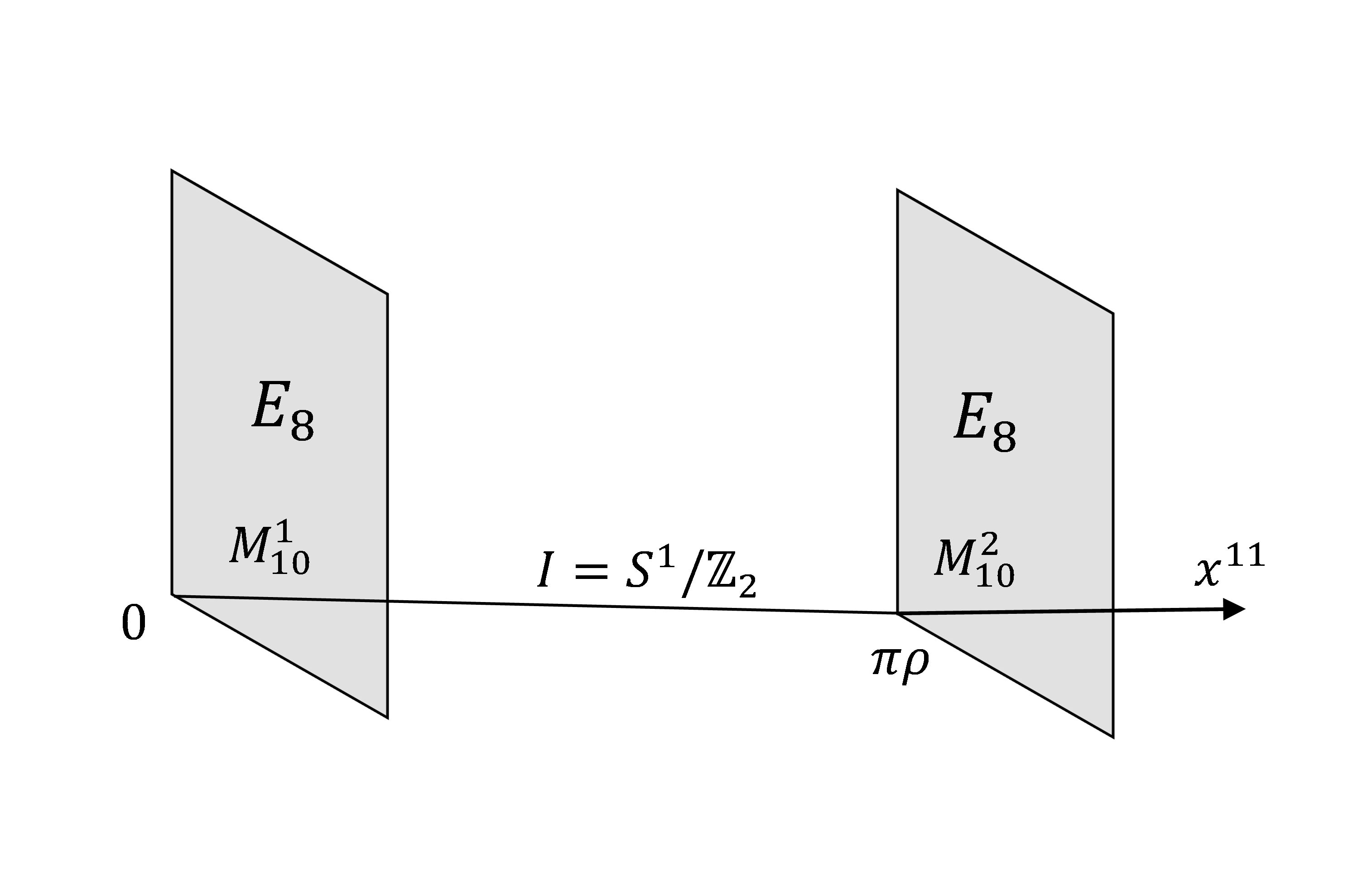}
\caption{Structure of the Hořava-Witten 11-dimensional vacuum. The 11D supergravity action, which is manifest in the bulk space, receives corrections at order $\kappa_{11}^{2/3}$, localized on the 10-dimensional 
orbifold planes $M_{10}^{1,2}$, located at $x^{11}=0,\pi \rho$ and separated across the 11th dimension. These corrections consist in the appearance of $E_8$ vector supermultiplets on each of the two 10-dimensional orbifold hyperplanes.}
  \label{fig:Vacum_structure2}
\end{figure}

As consequence, the supergravity action receives corrections at order $\kappa_{11}^{2/3}$, localized on the 10-dimensional 
orbifold planes. Restricting to terms at most quadratic
in derivatives, one finds that the action is given by~\cite{Lukas:1997fg}
\begin{equation}
\begin{split}
S&=S_{SG}+S_{YM}\\
&=\frac{1}{2\kappa_{11}^2} \left[\int_{\mathcal{M}_{11}}\sqrt{-g}
\left(-R-\tfrac{1}{2}| G|^2\right)-\frac{1}{6}\int_{\mathcal{M}_{11}} C\wedge G\wedge G
\right]\\
&\quad-\frac{1}{8\pi\kappa_{11}^2}\left(\frac{\kappa_{11}}{4\pi}  \right)^{2/3}\int_{M^1_{10}}\sqrt{-g}\>\text{tr}(F^{(1)})^2
-\frac{1}{8\pi\kappa_{11}^2}\left(\frac{\kappa_{11}}{4\pi}  \right)^{2/3}\int_{M^2_{10}}\sqrt{-g}\>\text{tr}(F^{(2)})^2\ .
\end{split}
\end{equation}
where $F^{(1,2)}_{IJ}$ are the field strengths of the two $E_8$ gauge fields $A^{(1,2)}_{I}$. Away from these orbifold planes, the theory is still supergravity. However, to preserve supersymmetry, it is necessary to add $\kappa_{11}^{2/3}$ corrections to
the Bianchi identity for $G_{\hat I\hat J \hat K \hat L}$, in the form of source terms localized on the hyperplanes, so that
\begin{equation}
(dG)_{11 IJ K L}=4\sqrt{2}\pi\left(\frac{\kappa_{11}^{2/3}}{4\pi}\right)\left[ J^{(0)}_{IJ K L}\delta(x^{11})
+J^{(1)}_{ IJ K L}\delta(x^{11}-\pi \rho)\right]\ ,
\end{equation}
where the sources are given by
\begin{equation}
J^{(0,1)}_{\hat I\hat J \hat K \hat L}=-\frac{1}{16\pi^2}\left(  \text{tr}\>F^{(1,2)}\wedge F^{(1,2)}-\frac{1}{2}R\wedge R \right)_{\hat I\hat J \hat K \hat L}\ .
\label{eq:sources}
\end{equation}

In addition, M-theory contains BPS
M2- and M5-branes, which can be constructed as 11D supergravity solutions. The heterotic $E_8 \times E_8$ string can be obtained from wrapping the
M2 membrane on the $S^1/\Z_2$ interval, as illustrated below.
\begin{figure}[H]
  \centering
  \includegraphics[width=0.6\textwidth]{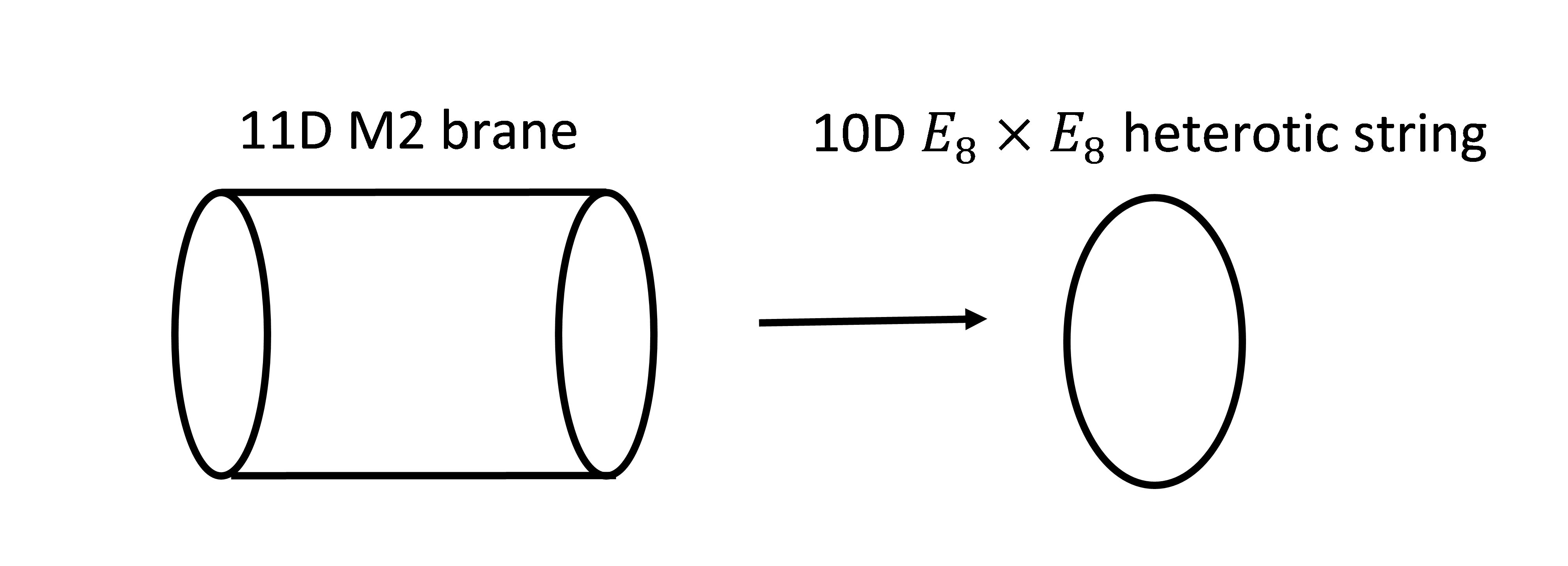}
  \label{fig:mbrane_string2}
\end{figure}
The unwrapped M2 brane is not $\Z_2$ invariant and does not correspond to a BPS brane. The M5-brane wrapped on $S^1/\Z_2$ is also not invariant, but
the M5-brane sitting at a point in $S^1/\Z_2$ does survive the projection and corresponds to
a heterotic NS5-brane. In principle, an arbitrary number of M5-branes can coexist in the interval between the 10-dimensional orbifold plane.

One consequence of adding a number $N$ of M5-branes to the system displayed in Figure \ref{fig:Vacum_structure2} is that the Bianchi identity gets modified:
\begin{equation}
\begin{split}
(dG)_{11 I J K  L}=4\sqrt{2}\pi\left(\frac{\kappa_{11}^{2/3}}{4\pi}\right)\Big[ &J^{(0)}\delta(x^{11})
+J^{(N+1)}\delta(x^{11}-\pi \rho)\\
&+\frac{1}{2}\sum_{n=1}^N J^{(n)}\left( \delta(x^{11}-x_n) +\delta(x^{11}+x_n) \right)
\Big]_{ IJ K L}\ ,
\label{bianchi_eq}
\end{split}
\end{equation}
The sources $J^{(0)}$ and $J^{(N+1)}$ arise on the orbifold fixed planes. Their form was given in eq. \eqref{eq:sources}, for $N=0$. The sources $J^{(n)}$, $n = 1, \dots ,N$ arise due to the presence of the five–branes. Their exact form has been 
computed in \cite{Brandle:2003uya}, but is not needed in the present discussion. Note that in the above expression, we summed over each of the five branes located at $x^n\in (0,\pi \rho)$, and their orbifold $\Z_2$ mirror images, located at $x^{n}\in(-\pi \rho,0)$.

The appearance of the boundary source terms in the Bianchi identity has a simple interpretation
by analogy with the theory of D-branes~\cite{Ibanez:2012zz}. $U(N)$ gauge fields describing the
theory of $N$ overlapping D$p$-branes encode the charges for lower-dimensional D-branes embedded
in the D$p$-branes. For example, the magnetic flux $\text{tr}F$ couples to the $p-1$-form RR
potential, so describes $D(p-2)$-brane charge, while $ \text{tr}{F}\wedge \text{tr}{F}$ describes
the embedding of D$(p-4)$-
brane charge. Furthermore, if the D$p$-brane is curved, then the
cohomology class $\text{tr}R\wedge R$ of the tangent bundle also induces D$(p-4)$-
brane charge. In eleven dimensions M five-branes are magnetic sources
for $G_{\hat I\hat J\hat K\hat L}$. Thus we can interpret the magnetic sources in the Bianchi identity as five-branes
embedded in the orbifold fixed planes.

\begin{figure}[t]
  \centering
  \includegraphics[width=0.76\textwidth]{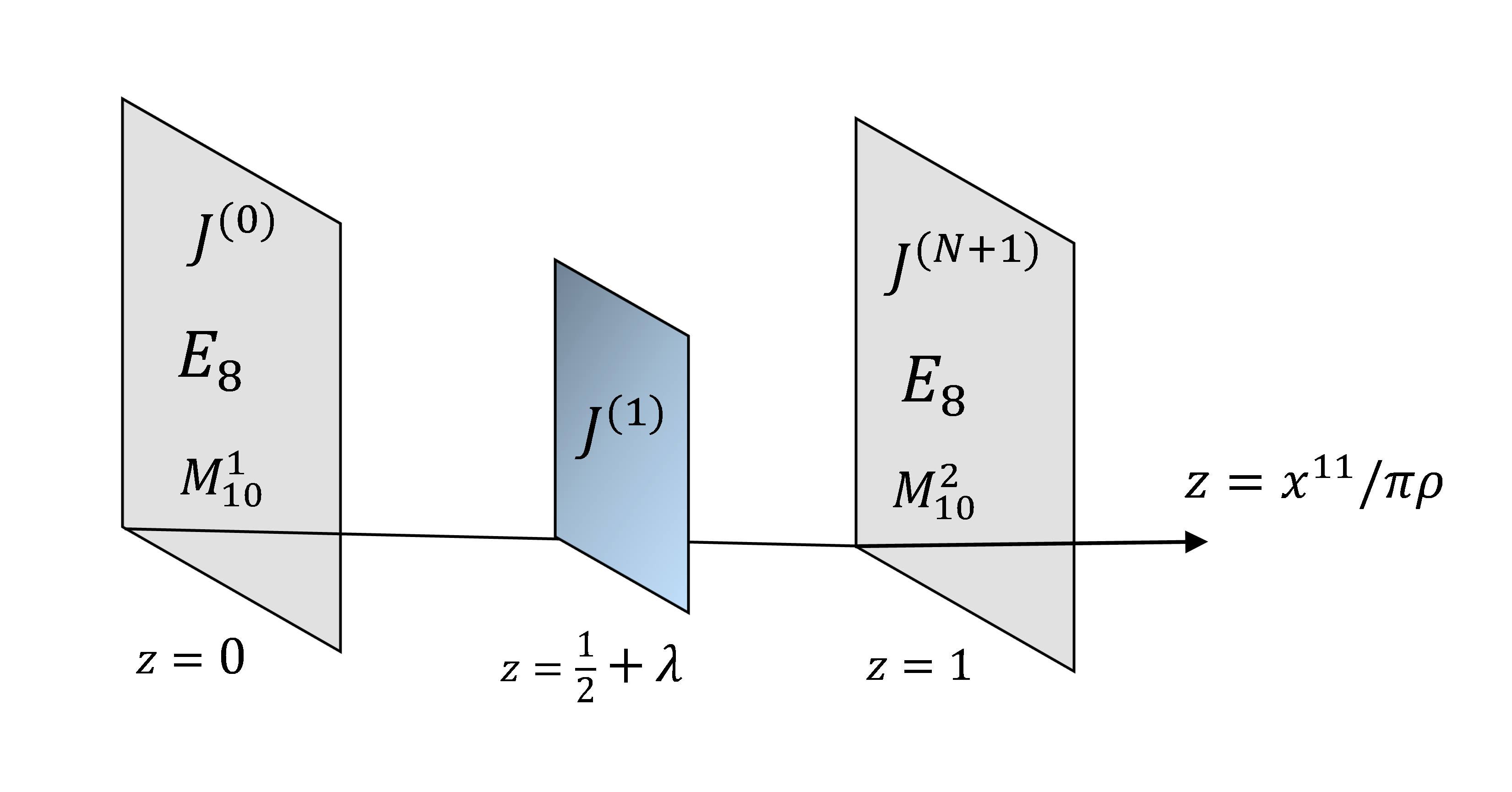}
\caption{Structure of the Hořava-Witten 11-dimensional vacuum. The supergravity action receives corrections at order $\kappa_{11}^{2/3}$, localized on the 10-dimensional 
orbifold planes $M_{10}^{1,2}$, located at $z=0,1$ and separated across the 11th dimension. These corrections consist in the appearance of $E_8$ vector supermultiplets on each of the orbifold plane. We allowed for the possibility of
introducing an M5-brane, located at a specific point $z=\frac{1}{2}+\lambda$ in the interval $z\in[0,1]$. The sources $J^{(0)}$ and $J^{(N+1)}$ arise on the orbifold fixed planes. Their form was given in eq. \eqref{eq:sources}, for $N=0$. The source $J^{(1)}$ is due to the presence of the five–brane.}
  \label{fig:Vacum_structure3}
\end{figure}

\section{Compactification Manifold}

To make contact with low-energy physics, one would like to consider compactifications of
the strongly-coupled theory which have $N = 1$ supersymmetry in four dimensions. There are several motivations to consider string compactifications preserving supersymmetry.
Firstly, they are
simpler to study from the theoretical point of view; regarding the compactification backgrounds
from the supergravity perspective, solutions to the (first-order) supersymmetry
conditions can be shown to automatically be solutions to the (much more involved, second order)
supergravity equations of motion. Also, compactifications which preserve supersymmetry are automatically stable and do not contain tachyons. Such compactifications are also appealing from a phenomenological perspective; supersymmetric
backgrounds open a playground to construct particle physics
models, with supersymmetry broken at low energy scales, potentially attainable by the LHC in the near future. 

To understand the properties of Calabi-Yau compactifications, we need to consider the supersymmetry transformation of the 11D gravitino
\begin{equation}
\delta \psi_{\hat I}=D_{\hat I}\eta+\frac{\sqrt{2}}{288}(\Gamma_{\hat I \hat J\hat K \hat L \hat M} 
-8g_{\hat I\hat J}\Gamma_{\hat K\hat L\hat M}) G^{\hat J\hat K\hat L\hat M}\eta+\dots ,
\end{equation}
where the dots indicate the omitted fermionic terms and $\eta$ is an eleven-dimensional Majorana
spinor. This spinor should be restricted by the condition
\begin{equation}
\eta(x^{11})=\Gamma_{11}\eta(-x^{11})
\end{equation}
for the supersymmetry variation to be compatible with the $\Z_2$ symmetry. This constraint
means that the theory has the usual 32 supersymmetries in the bulk but only 16 (chiral) supersymmetries on the 10-dimensional orbifold hyperplanes.

The condition for a spinor field $\eta$ to generate unbroken supersymmetry in a vacuum state is that the right
hand side vanishes (i.e. if the vacuum state is not invariant under supersymmetry, then supersymmetry is broken), 
\begin{equation}
\label{eq:vanish_eta}
D_{\hat I}\eta+\frac{\sqrt{2}}{288}(\Gamma_{\hat I \hat J\hat K \hat L \hat M} 
-8g_{\hat I\hat J}\Gamma_{\hat K\hat L\hat M}) G^{\hat J\hat K\hat L\hat M}\eta=0\ .
\end{equation}

At tree-level, the solution has a simple form. Without the corrections arising at linear order in $\kappa_{11}^{2/3}$, the Bianchi identity shown in \eqref{bianchi_eq} is simply
\begin{equation}
\begin{split}
(dG)_{11 I J K  L}=0\ .
\end{split}
\end{equation}
Therefore, both the Bianchi identity and the equation of motion 
\begin{equation}
\label{eq:eq_motion_G}
D_{\hat I}G^{\hat I \hat J \hat K\hat L}=0\ ,
\end{equation}
derived from the 11D supergravity action, can be solved by setting $G=0$. When this is the case, eq. \eqref{eq:vanish_eta}
is satisfied by picking a metric on $\mathcal{M}_{11}$ that admits
a covariantly constant spinor field, $D_{\hat I}\eta = 0$. On the other hand, when compactifying from eleven dimensions
to a four-dimensional flat space, on a manifold of the type $X \times S^1$, where $X$ is a six-dimensional manifold, 
it is the covariantly constant 6D spinors on $X$, which satisfy $D_A\eta_6=0$, that determine the unbroken supersymmetries in 4D.

Manifolds that admit covariantly constant spinors are Calabi-Yau manifolds. A Calabi-Yau manifold is a complex $2N$-dimensional manifold, which admits a Ricci flat
metric, or whose first Chern class $c_1(X) = 0$. The existence of a covariantly constant spinor field determines 
the $SU(N)$ holonomy of this manifold. In heterotic compactifications, we consider six-dimensional
Calabi-Yau manifolds, with $N=3$.

We have seen, however, that when compactifying on a $\Z_2$ invariant orbifold, the Bianchi identity receives correction terms, as shown in eq. \eqref{bianchi_eq}. These source terms have to be taken into account and, as consequence, the field strength solution $G=0$ is not possible. This effect further induces corrections to
the metric, which can be computed by requiring that $N = 1$ supersymmetry is preserved.

Without the order $\kappa_{11}^{2/3}$ corrections, the tree-level background metric has
the form
\begin{equation}
ds^2=g_{IJ}^{(0)}dx^{\hat I}dx^{\hat J}=\eta_{\mu \nu}dx^{\mu}dx^{\nu}
+g_{AB}dx^Adx^B+(dx^{11})^2\ ,
\end{equation}
where $g_{AB}$ is the metric of the Calabi-Yau space, which, in holomorphic coordinates, is related to the
K\"ahler form $\omega_{a\bar b}$ by $\omega_{a\bar b} = -ig_{a\bar b}$. Its K\"ahler form 
can be expanded in terms of the harmonic $(1,1)$-forms $\omega_{ia\bar b}, i = 1, \dots, h^{1,1}$ as
\begin{equation}
\omega_{a\bar b} = a^i\omega_{ia\bar b}\ .
\end{equation}
The coefficients $a^i = a^i(x^\mu)$, $\mu=0,\dots,3$ are the $(1, 1)$ moduli of the Calabi–Yau space. The Calabi–Yau volume
modulus $V = V (x^\mu)$ is defined by
\begin{equation}
V=\frac{1}{v}\int_X \sqrt{g}\ ,
\end{equation}
where $\sqrt{g}$ is the determinant of the Calabi–Yau metric $g_{AB}$. In order to make $V$ dimensionless we
have introduced a coordinate volume $v$ in this definition, which can be chosen for convenience. The
modulus $V$ then measures the Calabi-Yau volume in units of $v$. $ V$ is
not independent of the $(1, 1)$ moduli $a_i$, but it can be expressed as
\begin{equation}
V=\frac{1}{6}d_{ijk}a^ia^ja^k\ ,
\end{equation} 
where $d_{ijk}$ are the Calabi–Yau intersection numbers,
\begin{equation}
  d_{ijk} = \frac{1}{v}
  \int_X \omega_i \wedge \omega_j \wedge \omega_k\ .
\end{equation}

In the next chapter, we will use a particular Calabi--Yau threefold $X$, which is the fiber product of two rationally elliptic $\dd\mathbb{P}_{9}$ surfaces, that is, a self-mirror Schoen threefold \cite{Braun:2004xv}, quotiented with respect to a freely acting $\mathbb{Z}_{3} \times \mathbb{Z}_{3}$ isometry. Its Hodge data is $h^{1,1}=h^{1,2}=3$, so that it has three K\"ahler and three complex structure
moduli. 

However, adding the first-order corrections, of the $\kappa_{11}^{2/3}$ expansion, will distort this background solution. As mentioned previously, source terms appear at this order in
the Bianchi identity, shown in eq. \eqref{bianchi_eq}, and therefore, $G$ cannot be set to zero.

First note that since the l.h.s of the Bianchi Identity is an exact form, so must be the r.h.s. Therefore, the new Bianchi identity has a solution that preserves supersymmetry upon compactification if the source terms add up
to zero cohomologically, as an element of $H^{2,2}(X)$, that is
\begin{equation}
\label{eq:anomJ_canc}
\left[\sum_{n=0}^{N+1}J^{(n)}\right]=0\ .
\end{equation}
Physically speaking, this means that there can be no overall charge in a compact space, because there is nowhere for the flux to go. We can express this condition in a different form by defining the charges
\begin{equation}
\beta_i^{(n)}=\frac{1}{v^{1/3}}\int_X \omega_i\wedge J^{(n)}\ ,\quad i=1,\dots,h^{1,1}, \quad n=0,\dots, N+1\ .
\end{equation}
It is then possible to express the currents in terms of the $(1,1)$ harmonic forms,
\begin{equation}
*_{X}J^{(n)}=\frac{1}{v^{2/3}}\sum_{k=1}^{h^{1,1}}\beta_i^{(n)}\omega^i\ ,
\end{equation}
and therefore, the anomaly cancellation condition from eq. \ref{eq:anomJ_canc} becomes
\begin{equation}
\sum_{n=0}^{N+1}\beta_i^{(n)}=0\ .
\end{equation}

Having defined those conventions, the Bianchi identity in eq. \eqref{bianchi_eq} and the equation of motion from eq. \eqref{eq:eq_motion_G} can be solved in terms of a two form $\mathcal{B}$, defined by
\begin{equation}
-d\mathcal{B}^{(6)}=*G\ .
\end{equation}

It can then be shown~\cite{Brandle:2003uya} that the Bianchi identity can be written into the form of a Poisson equation
\begin{equation}
\begin{split}
(\Delta_X+\partial_{11}^2)\mathcal{B}=-\frac{\epsilon_S^\prime v^{2/3}}{\pi \rho}*_X\Big[ &J^{(0)}\delta(x^{11})
+J^{(N+1)}\delta(x^{11}-\pi \rho)\\
&+\frac{1}{2}\sum_{n=1}^N J^{(n)}\left( \delta(x^{11}-x_n) +\delta(x^{11}+x_n) \right)
\Big]\ ,
\end{split}
\end{equation}
where $\Delta_X$ denotes the Laplacian and $*X$ is the Hodge star operator, restricted to the
Calabi-Yau manifold X. We have also defined the strong-coupling expansion parameter $\epsilon_S^\prime$
\begin{equation}
  \epsilon_{S}^\prime= \pi\left(\frac{\kappa_{11}}{4\pi} \right)^{2/3}\frac{2\pi\rho}{v^{2/3}} \ ,
\end{equation}
which tracks the first order $\kappa_{11}^{2/3}$ corrections.

Next, after expanding the two-form $\mathcal B$ in terms of the $(1,1)$ harmonic forms,
\begin{equation}
\mathcal{B}=\sum_{i=1}^{h^{1,1}}b_i\omega^i\ ,
\end{equation}
it can be shown that the Poisson equation becomes
\begin{equation}
\begin{split}
\partial_{11}^2b_i=-\frac{\epsilon_S^\prime}{\pi \rho}\Big[ &\beta_i^{(0)}\delta(x^{11})
+\beta_i^{(N+1)}\delta(x^{11}-\pi \rho)\\
&+\frac{1}{2}\sum_{n=1}^N \beta_i^{(n)}\left( \delta(x^{11}-x_n) +\delta(x^{11}+x_n) \right)
\Big]\ ,
\end{split}
\end{equation}
which has linear solutions in $x^{11}$, of the type
\begin{equation}
b_i(z)=-\frac{\epsilon_S^\prime}{2}\left[\sum_{m=0}^{n}\beta^{(m)}_i(|z| -z_m )-\frac{1}{2}
\sum_{m=0}^{N+1}(z_m^2-2z_m)\beta_i^{(m)}\right]\ ,
\end{equation}
inside the interval $z_n\leq |z|, \leq z_{n+1}$,
where $z=x^{11}/\pi \rho$ and $z^n$, $n=0,\dots, N+1$ are the fixed position of the five-branes/orbifolds.

We then find the non-zero solutions to the Bianchi identity and the equation of motion, which appear at linear order in $\epsilon_S^\prime$
\begin{align}
G_{ABCD}&=\frac{1}{2}\epsilon_{ABCDEF}\partial_{11}\mathcal{B}^{EF}\\
G_{ABC,11}&=\frac{1}{2}\epsilon_{ABCDEF}\partial^D\mathcal{B}^{EF}\ .
\end{align}

\begin{figure}[t]
  \centering
  \includegraphics[width=0.76\textwidth]{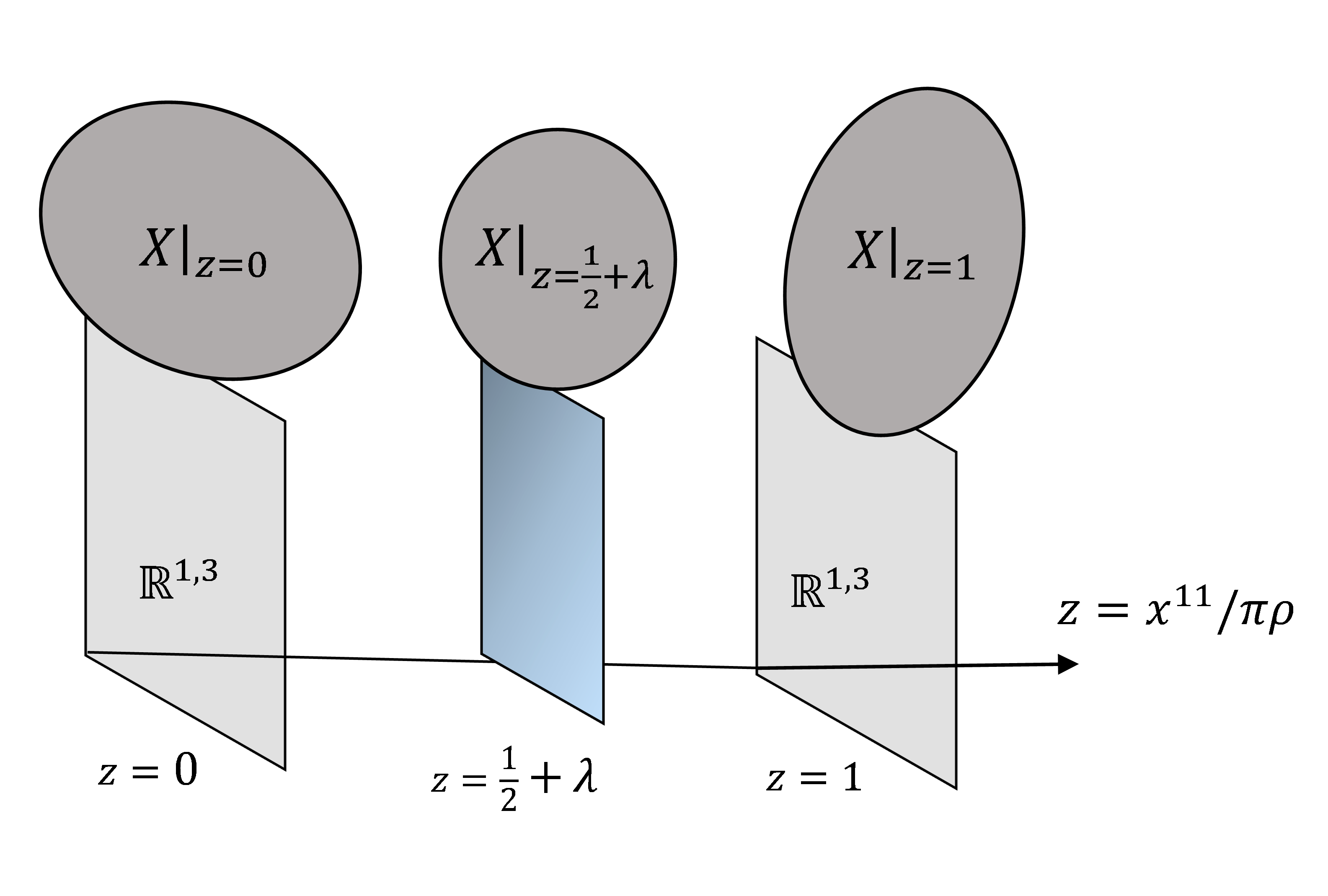}
\caption{Compactification of Hořava–Witten theory on a CY manifold $X$. We allowed for the possibility of
introducing an M5-brane wrapped on a holomorphic 2-cycle $\mathcal{C}_i$ and located at a specific point $z=\frac{1}{2}+\lambda$ in the interval $z\in[0,1]$. In the strongly coupled string regime, the size of the Calabi-Yau is smaller than the size across the 11th dimension. Therefore, on compactification, the universe appears first five-dimensional and then finally four-dimensional, after integrating over the orbifold separation. Also, note that the shape of the Calabi-Yau space changes across the 11th dimension of space. The shape change is induced by source terms from the two fixed orbifold planes and the five-brane, as shown in eq. \eqref{eq:metric_change}.}
  \label{fig:Vacum_structure4}
\end{figure}

Following~\cite{Lukas:1998hk}, it turns out that after solving the modified Bianchi identity, it is possible to find a solution for the metric that
solves the Einstein equations to order $\epsilon_S^\prime$. The source term corrections induce metric corrections of the form
\begin{equation}
\label{eq:metric_change}
ds^2=(1+b)\eta_{\mu \nu}+(g_{AB}+h_{AB})dx^Adx^B+(1+\gamma)(dx^{11})^2\ ,
\end{equation}
with
\begin{equation}
\begin{split}
&h_{AB}=2i\left( \mathcal{B}_{AB} -\frac{1}{3}\omega_{AB} \mathcal{B}\right)\ ,\\
&b=\frac{1}{3} \mathcal{B}_{AB}\omega^{AB}\ ,\quad \gamma=-\frac{2}{3} \mathcal{B}_{AB}\omega^{AB}\ .\\
\end{split}
\end{equation}

Subsequently, reducing to the 4D effective theory, we express the moduli of the theory in terms of their orbifold average functions defined as follows. For any arbitrary dimensionless function $f$ of the five $M_4
\times S^1/\Z_2$ coordinates, define its average over the $S^1/\Z_2$
orbifold interval as
\begin{equation}
\langle f \rangle_{11}=\frac{1}{\pi \rho}\int_0^{\pi\rho}{\dd x^{11}f} \ ,
\label{42}
\end{equation}
where $\rho$ is the reference length. Then $\langle f \rangle_{11}$ is
a function of the four coordinates $x^{\mu}$, $\mu=0,\dots,3$ of $M_4$
only.

\section{Vector Bundles}

Heterotic compactifications on Calabi-Yau manifolds $X$ necessarily have gauge field expectation values (vevs)
in the internal compact dimensions. This feature is a consequence of demanding that the total charge for the
Neveu-Schwarz form should vanish on the internal compact space. 
In addition to the holomorphic vector bundles on the observable and
hidden orbifold planes, the bulk space between these planes can
contain five-branes wrapped on two-cycles ${\cal{C}}_2^{(n)}$,
$n=1,\dots,N$ in $X$. Cohomologically, each such five-brane is
described by the $(2,2)$-form Poincar\'e dual to ${\cal C}_2^{(n)}$,
which we denote by $W^{(n)}$. Note that to preserve $N=1$
supersymmetry in the four-dimensional theory, these curves must be
holomorphic and, hence, each $W^{(n)}$ is an effective class.

In the previous section, we have seen that solutions to the new Bianchi identity, which preserve supersymmetry upon compactification, exist if the source terms add up
to zero cohomologically, as an element of $H^{2,2}(X)$, such that
\begin{equation}
\label{eq:anomaly_vB}
\left[\sum_{n=0}^{N+1}J^{(n)}\right]=0\ .
\end{equation}
The source currents have the forms
\begin{align}
    J^{(0)}=&\;\nonumber
    -\frac{1}{16 \pi^2}
    \Big( \tr_{E_8} F^{(1)} \wedge F^{(1)}
    -\frac{1}{2}\tr_{SO(6)} R \wedge R \Big) \\[1ex]
    J^{(n)}=&\;
    W^{(n)}, \quad n=1,\dots,N, \\[1ex]
    J^{(N+1)}=&\;
    -\frac{1}{16 \pi^2}
    \Big( \tr_{E_8} F^{(2)} \wedge F^{(2)}
    -\frac{1}{2}\tr_{SO(6)} R \wedge R \Big). \nonumber
  \end{align}
These relations clearly show that the existence of gauge field instantons in the internal dimensions of the compactification manifold is a requirement in a viable vacuum structure.

One of the interesting properties of these gauge field vevs on the compact dimensions concerns their
supersymmetry preserving properties. In the previous section, we studied the supersymmetry variation of the 11-dimensional gravitino $\Psi$, to find
how to preserve 
$N=1$ supersymmetry upon compactification on a six-dimensional manifold. Interesting relations and constraints can also be learned from the supersymmetry transformation of the 10-dimensional gauginos from the 
$E_8$ vector supermultiplets localized on the 10-dimensional planes $M^{1,2}_{10}$. For a gaugino $\lambda^\alpha$, where $\alpha$ runs over the adjoint representation of one of the $E_8$, this transformation has the form
\begin{equation}
D_\eta \lambda^\alpha=-\frac{1}{4g\sqrt{\phi}}\Gamma^{AB}F^\alpha_{AB}\eta+\dots\ ,
\end{equation}
where the dots represent the fermion terms. Therefore, in order to preserve
$N = 1$ supersymmetry in the four-dimensional effective theory upon turning an expectation value of the internal field strength, $F^\alpha$ on either of the two 10-dimensional orbifold planes, we need to ensure that
\begin{equation}
-\frac{1}{4g\sqrt{\phi}}\left(\Gamma^{ab}F^\alpha_{ab}+\Gamma^{\bar a\bar b}F^\alpha_{\bar a\bar b}+2\Gamma^{a\bar b}F^\alpha_{a\bar b}\right)\eta=0\ .
\end{equation}
We are led to the following set of equations
\begin{equation}
\label{eq:lalaHYM}
F^\alpha_{ab}=F^\alpha_{\bar a\bar b}=0\ ,\quad g^{a\bar b}F^\alpha_{a\bar b}=0\ .
\end{equation}
These are known as the Hermitian Yang-Mills equations, which must be satisfied on each of the two orbifold planes. The first set of relations, involving the purely holomorphic
and anti-holomorphic part of the field strength $F$, imply that a non-trivial solution to \eqref{eq:lalaHYM}
involves the construction of holomorphic vector bundles $V^{(1,2)}$ on the Calabi-Yau threefold, on each of the two fixed orbifolds, with structure groups $H^{(1,2)}\subset E_8$. Thus, to arrive at a complete vacuum, we have to specify a Calabi-Yau manifold and then
construct holomorphic vector bundles on the internal dimensions of each fixed orbifold.

Once a holomorphic gauge bundle is constructed, it needs to solve
\begin{equation}
 g^{a\bar b}F^{(1,2)}_{a\bar b}=0\ ,
\end{equation}
which can be written in the form
\begin{equation}
\omega\wedge \omega \wedge F^{(1,2)}=0\ .
\end{equation}
Note that we have contracted the field strengths with the generators $T^\alpha$ of either of the structure groups $H^{(1,2)}\subset E_8$. In general, these equations are difficult to solve. However, the Donaldson-Uhlenbeck-Yau theorem guarantees the existence of a solution to this equation
for gauge backgrounds satisfying the condition that the slope of a bundle $\cal{F}$, defined as
\begin{equation}
  \mu({\cal{F}})=
  \frac{1}{\rank({\cal{F}})v^{2/3}} 
  \int_X{c_1({\cal{F}})\wedge \omega \wedge \omega} \ ,
\end{equation}
is smaller than the slope of any of its sub-bundles $\cal{F}^\prime$, that is 
\begin{equation}
\mu({\cal{F}}^\prime)\> {\geq}\> \mu({\cal{F}}).
\end{equation}
Physically, it means that the gauge background does not split dynamically into lower energy
gauge configurations, corresponding to the sub-bundles.

A polystable bundle is one which can be written as the direct sum of stable bundles,
\begin{equation}
{\cal{F}}=\oplus_i{\cal{F}}_i\ ,
\end{equation}
 with the added condition that all the summand bundles have equal slope,
\begin{equation}
\mu({\cal{F}}_i) =\mu({\cal{F}})
\end{equation}
This definition indicates that stability implies polystability, but not the other way.

A simple structure of the boundary vector fields, which satisfies eq. \eqref{eq:anomaly_vB} is called the "standard embedding", in which no five-branes are present between the two fixed 10-dimensional walls,
and only one gauge instanton is turned on, with structure group $SU(3)\subset E_8$, identical to the structure of the internal manifold, such that
\begin{equation}
\tr_{E_8} F^{(1)} \wedge F^{(1)} =\tr_{SO(6)} R \wedge R\ .
\end{equation}

The $SU(3)$ instanton breaks the $E_8$ gauge group on the first orbifold plane to $E_6$ As a result, after integrating out the compact dimensions of the Calabi-Yau $X$, we are left with an $E_6$ gauge field on the first four-dimensional hyperplane and an
unbroken $E_8$ gauge field on the second four-dimensional hyperplane.

\section{Connection to Experiment}

 Depending on the choice of the Calabi-Yau threefold, as well as the specific gauge connections chosen on each orbifold
surface, there are seemingly endless possibilities for constructing viable vacuum structures. Given this large number of choices, 
it might be a good idea to narrow our search to the models that can reproduce phenomenologically realistic physics. That is, it is possible to
pick both the compactification geometry, as well as the vector bundles, so that we reproduce the observed Standard Model of
particles. In a series of papers~\cite{Braun:2005nv,Braun:2005bw,Braun:2005ux,Bouchard:2005ag,Anderson:2009mh,Braun:2011ni,Anderson:2011ns,Anderson:2012yf,Anderson:2013xka,Nibbelink:2015ixa,Nibbelink:2015vha,Braun:2006ae,Blaszczyk:2010db,Andreas:1999ty,Curio:2004pf}, a set of realistic vacua was obtained, entitled "heterotic standard models".

\begin{figure}[t]
  \centering
  \includegraphics[width=0.6\textwidth]{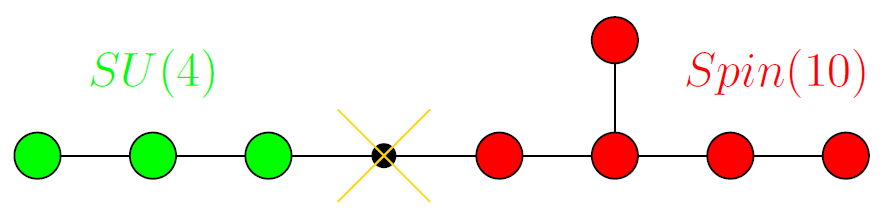}
\caption{The $SU(4)$ holomorphic vector bundle $V^{(1)}$, which is turned on inside the compact dimensions of the Calabi-Yau $X$, breaks the $E_8$ gauge group on the observable sector to the $Spin(10)\subset E_8$ gauge group in four dimensions. }
  \label{fig:SU(4)_diagram}
\end{figure}

In these models, the underlying geometry of the six-dimensional compactification manifold determines a significant amount of properties of the low energy theory. In particular, the Calabi--Yau manifold $X$ can be chosen to be a torus-fibered threefold
with fundamental group $\pi_1(X)=\Z_3 \times \Z_3$.
More specifically, the Calabi--Yau threefold $X$ is the fiber product of two rationally elliptic $\dd\mathbb{P}_{9}$ surfaces, that is, a self-mirror Schoen threefold \cite{Braun:2004xv}, quotiented with respect to a freely acting $\mathbb{Z}_{3} \times \mathbb{Z}_{3}$ isometry.

\begin{equation}
X=\frac{\tilde X}{\Z_3\times \Z_3}\ ,
\end{equation}
where 
\begin{equation}
\tilde X=\dd\mathbb{P}_{9}\times _{P_1}\dd\mathbb{P}_{9}\ .
\end{equation}
Each $\dd\mathbb{P}_{9}$ can be thought
of as an elliptic fibration over a base $P_1$. We take this base $P_1$ to be the same for each $\dd\mathbb{P}_{9}$. This restriction then means that the complex dimension of $\tilde X$ is $2 \times 2-1 = 3$, as required.
 Its Hodge data is $h^{1,1}=h^{1,2}=3$, so there are three K\"ahler and three complex structure
moduli. 

It is then possible to construct a slope-stable,
holomorphic vector bundle with vanishing slope on $X$ of the form
\begin{equation}
V^{(1)}=\frac{\tilde V^{(1)}}{\Z_3\times \Z_3}\ ,
\end{equation}
where $\tilde V^{(1)}$ has structure group $SU(4) \subset E_8 $, which is constructed by “extension” as
\begin{equation}
0\rightarrow V_1\rightarrow \tilde V^{(1)}\rightarrow V_2\rightarrow 0 \ .
\end{equation}
Each of $V_1 $ and $V_2$ is a specific tensor product of a line bundle with a rank two bundle pulled
back from a $\dd\mathbb{P}_{9}$ factor of $\tilde X$. In the heterotic M-theory picture, this bundle is associated with the "observable sector".

This $SU(4)$ holomorphic vector bundle $V^{(1)}$ breaks the $E_8$ gauge group on the observable sector to the $Spin(10)\subset E_8$ gauge group in four dimensions, as shown in Figure \ref{fig:SU(4)_diagram}. %
Although this model has a GUT-like gauge group on the observable sector, it does not contain the GUT-Higgs adjoint
multiplets necessary to break it to the Standard Model gauge group. In contrast with field
theory, string theory is a rigid structure and we cannot choose to introduce this multiplet by hand. However, for non-simply connected CY spaces (i.e with $\pi^1(X)\neq 0$), there exists an alternative of breaking the $Spin(10)$ group by turning
on non-trivial Wilson lines, thus bringing it closer to the SM gauge group.

\setlength{\unitlength}{.86cm}
\begin{figure}[!th]
\begin{center}
\scriptsize
\begin{picture}(5,10)(0,1)

\put(0.8,10){\color{blue}$\underline{SO(10)}$}
\put(-2,9.5){\color{red}\line(1,0){6}}
\put(4.3,9.3){\color{red} $M_{U}=M_{\chi_{B-L}}$}
\put(1.,9.4){\vector(0,-1){1.3}}
\put(1.2,9){$\chi_{B-L}$}

\put(-1.0,7.5){\color{blue}$\underline{SU(3)_C\otimes SU(2)_L\otimes SU(2)_R\otimes U(1)_{B-L}}$}
\put(-0.5,6.5){$L=(\textbf{1},\textbf{2},\textbf{1},-1)$}
\put(-0.5,5.9){$L^c=(\textbf{1},\textbf{1},\textbf{2},1)$}
\put(-0.5,5.3){$Q=(\textbf{3},\textbf{2},\textbf{1},1/3)$}
\put(-0.5,4.7){$Q^c=(\bar{\textbf{3}},\textbf{1},\textbf{2},-1/3)$}
\put(-0.5,3.9){$\mathcal{H}=(\textbf{1},\textbf{2},\textbf{2},0)$}
\put(-0.5,3.3){$H_C=(\textbf{3},\textbf{1},\textbf{1},2/3)$}
\put(-0.5,2.7){${\bar{H}}_C=(\bar{\textbf{3}},\textbf{1},\textbf{1},-2/3)$}

\put(-0.5,6.8){\line(-1,0){.2}}
\put(-0.5,4.6){\line(-1,0){.2}}
\put(-0.7,6.8){\line(0,-1){2.2}}
\put(-0.7,5.7){\line(-1,0){.2}}
\put(-1.4,5.6){\textbf{16}}
\put(3.,6.6){\oval(.4,.4)[tr]}
\put(3.2,6.6){\line(0,-1){.7}}
\put(3.4,5.9){\oval(.4,.4)[bl]}
\put(3.4,5.5){\oval(.4,.4)[tl]}
\put(3.2,5.5){\line(0,-1){.7}}
\put(3.,4.8){\oval(.4,.4)[br]}
\put(3.6,5.6){$\times 9$}

\put(-0.5,4.2){\line(-1,0){.2}}
\put(-0.5,2.6){\line(-1,0){.2}}
\put(-0.7,4.2){\line(0,-1){1.6}}
\put(-0.7,3.4){\line(-1,0){.2}}
\put(-1.4,3.3){\textbf{10}}
\put(3.1,4.2){\oval(.2,.2)[tr]}
\put(3.2,4.2){\line(0,-1){.1}}
\put(3.3,4.1){\oval(.2,.2)[bl]}
\put(3.3,3.9){\oval(.2,.2)[tl]}
\put(3.2,3.9){\line(0,-1){.1}}
\put(3.1,3.8){\oval(.2,.2)[br]}
\put(3.6,3.85){$\times 2$}

\put(-2,2.2){\color{red}\line(1,0){6}}
\put(4.3,2.1){\color{red}$M_{I}=M_{\chi_{3R}}$}
\put(1.,2.1){\vector(0,-1){1.}}
\put(1.2,1.7){$\chi_{3R}$}

\put(-1,.5){\color{blue}$\underline{SU(3)_C\otimes SU(2)_L\otimes U(1)_{3R}\otimes U(1)_{B-L}}$}
\put(-.5,-.5){$L=(\textbf{1},\textbf{2},0,-1)$}
\put(-.5,-1){$e^c=(\textbf{1},\textbf{1},1/2,1)$}
\put(-.5,-1.5){$\nu^c=(\textbf{1},\textbf{1},-1/2,1)$}
\put(-.5,-2){$Q=(\bar{\textbf{3}},\textbf{2},0,1/3)$}
\put(-.5,-2.5){$u^c=(\textbf{3},\textbf{1},-1/2,-1/3)$}
\put(-.5,-3.0){$d^c=(\textbf{3},\textbf{1},1/2,-1/3)$}
\put(-.5,-3.5){$H_u=(\textbf{1},\textbf{2},1/2,0)$}
\put(-.5,-4.0){$H_d=(\textbf{1},\textbf{2},-1/2,0)$}

\put(-.5,-.2){\line(-1,0){.2}}
\put(-.5,-3.1){\line(-1,0){.2}}
\put(-.7,-.2){\line(0,-1){2.9}}
\put(-.7,-1.7){\line(-1,0){.2}}
\put(-1.4,-1.8){\textbf{16}}
\put(3,-.4){\oval(.4,.4)[tr]}
\put(3.2,-.4){\line(0,-1){1.1}}
\put(3.4,-1.5){\oval(.4,.4)[bl]}
\put(3.4,-1.9){\oval(.4,.4)[tl]}
\put(3.2,-1.9){\line(0,-1){1}}
\put(3,-2.9){\oval(.4,.4)[br]}
\put(3.6,-1.8){$\times 3$}

\put(-.5,-3.2){\line(-1,0){.2}}
\put(-.5,-4.1){\line(-1,0){.2}}
\put(-.7,-3.2){\line(0,-1){.9}}
\put(-.7,-3.7){\line(-1,0){.2}}
\put(-1.4,-3.8){\textbf{10}}

\put(6.6,-1.5){MSSM}
\put(7,-2){+}
\put(5.3,-2.5){3 right-handed neutrino}
\put(6.0,-2.9){supermultiplets}

\end{picture}
\end{center}
 \vspace{5cm}
\caption{The particle spectra in the scaling regimes of the sequential Wilson line breaking pattern of $SO(10)$ in which 
$M_{\chi_{B-L}}=M_{U}~>~M_{\chi_{3R}}=M_{I}$.}
\label{fig:matterContent}
\end{figure}
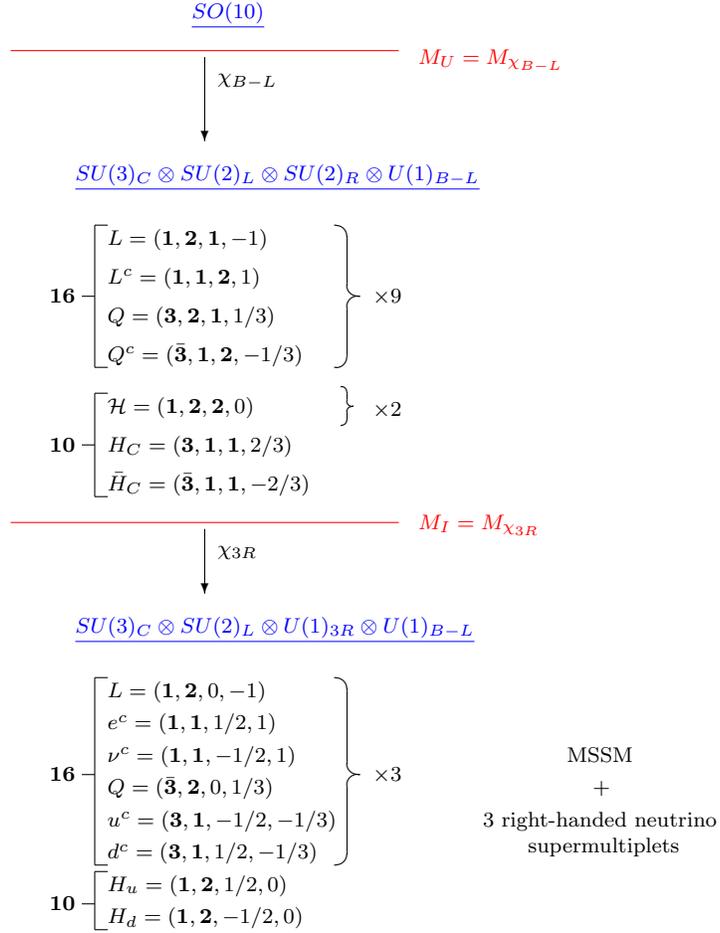

This is precisely what happens in the known constructions of the "heterotic standard models". Realistic low-energy spectra are obtained by turning on
two \emph{flat} Wilson lines $\chi_{B-L}$ and $\chi_{3R}$, each associated with a
different $\Z_3$ factor of the $\Z_3 \times \Z_3$ holonomy of $X$. Doing this preserves the $N=1$ supersymmetry of the effective theory, but breaks the 
observable gauge group down to
\begin{equation}
  Spin(10) 
  \to 
  SU(3)_C \times SU(2)_L \times U(1)_Y \times U(1)_{B-L} \ .
  \label{17}
\end{equation}
The result is the so-called $B-L$ MSSM. The low-energy gauge group, after turning on both $\mathbb{Z}_{3}$ Wilson lines, is that of the Standard Model augmented by an additional gauge $U(1)_{B-L}$ factor, and the particle content of the effective theory is precisely that of the MSSM, with three right-handed neutrino chiral multiplets and a single Higgs-Higgs conjugate pair, and no exotic fields. The complete breaking pattern is shown in Figure \ref{fig:matterContent}.

\section{Outline}

This thesis is structured in two parts. 

The first part follows how the $B-L$ MSSM is obtained from a "top-down" superstring analysis. We have already explained 
that such constructions involve turning on a holomorphic,
slope-stable $SU(4)$ vector bundle on the observable wall, which partially breaks the
observable sector $E_8$ gauge symmetry, leaving a $Spin(10)$ group unbroken in the 4D effective theory.
Two Wilson lines are then turned on to break the $Spin(10)$ further to the $SU(3)_C \times SU(2)_L \times U(1)_Y \times U(1)_{B-L}$ gauge group. On its own, however, this observable sector has no associated dynamical SUSY-breaking
mechanism. A typical proposal in heterotic compactifications is that the SUSY-breaking effects
occur in the hidden sector.

Following the work in \cite{Ashmore:2020ocb}, we show that for a viable heterotic vacuum,
this hidden sector must be consistent with a series of constraints. Previous attempts to build such a
hidden sector were valid only in the weakly coupled heterotic string regime. Unfortunately, in this
regime, one cannot obtain reasonable values for the unification scale or SM gauge couplings.
In Section 2.4, which follows \cite{Ashmore:2020ocb} we show how to rectify this problem. We propose a rank-two hidden sector gauge bundle $L\oplus L^{-1}$, characterized by a line bundle $L$, whose associated $U(1)$ is embedded into the hidden $E_8$
connection via the map $U(1)\rightarrow SU(2) \rightarrow E_8$. This bundle construction not only satisfies all of
the "vacuum constraints", but, in addition, corresponds to the strongly coupled
heterotic string. Notably, within a substantial region of the K\"ahler moduli space, the coupling is
strong yet under control and is large enough to yield the correct value for the observable sector
$SO(10)$ unification mass and gauge coupling.
In Section 2.5, we analyze other more general constructions of hidden sector vector
bundles, using multiple line bundles as the building blocks. We also discuss possible deformations of
the underlying hidden sector bundle within this context. 
Building on this work, in Section 2.6, which follows \cite{Ashmore:2020wwv}, we attach a possible SUSY-breaking mechanism, namely gaugino condensation in the hidden sector. The condensate induces non-zero F-terms in the 4D effective theory which break SUSY globally. This effect is
mediated by gravity and induces moduli-dependent soft SUSY-breaking terms in the observable
sector.

The second part treats the $B-L$ MSSM from a low-energy phenomenology perspective, following its current and future
detection efforts at the LHC. Within this model, the spontaneous breaking of R-parity leads to specific, and completely calculable, R-parity
violating (RPV) decays of the lightest supersymmetric particle (the LSP) into SM particles, thus
opening a new arena of the SUSY phenomenological landscape. Following~\cite{Dumitru:2018jyb}, we study the RPV decay
modes of two types of superparticles: charginos and neutralinos. Then, following \cite{Dumitru:2018nct} and \cite{Dumitru:2019cgf}, we analyze the
RPV decays of the Wino chargino, Wino neutralino, and Bino neutralino subspecies, in scenarios
in which they are the LSPs. We perform a statistical analysis of their decay modes, working with
a computer simulation that outputs sets of superparticle mass spectra that are entirely
consistent with the current experimental bounds. The findings suggest that these R-parity violating LSP decays
could be amenable to direct detection at the ATLAS and CMS detectors at the LHC. Detection of
these processes would not only be an explicit indication of "Beyond the Standard Model" physics,
but would also strongly hint at the existence of $N=1$ SUSY with spontaneously broken R-parity.

\chapter{The Strongly Coupled $E_8\times E_8$ Heterotic Vacuum}

\section{The \texorpdfstring{$B-L$}{B-L} MSSM Heterotic Standard Model}

The $B-L$ MSSM vacuum of heterotic M-theory was first introduced in \cite{Braun:2005bw,Braun:2005zv,Braun:2005nv} and various aspects of the theory were discussed in detail in \cite{Marshall:2014kea,Marshall:2014cwa,Dumitru:2018jyb,Dumitru:2018nct,Dumitru:2019cgf}. This phenomenologically realistic theory is obtained as follows. First, eleven-dimensional Hořava--Witten theory \cite{Horava:1995qa,Horava:1996ma} -- which is valid to order $\kappa_{11}^{2/3}$, where $\kappa_{11}$ is the eleven-dimensional Planck constant -- is compactified on a specific Calabi--Yau threefold $X$ down to a five-dimensional $M_{4} \times S^{1}/\mathbb{{Z}}_{2}$ effective theory, with $N=1, D=5$ supersymmetry in the bulk space and $N=1, D=4$ supersymmetry on the orbifold boundaries \cite{Lukas:1998yy, Lukas:1998tt}. By construction, this five-dimensional theory is also only valid to order $\kappa_{11}^{2/3}$. A BPS double domain wall vacuum solution of this theory was then presented \cite{Lukas:1998tt}. This BPS vacuum of the five-dimensional theory can, in principle, be computed to all orders as an expansion in $\kappa_{11}^{2/3}$ and used to dimensionally reduce to a four-dimensional, $N=1$ supersymmetric theory on $M_4$. However, since the five-dimensional effective theory is only defined to order $\kappa_{11}^{2/3}$, and since solving the BPS vacuum equations to higher-order for the Calabi--Yau threefold associated with the $B-L$ MSSM is very difficult, it is reasonable to truncate the BPS vacuum at order $\kappa_{11}^{2/3}$ as well. Dimensionally reducing with respect to this ``linearized'' solution to the BPS equations then leads to the four-dimensional $N=1$ supersymmetric effective Lagrangian for the $B-L$ MSSM vacuum of heterotic M-theory. By construction, this four-dimensional theory is also only valid to order $\kappa_{11}^{2/3}$ -- except for several quantities, specifically the dilaton, the gauge couplings of both the observable and hidden sectors, and the Fayet--Iliopoulos term associated with any $U(1)$ gauge symmetry of the hidden sector, which are well-defined to order $\kappa_{11}^{4/3}$ \cite{Lukas:1997fg, Lukas:1998tt, Lukas:1998hk, Ovrut:2018qog}. All geometric moduli are obtained by averaging the associated five-dimensional fields over the fifth dimension. 

Having discussed the basic construction elements of the $B-L$ MSSM heterotic vacuum, we will, in the rest of this section, present in detail the specific construction of this four-dimensional effective theory. The content of this section is based on the work done in \cite{Ashmore:2020ocb, Ashmore:2021xdm}.

\subsection{The Calabi--Yau Threefold}\label{sec:CY_sheon}

Let us start our discussion with the six-dimensional compactification manifold, whose underlying geometry determines a significant amount of properties of the low energy theory.
The Calabi--Yau manifold $X$ is chosen to be a torus-fibered threefold
with fundamental group $\pi_1(X)=\Z_3 \times \Z_3$. More specifically, the Calabi--Yau threefold $X$ is the fiber product of two rationally elliptic $\dd\mathbb{P}_{9}$ surfaces, that is, a self-mirror Schoen threefold \cite{Braun:2004xv}, quotiented with respect to a freely acting $\mathbb{Z}_{3} \times \mathbb{Z}_{3}$ isometry. Its Hodge data is $h^{1,1}=h^{1,2}=3$, so there are three K\"ahler and three complex structure
moduli. The complex structure moduli will play no role throughout this work. Relevant here is the degree-two Dolbeault cohomology group
\begin{equation}
  H^{1,1}\big(X,\C\big)=
  \Span_\C \{ \omega_1,\omega_2,\omega_3 \}  \ ,
  \label{1}
\end{equation}
where $\omega_i=\omega_{ia {\bar{b}}}$ are harmonic $(1,1)$-forms on
$X$ with the properties
\begin{equation}
  \omega_3\wedge\omega_3=0 \ ,\quad
  \omega_1\wedge\omega_3=3\,\omega_1\wedge\omega_1 \ ,\quad
  \omega_2\wedge\omega_3=3\,\omega_2\wedge\omega_2 \ .
  \label{2}
\end{equation}
Defining the intersection numbers as
\begin{equation}
  d_{ijk} = \frac{1}{v}
  \int_X \omega_i \wedge \omega_j \wedge \omega_k
   \quad i,j,k=1,2,3\ ,
  \label{3}
\end{equation}
where $v$ is a reference volume of dimension (length)$^6$,
it follows that
\begin{equation}\label{4}
  (d_{ijk}) = 
  \left(
    \begin{array}{ccc}
      (0,\tfrac13,0) & (\tfrac13,\tfrac13,1) & (0,1,0) \\
      (\tfrac13,\tfrac13,1) & (\tfrac13,0,0) & (1,0,0) \\
      (0,1,0) & (1,0,0) & (0,0,0)
    \end{array} 
  \right) \ .
\end{equation}
The $(i,j)$-th entry in the matrix corresponds to the triplet
$(d_{ijk})_{k=1,2,3}$.
The K\"ahler cone is the positive octant
\begin{equation}
  \Kcone = H^2_{+}(X,\R)
  \subset H^2(X,\R)\ .
\label{7}
\end{equation}
The K\"ahler form, defined to be $\omega_{a {\bar{b}}}=ig_{a
  {\bar{b}}}$, where $g_{a {\bar{b}}}$ is the Ricci-flat metric on $X$, can be
any element of $\Kcone$. That is, suppressing the Calabi--Yau indices,
the K\"ahler form can be expanded as
\begin{equation}
  \omega = a^i\omega_i 
  , \quad \text{where } a^i >0 \ .
\label{8}
\end{equation}
The real, positive coefficients $a^i$ are the three $(1,1)$ K\"ahler
moduli of the Calabi--Yau threefold. Here, and throughout this work,
upper and lower $H^{1,1}$ indices are summed unless otherwise
stated. The dimensionless volume modulus is defined by
\begin{equation}
  V=\frac{1}{v} \int_X \sqrt{g}
  \label{9}
\end{equation}
and, hence, the dimensionful Calabi--Yau volume is ${\bf{V}}=vV$. Using the
definition of the K\"ahler form and the intersection numbers \eqref{3}, $V$ can be written as
\begin{equation}
  V=\frac{1}{6v}\int_X
  \omega \wedge \omega \wedge \omega=
  \frac{1}{6} d_{ijk} a^i a^j a^k \ .
  \label{10}
\end{equation}
It is sometimes useful to express the three $(1,1)$ moduli in terms of $V$ and
two additional independent moduli. This can be accomplished by
defining the scaled shape moduli
\begin{equation}
  b^i=V^{-1/3}a^i \ , \qquad i=1,2,3 \ .
  \label{11}
\end{equation}
It follows from \eqref{10} that they satisfy the constraint
\begin{equation}
d_{ijk}b^ib^jb^k=6
\label{12}
\end{equation}
and, hence, represent only two degrees of freedom.

\subsection{The Observable Sector Bundle}

On the observable orbifold plane, the vector bundle $\Vvis$ on $X$
is chosen to be a specific holomorphic bundle with structure group $SU(4)\subset E_8$. The structure of this bundle was discussed in \cite{Braun:2005nv,Braun:2005bw,Gomez:2005ii,Braun:2006ae}. In order to preserve $N=1$ supersymmetry in the low-energy four-dimensional effective theory on $M_{4}$, this bundle must be both slope-stable and have vanishing slope~\cite{Braun:2005zv,Braun:2006ae}. Recall that the slope of any
bundle or sub-bundle $\cal{F}$ is defined as
\begin{equation}
  \mu({\cal{F}})=
  \frac{1}{\rank({\cal{F}})v^{2/3}} 
  \int_X{c_1({\cal{F}})\wedge \omega \wedge \omega} \ ,
  \label{50}
\end{equation}
where $\omega$ is the K\"ahler form in \eqref{8}. Since the first Chern class $c_1$ of any $SU(N)$ bundle must vanish, it follows immediately that  $\mu(\Vvis)=0$, as required.
However, demonstrating that our chosen bundle is slope-stable is non-trivial and was proven in detail in several papers \cite{Braun:2005nv,Braun:2005bw,Gomez:2005ii}. The $SU(4)$ vector bundle will indeed be slope-stable in a restricted, but large, region of the positive K\"ahler cone. As proven in detail in~\cite{Braun:2006ae}, this
will be the case in a subspace of the K\"ahler cone defined by seven
inequalities. In this region, all sub-bundles of $V^{(1)}$ will have a negative
slope. These can be slightly simplified into the statement that the moduli $a^{i}$, $i=1,2,3$, must satisfy at least one of the two inequalities
\begin{equation}\label{51}
  \begin{gathered}
    \left(
      a^1
      < 
      a^2
      \leq 
      \sqrt{\tfrac{5}{2}} a^1
      \quad\text{and}\quad
      a^3
      <
      \frac{
        -(a^1)^2-3 a^1 a^2+ (a^2)^2
      }{
        6 a^1-6 a^2
      } 
    \right)
    \quad\text{or}\\
    \left(
      \sqrt{\tfrac{5}{2}} a^1
      <
      a^2
      <
      2 a^1
      \quad\text{and}\quad
      \frac{
        2(a^2)^2-5 (a^1)^2
      }{
        30 a^1-12 a^2
      }
      <
      a^3
      <
      \frac{
        -(a^1)^2-3 a^1 a^2+ (a^2)^2
      }{
        6 a^1-6 a^2
      }
    \right) \ .
  \end{gathered}
\end{equation}
The subspace $\Kcone^s$ satisfying \eqref{51} is a full-dimensional
subcone of the K\"ahler cone $\Kcone$ defined in \eqref{7}. 
It is a cone because the inequalities are homogeneous. In other words, only the angular part of the K\"ahler moduli (the $b^i$) are constrained, but not the overall volume. 
\begin{figure}[t]
  \centering
  \includegraphics[width=0.9\textwidth]{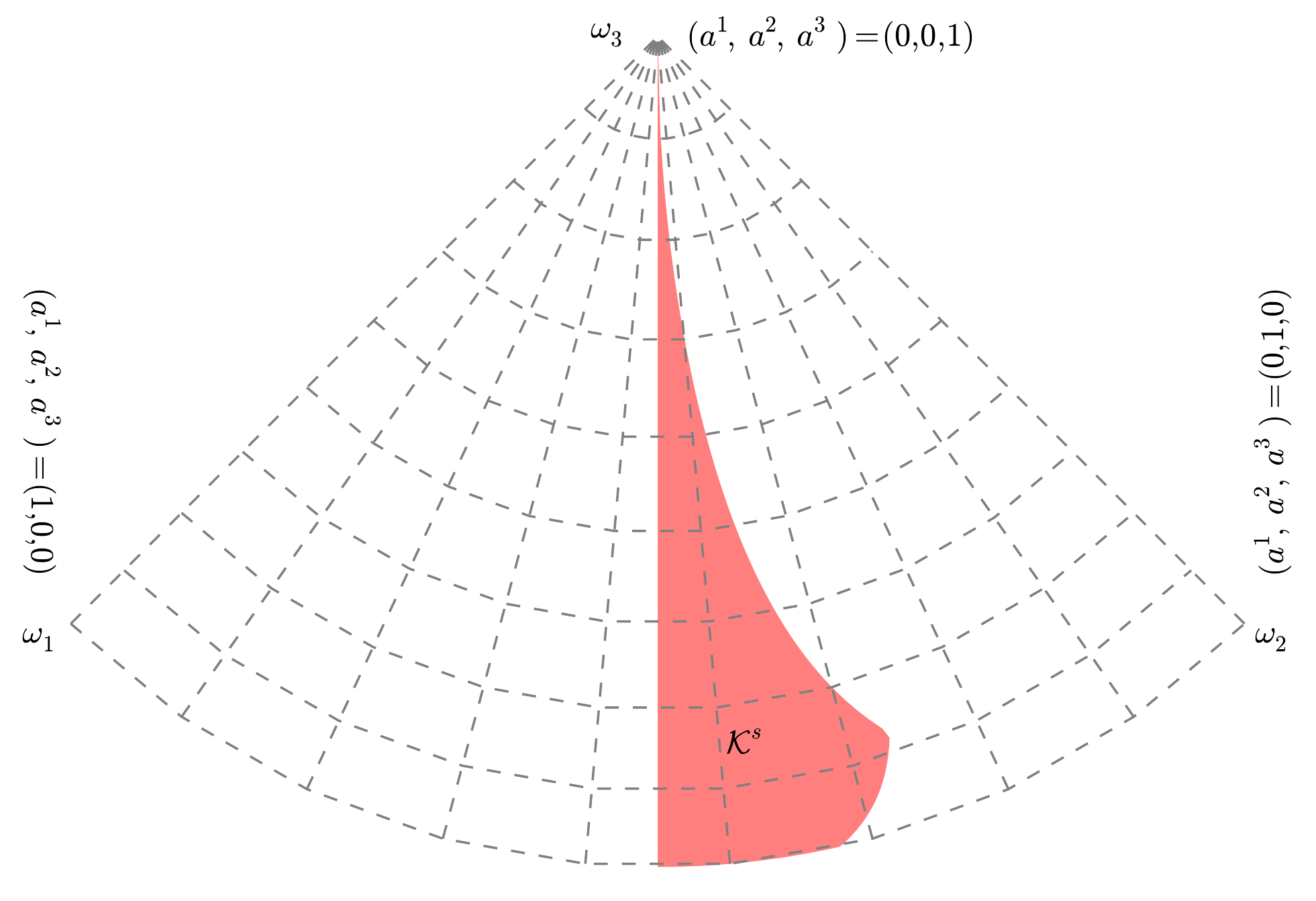}
  \caption{The observable sector stability region in the K\"ahler cone.}
  \label{fig:starmap}
\end{figure}
Hence, it is best displayed as a two-dimensional ``star map'' as seen by an
observer at the origin. This is shown in Figure \ref{fig:starmap}. For
K\"ahler moduli restricted to this subcone, the four-dimensional low-energy theory in the observable sector is $N=1$ supersymmetric.

Having discussed that our specific $SU(4)$ holomorphic vector bundle preserves four-dimensional $N=1$ supersymmetry, let us examine the physical content of the effective theory on $M_{4}$. To begin with, $SU(4) \times  Spin(10)$ is a maximal-rank subgroup of $E_{8}$.
Hence, $SU(4)$ breaks the $E_{8}$ group to
\begin{equation}
  E_8 \to Spin(10) \ .
  \label{13}
\end{equation}
However, to proceed further, one must break this $Spin(10)$ ``grand unified'' group down to the gauge group of the MSSM. This is accomplished by turning on
two \emph{flat} Wilson lines, each associated with a
different $\Z_3$ factor of the $\Z_3 \times \Z_3$ holonomy of $X$. Doing this preserves the $N=1$ supersymmetry of the effective theory, but breaks the 
observable gauge group down to
\begin{equation}
  Spin(10) 
  \to 
  SU(3)_C \times SU(2)_L \times U(1)_Y \times U(1)_{B-L} \ .
  \label{17}
\end{equation}
The mass scale associated with the Wilson lines can be approximately the same, or separated by up to one order of magnitude. Be that as it may, for energies below the lightest Wilson line mass, the particle spectrum of the $B-L$ MSSM is exactly that of the MSSM; that is, three families of quarks and leptons, including three right-handed neutrino chiral supermultiplets -- one per family -- and exactly one pair of Higgs-Higgs conjugate chiral superfields. There are no vector-like pairs of particles and no exotics of any kind. It follows from \eqref{17} however, that the gauge group is that of the MSSM plus an additional gauged $U(1)$ associated with the $B-L$ quantum numbers. The physics of this additional gauge symmetry -- which is broken far above the electroweak scale -- is discussed in detail in a number of papers \cite{Deen:2016vyh,Ambroso:2009jd,Ovrut:2012wg,Ovrut:2014rba,Ovrut:2015uea,Barger:2008wn,FileviezPerez:2009gr,FileviezPerez:2012mj} and is phenomenologically acceptable.

\subsection{The Hidden Sector Bundle}\label{sec:embedding}

In general, the hidden-sector vector bundle can have the generic form of a Whitney sum
\begin{equation}
V^{(2)}={\cal{V}}_{N} \oplus {\cal{L}}\ , \qquad {\cal{L}}=\bigoplus_{r=1}^R L_r \ ,
\label{dude1}
\end{equation}
where ${\cal{V}}_{N}$ is a slope-stable, non-abelian bundle and 
each $L_{r}$, $r=1,\dots,R$, is a holomorphic line bundle with  structure group $U(1)$. In this work we restrict our model to vector bundles consisting of the Whitney sum of line bundles only; that is
\begin{equation}
\qquad {V^{(2)}}={\cal{L}}=\bigoplus_{r=1}^R L_r \ .
\label{dude1aA}
\end{equation}
 Each line bundle $L_r$ is associated with a divisor of $X$ and is conventionally expressed as 
\begin{equation}
  L_r=\Ocal_X(l_r^1, l_r^2, l_r^3)  \ ,
\end{equation}
where the $l_r^i$ are integers satisfying the condition
\begin{equation}
  (l_r^1+l_r^2) \op{mod} 3 = 0 \ .
  \label{22}
\end{equation}
This additional constraint is imposed for these bundles to arise from $\Z_3 \times
\Z_3$ equivariant line bundles on the covering space of $X$.

The structure group is $U(1)^R$, where
each $U(1)$ factor has a specific embedding into the hidden sector
$E_8$ gauge group. It follows from the definition that
$\rank({\cal{L}})=R$ that the first Chern class is
\begin{equation}
  c_1({\cal{L}})
  =\sum_{r=1}^{R}c_1(L_r), \quad c_{1}(L_{r})=
  \frac{1}{v^{1/3}}  (l^1_r \omega_1 + l^2_r \omega_2 + l^3_r \omega_3) 
  .
\label{23}
\end{equation}
Note that since ${\cal{L}}$ is a sum of holomorphic line bundles,
$c_2({\cal{L}})=c_3({\cal{L}})=0$. However, the relevant quantity for the
hidden sector vacuum is the second Chern character defined in
\cite{Ovrut:2018qog}. For ${\cal{L}}$ this becomes
\begin{equation}
  ch_2({\cal{L}})
  = \sum_{r=1}^R ch_2(L_r) \ .
\label{24}
\end{equation}
Since $c_2(L_r)=0$, it follows that
\begin{equation}
  ch_2(L_r)=a_rc_1(L_r) \wedge c_1(L_r) 
  \label{25}
\end{equation}
where
\begin{equation}
  a_r=\tfrac{1}{4} \tr_{E_{8}} Q_r^2
  \label{26}
\end{equation}
with $Q_r$ the generator of the $r$-th $U(1)$ factor embedded into
the $\repd{248}$ adjoint representation of the hidden sector $E_8$, and the trace is taken over the $\repd{248}$ of $\Ex8$ (including a conventional factor of $1/30$).

In this section, following \cite{Ashmore:2020ocb}, we will further simplify the hidden sector vector bundle to consist of a {\it single} holomorphic line bundle $L$ with structure group $U(1) \subset E_{8}$ only. 
We embed a single line bundle into an $SU(2)$ subgroup of $E_8$, such that the low energy gauge group on the hidden sector is $E_7\times U(1)$. We obtain the $E_7\times U(1)$ subgroup after we break the 7th simple root of $\Ex 8$. In this case, $E_8$ first breaks to the maximal $E_7\times SU(2)$ subgroup, and then the $SU(2)$ is broken further to $U(1)$. 

Under $E_8\rightarrow E_7\times SU(2)$, the $\repd {248}$ adjoint representation of $E_8$ decomposes as
\begin{equation}
\label{eq:breakE8SU2}
\repd{248}=(\repd 1,\repd 3)+(\repd{56},\repd{2})+(\repd{133},\repd{1}).
\end{equation}
After breaking the $SU(2)$ group further to $S(U(1)\times U(1))\sim U(1)$, the above $\repd{248}$ representation decomposes under $E_7\times U(1)$ as
\begin{equation}
\label{eq:breakE8U1}
\repd{248}=\repd 1_{2}+\repd 1_{0}+\repd 1_{-2}+\repd{56}_{1}+\repd{56}_{-1}+\repd{133}_{0}\ .
\end{equation}

To find the $E_8$ connection which corresponds to this breaking pattern, we first build the $SU(2)$ connection at the decomposable locus. At the decomposable locus, the structure group of the $SU(2)$ bundle $V_{\repd 2}$ degenerates to $SU(2)\rightarrow S(U(1)\times U(1))\sim U(1)$. At this locus, we embed the $U(1)$ line bundle into the $SU(2)$ bundle $V_{\repd 2}$ by defining a group homomorphism
\begin{equation}
 U(1)\hookrightarrow S(U(1)\times U(1))\subset SU(2)\ .
\end{equation}
This map is in this case given by
\begin{equation}
 e^{i\phi}\hookrightarrow \left( \begin{matrix}e^{i\phi}&0\\0&e^{-i\phi}\end{matrix} \right)\ ,
\label{red5}
\end{equation}
where $\phi$ is the $U(1)$ phase. 
This embedding indicates to us that the $SU(2)$ bundle can be built from a $U(1)$ line bundle $L$ and its inverse alone, such that
\begin{equation}
V_{\repd 2}=L\oplus L^{-1}\ ,
\label{dude1cA}
\end{equation}
Equivalently, we can write the $SU(2)$ connection in terms of the $A_{U(1)}$ connection:
\begin{equation}
\label{eq:SU2U1}
A_{U(1)}\hookrightarrow A_{SU(2)} =\left( \begin{matrix}A_{U(1)}&0\\0&-A_{U(1)}\end{matrix} \right)\ .
\end{equation}
The $SU(2)$ connection can be embedded further into the $E_8$ connection. This $E_8$ connection commutes with the generators of $E_7$ on the hidden sector, as well as with itself and hence, it breaks the $E_8$ group to the $E_7\times U(1)$ low energy group. We derive the line bundle embedding into the $E_8$ from the decomposition \eqref{eq:breakE8U1}, by reading the $U(1)$ charges:
\begin{equation}
\label{eq:E8SU2U1}
{A}_{E_8} = A_{U(1)}\left( \begin{matrix} \left(\begin{matrix}2&&\\&0&\\&&-2\end{matrix}\right) & &\\&\left( \begin{matrix}1&0\\0&-1\end{matrix} \right)\otimes \id_{56}& \\ & & 0\times \id_{133}\end{matrix} \right)=A_{U(1)}Q\ .
\end{equation}
where $Q$ is the particular $U(1)$ generator which breaks $E_8$ to the low energy group. 

For the explicit choice of embedding, we find the numerical constant
\begin{equation}
  a=\tfrac{1}{4} \tr _{E_{8}}Q^2 =1\ ,
  \label{26A}
\end{equation}
where the trace $\tr$ includes a factor of $1/30$. This coefficient will enter several of the consistency conditions, such as the anomaly cancellation equation, required for an acceptable vacuum solution.

We note in passing that the four-dimensional effective theory associated with choosing this explicit embedding has gauge symmetry
\begin{equation}
H=E_{7} \times U(1) \ ,
\label{red7}
\end{equation}
where the second factor is an ``anomalous''  $U(1)$. It is identical to the structure group of $L$ and arises in the low-energy theory since $U(1)$ commutes with itself. This will be discussed in detail in Section \ref{sec:hidden_sector}.

An important comment can be now made about this bundle's stability. To begin with, we note that for the explicit bundle construction shown above, $ L$ and its inverse $L^{-1}$ are both sub-line bundles of the hidden sector $E_{8}$ gauge bundle. Since the connection of the hidden $E_8$ bundle vanishes, it can be poly-stable only iff
\begin{equation}
\mu(L)=\mu(L^{-1})=0\ .
\end{equation}
 The DUY theorem \cite{UY} states that the connection associated with this bundle, solves the HYM equation, and the solution is unique if $V$ is poly-stable. 

Generically, however, this will not be the case. It follows from \eqref{50} and \eqref{23} that the slope of $L$ is proportional to its first Chern class $c_{1}(L)=\frac{1}{v^{1/3}}(l_1\omega_{1}+l_2\omega_{2}+l_3\omega_{3})$ and, hence, its slope does not vanish anywhere in K\"ahler moduli space. We will discuss the bundle stability condition on the hidden sector in Section \ref{sec:hidden_sector}, where we also provide an explicit solution.

\subsection{Bulk Space Five-Branes}

In addition to the holomorphic vector bundles on the observable and
hidden orbifold planes, the bulk space between these planes can
contain five-branes wrapped on two-cycles ${\cal{C}}_2^{(n)}$,
$n=1,\dots,N$ in $X$. Cohomologically, each such five-brane is
described by the $(2,2)$-form Poincar\'e dual to ${\cal C}_2^{(n)}$,
which we denote by $W^{(n)}$. Note that to preserve $N=1$
supersymmetry in the four-dimensional theory, these curves must be
holomorphic and, hence, each $W^{(n)}$ is an effective class. In the following, we present the formalism associated with an arbitrary number $N$ of such five-branes. Later, in Section \ref{sec:constraints}, we will study a specific configuration, with a {\it single} five-brane. We denote its location in the bulk space by $z_{1}$, where 
$z_{1} \in [0,1]$. When convenient, we will re-express this five-brane location in terms of the parameter $\lambda=z_{1}-\frac{1}{2}$, where $\lambda \in [-\frac{1}{2},\frac{1}{2}]$.

\subsection{Anomaly Cancellation}\label{sec:anomaly}

As discussed in~\cite{Lukas:1998tt,Donagi:1998xe}, anomaly cancellation in heterotic M-theory requires that
\begin{equation}
  \sum_{n=0}^{N+1}J^{(n)}=0 ,
  \label{27}
\end{equation}
where 
\begin{equation}
  \label{28}
  \begin{split}
    J^{(0)}=&\;
    -\frac{1}{16 \pi^2}
    \Big( \tr_{E_8} F^{(1)} \wedge F^{(1)}
    -\frac{1}{2}\tr_{SO(6)} R \wedge R \Big) \\[1ex]
    J^{(n)}=&\;
    W^{(n)}, \quad n=1,\dots,N, \\[1ex]
    J^{(N+1)}=&\;
    -\frac{1}{16 \pi^2}
    \Big( \tr_{E_8} F^{(2)} \wedge F^{(2)}
    -\frac{1}{2}\tr_{SO(6)} R \wedge R \Big) \\
  \end{split}
\end{equation}
Note that the indices $n=0$ and $n=N+1$ denote the observable and hidden sector domain walls respectively, and {\it not} the location of a five-brane.
Using the definitions of the associated Chern characters, the anomaly cancellation condition can be
expressed as
\begin{equation}
  c_2(TX)-c_2(\Vvis)
  +\sum_{r=1}^R  a_r c_1(L_r) \wedge c_1(L_r) - W 
  = 0 ,
  \label{293434}
\end{equation}
where we have restricted the hidden sector bundle to be of the form \eqref{dude1aA} and $W=\sum_{n=1}^N W^{(n)}$ is the total five-brane class. For our particular compactification manifold and for our observable sector tangent bundle and gauge bundle of the $B-L$ MSSM given in \cite{Ovrut:2018qog}, we have that
\begin{equation}
  \frac{1}{v^{1/3}}\int_X \left(c_2(TX)-c_2(\mathcal{V}^{(1)})\right)\wedge \omega_i=\left( \tfrac{4}{3},\tfrac{7}{3},-4\right)_i\ .
\end{equation}
Furthermore, it follows from the properties of the Chern characters, and defining
\begin{equation}
  W_i = \frac{1}{v^{1/3}} \int_X W \wedge \omega_i \ ,
  \label{32}
\end{equation}
that the anomaly condition \eqref{293434} can be expressed as 
\begin{equation}
  W_i= \big( \tfrac{4}{3},\tfrac{7}{3},-4\big)\big|_i
  +\sum_{r=1}^R a_r d_{ijk} l^j_r l^k_r \geq 0 \ , \quad i=1,2,3  .
\label{33}
\end{equation}
The positivity constraint on $W$ follows from the requirement that it
be an effective class to preserve $N=1$ supersymmetry.

Finally, it is useful to define the charges
\begin{equation}
  \beta^{(n)}_i = 
  \frac{1}{v^{1/3}}
  \int_X J^{(n)} \wedge \omega_i \ , \quad i=1,2,3  .
\label{34}
\end{equation}
For example, when $n=0$, it follows from \eqref{28}, using results for the second Chern class of the observable sector gauge bundle given in \cite{Ovrut:2018qog} and the intersection numbers \eqref{3} and
\eqref{4}, we find that
\begin{equation}
  \beta^{(0)}_i = 
  \big( \tfrac{2}{3},-\tfrac{1}{3},4 \big)\big|_i \ .
  \label{35}
\end{equation}

The five-brane charges $\beta^{(n)}_i$, $n=1,\dots N$, satisfy the relation
\begin{equation}
W_i=\sum_{n=1}^N \beta_i^{(n)}\ .
\end{equation}

 Restricting this to a single hidden-sector line bundle $L$ and a single bulk-space five-brane, the anomaly cancellation equation can be simplified and then rewritten in the form
\begin{equation}
  W_i= \bigl( \tfrac{4}{3},\tfrac{7}{3},-4\big)\big|_i
  + a \, d_{ijk} l^j l^k \geq 0 \  \qquad i=1,2,3  \ ,
\label{33A}
\end{equation}
where the coefficient $a$ is defined in \eqref{26A}. The positivity constraint on $W$ follows from the requirement that the five-brane wraps an effective class to preserve $N=1$ supersymmetry.

\subsection{Going from 11D to 5D - Reduction Constraint}

In strongly coupled heterotic M-theory, there is a one-dimensional interval $S^{1}/{\mathbb{Z}}_{2}$ separating the observable and hidden orbifold planes. Denoting by $\rho$ an arbitrary reference radius of $S^{1}$, the reference length of this one-dimensional interval is given by $\pi \rho$. A real coordinate on this interval is written as $x^{11} \in [0, \pi \rho]$. As discussed the introduction as well as later in Section \ref{sec:5d_4d}, arbitrary dimensionless functions on $M_{4} \times S^{1}/{\mathbb{Z}}_{2}$ can be averaged over this interval, leading to moduli that are purely functions on $M_{4}$. Averaging the $b$ function in the five-dimensional metric,
$\dd s_{5}^{2} = \dots +b^{2}(\dd x^{11})^{2}$, defines a four-dimensional modulus
\begin{equation}
\frac{\Rhat}{2}=\langle   b \rangle _{11} \ .
\label{case1}
\end{equation}
The physical length of this orbifold interval is then given by $\pi \rho \Rhat$. It is convenient to define  a new coordinate $z$ by $z=\frac{x^{11}}{\pi \rho}$, which runs over the interval $z \in [0,1]$.

We now present the necessary condition for a consistent dimensional reduction on a Calabi--Yau threefold $X$ from the $d=11$ Hořava--Witten orbifold to five-dimensional heterotic M-theory. For this reduction to be viable, the averaged Calabi--Yau radius must, when calculated using the eleven-dimensional M-theory metric, be sufficiently smaller than the physical length of the $S^{1}/ \mathbb{Z}_{2}$ interval. That is, one must have
\begin{equation}
\frac{\pi \rho {\Rhat} V^{-1/3}}{(vV)^{1/6}} > 1 \ ,
\label{sun1}
\end{equation}
where the constant parameters $v$ and $\rho$ were introduced above and the moduli $V$ and $\Rhat$ are defined in \eqref{10} and \eqref{case1} respectively. The extra factor of $V^{-1/3}$ in the numerator arises because the $S^{1}/\mathbb{Z}_{2}$ interval length must be computed with respect to the eleven-dimensional metric. To see this, recall from \cite{Lukas:1998tt} that the eleven-dimensional metric ansatz for the reduction to five dimensions is given by
\begin{equation}
\dd s_{11}^{2}=V^{-2/3}g_{\alpha\beta} \dd x^{\alpha} \dd x^{\beta}+g_{AB} \dd x^{A}\dd x^{B}\ ,\label{eq:11d_metric}
\end{equation}
where $g_{\alpha\beta}$ is the five-dimensional metric and $g_{AB}$ is the metric of the Calabi–Yau threefold. Note that the factor of $V^{-2/3}$ is chosen so that $g_{\alpha\beta}$ is in the five-dimensional Einstein frame. To further reduce to four dimensions, one takes
\begin{equation}
g_{\alpha\beta} \dd x^{\alpha}\dd x^{\beta}=\Rhat^{-1}g_{\mu\nu} \dd x^{\mu} \dd x^{\nu}+\Rhat^{2}(\dd  x^{11})^{2} \ ,\label{eq:5d_metric}
\end{equation}
where $x^{11}$ runs from $0$ to $\pi\rho$ and $g_{\mu\nu}$ is the four-dimensional Einstein frame metric. As measured by the five-dimensional metric, the $S^{1}/ \mathbb{Z}_{2}$ orbifold interval has length $\pi\rho\Rhat$. However, if one wants to compare the scale of the orbifold interval with that of the  Calabi–Yau threefold, one must use the eleven-dimensional metric. Substituting (\ref{eq:5d_metric}) into (\ref{eq:11d_metric}) and averaging the value of $V$ over the orbifold interval, we find
\begin{equation}
\dd s_{11}^{2}=V^{-2/3}\Rhat^{-1}g_{\mu\nu} \dd x^{\mu} \dd x^{\nu}+V^{-2/3}\Rhat^{2}( \dd x^{11})^{2}+g_{AB} \dd x^{A} \dd x^{B}\ .\label{eq:11d_metric-1}
\end{equation}
From this we see that, in eleven dimensions, the orbifold interval has length $\pi\rho\Rhat V^{-1/3}$, as used in \eqref{sun1}. It is helpful to note that \eqref{sun1} can be written as
\begin{equation}
 \frac{\Rhat}{\epsilon_{R}V^{1/2}} > 1 \ ,\quad \text{where } \epsilon_{R}=\frac{v^{1/6}}{\pi \rho}  \ .
\label{soc1}
\end{equation}

\subsection{Going from 5D to 4D: The Linearized Double Domain Wall}\label{sec:5d_4d}

The five-dimensional effective theory of heterotic M-theory, obtained
by reducing Hořava--Witten theory on the above Calabi--Yau
threefold, admits a BPS double domain wall solution with five-branes
in the bulk space \cite{Lukas:1998yy,Donagi:1999gc,Lukas:1998tt,Lukas:1998hk,Lukas:1999kt,Lukas:1997fg}. This solution depends on 
the moduli $V$ and $b^i$ defined in the text, as well as the $a$, $b$ functions of the
five-dimensional metric
\begin{equation}
 \dd s_5^2=a^2\dd x^{\mu}\dd x^{\nu}\eta_{\mu\nu}+b^2(\dd x^{11})^{2} \ ,
  \label{37}
\end{equation}
all of which are dependent on the five coordinates $x^{\alpha}$,
$\alpha=0,\dots,3,11$ of $M_4 \times S^1/\Z_2$.
The detailed structure of the linearized double domain wall depends on the solution of three non-linear equations discussed in \cite{Lukas:1998tt}. These can be approximately solved by expanding to linear order in the quantity $\epsilon_S'\beta^{(0)}_i\big(z-\frac{1}{2}\big)$, where we define  $z={x^{11}}/{\pi\rho}$ with $z \in [0,1]$, $\beta^{(0)}_i$ is given in \eqref{35} and 
\begin{equation}
  \epsilon'_S = \pi \epsilon_{S}\ , \qquad
  \epsilon_{S}= \left(\frac{\kappa_{11}}{4\pi} \right)^{2/3}\frac{2\pi\rho}{v^{2/3}} \ .
  \label{40}
\end{equation}
It is also convenient to express the moduli of the theory in terms of orbifold average functions defined as follows. For an arbitrary dimensionless function $f$ of the five $M_4
\times S^1/\Z_2$ coordinates, define its average over the $S^1/\Z_2$
orbifold interval as
\begin{equation}
\langle f \rangle_{11}=\frac{1}{\pi \rho}\int_0^{\pi\rho}{\dd x^{11}f} \ ,
\label{42}
\end{equation}
where $\rho$ is the reference length. Then $\langle f \rangle_{11}$ is
a function of the four coordinates $x^{\mu}$, $\mu=0,\dots,3$ of $M_4$
only. The linearized solution is expressed in terms of orbifold
average functions
\begin{equation}
  V_0=\langle V \rangle_{11}\  , \quad 
  b^i_0=\langle b^i \rangle_{11} \ , \quad 
  \left(\frac{\Rhat_0}{2}\right)^{-\frac{1}{2}}=\langle a \rangle_{11}\  , \quad
  \frac{\Rhat_0}{2}=\langle b \rangle_{11} \ .
\label{42a}
\end{equation}
The fact that they are averaged is indicated by the subscript $0$.

The solution to these linearized equations depends on the number of five-branes located within the fifth-dimensional interval. Here, for simplicity, we will consider the vacuum to contain a single five-brane, wrapped on a holomorphic curve, and located at the fifth-dimensional coordinate $z_{1} \in [0,1]$. It was then shown in \cite{Lukas:1998tt} that the 
conditions for the validity of the linear approximation then break
into two parts. Written in terms of the averaged moduli, these are
\begin{equation}
  2\epsilon_S'\frac{\Rhat_0}{V_0}
  \left|
    \beta_i^{(0)} \big(z-\tfrac{1}{2}\big)
    -\frac{1}{2}\beta_i^{(1)}(1-z_1)^2
  \right|
  \ll 
  \left| d_{ijk} b_0^j b_0^k \right|
  \ , \quad z \in [0,z_1]\ ,
\label{45B}
\end{equation}
and 
\begin{equation}
  2\epsilon_S'\frac{\Rhat_0}{V_0}
  \left|
    (\beta_i^{(0)}+\beta_i^{(1)})
    \big(z-\tfrac{1}{2}\big)
    -\frac{1}{2}\beta_i^{(1)}z_1^2
  \right| 
  \ll 
  \left| d_{ijk} b_0^j b_0^k \right|
 \  , \quad z \in [z_1,1] \ .
\label{45C}
\end{equation}

When dimensionally reduced on this linearized BPS solution, the
four-dimensional functions $a_0^i$, $V_0$, $b_0^i$ and $\Rhat_0$
will become moduli of the $D=4$ effective heterotic M-theory. The
geometric role of $a_0^i$ and $V_0, b_0^i$ will remain the same as
above -- now, however, for the averaged Calabi--Yau threefold. For
example, the dimensionful volume of the averaged Calabi--Yau manifold
will be given by $vV_0$. The new dimensionless quantity $\Rhat_0$
will be the length modulus of the orbifold. The dimensionful length of
$S^1/\Z_2$ is given by $\pi \rho \Rhat_0$. Finally, since the
remainder of this section will be within the context of the $D=4$
effective theory, we will, for simplicity, {\it drop the subscript $``0"$
on all moduli} henceforth -- as well as everywhere in the text of this thesis. 

\subsection{The \texorpdfstring{$D=4$}{D=4} \texorpdfstring{$E_8 \times E_8$}{E8 x E8} Effective Theory}

When $d=5$ heterotic M-theory is dimensionally reduced to four
dimensions on the linearized BPS double domain wall with five-branes,
the result is an $N=1$ supersymmetric effective four-dimensional
theory with (potentially spontaneously broken) $E_8 \times E_8$ gauge group. The
Lagrangian will break into two distinct parts. The first contains
terms of order $\kappa_{11}^{2/3}$ in the eleven-dimensional Planck
constant $\kappa_{11}$, while the second consists of terms of order
$\kappa_{11}^{4/3}$.

\subsubsection{The \texorpdfstring{$\kappa_{11}^{2/3}$}{Order 2/3} Lagrangian}

This Lagrangian is well-known and was presented in~\cite{Lukas:1997fg}. Here we
revise only the relevant properties. In four
dimensions, the moduli must be organized into the lowest components of
chiral supermultiplets. Here, we need only consider the real part of
these components. Additionally, one specifies that these chiral
multiplets have \emph{canonical} K\"ahler potentials in the effective
Lagrangian. The dilaton is simply given by
\begin{equation}
  S=V+i\sigma \ ,
  \label{46}
\end{equation}
such that
\begin{equation}
  \re S=V \ .
  \label{46}
\end{equation}
Neither $a^i$ nor $b^i$ -- defined in \eqref{8} and \eqref{11} respectively -- have canonical kinetic energy. To
obtain this, one must define the re-scaled moduli
\begin{equation}
  t^i = 
  \Rhat b^i = 
  \Rhat V^{-1/3} a^i \ ,
  \label{47}
\end{equation}
where we have used \eqref{11} in the text, and choose the complex K\"ahler moduli
$T^i$ so that
\begin{equation}
\label{48}
T^i=t^i+2i\chi^i
\end{equation}
Denote the real modulus specifying the location of the $n$-th
five-brane in the bulk space by $z_n={x^{11}_n}/{\pi \rho}$ where
$n=1,\dots,N$. As with the K\"ahler moduli, it is necessary to define
the fields
\begin{equation}
   Z_n = \beta_i^{(n)} t^i z_n +2i\beta_i^{(n)}(-\eta^i\nu+\chi^iz) \ , \quad n=1,\cdots N\ .
\label{49}
\end{equation}
These rescaled $Z_n$ five-brane moduli have canonical kinetic
energy.

Note that the complex components of $S$, $T^{i}$ and $Z$ contain axionic scalars $\sigma$, $\chi^{i}$ and $\eta^i$, for $i=1,\dots,h^{1,1}$, respectively.

Furthermore, to this order, the gauge kinetic functions $f_1$ and $f_2$ on the hidden and the observable sector, respectively,
are equal,
\begin{equation}
f_1=f_2=S\ .
\end{equation} 

\subsubsection{The \texorpdfstring{$\kappa_{11}^{4/3}$}{Order 4/3} Lagrangian}

The terms in the BPS double domain wall solution proportional to
$\epsilon_S'$ lead to order $\kappa_{11}^{4/3}$ additions to the $D=4$
Lagrangian. These have several effects. The simplest is that the
five-brane location moduli now contributes to the definition of the
dilaton. Specifically, following~\cite{Brandle:2003uya,Anderson:2009nt} (see also~\cite{Weigand:2006yj} for the weakly coupled string result and \cite{Moore:2000fs} for a detailed derivation), the expressions for the complex scalar moduli fields, in the presence of five-branes, are
\begin{equation}
\begin{split}
\label{eq:def_scalar_intro}
& S=V+\frac{\epsilon_S^\prime}{2}\sum_{n=1}^N\beta^{(n)}_it^iz_n^2
+i\left[\sigma+ \epsilon_S^\prime \sum_{n=1}^N\beta^{(n)}_i\chi^iz_n^2\right] \ ,\\
&T^i=t^i+2i\chi^i\ ,\quad i=1,\dots,h^{1,1}\ ,\\
&Z_n=\beta^{(n)}_it^iz_n+2i\beta^{(n)}_i(-\eta^i\nu_n+\chi^iz_n)\ ,\quad n=1,\dots, N\ .
\end{split}
\end{equation}
These fields are the complex scalar components of the chiral superfields $\tilde S=(S,\psi_S,F_S)$, $\tilde T^i=(T^i,\psi_T^i, F_T^i)$, $\tilde Z_n=(Z_n,\psi_{Z_n},F_{Z_n})$.
The K\"ahler potentials associated with the moduli fields was presented in~\cite{Brandle:2003uya,Anderson:2009nt}. In the restricted case, in which there is a single five-brane between the observable and the hidden sector at $z=\frac{1}{2}+\lambda$, such that $W_i=\beta^{(1)}_i$, these are
\begin{equation}
\begin{split}
\label{eq:K_SandK_T}
&K_S=-\kappa_4^{-2}\ln\left(S+\bar S-\frac{\epsilon_S^\prime}{2} \frac{(Z+\bar Z)^2}{W_i(T^i+\bar T^i)}\right)\ , \\
&K_T=-\kappa_4^{-2}\ln\left(\frac{1}{48}d_{ijk}(T^i+\bar T^i)(T^j+\bar T^j)(T^k+\bar T^k)\right)\ .\\
\end{split}
\end{equation}

More profoundly, these $\kappa_{11}^{4/3}$ terms lead, first,  to threshold
corrections to the gauge kinetic functions, as computed
to order $\kappa_{11}^{4/3}$ in~\cite{Lukas:1998hk, Brandle:2003uya}. 
\begin{equation}
\begin{split}
f_1&=S-\frac{\epsilon_S^\prime}{2}\left(\beta_i^{(N+1)}T^i+2\sum_{n=1}^N Z^{(n)}\right)\ ,\\
f_2&=S+\frac{\epsilon_S^\prime}{2}\beta_i^{(N+1)}T^i\ .
\end{split}
\end{equation}

\subsection{Gauge Threshold Corrections}

 Written in terms of the fields $b^{i}$ defined in \eqref{11} and including five-branes in the bulk
space, the gauge threshold corrections on the observable and hidden sector are given by
\begin{equation}
  \frac{4\pi}{(g^{(1)})^2} \propto
  \text{Re}(f_1)=V \left(1+\epsilon_S' \frac{\Rhat}{2V} 
   \sum_{n=0}^{N}(1-z_n)^2 b^i \beta^{(n)}_i \right)
\label{62}
\end{equation}
and 
\begin{equation}
  \frac{4\pi}{(g^{(2)})^2} \propto
  \text{Re}(f_2)=V \left(1+\epsilon_S' \frac{\Rhat}{2V} 
  \sum_{n=1}^{N+1}z_n^2 b^i\beta^{(n)}_i\right)
  \label{63}
\end{equation}
respectively. The positive-definite constant of proportionality is identical for both gauge couplings and is not relevant to the present discussion. It is important to note that the effective parameter of the $\kappa_{11}^{2/3}$ expansion in \eqref{62} and \eqref{63}, namely $\epsilon_S'{\Rhat}/{V}$, is identical to the parameter appearing in \eqref{45B} and \eqref{45C}) for the validity of the linearized approximation with a single five-brane. That is, the effective strong coupling parameter of the $\kappa_{11}^{2/3}$ expansion is given by
\begin{equation}
\epsilon_{S}^{\rm eff}= \epsilon_S' \frac{\Rhat}{V} \ .
\label{pink1}
\end{equation}
We point out that this is, up to a constant factor of order one, precisely the strong coupling parameter presented in equation (1.3) of \cite{Banks:1996ss}.

Recall that $n=0$ and $n=N+1$ correspond to the observable and hidden sector domain walls -- not to five-branes. Therefore, $z_{0}=0$ and $z_{N+1}=1$.
Using \eqref{28} and  \eqref{34}, one can evaluate the $\beta^{(n)}_i$ coefficients in terms of the the $a^{i}, i=1,2,3$ K\"ahler moduli defined in \eqref{8}. Rewrite the above expressions in terms of these moduli using \eqref{10}, \eqref{11},
\eqref{24}, \eqref{25}, and redefine the five-brane moduli to be
\begin{equation}
  \lambda_n = 
  z_n-\tfrac{1}{2}
  \ , \qquad 
  \lambda_n \in \left[-\tfrac{1}{2},\tfrac{1}{2}\right]  \ .
\label{pink2}
\end{equation}
Furthermore, choosing our hidden sector bundle to be that given in \eqref{dude1aA}, we find that
\begin{equation}
  \label{64}
  \begin{split}
    \frac{4\pi}{(g^{(1)})^2} \propto \;&
    \frac{1}{6v}\int_X\omega \wedge \omega
    \wedge \omega 
    -\epsilon_S' \frac{\Rhat}{2V^{1/3}}
    \frac{1}{v^{1/3}}
    \\ &
    \times
    \int_X\omega \wedge 
    \left(
      -c_2(\Vvis) 
      +\tfrac{1}{2}c_2(TX)-\sum_{n=1}^{N}(\tfrac{1}{2}-\lambda_n)^2W^{(n)}  
    \right)
  \end{split}
\end{equation}
and 
\begin{equation}
  \label{65}
  \begin{split}
  \frac{4\pi}{(g^{(2)})^2} \propto \;&
  \frac{1}{6v}\int_X\omega \wedge \omega
  \wedge \omega 
  -\epsilon_S' \frac{\Rhat}{2V^{1/3}}
  \frac{1}{v^{1/3}}
  \\ &
  \times
  \int_X\omega \wedge 
  \left(
    \sum_{r=1}^{R} a_{r} c_{1}(L_{r}) \wedge c_{1}(L_{r})
    +\tfrac{1}{2}c_2(TX)-\sum_{n=1}^{N}(\tfrac{1}{2}+\lambda_n)^2W^{(n)}  
    \right)
  \end{split}
\end{equation}
where $a_r$ is given in \eqref{26}. The first term on the right-hand
side, that is, the volume $V$ defined in \eqref{10}, is the order
$\kappa_{11}^{2/3}$ result. The remaining terms are the $\kappa_{11}^{4/3}$
M-theory corrections first presented in~\cite{Lukas:1998hk}.

Consistency of the $D=4$ effective theory requires both
$(g^{(1)})^2$ and $(g^{(2)})^2$ to be positive. It follows that the
moduli of the four-dimensional theory are constrained to satisfy
\begin{multline}
  \frac{1}{v}\int_X\omega \wedge \omega \wedge \omega -3\epsilon_S'
  \frac{\Rhat}{V^{1/3}} \frac{1}{v^{1/3}}\int_X\omega \wedge
  \big(-c_2(\Vvis) 
  \\
  +\tfrac{1}{2}c_2(TX)-\sum_{n=1}^{N}(\tfrac{1}{2}-\lambda_n)^2W^{(n)}  \big) > 0
  \label{66}
\end{multline}
and 
\begin{multline}
  \frac{1}{v}\int_X\omega \wedge \omega \wedge \omega -3\epsilon_S'
  \frac{\Rhat}{V^{1/3}} \frac{1}{v^{1/3}}\int_X\omega \wedge
  \big(
    \sum_{r=1}^{R} a_{r} c_{1}(L_{r}) \wedge c_{1}(L_{r})
  \\
  +\tfrac{1}{2}c_2(TX)-\sum_{n=1}^{N}(\tfrac{1}{2}+\lambda_n)^2W^{(n)}  \big) > 0 .
  \label{67}
\end{multline}
One can use \eqref{3}, \eqref{4},
\eqref{8}, \eqref{23}  and \eqref{32} to rewrite
these expressions as
\begin{equation}
  \label{68}
  \begin{split}
    d_{ijk} a^i a^j a^k- 3 \epsilon_S' \frac{\Rhat}{V^{1/3}} \Big(
    -(\tfrac83 a^1 + \tfrac53 a^2 + 4 a^3)
    + \qquad& \\
    + 2(a^1+a^2) -\sum_{n=1}^{N}(\tfrac{1}{2}-\lambda_n)^2 a^i \;W^{(n)}_i
    \Big) &> 0 
  \end{split}
\end{equation}
and
\begin{equation}
  \label{69}
  \begin{split}
    d_{ijk} a^i a^j a^k- 3 \epsilon_S' \frac{\Rhat}{V^{1/3}}
    \Big(d_{ijk}a^i \sum_{r=1}^{R}a_r l^j_r l^k_r
    + \qquad& \\
    + 2(a^1+a^2) -\sum_{n=1}^{N}(\tfrac{1}{2}+\lambda_n)^2 a^i
    \;W^{(n)}_i \Big) &> 0
  \end{split}
\end{equation}
respectively.

We now restrict those results to the case of a hidden-sector bundle constructed from of a single line bundle $L$ and a single five-brane located at $\lambda= z_{1}-\frac{1}{2} \in \left[-\tfrac{1}{2},\tfrac{1}{2}\right]$. The charges $\beta_i^{(0)}$ and $\beta_i^{(1)}$, and the constant coefficient $\epsilon'_S$ are 
\begin{equation}
\beta_i^{(0)}=\left(\tfrac{2}{3},-\tfrac{1}{3},4\right)_{i} \ , \qquad \beta_i^{(1)}=W_{i}  \ ,
\label{ruler1}
\end{equation}
and 
\begin{equation}
  \epsilon'_S = \pi \epsilon_{S} \ , \qquad
  \epsilon_{S}= \left(\frac{\kappa_{11}}{4\pi} \right)^{2/3}\frac{2\pi\rho}{v^{2/3}} \ .
  \label{40AA}
\end{equation}
The parameters $v$ and $\rho$ are defined above and $\kappa_{11}$ is the eleven-dimensional  Planck constant. Written in terms of the K\"ahler moduli $a^{i}$ using \eqref{11}, 
the constraints that $(g^{(1)})^{2}$ and $(g^{(2)})^{2}$ be positive definite are then given by
\begin{equation}
  \label{68AA}
  \begin{split}
    d_{ijk} a^i a^j a^k- 3 \epsilon_S' \frac{\Rhat}{V^{1/3}} \bigl(
    -(\tfrac{8}{3} a^1 + \tfrac{5}{3} a^2 + 4 a^3)
    \qquad &\\
    + 2(a^1+a^2) -(\tfrac{1}{2}-\lambda)^2 a^i \,{W}_i
    \bigr) &> 0 \ ,
  \end{split}
\end{equation}
and
\begin{equation}
  \label{69AA}
  \begin{split}
    d_{ijk} a^i a^j a^k- 3 \epsilon_S' \frac{\Rhat}{V^{1/3}}
    \bigl(a\,d_{ijk}a^i l^j l^k
    \qquad &\\
    + 2(a^1+a^2) -(\tfrac{1}{2}+\lambda)^2 a^i
    \,{W}_i \bigr) &> 0 \ ,
  \end{split}
\end{equation}
respectively. The Calabi--Yau volume modulus $V$ is defined in terms of the $a^{i}$ moduli in \eqref{10}, and $\Rhat$ is the independent $S^{1}/{\mathbb{Z}}_{2}$ length modulus defined in \eqref{case1}. Note that the coefficient $a$ defined in \eqref{26A} enters both expressions via the five-brane class $ {W}_i$ and independently occurs in the second term of \eqref{69AA}.

\section{Matching with Observations - Phenomenological Constraints}\label{sec:observ}

In the previous section, we explored the mathematical constraints required for the theory to be anomaly free with a hidden sector containing a single line bundle $L$ and a single bulk space five-brane. Furthermore, the K\"ahler moduli space constraints were presented so that a specific $SU(4)$ holomorphic vector bundle in the observable sector would be slope-stable and preserve $N=1$ supersymmetry. Important phenomenological properties of the resultant effective theory were presented; specifically, that the low-energy gauge group, after turning on both $\mathbb{Z}_{3}$ Wilson lines, is that of the Standard Model augmented by an additional gauge $U(1)_{B-L}$ factor, and that the particle content of the effective theory is precisely that of the MSSM, with three right-handed neutrino chiral multiplets and a single Higgs-Higgs conjugate pair, and no exotic fields.

That being said, for the $B-L$ MSSM to be completely realistic there are additional low-energy properties that it must possess. These are: 1) spontaneous breaking of the gauged $B-L$ symmetry at a sufficiently high scale, 2) spontaneous breaking of electroweak symmetry with the measured values of the $W^{\pm}$ and $Z^0$ masses, 3) the Higgs mass must agree with its measured value, and 4) all sparticle masses must exceed their current experimental lower bounds. In a series of papers \cite{Ovrut:2014rba,Ovrut:2015uea,Deen:2016vyh}, using generic soft supersymmetry breaking terms added to the effective theory, scattering the initial values of their parameters statistically over various physically interesting regions and running all parameters of the effective theory to lower energy using an extensive renormalization group analysis, it was shown that there is a wide range of initial conditions that completely solve all of the required phenomenological constraints. These physically acceptable initial conditions are referred to as ``viable black points''. More information about those physically viable sets of points can be found in the second part of this thesis, where the main topic is precisely the phenomenology of the B-L MSSM. 

 Relevant to this theory is the fact that for two distinct choices of the mass scales of the two $\mathbb{Z}_{3}$ Wilson lines, the four gauge parameters associated with the $SU(3)_{C} \times SU(2)_{L} \times U(1)_{Y} \times U(1)_{B-L}$ group were shown to grand unify -- albeit at different mass scales. Let us discuss these two choices in turn.

\subsubsection{Split Wilson Lines}
\begin{figure}
\centering
\includegraphics[scale=0.9]{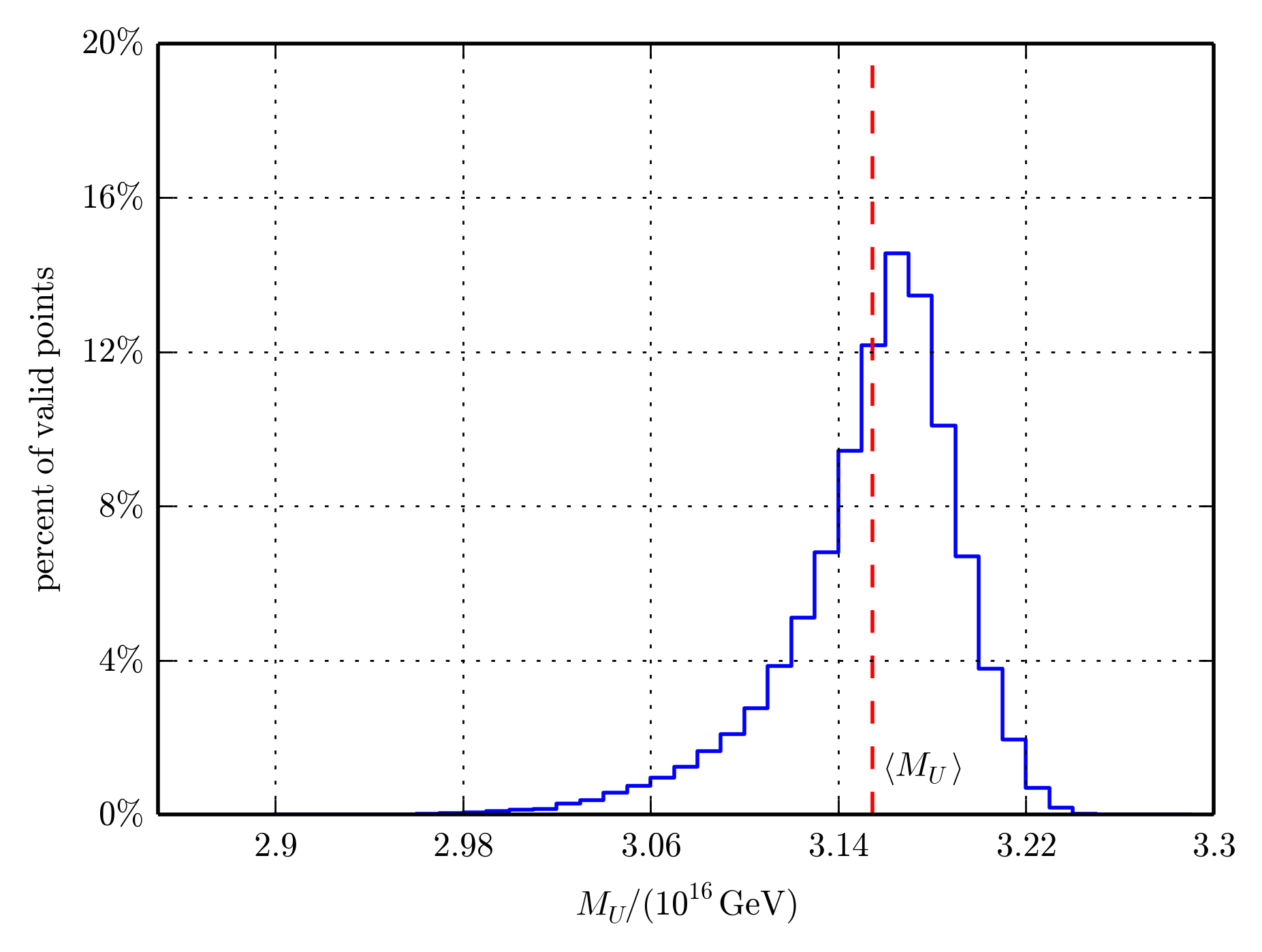}
\caption{A histogram of the unification scale for the 53,512 phenomenologically viable black points in the split Wilson line ``left-right'' unification scheme. The average unification scale is $\langle M_U\rangle=3.15\times10^{16}~\text{GeV}$.}
\label{fig:hist_uni_a}
\end{figure}

\begin{figure}
\centering
\includegraphics[scale=0.9]{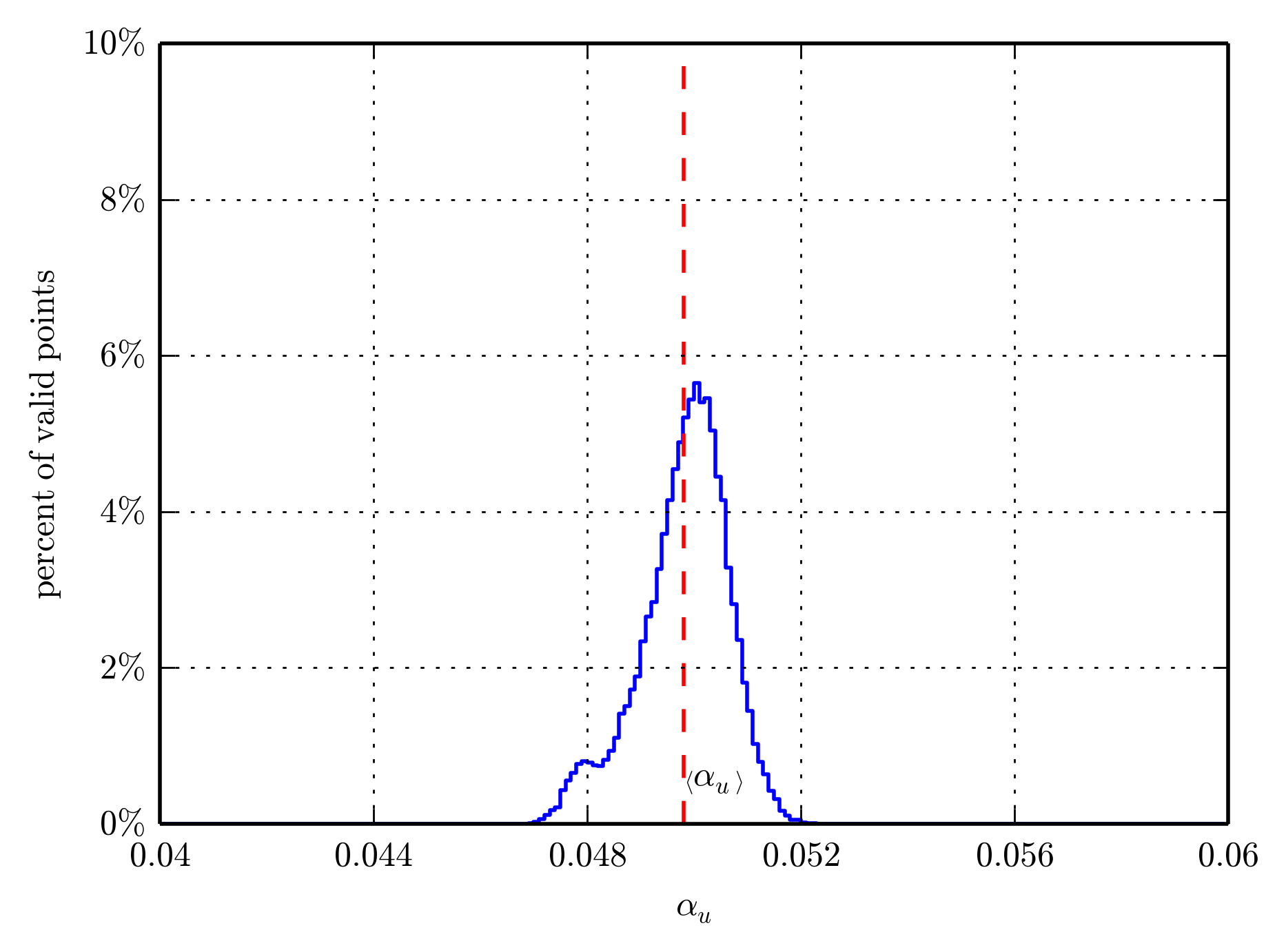}
\caption{A histogram of the unification scale for the 53,512 viable black points in the split Wilson line ``left-right'' unification scheme. The average value of the unified gauge coupling is $\langle \alpha_u\rangle=0.0498=\frac{1}{20.08}$.}
\label{fig:hist_uni_b}
\end{figure}

The first scenario involved choosing one of the Wilson lines to have a mass scale identical to the $Spin(10)$ breaking scale and to fine-tune the second Wilson line to have a somewhat lower scale, chosen to give exact gauge coupling unification. The region between the two Wilson line mass scales can exhibit either a ``left-right'' scenario or a Pati--Salam scenario depending on which Wilson line is chosen to be the lightest.  We refer the reader to \cite{Ovrut:2012wg} for details. Here, to be specific, we will consider the ``left-right'' split Wilson line scenario. For a given choice of viable black point, the gauge couplings unify at a specific mass scale $M_{U}$ with a specific value for the unification parameter $\alpha_{u}$. It was shown in \cite{Deen:2016vyh} that there were 53,512 phenomenologically viable black points. The results for $M_{U}$ and the associated gauge parameter $\alpha_{u}$ are plotted statistically over these viable black points in Figures \ref{fig:hist_uni_a} and \ref{fig:hist_uni_b} respectively. The average values for the unification scale and gauge parameter, $\langle M_U\rangle$ and $\langle \alpha_u \rangle$ respectively, are indicated.
\\
\indent The results presented in Figures \ref{fig:hist_uni_a} and \ref{fig:hist_uni_b} lead us to postulate two new ``phenomenological'' constraints on our $B-L$ MSSM vacuum. The first constraint is that
\begin{equation}
\langle M_{U}\rangle=3.15 \times 10^{16}~\text{GeV} \equiv\frac{1}{\boldsymbol{V}^{1/6}}=\frac{1}{v^{1/6}V^{1/6}} \ .
\label{jack145}
\end{equation}

 To elucidate the second constraint, we must present the explicit expression for the $D=4$ effective Lagrangian for the observable and hidden sector gauge field kinetic terms. This was calculated in \cite{Lukas:1997fg,Lukas:1998yy,Lukas:1998tt} and, ignoring gravitation, was found to be
\begin{equation}
\mathcal{L}=\dots-\frac{1}{16\pi\hat{\alpha}_{\text{GUT}}}(\re f_{1}  \tr_{E_{8}}F_{1}^{\mu\nu}F_{1\mu\nu}+\re f_{2}  \tr_{E_{8}}F_{2}^{\mu\nu}F_{2\mu\nu})+\dots\label{eq:het_lagrangian}
\end{equation}
where $\hat{\alpha}_{\text{GUT}}$ is a parameter given by\footnote{As discussed in \cite{Conrad:1997ww}, the expression for $\ah$ presented here is two times larger than the result given in \cite{Lukas:1997fg,Lukas:1998yy,Lukas:1998tt}.}
\begin{equation}
\hat{\alpha}_{\text{GUT}}=\frac{\kappa_{11}^{2}}{v}\left( \frac{4\pi}{\kappa_{11}} \right)^{2/3} \ .
\label{bag1}
\end{equation}
 It then follows from Figure \ref{fig:hist_uni_b} and \eqref{eq:het_lagrangian} that
\begin{equation}
\langle \alpha_{u} \rangle = \frac{1}{20.08}=\frac{\hat{\alpha}_{\text{GUT}}}{\re f_{1}}  \ .
\label{bag3}
\end{equation}
Furthermore, from the dimensionality reduction, we know that
\begin{equation}
\kappa_{4}^{2}= \frac{8\pi}{M_{P}^{2}} =\frac{\kappa_{11}^{2}}{v2\pi\rho} \ ,
\label{bag4}
\end{equation}
where $\kappa_{4}$ and $M_{P}=1.221 \times 10^{19}~\text{GeV}$ are the four-dimensional Newton's constant and Planck mass respectively. It then follows that
\begin{equation}
 \rho=\left( \frac{\hat{\alpha}_{GUT}}{8 \pi^{2}} \right)^{3/2}v^{1/2}M_{P}^{2} \ .
\label{bag5}
\end{equation}
Finally, using these relations, the expression for $\epsilon_S'$ in  \eqref{40AA} can be rewritten as 
\begin{equation}
 \epsilon_S' =\frac{2\pi^{2}\rho^{4/3}}{v^{1/3}M_{P}^{2/3}} \ .
\label{soc2}
\end{equation}

That is, using the physical values of the constants $M_U$ and $\alpha_{u}$ determined in this section, one can determine {\it all} constant parameters of the theory -- that is, $v$, $\hat{\alpha}_{GUT}$, $\rho$, $ \epsilon_S^\prime$ and $\epsilon_{R}$, at any given fixed point $(a^1,a^2,a^3)$ inside the K\"ahler moduli space.

\subsubsection{Simultaneous Wilson lines}

\begin{figure}
\centering
\includegraphics[scale=0.9]{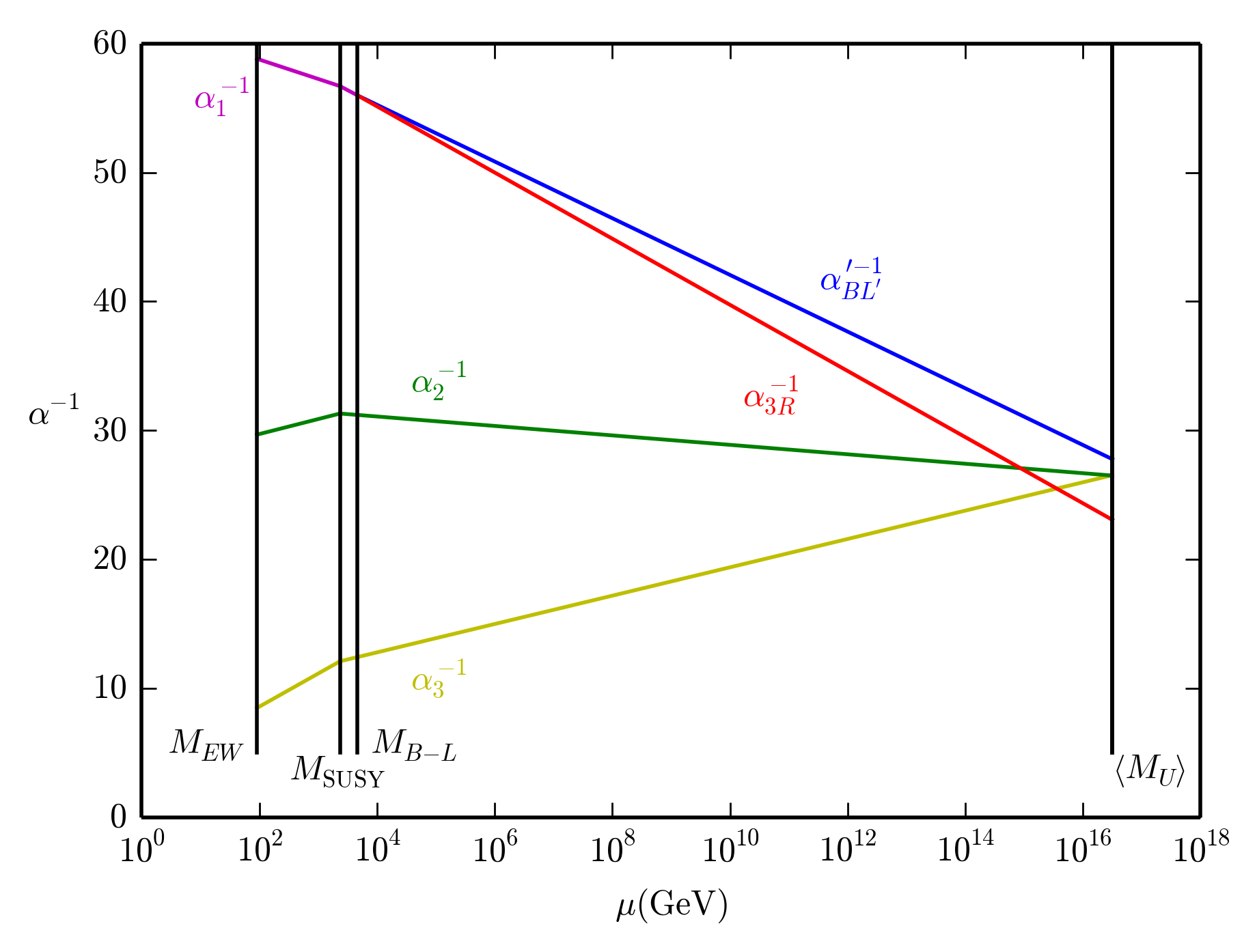}
\caption{Running gauge couplings for a sample ``valid black point''  with $M_{SUSY}=2350$ GeV, $M_{B-L}=4670$ GeV and $\sin^{2}\theta_R = 0.6$. In this example, $\alpha_3(\langle M_U \rangle)=0.0377$, $\alpha_2(\langle M_U \rangle )=0.0377$, $\alpha_{3R}(\langle M_U \rangle)=0.0433$, and $\alpha_{BL^\prime}(\langle M_U \rangle)=0.0360$.}
\label{fig:gauge_uni}
\end{figure}

In the previous subsection, we presented the phenomenological constraints for the ``left-right'' split Wilson line scenario. Here, we will again discuss the two phenomenological constraints, but this time in the scenario where the mass scales of the two Wilson lines and the ``unification'' scale are approximately degenerate. Although somewhat less precise than the split Wilson line scenario, this ``simultaneous'' Wilson line scenario is more natural in the sense that less fine-tuning is required. We refer the reader to \cite{Deen:2016vyh} for details. In this new scenario, we continue to use the previous mass scale $\langle M_{U}\rangle=3.15 \times 10^{16}~\text{GeV}$ as the $SO(10)$ ``unification'' scale -- since its mass is set by the scale of the gauge bundle -- even though when the Wilson lines are approximately degenerate the low-energy gauge couplings no longer unify there. Rather, they are split at that scale by individual ``threshold'' effects. Since the full $B-L$ MSSM low energy theory now exists at $\langle M_{U} \rangle$, we will assume that soft supersymmetry breaking also occurs at that scale. As shown in \cite{Deen:2016vyh}, we find that 44,884 valid black points satisfy all low-energy physical requirements -- including the correct Higgs mass. Rather than statistical plots over the set of all phenomenological black points, as we did in Figures \ref{fig:hist_uni_a} and \ref{fig:hist_uni_b} for the previous scenario, here we present a single figure showing the running of the inverse $\alpha$ parameters for the $SU(3)_{C}$, $SU(2)_{L}$, $U(1)_{3R}$ and $U(1)^{'}_{B-L}$  gauge couplings. This is presented in Figure \ref{fig:gauge_uni}. 
Note that in the analysis of Figure \ref{fig:gauge_uni}, we use the $U(1)_{3R}$ gauge group instead of $U(1)_{Y}$ and $U(1)^{'}_{B-L}$ instead of  $U(1)_{B-L}$, which is a minor redefinition of the $B-L$ charges, since this simplifies the renormalization group analysis. However, the averages over their gauge thresholds differ only minimally from the basis used here. Furthermore, we will augment the results of Figure \ref{fig:gauge_uni} with a more detailed discussion below which uses our standard basis.

As discussed in the previous paragraph, the first constraint in this new scenario is identical to constraint \eqref{jack145} above. That is,
\begin{equation}
\langle M_{U}\rangle=3.15 \times 10^{16}~\text{GeV} \equiv\frac{1}{\boldsymbol{V}^{1/6}}=\frac{1}{v^{1/6}V^{1/6}} \ .
\label{jack1A}
\end{equation}
To elucidate the second phenomenological constraint in this scenario, however, requires further analysis. First note from Figure \ref{fig:gauge_uni}, which is computed for a {\it single} initial valid black point, that at $ \langle M_{U} \rangle$ the values of the $\alpha$ parameters for each of the four gauge couplings are given by
\begin{gather}
\alpha_3(\langle M_U\rangle)=0.0377\ ,\qquad \alpha_2(\langle M_U\rangle)=0.0377\ , \\
\alpha_{3R}(\langle M_U\rangle)=0.0433\ ,\qquad \alpha_{BL^\prime}(\langle M_U\rangle)=0.0360 \ ,
 \label{JFK1}
 \end{gather}
 respectively. Taking the average over these parameters, we find that for that specific valid black point,
 \begin{equation}
 \alpha_{u}^{\rm avg}= \frac{1}{25.87} \ .
 \label{JFK2}
 \end{equation}
 However, to get a more generic value for the average $\langle \alpha_{u} \rangle$ at the unification scale $\langle M_{U} \rangle$, one can either: 1) repeat the same analysis as in Figure 2.4, statistically calculating over all valid black points and finding the average of the results, or 2) use the following technique, which is unique to a string theory analysis. Since our observable sector comes from an $E_{8}\times E_{8}$ heterotic string theory in ten dimensions, we will use the second analysis for simplicity.
 
 To do this, we note that, at the string tree level, the gauge couplings are expected to grand 
 unify to a single parameter $g_{\rm string}$ at a ``string unification'' scale
\begin{equation}
M_{\rm string}=g_{\rm string} \times 5.27 \times 10^{17}~\mbox{GeV} \ .
\end{equation}
The string coupling parameter $g_{\rm string}$ is set by the value of the dilaton, and is typically of ${\cal{O}}(1)$. A common value in the literature, see for example \cite{Dienes:1996du,Bailin:2014nna,Nilles:1998uy}, is $g_{\rm string}= 0.7$ which, for specificity, we will use henceforth. Therefore, we take $\alpha_{\rm string}$ and the string unification scale to be 
\begin{equation}
\alpha_{\rm string}=\frac{g_{\rm string}^{2}}{4\pi} = 0.0389 \ , \qquad    M_{\rm string}=3.69 \times 10^{17}~ \mbox{GeV} \ ,
\label{hani4}
\end{equation}
respectively. Note that $ M_{\rm string}$ is approximately an order of magnitude larger than $\langle M_{U}\rangle$. Below $M_{\rm string}$ however, the couplings evolve according to the renormalization group equations of $B-L$ MSSM effective field theory. This adds another scaling regime, $\langle M_{U}\rangle \rightarrow M_{\rm string}$, to those discussed previously.
The effective field theory in this regime remains that of the $B-L$ MSSM, with the same renormalization group factors  
as between the $B-L$ breaking scale and  $\langle M_{U}\rangle $.
However, the gauge coupling renormalization group equations are now altered to
\begin{equation}
4\pi {\alpha_{a}}^{-1}( p)=4\pi \alpha_{\rm string }^{-1}-b_{a}\ln\left(\frac{p^2}{M_{\rm string}^{2}}\right) \ ,
\label{hani6}
\end{equation}
where the index $a$ runs over $SU(3), SU(2), 3R, B-L$, the coefficients $b_{a}$ are given in \cite{Ovrut:2015uea} and, for simplicity, we have ignored the ``string threshold'' corrections calculated in \cite{Deen:2016vyh}.
Note that the one-loop running gauge couplings do not unify exactly at $\langle M_U\rangle $. Rather, they are ``split'' by dimensionless threshold effects. Using \eqref{hani4} and taking $p^2=\langle M_U\rangle ^{2}$, one can evaluate the $\alpha_{a}$ parameter for each of the four gauge couplings at the  scale $\langle M_U\rangle $. We find that
\begin{gather}
\alpha_{SU(3)}(\langle M_U\rangle )=0.0430\ ,\qquad\alpha_{SU(2)}(\langle M_U\rangle )=0.0383\ ,\\
\alpha_{3R}(\langle M_U\rangle )=0.0351\ ,\qquad\alpha_{B-L}(\langle M_U\rangle )=0.0356\ ,
\label{tr1}
\end{gather}
and, hence, the average ``unification'' parameter at $\langle M_U\rangle $ is given by
\begin{equation}
\langle  \alpha_{u}\rangle =\frac{1}{26.46} \ .
\label{tr2}
\end{equation}
It follows that for the ``simultaneous'' Wilson line scenario, the second phenomenological constraint is altered to become
\begin{equation}
\langle  \alpha _{u}\rangle =\frac{1}{26.46}=\frac{\hat{\alpha}_{GUT}}{\re f_{1}}  \ .
\label{tr3}
\end{equation}
As with the ``left-right'' Wilson line scenario in the previous subsection, given the values for $v$ and $\hat{\alpha}_{\text{GUT}}$ from \eqref{jack1A} and \eqref{tr3}, one can then compute the parameters $\rho$, $\epsilon_S'$ and $\epsilon_{R}$ using \eqref{bag5}, \eqref{soc2} and \eqref{soc1} respectively.

\section{Solving the \texorpdfstring{$B-L$}{B-L} MSSM Vacuum Constraints}\label{sec:constraints}
  
In this section, we will present a simultaneous solution to all of the $B-L$ MSSM vacuum constraints listed above. We briefly review them since they impose important constraints on the physically allowed region of K\"ahler and $\hat{R}$ moduli space.  We study a specific configuration, with a {\it single} five-brane. We denote its location in the bulk space by $z$, where 
$z \in [0,1]$. When convenient, we will re-express this five-brane location in terms of the parameter $\lambda=z-\frac{1}{2}$, where $\lambda \in [-\frac{1}{2},\frac{1}{2}]$.

\begin{enumerate}
	
\item In order to preserve $N=1$ supersymmetry in the four-dimensional effective theory, the $SU(4)$ bundle must be both slope-stable and have vanishing slope~\cite{Braun:2005zv,Braun:2006ae}. As proven in detail in~\cite{Braun:2006ae}, the $SU(4)$ bundle in the observable sector of the $B-L$ MSSM is slope-stable, and, hence, admits a connection that satisfies the Hermitian Yang--Mills (HYM) equations, in the regions of the positive K\"ahler cone defined by
\begin{equation}  \label{51}
  \begin{gathered}
    \left(
      a^1
      < 
      a^2
      \leq 
      \sqrt{\tfrac{5}{2}} a^1
      \quad\text{and}\quad
      a^3
      <
      \frac{
        -(a^1)^2-3 a^1 a^2+ (a^2)^2
      }{
        6 a^1-6 a^2
      } 
    \right)
    \quad\text{or}  \\
    \left(
      \sqrt{\tfrac{5}{2}} a^1
      <
      a^2
      <
      2 a^1
      \quad\text{and}\quad
      \frac{
        2(a^2)^2-5 (a^1)^2
      }{
        30 a^1-12 a^2
      }
      <
      a^3
      <
      \frac{
        -(a^1)^2-3 a^1 a^2+ (a^2)^2
      }{
        6 a^1-6 a^2
      }
    \right) \ .
  \end{gathered}
\end{equation}
\item  The four-dimensional effective theory is derived by first compactifying on a Calabi--Yau threefold to give a five-dimensional theory, and then reducing further on the $S^{1}/{\mathbb{Z}}_{2}$ interval. For this to be consistent, we require that the length of the interval is sufficiently large compared to the average Calabi--Yau radius. This condition takes the form
\begin{equation}
\frac{\pi \rho {\Rhat} V^{-1/3}}{(vV)^{1/6} } > 1 \ .
\label{seb3}
\end{equation}

\item  The reduction on the orbifold interval from 5D to 4D uses a linearized approximation to the five-dimensional BPS solution of heterotic M-theory.
\begin{equation}
 2\epsilon_S'\frac{\Rhat}{V^{1/3}}
  \left|
    \beta_i^{(0)} \big(z-\tfrac{1}{2}\big)
    -\tfrac{1}{2}W_i(\tfrac{1}{2}-\lambda)^2
  \right|
  \ll 
  | d_{ijk} a^j a^k |
  \ , \quad z \in [0, \lambda + \tfrac{1}{2}] \label{83} \ ,
  \end{equation}
\begin{equation}
2\epsilon_S'\frac{\Rhat}{V^{1/3}}
  \left|
    (\beta_i^{(0)}+W_i)
    \big(z-\tfrac{1}{2}\big)
    -\tfrac{1}{2}W_i(\tfrac{1}{2}+\lambda)^2
  \right| 
  \ll 
  | d_{ijk} a^j a^k |
  \ , \quad z \in [\lambda + \tfrac{1}{2},1] \ .
  \label{84}
\end{equation}

\item The squares of the ``unified'' gauge couplings in both the observable and hidden sectors must be positive-definite.
As shown in the previous section, these conditions can be written as
\begin{align}
d_{ijk}a^ia^ja^k+3\frac{\epsilon_S^\prime \hat R}{V^{1/3}}\left(\tfrac{2}{3}a^1-\tfrac{1}{3}a^2+4a^3+(\tfrac{1}{2}-\lambda)^2W_ia^i\right)&>0 \ , \label{tree1}\\
d_{ijk}a^ia^ja^k-3\frac{\epsilon_S^\prime \hat R}{V^{1/3}}\left(\tfrac{2}{3}a^1-\tfrac{1}{3}a^2+4a^3+(1-(\tfrac{1}{2}+\lambda)^2)W_ia^i\right)&>0 \ ,
\label{eq:positive2}
\end{align}
where $z=\tfrac{1}{2}+\lambda$ gives the position of the five-brane in the interval.

%

\item We want our top-down model to give reasonable four-dimensional physics, which leads to a number of ``phenomenological'' constraints. One such constraint is that the $Spin(10)$ grand unification scale, $\langle M_{U} \rangle$, and the associated unified gauge coupling in the observable sector, $\langle \alpha_{u}\rangle=\langle g^{(1)}\rangle^2/4\pi$, be consistent with phenomenologically acceptable values for these quantities. As discussed in the previous section, reasonable choices for these quantities are
\begin{equation}
\langle M_{U}\rangle=3.15 \times 10^{16}~\text{GeV} \ ,\qquad \langle \alpha_{u} \rangle = \frac{1}{20.08} \ .
\label{jack1}
\end{equation}
for the split Wilson lines scenario, and
\begin{equation}
\langle M_{U}\rangle=3.15 \times 10^{16}~\text{GeV} \ ,\qquad \langle \alpha_{u} \rangle = \frac{1}{26.46} \ .
\label{jack11}
\end{equation}
in the simultaneous Wilson lines scenario.

\end{enumerate}

 The slope-stability condition for the $SU(4)$ observable sector gauge bundle is independent of the choice of the hidden-sector gauge bundle and any bulk-space five-branes. However, the remaining constraints depend strongly upon the specific choice of the line bundle $L$ in the hidden sector, its exact embedding in the hidden sector $E_{8}$ gauge group, and, finally, on the location $\lambda$ and the effective class of the five-brane in the $S^{1}/{\mathbb{Z}}_{2}$ interval.

We propose the following line bundle $L$ as a solution for all these constraints
\begin{equation}
  L=\Ocal_X(2, 1, 3)  \ .
  \label{red1}
\end{equation}
The reason why a line bundle with these divisor values is an appealing choice will be explained later. Note that each entry is an integer and that $l^1=2$ and $l^2=1$ satisfy the equivariance condition \eqref{22}, as they must. The reason for this choice of line bundle, and the presentation of several other line bundles that lead to acceptable results, will be discussed below. Generically, there are numerous distinct embeddings of a given arbitrary line bundle into an $E_{8}$ gauge group, each with its commutant subgroup. For concreteness, we choose the particular embedding of $L=\Ocal_X(2, 1, 3)$ into $E_{8}$ discussed in Section 2.1.3, for which $a=1$. For this line bundle choice, using eq. \eqref{33A} we find that
\begin{equation}
W_{i}=(9 ,17 , 0)|_{i} \geq 0 \quad\text{for each } i=1,2,3  \ .
\label{red8}
\end{equation}
Hence, the anomaly cancellation condition is satisfied and the five brane class is effective, preserving supersymmetry.

\subsection{Moduli Scaling: Simplified Gauge Parameter Constraints}

Before presenting solutions to the rest of the constraints shown above, we observe the following important fact. The form of those constraints remains invariant under the scaling
\begin{equation}
a^{i} \to \mu a^{i} \ ,\qquad  \epsilon'_{S}\Rhat \to \mu^{3} \epsilon'_{S}\Rhat \ ,
\label{wall1}
\end{equation}
where $\mu$ is any positive real number. It follows that the coefficient $\epsilon_S'\Rhat/V^{1/3}$ in front of the $\kappa_{11}^{4/3}$ terms in each of the two constraint equations can be set to unity by choosing the appropriate constant $\mu$; that is
\begin{equation}
\epsilon_S'\frac{\Rhat}{V^{1/3}} \to 1 \ .
\label{wall2}
\end{equation}
We will refer to this choice of $\epsilon_S'\Rhat/V^{1/3}=1$ as the ``unity'' gauge. 

This scalling invariance property is obvious for the $SU(4)$ slope-stability condition discussed above, but can be verified for all the other conditons enumerated above. For example, under the $a^{i} \to \mu a^{i}$ scaling, the gauge coupling positivity constraints \eqref{eq:positive2} simplify to %
\begin{equation}
  \label{68A}
  \begin{split}
    d_{ijk} a^i a^j a^k- 3\bigl(
    -(\tfrac83 a^1 + \tfrac53 a^2 + 4 a^3)
    + 
    + 2(a^1+a^2) -(\tfrac{1}{2}-\lambda)^2 a^i \,{W}_i 
    \bigr) &> 0  \ ,
  \end{split}
\end{equation}
and
\begin{equation}
  \label{69A}
  \begin{split}
    d_{ijk} a^i a^j a^k- 3
    \bigl(a\,d_{ijk}a^i l^j l^k
    + 
    + 2(a^1+a^2) -(\tfrac{1}{2}+\lambda)^2 a^i
    \,{W}_i \bigr) &> 0 \ .
  \end{split}
\end{equation}

For the explicit choice of line bundle $L=\Ocal_X(2, 1, 3)$, these constraints become
\begin{equation}
  \begin{split}
	({a^1})^2a^2+a^1({a^2})^2+6a^1a^2a^3&+2a^1-a^2+12a^3+\\
	&+3\left(\frac{1}{2}-\lambda \right)^2(9a^1+17a^2)>0
  \end{split}
\label{clip1}
\end{equation}
and 
\begin{equation}
  \begin{split}
	({a^1})^2a^2+a^1({a^2})^2+6a^1a^2a^3&-29a^1-50a^2-12a^3+\\
	&+3\left(\frac{1}{2}+\lambda \right)^2(9a^1+17a^2)>0.
  \end{split}
\label{clip2}
\end{equation}

We will also explicitly fix the location of the bulk space five-brane by choosing its location parameter $\lambda$. As can be seen in \eqref{clip2}, the condition $(g^{(2)})^2>0$ is most easily satisfied when the value of $\lambda$ is as large as possible; that is, for the five-brane to be near the hidden wall.
For concreteness, we will take 
\begin{equation}
\lambda=0.49 \ .
\label{cup1}
\end{equation}
Note that we do not simply set $\lambda=\frac{1}{2}$, to avoid unwanted ``small instanton'' transitions of the hidden sector~\cite{Ovrut:2000qi}; that is, to keep the five-brane as an independent entity.

\subsection{Solution Space}

Let us first discuss the physical constraints. We demand that \emph{at every point} in the region of K\"ahler moduli space, all parameters of the theory are adjusted so that $\langle  M_U \rangle $  and $\langle  \alpha_u \rangle $ are fixed at the physical values presented in Section \ref{sec:observ} -- that is, the unification scale is always set to $\langle M_U\rangle =3.15\times 10^{16}~\text{GeV}$, whereas for $SO(10)$ breaking with split Wilson lines, $\langle  \alpha_{u}\rangle =1/20.08$, while for $SO(10)$ breaking with simultaneous Wilson lines $\langle  \alpha_{u}\rangle =1/26.46$. It follows from the physical constraint equations \eqref{jack1}, and \eqref{bag3} and \eqref{tr3} that the values of $v$ and ${\hat{\alpha}}_{GUT}$ -- and, hence, the remaining parameters $\rho$, $\epsilon_S^\prime$ and $\epsilon_R$ -- can be always be chosen so as to obtain the required values of $\langle M_U\rangle $ and $\langle  \alpha_{u}\rangle $.  To make this explicit, one can invert constraint equations \eqref{jack1}, \eqref{bag3} and \eqref{tr3} so as to express $v$ and ${\hat{\alpha}}_{GUT}$ explicitly as functions of $\langle M_{U}\rangle $, $\langle  \alpha_{u}\rangle $ and the K\"ahler moduli. That is, expression \eqref{jack1} can be inverted to give
\begin{equation}
v=\frac{1}{\langle M_U\rangle ^{6}V} \ ,
\label{door1}
\end{equation}
while \eqref{bag3} and \eqref{tr3} give
\begin{equation}
{\hat{\alpha}}_{GUT}=\frac{1}{\langle  \alpha_{u} \rangle \re f_{1}} \ .
\label{door2}
\end{equation}
 Inserting these expressions into \eqref{bag5}, \eqref{soc2} and \eqref{soc1}, one obtains the following expressions for $\rho$, $\epsilon_S^\prime$ and $\epsilon_R$ respectively. We find that
\begin{equation}
 \rho= \left(\frac{\langle  \alpha_{u} \rangle}{16\pi^2}\right)^{3/2}\frac{M_P^2}{\langle M_{U}\rangle ^3}  \frac{ ( \re f_{1} )^{3/2}}{V^{1/2}}
\label{seb6}
\end{equation}
and
\begin{equation}
 \epsilon_S' =\frac{\langle  \alpha_{u} \rangle^2}{128\pi^2}\frac{M_P^2}{\langle M_U\rangle ^2}\frac{(\re f_{1})^2}{V^{1/3}}\ ,\qquad \epsilon_{R}=\frac{64\pi^{2}}{\langle \alpha_{u}\rangle ^{3/2} }\frac{\langle M_{U}\rangle^{2}}{M_{P}^{2}}\frac{V^{1/3}}{(\re f_{1})^{3/2}}  \ .
\label{seb7}
\end{equation}
Therefore, these parameters can be calculated at any point $(a^1,a^2,a^3)$ in the moduli space.

Next, we discuss the dimensional reduction constraint. We require that \eqref{sun1} be valid; that is, the length of the $S^{1}/{\mathbb{Z}}_{2}$ orbifold interval should be larger than the average Calabi--Yau radius 
\begin{equation}
\frac{\pi \rho {\Rhat} V^{-1/3}}{(vV)^{1/6} } > 1 \ .
\label{seb3}
\end{equation}
This condition can be verified for any $(a^1,a^2,a^3)$ point in the K\"ahler moduli space, using the parameter expressions given above. Also, note that
in "unity gauge",
\begin{equation}
\Rhat=\frac{V^{1/3}}{\epsilon'_{S}} \ .
\label{exam1}
\end{equation}
It is possible to use "unity gauge", because equation \eqref{seb3}, as well as all the ``vacuum'' constraints enumerated above, remain invariant under the scaling
\begin{equation}
a^{i} \rightarrow \mu a^{i}\ , \qquad\epsilon'_{S} \Rhat \rightarrow \mu^{3} \epsilon'_{S}\Rhat \ , 
\label{er4}
\end{equation}
where $\mu > 0$. It follows from this that if any point \{$a^{i}$\} so will any point \{$\mu a^{i}$\}. Do any of these ``scaled'' moduli carry new information concerning both the reduction and the physical constraints? The answer to this is no, as we will now demonstrate.

Let us pick any point \{$a^{i}$\} assume that the values of the parameters at that point are given by $v$, ${\hat{\alpha}}_{GUT}$, $\rho$ and $\epsilon'_{S}$ obtained from expressions \eqref{door1}, \eqref{door2}, \eqref{seb6} and \eqref{seb7} respectively evaluated at this point. We now want to determine how each of these parameters changes under the $\mu$ scaling given in \eqref{er4}. To do this, one can again use the same equations  \eqref{door1}, \eqref{door2}, \eqref{seb6} and \eqref{seb7}, but now scaling the original point as in \eqref{er4}. To do this, we must know the scaling behavior of both $V$ and $\re f_{1}$ respectively. It follows from \eqref{10} that $V \rightarrow \mu^{3} V$. However, to obtain the scaling behavior of $\re f_{1}$, one must go back to \eqref{69AA} and recall that the terms in $Ref_{1}$ linear in the K\"ahler moduli are, generically, multiplied by the factor $\epsilon'_{S} \Rhat/ V^{1/3}$, which scales as $\mu^2$. Hence, it follows from \eqref{er4} that under $\mu$ scaling $\re f_{1} \rightarrow \mu^{3} \re f_{1}$. Using these results, we find that
\begin{gather}
v \rightarrow \mu^{-3}v\ ,\quad{\hat{\alpha}}_{GUT} \rightarrow \mu^{3}{\hat{\alpha}}_{GUT}\ ,\quad\rho \rightarrow \mu^{3} \rho\ ,\\
\epsilon'_{S} \rightarrow  \mu^{5}\epsilon'_{S}\ ,\quad\epsilon_{R} \rightarrow \mu^{-7/2} \epsilon_{R} \ ,
\label{sf1}
\end{gather}
as well as  $\epsilon'_{S}\Rhat \rightarrow \mu^{3} \epsilon'_{S}\Rhat$ and
\begin{equation}
\Rhat \rightarrow \mu^{-2} \Rhat \ .
\label{sf2}
\end{equation}
It is now straightforward to insert these results into the expression for the ratio of the orbifold interval length/average Calabi--Yau radius. We find that the scaling of the individual parameters and moduli {\it exactly cancel}. That is, under the scaling given in \eqref{er4} and \eqref{sf1}
\begin{equation}
\frac{\pi \rho {\Rhat} V^{-1/3}}{(vV)^{1/6} }  \rightarrow \frac{\pi \rho {\Rhat} V^{-1/3}}{(vV)^{1/6} } \ .
\label{sf3}
\end{equation}
We conclude from this that the $\mu$-scaled point $\{\mu a^{i}\}$ of any point $\{a^{i}\}$ is subject to the same ``vacuum'' and ``phenomenological'' constraints.

\begin{figure}[t]
   \centering
\begin{subfigure}[c]{0.47\textwidth}
\includegraphics[width=1.0\textwidth]{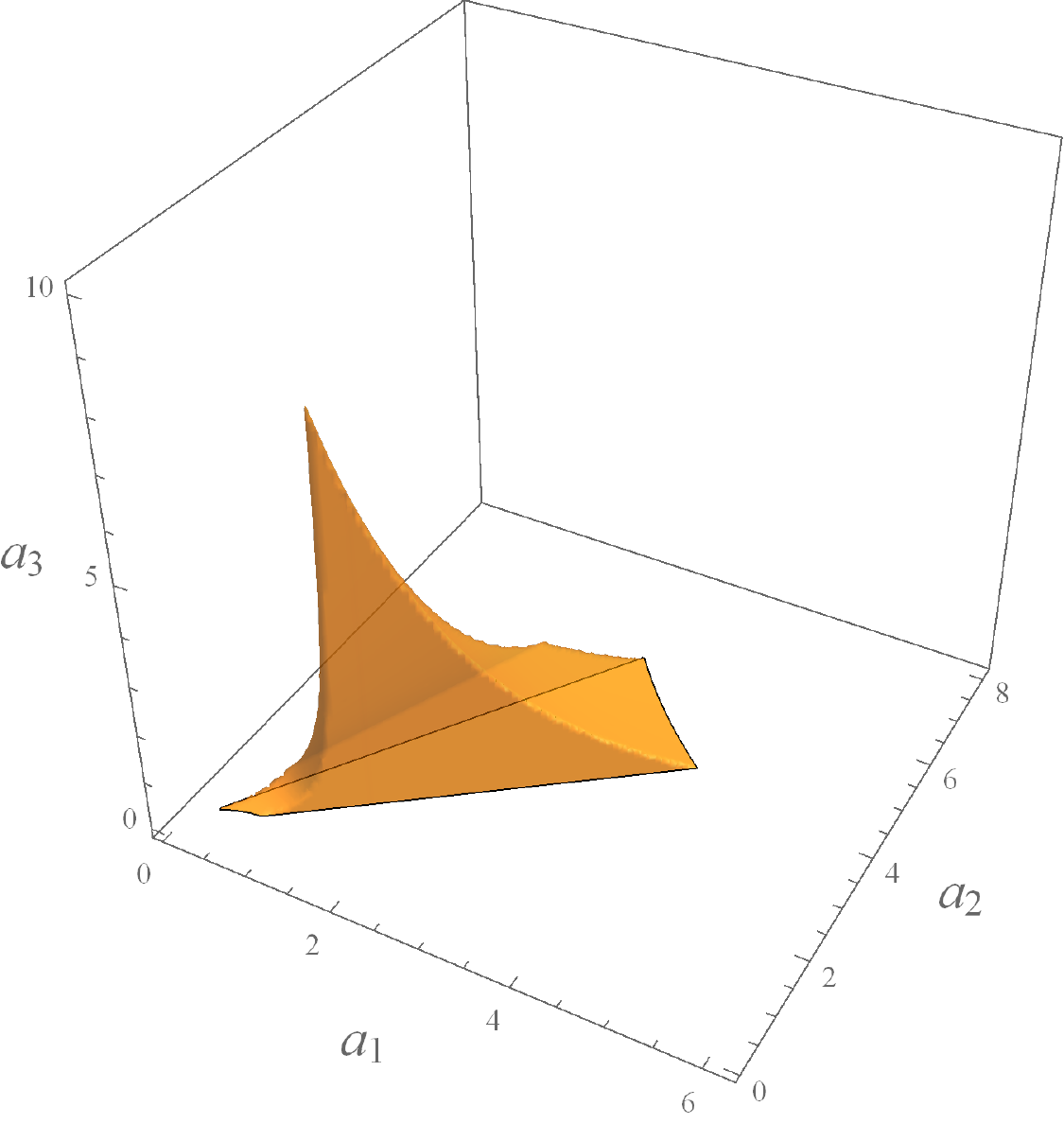}
\caption{$\langle  \alpha_{u}\rangle =\frac{1}{20.08}$}
\end{subfigure}
\begin{subfigure}[c]{0.47\textwidth}
\includegraphics[width=1.0\textwidth]{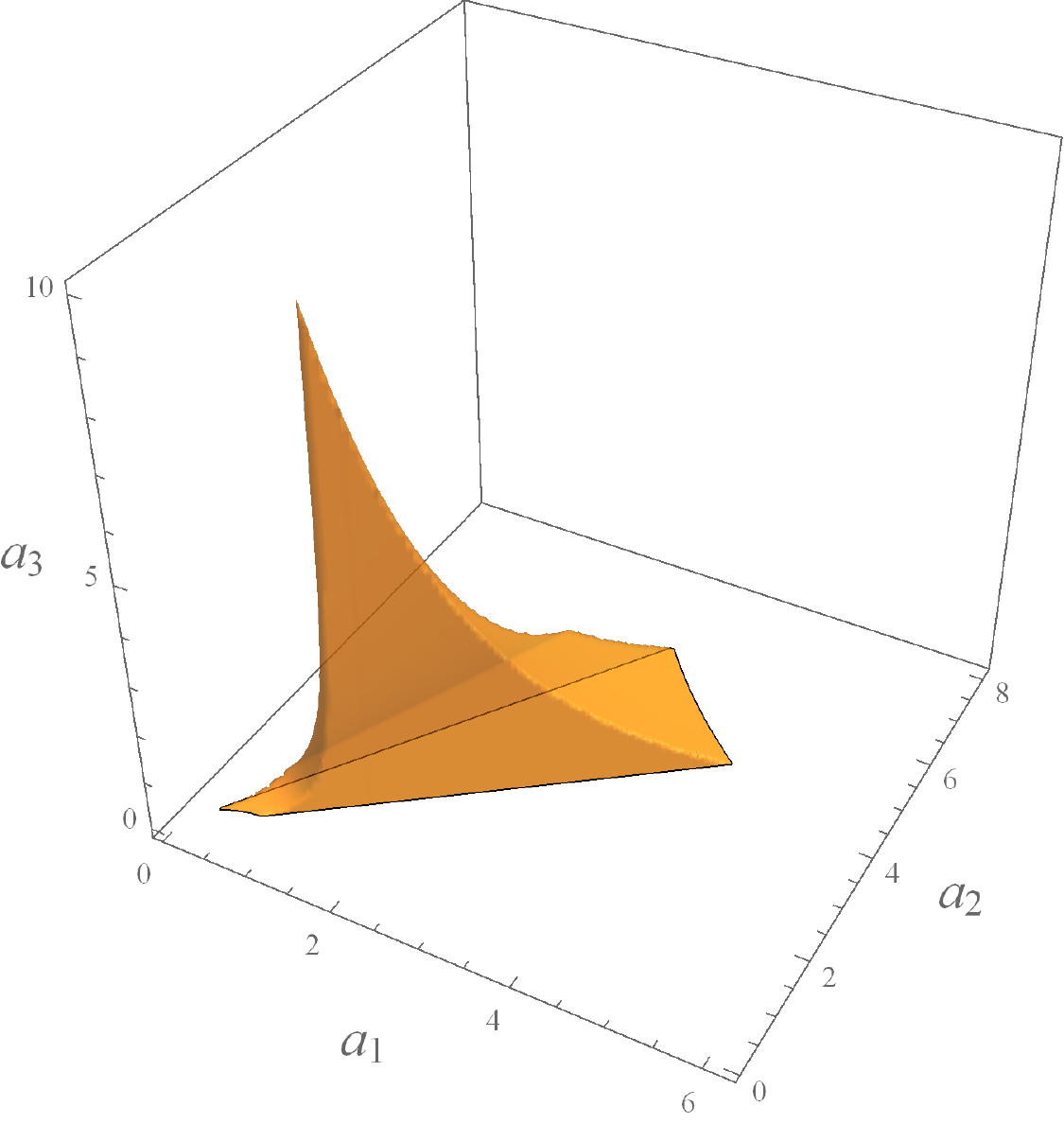}
\caption{$\langle  \alpha_{u}\rangle =\frac{1}{26.46}$}
\end{subfigure}
\caption{The region of K\"ahler moduli space where the $SU(4)$ slope-stability conditions, the anomaly cancellation constraint and the positive squared gauge coupling constraint are satisfied, in addition to the dimensional reduction and the phenomenological constraints. The results are valid for a hidden sector line bundle $L=\Ocal_X(2, 1, 3)$  with $a=1$ and for a single five-brane located at $\lambda=0.49$. We study both cases of split Wilson lines, with $\langle  \alpha_{u}\rangle =\frac{1}{20.08}$, and simultaneous Wilson lines with $\langle  \alpha_{u}\rangle =\frac{1}{26.46}$. Note that reducing the size of $\langle  \alpha_u\rangle $ increases the space of solutions.}
\label{fig:PhysContraint}
\end{figure}

With this insight, we will be able to find the subspace of that region of K\"ahler moduli space in which the vacuum constraints are satisfied. We will relax, however, the constraints shown in \eqref{83} and \eqref{84} that control the accuracy of the linearized approximation to the five-dimensional BPS solution. Scanning over all points $\{a^i\}$ in the K\"ahler cone, we find, of course, two such regions -- one corresponding to the ``split'' Wilson line scenario and a second corresponding to the ``simultaneous'' Wilson line scenario. These regions are shown as the orange subspaces of Figure \ref{fig:PhysContraint} (a) and (b) respectively.

One can go further and, by scanning over the brown subspace associated with each Wilson line scenario, find the numerical range of the ratio $\frac{\pi \rho {\Rhat} V^{-1/3}}{(vV)^{1/6}}$ in each case. We find that
\begin{equation}\
1 \lesssim \frac{\pi \rho {\Rhat} V^{-1/3}}{(vV)^{1/6}} \lesssim 17.4
\label{er1}
\end{equation}
for the split Wilson line scenario and 
\begin{equation}\
1 \lesssim \frac{\pi \rho {\Rhat} V^{-1/3}}{(vV)^{1/6}} \lesssim 19.8
\label{er2}
\end{equation}
for the simultaneous Wilson lines.

\section{The Hidden Sector - Slope Stability, Supersymmetry, Matter Content}\label{sec:hidden_sector}

Thus far, we have found the region of K\"ahler moduli space in which the $\text{\ensuremath{\SU 4}}$ bundle is slope-stable with vanishing slope, the five-brane class is effective, the squares of both gauge couplings are positive, the length of the orbifold is larger than the characteristic length scale of the Calabi--Yau threefold and the vacuum is consistent with both the mass scale and gauge coupling of $SO(10)$ grand unification in the observable sector. 

Importantly, however, we must satisfy two remaining conditions. First, it is necessary that the gauge connection associated with the hidden sector line bundle solves the Hermitian Yang--Mills (HYM) equations~\cite{UY,Donaldson} and, second, that the line bundle is such that the low-energy effective theory admits an $N=1$ supersymmetric vacuum. We will now analyze both of these remaining constraints.  To carry out these analyses, it is first necessary to introduce the Fayet--Iliopoulos term associated with the hidden sector $U(1)$ gauge group and to discuss the $\kappa_{11}^{4/3}$ correction to both the Fayet--Iliopoulos term and to the slope.

\subsection{The anomalous $U(1)$}

In the previous section, we have shown that line the bundle $L$ and its embedding into $E_8$ are chosen such that the 4D low energy group is $E_7\times U(1)$. In particular, the low-energy $U(1)$ structure group leads to an anomalous gauge 3-point function. As it is well-known, in string theory this anomaly can be removed by the Green-Schwarz mechanism~\cite{Green:1984sg}. The associated $U(1)$ structure group of $L$ is then referred to as an ``anomalous'' $U(1)$ and has several important properties. The most relevant for the present analysis is the fact that the Green-Schwarz mechanism induces
an {\it inhomogenous} transformation in both the dilaton, the K\"ahler moduli, and the five-brane modulus discussed above, even though they are uncharged under $U(1)$. The Green-Schwarz mechanism for an anomalous $U(1)$ and the resulting inhomogenous transformations for the dilaton, K\"ahler moduli and five-brane modulus are shown in Appendix A. Here we simply state the results.
We found that under an infinitesimal anomalous $U(1)$ gauge transformation, the scalar parts of these moduli transform as

\begin{equation}
\begin{split}
\label{eq:def_scalar_transform_intro}
& \delta_\theta S=-2i\pi a\epsilon_S^2\epsilon_R^2\left(\tfrac{1}{2} \beta^{(2)}_i l^i + W_il^iz^2\right)\theta\ \equiv k_S\theta,\\
&\delta_\theta T^i=-2i a\epsilon_S\epsilon_R^2l^i\theta \equiv k_T^i\theta\ ,\quad i=1,\dots,h^{1,1}\ ,\\
&\delta_\theta Z=-2ia\epsilon_S\epsilon_R^2W_il^iz\theta=k_Z\theta\ .\\
\end{split}
\end{equation}
where $\theta$ is an infinitesimal parameter. 

We note in passing, that the hidden sector low energy effective theory contains matter chiral superfields $(C_L,\psi_L)$ whose fermions $\psi_L$ are associated with the zero-modes of the 6D Dirac operator. These can be determined, for example, using the Euler characteristic of powers of the associated bundle, as discussed in~\cite{Green:1987mn}. We will explicitly compute these modes in Section \ref{sec:hidden_matter_cont}, for our specific line bundle hidden sector. Unlike the moduli superfields, these matter superfields generically transform {\it homogeneously} under both the $E_7$ and 
$U(1)$ subfactors of the low energy gauge group. For example, the scalar fields $C^L$, with $Q^L$ charge under the $U(1)$ charges by $Q^L$. That is, under an infinitesimal $U(1)$ gauge transformation
\begin{equation}
\label{eq:def_matter_transform_intro}
\delta_\theta C^L=-iQ^LC^L\theta\equiv k_C^L \theta\ , \quad L=1, \dots, \mathcal{N}\ ,
\end{equation}
Finally, as presented in~\cite{Brandle:2003uya,Lukas:1998tt}, the K\"ahler potential associated with the matter superfields is given by
\begin{equation}
\label{eq:K_matter_int}
K_{\text{matter}}=e^{ \kappa^2_4K_T/3}\mathcal{G}_{L\bar M}C^L\bar C^{\bar M}\ ,
\end{equation}
where $\mathcal{G}_{L\bar M}$ is an unspecified, generically moduli dependent, Hermitian matrix on the $H^1$ cohomologies associated with the $C^L$ matter fields in the hidden sector. In this analysis, we will, for simplicity, assume that $\mathcal{G}_{L\bar M}$ is a constant matrix.

The explicit vectors $k_S$, $k_T^i$, $k_{Z}$ and $k_C^L$, defined by the transformations above, that is
\begin{align}
\label{rain2}
&\nonumber k_S=-2i\pi a\epsilon_S^2\epsilon_R^2\left(\tfrac{1}{2} \beta^{(2)}_i l^i + W_il^iz^2\right) \ ,\\
&  k_T^i =-2i a\epsilon_S\epsilon_R^2l^i\ ,\\
&\nonumber k_Z=-2ia\epsilon_S\epsilon_R^2W_il^iz\ ,\\
&\nonumber  k_C^L=-iQ^LC^L
\end{align}
can be shown to be the Killing vectors on the space of all moduli and the matter chiral superfields.

\subsection{\texorpdfstring{$D=4$}{D = 4} Effective Lagrangian and the Anomalous \texorpdfstring{$U(1)$}{U(1)} Mass}

Before proceeding to the discussion of $N=1$ supersymmetry, it will be useful to present the $D=4$ effective theory for the hidden sector and to explicitly compute the anomalous mass of the $U(1)$ gauge boson. We present the results for a generic hidden sector line bundle $L=\mathcal{O}_{X}(l^1,l^2,l^3)$ with an arbitrary embedding into the hidden sector $E_{8}$. However, we conclude this subsection by computing the anomalous mass associated with the specific line bundle $L=\mathcal{O}_{X}(2,1,3)$  embedded  into $E_{8}$ as in \eqref{red5} with $a=1$.  

\subsubsection{\texorpdfstring{$D=4$}{D = 4} Effective Lagrangian}
Following the conventions of \cite{Brandle:2003uya,Freedman:2012zz}, the relevant terms in the four-dimensional effective action for the hidden sector of the strongly coupled heterotic string are
\begin{equation}
\begin{split}
 \label{eq:het_lagrangian-4}
\mathcal{L}\supset -
& g_{S\bar S}D_\mu S D^\mu \bar S-g_{T^i\bar T^j}D_\mu T^i D^\mu \bar T^{\bar j}-g_{Z\bar Z}D_\mu Z D^\mu \bar Z- g_{C^L\bar C^{\bar M}}D_\mu C^L D^\mu \bar C^{\bar M}\\
(-& g_{S\bar T^i}D_\mu S D^\mu \bar T^i- g_{S\bar Z}D_\mu S D^\mu \bar Z- g_{T^i\bar C^L}D_\mu T^i D^\mu \bar C^L
 - g_{Z\bar T^i}D_\mu Z D^\mu \bar T^i+hc)\\
&-\frac{4a\re f_{2}}{16\pi\ah}F_{2}^{\mu\nu}F_{2\mu\nu}-\frac{\pi\ah}{2a\re f_{2}}D_{\Uni 1}^{2}\ , 
\end{split}
\end{equation}
where $C^L$ denote the scalar components of the charged zero-mode chiral superfields, generically with different $U(1)$ charges $Q^{L}$ discussed in the previous subsection,  $F_{2\mu\nu}$ is the hidden sector four-dimensional $\Uni 1$ field strength. The K\"ahler metrics $g_{S\bar S}=\partial_S\partial_{\bar S} K$, $g_{T^i\bar T^j}=\partial_{T^i}\partial_{\bar T^j} K$, $g_{C^L\bar C^M}=\partial_{C^L}\partial_{\bar C^M} K$, etc. are defined in terms of the K\"ahler potential for $S$, $T^i$, $Z$ and $C^L$ fields
\begin{equation}
\label{blue1}
K=K_S+K_T+K_{\text{matter}}\ ,
\end{equation}
where $K_S$, $K_T$ and $K_{\text{matter}}$ were given in \eqref{eq:K_SandK_T} and \eqref{eq:K_matter_int}. As we will see below, the exact form of $g_{C^L\bar C^M}$ is not important in this analysis, whereas the exact form of $g_{ij}\equiv g_{T^i\bar T^j}$ will be essential in the calculation of the anomalous $U(1)$ vector superfield mass. An explicit calculation of $g_{ij}$ is presented in Appendix C.
The covariant derivatives for the scalar components 
\begin{equation}
\begin{split}
&{D}_\mu S=\partial_\mu S-A_\mu  k_S\ ,\\
&D_\mu T^i=\partial_\mu T^i-A_\mu k_T^i\ ,\quad i=1,\dots,h^{1,1}\ ,\\
&D_\mu Z=\partial_\mu Z-A_\mu k_Z\ ,\\
&D_\mu C^L=\partial_\mu C^L-A_\mu k_C^L\ ,\\
\end{split}
\end{equation}
are defined to ensure that the kinetic terms are invariant under the anomalous $U(1)$ transformations.

Also note that we have written the kinetic term for the hidden sector gauge field as a trace over $\Uni1$ instead of $\Ex8$, so that $\tr_{E_{8}} F_2^{\mu\nu} F_{2\mu\nu}=4a\,F_2^{\mu\nu} F_{2\mu\nu}$. The final term in \eqref{eq:het_lagrangian-4} is the potential energy, where $D_{\Uni 1}$ is proportional to the solution of the auxiliary D-field equation of motion and is given by
\begin{equation}
D_{\Uni 1}=ik_S\frac{\partial K}{\partial S}+ik_T^i\frac{\partial K}{\partial T^i}+ik_Z\frac{\partial K}{\partial Z}
+ik_C^L\frac{\partial K}{\partial C^L}\ . \ .
\label{again1}
\end{equation}
Therefore, using the Killing vector expressions from \eqref{rain2}, we find that
\begin{equation}
\begin{split}
\label{rain3}
D_{U(1)}&={-\frac{a\epsilon_S\epsilon_R^2}{2\kappa_4^2\hat RV^{2/3}}\left( \mu(L)+\frac{\pi \epsilon_S\hat R}{V^{1/3}}\left(\beta_i^{(2)}+z^2W_i\right)l^i    \right)}\\
&\qquad\qquad+\mathcal{G}_{L\bar M} e^{\kappa_4^2K_T/3}{\left(   1+   \frac{a\epsilon_S\epsilon_R^2}{12Q^L\hat RV^{2/3}} {\mu(L)}\right)}  Q^L C^L\bar C^{\bar M}\\
&={-\frac{a\epsilon_S\epsilon_R^2}{2\kappa_4^2\hat RV^{2/3}}\left( \mu(L)+\frac{\pi \epsilon_S\hat R}{V^{1/3}}\left(\beta_i^{(2)}+z^2W_i\right)l^i    \right)} + Q^LG_{L\bar M} C^L\bar C^{\bar M}.
\end{split}
\end{equation}
In the above expression,
\begin{equation}
\mu(L)=
  \frac{1}{\rank({{L}})v^{2/3}} 
  \int_X{c_1({{L}})\wedge \omega \wedge \omega} = d_{ijk}l^ia^ja^k 
\label{rain4A}
\end{equation}
is the \emph{tree-level} expression for the slope of the line bundle $L$ and we have defined the metric
\begin{equation}
\label{def_GLM}
G_{L\bar M}=\mathcal{G}_{L\bar M} e^{\kappa_4^2K_T/3}{\left(   1+   \frac{a\epsilon_S\epsilon_R^2}{12Q^L\hat RV^{2/3}} {\mu(L)}\right)} \ .
\end{equation}
The complex scalar fields $C^{L}$ enter the expression for $D_{U(1)}$ since they transform linearly under $U(1)$ with charge $Q^{L}$. 

For the vacuum to be $N=1$ supersymmetric, it is necessary that the vacuum expectation values of the fields $S$, $T^i$, $Z$ and $C^L$ satisfy the D-flatness condition
\begin{equation}
V=V_D=0\quad \Rightarrow \quad \langle D_{U(1)} \rangle=0
\end{equation} 
At this point, it is useful to define the so-called Fayet-Iliopoulos (FI) term within our present context. The FI term is defined to be
\begin{align}
\text{FI}&=i\langle k_S\frac{\partial K}{\partial S}\rangle+i\langle k_T^i\frac{\partial K}{\partial T^i}\rangle+i\langle k_Z\frac{\partial K}{\partial Z}\rangle \nonumber \\
&= -\frac{a\epsilon_S\epsilon_R^2}{2\kappa^2_4V^{2/3}\hat R}\left( \mu(L) +\frac{\pi\epsilon_S \hat R}{V^{1/3}}
\left(\beta_i^{(2)} +  W_iz^2  \right)  l^i  \right)\ ,
\label{rain5}
\end{align}
that is, the part of  $D_{U(1)} $ that is independent of the matter fields $ C^{L}$. The reader should be aware that we \emph{dropped} the VEV bracket notation for brevity. Note that the moduli-dependent parameters from the expression for the FI term, such as $V$ and $\hat R$, have fixed values inside the vacuum we have just described. The string one-loop corrected Fayet--Iliopoulos (FI) term was also
also computed in~\cite{Blumenhagen:2005ga}, within the context of the weakly coupled heterotic string, and in \cite{Anderson:2009sw,Anderson:2009nt} for a specific embedding of $U(1)_{r}$ into $E_8$ in strongly coupled heterotic $M$-theory.  In conclusion, $N=1$ supersymmetry is preserved in the hidden sector iff
\begin{equation}
D_{U(1)}=\text{FI}+Q^L\langle G_{L\bar M} C^L\bar C^{\bar M}\rangle =0 \ .
\label{rain4A}
\end{equation}
The D-term stabilization condition above defines two types of vacua:
\begin{itemize}
\item With vanishing FI term, FI$=0$. In this case, the $C^L$ matter fields do not need to obtain non-zero VEVs $\langle C^L\rangle=0$;

\item With non-vanishing FI term, FI$\neq 0$. In this case, the $C^L$ matter fields must obtain non-zero VEVs $\langle C^L \rangle \neq 0$ to cancel the FI term, assuming they have charges $Q_L$ with the right sign to allow such a cancellation. 
\end{itemize}
We will continue our discussion about these two types of vacua in Section \ref{sec:susy_vacua_4d}, within a specific context.

\subsubsection{A Fayet--Iliopoulos Term and the \texorpdfstring{$\kappa_{11}^{4/3}$}{kappa 4/3} Slope Correction}

The value of the FI term, whether it is vanishing or non-vanishing, will play an important role below in categorizing specific vacua. Given its importance in our analysis, let us now discuss this term in more detail, and how it appeared in literature in more general contexts.

In the heterotic standard model vacuum, the observable sector vector
bundle $\Vvis$ has structure group $SU(4)$. Hence, it does not lead
to an anomalous $U(1)$ gauge factor in the observable sector of the
low energy theory. However, the hidden sector structure group can contain such abelian subgroups. Let us consider again the general case, in which the hidden sector bundle ${V^{(2)}}=\bigoplus_{r=1}^R L_r$
consists of a sum of line bundles with the additional structure group $U(1)^{R}$. Each $U(1)$ factor leads to
an anomalous $U(1)$ gauge group in the four-dimensional effective
field theory of the hidden sector and, hence, an associated FI-term.  Let $L_r$ be
any one of the irreducible line bundles of $\Vhid$. Comparing various results in the literature with our above results, it is straightforward to show that to order $\kappa_{11}^{4/3}$ the one-loop corrected FI-term for $L_{r}$ in the strongly coupled heterotic string is 
\begin{equation}
 \label{54}
  \text{FI}_r =
  \frac{a_r}{2} \frac{ \epsilon_S \epsilon_R^2}{\kappa_{4}^{2}}
  \frac{1} {\Rhat V^{2/3}}
  \Big( \mu(L_r) + 
     \epsilon_S' \frac{\Rhat}{V^{1/3}} 
    \int_X c_1(L_r) \wedge 
   \big( J^{(N+1)}+\sum_{n=1}^{N} z_n^2 J^{(n)} \big) \Big) , 
\end{equation}
where $a_r$ is a group-theoretical coefficient, defined in \eqref{26}, determined by how the $\Uni1$ structure group of $L_r$ embeds in $\Ex8$, and $\mu(L_r)$ is given in \eqref{50}. We note that the $\kappa_{11}^{2/3}$ part of this expression is identical to that derived in \cite{Anderson:2009nt}.
Insert \eqref{28}  and, following the
conventions of~\cite{Blumenhagen:2005ga,Weigand:2006yj}, redefine the five-brane moduli as in \eqref{pink2}. Furthermore, choosing our hidden sector bundle to be the sum of line bundles   ${\cal{L}} = \bigoplus_{r=1}^R L_r,\>
  L_r=\Ocal_X(l^1_r, l^2_r, l^3_r)$, we find that
the FI-term becomes
\begin{multline}
  \label{56}
  \text{FI}_r = 
   \frac{a_r}{2} \frac{ \epsilon_S \epsilon_R^2}{\kappa_{4}^{2}}
  \frac{1} {\Rhat V^{2/3}}
  \Big( \mu(L_r) - 
  \epsilon_S' \frac{\Rhat}{V^{1/3}} 
  \\
  \int_X c_1(L_r)\wedge
  \big(\sum_{s=1}^{R} a_{s} c_{1}(L_{s}) \wedge c_{1}(L_{s})
  +\tfrac{1}{2} c_2(TX) 
  -\sum_{n=1}^{N}(\tfrac{1}{2}+\lambda_n)^2 W^{(n)} \big) \Big) \ ,
\end{multline}
The first term on the right-hand
side, that is, the slope of $L_r$, is the
order $\kappa_{11}^{2/3}$ result. The remaining terms are the
$\kappa_{11}^{4/3}$ M-theory corrections first presented in~\cite{Lukas:1998hk}.
Note that the dimensionless parameter $ \epsilon_S'
\frac{\Rhat}{V^{1/3}}$ of the $\kappa_{11}^{4/3}$ term is identical to
the expansion coefficient of the linearized solution -- when expressed
in terms of the $a^i$ moduli.
Finally, recalling definition \eqref{50} of the slope, 
using \eqref{3}, \eqref{4}, \eqref{8},
\eqref{23}, \eqref{32} and the properties of the second Chern character, it follows that for each $L_r$ the associated Fayet--Iliopoulos factor $FI_{r}$ in  \eqref{56} can be written as
\begin{multline}
  \label{59} 
\text{FI}_{r}= \frac{a_r}{2} \frac{ \epsilon_S \epsilon_R^2}{\kappa_{4}^{2}}
  \frac{1} {\Rhat V^{2/3}} \Big(d_{ijk} l_r^i a^j a^k - 
  \epsilon_S' \frac{\Rhat}{V^{1/3}} 
  \\
 \big(d_{ijk}l_r^i\sum_{s=1}^{R}a_{s}l_{s}^jl_{s}^k 
  + l^i_r(2,2,0)|_i
  -\sum_{n=1}^{N}\left(\tfrac{1}{2}+\lambda_n\right)^2l_r^iW^{(n)}_i\big) \Big)\ ,
\end{multline}

 Returning to our chosen configuration, it follows from \eqref{59} that the Fayet--Iliopoulos term associated with a generic single line bundle $L=\Ocal_X(l^{1}, l^{2}, l^{3})$, embedded in the hidden $E_8$ such that $a=1$, and a single five-brane located at $\lambda\in[-1/2,1/2]$ is given in ``unity'' gauge $\epsilon_S\hat R/V^{1/3}=1$ by
\begin{equation}
  \label{again2A} 
\text{FI}= \frac{a}{2} \frac{ \epsilon_S \epsilon_R^2}{\kappa_{4}^{2}}
  \frac{1} {\Rhat V^{2/3}} \left(d_{ijk} l^i a^j a^k - a\,d_{ijk}l^il^jl^k 
  - l^i(2,2,0)|_i
  +\left(\tfrac{1}{2}+\lambda\right)^2l^iW_i \right) \ ,
\end{equation}
with the volume modulus $V$ and $W_{i}$ presented in \eqref{10} and \eqref{33A} respectively and $\Rhat$ defined in \eqref{case1}.  

We can use the "unity gauge" in which $\epsilon_S\hat R/V^{1/3}=1$, because under the scaling $a^i\rightarrow \mu a^i$, the FI term remains invariant, as can be checked using the results given in \eqref{sf1}.

It is important to note -- using \eqref{3}, \eqref{50} and \eqref{23} -- that the ``classical'' slope of the line bundle $L=\Ocal_X(l^{1}, l^{2}, l^{3})$ is given by\footnote{Note that this is not the same as the scaling factor $\mu$ of the previous section. From here onwards, $\mu$ will denote the slope.}
\begin{equation}
\mu(L)=d_{ijk} l^ia^j a^k \ ,
\label{river1}
\end{equation}
that is, the first term in the bracket of \eqref{again2A}. It follows that the remaining terms in the bracket, specifically
\begin{equation}
- d_{ijk}l^il^jl^k - l^i(2,2,0)|_i+\left(\tfrac{1}{2}+\lambda\right)^2l^iW_i \ ,
\label{river2}
\end{equation}
are the strong coupling $\kappa_{11}^{4/3}$ corrections to the slope of $L$. For the remainder of this calculation, we will take the slope of the line bundle $L=\Ocal_X(l^{1}, l^{2}, l^{3})$ to be the 
$\kappa_{11}^{4/3}$, genus-one corrected expression
\begin{equation}
\mu(L)=d_{ijk} l^ia^j a^k- d_{ijk}l^il^jl^k - l^i(2,2,0)|_i+\left(\tfrac{1}{2}+\lambda\right)^2l^iW_i \ .
\label{trenton1}
\end{equation}

\subsubsection{The Anomalous \texorpdfstring{$U(1)$}{U(1)} Mass}

As is commonly known, a $\Uni 1$ symmetry that appears in both the internal and four-dimensional gauge groups is generically anomalous~\cite{Dine:1986zy,Dine:1987xk,Lukas:1999nh,Blumenhagen:2005ga}. Hence, there must be a Green–Schwarz mechanism in the original heterotic M-theory that will cancel this anomaly in the effective field theory. Importantly, however, in addition to canceling this anomaly, the Green--Schwarz mechanism will give a mass for the $U(1)$ vector superfield~\cite{Green:1984sg}. The way this mechanism works was reproduced in Appendix B. Here we will apply the conclusions of Appendix B to the specific field content of our theory. 

We mentioned previously that the value of the FI term characterizes two sets of vacua, with very different properties, depending on if the FI is zero or not. Assuming that the D-term stabilization condition sets $\text{FI}=0$, then one $U(1)$ $N=1$ supervector multiplet becomes massive,
with components
\begin{equation}
\label{late2}
(\phi, A_{\mu}, \Psi ).
\end{equation}
To linear order in $\epsilon_S$, the scalar component of this massive gauge vector supermultiplet is given by
\begin{equation}
 \phi=\frac{ \frac{1}{8V^2}\pi \epsilon_S\left( \beta^{(2)}_i+  W_iz^2\right) l^i\delta V +\kappa_4^2g^T_{i\bar j}l^{\bar j}\delta t^i}
{\sqrt{\kappa_4^2g^{T}_{i \bar{j}}l^{i}l^{\bar{j}} } }+\mathcal{O}(\epsilon_S^2)\ ,
\end{equation}
which represents a linear combination of the moduli scalar perturbations $\delta V$ and $\delta t^i$ around the D-flat vacuum defined by
$\langle D_{U(1)}\rangle =\text{FI}=0$.
On the other hand, the corresponding linear combination of moduli axions,
\begin{equation}
\eta=\frac{ \frac{1}{8V^2}\pi \epsilon_S\left( \beta^{(2)}_i+  W_iz^2\right) l^i\sigma +2\kappa_4^2g^T_{i\bar j}l^{\bar j}\chi^{i}}
{\sqrt{\kappa_4^2g^{T}_{i \bar{j}}l^{i}l^{\bar{j}} }}+\mathcal{O}(\epsilon_S^2) \ ,
\end{equation}
 forms the longitudinal degree of freedom of the massive $U(1)$ vector boson $A_{\mu}$. Specifically, it is absorbed by redefining the gauge field as
\begin{equation}
A_\mu^\prime = A_\mu-\frac{\partial_\mu \eta}{\sqrt{\langle g_{B\bar C}k^B\bar k^{\bar C}\rangle}}\ .
\end{equation}
Note that, for simplicity,  we have \emph{dropped the VEV bracket notation} from these final expressions for all moduli-dependent quantities. We will continue to do this in every {\it final} result throughout the rest of this work. Note that at order $\epsilon_S^2$, the five brane axion and the five-brane scalar perturbation $\delta z$ enter the definitions of $\phi$ and $\eta$ as well.

Similarly, the fermion $\psi^1_{\xi}$ is a linear combination of the fermionic parts of the chiral moduli multiplets. Using the expressions for the Killing fields $k_{S}$, $k_{T}^{i}$, $k_Z$ given in \eqref{rain2} and for the moduli space metrics presented in Appendix C, now restricted to $i,j=1,2,3$, we find that
\begin{equation}
 \psi_\xi^1=\frac{ \frac{1}{4V^2}\pi \epsilon_S\left( \beta^{(2)}_i+  W_iz^2\right) l^i\psi_S +\kappa_4^2g^T_{i\bar j}l^{\bar j}\psi_T^i}
{\sqrt{\kappa_4^2g^{T}_{i \bar{j}}l^{i}l^{\bar{j}}} }+\mathcal{O}(\epsilon_S^2)\ .
\end{equation}
As discussed in Appendix B,  $\psi^1_{\xi}$ combines with the $U(1)$ gaugino $\lambda$ to form a massive Dirac fermion
\begin{equation}
\label{late1}
\Psi= \dbinom{\lambda_{2}^{\dagger}/\langle g_2\rangle}{\psi_{\xi}^{1}} \ .
\end{equation}
 Since $N=1$ supersymmetry is protected by the D-flatness condition, the masses of vector supermultiplet components, $\phi$, $ A_{\mu}$, and $ \Psi $ are all identical. Once again, using the results of the Appendix B, and the metrics given in Appendix C, we find that the squared mass of this vector superfield can be expressed as
\begin{equation}
\begin{split}
m_{A}^{2} =\frac{\pi \hat{\alpha}_{\text{GUT}}}{a\text{Re} f_{2} }\>8a^{2}\epsilon_{S}^{2}\epsilon_{R}^{4} \Big[& g^{T}_{i \bar{j}}l^{i}l^{\bar{j}} +\frac{\pi^2\epsilon_S^2}{8V^2}\left(  \beta^{(2)}_il^i-
W_il^i z^2\right)^2\Big]\ .
\label{late4}
\end{split}
\end{equation}
Using the $g_{ij}$ metric expression given in Appendix C, we get
for $m_A^2$ of the form
\begin{equation}
m_A^2 =\frac{4\pi\hat\alpha_{GUT}}{a \re f_2}\frac{a^2\epsilon_S^2\epsilon_R^4}{\kappa_{4}^{2}\Rhat^{2}}
\left(\frac{1}{8V^{4/3}}\mu(L)^{2}-\frac{1}{2V^{1/3}}d_{ijk}l^{i}l^{j}a^{k} +\frac{\pi^2\epsilon_S^2}{8V^2}\left(  \beta^{(2)}_il^i-
W_il^i z^2\right)^2\right) \ ,
\label{eq:anomalous_massA}
\end{equation}
which is valid for a generic line bundle $L=\mathcal{O}_{X}(l^1,l^2,l^3)$ embedded arbitrarily into the hidden sector $E_{8}$. 
%
%


\subsection{Slope-Stability of the Hidden Sector Bundle \texorpdfstring{$L=\Ocal_X(2, 1, 3)$}{L = O(2,1,3)} } 

Although any line bundle $L$ is automatically slope-stable, since it has no sub-bundles, in order for its gauge connection to ``embed'' into the hidden sector $E_{8}$ gauge connection it is necessary to extend the bundle to $L \oplus L^{-1}$, as discussed in subsection \ref{sec:embedding}. However, even though the connection associated with the bundle $L \oplus L^{-1}$ can, in principle, embed properly into the $\bf 248$ gauge connection of the hidden sector $E_{8}$, it remains necessary to show that $L \oplus L^{-1}$ is ``slope-stable''; that is, that its associated connection satisfies the Hermitian Yang--Mills equations. More properly stated, since $L \oplus L^{-1}$ is the Whitney sum of two line bundles, it was shown in \cite{UY, Donaldson} that it will admit a connection that uniquely satisfies the Hermitian Yang--Mills equations if and only if it is ``polystable''; that is, if and only if
\begin{equation}
\mu(L)=\mu(L^{-1})=\mu(L \oplus L^{-1}) \ .
\label{poly1}
\end{equation}
Since $\mu(L \oplus L^{-1})$ must vanish by construction, it follows that $L \oplus L^{-1}$ is polystable if and only if $\mu(L)=0$.

\begin{figure}[t]
   \centering
\begin{subfigure}[c]{0.6\textwidth}
\caption*{}
\end{subfigure}\\
\begin{subfigure}[c]{0.49\textwidth}
\includegraphics[width=1.0\textwidth]{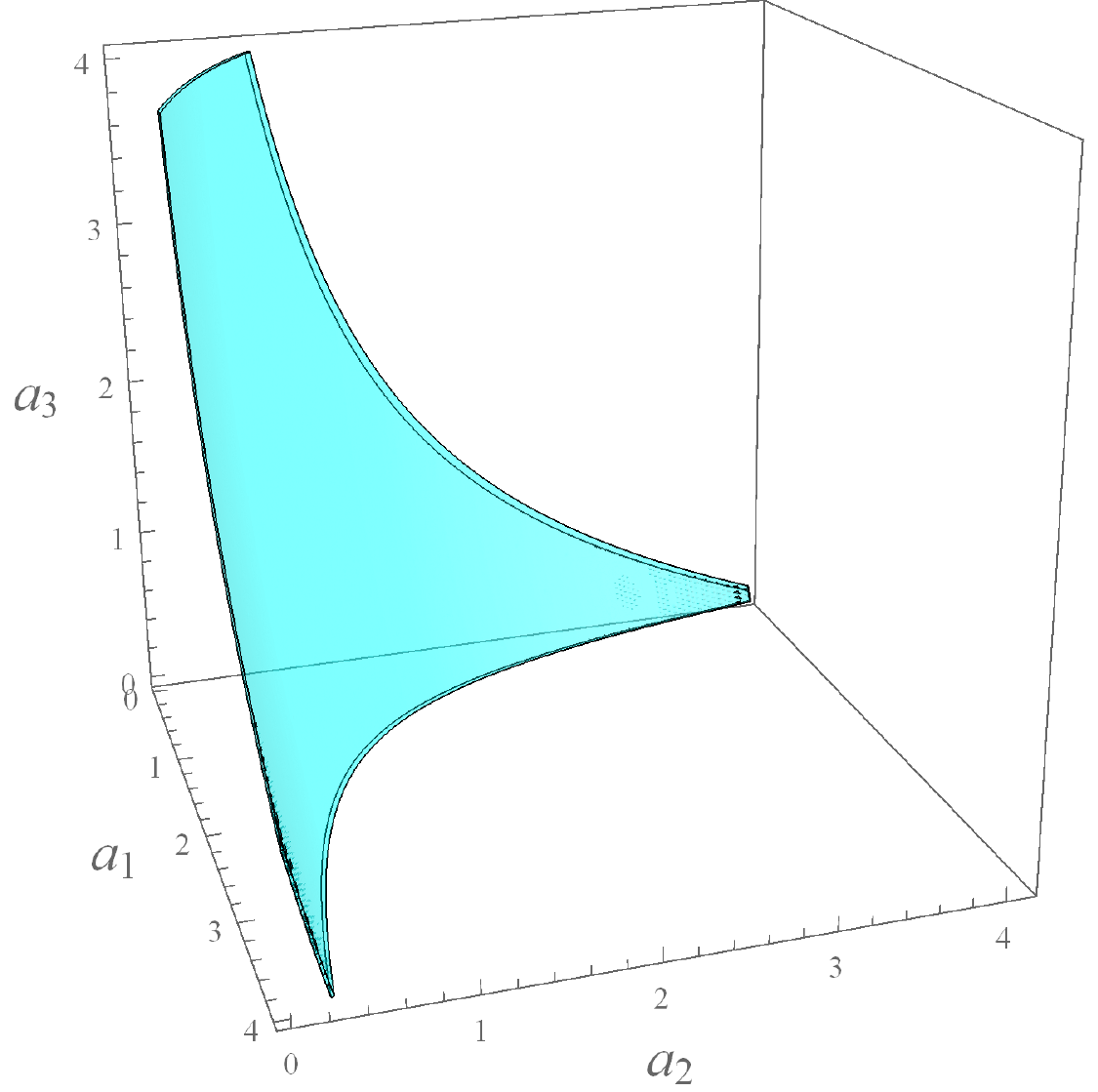}
\end{subfigure}
\caption{The surface in K\"ahler moduli space where the genus-one corrected slope of the hidden sector line bundle $L=\Ocal_X(2, 1, 3)$ vanishes.}
\label{fig:ZeroSlope}
\end{figure}

Let us now consider the specific line bundle $L=\Ocal_X(2, 1, 3)$ embedded into $SU(2) \subset E_{8}$ as in subsection \ref{sec:embedding}, with coefficient $a=1$, and take $\lambda=0.49$. It follows from \eqref{trenton1} that the associated genus-one corrected slope is
\begin{equation}
\mu(L)= \tfrac{1}{3}(a^1)^{2}+\tfrac{2}{3}(a^2)^{2} +8a^1a^2+4a^2a^3 +2a^1a^3 -13.35 \ .
\label{trenton3}
\end{equation}
Hence, this specific hidden sector bundle will be slope polystable -- and, therefore, admit a gauge connection satisfying the corrected Hermitian Yang--Mills equations -- if and only if the K\"ahler moduli $a^{i}, i=1,2,3$ satisfy the condition that
\begin{equation}
 \tfrac{1}{3}(a^1)^{2}+\tfrac{2}{3}(a^2)^{2} +8a^1a^2+4a^2a^3 +2a^1a^3 -13.35 = 0 \ .
\label{trenton4}
\end{equation}
The region of K\"ahler moduli space satisfying this condition is the two-dimensional surface displayed in Figure \ref{fig:ZeroSlope}. 

\begin{figure}[t]
   \centering
\begin{subfigure}[c]{0.6\textwidth}
\caption*{}
\end{subfigure}\\
\begin{subfigure}[c]{0.49\textwidth}
\includegraphics[width=1.0\textwidth]{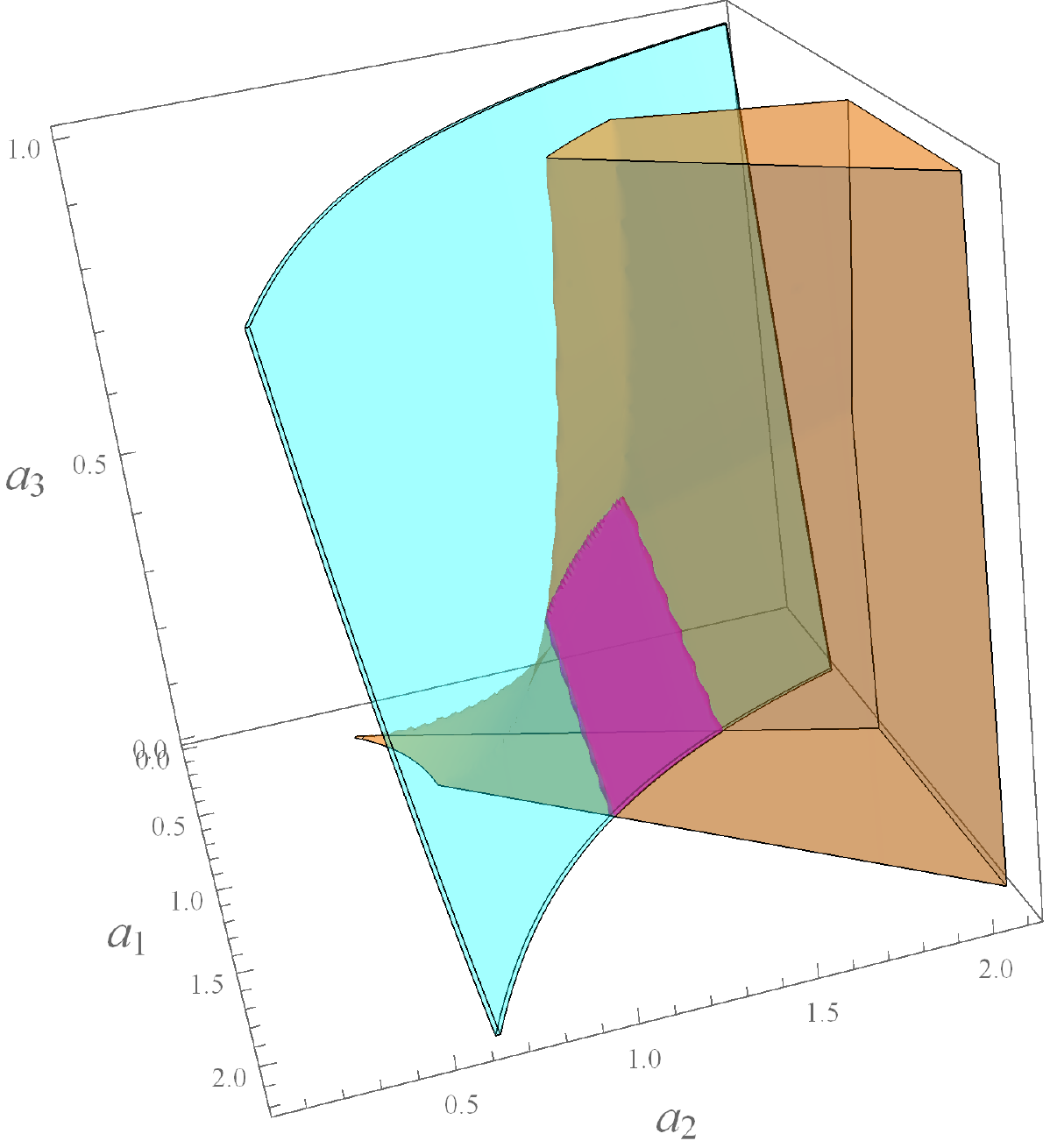}
\end{subfigure}
\caption{ The magenta region shows the intersection between the orange regions of Figure \ref{fig:PhysContraint} (a) and (b) and the two-dimensional cyan surface in Figure \ref{fig:ZeroSlope}. Therefore, the magenta region represents the sub-region of the vanishing, genus-one corrected slope surface, each point of which satisfies all the necessary constraints discussed in Section \ref{sec:observ}. The size of the magenta region is the same for both the split and simultaneous Wilson lines scenarios. However, the values of the coupling parameters differ slightly for these two cases, at any point in this intersection subspace. }
\label{fig:Intersection}
\end{figure}

Recall that for hidden sector line bundle $L=\Ocal_X(2, 1, 3)$  with $a=1$ and a single five-brane located at $\lambda=0.49$, the region of K\"ahler moduli space satisfying all {\it previous} constraints -- that is, the $SU(4)$ slope-stability conditions, the anomaly cancellation constraint, the positive squared gauge coupling constraints, in addition to the dimensional reduction and the phenomenological constraints -- are shown as the orange regions in Figure  \ref{fig:PhysContraint} (a) and (b), for the split Wilson lines and the simultaneous Wilson line scenarios respectively. It follows that the intersection of the orange regions of Figure \ref{fig:PhysContraint} (a) and (b) with the two-dimensional surface in Figure \ref{fig:ZeroSlope} will further constrain our theory so that the hidden sector gauge connection satisfies the corrected Hermitian Yang--Mills equations -- as it must. The regions of intersection are displayed graphically in Figure \ref{fig:Intersection}. We emphasize that although the brown regions of Figure \ref{fig:PhysContraint} (a) and (b) overlap in this region of K\"ahler moduli space, each point in their overlap region has a somewhat different set of parameters associated with it. Hence, in discussing a point in the magenta region of Figure \ref{fig:Intersection}, for example, it is necessary to state whether it is arising from the split Wilson line or simultaneous Wilson line scenario.

\begin{figure}[t]
   \centering
\begin{subfigure}[c]{0.49\textwidth}
\includegraphics[width=1.0\textwidth]{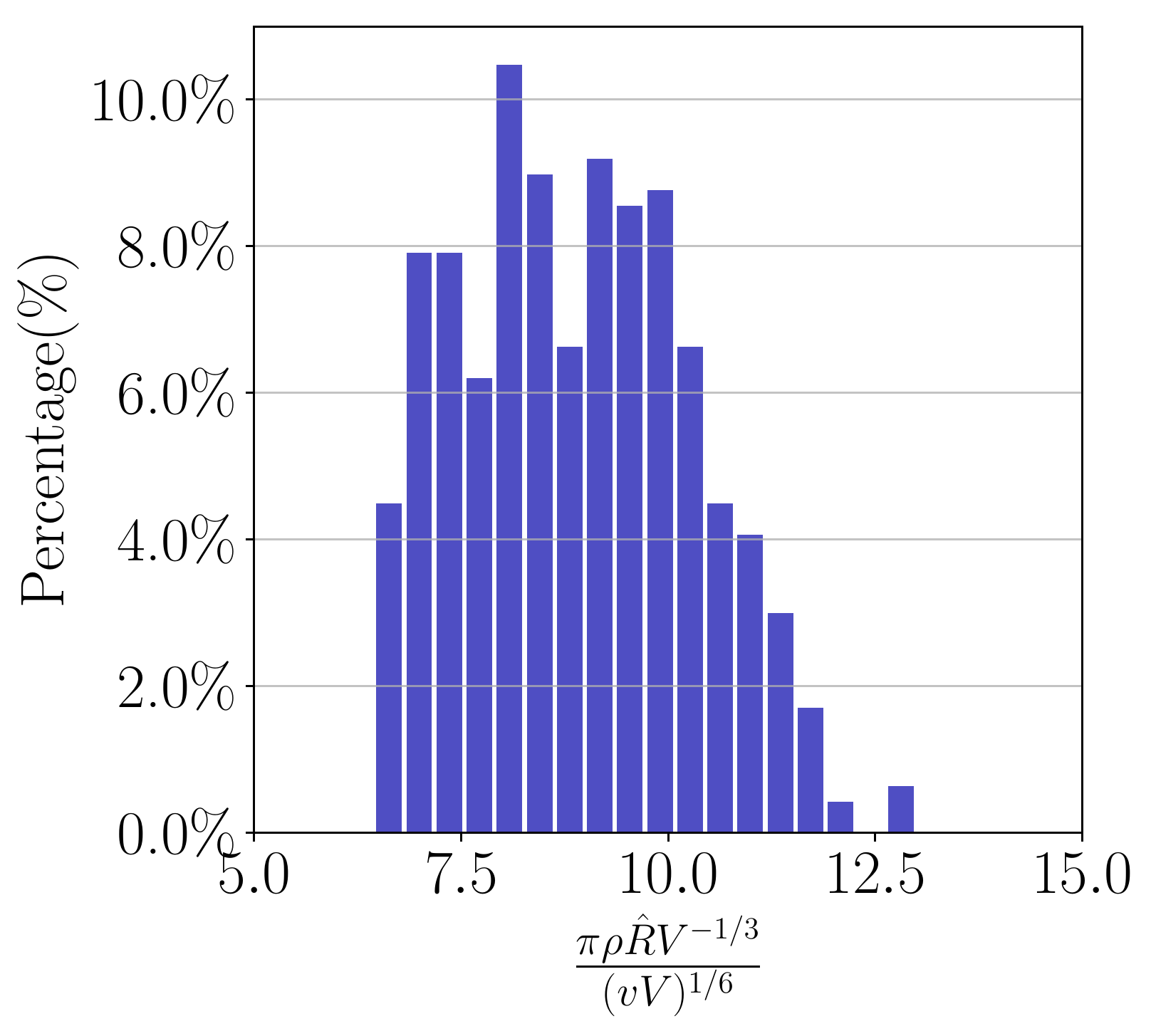}
\caption{Split Wilson lines: $\langle  \alpha_{u}\rangle =\frac{1}{20.08}$.}
\end{subfigure}
\begin{subfigure}[c]{0.49\textwidth}
\includegraphics[width=1.0\textwidth]{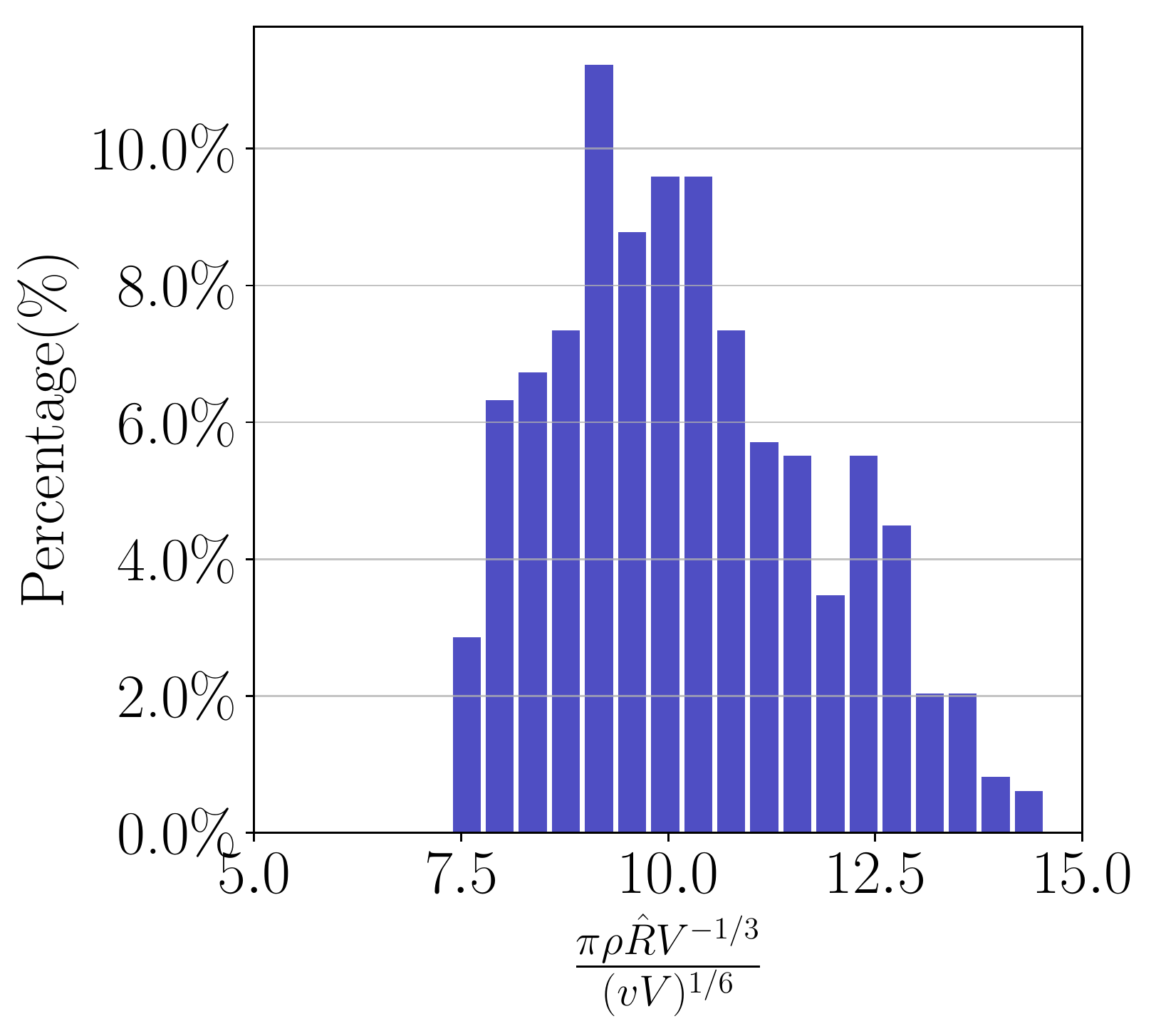}
\caption{Simultaneous Wilson lines: $\langle  \alpha_{u}\rangle =\frac{1}{26.46}$.}
\end{subfigure}
\caption{Plots of the percentage of occurrence versus the ratio of the orbifold interval length to the average Calabi--Yau radius; for the split Wilson line scenario in (a) and for the simultaneous Wilson line scenario in (b). The results shown in (a) and (b) represent a scan over the magenta region of K\"ahler moduli space displayed in Figure \ref{fig:Intersection}, where all vacuum, reduction and physical constraints are satisfied and the line bundle  $L=\Ocal_X(2, 1, 3)$ is slope polystable.}
\label{fig:2Dplots}
\end{figure}

As we did previously for the orange regions presented in Figure \ref{fig:PhysContraint} (a) and (b), it is of interest to scan over the magenta subspace of Figure \ref{fig:Intersection} to find the numerical range of the ratio $\frac{\pi \rho {\Rhat} V^{-1/3}}{(vV)^{1/6}}$. We find that
\begin{equation}\
6.4 \lesssim \frac{\pi \rho {\Rhat} V^{-1/3}}{(vV)^{1/6}} \lesssim 12.9
\label{er1A}
\end{equation}
for the split Wilson line scenario and 
\begin{equation}\
7.3 \lesssim \frac{\pi \rho {\Rhat} V^{-1/3}}{(vV)^{1/6}} \lesssim 14.7
\label{er2A}
\end{equation}
for simultaneous Wilson lines. In fact, one can go further and present a histogram of the percentage versus the ratio $\frac{\pi \rho {\Rhat} V^{-1/3}}{(vV)^{1/6}}$ for each scenario. These histograms are shown in Figure \ref{fig:2Dplots} (a) and (b) for the split Wilson line and simultaneous Wilson line scenarios respectively.

Before proceeding to the discussion of $N=1$ supersymmetry in the $D=4$ effective theory, it will be useful to present the formalism for computing the low-energy matter spectrum associated with a given hidden sector line bundle. We do this in the next subsection, displaying the formalism and low energy spectrum within the context of the line bundle $L=\Ocal_X(2, 1, 3)$ for specificity.

\subsection{The Hidden Sector Matter Spectrum of the \texorpdfstring{$D=4$}{D = 4} Effective Theory} \label{sec:hidden_matter_cont}

Having found the explicit sub-region of K\"ahler moduli space that satisfies all required constraints for the $L=\Ocal_X(2, 1, 3)$ line bundle, in this subsection, we will discuss the computation of the vector and chiral matter content of the $D=4$ low-energy theory of the hidden sector. Generically,
the low-energy matter content depends on the precise hidden sector line bundle under consideration, as well as its embedding into $E_{8}$. In this subsection, we again choose the line bundle to be $L=\mathcal{O}_{X}(2,1,3)$,  embedded  into $E_{8}$ as in \eqref{red5} with $a=1$. However, the formalism presented applies to any line bundle with any embedding into $E_{8}$. The commutant of the $U(1)$ structure group of our specific embedding is $\Uni1 \times E_{7}$. As discussed in Section \ref{sec:embedding}, the $\repd{248}$ decomposes under $U(1) \times\Ex 7$ as
\begin{equation}
\repd{248} \to 
(0, \repd{133}) \oplus 
\bigl( (1, \repd{56}) \oplus (-1, \repd{56})\bigr) \oplus 
\bigl( (2, \repd{1}) \oplus (0, \repd{1}) \oplus (-2, \repd{1}) \bigr)\ .
\label{red55}
\end{equation}
The $(0,\repd{133})$ corresponds to the adjoint representation of $\Ex 7$, while the $(\pm1,\repd{56})$ give rise to chiral matter superfields with  $\pm1$ $U(1)$ charges transforming in the $\underline{\bf56}$ representation of $\Ex 7$ in four dimensions. 
The $(\pm2,\repd 1)$ are $E_7$ singlet chiral superfields fields with charges $\pm 2$ under $U(1)$. Finally, the $(0,\repd{1})$ gives the one-dimensional adjoint representation of the $\Uni 1$ gauge group. The embedding of the line bundle is such that fields with $U(1)$ charge $-1$ are counted by $H^{*}(X,L)$, charge $-2$ fields are counted by $H^{*}(X,L^{2})$ and so on.\footnote{This is due to the form of the gauge transformation of the matter fields. This was chosen so as to agree with \cite{Wess:1992cp,Anderson:2009nt}.}

The low-energy massless spectrum can be determined by examining the chiral fermionic zero-modes of the Dirac operators for the various representations in the decomposition of the $\repd{248}$. Generically, the Euler characteristic $\chi(\mathcal{F})$ counts $n_{\text{R}}-n_{\text{L}}$, where $n_{R}$ and $n_{L}$ are the number of right- and left-chiral zero-modes respectively transforming under the representation associated with the bundle $\mathcal{F}$. With the notable exception of $\mathcal{F}=\mathcal{O}_{X}$, which is discussed below, paired right-chiral and left-chiral zero-modes are assumed to form a massive Dirac fermion and are integrated out of the low-energy theory. Therefore, it is precisely the difference of the number of right-chiral fermions minus the left-chiral fermions, counted by the Euler characteristic $\chi$, that give the massless zero-modes of the $D=4$ theory. On a Calabi–Yau threefold $X$, $\chi(\mathcal{F})$ can be computed by the Atiyah–Singer index theorem as
\begin{equation}
\chi(\mathcal{F})=\sum_{i=0}^{3}(-1)^{i}h^{i}(X,\mathcal{F})=\int_{X}\op{ch}(\mathcal{F})\wedge\op{Td}(X)\ ,
\end{equation}
where $h^{i}$ are the dimensions of the $i$-th cohomology group, $\op{ch}(\mathcal{F})$ is the Chern character of $\mathcal{F}$, and $\op{Td}(X)$ is the Todd class of the tangent bundle of $X$. When $\mathcal{F}=L=\mathcal{O}_{X}(l^{1},l^{2},l^{3})$ is a line bundle, this simplifies to
\begin{equation}\label{eq:chi}
\chi(L)=\tfrac{1}{3}(l^{1}+l^{2})+\tfrac{1}{6}d_{ijk}l^{i}l^{j}l^{k}\ .
\end{equation}
Unlike the case of an $\SU N$ bundle, when $L$ is a line bundle with non-vanishing first Chern class, $\chi$ can receive contributions from all \emph{four} $h^i$, $i=0,1,2,3$. For example, $h^1(X,L)+h^3(X,L)$ then counts the number of (left-handed) chiral multiplets while $h^0(X,L)+h^2(X,L)$ counts (right-handed) anti-chiral multiplets, both transforming in the $(-1,\repd{56})$ representation. Note that the multiplets counted by $h^0(X,L)+h^2(X,L)$ are simply the CPT conjugate partners of those already counted by $h^1(X,L^{-1})+h^3(X,L^{-1})$. Since it is conventional to give a supersymmetric matter spectrum in terms of (left-handed) chiral supermultiplets, it is sufficient to compute $h^1+h^3$ for the various bundles under consideration.

Using \eqref{eq:chi}, it is straightforward to compute the value of $\chi$ for the powers of $L$ associated with the decomposition \eqref{red55}. These are presented in Table \ref{tab:chiral_spectrum}. Having done this, let us discuss the spectrum in more detail.\footnote{See \cite{Braun:2005ux}, for example, for a similar discussion of the hidden-sector spectrum for an $\SU 2$ bundle.} 
\begin{table}
	\noindent \begin{centering}
		\begin{tabular}{rrr}
			\toprule 
			$U(1) \times \Ex 7$ & Cohomology & Index $\chi$\tabularnewline
			\midrule
			\midrule 
			$(0,\repd{133})$ & $H^{*}(X,\mathcal{O}_{X})$ & $0$\tabularnewline
			\midrule 
			$(0,\repd 1)$ & $H^{*}(X,\mathcal{O}_{X})$ & $0$\tabularnewline
			\midrule 
			$(-1,\repd{56})$ & $H^{*}(X,L)$ & $8$\tabularnewline
			\midrule 
			$(1,\repd{56})$ & $H^{*}(X,L^{-1})$ & $-8$\tabularnewline
			\midrule 
			$(-2,\repd 1)$ & $H^{*}(X,L^{2})$ & $58$\tabularnewline
			\midrule 
			$(2,\repd 1)$ & $H^{*}(X,L^{-2})$ & $-58$\tabularnewline
			\bottomrule
		\end{tabular}
		\par\end{centering}
	\caption{The chiral spectrum for the hidden sector $\protect\Uni 1\times\protect\Ex 7$ with a single line bundle $L=\mathcal{O}_{X}(2,1,3)$. The Euler characteristic (or index) $\chi$ gives the difference between the number of right- and left-chiral fermionic zero-modes transforming in the given representation. We denote the line bundle dual to $L$ by $L^{-1}$ and the trivial bundle $L^{0}$ by $\mathcal{O}_{X}$.\label{tab:chiral_spectrum}}
	
\end{table}
\begin{itemize}
	\item The index of the bundle $\mathcal{O}_X$ associated with the $(0,\repd{133})$ and $(0,\repd{1})$ representations vanishes, so the corresponding fermionic zero-modes must be \emph{non-chiral}. As discussed in \cite{Braun:2013wr}, since the trivial bundle $\mathcal{O}_{X}$ has $h^{0}(X,\mathcal{O}_{X})=h^{3}(X,\mathcal{O}_{X})=1$ and zero otherwise, there is a single right-chiral fermionic zero-mode (counted by $h^{0}$) and a single left-chiral fermionic zero-mode (counted by $h^{3}$), which combine to give the conjugate gauginos in a massless vector supermultiplet. In other words, the low-energy theory has one vector supermultiplet transforming in the $(0,\repd{133})$ adjoint representation of $E_{7}$ and one vector supermultiplet in the $(0,\repd{1})$ adjoint representation of $\Uni 1$. 
	
	\item The $(1, \repd{56})$ multiplets are counted by $H^{*}(X,L^{-1})$. Since $\chi(L^{-1})=-8$, there are 8 unpaired left-chiral fermionic zero-modes that contribute to 8 chiral matter supermultiplets transforming in the $(1,\repd{56})$ of $U(1) \times E_{7}$.
        \item Similarly, the $(-1, \repd{56})$ multiplets are counted by $H^{*}(X,L)$. Since $\chi(L)=8$, there are 8 unpaired right-chiral fermionic zero-modes that contribute to 8 anti-chiral matter supermultiplets transforming in the $(-1,\repd{56})$ of $U(1) \times E_{7}$.  However, these do not give extra fields in the spectrum: they are (right-handed) anti-chiral $(-1, \repd{56})$ supermultiplets which are simply the CPT conjugate partners of the 58 chiral $(1, \repd{56})$ supermultiplets already counted above~\cite{Green:1987mn}.
	\item Since $\chi(L^{-2})=-58$, there are 58 unpaired left-chiral fermionic zero-modes that contribute to 58 chiral matter supermultiplets transforming in the $(2,\repd{1})$ representation of $U(1) \times E_{7}$.

	\item Similarly, the  $(-2,\repd{1})$ multiplets are counted by $H^{*}(X,L^{2})$. Since $\chi(L^{2})=58$, there are 58 unpaired right-chiral fermionic zero-modes that contribute to 58 charged anti-chiral matter supermultiplets transforming in the $(2,\repd{1})$ representation of $U(1) \times E_{7}$. However, as discussed above, these do not give extra fields in the spectrum: they are (right-handed) anti-chiral $(-2,\repd{1})$ supermultiplets which are simply the CPT conjugate partners of the 58 chiral $(2,\repd{1})$ supermultiplets already counted above. 
	
\end{itemize}
In summary, the $\Uni 1\times\Ex 7$ hidden sector massless spectrum for $L=\mathcal{O}_{X}(2,1,3)$ is
\begin{equation}
1\times(0,\repd{133})+1\times(0,\repd{1})+8\times(1,\repd{56})+58\times(2,\repd{1})\ ,\label{eq:matter}
\end{equation}
corresponding to one vector supermultiplet transforming in the adjoint representation of $\Ex7$, one $\Uni1$ adjoint representation vector supermultiplet, eight chiral supermultiplets transforming as $(1,\repd{56})$ and 58 chiral supermultiplets transforming as $(2,\repd{1})$. 

Note that since we have a chiral spectrum charged under $\Uni1$ with all positive charges, the $\Uni1$ gauge symmetry will be anomalous. As we discuss in the next subsection, this anomaly is canceled by the four-dimensional version of the Green--Schwarz mechanism which, in addition, gives a non-zero mass to this ``anomalous'' hidden sector $U(1)$.

\subsection{Supersymmetric Vacuum Solutions in Four Dimensions}\label{sec:susy_vacua_4d}

In this subsection, the generic form of the $U(1)$ D-term potential in the four-dimensional effective theory for an arbitrary hidden sector line bundle $L=\Ocal_X(l^{1}, l^{2}, l^{3})$ will be presented.
Using this result, we will then discuss the conditions for unbroken $N=1$ supersymmetry in the four-dimensional theory.
In the next subsection, we will be more specific, focusing on the example of $L=\mathcal{O}_X(2,1,3)$ with $a=1$ discussed above.

$N=1$ supersymmetry will be preserved in the $D=4$ effective theory only if the $D_{\Uni 1}$ term presented in \eqref{rain4A} vanishes at the minimum of the potential energy; that is
\begin{equation}
\langle D_{\Uni 1} \rangle=0 \ .
\end{equation}
Whether or not the $D=4$ effective theory can satisfy this condition, and the exact details as to how it does so depends strongly on the value of the Fayet--Iliopoulos term. There are two generic possibilities.
\begin{enumerate}
\item[](i) The genus-one corrected FI-term vanishes. In this case, the VEVs of the scalar fields either all vanish or the VEVs of those fields with opposite charges, should they exist, cancel against each other.
\item[](ii) The genus-one corrected FI-term is non-vanishing. In this case, non-zero VEVs of the scalar fields $C^L$ with the same sign as FI turn on to cancel the non-vanishing FI-term.
\end{enumerate}
Each of the two scenarios comes with its conditions which have to be met. In the first case, to obtain a vanishing FI-term, the strong coupling $\kappa_{11}^{2/3}$ corrections to the slope need to cancel the tree-level ``classical'' slope in \eqref{again2}. For that to happen, one needs to be in a very strongly coupled regime, in which working only to order $\kappa_{11}^{2/3}$ may be a poor approximation. In the second case, the low-energy spectrum needs to contain scalars $C^L$ with the correct charge $Q^L$ under $U(1)$, such that their VEVs can cancel the non-zero $\text{FI}$ contribution. In such a scenario, one can move in K\"ahler moduli space--while still satisfying all the vacuum and phenomenological constraints (that is, outside the magenta region of Figure \ref{fig:Intersection} while remaining within the brown region)--to a less strongly coupled regime in which the first-order expansion to $\kappa_{11}^{2/3}$ is more accurate. However, we will show that the VEVs of these scalar fields will deform the hidden sector line bundle to an $SU(2)$ bundle, which might not be slope-stable. 

\subsubsection{Vanishing FI}\label{sec:vanishing_FI}

Let us start by analyzing the first case. A simple way to ensure unbroken $N=1$ supersymmetry in $D=4$ and slope stability of the hidden sector bundle is to require that
\begin{equation}
\text{FI}=0 \ .
\label{lamp1}
\end{equation}
There are then two scenarios in which supersymmetry can remain unbroken in the low energy theory. These are the following:
\begin{enumerate}
	\item The first, and simplest possibility is that the charges $Q^{L}$ of the scalar fields $C^L$ are all of the same sign. It follows that the potential energy will set all VEVs to zero, $\langle C^{L} \rangle=0$, and hence $D_{\Uni 1}$ will vanish at this {\it stable} vacuum. Thus $N=1$ supersymmetry will be unbroken.
	\item A second possibility is that some of the $Q^{L}$ signs may differ. This will lead to unstable, flat directions where both $D_{\Uni 1}$ and the potential energy vanish. If one is at an arbitrary point, away from the origin, in a flat direction, then at least two VEVs will be non-vanishing $\langle C^{L} \rangle \neq 0$ and, hence, although preserving $N=1$ supersymmetry, such a vacuum would also spontaneously break the $U(1)$ symmetry. 
\end{enumerate}

Having discussed this second possibility, we note again that the associated potential energy must have at least one flat direction and, hence, is an {\it unstable} vacuum state. For this reason, we will ignore such vacua. However, the first scenario is easily satisfied, as we now demonstrate with an explicit example.

\subsubsection*{$\text{FI}=0$ Example: $N=1$ Supersymmetry for $L=\mathcal{O}_X(2,1,3)$}

We now discuss $N=1$ supersymmetry in the example introduced above, where the line bundle is taken to be $L=\mathcal{O}_X(2,1,3)$ and embedded into $SU(2) \subset E_{8}$ as in \eqref{red5} with coefficient $a=1$, and the location of the single five-brane is at $\lambda=0.49$.
In this case, for the bundle $L \oplus L^{-1}$ to be polystable, it was necessary to restrict this region of moduli space to points that set the genus-one corrected slope \eqref{trenton3} -- and, hence, the FI-term--to zero. Furthermore, the low-energy scalar spectrum carrying non-vanishing $U(1)$ charge was shown in Table \ref{tab:chiral_spectrum}. It was shown there that the low-energy scalar spectrum of the hidden sector -- specifically $8 \times (1,\repd{56})+58 \times (2,\repd{1})$ -- each had charges $Q^{L}$ of the same sign. It then follows from the above discussion that the potential energy must have a unique minimum where the VEVs vanish, $\langle C^{L} \rangle=0$, such that $\langle D_{\Uni 1} \rangle=0$ at this minimum. Hence, $N=1$ supersymmetry is {\it unbroken} in the vacuum state of the $D=4$ effective theory. As discussed above, since the $U(1)$ symmetry is anomalous, the mass $m_{A}$ presented in \eqref{eq:anomalous_massA} is non-vanishing.

We would like to point out that $L=\mathcal{O}_X(2,1,3)$ is not the only hidden sector line bundle which, if embedded into $SU(2) \subset E_{8}$ as in \eqref{red5} with $a=1$, has a region of K\"ahler moduli space where all required constraints are satisfied, $\text{FI}=0$ and the $D=4$ vacuum preserves $N=1$ supersymmetry. However, any such line bundle $L$ must be ``ample''--that is, each of its defining integers $l^{i}, i=1,2,3$ where $l^{1}+l^{2}=0~ mod ~3$ must either be all positive or all negative. The reason is that for the Schoen manifold defined in Section \ref{sec:CY_sheon} , one can show that the genus-one corrected Fayet-Iliopoulos term can vanish, that is, $\text{FI}=0$, if and only if $L$ is ample. Restricting to ample line bundles, one can indeed find a significant number satisfying all required constraints. However, of these, many have a large number of equal sign zero-mode chiral multiplets -- some with large charges $Q^{L}$--making them incompatible with spontaneous supersymmetry breaking via gaugino condensation. While potentially of physical interest, we wish to focus on the subset of ample line bundles that have a sufficiently small zero-mode chiral spectrum, with sufficiently small charges, to be compatible with supersymmetry breaking via $E_{7}$ gaugino condensation. These line bundles are specified by
	\begin{gather}\label{many}
		\mathcal{O}_X(2,1,3)\ , \qquad \mathcal{O}_X(1,2,3)\ ,\qquad \mathcal{O}_X(1,2,2)\ ,  \qquad\mathcal{O}_X(2,1,2)\ , \nonumber \\
		\mathcal{O}_X(2,1,1)\ , \qquad\mathcal{O}_X(1,2,1)\ ,  \qquad\mathcal{O}_X(2,1,0) \ ,
	\end{gather}
and their duals; that is, for example, $\mathcal{O}_X(-2,-1,-3)$. Spontaneous supersymmetry breaking via $E_{7}$ gaugino condensation in this context will be explored in Section \ref{sec:low_energy_gaugino}.

As discussed at the beginning of this section, although this hidden sector vacuum satisfies all required physical and phenomenological constraints, setting the FI-term to zero necessitates exact the cancellation of the genus-one corrected slope against the tree-level slope of the hidden sector line bundle. Unsurprisingly, this fine-tuning can only be carried out in a relatively strongly coupled regime of heterotic M-theory -- thus making the validity of the linearized approximation used in our analysis uncertain. It is, therefore, of some interest to explore vacua for which the genus-one corrections to the slope are significantly smaller than the tree-level slope of the hidden sector bundle. In this case, one expects the effective coupling parameter to be smaller than in the previous scenario and, hence, the linearized results to be a better approximation. For this reason, we now consider hidden sector vacua where the FI-term does not vanish.

\subsubsection{Non-vanishing FI}

We now consider what happens when the $\kappa_{11}^{2/3}$ correction to the tree-level slope is small and so cannot be used to set the FI-term to zero. The question then is, given a non-vanishing FI-term
\begin{equation}
	\text{FI}\neq0\ ,
\end{equation}
can one still preserve $N=1$ supersymmetry in the four-dimensional effective theory? 

Recall that the conditions for a supersymmetric vacuum are the vanishing of the F- and D-terms. If we insist that the vacuum has an unbroken $\Ex 7$ gauge symmetry, the $\Ex 7$ D-terms will vanish by setting the VEVs of the non-abelian $(\pm1,\repd{56})$ matter fields to be zero. Any F-terms involving the non-abelian matter fields will then also vanish. As discussed in \cite{Anderson:2010ty}, the F-term conditions for the $(\pm2,\repd 1)$ matter fields permit us to give VEVs to only one set of fields, that is, either $(2,\repd 1)$ or $(-2,\repd 1)$ but not both. The remaining condition for the vacuum solution to be supersymmetric is the vanishing of the $\Uni 1$ D-term, $\langle D_{\Uni 1}\rangle=0$. Since the FI term does \emph{not} vanish for any choice of line bundle when the $\kappa_{11}^{2/3}$ correction is small, one is forced to cancel the FI term against the VEVs of the charged singlet fields. In other words, we want
\begin{equation}\label{eq:vevs}
	Q^{L}\langle C^{L}\rangle G_{LM}\langle\bar{C}^{M}\rangle=\text{FI}\quad\Rightarrow\quad\langle D_{\Uni 1}\rangle=0\ .
\end{equation}
Obviously, such cancellation will depend on the relative sign of FI and the charges of the scalars $C^{L}$. For example, if the FI term is \emph{positive}, one needs at least one zero-mode chiral supermultiplet whose scalar component is a singlet under the non-abelian group and has \emph{positive} $\Uni 1$ charge. Whether or not such scalar fields are present will depend on the specific line bundle studied.

The embedding of $\Uni 1$ inside $\Ex 8$ that we have considered for much of this analysis factors through the $\SU 2$ subgroup of $\Ex 8$ that commutes with $\Ex 7$. The $\Uni 1$ gauge connection $A$ for the line bundle $L$ can be thought of as defining an $\Ex 8$ connection by first embedding in $\SU 2$ as $\text{diag}(A,-A)$ and then embedding $\SU 2$ in $\Ex 8$. 

First note that the connection $\text{diag}(A,-A)$ is a connection for an $\SU 2$ bundle
which splits as a direct sum 
\begin{equation}
	\mathcal{V}=L\oplus L^{-1}
\end{equation}
of line bundles. How does this relate to supersymmetry? The induced $\Ex 8$ connection will solve the (genus-one corrected) Hermitian Yang–Mills equation, and so give a supersymmetric solution, if the $\SU 2$ connection itself solves the (genus-one corrected) Hermitian Yang–Mills equation. This is guaranteed if the rank two $L\oplus L^{-1}$ bundle is polystable with a vanishing slope.\footnote{Here, the slope is taken to mean the genus-one corrected slope. The same comments apply if one considers only the tree-level expression.} Since $\mu(\mathcal{V})=0$ by construction, the remaining conditions for polystability are
\begin{equation}
	\mu(L)=\mu(L^{-1})=0\ .
\end{equation}
This is exactly the vanishing $FI$ case studied in \ref{sec:vanishing_FI}, where the corrected slope of $L$ is set to zero and the VEVs of the charged singlet matter fields vanish.

When $\mu(L)\neq0$, the $\SU 2$ bundle $\mathcal{V}$ is no longer polystable and so its connection does not solve the Hermitian Yang–Mills equation. The four-dimensional consequence of this is that the FI term no longer vanishes. However, we might be able to turn on VEVs for appropriate charged singlet matter fields to cancel the FI term and set the D-term to zero, thus preserving supersymmetry. One might wonder: what is the bundle interpretation of turning on VEVs for these charged singlet matter fields? As discussed in \cite{0905.1748,1012.3179,1506.00879}, these VEVs should be seen as deforming the gauge bundle away from its split form $\mathcal{V}=L\oplus L^{-1}$ to a \emph{non-split} $\SU 2$ bundle $\mathcal{V}'$ which admits a connection that \emph{does} solve the Hermitian Yang--Mills equations.

Consider the case where $\mu(L)>0$ (equivalent to $\text{FI}>0$) in some region of K\"ahler moduli space where the constraints of Section \ref{sec:constraints} are all satisfied. From \eqref{eq:vevs} we see that one can set $\langle D_{\Uni 1}\rangle=0$ provided we have charged scalars $C^{L}$ with positive charge, $Q^{L}>0$. From the cohomologies in Table \ref{tab:chiral_spectrum}, the required scalars are those transforming in $(2,\repd 1)$, with the chiral superfields which contain these scalars counted by $h^{1}(X,L^{-2})+h^{3}(X,L^{-2})$. Hence, giving VEVs to $(2,\repd 1)$ scalars corresponds to allowing non-trivial elements of $H^{1}(X,L^{-2})\oplus H^{3}(X,L^{-2})$. The first summand has an interpretation as the space of extensions of $L^{-1}$ by $L$, with the exact sequence
\begin{equation}
	0\to L^{-1}\to\mathcal{V}'\to L\to0
\end{equation}
defining an $\SU 2$ bundle $\mathcal{V}'$. This extension can be non-trivial ($\mathcal{V}\neq\mathcal{V}'$) provided
\begin{equation}
	\op{Ext}^{1}(L^{-1},L) = H^{1}(X,L^{-2})\neq0\ .
\end{equation}
Choosing a non-zero element of this space then corresponds to turning on VEVs for some set of $(2,\repd 1)$ scalars. Thus we see that giving VEVs to positively charged singlet scalars arising from $H^1(X,L^{-2})$ amounts to deforming the induced $L \oplus L^{-1}$ bundle $\mathcal{V}$ to the $\SU 2$ bundle $\mathcal{V}'$. Note that if $H^1(X,L^{-2})=0$, the VEVs of positively charged matter coming from $H^3(X,L^{-2})$ cannot be interpreted as deforming to a new $SU(2)$ bundle. Hence, the bundle remains  $L \oplus L^{-1}$ which is unstable and, therefore, its gauge connection does not solve the Hermitian Yang--Mills equation.

Assuming one can show for a given line bundle $L$ that $H^1(X,L^{-2}) \neq 0$, it might seem that we are done – the $\Uni 1$ D-term vanishes, and supersymmetry appears to have been restored. However, the four-dimensional analysis is insensitive to whether the new bundle $\mathcal{V}'$ is slope stable and thus actually admits a solution to Hermitian Yang–Mills. Unfortunately, checking slope stability is a difficult calculation that one must do explicitly for each example. As a preliminary check, one can first see whether $\mathcal{V}'$ satisfies some simpler \emph{necessary} conditions for slope stability. First, the obvious subbundle $L^{-1}$ should not destabilize $\mathcal{V}'$. In our case, this is guaranteed as we have assumed $\mu(L)>0$, so that $L^{-1}$ has a negative slope.\footnote{If instead $\mu(L) <0$, one simply swaps the roes of $L$ and $L^{-1}$ in the above discussion and instead considers the extension of $L^{-1}$ by $L$.} Second, $\mathcal{V}'$ must satisfy the Bogomolov inequality~\cite{huybrechts2010geometry}. For a bundle with vanishing first Chern class, this states that if $\mathcal{V}'$ is slope stable with respect to some choice of K\"ahler class $\omega=a^{i}\omega_{i}$, then
\begin{equation}
	\int_{X}c_{2}(\mathcal{V}')\wedge\omega\geq0\ .
\end{equation}
Since $\mathcal{V}'$ is constructed as an extension of line bundles, we have
\begin{equation}
	c_{2}(\mathcal{V}')\equiv c_{2}(L\oplus L^{-1})=-\tfrac{1}{2}c_{1}(L)\wedge c_{1}(L)\ ,
\end{equation}
with $c_{1}(L)={v^{-1/3}}l^{i}\omega_{i}$. Thus if $\mathcal{V}'$ is to be slope stable, we must be in a region of K\"ahler moduli space where
\begin{equation}\label{eq:Bolg}
	-\tfrac{1}{2}\int_{X}c_{1}(L)\wedge c_{1}(L)\wedge\omega\geq0\quad\Rightarrow\quad d_{ijk}l^{i}l^{j}a^{k}\leq0\ .
\end{equation}
Note that this is a necessary but not sufficient condition. However, it is often the case that the Bogomolov inequality is the only obstruction to finding stable bundles~\cite{Braun:2005zv}.

We thus have a new set of necessary conditions on $L$ (in addition to the physically and mathematically required constraints presented in Section \ref{sec:constraints}) for there to be a supersymmetric vacuum after turning on the VEVs to cancel the FI-term. These are
\begin{enumerate}
	\item Singlet matter with the correct charge must be present so that FI can be canceled and the D-term set to zero.
	\item $H^1(X,L^{-2})$ must not vanish.
	\item The Bogomolov inequality, $d_{ijk}l^{i}l^{j}a^{k}\leq0$, must be satisfied.
\end{enumerate}

Does our previous choice of $L=\mathcal{O}_{X}(2,1,3)$ satisfy these conditions? Note that $\mu(L)>0$ everywhere in the K\"ahler cone for this line bundle. From the low-energy spectrum in Table \ref{tab:chiral_spectrum}, we see we have 58 massless positively charged singlets transforming in the $(2,\repd 1)$ representation, and so we do indeed have the correct matter to cancel the FI-term and set the D-term to zero. However, as discussed in \cite{Braun:2013wr}, if $L$ is an {\it ample} line bundle then
\begin{equation}
 H^1(X,L)=0 \ .
 \label{snow1}
 \end{equation}
 Since $L=\mathcal{O}_{X}(2,1,3)$--and, hence, $L^{-2}$--is ample, it follows that $H^1(X,L^{-2})=0$. Therefore, condition 2 above implies that $L \oplus L^{-1}$ cannot admit an extension to an $SU(2)$ bundle $\mathcal{V}'$.
Ignoring this for a moment, and assuming that there did exist an $SU(2)$ extension, we would still have to check whether or not the Bogomolov inequality, a necessary condition for $\mathcal{V}'$ to be slope stable, is satisfied.
However, from \eqref{eq:Bolg} and the positivity of the K\"ahler moduli, we see that it is {\it impossible} to satisfy this inequality, implying (again) that the split bundle cannot be deformed to admit a solution to the Hermitian Yang–Mills equation. Moreover, we see the same will be true for any ample line bundle -- the $l^{i}$ are positive and $d_{ijk}l^{i}l^{j}a^{k}\leq0$ is not satisfied anywhere in the K\"ahler cone.

What about other choices of line bundle? It turns out that of the three conditions, the Bogomolov inequality is the more difficult to satisfy. Scanning over different choices of $L$, one finds that in the region of K\"ahler moduli space where the $\SU 4$ bundle is stable, the only line bundles that are equivariant with $\mu(L)>0$,\footnote{We restrict to $\mu(L)>0$ in our scan to match our analysis above. Including bundles with $\mu(L)<0$ would give the reverse extension sequence with the bundle and its dual swapped, leading to the same $\SU 2$ bundles that were already captured by restricting to a positive slope.} allow for anomaly cancellation and satisfy the Bogomolov inequality are
\begin{equation}
	\mathcal{O}_{X}(1,2,-1)\ ,\qquad\mathcal{O}_{X}(2,1,-1)\ ,\qquad\mathcal{O}_{X}(7,2,-2)\ ,\qquad\mathcal{O}_{X}(7,5,-3)\ .
\end{equation}
Do any of these have positively charged singlet matter in their low-energy spectrum to allow for a non-trivial extension? That is, do we have $H^{1}(X,L^{-2})>0$ for any of these candidate line bundles? For $\mathcal{O}_{X}(1,2,-1)$ and $\mathcal{O}_{X}(2,1,-1)$, it is simple to show using a Leray spectral sequence that the answer is no. For a definitive answer in the remaining two cases, one must extend the analysis of Appendix A of \cite{Braun:2005zv} to higher degree line bundles on dP$_9$. This is beyond the scope of the present work. Therefore, for now, we content ourselves with noting that $\chi(L^{-2})$ is positive for both remaining line bundles, which is consistent with $H^{1}(X,L^{-2})=0$ and the absence of an extension to a $SU(2)$ bundle $\mathcal{V}'$.

As exploited by a number of other works~\cite{Anderson:2010ty,Anderson:2012yf,Nibbelink:2015ixa}, moving from a single line bundle to two or more such bundles provides a richer low-energy spectrum, making it much easier to find examples that possess the correct charged matter and satisfy both the phenomenological constraints and the Bogomolov inequality. We analyze this case in the next section.

So far, we have explicitly chosen the hidden sector line bundle $\mathcal{O}_X(2,1,3)$, embedded in a specific way with embedding coefficient $a=1$ into the $E_{8}$ gauge group, and studied its phenomenological properties. This choice of the hidden sector was shown to satisfy all ``vacuum'' constraints required to be consistent with present low-energy phenomenology, as well as both the ``reduction'' and ``physical'' constraints required to be a ``strongly-coupled'' heterotic vacuum consistent with both the mass scale and gauge coupling of a unified $SO(10)$ theory in the observable sector. Additionally, we showed that the induced $\SU2$ bundle $L \oplus L^{-1}$ is polystable after including genus-one corrections, and that the effective low-energy theory admits an $N=1$ supersymmetric vacuum. We pointed out that there are actually a large number of different line bundles that one could choose, and a large number of inequivalent embeddings of such line bundles into $E_{8}$. An alternative choice of hidden sector bundle could lead to: 1) a different commutant subgroup $H$ and hence a different low-energy gauge group, 2) a different spectrum of zero-mass particles transforming under $H \times U(1)$, 3) a different value for the associated Fayet--Iliopoulos term and, hence, a different D-term mass for the $U(1)$ vector superfield, and so on. We will explore an alternative type of hidden sector in the next section.

\section{Hidden Sectors with two Line Bundle Embedding}\label{2 line bundles embedding}

In Section 2.3 we showed that vacuum configurations with hidden sectors built from a single line bundle require that the genus-one corrected slope of the line bundle vanishes. It seems that such a configuration always pushes us into a strongly coupled regime in which the accuracy of the linear approximation used to derive the effective four-dimensional theory is uncertain. Here, we begin with a general overview of constructing Abelian line bundle backgrounds, before specializing our discussion on hidden sectors built from {\it two} line bundles embedded into the hidden $E_8$ gauge group. As an example of how this works, we will then focus on embeddings which lead to the breaking pattern $E_8\to E_6\times U(1)\times U(1)$. 

\subsection{Line Bundle Embeddings}\label{sec:embeddings}

A particularly simple set of hidden sector bundles are those constructed from line bundles. These are defined by a set of line bundles and the embedding of their corresponding $\Uni 1$ groups into $\Ex 8$. Given a set of line bundles, there are multiple inequivalent ways to embed their Abelian gauge connections into the ten-dimensional hidden $\Ex 8$ connection. A particularly useful formalism for describing these embeddings is using ``line bundle vectors''. Following \cite{Nibbelink:2016wms}, the Abelian gauge connections can be embedded in the hidden sector $\Ex 8$ by expanding the curvature $F_{E_{8}}$ as
\begin{equation}
	\frac{F_{E_{8}}}{2\pi}=\frac{1}{v^{1/3}}\omega_{i}H_{i}\ ,\label{eq:flux-1}
\end{equation}
where the coefficients $H_{i}$ are matrices valued in the Lie algebra of $\Ex 8$. As in \cite{Ashmore:2020ocb,Ashmore:2020wwv}, the $\omega_{i}$ are the three harmonic $(1,1)$-forms that span the $H^{1,1}(X,\mathbb{C})$ cohomology on the Schoen threefold $X$. Since the background is Abelian, one can expand the coefficients as
\begin{equation}
	H_{i}=V_{i}^{I}H_{I}\ ,
\end{equation}
where $I=1,\ldots,8$ runs over the Cartan subalgebra of the hidden $\Ex 8$. Here the $H_{I}$ denote the Cartan generators of the $\SO{16}\subset\Ex 8$ maximal subgroup, normalized so that
\begin{equation}\label{gen_norm}
	\tr H_{I}H_{J}=2\,\delta_{IJ}\ ,
\end{equation}
where the trace is $1/30$ of the trace over the $\repd{248}$ of $E_8$ or, equivalently, taken in the fundamental $\repd{16}$ representation of $\SO{16}$.\footnote{This agrees with the normalisation in \cite{Nibbelink:2016wms} after noting that their trace is taken in the fundamental of an SU group rather than SO. See \cite[Appendix A]{Blaszczyk:2015zta} for more details.} The eight-component vectors $\boldsymbol{V}_{i}=V_{i}^{I}$ are known as \emph{line bundle vectors}. Given a choice of Cartan generators, an Abelian hidden sector bundle is completely specified by a choice of three line bundles vectors $\boldsymbol{V}_{i}$, $i=1,2,3$. For example, $\boldsymbol{V}_{1}$ then encodes how much $\omega_{1}$ contributes to the curvature $F_{E_8}$. This is somewhat abstract at the moment, but we will see how this works with an explicit example later.

As noted in \cite{Nibbelink:2015ixa}, the flux $F_{E_{8}}$ has to be quantized when evaluated on a string state and integrated over any curve dual to a divisor defined by a sum of the $\omega_{i}$. Since the string states are characterized by weight vectors that lie on the $\Ex 8$ root lattice $\Lambda$, the line bundle vectors must also lie on the root lattice, that is $\boldsymbol{V}_{i}\in\Lambda.$ Following the conventions of \cite{Feger:2012bs,Feger:2019tvk}, the $\Ex 8$ root lattice is given by the set of points $\Lambda\in\bR^{8}$ such that all eight coordinates are integers or half-integers (but not a mix of the two), and the coordinates sum to an even integer. This constrains the form of the line bundle vectors and ensures that the curvature of the resulting hidden sector bundle obeys flux quantization.

The second Chern character of the hidden sector bundle $\mathcal{V}^{(2)}$ constructed from the line bundles is given by
\begin{equation}
	\text{ch}_{2}(\mathcal{V}^{(2)})=\frac{1}{16\pi^{2}}\tr F_{E_{8}}\wedge F_{E_{8}}=\tfrac{1}{2}(\boldsymbol{V}_{i}\cdot\boldsymbol{V}_{j})\,\frac{1}{v^{2/3}}\omega_{i}\wedge\omega_{j}\ ,\label{eq:chern2-1}
\end{equation}
where $\boldsymbol{V}_{i}\cdot\boldsymbol{V}_{j}=V_{i}^{I}V_{j}^{I}$ is the Euclidean scalar product between the $i^{\text{th}}$ and $j^{\text{th}}$ line bundle vectors. Since $c_{1}(\mathcal{V})=0$ -- following from the fact that the generators $H_{I}$ are traceless -- the second Chern class is given by $c_{2}(\mathcal{V}^{(2)})=-\text{ch}_{2}(\mathcal{V}^{(2)})$. Wedging with $\omega_{i}$ and integrating over $X$, one finds that the second Chern numbers of $\mathcal{V}^{(2)}$ are
\begin{equation}
c_{2,i}(\mathcal{V}^{(2)})\equiv\frac{1}{v^{1/3}}\int_X c_2(\mathcal{V}^{(2)})\wedge\omega_i	 = -\tfrac{1}{2}d_{ijk}(\boldsymbol{V}_{j}\cdot\boldsymbol{V}_{k})\ .
\end{equation}

The unbroken gauge group $G$ in four dimensions is given by the commutant of the structure group of $\mathcal{V}^{(2)}$ with the ten-dimensional gauge group. The \emph{non-Abelian} part of $G$ can be computed by finding all roots\footnote{The 240 roots are given by vectors $\boldsymbol{r}$ that lie on the root lattice with length squared equal to 2; $\{\boldsymbol{r}\in\Lambda\;|\;\boldsymbol{r}\cdot \boldsymbol{r}=2\}$.} $\boldsymbol{r}$ of $\ex 8$ that are orthogonal to all of the line bundle vectors:
\begin{equation}
	H_{i}(\boldsymbol{r})\equiv\boldsymbol{V}_{i}\cdot \boldsymbol{r}=0\ ,\quad\text{for all }i=1,\ldots,h^{1,1}\ .\label{eq:unbroken_condition}
\end{equation}
This ensures that the components of the $\Ex 8$ connection that form the four-dimensional connection are uncharged with respect to $G$. Since we are considering bundles with an Abelian structure group, there may also be $\Uni 1$ factors in $G$ (since they commute with themselves).

One can also calculate the chiral part of the matter spectrum in the resulting four-dimensional theory by computing the Euler characteristic $\chi$ for the bundles in which the various matter fields transform. This can be done using the multiplicity operator $\mathcal{N}$~\cite{Nibbelink:2007rd}. More precisely, the Euler characteristic $\chi(V)$ for a bundle $V$ whose sections transform in the representation $\repd R$ is calculated by $\mathcal{N}(\boldsymbol{r})$, where $\boldsymbol{r}$ is the $\ex 8$ root corresponding to the highest weight of $\repd R$. That is, given a decomposition of the $\repd{248}$ into representations $\repd R$, each $\repd R$ is characterized by some highest weight, which corresponds to some root of $\ex 8$ (since the roots are the weights of the $\repd{248}$). To compute $\chi(V)$, one simply finds the $\ex 8$ root $\boldsymbol{r}$ that the highest weight of $\repd R$ corresponds to and then evaluates $\mathcal{N}(\boldsymbol{r})$ as
\begin{equation}
	\chi(V)=\mathcal{N}(\boldsymbol{r})\equiv\tfrac{1}{12}c_{2,i}(X)H_{i}(\boldsymbol{r})+\tfrac{1}{6}d_{ijk}H_{i}(\boldsymbol{r})H_{j}(\boldsymbol{r})H_{k}(\boldsymbol{r})\ ,\label{eq:mult-1}
\end{equation}
where the second Chern numbers of the Schoen threefold are $c_{2,i}(X)=(4,4,0)_i$. Note that we will often abuse notation and write $\chi(\repd R)$ for the Euler characteristic of the bundle $V$ transforming in the $\repd R$ representation. With these conventions, a left chiral fermion zero-mode in four dimensions has $\mathcal{N}(\boldsymbol{r})<0$.

\subsection{Embedding Constraints for Two Line Bundles}

As mentioned above, since our Schoen threefold has $h^{1,1}=3$, we need to specify three line bundle vectors $\boldsymbol{V}_{i}$. To be concrete, we now consider the explicit example of a line bundle background that breaks $\Ex 8$ to $\Ex 6\times\Uni 1\times\Uni 1$. We will describe this bundle using both the line bundle vector description above and a more standard description. 

It is useful to decompose the $\boldsymbol{V}_{i}$ into a set of line bundle data $(m^{i},n^{i})$ and a set of linearly independent, eight-component basis vectors $(\boldsymbol{t}_{1},\boldsymbol{t}_{2})$. The line bundle vectors can then be written as
\begin{equation}\label{line_vecs}
	\boldsymbol{V}_{i}=m^{i}\boldsymbol{t}_{1}+n^{i}\boldsymbol{t}_{2}\ .
\end{equation}
To match with our previous conventions in the case of the single line bundle, we take the generators of the $\Uni 1\times\Uni 1$ structure group to be $(-\boldsymbol{t}_{1},-\boldsymbol{t}_{2})$. One then finds that matter fields transform according to
\begin{equation}\label{eq:bundle_def}
	\repd 1_{-1,0}\sim\mathcal{O}_{X}(m^{1},m^{2},m^{3})=L_{1}\ ,\qquad\repd 1_{0,-1}\sim\mathcal{O}_{X}(n^{1},n^{2},n^{3})=L_{2}\ ,
\end{equation}
From this we see that $(m^{i},n^{i})$ specify the line bundles, while $(\boldsymbol{t}_{1},\boldsymbol{t}_{2})$ give the embedding of $\Uni 1\times\Uni 1$ into the hidden sector $\Ex 8$. If we want to consider the embedding of a single $\Uni 1$ into $\Ex 8$, one takes $\boldsymbol{t}_{2}=\boldsymbol{0}$, while more $\Uni 1$s could be embedded by including more basis vectors. This should be compared with the discussion in \cite[Section 7.2]{Nibbelink:2016wms}, where they give an example of a $U(1) \times U(1)$ bundle with three line bundle vectors and a single relation between them, implying that they can be written in a basis with two linearly independent generators.

More details on the generators and our conventions can be found in Appendix \ref{sec:Conventions}. The eight-component basis vectors are taken to be
\begin{equation}\label{eq:u1_vectors}
	\boldsymbol{t}_{1}=(0,0,0,0,0,-1,1,0)\ ,\qquad\boldsymbol{t}_{2}=(0,0,0,0,0,-1,-1,-2)\ ,
\end{equation}
where the two $\Uni 1$ groups are generated by $(-\boldsymbol{t}_{1},-\boldsymbol{t}_{2})$. Note that this choice obeys flux quantisation when the entries of $m^{i}$ and $n^{i}$ are integers. Note also that one must have $m^{1}+m^{2}\mod3=0$ for equivariance of the line bundle $L_{1}$, with the same condition for the $n^{i}$ as well.

It is easy to see which simple roots of $\Ex 8$ are broken by this choice – one simply takes the inner product of each $\boldsymbol{V}_{i}$ with the simple roots in (\ref{eq:roots}). In particular, $\boldsymbol{t}_{1}$ breaks a combination of $\alpha_{6}$ and $\alpha_{7}$, while $\boldsymbol{t}_{2}$ breaks $\alpha_{6}$ alone. Together, they break both $\alpha_{6}$ and $\alpha_{7}$, suggesting that the unbroken gauge group will be $\Ex 6$. One can check this explicitly using (\ref{eq:unbroken_condition}) from which one sees that 72 roots of $\Ex 8$ are annihilated by the line bundle vectors, which then form the 72 roots of the unbroken $\Ex 6$ which commutes with $\Uni 1\times\Uni 1$ inside $\Ex 8$. One can also see this by transforming the basis vectors to the $\omega$-basis, where they are given by $(0,0,0,0,0,-1,2,0)$ and $(0,0,0,0,0,-1,0,0)$. Using the \emph{Mathematica }package LieART~\cite{Feger:2012bs,Feger:2019tvk}, one can then find the decomposition of the adjoint representation of $\Ex 8$:
\begin{equation}
	\label{eq:decompE8SU3big}
	\begin{split}
		\repd{248}&=\repd{78}_{0,0}+2\times\repd{1}_{0,0}+\repd{1}_{2,0}+\repd{1}_{-2,0}+\repd{1}_{1,3}+\repd{1}_{-1,3}+
		\repd{1}_{1,-3}+\repd{1}_{-1,-3}\\
		&\eqspace+\repd{27}_{1,-1}+\repd{27}_{-1,-1}+\repb{27}_{1,1}+\repb{27}_{-1,1}+\repd{27}_{0,2}+\repb{27}_{0,-2}\ .\\
	\end{split}
\end{equation}



This particular breaking pattern can be obtained more conventionally by first breaking $E_8\rightarrow E_6\times SU(3)$, under which the adjoint representation of $E_8$ decomposes as
\begin{equation}
\label{eq:E8SU3first}
\bf{248}=(\repd{78},\repd{1})+(\repd{1},\repd{8})+(\repd{27},\repb{3})+(\repb{27},\repd{3})\ .
\end{equation}
Breaking $SU(3)$ further to $SU(2)\times U(1)$, the $SU(3)$ representations that appear above decompose as
\begin{equation}
\label{eq:decompSU3}
\begin{split}
\repd{3}&=\repd{2}_{1}+\repd{1}_{-2}\ ,\\
\repb{3}&=\repd{2}_{-1}+ \repd{1}_{2}\ ,\\
\repd{8}&=\repd{3}_{0}+\repd{2}_{3}+\repd{2}_{-3}+\repd{1}_0\ .\\
\end{split}
\end{equation}
Finally, we break the $SU(2)$ further to $U(1)$, with the $SU(2)$ representations above decomposing as
\begin{align}
\repd{2}&=\repd{1}_{1}+\repd{1}_{-1}\ ,\\
\repd{3}&=\repd{1}_2+\repd{1}_0+\repd{1}_{-2}\ .
\end{align}
Putting this together, one sees that the adjoint representation of $E_8$ decomposes under this $E_6\times U(1)\times U(1)$ exactly as in \eqref{eq:decompE8SU3big} above. We thus have two equivalent descriptions of the embedding of this $U(1)\times U(1)$ in $E_8$, either via the line bundle vectors defined by \eqref{eq:u1_vectors} or via a chain of subgroups starting from the maximal subgroup $E_6\times SU(3)$.

We can find the $E_8$ connection that corresponds to this breaking pattern by first building the $SU(3)$ connection induced by the breaking $SU(3)\rightarrow S(U(1)\times U(1)\times U(1))\simeq U(1)\times U(1)$ and two $U(1)$ connections, $A^{(1)}_{U(1)}$ and $A^{(2)}_{U(1)}$. The two $U(1)$s embed into $SU(3)$ as
\begin{equation}
( \ee^{\ii\phi_1},\ee^{\ii\phi_2})\hookrightarrow  \begin{pmatrix}\ee^{-2\ii\phi_2}&0\\0&\ee^{\ii\phi_2}\begin{pmatrix}\ee^{\ii\phi_1}&0\\0&\ee^{-\ii\phi_1}\end{pmatrix}\end{pmatrix} \ .
\end{equation}
Using this, we can build a connection associated with a rank three bundle $V_{\repd 3}$ as
\begin{equation}
	\begin{split}
		\label{eq:SU3U1U1}
		A_{SU(3)} = \begin{pmatrix}-2A^{(2)}_{U(1)}&0&0\\0&A^{(1)}_{U(1)}+A^{(2)}_{U(1)}&0\\0&0&-A^{(1)}_{U(1)}+A^{(2)}_{U(1)}\end{pmatrix}\ ,\\
	\end{split}
\end{equation}
where $A^{(1)}_{U(1)}$ and $A^{(2)}_{U(1)}$ are the $L_1$ and $L_2$  line bundle connections respectively. The form of connection \eqref{eq:SU3U1U1} implies that the rank three bundle is the Whitney sum
\begin{equation}
\label{eq:defineBundle2}
\begin{split}
V_{\repd 3}&= L_2^{-2}\oplus\left((L_1\oplus L_1^{-1})\otimes L_2\right)
\\&= L_2^{-2}\oplus L_1L_2\oplus L_1^{-1}L_2\equiv\mathcal{F}\oplus\mathcal{K}\oplus\mathcal{E}\ ,
\end{split}
\end{equation}
where we have defined
\begin{equation}\label{eq:FKL}
\mathcal{F}=L_2^{-2}\ , \qquad \mathcal{K}=L_1 L_2\ , \qquad \mathcal{E}=L_1^{-1} L_2\ .
\end{equation}
From the form of the $SU(3)$ connection in \eqref{eq:SU3U1U1}, one can read off that the $U(1)$ connections associated with the line bundles $\mathcal{F}$, $\mathcal{K}$ and $\mathcal{E}$ are 
\begin{equation}
	A^{\mathcal{F}}_{U(1)}=-2A^{(2)}_{U(1)}\ ,\qquad A^{\mathcal{K}}_{U(1)}=A^{(1)}_{U(1)}+A^{(2)}_{U(1)}\ , \qquad A^{\mathcal{E}}_{U(1)}=-A^{(1)}_{U(1)}+A^{(2)}_{U(1)}
\end{equation}
respectively.  Note that $A^{\mathcal{E}}_{U(1)}+A^{\mathcal{F}}_{U(1)}+A^{\mathcal{K}}_{U(1)}=0$ and, therefore, that
\begin{equation}
c_1(\mathcal{F})+c_1(\mathcal{K})+c_1(\mathcal{E})=0\ ,
\end{equation}
which is simply the condition that $V_{\repd 3}$ is an $SU(3)$ bundle.

The DUY theorem \cite{Donaldson, UY} implies that there exists a connection $A_{SU(3)}$ that solves the HYM equation if $V_{\repd 3}=\mathcal{F}\oplus\mathcal{K}\oplus\mathcal{E}$ is poly-stable. Since the slope of any $SU(N)$ bundle such as $V_{\repd 3}$ vanishes, it follows that $V_{\repd 3}$ can be poly-stable only if the slopes of each of its subbundles vanish as well. Hence, the hidden sector bundle $V_{\repd 3}$ given in 
\eqref{eq:defineBundle2} will be slope poly-stable if
\begin{equation}
\mu(\mathcal{F})=\mu(\mathcal{K})=0 \ .
\end{equation}
Note that the slope of $\mathcal{E}$ vanishes automatically if the slopes of $\mathcal{F}$ and $\mathcal{K}$ do, so the condition above is sufficient.
%

The $SU(3)$ connection in \eqref{eq:SU3U1U1} embeds further into an $E_8$ connection such that it commutes with $E_6$ as in \eqref{eq:E8SU3first}. 
%
The embedding of the line bundle connections into the hidden $E_8$ is then given by 
\begin{equation}\label{eq:E8_connection}
(A^{(1)}_{U(1)},A^{(2)}_{U(1)})\hookrightarrow A_{E_8}=A^{(1)}_{U(1)}Q_1+A^{(2)}_{U(1)}Q_2\ ,
\end{equation}
where $Q_1$ and $Q_2$ are elements of the $E_8$ algebra whose traces obey
\begin{equation}\label{eq:traces}
	\tfrac{1}{4}\tr Q_1^2=1\ , \qquad \tfrac{1}{4}\tr Q_2^2=3\ , \qquad \tr Q_1Q_2=0\ ,
\end{equation}
which can also be read off from the decomposition in \eqref{eq:decompE8SU3big}. We see that $Q_1$ and $Q_2$ contain the charges associated with each of the two $U(1)$ subgroups.

We can also see this from the line bundle vector description as follows. Since the curvature can be expanded in the Cartan generators as in \eqref{eq:flux-1}, we can write
\begin{equation}
	F_{E_8} = \frac{2\pi}{v^{1/3}}\omega_i H_i = \frac{2\pi}{v^{1/3}} \omega_i (m^i t^I_1 + n^i t^I_2)H_I = (F_{U(1)}^{(1)} t^I_1 + F_{U(1)}^{(2)} t^I_2)H_I\ ,
\end{equation}
where we have identified $F_{U(1)}^{(1)}=2\pi v^{-1/3} m^i\omega_i$ as the curvature of the line bundle $L_1=\mathcal{O}_X(m^1,m^2,m^3)$ defined in \eqref{eq:bundle_def}, and similarly for $L_2$. Comparing with \eqref{eq:E8_connection}, we read off that $Q_1 = t_1^I H_I$ and $Q_2 = t_2^I H_I$. It is then simple to repeat the calculation of the traces to give
\begin{equation}\label{eq:traces_line}
	\tfrac14\tr Q_1^2 = \tfrac12\boldsymbol{t}_1\cdot \boldsymbol{t}_1 = 1 \ , \qquad \tfrac14\tr Q_2^2 = \tfrac12\boldsymbol{t}_2\cdot \boldsymbol{t}_2 = 3\ , \qquad \tfrac14\tr Q_1 Q_2 = \tfrac12\boldsymbol{t}_1\cdot \boldsymbol{t}_2 = 0 \ ,
\end{equation}
where we have used the normalisation of the generators in \eqref{gen_norm}. We see that this agrees with \eqref{eq:traces} upon inserting the definitions from \eqref{line_vecs} and \eqref{eq:u1_vectors}.
%
%

\subsection*{Anomaly Condition}

As discussed in Section \ref{sec:anomaly}, anomaly cancellation in heterotic M-theory requires that
\begin{equation}
  c_2(TX)-c_2(\mathcal{V}^{(1)})
  -\op{ch}_2(\mathcal{V}^{(2)}) - W 
  = 0 \ ,
  \label{29}
\end{equation}
where $\mathcal{V}^{(1)}$ is the observable $SU(4)$ bundle, $\mathcal{V}^{(2)}$ is the hidden sector bundle, $TX$ is the tangent bundle of the compactification threefold, while $W$ is the effective class of the single five-brane between the hidden and observable sector. 

The second Chern character of the hidden sector bundle is
\begin{equation}
\op{ch}_{2}(\mathcal{V}^{(2)})=\frac{1}{16\pi^{2}}\tr F_{E_8}\wedge F_{E_8},
\end{equation}
where $F_{E_8}$ is the curvature of the $E_8$ connection induced from the two line bundle connections in \eqref{eq:E8_connection}.
%
Expanding out, we find
\begin{equation}
\op{ch}_{2}(\mathcal{V}^{(2)})=\frac{1}{16\pi^{2}}\left(\tr Q_1^2F_{U(1)}^{(1)}\wedge F_{U(1)}^{(1)}+\tr Q_2^2F_{U(1)}^{(2)}\wedge F_{U(1)}^{(2)}+
2\tr Q_1Q_2F_{U(1)}^{(1)}\wedge F_{U(1)}^{(2)}\right) \ ,
\end{equation}
where the $F_{U(1)}^{(i)}$ are given in terms of the first Chern class of each line bundle as $ c_{1}^i=F^{(i)}_{U(1)}/2\pi$. Again denoting the $L_{1}$ and $L_{2}$ line bundles by 
\begin{equation}\label{eq:L1_L2_def}
L_1=\mathcal{O}_X(m^1,m^2,m^3)\ , \qquad L_2=\mathcal{O}_X(n^1,n^2,n^3)\ ,
\end{equation}
the second Chern character of $\mathcal{V}^{(2)}$ is then simply
\begin{equation}
\begin{split}\label{ch2}
\op{ch}_{2}(\mathcal V^{(2)})&=\,c_{1}(L_1)\wedge c_{1}(L_1)+3c_{1}(L_2)\wedge c_{1}(L_2)\\
&=\frac{1}{v^{2/3}}(m^1\omega_{1}+m^2\omega_{2}+m^3\omega_{3})^{2}+\frac{3}{v^{2/3}}(n^1\omega_{1}+n^2\omega_{2}+n^3\omega_{3})^{2}\ ,
\end{split}
\end{equation}
where we used the trace relations from \eqref{eq:traces}. Remember that for our particular configuration,
\begin{equation}
  \frac{1}{v^{1/3}}\int_X \left(c_2(TX)-c_2(\mathcal{V}^{(1)})\right)\wedge \omega_i=\left( \tfrac{4}{3},\tfrac{7}{3},-4\right)_i\ .
\end{equation}
If we define 
\begin{equation}
W_{i}= \frac{1}{v^{1/3}}\int_X{W \wedge \omega_{i}} \ ,
\label{sun2}
\end{equation}
it then follows that the anomaly condition is given by
\begin{equation}
\label{eq:anomaly_modified}
W_i=\left(\tfrac{4}{3},\tfrac{7}{3},-4  \right)_i+d_{ijk}m^jm^k+3d_{ijk}n^jn^k\ .
\end{equation}
In order to preserve $N=1$ supersymmetry $W$ must be an effective class; that is, each component $W_{i}$,  $i=1,2,3$ must be non-negative.

Again we can compare this with the line bundle vector formalism. We have already given the second Chern character in \eqref{eq:chern2-1}. Expanding this out and using the trace relations in \eqref{eq:traces_line}, one can check that it reproduces \eqref{ch2} above. Furthermore, the anomaly cancellation condition can then be written as
\begin{equation}
	W_{i}=(\tfrac{4}{3},\tfrac{7}{3},-4)_{i}+\tfrac{1}{2}d_{ijk}(\boldsymbol{V}_{j}\cdot\boldsymbol{V}_{k})\ ,
\end{equation}
which agrees with \eqref{eq:anomaly_modified} upon using the definitions from \eqref{line_vecs} and \eqref{eq:u1_vectors}.

\subsection*{Low-Energy Fields}

As usual, low-energy matter superfields fields arising from the decomposition in eq.~\eqref{eq:decompE8SU3big} are associated with bundle-valued cohomologies on the Calabi--Yau threefold. Using the identification \eqref{eq:FKL}, we find that the $E_6$ singlets with non-zero $U(1)$ charges are associated with
\begin{align}
	\repd 1_{0,2}&\sim H^\bullet(X,\mathcal{F})=H^\bullet(X,L_2^{-2})\ ,\\
	 \repd 1_{-1,-1}&\sim H^\bullet(X,\mathcal{K})=H^\bullet(X,L_1 L_2)\ ,\\
	 \repd 1_{1,-1}&\sim H^\bullet(X,\mathcal{E})=H^\bullet(X,L_1^{-1}L_2)\ .
\end{align}
%
To find the cohomologies for the other representations in the decomposition, we can either use the fact that all $SU(3)$ representations can be obtained from the fundamental $\repd 3$ and its conjugate $\repb 3$, and then use the decomposition of the $\repd3$ in terms of $U(1)\times U(1)$, or we can just count the charges in \eqref{eq:decompE8SU3big} and use the above identifications to find the corresponding cohomology. 
The representations we obtain for the $E_6\times U(1)\times U(1)$ low-energy group, as well as their corresponding cohomologies, are shown in Table \ref{tab:4thTable}. Note that each representation $\repd R$ in Table \ref{tab:4thTable} has an associated cohomology of the form
\begin{equation}
H^\bullet(X,L_{\repd R})=H^\bullet(X,L_1^{-q_{\repd R}}\otimes L_2^{-p_{\repd R}})\ ,
\end{equation}
where $q_{\repd R}$ and $p_{\repd R}$ are the charges of $\repd R$ for each of the two $U(1)$ groups.

Let us now consider the low-energy matter spectrum. For fields in the $\repd R$ representation, with associated line bundle $L_{\repd R}$, the Euler characteristic $\chi(L_{\repd R})$ counts the chiral asymmetry. For a line bundle of the form $L_{\repd R}=\mathcal{O}_{X}(l^1_{\repd R},l^2_{\repd R},l^3_{\repd R})$, the Euler characteristic is given by
\begin{equation}
\chi(L_{\repd R})=\sum_{i=0}^{3}(-1)^ih^i(X,L_{\repd R})=\int_X \op{ch}(L_{\repd R})\wedge \op{Td}(X)\ ,
\label{green1}
\end{equation}
where  $\op{ch}(L_{\repd R})$ is the Chern character of $L_{\repd R}$ and $\op{Td}(X)$ is the Todd class of the tangent bundle of $X$. On the Schoen manifold we are considering, this simplifies to 
\begin{equation}
\label{eq:Euler_ch_li}
\chi(L_{\repd R})=\tfrac{1}{3}(l^1_{\repd R}+l^2_{\repd R})+\tfrac{1}{6}d_{ijk}l^i_{\repd R}l^j_{\repd R}l^k_{\repd R}\ .
\end{equation}
Using the intersection numbers $d_{ijk}$ given in \eqref{4}, this expression becomes
\begin{equation}
\chi(L_{\repd R})=\tfrac{1}{6}\left(l^1_{\repd R}l^2_{\repd R}(l^1_{\repd R}+l^2_{\repd R}+6l^3_{\repd R})+2l^1_{\repd R}+2l^2_{\repd R}\right)\ .
\label{hope2}
\end{equation}
The numbers $l_{\repd R}^i$ characterizing the line bundle $L_{\repd R}$ depend on the low-energy representation $\repd R$. For our line bundle embedding and a representation $\repd R$ with $U(1)$ charges $q_{\repd R}$ and $p_{\repd R}$, the Euler characteristic is given by $\chi(L_{\repd R})=\chi({L_1^{-q_{\repd R}}L_2^{-p_{\repd R}}})$. Defining $L_1$ and $L_2$ as in \eqref{eq:L1_L2_def}, one finds that 
\begin{equation}
 l_{\repd R}^i=-q_{\repd R}m^i-p_{\repd R}n^i\ , \quad i=1,2,3\ .
\end{equation}
It is then simple to evaluate the Euler characteristic for each representation by substituting the
values of $l_{\repd R}^i$ into \eqref{hope2}.
For example, for the line bundle $L_1^{-1}L_2^{-3}$ associated with the representation $\repd 1_{1,3}$ in 
Table \ref{tab:4thTable}, the $l^i$ are
\begin{equation}
l^i=-m^i-3n^i\ .
\end{equation}
The Euler characteristic is then given by
\begin{equation}\label{chi}
\chi(\repd{1}_{1,3})\equiv\chi(L_1^{-1}L_2^{-3}) =\tfrac{1}{6}\left(l^1l^2(l^1+l^2+6l^3)+2l^1+2l^2\right)\ .
\end{equation}
It is straightforward to check that this agrees with the Euler characteristic as computed using the multiplicity operator in \eqref{eq:mult-1}. For example, the highest weight of $\repd{1}_{1,3}$ can be obtained by projecting the root $\boldsymbol{r}=(0,0,0,0,0,1,0,1)$ of $\mathfrak{e}_8$. Using this in \eqref{eq:mult-1}, one finds that $\mathcal{N}(\boldsymbol{r})$ evaluates to \eqref{chi}.

The matter fields associated with any given representation and cohomology in Table \ref{tab:4thTable} are either 1) chiral superfields if the corresponding Euler characteristic $\chi <0$ or 2) anti-chiral superfields if $\chi >0$. For any given entry in Table \ref{tab:4thTable}, the sign of the Euler characteristic depends on the choice of line bundles $L_{1}$ and $L_{2}$. Hence, the same entry in the Table can correspond to either a chiral superfield or an anti-chiral superfield depending on the circumstances of the solution. Note, however, that if a representation, such as ${\underline{\bf 1}}_{1,3}$, corresponds to chiral superfields then the conjugate representation, such as ${\underline{\bf 1}}_{-1,-3}$, corresponds to anti-chiral superfields. With this in mind, in the last column of the table, we have assigned a specific symbol to the matter supermultiplet of each representation. Having done that, we note that each such superfield contains a complex scalar field component. In the following, it is convenient to abuse notation and denote each superfield and its complex scalar field component using the same symbol. Whether we are referring to the full supermultiplet or its scalar component will be clear from the context.

\begin{table}[t]
	\noindent \begin{centering}
		\begin{tabular}{clc}
			\toprule 
			$E_6\times U(1)\times U(1) $ & Cohomology & Field Name\tabularnewline
			\midrule
			\midrule 
			$\repd{1}_{1,3}$ & $H^{\bullet}(X,\mathcal{F}\otimes \mathcal{K}^*=L_1^{-1}L_2^{-3})$ &$C_1$\tabularnewline
			\midrule 
			$\repd{1}_{-1,-3}$ & $H^{\bullet}(X,\mathcal{F}^*\otimes \mathcal{K}=L_1L_2^{3})$& $\tilde C_1$\tabularnewline
			\midrule 
			$\repd{1}_{-1,3}$ & $H^{\bullet}(X,\mathcal{F}\otimes \mathcal{E}^*=L_1L_2^{-3})$ & $C_2$\tabularnewline
			\midrule 
			$\repd{1}_{1,-3}$ & $H^{\bullet}(X,\mathcal{F}^*\otimes \mathcal{E}=L_1^{-1}L_2^{3})$ &$\tilde C_2$\tabularnewline
			\midrule 
			$\repd{1}_{2,0}$ & $H^{\bullet}(X,\mathcal{K}^*\otimes \mathcal{E}=L_1^{-2})$ &$C_3$\tabularnewline
			\midrule 
			$\repd{1}_{-2,0}$ & $H^{\bullet}(X,\mathcal{K}\otimes \mathcal{E}^*=L_1^{2})$ & $\tilde C_3$ \tabularnewline
			\midrule 
			$\repd{27}_{0,-2}$ & $H^{\bullet}(X,\mathcal{F}^*=L_2^{2})$  & $f_1$\tabularnewline
			\midrule
			$\repd{27}_{1,1}$ & $H^{\bullet}(X,\mathcal{K}^*=L_1^{-1}L_2^{-1})$  & $f_2$\tabularnewline
			\midrule
			$\repd{27}_{-1,1}$ & $H^{\bullet}(X,\mathcal{E}^*=L_1L_2^{-1})$  & $f_3$\tabularnewline
			\midrule
			$\repd{27}_{0,2}$ & $H^{\bullet}(X,\mathcal{F}=L_2^{-2})$  & $\tilde f_1$\tabularnewline
			\midrule
			$\repd{27}_{-1,-1}$ & $H^{\bullet}(X,\mathcal{K}=L_1L_2)$  & $\tilde f_2$\tabularnewline
			\midrule
			$\repd{27}_{1,-1}$ & $H^{\bullet}(X,\mathcal{E}=L_1^{-1}L_2)$  & $\tilde f_3$\tabularnewline
			\bottomrule
		\end{tabular}
		\par\end{centering}
		\caption{Low-energy representations of $E_6\times U(1)\times U(1)$ and their associated cohomologies. $L_1$ is a line bundle of the form $L_1=\mathcal{O}_X(m^1,m^2,m^3)$, while we write $L_2=\mathcal{O}_X(n^1,n^2,n^3)$ for $L_2$ . The entries in the third column correspond to either a chiral or an anti-chiral supermultiplet if the Euler characteristic of the associated line bundle is negative or positive respectively. Hence, the supermultiplets corresponding to a line bundle and its inverse bundle are conjugates of each other.
Note that the fields $C_i$ and $\tilde C_i$ are singlets under $E_6$. Deforming the bundle $V_{\repd 3}$ away from the decomposable locus is equivalent to turning on different combinations of VEVs for the scalar components of these supermultiplets in the effective theory.}
		\label{tab:4thTable}
\end{table}

\subsection*{Genus-One Corrected FI Terms}\label{app:FI}

As previously explained, a $\Uni 1$ symmetry that appears in both the internal and four-dimensional gauge generates a D-term potential proportional to an FI term associated with the $U(1)$ bundle
\begin{equation}
\label{eq:B1_App}
\text{FI}=\frac{a_L}{2}\frac{\epsilon_S\epsilon_R^2}{\kappa_4^2}\frac{1}{\hat R V^{2/3}}\left[\mu(L)+\frac{\epsilon_S^\prime \hat R}{V^{1/3}}\int_X c_1(L)\wedge \left(  J^{(N+1)}+\sum^N_{n=1}z_n^2J^{(n)}  \right)\right]\ ,
\end{equation} 
where the complex two-forms $J$ are defined in  \cite{Ovrut:2015uea} and $n$ runs over all five-branes in the bulk interval. In our setup, we only have one five brane at position $z=\lambda +\frac{1}{2}$  with the source term given by $J^1=W$. The coefficient $a_L$ depends on the exact embedding of the line bundle $L$ associated with the FI term into the hidden sector $E_8$. 
In the case of a hidden sector with a single line bundle, there is one FI term associated with it. In this case, the coefficient $a_L$ is simply equal to the coefficient $a$ derived for the second Chern character. For the particular embedding $U(1)\rightarrow SU(2)\rightarrow E_8$ that we studied in the previous sections, $a=1$.

In the case of a hidden sector with two line bundles, we have two FI terms. Each is  associated with one of the two line bundles $\mathcal{F}$ and $\mathcal{K}$ defined in the decomposition 
\begin{equation}
V_{\repd 3}=\mathcal{F}\oplus \mathcal{K} \oplus \mathcal{E}
\end{equation}
of the $SU(3)$ bundle $V_{\repd 3}$ at the decomposable locus.
Note that the line bundle $\mathcal{E}$ depends on $\mathcal{K}$ and $\mathcal{F}$ such that 
$c_1(\mathcal{E})=-c_1(\mathcal{F})-c_1(\mathcal{K})$. Hence, $V_{\repd 3}$ has the structure group
$S(U(1)\times U(1)\times U(1))\sim U(1)\times U(1)$ at the stability wall. The genus-one corrected FI terms in this case are
\begin{equation}
\begin{split}
\label{eq:B3_App}
&\text{FI}_{\mathcal{F}}=\frac{a_{\mathcal{F}}}{2}\frac{\epsilon_S\epsilon_R^2}{\kappa_4^2}\frac{1}{\hat R V^{2/3}}\left[\mu(\mathcal{F})+\frac{\epsilon_S^\prime \hat R}{V^{1/3}}\int_X c_1(\mathcal{F})\wedge \left(  J^{(N+1)}+\sum^N_{n=1}z_n^2J^{(n)}  \right)\right]\ , \\
&\text{FI}_{\mathcal{K}}=\frac{a_{\mathcal{K}}}{2}\frac{\epsilon_S\epsilon_R^2}{\kappa_4^2}\frac{1}{\hat R V^{2/3}}\left[\mu(\mathcal{K})+\frac{\epsilon_S^\prime \hat R}{V^{1/3}}\int_X c_1(\mathcal{K})\wedge \left(  J^{(N+1)}+\sum^N_{n=1}z_n^2J^{(n)}  \right)\right] \ .
\end{split}
\end{equation} 
Note that the FI terms in the effective theory are associated with the line bundles $\mathcal{F}=L_2^{-2}$ and $\mathcal{K}=L_1L_2$ and \emph{not} with the bundles $L_1=\mathcal{O}_X(m^1,m^2,m^3)$ and $L_2=\mathcal{O}_X(n^1,n^2,n^3)$. 
Hence, the coefficients $a_{\mathcal{F}}$ and $a_{\mathcal{K}}$ in front of the expressions in \eqref{eq:B3_App} depend on the how the bundles 
$\mathcal{F}$ and $\mathcal{K}$ embed into the $E_8$ connection. This calculation was trivial in the single bundles case, because we parametrized directly all the equations in terms of the bundle $L$ associated with the $FI$ term. For the two line bundle case of interest, one can read off the generators $Q_{\mathcal{F}}$ and $Q_{\mathcal{K}}$ associated with their embedding into $E_8$ from Table \ref{tab:4thTable}. These are given by
\begin{align}
&Q_{\mathcal{F}}=(1,-1,2,-2,-1,1,-\id_{27},0\times \id_{27}, \id_{27},\id_{27},0\times \id_{27}, -\id_{27})\ ,\\
&Q_{\mathcal{K}}=(-1,1,1,-1,-2,2,0\times \id_{27},- \id_{27}, \id_{27},0\times\id_{27},\id_{27}, -\id_{27})\ .
\end{align}
Hence, we get
\begin{align}
&a_{\mathcal{F}}=\frac{1}{4} \tr Q_{\mathcal{F}}^2=1,\\
&a_{\mathcal{K}}=\frac{1}{4} \tr Q_{\mathcal{K}}^2=1\ .
\end{align}

\subsection*{Extension Bundle}

In this section, we study how to deform the Whitney sum bundle defined in eq.~\eqref{eq:defineBundle2} away from the decomposable locus to construct an irreducible $\SU3$ bundle. We will also discuss how this construction appears in the effective field theory.

Consider the pair of exact sequences
\begin{equation}
\begin{split}
\label{eq:sequences}
0&\rightarrow \mathcal{F}\rightarrow W\rightarrow \mathcal{E}\rightarrow  0\ ,\\
0&\rightarrow \mathcal{K}\rightarrow V^{\prime}_{\repd 3} \rightarrow W \rightarrow 0\ .
\end{split}
\end{equation}
\setcounter{footnote}{0}
These define an $SU(3)$ bundle $V^{\prime}_{\repd 3}$, since $c_1(V^{\prime}_{\repd 3})=0$. The first extension can be non-trivial -- that is, $W$ is not simply $\mathcal{F}\oplus\mathcal{E}$ -- if and only if the Ext group $\op{Ext}^1(\mathcal{E},\mathcal{F})=H^1(X, \mathcal{F}\otimes \mathcal{E}^*)$ is non-trivial. Similarly, the second extension can be non-trivial if and only if $\op{Ext}^1(W,\mathcal{K})=H^1(X,\mathcal{K}\otimes W^*)$ is non-trivial. It is relatively easy to see that a non-trivial extension class for the first extension sequence corresponds to turning on a VEV for the field $C_2$ in the effective theory. Indeed, from Table \ref{tab:4thTable} we see that $H^1(X, \mathcal{F}\otimes \mathcal{E}^*)$ counts the number of $C_2$ fields. It is straightforward, if not obvious, to see that the extension class for the second sequence is equivalent to turning on a VEV for the field $\tilde C_3$. To prove this, note that one can show that $\op{Ext}^1(W,\mathcal{K})=H^1(X,\mathcal{K}\otimes W^*)=H^1(X,\mathcal{K}\otimes \mathcal{E}^*)$~\cite{Anderson:2010ty}.\footnote{Specifically, in Section 3.3 of \cite{Anderson:2010ty} it is
shown that $H^1(X,\mathcal{K}\otimes W^*)=H^1(X,\mathcal{K}\otimes \mathcal{E}^*)+\kernel \delta$, where $\kernel \delta\in H^1(X,\mathcal{K}\otimes \mathcal{F}^*)$. Note that this branch is confined to the zero element of $H^1(X,\mathcal{K}\otimes \mathcal{F}^*)$.}  It follows from Table \ref{tab:4thTable} that $H^1(X, \mathcal{K}\otimes \mathcal{E}^*)$ counts the number of $\tilde C_3$ fields.
We learn that the pair of non-trivial extensions in \eqref{eq:sequences} corresponds to turning on VEVs for the fields $ C_2$ and $\tilde C_3$ in the effective theory,
\begin{equation}
\langle  C_2\rangle \neq 0\ , \qquad \langle \tilde C_3\rangle \neq 0 \ ,
\end{equation}
while the VEVs of the other charged matter fields are set to zero. Having allowed for these VEVs, one should check that the F-term and D-term constraints are satisfied so that the vacuum of the four-dimensional theory preserves supersymmetry. As shown in \cite{Anderson:2010ty}, one can turn on this combination of VEVs while still having the superpotential and its derivative vanish -- that is, the effective theory satisfies the F-flatness requirement. Furthermore, using the generic expression for a D-term given in \cite{Anderson:2010ty}, we see that fields $C_2$ and $\tilde C_3$ have the correct charges to cancel the FI terms associated with the two $U(1)$ bundles $\mathcal{F}$ and $\mathcal{K}$, as defined in eq. \eqref{eq:B3_App}. Therefore,
the theory is also D-flat.

Note that there is another pair of extension sequences which correspond to the same non-zero elements of $H^1(X,\mathcal{F}\otimes \mathcal{E}^*)$ and $H^1(X,\mathcal{K}\otimes \mathcal{E}^*)$, namely
\begin{align}\label{eq:sequences_disc2}
	\begin{split}
		 0&\rightarrow \mathcal{K}\rightarrow W^\prime\rightarrow \mathcal{E}\rightarrow  0\ ,\\
			0&\rightarrow \mathcal{F}\rightarrow \mathcal{V}^{\prime}_{\repd 3} \rightarrow W^\prime \rightarrow 0\ ,
	\end{split}
	\begin{split}
		\op{Ext}^1(\mathcal{E},\mathcal{K})&=H^1(X,\mathcal{K}\otimes \mathcal{E}^*)\neq 0\ ,\\
		\op{Ext}^1(W^\prime,\mathcal{F})&=H^1(X,\mathcal{F}\otimes \mathcal{E}^*) \neq 0\ .
	\end{split}
\end{align}
This seemingly gives us a different $SU(3)$ bundle $\mathcal{V}^{\prime}_{\repd 3}$ for the same set of VEVs. However, as shown in \cite{Anderson:2010ty}, this new bundle is isomorphic to the above, that is
\begin{equation}
\mathcal{V}^{\prime}_{\repd 3}\simeq V^{\prime}_{\repd 3}\ .
\label{cold1}
\end{equation}
From an effective theory perspective, both pairs of sequences correspond to turning on VEVs for the low-energy fields $\tilde{C}_2$ and $C_3$. Roughly speaking, the choice between \eqref{eq:sequences} or \eqref{eq:sequences_disc2} corresponds to turning on VEVs for $C_2$ first and then $\tilde{C}_3$ or vice versa, respectively.

As we will discuss next, there are actually six inequivalent \emph{branches} along which one can, in principle, deform the $SU(3)$ bundle away from the decomposable locus. From the low-energy perspective, deforming the bundle along one branch or another comes from turning on different pairs of VEVs of the $E_6$ singlet fields $C_i$ and $\tilde C_i$ which are both F-flat and D-flat. The extensions in \eqref{eq:sequences} or \eqref{eq:sequences_disc2} give one possible branch. 
We show the full branch structure for a hidden sector built from two line bundles in Table \ref{tab:branches}, but for the moment let us continue focusing on the $\langle  C_2\rangle,\langle \tilde C_3\rangle \neq 0$ branch. We have seen that a non-trivial extension is possible if and only if the cohomology groups $H^1(X,\mathcal{F}\otimes \mathcal{E}^*)$ and $H^1(X,\mathcal{K}\otimes \mathcal{E}^*)$ are non-empty, that is
\begin{equation}
h^1(X,\mathcal{F}\otimes \mathcal{E}^*)>0\ , \qquad h^1(X,\mathcal{K}\otimes \mathcal{E}^*)>0 \ .
\end{equation}
It is useful to observe that the case where either of the line bundles $\mathcal{F}\otimes \mathcal{E}^*$ or $\mathcal{K}\otimes \mathcal{E}^*$ is ample is eliminated from the start, since ample line bundles have vanishing first cohomology groups on a Calabi--Yau manifold. Hence, one must choose the line bundles $L_{1}$, $L_{2}$ so that $\mathcal{F}\otimes \mathcal{E}^*$ and $\mathcal{K}\otimes \mathcal{E}^*$ are {\it not} ample. That being said, computing the cohomology groups $H^1(X,\mathcal{F}\otimes \mathcal{E}^*)$ and $H^1(X,\mathcal{K}\otimes \mathcal{E}^*)$ directly on our Schoen manifold $X$ is a difficult task, which will not be included in this work.

\subsection*{Bundle Stability}

The conditions above ensure that an extension is possible, which corresponds to the existence of non-zero VEVs of certain charged matter fields in the four-dimensional effective theory. The existence of the deformed $SU(3)$ bundle is not sufficient to ensure supersymmetry, however. One must also require that the new bundle admits a connection that solves the HYM equation; that is, the bundle must be slope-stable. Therefore, the next question one must ask, assuming the extension exists, is whether the resulting bundle is slope-stable. As we have emphasized, checking the stability of a bundle is generally a difficult calculation which, at the moment, cannot be done algorithmically on our particular Schoen threefold. Instead, we will focus on some necessary conditions; specifically that some obvious subbundles have negative slope and that the Bogomolov inequality is satisfied. This last requirement is significant  since often the Bogomolov inequality is the only obstruction to finding a slope-stable bundle~\cite{Braun:2006ae}. 

From the first sequence in \eqref{eq:sequences}, we learn that there is an embedding
\begin{equation}
\mathcal{K}\hookrightarrow V^\prime_{\repd 3}\ .
\end{equation}
That is, the line bundle $\mathcal{K}$ injects into $V_{\repd 3}^\prime$. Since $V_{\repd 3}^\prime$ has a vanishing slope, the first necessary condition for stability is that $\mathcal{K}$ has a negative tree-level slope:
\begin{equation}
\label{eq:slopecond1}
\mu(\mathcal{K})=\mu(L_1L_2)<0\quad \Rightarrow \quad d_{ijk}(m^i+n^j)a^ja^k<0\ ,
\end{equation}
Furthermore, it would appear from \eqref{eq:sequences_disc2} and \eqref{cold1}, and is proven in Appendix \ref{app:subbundles}, that the bundle $\mathcal{F}$ also injects into $ V^\prime_{\repd 3}$,
\begin{equation}
\mathcal{F}\hookrightarrow V^\prime_{\repd 3}\ ,
\end{equation}
and so the slope of $\mu(\mathcal{F})$ must also be negative:
\begin{equation}
\label{eq:slopecond2}
\mu(\mathcal{F})=\mu(L_2^{-2})<0\quad \Rightarrow \quad -2d_{ijk}n^ja^ja^k<0\ .
\end{equation}

Given a choice of $L_1$ and $L_2$ such that \eqref{eq:slopecond1} and \eqref{eq:slopecond2} are satisfied, the final necessary condition that we impose is that $V_{\repd 3}^{\prime}$ satisfies the Bogomolov inequality
\begin{equation}
\int c_{2}(V_{\repd 3}^{\prime})\wedge\omega\geq0\ .
\label{train1}
\end{equation}
A bundle is stable only if it satisfies this highly non-trivial constraint. Thankfully, since $c_{2}$ is topological, this can be computed from the data of the Whitney sum bundle $V_{\repd 3}=\mathcal{F}\oplus\mathcal{K}\oplus\mathcal{E}$. The total Chern class is
\begin{equation}
\begin{split}
c(V_{\repd 3}^{\prime})\equiv c(V_{\repd 3}) & =c( L_2^{-2}\oplus L_1L_2\oplus L_1^{-1}L_2)\\
 &=1-3c_1^2(L_2)-c_1^2(L_1)+\dots
 \end{split}
 \end{equation}
from which we see that $c_2(V_{\repd 3}^\prime)=3c_1^2(L_2)-c_1^2(L_2)$. It follows that condition \eqref{train1} becomes
\begin{equation}
\label{eq:bogomolov_2lines}
d_{ijk}m^im^ja^k+3 d_{ijk}n^in^ja^k\leq 0\ .
\end{equation}
We emphasize that since the K\"ahler form $\omega$ appears in the calculation of the two slopes and the Bogomolov inequality, all three of these conditions depend on where one is in the K\"ahler cone, which is also restricted by the physical constraints discussed in Section \ref{sec:constraints}


\subsection{Line Bundles Scan}\label{sec:first_survey}

We are now in a position to look for an appropriate hidden sector bundle, which has a non-trivial extension $V_{\repd 3}^\prime$ that might be stable away from the decomposable locus. In addition to the physical constraints  discussed in Section \ref{sec:constraints}, we impose the stability constraints derived earlier. Specifically, we want to find two line bundles $L_1=\mathcal{O}_X(m^1,m^2,m^3)$ and $L_2=\mathcal{O}_X(n^1,n^2,n^3)$
such that

\begin{enumerate}
\item Both $L_1$ and $L_2$ are equivariant:
\begin{equation}
(m_1+m_2)\op{mod}3=0\ , \quad (n_1+n_2)\op{mod}3=0\ .
\end{equation}

\item The five-brane class is effective:
\begin{equation}
W_i=\left(\tfrac{4}{3},\tfrac{7}{3},-4  \right)_i+d_{ijk}m^jm^k+3d_{ijk}n^jn^k\geq0\ .
\end{equation}

\item The cohomology groups $H^1(X, \mathcal{F}\otimes \mathcal{E}^*)=H^1(X, L_1L_2^{-3})$ and $H^1(X, \mathcal{K}\otimes \mathcal{E}^*)=H^1(X, L_1^2)$ are non-zero. In this scan we will simply impose the necessary condition that $\mathcal{F}\otimes \mathcal{E}^*=L_1L_2^{-3}$ and $ \mathcal{K}\otimes \mathcal{E}^*=L_1^2$ are not ample.
%
%
%

\item The extended bundle $V_{\repd 3}^\prime$ is stable. The necessary conditions that we impose are:

(i) The slopes of the line bundles $\mathcal{F}$ and $\mathcal{K}$ are negative:
\begin{equation}
\label{eq:const_2slopes}
 d_{ijk}(m^i+n^i)a^j a^k<0 \ , \qquad -2d_{ijk}n^i a^j a^k<0\ .
\end{equation}

(ii) The extension bundle $V_{\repd 3}^\prime$ satisfies the Bogomolov inequality
\begin{equation}
\label{eq:bogolomov_cons2}
d_{ijk}m^im^ja^k+3 d_{ijk}n^in^ja^k\leq 0\ .
\end{equation}

\end{enumerate}

We performed a systematic scan over all possible pairs of line bundles, with $|m^{i}|\leq 15$, $|n^{i}|\leq 15$ for $i=1,2,3$. The values of $(a^1,a^2,a^3)$ we sample sit inside the ``orange'' subspaces of the K\"ahler cone shown in Figure \ref{fig:PhysContraint}. 
We find a number of pairs of line bundles that lead to a solution satisfying all of these constraints. These include the line bundles
\begin{equation}
L_1=\mathcal{O}_X(-5, -1, 1)\ , \qquad  L_2=\mathcal{O}_X(2,1, -1).
\label{door1}
\end{equation}
 Before giving the other examples, let us analyze this case in more detail. For this pair of line bundles, we find that the class of the five-brane is
\begin{equation}
W_i=(2,  0, 18)_i\ ,
\end{equation}
which is indeed effective ($W_i\geq 0$).
Furthermore, the bundles $L_1L_2^{-3}=\mathcal{O}_X(-11,-4,4)$ and $L_1^2=\mathcal{O}_X(-10,-2,2)$, associated with the $\repd 1_{-1,3}$ and the $\repd 1_{-2,0}$ representations respectively, are clearly non-ample. However, we do not know if the extensions in \eqref{eq:sequences} exist without computing the cohomologies
\begin{align}
\label{eq:cohomology}
&H^1(X, \mathcal{F}\otimes \mathcal{E}^*)=H^1(X, L_1L_2^{-3})=H^1(X, \mathcal{O}_X(-11,-4,4)).\\
&H^1(X, \mathcal{K}\otimes \mathcal{E}^*)\equiv H^1(X, L_1^{2})=H^1(X, \mathcal{O}_X(-10,-2,2))\ .
\end{align}

\subsection{Different Extension Branches}\label{sec:more_branches}

\begin{table}[t]
	\noindent \begin{centering}
		\begin{tabular}{cccc}
			\toprule 
			Branch & Field VEVs & Sequences & Ext\tabularnewline
						\midrule
			\midrule 
			1& $\langle C_2\rangle,\langle \tilde C_3 \rangle \neq 0$&
			 $\begin{matrix} 0\rightarrow \mathcal{F} \rightarrow W\rightarrow \mathcal{E}\rightarrow 0\\ 0\rightarrow \mathcal{K} \rightarrow V_{\repd 3}^\prime \rightarrow W\rightarrow 0 \end{matrix} $  &  
			 $\begin{matrix}H^1(X,\mathcal{F}\otimes \mathcal{E}^*)\\H^1(X,\mathcal{K}\otimes \mathcal{E}^*)\end{matrix}$\tabularnewline
			 \midrule
			 	2& $\langle C_3\rangle,\langle \tilde C_2 \rangle \neq 0$&
			 $\begin{matrix} 0\rightarrow \mathcal{E} \rightarrow W^{(2)}\rightarrow \mathcal{K}\rightarrow 0\\ 0\rightarrow W^{(2)} \rightarrow {V_{\repd 3}^\prime}^{(2)} \rightarrow \mathcal{F}\rightarrow 0 \end{matrix} $  &  
			 $\begin{matrix}H^1(X,\mathcal{E}\otimes \mathcal{K}^*)\\H^1(X,\mathcal{E}\otimes \mathcal{F}^*)\end{matrix}$\tabularnewline
			 \midrule
			 	3& $\langle C_2\rangle,\langle  C_1 \rangle \neq 0$&
			 $\begin{matrix} 0\rightarrow \mathcal{F} \rightarrow W^{(3)}\rightarrow \mathcal{E}\rightarrow 0\\ 0\rightarrow W^{(3)} \rightarrow {V_{\repd 3}^\prime}^{(3)} \rightarrow \mathcal{K}\rightarrow 0 \end{matrix} $  &  
			 $\begin{matrix}H^1(X,\mathcal{F}\otimes \mathcal{E}^*)\\H^1(X,\mathcal{F}\otimes \mathcal{K}^*)\end{matrix}$\tabularnewline
			 \midrule
			 	4& $\langle \tilde C_1\rangle,\langle \tilde C_2 \rangle \neq 0$&
			 $\begin{matrix} 0\rightarrow \mathcal{K} \rightarrow W^{(4)}\rightarrow \mathcal{F}\rightarrow 0\\ 0\rightarrow \mathcal{E} \rightarrow {V_{\repd 3}^\prime}^{(4)} \rightarrow W^{(4)}\rightarrow 0 \end{matrix} $  &  
			 $\begin{matrix}H^1(X,\mathcal{K}\otimes \mathcal{F}^*)\\H^1(X,\mathcal{E}\otimes \mathcal{F}^*)\end{matrix}$\tabularnewline
			 \midrule
			 	5& $\langle C_3\rangle,\langle C_1 \rangle \neq 0$&
			 $\begin{matrix} 0\rightarrow \mathcal{E} \rightarrow W^{(5)}\rightarrow \mathcal{K}\rightarrow 0\\ 0\rightarrow \mathcal{F} \rightarrow {V_{\repd 3}^\prime}^{(5)} \rightarrow W^{(5)}\rightarrow 0 \end{matrix} $  &  
			 $\begin{matrix}H^1(X,\mathcal{E}\otimes \mathcal{K}^*)\\H^1(X,\mathcal{F}\otimes \mathcal{K}^*)\end{matrix}$\tabularnewline
			 \midrule
			 	6& $\langle \tilde C_1\rangle,\langle \tilde C_3 \rangle \neq 0$&
			 $\begin{matrix} 0\rightarrow \mathcal{K} \rightarrow W^{(6)}\rightarrow \mathcal{F}\rightarrow 0\\ 0\rightarrow W^{(6)} \rightarrow {V_{\repd 3}^\prime}^{(6)} \rightarrow \mathcal{E} \rightarrow 0 \end{matrix} $  &  
			 $\begin{matrix}H^1(X,\mathcal{K}\otimes \mathcal{F}^*)\\H^1(X,\mathcal{K}\otimes \mathcal{E}^*)\end{matrix}$\tabularnewline
			\bottomrule
		\end{tabular}
		\par\end{centering}
		\caption{The six extension branches of the split Whitney sum bundle $V_{\repd 3}=\mathcal{F}\oplus \mathcal{K}\oplus
		\mathcal{E}$. For each branch, there is also a second pair of sequences which corresponds to switching the order of the extensions. The resulting bundles can be shown to be isomorphic~\cite{Anderson:2010ty} and, hence, we do not display them.}
		\label{tab:branches}
\end{table}

In the example in the previous section, we showed it might be possible to extend the decomposable bundle $V_{\repd 3}$, defined in \eqref{eq:defineBundle2}, to a non-trivial stable $SU(3)$ bundle $V_{\repd 3}^\prime$ via the extension sequences defined in \eqref{eq:sequences}. As we discussed, this extension is equivalent to turning on VEVs for the fields $C_2$ and $\tilde C_3$ in the four-dimensional effective theory. We now ask if we can still solve the system of vacuum constraints if we chose a different extension sequence or, equivalently, if we chose to turn on VEVs for different combinations of $C$ and $\tilde C$ fields. 

First of all, not all combinations of pairs of VEVs are allowed. The F-flatness conditions, coming from the vanishing of the superpotential and its first derivative, reduce the fifteen combinations of pairs of VEVs to six~\cite{Anderson:2010ty}. In Table \ref{tab:branches} we give the allowed VEVs for each branch and the corresponding extension sequences. The branch we studied in the previous section corresponds to branch one in this table. 

\subsection{Possible Generalization}

We finish by pointing out that the analyses and methods developed in this work can be extended to models with hidden sectors built using different embeddings into $E_{8}$ or from three or more line bundles. In particular, the method of line bundle vectors~\cite{Nibbelink:2016wms} reviewed in Section \ref{sec:embeddings} is ideally suited for this problem. One could imagine algorithmically generating hidden sectors built from line bundles. As a taste of how this would work, we finish with two such examples that should make clear that there are a large number of models to explore.

Consider for instance another example built from two line bundles. We use two linearly independent generators $(\boldsymbol{t}_{1},\boldsymbol{t}_{2})$ given by
\begin{align}
	\boldsymbol{t}_{1} & =(0,0,0,0,0,0,0,-2)\ ,\\
	\boldsymbol{t}_{2} & =(0,0,0,0,0,0,-2,0)\ ,
\end{align}
with the two line bundle vectors given by
\begin{equation}
\boldsymbol{V}_{i}=m^{i}\boldsymbol{t}_{1}+n^{i}\boldsymbol{t}_{2}\ .
\end{equation}
The non-Abelian commutant of $U(1)^{2}$ inside $E_{8}$ is then $SO(12)$. 

 Using the \emph{Mathematica }package LieART~\cite{Feger:2012bs,Feger:2019tvk}, one can then find the decomposition of the adjoint representation of $\Ex 8$:
\begin{equation}
	\label{eq:decompSO12big}
	\begin{split}
		\repd{248}&=\repd{66}_{0,0}+2\times\repd{1}_{0,0}+\repd{1}_{2,0}+\repd{1}_{-2,0}+\repd{1}_{0,2}+\repd{1}_{0,-2}+
		\repd{32}_{1,0}+\repd{32}_{-1,0}+\repd{32}_{1,1}+\repd{32}_{1,-1}\\
		&\eqspace+\repd{12}_{1,1}+\repd{12}_{-1,1}+\repd{12}_{1,-1}+\repd{12}_{-1,1}\ .\\
	\end{split}
\end{equation}

This breaking pattern is obtained by first breaking $E_8$ to $E_7\times SU(2)$, under which the adjoint representation of $E_8$ decomposes as
\begin{equation}
\label{eq:SO12first}
\bf{248}=(\repd{133},\repd{1})+(\repd{56},\repd{2})+(\repd{1},\repd{3})\ .
\end{equation}
Breaking $E_7$ further to $SO(12)\times SU(2)$, the $E_7$ representations that appear above decompose as
\begin{equation}
\label{eq:SO12second}
\begin{split}
\repd{56}&=(\repd{12},\repd{2})+(\repd{32},\repd{1})\ ,\\
\repd{133}&=(\repd{66},\repd{1})+(\repd{32},\repd{2})+(\repd{1},\repd{3})\ ,\\
\repd{1}&=(\repd{1},\repd{1})\ .\\
\end{split}
\end{equation}
Finally, we break both $SU(2)$ groups down to $U(1)$s. The fundamental representation of $SU(2)$ decomposes as
\begin{align}
\repd{2}&=\repd{1}_{1}+\repd{1}_{-1}\ ,\\
\end{align}
Putting this together, one sees that the adjoint representation of $E_8$ decomposes under $SO(12)\times U(1)\times U(1)$ exactly as in \eqref{eq:decompSO12big} above.

Each of the two line bundles $L_1$ and $L_2$ associated with the two $U(1)$ factors are embedded into an $SU(2)$ bundle. At the decomposable locus we produce a diagonal $SU(2)\times SU(2)$ bundle
\begin{equation}
\label{eq:defineBundlesSU222}
\begin{split}
V_{2\times 2}=( L_1^{-1}\oplus L_1)\oplus( L_2^{-1}\oplus L_2)\ .
\end{split}
\end{equation}
with $U(1)\times U(1)$ structure group

In this case, the only singlets under the non-Abelian factor that are charged under the two $U(1)$s are $(\repd 1)_{2,0}$, $(\repd 1)_{0,2}$ and their conjugates. These give charged singlet matter fields whose VEVs may be used to set the D-terms associated with the low-energy $U(1)$s to zero. Turning a combination of two VEVs would correspond to deforming the hidden sector bundle with $U(1)\times U(1)$ structure group into one with $\SU 2\times \SU 2$ instead.

One could also go beyond the two line bundle embedding scenario and build a hidden sector from three line bundles. We use three linearly independent generators $(\boldsymbol{t}_{1},\boldsymbol{t}_{2},\boldsymbol{t}_{3})$ given by
\begin{align}
	\boldsymbol{t}_{1} & =(0,0,0,0,0,0,0,-2)\ ,\\
	\boldsymbol{t}_{2} & =(0,0,0,0,0,-1,-1,-2)\ ,\\
	\boldsymbol{t}_{3} & =(0,0,0,0,0,0,-1,-1)\ ,
\end{align}
with the three line bundle vectors given by
\begin{equation}
\boldsymbol{V}_{i}=m^{i}\boldsymbol{t}_{1}+n^{i}\boldsymbol{t}_{2}+p^{i}\boldsymbol{t}_{3}\ .
\end{equation}
Using \eqref{eq:unbroken_condition}, one can then show that the non-Abelian commutant of $U(1)^{3}$ inside $E_{8}$ is $SO(10)$, so that the low-energy gauge group is $SO(10)\times U(1)^{3}$. It is then simple to find the decomposition of the $\repd{248}$:
\begin{align}
	\repd{248} & \to\repd 1_{2,3,4}+\repd 1_{2,3,1}+\repd 1_{2,1,1}+\repd 1_{2,1,0}+\repd 1_{0,2,1}+\repd 1_{0,0,1}+\text{c.c.}\\
	& \eqspace+3\repd 1_{0,0,0}+\repd{45}_{0,0,0}\\
	& \eqspace+\repd{10}_{2,2,1}+\repd{10}_{0,1,1}+\repd{10}_{0,1,0}+\text{c.c.}\\
	& \eqspace+\repd{16}_{1,2,1}+\repd{16}_{1,0,0}+\repd{16}_{-1,-1,0}+\repd{16}_{-1,-1,-1}+\text{c.c.}
\end{align}

This pattern corresponds to first breaking the $E_8$ group to the maximal subgroup $SO(16)$, and then breaking further the $SO(16)$ to $SU(4)\times SO(10)$. One could then embed the $U(1)\times U(1)\times U(1)$ structure group into the $SU(4)$ connection to obtain a diagonal $V_{\repd 4}$ bundle at the decomposable locus of the type
\begin{equation}
\label{eq:defineBundlesSU223}
\begin{split}
V_{\repd 4}=&	L_3^{-3}\oplus\left(L_2^{-2}\oplus(L_1^{-1}\oplus L_1)\otimes L_2\right)\otimes L_3\ .\\
=&L_3^{-3}\oplus L_2^{-2}L_3\oplus L_1L_2L_3\oplus L_1^{-1}L_2L_3\ .
\end{split}
\end{equation}
In the expression above, $L_1$, $L_2$ and $L_3$ are the line bundles associated with the three $U(1)$ groups.

We see there are multiple $SO(10)$ singlets that are charged under the three $U(1)$s. These are the charged singlet matter fields whose VEVs may be used to set the D-terms for the three $U(1)$s to zero. From a ten-dimensional perspective, these VEVs correspond to deforming the hidden sector bundle constructed from the three line bundles to a non-trivial $\SU 4$ extension bundle. One can then impose conditions on these line bundles, following the analysis laid out in this section.

Clearly, there are a large number of hidden sectors that one can construct using different numbers of line bundles and various embeddings into $E_{8}$.

\section{Breaking Supersymmetry at Low Energy}\label{sec:low_energy_gaugino}

In the previous sections, we presented the constraints imposed on a heterotic M-theory vacuum whose hidden sector is defined by a single line bundle $L$ with its $U(1)$ structure group embedded into the $SU(2)$ subgroup of $SU(2) \times E_{7} \subset E_8$ via the induced vector bundle $L \oplus L^{-1}$. We demanded that $d=4$,  $N=1$ supersymmetry be exactly preserved.  In this section, however, we will analyze how spontaneous supersymmetry breaking in four dimensions can occur due to gaugino condensation of $E_{7}$ in the hidden sector.

As is well known \cite{Dine:1985rz,Lukas:1997rb,Nilles:1998sx}, if the $E_{7}$ gauge group of the hidden sector becomes strongly coupled below the compactification scale, then the associated gauginos condense and produce an effective superpotential for the relevant geometric moduli of the theory -- in our case, the dilaton $S$ and the complexified K\"ahler moduli $T^{i}$, $i=1,2,3$. The form of this gaugino condensate superpotential is given by~\cite{Dine:1985rz,Lukas:1997rb,Lukas:1999kt}
\begin{equation}
W=\langle M_{U} \rangle ^3 \exp\left({-\frac{6\pi}{b_L \hat{\alpha}_{\text{GUT}}}f_2}\right)\ ,
\label{sup1}
\end{equation}
where $\hat{\alpha}_{\text{GUT}}$ was defined in \eqref{bag1}, and $b_{L}$ is a real number associated with the beta-function of the $E_{7}$ gauge coupling $g^{(2)}$ for a choice of hidden sector line bundle $L$. Both $b_{L}$  and the associated ``condensation scale'' $\Lambda$ will be defined below.

Remember that to order $\kappa_{11}^{4/3}$, the general expressions for the gauge kinetic functions are
 in~\cite{Lukas:1998hk, Brandle:2003uya}
\begin{equation}
\begin{split}
\label{eq:2fs}
f_1&=S-\frac{\epsilon_S^\prime}{2}\left(\beta_i^{(N+1)}T^i+2\sum_{n=1}^N Z^{(n)}\right)\ ,\\
f_2&=S+\frac{\epsilon_S^\prime}{2}\beta_i^{(N+1)}T^i\ .
\end{split}
\end{equation}
The real parts of the gauge kinetic functions are related to the gauge couplings on the two sectors
\begin{equation}
  \frac{4\pi}{(g^{(1)})^2} \propto
  \text{Re}(f_1)=V \left(1+\epsilon_S' \frac{\Rhat}{2V} 
   \sum_{n=0}^{N}(1-z_n)^2 b^i \beta^{(n)}_i \right)
\label{62o}
\end{equation}
and 
\begin{equation}
  \frac{4\pi}{(g^{(2)})^2} \propto
  \text{Re}(f_2)=V \left(1+\epsilon_S' \frac{\Rhat}{2V} 
  \sum_{n=1}^{N+1}z_n^2 b^i\beta^{(n)}_i\right)
  \label{63o}
\end{equation}
respectively. For a single  line bundle $L=\mathcal{O}_X(l^1,l^2,l^3)$  in the hidden sector and a single five-brane, we find that
\begin{equation}
f_2=S+\frac{\epsilon_S^\prime}{2}\bigl( -(2,2,0)_{i}-d_{ijk}l^{j}l^{k}\bigr)T^{i} \ ,
\label{sup2}
\end{equation}
where $T^{i}=t^{i}+\ii 2\chi$,  $i=1,2,3$ and the strong coupling parameter $\epsilon_S^\prime$ was defined in \eqref{soc2}.
We want to emphasize that expressions \eqref{sup1} and \eqref{sup2} for the condensate superpotential and $f_{2}$ respectively are known only to linear order in $\epsilon_S^\prime$~\cite{Lukas:1997fg,Lukas:1998tt,Brandle:2001ts,Brandle:2003uya}.

An important point is that the potential induced by the gaugino superpotential is actually runaway and so does not fix the moduli fields at a minimum. Instead, we simply assume that the moduli can be fixed to some values (such as by picking a point in the viable region of K\"ahler moduli space) and then analyze the supersymmetry breaking due to the gaugino superpotential at those values of the moduli. In a complete model, one would be able to fix all of the moduli fields dynamically. However, this is far outside the scope of the present work.\footnote{One could use gaugino condensation in multiple gauge groups to generate a potential with a determined minimum and thus fix some of the moduli. However, this would still leave bundle moduli, complex structure moduli, and so on, to fix~\cite{Anderson:2010mh, Anderson:2011cza, Anderson:2011ty,Correia:2007sv}. To the knowledge of the authors, within the context of phenomenologically realistic models, there are \emph{no} models that fix all moduli.}

The observable sector $Spin(10)$ gauge coupling $g^{(1)}$ is fixed at the unification scale $\langle M_U \rangle$ based on phenomenological data. As discussed in \cite{Ashmore:2020ocb} and Section 2, we find that $\langle \alpha_{u}\rangle=\frac{1}{20.08}$ in the split Wilson lines scenario and $\langle \alpha_{u}\rangle=\frac{1}{26.64}$ in the simultaneous Wilson lines scenario (with $\alpha=g^{(1)2}/4\pi$). We cannot determine the value of the hidden sector gauge coupling $g^{(2)}$ at $\langle M_U \rangle$ based on direct observation. However, the two gauge couplings $g^{(1)}$ and $g^{(2)}$ both occur in the $d=10$ Hořava--Witten theory and, hence, it is possible to find an expression relating them to couplings in the $d=4$ effective theory after compactification. As shown in eq. \eqref{bag1}, in the observable sector the $Spin(10)$ gauge coupling at the unification scale is given by~\cite{Banks:1996ss,Lukas:1997fg}
\begin{equation}
\langle \alpha_{u}\rangle=\frac{\hat \alpha_{\text{GUT}}}{\re f_1}\ ,
\label{need2}
\end{equation}
where $f_{1}$ is the gauge kinetic function on the observable sector. Similarly, the $E_7$ hidden sector gauge coupling at the unification scale $\langle M_{U} \rangle$ is defined to satisfy
\begin{equation}
\langle \alpha^{(2)}_{u}\rangle=\frac{\hat \alpha_{\text{GUT}}}{\re f_2}\ ,
\label{need3A}
\end{equation}
where we denote the hidden sector gauge parameter by $\alpha^{(2)}=\frac{g^{(2)2}}{4\pi}$. The moduli dependent function $f_{2}$ was given in \eqref{sup2}. Using the relations 
\begin{equation}
\re S=V+\frac{\epsilon_{S}'}{2}\bigl(\tfrac{1}{2}+\lambda\bigr)^{2}W_{i}t^{i} 
\label{cl1}
\end{equation}
and
\begin{equation}
\re T^{i}=t^{i}=\frac{\hat{R}}{V^{1/3}} \ ,
\label{cl2}
\end{equation}
it follows from \eqref{sup2} that 
\begin{equation}
\re f_{2}=V+\epsilon_{S}'\frac{\hat{R}}{V^{1/3}}\left(\bigl(-(1,1,0)_{i}-\tfrac{1}{2}d_{ijk}l^{j}l^{k}\bigr)a^{i}+\bigl(\tfrac{1}{2}+\lambda\bigr)^{2}W_{i}a^{i}\right) \ .
\label{cl3}
\end{equation}
Finally, using \eqref{need2} and \eqref{need3A}, we find that
\begin{equation}
\langle \alpha^{(2)}_{u}\rangle=\frac{\re f_1}{\re f_2}\langle \alpha_{u}\rangle ,
\end{equation}
allowing one to solve for $\langle \alpha^{(2)}_{u}\rangle$ at any point in moduli space. Remember that inside the orange subspaces of the K\"ahler cone shown in Figure \ref{fig:PhysContraint}, ${\rm{Re}}f_{1}$ and ${\rm{Re}}f_{2}$ have positive definite values. Hence, $\langle \alpha^{(2)}_{u}\rangle$ is well defined within that subspace.

For an arbitrary momentum $p$ below the unification scale, the renormalization group equation for the hidden sector gauge parameter $\alpha^{(2)}$ is given by
\begin{equation}
\alpha^{(2)}(p)^{-1}=\langle \alpha_{u}^{(2)}\rangle^{-1}-\frac{b_L}{2\pi} \ln \left(\frac{ \langle M_U \rangle}{p}\right)\ .
\label{tea1}
\end{equation} 
Note that for $b_{L} < 0$, the value of $\alpha^{(2)}(p)^{-1}$ is identical to the perturbative gauge coupling $\langle \alpha_{u}^{(2)}\rangle^{-1}$ for $p= \langle M_U \rangle$ and only becomes more weakly coupled for $p< \langle M_U \rangle$. It follows that the $E_{7}$ gauge group never becomes strongly coupled in the effective theory and, therefore, gaugino condensation can never occur. Therefore, we will only consider line bundles $L$ for which 
\begin{equation}
b_{L}>0. 
\label{bl1}
\end{equation}
In this case,
roughly speaking, the hidden sector $E_{7}$ gauge theory becomes strongly coupled and, hence, its gauginos condense, at a momentum $p \approx \Lambda$ where $ \alpha^{(2)}(\Lambda)^{-1}$ can be  well approximated by $0$. It then follows from \eqref{tea1} that, at this scale, 
\begin{equation}
\langle \alpha_{u}^{(2)}\rangle^{-1}=\frac{b_L}{2\pi} \ln \left(\frac{\langle M_U\rangle}{\Lambda}\right)\ .
\label{tea2}
\end{equation}
The condensation scale $\Lambda$ can then be expressed as 
\begin{equation}
\label{eq:condesation_scale}
\Lambda=\langle M_U\rangle \ee^{\frac{-2\pi}{b_L} \langle \alpha^{(2)}_{u}\rangle^{-1}}=\langle M_U \rangle \ee^{\frac{-2\pi}{b_L}\frac{\re f_2}{\re f_1\langle \alpha_u \rangle}} \ .
\end{equation}
$\Lambda$, like $W$ defined in \eqref{sup1}, is a function of the K\"ahler moduli. It remains to compute the coefficient $b_L$. 

For any line bundle $L=\mathcal{O}_X(l^1,l^2,l^3)$ in the hidden sector satisfying all constraints in Section 2, the beta-function coefficient $b_L$ for the $E_{7}$ gauge coupling is given by
\begin{equation}
b_L=3\,T(\Rep{133})-\sum_{\Rep{r}}n_{\Rep r}T(\Rep{r}) \ .
\label{hope1}
\end{equation}
Here, the sum is over the $E_{7}$ representations $\Rep{r}$ that arise in the decomposition of the of the adjoint representation $\Rep{248}$ of $E_{8}$ with respect to the low-energy hidden sector gauge group $U(1) \times E_{7}$, and the coefficients $n_{\Rep{r}}$ are the number of light chiral matter fields which transform as $\Rep{r}$.  $T(\Rep{r})$ denotes the Dynkin index of the representation $\Rep r$, defined by
\begin{equation}
T(\Rep{r})=\text{tr}_{\Rep{r}}{\mathrm T}_a^2 \ ,
\end{equation}
where ${\mathrm T}_a$ is an arbitrary Lie algebra generator of $E_{7}$ in the representation $\Rep{r}$. See, for example, \cite{Yamatsu:2015npn}. It follows that calculating the beta-function coefficient $b_{L}$ for a specific line bundle $L$ requires one to explicitly compute the particle content of its low-energy theory. To do this, we first recall that all line bundles under consideration have their $U(1)$ structure group embed into the $SU(2)$ subgroup of $SU(2) \times E_{7} \subset E_{8}$. It follows that for all such line bundles, the adjoint $\Rep{248}$ representation of the hidden sector $E_8$ decomposes under $U(1) \times\Ex 7$ as
\begin{equation}
\Rep{248} \to 
(0, \Rep{133}) \oplus 
\bigl( (1, \Rep{56}) \oplus (-1, \Rep{56})\bigr) \oplus 
\bigl( (2, \Rep{1}) \oplus (0, \Rep{1}) \oplus (-2, \Rep{1}) \bigr)\ .
\label{red55}
\end{equation}
The $(0,\Rep{133})$ corresponds to the adjoint representation of $\Ex 7$, while the $(\pm1,\Rep{56})$ give rise to chiral matter superfields with  $\pm1$ $U(1)$ charges transforming in the $\underline{\bf56}$ representation of $\Ex 7$ in four dimensions. 
The $(\pm2,\Rep 1)$ are $E_7$ singlet chiral superfields fields with charges $\pm 2$ under $U(1)$. Finally, the $(0,\Rep{1})$ gives the one dimensional adjoint representation of the $U(1)$ gauge group. The embedding of the line bundle is such that fields with $U(1)$ charge $-1$ are counted by $H^{*}(X,L)$, charge $-2$ fields are counted by $H^{*}(X,L^{2})$ and so on.

The low-energy massless spectrum can be determined by examining the chiral fermionic zero-modes of the Dirac operators for the various representations in the decomposition of the $\Rep{248}$. The Euler characteristic $\chi(\mathcal{F})$ counts $n_{\text{R}}-n_{\text{L}}$, where $n_{\text R}$ and $n_{\text L}$ are the number of right- and left-chiral zero-modes respectively transforming under the group representation associated with the bundle $\mathcal{F}$. With the notable exception of $\mathcal{F}=\mathcal{O}_{X}$, which corresponds  to the massless vector superfields of both the  $(0, \Rep{133})$ and $(0, \Rep{1})$ adjoint representations of $E_{7}$ and $U(1)$ respectively, right-chiral and left-chiral zero modes pair up and form massive fermion states, which can then be integrated out of the low-energy theory. However, the remaining unpaired zero modes of the $d=4$ effective theory remain massless and are precisely those counted by the Euler characteristic for each representation. On a Calabi--Yau threefold $X$,  $\chi(\mathcal{F})$ can be computed using the Atiyah--Singer index theorem and is given by
\begin{equation}
\chi(\mathcal{F})=\int_{X}{\rm {ch}}(\cal{F}) \wedge {\rm {Td}}({\rm X}) \ ,
\label{ni1}
\end{equation}
where ${\rm {ch}}(\cal{F})$ is the Chern character of $\mathcal{F}$ and ${\rm {Td}}({\rm X})$ is the Todd class of the tangent bundle of $X$.
It is useful to note that for a line bundle of the form ${\cal{F}}=\mathcal{O}_X(f^1,f^2,f^3)$, expression \eqref{ni1} simplifies to
\begin{equation}
\chi({\cal{F}})= \frac{1}{3}(f^{1}+f^{2})+\frac{1}{6}d_{ijk}f^{i}f^{j}f^{k} \ .
\label{st1}
\end{equation}
For the decomposition of the $\Rep{248}$ presented in \eqref{red55}, the bundles ${\cal{F}}$ corresponding to the $U(1) \times E_7$ representations $(0, \Rep{133})$, $(\pm1, \Rep{56})$, $(\pm2, \Rep{1})$ and 
$(0, \Rep{1})$ are $\mathcal{F}=\mathcal{O}_{X}, L^{\mp1}, L^{\mp2}$ and $\mathcal{O}_{X}$ respectively. Using the fact that the $E_{7}$ singlet states $(\pm2, \Rep{1})$ and $(0, \Rep{1})$ cannot contribute to $b_{L}$, that is, 

\begin{equation}
T({\Rep1})=0 \ ,
\label{no1}
\end{equation}
we need only consider $(0, \Rep{133})$ and$(\pm1, \Rep{56})$. Furthermore, since the $E_{7}$ adjoint representation can occur only once in the spectrum, it follows that we need to determine only the number of states $(\pm1, \Rep{56})$. Of these, the low-energy spectrum consists of the left chiral states only, which correspond to $(+1, \Rep{56})$. They are counted by the Euler characteristic of the line bundle $L^{-1}$. It follows from \eqref{st1} that for $L=\mathcal{O}_X(l^1,l^2,l^3)$, 
\begin{equation}
\chi(L^{-1})=-\frac{1}{6}\bigl(2l^1+2l^2+({l^1})^2l^2+l^{1}({l^2})^2+6l^{1}l^2l^3\bigr)\ .
\label{tr1}
\end{equation}

With this information about the low-energy spectrum, we can now compute the $b_L$ coefficient for any line bundle $L=\mathcal{O}_X(l^1,l^2,l^3)$ in the hidden sector satisfying all constraints in Section 2. 
Noting \cite{Yamatsu:2015npn} that the Dynkin indices satisfy \eqref{no1} and 
 \begin{equation}
  T(\Rep{133})=18 \quad T(\Rep{56})=6 \ ,
  \label{cut1}
 \end{equation}
we find, using \eqref{hope1}, that
 \begin{equation}
 \label{beta_expr}
 b_{L}=3\times T(\Rep{133})-|\chi(L^{-1})|\times T(\Rep{56})=54-6|\chi(L^{-1})| 
 \end{equation} 
 and, hence, from \eqref{tr1} that
 \begin{equation}
 b_{L}=54-| 2 l^{1} +2 l^{2}+(l^{1})^2 l^2+l^{1}(l^{2})^2+6l^{1}l^{2}l^{3}| \ .
 \label{cut2}
 \end{equation} 
 We can now explain why, out of all the line bundles satisfying the constraints in Section \ref{sec:susy_vacua_4d}, only the seven ample line bundles listed in \eqref{many} can exhibit $E_{7}$ gaugino condensation. Furthermore, we can now justify the preference for the line bundle $L=\mathcal{O}_X(2,1,3)$ in our previous sections and the rest of the present work.

In Section \ref{sec:susy_vacua_4d} we introduced a new constraint; that is, we demanded that the line bundle $L$ be such that $N=1$ supersymmetry is spontaneously broken by the condensation of the $E_{7}$ gauginos. However, as discussed above, for momenta below the unification scale $\langle M_{U} \rangle$, the $E_7$ gauge theory becomes strongly coupled if and only if condition \eqref{bl1} is satisfied. Therefore, we now impose, using \eqref{cut2}, the extra condition that
\begin{equation}
54-\left|2 l^{1} +2 l^{2}+(l^{1})^2 l^2+l^{1}(l^{2})^2+6l^{1}l^{2}l^{3}\right|>0\ .
\label{al1}
\end{equation}
We find that the only line bundles that satisfy all the constraints in Section 2, as well as this new constraint, are 
\begin{gather}
\mathcal{O}_X(2,1,3)\ , \qquad \mathcal{O}_X(1,2,3)\ , \qquad \mathcal{O}_X(1,2,2)\ ,  \qquad\mathcal{O}_X(2,1,2)\ , \nonumber \\
\mathcal{O}_X(2,1,1)\ ,  \qquad \mathcal{O}_X(1,2,1)\ ,  \qquad\mathcal{O}_X(2,1,0) \ ,
\label{many1A}
\end{gather}
which are exactly those presented in \eqref{many}. Calculating the $b_{L}$ coefficient for each of these bundles using \eqref{cut2}, we find that
\begin{itemize}
\item{$\mathcal{O}_X(2,1,3)$ and $\mathcal{O}_X(1,2,3)$: $b_{L}=6$;}
\item{$\mathcal{O}_X(2,1,2)$ and $\mathcal{O}_X(1,2,2)$: $b_{L}=18$;}
\item{$\mathcal{O}_X(2,1,1)$ and $\mathcal{O}_X(1,2,1)$: $b_{L}=30$;}
\item{ $\mathcal{O}_X(2,1,0)$: $b_{L}=42$.}
\end{itemize}
For any other line bundle satisfying all the constraints given in Section 2, one can show that $b_{L}<0$. For example, 
for $\mathcal{O}_X(2,1,4)$ and $\mathcal{O}_X(1,2,4)$, $b_{L}$ has already become negative; that is $b_{L}=-6$. 

So which of the above line bundles is, from the point of view of low-energy phenomenology, most interesting to study? It is clear from expression \eqref{eq:condesation_scale} that the gaugino condensation scale $\Lambda$, at any fixed point in K\"ahler moduli space, will be the smallest for the line bundle $L$ with the lowest value of $b_{L}$. For this reason, we will focus on the two line bundles $\mathcal{O}_X(2,1,3)$ and $\mathcal{O}_X(1,2,3)$, each of which was shown above to have $b_{L}=6$. We also find that the ``viable'' region of K\"ahler moduli space, is considerably larger for the first of these two bundles. Therefore, we will focus on the line bundle $\mathcal{O}_X(2,1,3)$.
For completeness, in Table \ref{tab:chiral_spectrum2} we present the complete low energy spectrum, the Euler characteristics and the Dynkin coefficients for the line bundle $L=\mathcal{O}_{X}(2,1,3)$.
 
 \begin{table}
	\noindent \begin{centering}
		\begin{tabular}{rrrr}
			\toprule 
			$U(1) \times \Ex 7$ & Cohomology & Index $\chi$ &$T(\Rep{r})$\tabularnewline
			\midrule
			\midrule 
			$(0,\Rep{133})$ & $H^{*}(X,\mathcal{O}_{X})$ & $0$&18\tabularnewline
			\midrule 
			$(0,\Rep 1)$ & $H^{*}(X,\mathcal{O}_{X})$ & $0$&0\tabularnewline
			\midrule 
			$(-1,\Rep{56})$ & $H^{*}(X,L)$ & $8$&6\tabularnewline
			\midrule 
			$(1,\Rep{56})$ & $H^{*}(X,L^{-1})$ & $-8$&6\tabularnewline
			\midrule 
			$(-2,\Rep 1)$ & $H^{*}(X,L^{2})$ & $58$&0\tabularnewline
			\midrule 
			$(2,\Rep 1)$ & $H^{*}(X,L^{-2})$ & $-58$&0\tabularnewline
			\bottomrule
		\end{tabular}
		\par\end{centering}
	\caption{The chiral spectrum for the hidden sector $\protect\Uni 1\times\protect\Ex 7$ with a single line bundle $L=\mathcal{O}_{X}(2,1,3)$. The Euler characteristic (or index) $\chi$ gives the difference between the number of right- and left-chiral fermionic zero-modes transforming in the given representation. We denote the line bundle dual to $L$ by $L^{-1}$ and the trivial bundle $L^{0}$ by $\mathcal{O}_{X}$.\label{tab:chiral_spectrum2}}
\end{table}

\begin{figure}[t]
   \centering
     \begin{subfigure}[b]{0.495\textwidth}
\includegraphics[width=1.0\textwidth]{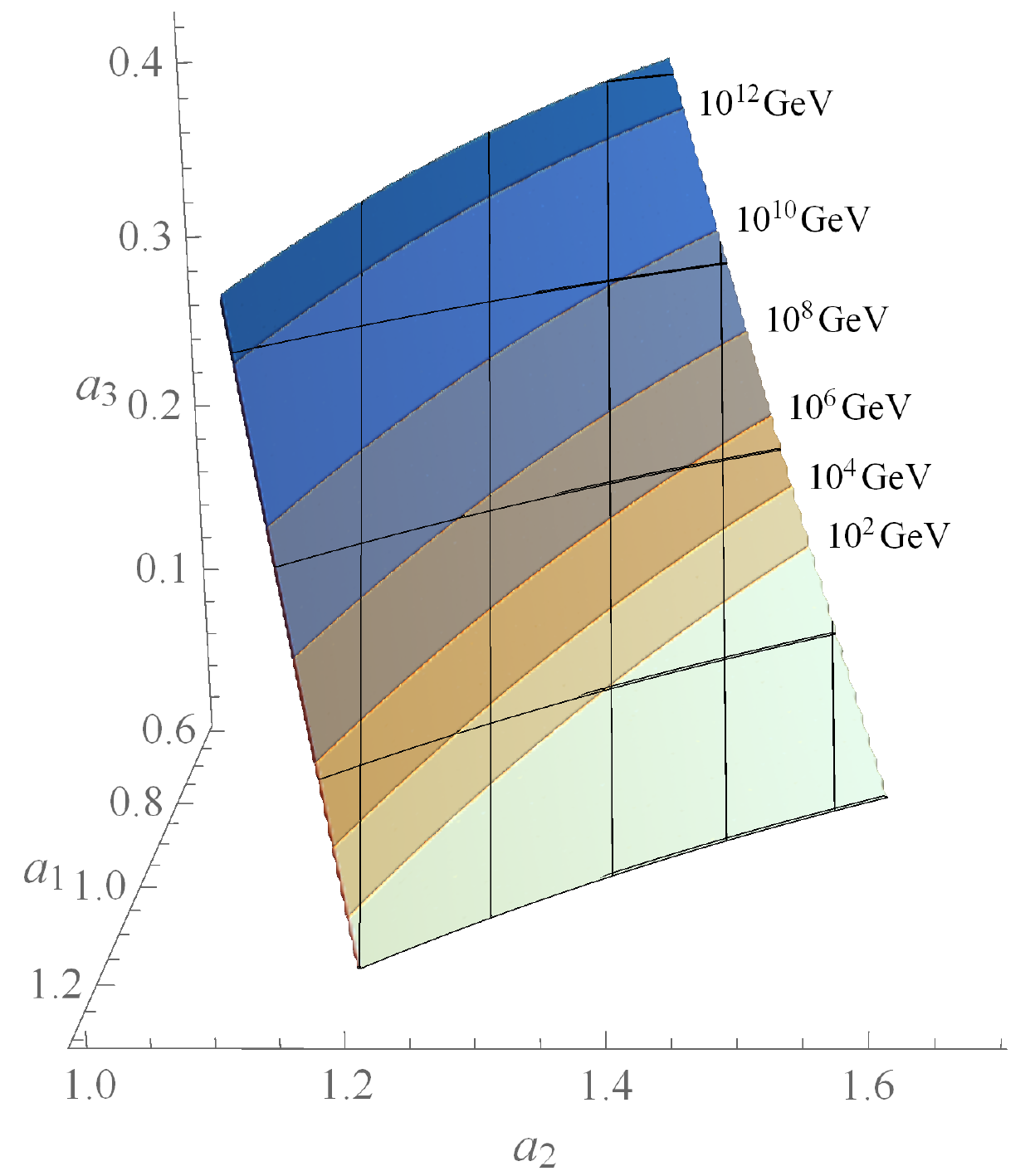}
\caption{$\langle \alpha_{u} \rangle=\frac{1}{20.08}$}
\end{subfigure}
     \begin{subfigure}[b]{0.495\textwidth}
\includegraphics[width=1.0\textwidth]{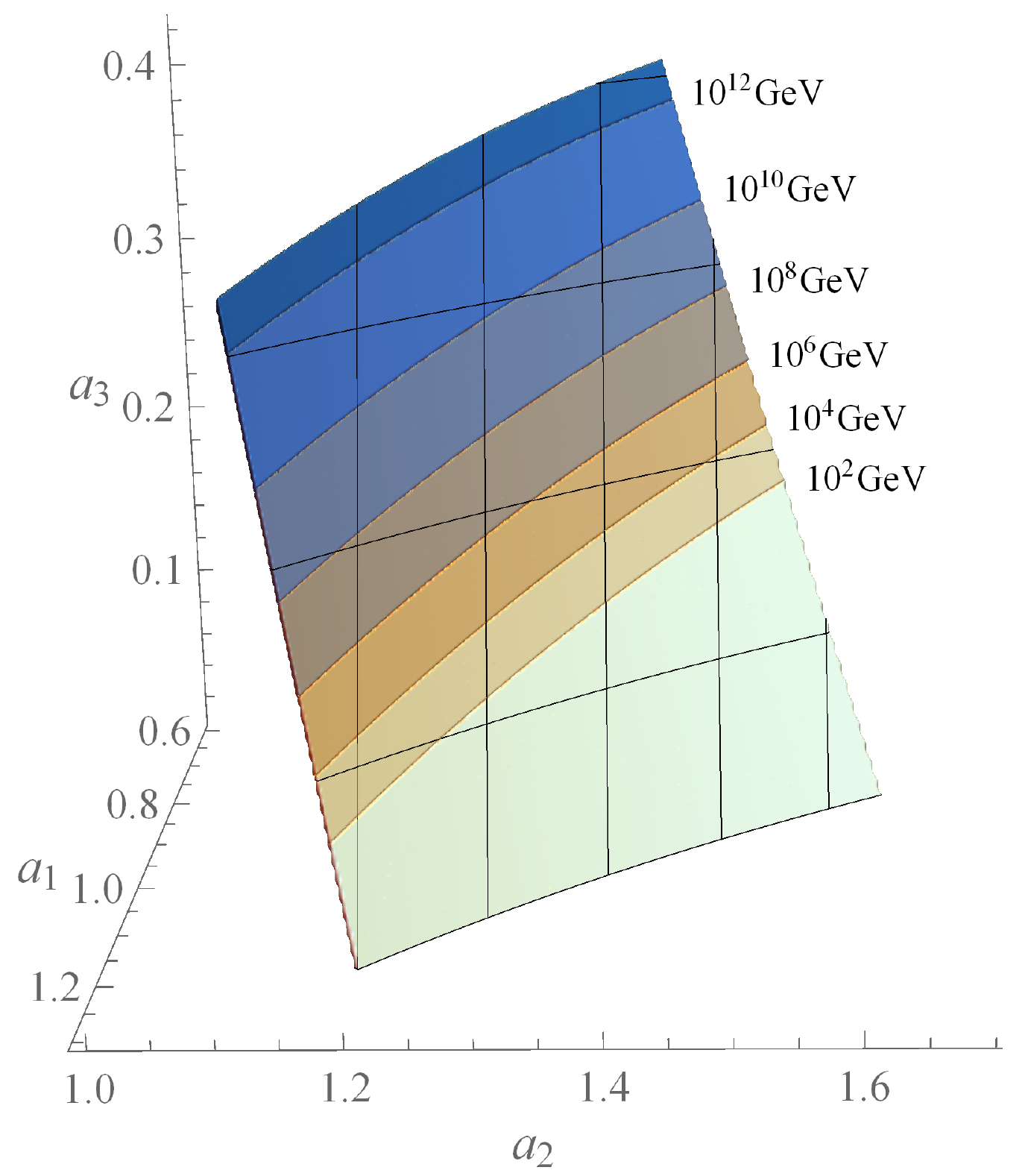}
\caption{$\langle \alpha_{u} \rangle=\frac{1}{26.46}$}
\end{subfigure}\\
\caption{Variation of the mass scale $m_{\text{susy}}\sim 8\pi\Lambda^3/M_p^2$ of the soft breaking terms across the ``viable'' region of K\"ahler moduli space displayed with magenta in Figure \ref{fig:Intersection}, for the line bundle $L= \mathcal{O}_X(2,1,3)$. The numbers indicate the $m_{\text{susy}}$ value corresponding to each contour.  Figures (a) and (b) respectively, show the results for both the split and the simultaneous Wilson lines scenarios.  $m_{\text{susy}}$ scales below the EW scale $\sim10^2$ GeV become unphysically small and, therefore, are not displayed.}
\label{fig:susyScale_both}
\end{figure}

In a following section, we will explicitly calculate the soft supersymmetry breaking terms in the observable matter sector. Here, however, we can use what we have learned so far to predict the scale of those soft SUSY breaking terms. 
Supersymmetry breaking first occurs in the $S$ and $T^{i}$ moduli via the gaugino condensate superpotential $W$ given in \eqref{sup1}. Here, it is useful to note that writing $f_{2}=\re f_{2}+\ii \imag f_{2}$, and using \eqref{need3A} and \eqref{tea2}, it follows that 
\begin{equation}
W=\Lambda^{3}  \ee^{-\ii \frac{6 \pi}{b_{L}\hat{\alpha}_{\text GUT}}\imag f_{2}} \ .
\label{mk1}
\end{equation}
The supersymmetry breaking in the $S$ and $T^{i}$ moduli is then gravitationally mediated to the observable matter sector. The scale of SUSY breaking in the low-energy observable matter sector is then of order
\begin{equation}
m_{\rm{susy}} \sim \kappa_{4}^{2} \Lambda^3=8 \pi \frac{\Lambda^{3}}{M_{P}^{2}} \ ,
\label{ew1}
\end{equation}
where $\Lambda$ is the condensation scale given in \eqref{eq:condesation_scale} and we have used the fact that $\kappa_{4}^{2}=8\pi/M_{P}^{2}$.

For the configuration we studied in the previous section, with a hidden sector bundle defined by the line bundle $L=\mathcal{O}_{X}(2,1,3)$, and a single five brane between the observable and the hidden sector, at $\lambda=0.49$, the above relations become
\begin{equation}
\re f_{1}=V+\frac{1}{3}a^{1}-\frac{1}{6}a^{2}+2a^{3}+\frac{1}{2}\bigl(\tfrac{1}{2}-\lambda\bigr)^{2}(9a^{1}+17a^{2})
\label{mw1}
\end{equation}
and
\begin{equation}
\re f_{2}=V-\frac{29}{6}a^{1}-\frac{25}{3}a^{2}-2a^{3}+\frac{1}{2}\bigl(\tfrac{1}{2}+\lambda\bigr)^{2}(9a^{1}+17a^{2})\ ,
\label{mw2}
\end{equation}
where we have used the "unity gauge" notation $\epsilon_S^\prime\hat R/V^{1/3}=1$.

Using these results, as well as $\langle M_{U}\rangle$ given in \eqref{jack1}, $b_{L}=6$ and the values for $\langle \alpha_{u} \rangle$ for the split and unified Wilson lines scenarios respectively, one can compute $\Lambda$ in \eqref{eq:condesation_scale} and, hence, using \eqref{ew1} the value of $m_{\rm{susy}}$. Clearly, the result is a function of the point in moduli space where $m_{\rm{susy}}$ is evaluated. Plots of $m_{\rm{susy}}$ evaluated over the ``viable'' region of K\"ahler moduli space displayed with magenta in Figure \ref{fig:Intersection}, are shown in Figure \ref{fig:susyScale_both}.

\subsection{Spontaneous Supersymmetry Breaking in the Low-Energy Effective Theory}

Spontaneous $N=1$ supersymmetry breaking, induced by gaugino condensation in the hidden sector, appears in the low-energy effective theory as potentially non-vanishing moduli $F$-terms in the chiral superfields of the dilaton $S$, the complexified K\"ahler moduli $T^{i}$ and the single five-brane modulus $Z$~\cite{Choi:1997cm,Kaplunovsky:1993rd,Horava:1996vs,Lukas:1997rb,Nilles:1998sx,Binetruy:1996xja,Antoniadis:1997xk,Minasian:2017eur,Gray:2007qy,Lukas:1999kt,Font:1990nt}. We will assume that the resulting moduli-mediated soft terms~\cite{Soni:1983rm,Kaplunovsky:1993rd,Louis:1994ht,Choi:1997cm,Brignole:1998dxa} dominate over any anomaly-mediated SUSY breaking effects~\cite{hep-th/9810155,hep-ph/9810442,hep-th/9911029,hep-ph/0011081}. Defining the index $a$ which runs over $(S, T^{1}, T^{2}, T^{3}, Z)$, the moduli $F$-terms are then given by
\begin{equation}
\bar F^{\bar b}=\kappa_4^2 \ee^{\hat K/2}\hat K^{\bar b a}(\partial_a W+W\partial_a \hat K) \ ,
\label{f1}
\end{equation}
where $\kappa_{4}^{2}=8\pi/M_{P}^{2}$ and $W$ is the gaugino condensate superpotential given in \eqref{sup1}. In addition, these $F$-terms depend strongly on the dimensionless K\"ahler potential $\hat{K}$. Before continuing, we note that since the scalar fields of all chiral multiplets have been chosen to be dimensionless, it follows that all $F$ auxiliary fields have dimension one.  Since $\kappa_{4}^{2}~ W \propto \Lambda^{3}/M_{P}^{2}$, it follows that the mass scale of each $F$-term is ${\cal{O}}(8\pi\Lambda^{3}/M_{P}^{2})$. Thus, although gaugino condensation and the superpotential $W$ occur in the hidden sector, their supersymmetry breaking effects on the $S$, $T^{i}$ and $Z$ moduli are mediated via gravitational interactions and thus are Planck mass suppressed. The  K\"ahler potential $\hat{K}$ was defined in \cite{Lukas:1999kt} and is given by
\begin{equation}
\hat K=\tilde K_S+K_T\ ,
\label{sum1}
\end{equation}
where
\begin{equation}
\begin{split}
\tilde K_S=&-\ln \left(  S+\bar S-\frac{\epsilon_S^\prime}{2}\frac{(Z+\bar Z)^2}{W_{i} (T+\bar T)^i} \right)\\
\simeq&-\ln(S+\bar S)+\frac{\epsilon_S^\prime}{2}\frac{(Z+\bar Z)^2}{(S+\bar S)W_i(T+\bar T)^i}\ ,
\label{split2}
\end{split}
\end{equation}
\begin{equation}
Z=W_it^iz+2iW_i(-\eta^i\nu+\chi^iz), \quad \text{with} \quad t^i=\tfrac{1}{2}(T+\bar T)^i,
\end{equation}
and 
\begin{equation}
K_T=-\ln \left( \tfrac{1}{48}d_{ijk}(T+\bar T)^i(T+\bar T)^j(T+\bar T)^k \right)
\label{tel1}
\end{equation}

\subsection{Soft Supersymmetry Breaking Terms in the Observable Sector}

The $B-L$ MSSM, the $Spin(10)$ group of the observable sector is broken near the unification scale $\langle M_{U} \rangle$ to the low energy gauge group $SU(3)_{C}\times SU(2)_{L}\times U(1)_{3R}\times U(1)_{B-L}$ by two independent Wilson lines associated with the $\mathbb{Z}_{3} \times \mathbb{Z}_{3}$ homotopy group of the Calabi--Yau threefold. 
The particle content of the resulting effective theory
is precisely that of the MSSM, with three right-handed neutrino chiral multiplets and
a single Higgs-Higgs conjugate pair, with no exotic fields.  

The superpotential of the observable sector evaluated at $\langle M_{U} \rangle$ is given by
\begin{equation}
W=\mu H_{u} H_{d}+Y_{u}QH_{u}u^{c}-Y_{d}QH_{d}d^{c}-Y_{e}QH_{d}\ee^{c}+Y_{\nu}QH_{u}\nu^{c} \ ,
\label{con1}
\end{equation}
where flavor and gauge indices have been suppressed and the Yukawa couplings are three-by-three matrices in flavor space. The observed smallness of the three CKM mixing angles and the CP-violating phase dictate that the quark and lepton Yukawa matrices should be nearly diagonal and real. Furthermore, the smallness of the first and second family fermion masses implies that all components of the up/down quark and lepton Yukawa couplings, with the exception of the top and bottom quarks, and the tau lepton, can be neglected for the this work. Similarly, the very light neutrino masses imply that the neutrino Yukawa couplings can also be neglected in our analysis. The $\mu$-parameter can be chosen to be real, but not necessarily positive, without loss of generality. We do not attempt to solve the ``$\mu$-problem''. Hence, we allow the dimension-one parameter $\mu$ to take any value required to obtain the correct $Z$-boson mass.

To align with the formalism presented in \cite{Kaplunovsky:1993rd} and \cite{Lukas:1999kt}, where the general form of the soft supersymmetry breaking terms in the Lagrangian which we use in this analysis were computed, we rewrite the superpotential \eqref{con1} in the form
\begin{equation}
W= \frac{1}{2}\hat{\mu}_{IJ}C^{I}C^{J}+\frac{1}{3}\hat{Y}^{IJK}C^{I}C^{J}C^{K} \ ,
\label{con2}
\end{equation}
where $C^{I}$ are the chiral superfields associated with the top and bottom quarks, the tau lepton and the up- and down-Higgs particles. Comparing \eqref{con1} and \eqref{con2}, it follows that  $\hat{\mu}_{IJ}$ is symmetric and all components vanish with the exception of $\hat{\mu}_{H_{u}H_{d}}=\mu$. Similarly, the coefficients $\hat{Y}_{IJK}$ are completely symmetric and, following the formalism in \cite{Lukas:1999kt}, are given by
\begin{equation}
\hat{Y}_{IJK}=2\sqrt{2\pi\hat{\alpha}_{GUT}} \>y_{IJK} \ ,
\label{con3}
\end{equation}
where $y_{IJK}$ are the physical Yukawa parameters for the top, bottom and tau particles scaled up to $\langle M_{U} \rangle$.

The soft supersymmetry breaking terms associated with the observable sector of the $B-L$ MSSM are of the form
\begin{align}
 \label{eq:6iii}
\begin{split}
	-\mathcal L_{\mbox{\scriptsize soft}}  &= 
	\left(
		\frac{1}{2} M_3 \tilde g^2+ \frac{1}{2} M_2 \tilde W^2+ \frac{1}{2} M_R \tilde W_R^2+\frac{1}{2} M_{BL} \tilde {B^\prime}^2
	\right.
		\\
	& \eqspace\left.
		\hspace{0.4cm} +a_u \tilde Q H_u \tilde u^c - a_d \tilde Q H_d \tilde d^c - a_e \tilde L H_d \tilde e^c
		+ a_\nu \tilde L H_u \tilde \nu^c + b H_u H_d + \text{h.c.}
	\right)
	\\
	&\eqspace + m_{\tilde Q}^2|\tilde Q|^2+m_{\tilde u^c}^2|\tilde u^c|^2+m_{\tilde d^c}^2|\tilde d^c|^2+m_{\tilde L}^2|\tilde L|^2
	+m_{\tilde \nu^c}^2|\tilde \nu^c|^2+m_{\tilde e^c}^2|\tilde e^c|^2  \\  
	&\eqspace+m_{H_u}^2|H_u|^2+m_{H_d}^2|H_d|^2 \ .
\end{split}
\end{align}
The $b$ parameter can be chosen to be real and positive without loss of generality. The gaugino soft masses can, in principle, be complex. This, however, could lead to CP-violating effects that are not observed. Therefore, we proceed by assuming they all are real. The $a$-parameters and soft scalar masses can, in general, be Hermitian matrices in family space. Again, however, this could lead to unobserved flavor and CP violation. Therefore, we will assume they all are diagonal and real. For more explanation of these assumptions, see \cite{Ovrut:2015uea}.

As we did for the superpotential, to compare with the formalism presented in \cite{Kaplunovsky:1993rd,Lukas:1999kt}, we will rewrite the soft supersymmetry breaking terms in \eqref{eq:6iii} as
\begin{equation}
-\mathcal L_{\mbox{\scriptsize soft}}= \left( \frac{1}{2}M_{i}(\lambda^{i})^{2}+\frac{1}{3}a_{IJK}\tilde{C}^{I}\tilde{C}^{J}\tilde{C}^{K} +\frac{1}{2}B_{IJ}\tilde{C}^{I}\tilde{C}^{J}+\text{h.c.} \right)+m_{I\bar{J}}^{2}\tilde{C}^{I}\bar{\tilde{C}}^{\bar{J}} \ ,
\label{con4}
\end{equation}
where $\lambda^{i}$ are the gauginos for $i=3,2,3R,BL$, and $\tilde{C}^{I}$ are the scalar components of the chiral superfields associated with the top and bottom quarks, the tau lepton and up- and down-Higgs particles. Comparing \eqref{con4} with \eqref{eq:6iii}, it follows that a) the matrix $m_{I\bar{J}}^{2}$ is diagonal with its $I,\bar{J}$ indices running over stop, sbottom, stau, and Higgs-up and Higgs-down scalars, b) $B_{IJ}$  is symmetric, all of whose terms vanish except $B_{H_{u}H_{d}}=b$ and c) $a_{IJK}$ is totally symmetric where, as with $m_{I\bar{J}}^{2}$, its indices $I,J,K$ run over stop, sbottom, stau and Higgs-up and Higgs-down scalars only, and are associated with the $a$-parameters in \eqref{eq:6iii} accordingly.

The parameters for each of these soft supersymmetry breaking terms -- including the induced gravitino mass which enters some of the soft breaking coefficients -- can be explicitly computed for any spontaneous SUSY breaking mechanism which results in moduli-dominated soft terms. Using the notation of \cite{Kaplunovsky:1993rd}, the generic expressions for these parameters are the following:
\begin {enumerate}
\item The gravitino mass:

$m_{3/2}=\kappa_{4}^{2}\ee^{\hat{K}/2}|W|. $  

\item The gaugino masses:

$M_{i}=\frac{1}{2}F^{a}\partial_{a}\ln g^{-2}_{i}.$    

Note that the gaugino mass is, in general not ``universal'' -- that is, it is not necessarily the same for all gauginos. However, for the present analysis of the $B-L$ MSSM, the gaugino masses will turn out to be identical. We explain why this is the case in the discussion to follow.

\item The quadratic scalar masses:

$m_{I\bar{J}}^{2}=m^{2}_{3/2}Z_{I\bar{J}}-F^{a}\bar{F}^{\bar{b}}R_{a \bar{b} I\bar{J}}\ .$   

Here $Z_{I\bar{J}}$ is defined by $K_{\text {matter}}=Z_{I\bar{J}}\tilde{C}^{I}{\bar{\tilde{C}}}{}^{\bar{J}}$ for generic observable-sector scalar fields $\tilde{C}^{I}$.
The explicit forms for $Z_{I\bar{J}}$ and $R_{a \bar{b} I\bar{J}}$ are presented in the discussion below.

\item The cubic scalar coefficients:

$a_{IJK}=F^{a}\big( \partial_{a}Y_{IJK}+\frac{1}{2}\hat{K}_{a}Y_{IJK}-3\Gamma^{N}_{a(I}Y_{JK)N} \big)\ .$ 

The parameters $Y_{IJK}$ and $\Gamma^{N}_{aI}$ will be given in the following analysis. 

\item The holomorphic quadratic coefficient:

$ B_{IJ}=F^{a}\bigl(\partial_{a}\mu_{IJ}+\frac{1}{2}(\partial_{a}\hat{K})-3\Gamma^{N}_{a(I}\mu_{J)N} \bigr)- m_{3/2}\mu_{IJ} \ .$

The parameter $\mu_{IJ}$ will be discussed below.

\end{enumerate}

The above expressions for the soft supersymmetry breaking terms are generic; that is, they can arise from any vacuum state which spontaneously breaks SUSY via non-vanishing $F$-terms. However, for the remainder of this section, we will consider supersymmetry breaking to occur explicitly from a ``gaugino condensate'' in the hidden sector of the $B-L$ MSSM theory.

\subsection*{Gravitino Mass}

The gravitino mass is simply defined to be
\begin{equation}
m_{3/2}=\kappa_{4}^{2}\ee^{\hat{K}/2}|W| \ ,
\label{van1}
\end{equation}
where $\kappa_{4}^{2}=8 \pi/M_{P}^{2}$, $\hat{K}$ is defined in \eqref{sum1} and $W$ is the gaugino condensate superpotential presented in \eqref{sup1}.

\subsection*{Gaugino Mass}

The generic expression for the gaugino mass  associated with the $i$-th factor of an observable sector gauge group of the form $G=\Pi_{i} G_{i}$ is given by
\begin{equation}
M_i=\frac{1}{2}F^a\partial_a \ln g_i^{-2}\qquad a=S,T^1,T^2,T^3,Z \ .
\label{g1}
\end{equation}
In our case, the index $i$ spans the factors in the $d=4$ low-energy gauge group $G=SU(3)_{C}\times SU(2)_{L}\times U(1)_{3R}\times U(1)_{B-L}$ of the  $B-L$ MSSM in the observable sector. That is,
\begin{equation}
i=3,\>2,\>3R,\> B-L \ .
\end{equation}
As discussed in \cite{Ashmore:2020ocb}, in the simultaneous Wilson lines scenario each gauge coupling $g_{i}^{2}$ is related to its average value $\langle g_{u}^{2}\rangle$  at the unification scale $\langle M_{U} \rangle=3.15 \times 10^{16}$ GeV by
\begin{equation}
g_i^2=c_i\langle g_u^2\rangle
\end{equation}
for some constant coefficient $c_{i}$.
Then
\begin{equation}
M_i=\frac{1}{2}F^a\partial_a \ln g_i^{-2}=\frac{1}{2}F^a\frac{1}{g_i^{-2}}\partial_a g_i^{-2}\\
=\frac{1}{2}F^{a} c_{i} \langle g_u^{2} \rangle \partial_a \frac{1}{c_{i} \langle g_u^{2} \rangle}
 \end{equation}
Now  \eqref{need2} implies that 
\begin{equation}
\langle g_u^2 \rangle=\frac{4\pi\hat \alpha_{\text{GUT}}}{\re  f_1} \ ,
\end{equation}
where $ \hat{\alpha}_{\text{GUT}}$ is a constant parameter. It follows that the constants $c_{i}$ and ${ \hat{\alpha}}_{\text{GUT}}$ drop out
and, hence, the gaugino masses defined \eqref{g1} are all identical. Defining this unique parameter to be $M_{1/2}$, we find that
\begin{equation}
M_{1/2}=\frac{1}{2\re f_1}F^a\partial_a \re  f_1, 
\end{equation}
as presented previously. A similar argument can be made for the split Wilson lines scenario. 

From eq. \label{eq:2fs} we learn that the expression for the real part of $f_{1}$, can be then be written as
\begin{multline}
\re f_{1}=\frac{S+\bar{S}}{2}-\tfrac{1}{4}\epsilon_{S} ' \Big( \beta_1^{(N+1)}(T^{1}+\bar{T}^{1}) \\
+\beta_2^{(N+1)}(T^{2}+\bar{T}^{2}) +\beta^{(N+1)}_3(T^{3}+\bar{T}^{3}) +2(Z+\bar{Z}) \Big) \ .
\label{need0}\
\end{multline}
It follows that 
\begin{gather}
\partial_{S}\re f_{1}=\frac{1}{2}, \quad \partial_{T^{1}}\re f_{1}=- \epsilon_{S} ' \frac{\beta_1^{(N+1)}}{4},\quad\partial_{T^{2}}\re f_{1}=- \epsilon_{S} '\frac{\beta_2^{(N+2)}}{4} ,  \\
\qquad \qquad \partial_{T^{3}}\re f_{1}=-\epsilon_{S} '\frac{\beta_3^{(N+1)}}{4} ,\quad \partial_{Z}\re f_{1}=-\frac{1}{2}\epsilon_{S}'  \ .
\label{need00}
\end{gather}
Putting everything together, the universal gaugino soft supersymmetry breaking coefficient is given, in unity gauge, by
{\small
\begin{equation}
\begin{split}
M_{1/2}&=\frac{1}{2\text{Re}\>f_1}  \left[ \frac{1}{2}F^{S}- \epsilon_{S} ' \frac{\beta_1^{(N+1)}}{4}F^{T^{1}} -\epsilon_{S} ' \frac{\beta_2^{(N+1)}}{4}F^{T^{2}} -\epsilon_{S} '\frac{\beta_3^{(N+1)}}{4} F^{T^{3}} - \epsilon_{S} ' \frac{1}{2}F^{Z}  \right] \ .
\end{split}
\label{need523}
\end{equation}
}
In unity gauge, the real part of $f_{1}$ for the line bundle $L=\mathcal{O}_{X}(2,1,3)$ was found to be
\begin{equation}
{\rm {Re}} f_1=V+\frac{1}{3}a^1-\frac{1}{6}a^2+2a^3+\frac{1}{2}\left( \tfrac{1}{2}-\lambda \right)^2W_ia^i \ ,
\label{need3}
\end{equation}
where 
\begin{equation}
W_{i}=(9,17,0) \ .
\end{equation}
and $\lambda=0.49$, as done in \cite{Ashmore:2020ocb}. Therefore, one can compute the value of the universal gaugino soft supersymmetry breaking coefficient at any point in the physical ``viable'' subspace of K\"ahler moduli space for $L=\mathcal{O}_{X}(2,1,3)$, as we did above for both $m_{\rm{susy}}$ and $m_{3/2}$
\begin{equation}
\begin{split}
M_{1/2}&=\frac{1}{2\big( V+\frac{1}{3}a^1-\frac{1}{6}a^2+2a^3+\frac{1}{2}\left( \frac{1}{2}-\lambda \right)(9a^{1}+17a^{2}) \big)}  \\
& \times \left[ \frac{1}{2}F^{S}+ \epsilon_{S} '  \frac{29}{12}F^{T^{1}} +\epsilon_{S} '  \frac{25}{6}F^{T^{2}} +\epsilon_{S} ' F^{T^{3}} - \epsilon_{S} ' \frac{1}{2}F^{Z}  \right] \ .
\end{split}
\label{need5}
\end{equation}

\subsection*{Quadratic Scalar Masses}

The generic form for the quadratic scalar mass coefficients is given by~\cite{Kaplunovsky:1993rd,Brignole:1998dxa,Lukas:1997rb,Lukas:1999kt}
\begin{equation}
m_{I\bar{J}}^{2}=m^{2}_{3/2}Z_{I\bar{J}}-F^{a}\bar{F}^{\bar{b}}R_{a \bar{b} I\bar{J}} \ .
\label{q1}
\end{equation}
%
To linear order in $\epsilon_S^\prime$, it was shown in \cite{Lukas:1999kt} that, for a {\it single} five-brane located at $z \in [0,1]$,
\begin{equation}
Z_{I\bar{J}}=\ee^{K_{T}/3}\left[K_{BI\bar{J}}-\frac{\epsilon_S^\prime}{2(S+\bar{S})}\tilde{\Gamma}^{i}_{BI\bar{J}}\big(\beta_{i}^{(0)}+(1-z)^{2}W_{i}  \big)   \right] \ ,
\label{q2} 
\end{equation}
where $\beta_{i}^{(0)}$ is the ``charge'' on the observable wall and $W_{i}$ is the five-brane, which for the $B-L$ MSSM vacuum, is
\begin{equation}
\beta_{i}^{(0)}=\left(\frac{2}{3},-\frac{1}{3},4 \right)_{i} \ .
\label{l1}
\end{equation}
The $T^i$-dependent K\"ahler potential was presented in \eqref{tel1} and $K_{BI\bar{J}}$ is defined by
\begin{equation}
K_{BI\bar{J}}=G_{I\bar{J}} \ ,
\label{kn1}
\end{equation}
where $G_{I\bar{J}}$ is a positive-definite Hermitian metric on the $H^{1}$ cohomologies associated with the $\tilde{C}^{I}$ matter scalars in the observable sector \cite{Lukas:1997rb,Lukas:1999kt}. Generically, $G_{I\bar{J}}$ is moduli dependent.  The quantity  $\tilde{\Gamma}^{i}_{BI\bar{J}}$ is given by
\begin{equation}
\tilde{\Gamma}^{i}_{BI\bar{J}}={\Gamma}^{i}_{BI\bar{J}}-(T^{i}+\bar{T}^{i})K_{BI\bar{J}}-\tfrac{2}{3}(T^{i}+\bar{T}^{i})(T^{k}+\bar{T}^{k})K_{Tkj}{\Gamma}^{j}_{BI\bar{J}} \ ,
\label{kn2}
\end{equation}
with
\begin{equation}
\Gamma^{i}_{BI\bar{J}}=K_{T}^{ij} \frac{\partial K_{BI\bar{J}}}{\partial T^{j}}\ ,
\label{kn3}
\end{equation}
and $K_{T}^{ij}$ is the inverse of the matrix 
%
\begin{equation}
K_{Tij}=\frac{\partial^{2}K_{T}}{\partial T^{i} \partial T^{j}} =-\frac{d_{lmi}a^{i}}{4\hat{R}^{2}V^{1/3}} +\frac{d_{mij}a^{i}a^{j}d_{lpq}a^{p}a^{q}}{16\hat{R}^{4}V^{2/3}}\ .
\label{kn4}
\end{equation}

Next, we consider the tensor $R_{a \bar{b} I\bar{J}}$ in \eqref{q1}. It is defined to be
\begin{equation}
R_{a \bar{b} I\bar{J}}=\partial_{a}\partial_{\bar{b}}Z_{I\bar{J}}-\Gamma^{N}_{aI}Z_{N\bar{L}}{\bar{\Gamma}}^{\bar{L}}_{\bar{b}\bar{J}}\ ,
\label{r1}
\end{equation}
where
\begin{equation}
\Gamma^{N}_{aI}=Z^{N\bar{J}}\partial_{a}Z_{\bar{J}I} \ .
\label{r2}
\end{equation}
Writing the generic expression for $Z_{I\bar{J}}$ in \eqref{q2} as
\begin{equation}
Z_{I\bar{J}}=Z_{I\bar{J}}^{(0)}+Z_{I\bar{J}}^{(\epsilon_{S}')} \ ,
\label{r3}
\end{equation}
it follows that
\begin{equation}
\Gamma^{N}_{aI}=\Gamma^{(0)N}_{aI}+\Gamma^{(\epsilon_{S}')N}_{aI}\ ,
\end{equation}
where
\begin{equation}
\Gamma^{(0)N}_{aI}=Z^{(0)N\bar{J}}\partial_{a}Z^{(0)}_{\bar{J}I}, \qquad \Gamma^{(\epsilon_{S}')N}_{aI}=Z^{(0)N\bar{J}}\partial_{a}Z^{(\epsilon_{S}')}_{\bar{J}I}+Z^{(\epsilon_{S}')N\bar{J}}\partial_{a}Z^{(0)}_{\bar{J}I} \ .
\end{equation}
Inserting these expressions into \eqref{r1}, we find that
\begin{equation}
R_{a\bar{b}I\bar{J}}= R^{(0)}_{a\bar{b}I\bar{J}}+R^{(\epsilon_{S}')}_{a\bar{b}I\bar{J}}\ ,
\label{bc2}
\end{equation}
with
\begin{equation}
R^{(0)}_{a\bar{b}I\bar{J}}=\partial_{a}\partial_{\bar{b}}Z^{(0)}_{I\bar{J}} - \Gamma^{(0)N}_{aI}Z^{(0)}_{N\bar{L}}\bar{\Gamma}^{(0)\bar{L}}_{\bar{b}\bar{J}} \ ,
\label{bc3}
\end{equation}
and
\begin{equation}
R^{(\epsilon_{S}')}_{a\bar{b}I\bar{J}}=\partial_{a}\partial_{\bar{b}}Z^{(\epsilon_{S}')}_{I\bar{J}} - \Gamma^{(\epsilon_{S}')N}_{aI}Z^{(0)}_{N\bar{L}}\bar{\Gamma}^{(0)\bar{L}}_{\bar{b}\bar{J}} - \Gamma^{(0)N}_{aI}Z^{(\epsilon_{S}')}_{N\bar{L}}\bar{\Gamma}^{(0)\bar{L}}_{\bar{b}\bar{J}} - \Gamma^{(0)N}_{aI}Z^{(0)}_{N\bar{L}}\bar{\Gamma}^{(\epsilon_{S}')\bar{L}}_{\bar{b}\bar{J}} \ .
\label{bc4}
\end{equation}

Putting everything together, one can express the quadratic scalar soft coefficients in \eqref{q1} as 
\begin{equation}
m_{I\bar{J}}^{2} = m^{(0)2}_{I\bar{J}} +m^{(\epsilon_{S}')2}_{I \bar{J} } \ ,
\label{bc5}
\end{equation}
where
\begin{equation}
m^{(0)2}_{I\bar{J}}=m_{3/2}^{2}Z^{(0)}_{I\bar{J}}-F^{a}\bar{F}^{\bar{b}}R^{(0)}_{a\bar{b}I\bar{J}}\ ,
\label{bc6}
\end{equation}
and
\begin{equation}
m^{(\epsilon_{S}')2}_{I\bar{J}}=m_{3/2}^{2}Z^{(\epsilon_{S}')}_{I\bar{J}}-F^{a}\bar{F}^{\bar{b}}R^{(\epsilon_{S}')}_{a\bar{b}I\bar{J}} \ ,
\label{bc7}
\end{equation}
with $Z^{(0)}_{I\bar{J}}$ and  $Z^{(\epsilon_{S}')}_{I\bar{J}}$ defined in \eqref{r3}, and  $R^{(0)}_{a\bar{b}I\bar{J}}$ and $R^{(\epsilon_{S}')}_{a\bar{b}I\bar{J}}$ given in \eqref{bc3} and \eqref{bc4} respectively.

Let us now compute these quantities explicitly. First of all, we note that there is currently no known method to explicitly compute $G_{I\bar J}$.\footnote{This \emph{should} be computable using numeric metrics on Calabi--Yau threefolds~\cite{Headrick:2005ch,Douglas:2006rr,Braun:2007sn,Headrick:2009jz,Douglas:2006hz,Anderson:2011ed,Anderson:2010ke,Cui:2019uhy,Anderson:2020hux,Douglas:2020hpv} and their moduli spaces~\cite{Keller:2009vj}, and the corresponding eigenmodes of the Laplacian~\cite{Braun:2008jp,Ashmore:2020ujw}.} With this in mind, for the rest of this work we shall assume that $K_{BI\bar{J}}=G_{I\bar{J}}$ is {\it moduli independent}, that is, simply an Hermitian matrix of numbers. We will denote this choice by
\begin{equation}
G_{I\bar{J}}={\cal{G}}_{I\bar{J}} \ .
\label{s1}
\end{equation}
We assume this to be the case henceforth. It then follows from \eqref{kn3} that $\Gamma^{i}_{BI\bar{J}}=0$ and, hence,
\begin{equation}
\tilde{\Gamma}^{i}_{BI\bar{J}}=-(T+\bar{T} )^{i}{\cal{G}}_{I\bar{J}} \ .
\label{s2}
\end{equation}
Then, using the metric \eqref{s1}, expression \eqref{q2} for $Z_{I\bar{J}}$ simplifies to 
\begin{equation}
Z_{I\bar{J}}=Z_{I\bar{J}}^{(0)}+Z_{I\bar{J}}^{(\epsilon_{S}')}
\label{s3}
\end{equation}
where
\begin{equation}
Z_{I\bar{J}}^{(0)}=\ee^{K_{T}/3} {\cal{G}}_{I\bar{J}}
\label{s4}
\end{equation}
%
%
\begin{equation}
Z_{I\bar{J}}^{(\epsilon_{S}')}=\frac{\epsilon_S^\prime}{2}\ee^{K_{T}/3}  \frac{(T+\bar{T})^{i}}{S+\bar{S}} \left[ \left( \frac{2}{3}, -\frac{1}{3},4 \right)_{i}+\left(1-\frac{Z+\bar{Z}}{ W_{l}(T+\bar{T})^{l}} \right) ^{2} W_{i} \right] {\cal{G}}_{I\bar{J}} \ .
\label{s5}
\end{equation}
Note that we have rewritten the last term so as to be able to differentiate this expression with respect to $Z$. This will be necessary in order to compute $R^{(\epsilon_{S}')}_{a\bar{b}I\bar{J}}$ below. It is also useful to rewrite the expressions for $Z_{I\bar{J}}^{(0)}$ and $Z_{I\bar{J}}^{(\epsilon_{S}')}$ in terms of the $\hat{R}$, $a^{i}$, $V$ and $\lambda$. Doing this, we find
\begin{equation}
Z_{I\bar{J}}^{(0)}=\frac{1}{\hat{R}}{\cal{G}}_{I\bar{J}}
\label{s4A}
\end{equation}
and
\begin{equation}
Z_{I\bar{J}}^{(\epsilon_{S}')}=\epsilon_S^\prime \frac{a^{i}}{2V^{4/3}} \left[ \left( \frac{2}{3}, -\frac{1}{3},4 \right)_{i}+ (\frac{1}{2}-\lambda)^{2} W_{i} \right] {\cal{G}}_{I\bar{J}} \ .
\label{s5A}
\end{equation}
Recall from \eqref{wall1} that unity gauge is defined by setting $\epsilon_S'\frac{\Rhat}{V^{1/3}} = 1 $.
It follows that in unity gauge the expression for $Z_{I\bar{J}}^{(\epsilon_{S}')}$ becomes
\begin{equation}
Z_{I\bar{J}}^{(\epsilon_{S}')}=\frac{a^{i}}{2\hat{R}V} \left[ \left( \frac{2}{3}, -\frac{1}{3},4 \right)_{i}+ (\frac{1}{2}-\lambda)^{2} W_{i} \right] {\cal{G}}_{I\bar{J}} \ .
\label{s5AA}
\end{equation}
Finally, evaluating this expression for the specific line bundle $L=\mathcal{O}_{X}(2,1,3)$ with $W_{i}=(9,17,0)$ discussed above, we find that
\begin{equation}
Z_{I\bar{J}}^{(\epsilon_{S}')}=\frac{1}{2\hat{R}V} \left[ \left( \frac{2}{3}+9 (\frac{1}{2}-\lambda)^{2} \right) a^{1}+ \left(-\frac{1}{3}+17(\frac{1}{2}-\lambda)^{2} \right) a^{2} +4a^{3} \right]{\cal{G}}_{I\bar{J}} \ .
\label{s5AAA}
\end{equation}

Let us now compute $R^{(0)}_{a\bar{b}I\bar{J}}$ and $R^{(\epsilon_{S}')}_{a\bar{b}I\bar{J}}$ using expressions \eqref{bc3} and \eqref{bc4} respectively. We begin with $R^{(0)}_{a\bar{b}I\bar{J}}$. Using the fact that
\begin{equation}
Z^{(0)N\bar{J}}=\ee^{-K_{T}/3}{ \cal{G}}^{N\bar{J}} \ ,
\label{s6}
\end{equation}
it follows that 
\begin{equation}
\Gamma^{(0)N}_{aI}=\frac{1}{3}(\partial_{a}K_{T})\delta^{N}_{I}
\label{s7}
\end{equation}
and, hence
\begin{equation}
-\Gamma^{(0)N}_{aI}Z^{(0)}_{N\bar{L}}\bar{\Gamma}^{(0)\bar{L}}_{\bar{b}\bar{J}} = \frac{1}{9}\ee^{K_{T}/3} (\partial_{a}K_{T})(\partial_{\bar{b}}K_{T}){\cal{G}}_{I\bar{J}} \ .
\label{s8}
\end{equation}
It is straightforward to show from \eqref{s4} that
\begin{equation}
\partial_{a}\partial_{\bar{b}}Z^{(0)}_{I\bar{J}}= \ee^{K_{T}/3}\left(\frac{1}{9}(\partial_{a}K_{T})(\partial_{\bar{b}}K_{T})+\frac{1}{3}(\partial_{a}\partial_{\bar{b}}K_{T}) \right){\cal{G}}_{I\bar{J}} \ .
\label{s9}
\end{equation}
Using \eqref{s8} and \eqref{s9}, expression \eqref{bc3} becomes
\begin{equation}
R^{(0)}_{a\bar{b}I\bar{J}}=\frac{\ee^{K_{T}/3}}{3}(\partial_{a}\partial_{\bar{b}}K_{T}){\cal{G}}_{I\bar{J}} \ .
\label{s10}
\end{equation}
Note that this vanishes if index $a$ and/or ${b}$ is $S, Z$. For  $a=i$, ${b}=j$ for $i,j=1,2,3$, $\partial_{i}\partial_{{j}}K_{T}$ is given by \eqref{kn4}.
Let us now compute $R^{(\epsilon_{S}')}_{a\bar{b}I\bar{J}}$. It follows from \eqref{bc4} that, in addition to  the inverse of $Z^{(0)}_{I\bar{J}}$ given in \eqref{s6}, one also needs to know the inverse $Z^{(\epsilon_{S}'))}_{I\bar{J}}$ in \eqref{s5}. Calculating this to linear order in $\epsilon_{S}'$ is straightforward. It is found to be
\begin{equation}
Z^{(\epsilon_{S}')N\bar{J}}=-\frac{\epsilon_S^\prime}{2}\ee^{-K_{T}/3}  \frac{(T+\bar{T})^{i}}{S+\bar{S}} \left[ \left( \frac{2}{3}, -\frac{1}{3},4 \right)_{i}+\left(1-\frac{Z+\bar{Z}}{ W_{l}(T+\bar{T})^{l}} \right) ^{2} W_{i} \right]{ \cal{G}}^{N\bar{J}} \ .
\label{ep1}
\end{equation}
To continue, recall from \eqref{s7} that
\begin{equation}
\Gamma^{(0)N}_{aI}=\frac{1}{3}(\partial_{a}K_{T})\delta^{N}_{I} \ .
\label{ep2}
\end{equation}
Furthermore, using \eqref{s5} and the inverse \eqref{ep1} one can show that
\begin{equation}
\Gamma^{(\epsilon_S^\prime)N}_{aI}=\frac{\epsilon_S^\prime}{2} \partial_{a}\left( \frac{(T+\bar{T})^{i}}{S+\bar{S}}[X]_{i}\right)    \delta^{N}_{I} \ ,
\label{ep3}
\end{equation}
where we have introduced 
\begin{equation}
[X]_{i}=\left( \frac{2}{3}, -\frac{1}{3},4 \right)_{i}+\left(1-\frac{Z+\bar{Z}}{W_{l}(T+\bar{T})^{l}} \right) ^{2} W_{i}
\label{ep4}
\end{equation}
to simplify the notation. Using \eqref{ep3} and \eqref{ep4} it is straightforward to show that the last three terms in \eqref{bc4} are given by
\begin{align}
& - \Gamma^{(\epsilon_{S}')N}_{aI}Z^{(0)}_{N\bar{L}}\bar{\Gamma}^{(0)\bar{L}}_{\bar{b}\bar{J}} - \Gamma^{(0)N}_{aI}Z^{(\epsilon_{S}')}_{N\bar{L}}\bar{\Gamma}^{(0)\bar{L}}_{\bar{b}\bar{J}} - \Gamma^{(0)N}_{aI}Z^{(0)}_{N\bar{L}}\bar{\Gamma}^{(\epsilon_{S}')\bar{L}}_{\bar{b}\bar{J}}\nonumber \\
&=-\frac{\epsilon_{S}'}{6}\ee^{K_{T}/3}  \Big(  \frac{1}{3} \frac{(T+\bar{T})^{i}}{S+\bar{S}} [X]_{i}(\partial_{a}K_{T})(\partial_{\bar{b}}K_{T})  \label{ep5}     \\
&\eqspace +\partial_{a} ( \frac{(T+\bar{T})^{i}}{S+\bar{S}} [X]_{i} ) (\partial_{\bar{b}}K_{T})  + (\partial_{a}K_{T})\partial_{\bar{b}} ( \frac{(T+\bar{T})^{i}}{S+\bar{S}} [X]_{i} ) \Big){ \cal{G}}_{I\bar{J}} \ . \nonumber
\end{align}
Similarly, using $Z_{I\bar{J}}^{(\epsilon_{S}')}$ in \eqref{s5}, it is tedious but straightforward to show that
\begin{align}
\partial_{a}\partial_{\bar{b}}Z_{I\bar{J}}^{(\epsilon_{S}')}&= \frac{\epsilon_{S}'}{6}\ee^{K_{T}/3}  \Big( \frac{1}{3} \frac{(T+\bar{T})^{i}}{S+\bar{S}} [X]_{i}(\partial_{a}K_{T})(\partial_{\bar{b}}K_{T}) \label{ep6} \\ 
& \eqspace+\partial_{a} ( \frac{(T+\bar{T})^{i}}{S+\bar{S}} [X]_{i} ) (\partial_{\bar{b}}K_{T})  + (\partial_{a}K_{T})\partial_{\bar{b}} ( \frac{(T+\bar{T})^{i}}{S+\bar{S}} [X]_{i} ) \nonumber \\
&\eqspace+(\partial_{a}\partial_{\bar{b}}K_{T}) \frac{(T+\bar{T})^{i}}{S+\bar{S}} [X]_{i} +3\partial_{a}\partial_{b} (\frac{(T+\bar{T})^{i}}{S+\bar{S}} [X]_{i}) \Big) {\cal{G}}_{I\bar{J}} \ . \nonumber
\end{align}
Adding \eqref{ep5} and \eqref{ep6}, we see that the first three terms in each expression exactly cancel and, hence, it follows from \eqref{bc4} that
\begin{equation}
R^{(\epsilon_{S}')}_{a\bar{b}I\bar{J}}=\frac{\epsilon_{S}'}{6}\ee^{K_{T}/3}  
\Big((\partial_{a}\partial_{\bar{b}}K_{T}) \frac{(T+\bar{T})^{i}}{S+\bar{S}} [X]_{i} +3\partial_{a}\partial_{b} (\frac{(T+\bar{T})^{i}}{S+\bar{S}} [X]_{i}) \Big) {\cal{G}}_{I\bar{J}} \ .
\label{ep7}
\end{equation}
Having presented the generic expression for $R^{(0)}_{a\bar{b}I\bar{J}}$ and $R^{(\epsilon_{S}')}_{a\bar{b}I\bar{J}}$ in \eqref{s10} and \eqref{ep7} respectively, it is again useful to rewrite them  in terms of the $\hat{R}$, $a^{i}$, $V$. For $R^{(0)}_{a\bar{b}I\bar{J}}$ we find that
\begin{equation}
R^{(0)}_{a\bar{b}I\bar{J}}=\frac{1}{3\hat{R}}(\partial_{a}\partial_{\bar{b}}K_{T}){\cal{G}}_{I\bar{J}} \ ,
\label{s10A}
\end{equation}
where $\partial_{a}\partial_{\bar{b}}K_{T}$ is given in \eqref{kn4}. The expression for $R^{(\epsilon_{S}')}_{a\bar{b}I\bar{J}}$, however, is considerably more complicated. In order to simplify a long calculation, we will present the components of this quantity, not only in terms of the variables $\hat{R}$, $a^{i}$, $V$ and $\lambda$, but will further restrict the result to the line bundle $L=\mathcal{O}_{X}(2,1,3)$ with $W_{i}=(9,17,0)$ discussed above only. Moreover, we will present the results in the unity gauge, setting $\epsilon_S^\prime\hat R/V^{1/3}$. For this specific case, defining the indices $a=S,T^{1},T^{2},T^{3},Z$ and $\bar{b}=\bar{S},\bar{T}^{1},\bar{T}^{2},\bar{T}^{3},\bar{Z}$, we find that
\begin{equation}
R^{(\epsilon_{S}')}_{a\bar{b}I\bar{J}}= M^{(\epsilon_{S}')}_{a\bar{b}}{\cal{G}}_{I\bar{J}}   \ ,
\label{fi1}
\end{equation}
where $M^{(\epsilon_{S}')}_{a\bar{b}}$ is the real, symmetric matrix specified by
\begin{align}
\noindent \bar{S}~ {\rm Terms}:~~ M_{S\bar{S}}&=\frac{1}{4\hat{R}V^{3}} \left( \frac{2a^{1}}{3}-\frac{a^{2}}{3}+4a^{3}+\left(\tfrac{1}{2}-\lambda\right)^{2}(9a^{1}+17a^{2}) \right) \label{help1}\ , \\
M_{T^{1}\bar{S}}&=-\frac{1}{8\hat{R}^{2}V^{5/3}} \left( \frac{2}{3}+9\left(\tfrac{1}{2}-\lambda\right)^{2} \right)\ , \\
M_{T^{2}\bar{S}}&=-\frac{1}{8\hat{R}^{2}V^{5/3}} \left( -\frac{1}{3}+17\left(\tfrac{1}{2}-\lambda\right)^{2} \right) \ ,\\
M_{T^{3}\bar{S}}&=-\frac{1}{2\hat{R}^{2}V^{5/3}} \ ,\\
M_{Z\bar{S}}&=\frac{1}{\hat{R}V}\left(\tfrac{1}{2}-\lambda\right)\left(1-\frac{1}{2V}\left(\lambda+\tfrac{1}{2}\right)(9a^{1}+17a^{2})  \right)\ ,
\label{res1}\\
\bar{T}^{1}~ {\rm Terms}:~~ M_{T^{1}\bar{T}^{1}}&=\frac{1}{6 \hat{R} V} (\partial_{T^{1}}\partial_{\bar{T}^{1}}K_{T})\biggl(  \frac{2a^{1}}{3}-\frac{a^{2}}{3}+4a^{3}  \\
&\eqspace+\left(\tfrac{1}{2}-\lambda\right)^{2}(9a^{1}+17a^{2}) \biggr)\ ,  \nonumber\\
M_{T^{2}\bar{T}^{1}}&=\frac{1}{6 \hat{R} V} (\partial_{T^{2}}\partial_{\bar{T}^{1}}K_{T})\biggl(  \frac{2a^{1}}{3}-\frac{a^{2}}{3}+4a^{3} \\
&\eqspace+\left(\tfrac{1}{2}-\lambda\right)^{2}(9a^{1}+17a^{2}) \biggr)\ ,  \nonumber\\
M_{T^{3}\bar{T}^{1}}&=\frac{1}{6 \hat{R} V} (\partial_{T^{3}}\partial_{\bar{T}^{1}}K_{T})\biggl(  \frac{2a^{1}}{3}-\frac{a^{2}}{3}+4a^{3} \\
&\eqspace+\left(\tfrac{1}{2}-\lambda\right)^{2}(9a^{1}+17a^{2}) \biggr)\ ,  \nonumber \\
M_{Z\bar{T}^{1}}&=-\frac{9}{4\hat{R^{3}} V^{1/3}}  \frac{\left(\frac{1}{2}-\lambda\right)}{(9a^{1}+17a^{2})}\ ,
\label{res2}\\
\bar{T}^{2}~ {\rm Terms}:~~ M_{T^{2}\bar{T}^{2}}&=\frac{1}{6 \hat{R} V} (\partial_{T^{2}}\partial_{\bar{T}^{2}}K_{T})\biggl(  \frac{2a^{1}}{3}-\frac{a^{2}}{3}+4a^{3} \\
&\eqspace+\left(\tfrac{1}{2}-\lambda\right)^{2}(9a^{1}+17a^{2}) \biggr)\ ,  \nonumber\\
M_{T^{3}\bar{T}^{2}}&=\frac{1}{6 \hat{R} V} (\partial_{T^{3}}\partial_{\bar{T}^{2}}K_{T})\biggl(  \frac{2a^{1}}{3}-\frac{a^{2}}{3}+4a^{3} \\
&\eqspace+\left(\tfrac{1}{2}-\lambda\right)^{2}(9a^{1}+17a^{2}) \biggr)\ , \nonumber\\
M_{Z\bar{T}^{2}}&=-\frac{17}{4\hat{R}^{3} V^{1/3}}  \frac{\left(\tfrac{1}{2}-\lambda\right)}{(9a^{1}+17a^{2})} \ ,
\label{res3}\\
\bar{T}^{3}~ {\rm Terms}:~~ M_{T^{3}\bar{T}^{3}}&=\frac{1}{6 \hat{R} V} (\partial_{T^{3}}\partial_{\bar{T}^{3}}K_{T})\biggl(  \frac{2a^{1}}{3}-\frac{a^{2}}{3}+4a^{3} \\
&\eqspace+\left(\tfrac{1}{2}-\lambda\right)^{2}(9a^{1}+17a^{2}) \biggr)\ ,  \nonumber\\
M_{Z\bar{T}^{2}}&=0\ ,
\label{res4}\\
\bar{Z}~ {\rm Terms}:~~ M_{Z\bar{Z}}&= \frac{1}{8\hat{R}^{3}V^{1/3}}  \frac{1}{(9a^{1}+17a^{2})}  \ . 
\label{res5}
\end{align}
For completeness, we restate that
\begin{equation}
\partial_{T^{m}}\partial_{\bar{T}^{l}}K_{T}=\frac{-d_{lmi}a^{i}}{4\hat{R}^{2}V^{1/3}} +\frac{d_{mij}a^{i}a^{j}d_{lpq}a^{p}a^{q}}{16\hat{R}^{4}V^{2/3}} \ ,
\label{kn4A}
\end{equation}
where
\begin{align}
d_{1jk}a^{j}a^{k}&=\frac{2}{3}a^{1}a^{2}+\frac{1}{3}(a^{2})^{2}+2a^{2}a^{3}\ , \\
d_{2jk}a^{j}a^{k}&=\frac{(a^{1})^{2}}{3}+\frac{2}{3}a^{1}a^{2}+2a^{1}a^{3}\ , \\
d_{3jk}a^{j}a^{k}&=2a^{1}a^{2} \ .
\label{ing1}
\end{align}

Using the above results, one can now calculate the coefficients of the quadratic scalar soft supersymmetry breaking terms to linear order in $\epsilon_{S}'$. Recall from \eqref{bc5} that
\begin{equation}
m_{I\bar{J}}^{2} = m^{(0)2}_{I\bar{J}} +m^{(\epsilon_{S}')2}_{I \bar{J} } \ .
\label{bc5A}
\end{equation}
Then it follows from \eqref{bc6}, \eqref{s4} and \eqref{s10A} that 
\begin{equation}
m^{(0)2}_{I\bar{J}}=\frac{1}{\hat{R}} \left( m_{3/2}^{2}-\frac{1}{3}F^{a}\bar{F}^{\bar{b}}(\partial_{a}\partial_{\bar{b}}K_{T})  \right) {\cal{G}}_{I\bar{J}} \ ,
\label{bc6AA}
\end{equation}
and from \eqref{bc7}, \eqref{s5AAA} and \eqref{fi1} that, in unity gauge for $L=\mathcal{O}_{X}(2,1,3)$ with $W_{i}=(9,17,0)$,
\begin{align}
\label{eq:squared_scalar_mass}
m^{(\epsilon_{S}')2}_{I\bar{J}}=&\biggl(\frac{m_{3/2}^{2}}{2\hat{R}V} \left[ \left( \frac{2}{3}+9 \left(\tfrac{1}{2}-\lambda\right)^{2} \right) a^{1}+ \left(-\frac{1}{3}+17\left(\tfrac{1}{2}-\lambda\right)^{2} \right) a^{2} +4a^{3} \right] \\
& \eqspace -F^{a}\bar{F}^{\bar{b}}M^{(\epsilon_{S}')}_{a\bar{b}} \biggr) {\cal{G}}_{I\bar{J}}\ , \nonumber
\end{align}
with the coefficients of $M^{(\epsilon_{S}')}_{a\bar{b}}$ given in \eqref{help1} -- \eqref{res5}.
Adding \eqref{bc6AA} and \eqref{eq:squared_scalar_mass}, the scalar masses squared $m_{I\bar{J}}^{2}$ can be put in the simple form
\begin{equation}
m_{I\bar{J}}^{2} =m^2_s(a^1,a^2,a^3)\mathcal{G}_{I\bar J}\ ,
\end{equation}
 where $m_s^2$ is a moduli-dependent function which is independent of the $I,\bar{J}$ indices. This function can be computed at any point inside the ``viable'' region of K\"ahler moduli space associated with $L=\mathcal{O}_{X}(2,1,3)$. Recall that we have assumed that $\mathcal{G}_{I\bar J}$ is a moduli-independent matrix, with numerical entries.

\subsection*{Cubic Scalar Coefficients}

The generic form for the mass-dimension-one coefficients of the cubic scalar soft supersymmetry breaking terms was shown in \cite{Soni:1983rm,Kaplunovsky:1993rd,Louis:1994ht,Brignole:1998dxa} to be 
\begin{equation}
a_{IJK}=F^{a}\left( \partial_{a}Y_{IJK}+\tfrac{1}{2}(\partial_{a}\hat{K})Y_{IJK}-3\Gamma^{N}_{a(I}Y_{JK)N} \right) \ ,
\label{night1}
\end{equation}
where $\hat{K}$ is given in \eqref{sum1}, \eqref{split2} and \eqref{tel1}, and $Y_{IJK}$ is
\begin{equation}
Y_{IJK}=\ee^{\hat{K}/2}\hat{Y}_{IJK}=\ee^{\hat{K}/2} 2\sqrt{2\pi\hat{\alpha}_{\text{GUT}}} ~y_{IJK}\ ,
\label{night2}
\end{equation}
with $\hat{\alpha}_{\text{GUT}}$ defined in \eqref{bag1}, $y_{IJK}$ the Yukawa couplings at mass scale $\langle M_{U} \rangle$ and, as defined in \eqref{r2},
\begin{equation}
\Gamma^{N}_{aI}=Z^{N\bar{J}}\partial_{a}Z_{\bar{J}I} \ .
\label{r2A}
\end{equation}
Noting that \eqref{night2} implies 
\begin{equation}
\partial_{a}Y_{IJK}=\frac{1}{2}(\partial_{a}\hat{K})Y_{IJK} \ ,
\label{night3}
\end{equation}
it follows that expression \eqref{night1} can be simplified to
\begin{equation}
a_{IJK}=F^{a}\left( (\partial_{a}\hat{K})Y_{IJK}-3\Gamma^{N}_{a(I}Y_{JK)N} \right) \ .
\label{night1A}
\end{equation}
As discussed previously, subject to the assumption that
\begin{equation}
Z_{I\bar{J}}^{(0)}=\ee^{K_{T}/3}{ \cal{G}}_{I\bar{J}} \ ,
\label{doc1}
\end{equation}
it follows from \eqref{ep2} and \eqref{ep3} that
\begin{equation}
\Gamma^{N}_{aI}=\Gamma^{(0)N}_{aI} +\Gamma^{(\epsilon_{S}')N}_{aI}=\left( \frac{1}{3}(\partial_{a}K_{T}) +\frac{\epsilon_{S}'}{2} \partial_{a}\big( \frac{(T+\bar{T})^{i}}{(S+\bar{S})}[X]_i \big)  \right)\delta^{N}_{I} \ ,
\label{doc2}
\end{equation}
where $[X]_i$ is defined in \eqref{ep4}. Furthermore, the symmetry of the cubic couplings $Y_{IJK}$ implies that
\begin{equation}
\Gamma^{N}_{a(I}Y_{JK)N}=\left[ \frac{1}{3}(\partial_{a}K_{T}) +\frac{\epsilon_{S}'}{2} \partial_{a}\left( \frac{(T+\bar{T})^{i}}{(S+\bar{S})} [X]_i \right) \right] Y_{IJK} \ .
\label{doc3}
\end{equation}
Inserting this into \eqref{night1A} and using \eqref{sum1}, it follows that
\begin{equation}
a_{IJK}=F^{a}\partial_{a} \left( \tilde{K}_{S}-\frac{3}{2} \epsilon_{S}' \frac{(T+\bar{T})^{i}}{(S+\bar{S})} [X]_i \right) Y_{IJK}\ .
\label{doc4}
\end{equation}

To compare this result to the formalism for the soft terms in the $B-L$ MSSM presented in \cite{Ovrut:2012wg}, it is convenient to write \eqref{doc4} in the form
\begin{equation}
\begin{split}
a_{IJK}&= \mathcal{A}(S,T^{i},Z) ~y_{IJK} \ ,\\
\end{split}
\label{doc5}
\end{equation}
where $\mathcal{A}$ is the specific function of the moduli given by
\begin{equation}
\mathcal{A}(S,T^{i},Z)=2 \sqrt{2\pi\hat{\alpha}_{\text{GUT}}}\ee^{\hat{K}/2}F^{a}\partial_{a} \left( \tilde{K}_{S}-\frac{3}{2} \epsilon_{S}' \frac{(T+\bar{T})^{i}}{(S+\bar{S})} [X]_i \right)\ .
\label{doc6}
\end{equation}
As discussed in detail in \cite{Ovrut:2014rba,Ovrut:2015uea}, in the renormalization analysis of the $B-L$ MSSM the experimental values of the quark and lepton Yukawa parameters $y_{IJK}$ are entered into the theory at the electroweak scale. These parameters are then run-up using the RGEs to give precise values for the Yukawa couplings $ y_{IJK}$ at the unification scale $\langle M_{U} \rangle$. Hence, the $y_{IJK}$ parameters in the above analysis and in \eqref{doc5} are completely specified. Hence, the only unknown part of the soft supersymmetry breaking cubic parameters is the {\it universal} moduli function $\cal{A}$ defined in \eqref{doc6}. Its exact value will depend on where it is evaluated in moduli space. We note, for completeness, that the renormalization group equation used in \cite{Ovrut:2014rba,Ovrut:2015uea} sets all Yukawa parameters to zero except for the top and bottom quarks and for the tau lepton, including in the soft supersymmetry breaking terms. As shown in earlier work, the remaining Yukawa parameters are too small to lead to significant effects and are hence ignored. 

The $\partial_{a} \big( \tilde{K}_{S}-\frac{3}{2} \epsilon_{S}' \frac{(T+\bar{T})^{i}}{(S+\bar{S})} [X]_i \big)$ factors in the universal soft supersymmetry breaking cubic coefficient \eqref{doc6} can be explicitly calculated as functions of the moduli using the expression for $[X]_i$ in \eqref{ep4}. However, the generic results are not particularly enlightening. As we did in previous sections, we will present each of these quantities written in terms of the variables $\hat{R}$, $a^{i}$, $V$ and $\lambda$, and restricted to the case of the line bundle $L=\mathcal{O}_{X}(2,1,3)$ discussed above. Furthermore, we will present the results in unity gauge $\epsilon_S^\prime \hat R/V^{1/3}=1$. For this specific case, defining the indices $a=S,T^{1},T^{2},T^{3},Z$, we find that 
\begin{align}
\partial_{S} \left( \tilde{K}_{S}-\frac{3}{2} \epsilon_{S}' \frac{(T+\bar{T})^{i}}{(S+\bar{S})} [X]_i \right)=&-\frac{1}{2V}\biggl(1-\frac{1}{2V}(2a^{1}-a^{2}+12a^{3}) \label{toi1}\\
&\eqspace-\frac{3}{2V}\left(\lambda-\tfrac{1}{2}\right)^{2} (9a^{1}+17a^{2}) \biggr)\ , \nonumber \\
\partial_{T^{i}} \left( \tilde{K}_{S}-\frac{3}{2} \epsilon_{S}' \frac{(T+\bar{T})^{l}}{(S+\bar{S})} [X]_l \right)=& -\frac{1}{4\hat{R}V^{2/3}} \biggl[(2,-1,12)_{i} \label{toi2} \\
& \eqspace +\left(3-2\left(\lambda+\tfrac{1}{2}\right)^{2}\right)(9,17,0)_{i} \biggr]\ ,  \nonumber \\
\partial_{Z} \left( \tilde{K}_{S}-\frac{3}{2} \epsilon_{S}' \frac{(T+\bar{T})^{i}}{(S+\bar{S})} [X]_i \right)=&\frac{1}{2\hat{R}V^{2/3}}\left(3-2\left(\lambda-\tfrac{1}{2} \right) \right) \label{toi3} \ .
\end{align}
Putting these results into \eqref{doc6} and computing the associated F-term fields $F^{a}$, it follows from \eqref{doc6} that one can compute the universal $\mathcal{A}$ coefficient for any given point in the ``viable'' region of K\"ahler moduli space associated with the line bundle $L=\mathcal{O}_{X}(2,1,3)$.

\subsection*{The Holomorphic Quadratic Term}

The generic form for the mass-dimension-two coefficient of the quadratic scalar soft supersymmetry breaking terms $C_{I}C_{J}+\text{h.c.}$ was shown in \cite{Soni:1983rm,Kaplunovsky:1993rd,Louis:1994ht} to be
\begin{equation}
B_{IJ}=F^{a}\left(\partial_{a}\mu_{IJ}+\tfrac{1}{2}(\partial_{a}\hat{K})-3\Gamma^{N}_{a(I}\mu_{J)N} \right)- m_{3/2}\mu_{IJ}
\label{naan1}
\end{equation}
where $\hat{K}$ is given in \eqref{sum1}, \eqref{split2} and \eqref{tel1}, $m_{3/2}$ is defined in \eqref{van1} and 
\begin{equation}
\mu_{IJ}=\ee^{\hat{K}/2}\hat{\mu}_{IJ}\ ,
\label{naan2}
\end{equation}
with $\hat{\mu}_{IJ}$ the dimension-one parameter of the $C^{I}C^{J}$ holomorphic term in the superpotential.

Note that the three terms in the brackets in \eqref{naan1} are exactly of the same form as the expression for $a_{IJK}$ given in \eqref{night1} in the previous subsection, with $Y_{IJK}$ now replaced by $\mu_{IJ}$. Using the fact that $\mu_{IJ}$ is symmetric in $IJ$, and following the same procedure as was used to evaluate $a_{IJK}$ above, we find that 
\begin{equation}
F^{a}\left(\partial_{a}\mu_{IJ}+\tfrac{1}{2}(\partial_{a}\hat{K})-3\Gamma^{N}_{a(I}\mu_{J)N} \right)=F^{a}\partial_{a} \left( \tilde{K}_{S}-\frac{3}{2} \epsilon_{S}' \frac{(T+\bar{T})^{i}}{(S+\bar{S})} [X]_i \right) \mu_{IJ} \ ,
\label{naan3}
\end{equation}
where $\tilde{K}_{S}$ is defined in \eqref{split2} and $[X]_i$ is presented in \eqref{ep4}. It follows that 
\begin{equation}
B_{IJ}=\ee^{\hat{K}/2}\left[F^{a}\partial_{a} \left( \tilde{K}_{S}-\frac{3}{2} \epsilon_{S}' \frac{(T+\bar{T})^{i}}{(S+\bar{S})} [X]_i \right)-m_{3/2}   \right]\hat  \mu_{IJ} \ .
\label{naan4}
\end{equation}
Restricting these expression to the line bundle $L=\mathcal{O}_{X}(2,1,3)$ discussed above and expressing the results in the variables $\hat{R}$, $a^{i}$, $V$ and $\lambda$ in unity gauge, $B_{IJ}$ can then be evaluated using expressions in \eqref{toi1}, \eqref{toi2}, \eqref{toi3} and \eqref{van1}. As was the case for the $a_{IJK}$ coefficients above, it is useful to write $B_{IJ}$ in the form
\begin{equation}
B_{IJ}={\mathcal{B}}(S, T^{i}, Z) {\hat{\mu}}_{IJ}\ ,
\label{pa1}
\end{equation}
where $\mathcal{B}$ is a function of the moduli given by
\begin{equation}
{\mathcal{B}}(S, T^{i}, Z) =\ee^{\hat{K}/2}\left[F^{a}\partial_{a} \left( \tilde{K}_{S}-\frac{3}{2} \epsilon_{S}' \frac{(T+\bar{T})^{i}}{(S+\bar{S})} [X]_i \right)-m_{3/2}   \right] \ .
\label{pa2}
\end{equation}
To compare this result to the superpotential and the soft terms in the $B-L$ MSSM presented in \eqref{con2} and \eqref{eq:6} respectively, we note that 
\begin{equation}
	{\hat{\mu}}_{IJ}=   \begin{cases}
		\mu & \text{for }  I=H_{u}, J=H_{d} \text{ or } I=H_{d}, J=H_{u} \ , \\
		0 & \text{for other choices of }I,J \ ,
	\end{cases}                               
	\label{pa3}  
\end{equation}
and, therefore, that
\begin{equation}
	B_{IJ}=   \begin{cases}
		b & \text{for }  I=H_{u}, J=H_{d} \text{ or } I=H_{d}, J=H_{u} \ , \\
		0 & \text{for other choices of }I,J \ ,
	\end{cases}                               
	\label{pa4}  
\end{equation}
where 
\begin{equation}
b={\mathcal{B}}(S, T^{i}, Z)\, \mu \ .
\label{pa5}
\end{equation}
As in \cite{Ovrut:2014rba,Deen:2016zfr,Dumitru:2018jyb,Dumitru:2018nct,Dumitru:2019cgf}, we make no attempt to solve the ``${\mu}$ problem" \cite{Martin:1997ns}. Therefore, the value of the parameter $\mu$ at the unification scale is unconstrained. However, the ratio between $b$ and $\mu$
\begin{equation}
\frac{b}{\mu}={\mathcal{B}}(S, T^{i}, Z)
\label{pa6}
\end{equation}
{\it is} constrained at the unification scale for any given viable point in moduli space.

\newpage

\chapter{Searching for the $B-L$ MSSM at the LHC}

\section{The Phenomenology of the $B-L$ MSSM}

\subsection{The B-L MSSM}
\label{sec:2}

In this chapter, we review the contents of the $B-L$ MSSM theory relevant from a phenomenological standpoint. Within this context, we analyze $R$-parity violating decays of various superparticles and their potential signatures at the LHC.

The low energy manifestation of the ``heterotic standard model'', that is, the $B-L$ MSSM, arises from the breaking of an $SO(10)$ GUT theory via two independent Wilson lines, denoted by $\chi_{3R}$ and $\chi_{B-L}$, associated with the diagonal $T_{3R}$ generator of $SU(2)_{R}$ and the generator $T_{B-L}$ of $U(1)_{B-L}$ respectively. These specific generators  are chosen since it can be shown that there is no kinetic mixing of their respective Abelian gauge kinetic terms at any energy scale-- thus simplifying the RG calculations \cite{Ovrut:2012wg}. However, identical physical results will be obtained for any linear combination of these generators. Associated with these Wilson lines are two mass scales, $M_{\chi_{3R}}$ and $M_{\chi_{B-L}}$, with three possible relations between them; 1) $M_{\chi_{B-L}} > M_{\chi_{3R}}$, 2) $M_{\chi_{3R}} > M_{\chi_{B-L}}$ and 3) $M_{\chi_{3R}} = M_{\chi_{B-L}}$. As discussed in \cite{Ovrut:2012wg}, the masses in the first two relations can be adjusted so as to enforce exact unification at one loop of all gauge couplings at the $SO(10)$ unification scale $M_{U}$, whereas gauge unification cannot occur for the third mass relationship without accounting for threshold effects at the unification scale or the SUSY scale \cite{Deen:2016vyh,Ovrut:2012wg,Fundira:2017vip}. For this reason, we will not consider the third option. The gauge coupling RG equations associated with each of the first two mass relations were discussed in detail in \cite{Ovrut:2012wg} and, as far as low energy LHC phenomenology is concerned, give almost identical results. For specificity, therefore, we will focus on the first relationship and, without loss of accuracy, choose $M_{\chi_{B-L}}=M_{U}$. The lower scale $M_{\chi_{3R}}$, which we henceforth denote by $M_{I}$, is adjusted to obtain exact gauge coupling unification. We emphasize, however, that the low energy results predicted for the LHC are almost unchanged even if $M_{I}$ is chosen to yield only ``approximate'' gauge unification --- with moderate-sized gauge ``thresholds''. Conventionally, the scale of supersymmetry breaking is defined to be
\begin{equation}
M_{SUSY} = \sqrt{m_{{\tilde{t}}_{1}}m_{{\tilde{t}}_{2}}} \ ,
\label{cham1}
\end{equation}
where $m_{{\tilde{t}}_{1}}$ and $m_{{\tilde{t}}_{2}}$ are the lightest and heaviest stop masses respectively; see ,for example, \cite{Ovrut:2015uea}.
Suffice it here to say that for supersymmetry breaking to occur between the electroweak scale and 10~TeV, which will be the case in this analysis, the unification scale $M_{U}$ is found to be $\cal{O}$$(3 \times 10^{16}~{\rm GeV})$. Over the same range of supersymmetry breaking, however, the  intermediate scale $M_{I}$ changes from $\cal{O}$$(2 \times 10^{16}~{\rm GeV})$ to $\cal{O}$$(3 \times 10^{15}~{\rm GeV})$ respectively \cite{Deen:2016vyh}.

The details of the symmetry breaking and the respective mass spectra for this choice of Wilson line hierarchy were given in \cite{Ovrut:2012wg}. Here, we simply note that in the mass regime between $M_{U}$ and $M_{I}$, the gauge group is broken from $SO(10)$ to $SU(3)_{C} \times SU(2)_{L} \times SU(2)_{R} \times U(1)_{B-L}$ with the spectrum shown in Figure \ref{fig:matterContent}. This theory is referred to as the ``left-right'' model \cite{Senjanovic:1975rk} \cite{Grimus:1993fx}. As discussed above, for the supersymmetry breaking scales of interest in this analysis, this mass regime will on average be considerably smaller than one order of magnitude in GeV. 

At the ``intermediate'' scale $M_{I}$, the second Wilson line breaks this ``left-right'' model down to the exact $B-L$ MSSM. This theory has the $SU(3)_{C} \times SU(2)_{L} \times U_{Y}(1)$ gauge group of the standard model augmented by an additional $U(1)_{B-L}$ Abelian symmetry. As mentioned above, it is convenient-- and equivalent --to use the Abelian group $U(1)_{3R}$ with the generator 
\begin{equation}
T_{3R}=Y-\frac{B-L}{2}
\label{eq:1}
\end{equation}
in the RGE's since the associated gauge kinetic term cannot mix  with the  gauge kinetic energy of $U(1)_{B-L}$. That is, the $B-L$ MSSM gauge group is chosen, for computational convenience, to be
\begin{equation}
	SU(3)_C\otimes SU(2)_L\otimes U(1)_{3R}\otimes U(1)_{B-L} \ .
	\label{eq:2}
\end{equation}
The associated gauge couplings will be denoted by $g_3$, $g_2$, $g_{R}$ and $g_{BL}$.
The spectrum, as shown in Figure \ref{fig:matterContent}, is exactly that of the MSSM with three right-handed neutrino chiral multiplets, one per family; that is, three generations of matter superfields
\begin{eqnarray}
	Q=\trix{c}u\\d\notrix\sim({\bf 3}, {\bf 2}, 0, \frac{1}{3}) & \begin{array}{rl}u^c\sim&(\bar{\bf 3}, {\bf 1}, -1/2, -\frac{1}{3}) \\
	d^c\sim&(\bar{\bf 3}, {\bf 1}, 1/2, -\frac{1}{3})\end{array} \ , \nonumber\\
	L=\trix{c}\nu\\e\notrix\sim({\bf 1}, {\bf 2}, 0, -1)&\begin{array}{rl}\nu^c\sim&({\bf 1}, {\bf 1}, -1/2, 1)\\
	e^c\sim&({\bf 1}, {\bf 1}, 1/2, 1)\end{array} \ ,
	\label{eq:3}
\end{eqnarray}
along with two Higgs supermultiplets
\begin{equation}
H_u=\trix{c}H_u^+\\H_u^0\notrix\sim({\bf 1}, {\bf 2}, 1/2, 0) ~~, ~ H_d=\trix{c}H_d^0\\H_d^-\notrix\sim({\bf 1}, {\bf 2}, -1/2, 0) \ .
\label{eq:4}
\end{equation}

The superpotential of the $B-L$ MSSM is given by
\begin{eqnarray}
	W=Y_u Q H_u u^c - Y_d Q H_d d^c -Y_e L H_d e^c +Y_\nu L H_u \nu^c+\mu H_u H_d \ ,
	\label{eq:5}
\end{eqnarray}
where flavor and gauge indices have been suppressed and the Yukawa couplings are three-by-three matrices in flavor space. In principle, the Yukawa matrices are arbitrary complex matrices. However, the observed smallness of the three CKM mixing angles and the CP-violating phase dictate that the quark Yukawa matrices be taken to be nearly diagonal and real. The charged lepton Yukawa coupling matrix can also be chosen to be diagonal and real. This is accomplished  by moving the rotation angles and phases into the neutrino Yukawa couplings which, henceforth, must be complex matrices. Furthermore, the smallness of the first and second family fermion masses implies that all components of the up, down quark and charged lepton Yukawa couplings-- with the exception of the (3,3) components --can be neglected for the purposes of the RG running. Similarly, the very light neutrino masses imply that the neutrino Yukawa couplings are sufficiently small so as to be neglected for the purposes of RG running. However, the $Y_{\nu i3}$, $i=1,2,3$ neutrino Yukawa couplings cannot be neglected for the calculations of the neutralino, neutrino and chargino mass matrices, as well as in decay rates/branching ratios.
The $\mu$-parameter can be chosen to be real, but not necessarily positive, without loss of generality. We implement these constraints in the remainder of our analysis.

Spontaneous supersymmetry breaking is assumed to occur in a hidden sector-- a natural feature of both strongly and weakly coupled $E_{8} \times E_{8}$ heterotic string theory --and be transmitted through gravitational mediation to the observable sector and, hence, to the $B-L$ MSSM. Since the $B-L$ MSSM first manifests itself at the scale $M_{I}$, we will begin our analysis by presenting the most general soft supersymmetry breaking interactions at that scale. That is, at scale $M_{I}$,
the soft supersymmetry breaking Lagrangian is given by
\begin{align}
 \label{eq:6}
\begin{split}
	-\mathcal L_{\mbox{\scriptsize soft}}  = &
	\left(
		\frac{1}{2} M_3 \tilde g^2+ \frac{1}{2} M_2 \tilde W^2+ \frac{1}{2} M_R \tilde W_R^2+\frac{1}{2} M_{BL} \tilde {B^\prime}^2
	\right.
		\\
	& \left.
		\hspace{0.4cm} +a_u \tilde Q H_u \tilde u^c - a_d \tilde Q H_d \tilde d^c - a_e \tilde L H_d \tilde e^c
		+ a_\nu \tilde L H_u \tilde \nu^c + b H_u H_d + h.c.
	\right)
	\\
	& + m_{\tilde Q}^2|\tilde Q|^2+m_{\tilde u^c}^2|\tilde u^c|^2+m_{\tilde d^c}^2|\tilde d^c|^2+m_{\tilde L}^2|\tilde L|^2
	+m_{\tilde \nu^c}^2|\tilde \nu^c|^2+m_{\tilde e^c}^2|\tilde e^c|^2  \\  
	&+m_{H_u}^2|H_u|^2+m_{H_d}^2|H_d|^2 \ .
\end{split}
\end{align}
The $b$ parameter can be chosen to be real and positive without loss of generality. The gaugino soft masses can, in principle, be complex. This, however, could lead to CP-violating effects that are not observed. Therefore, we proceed by assuming they all are real. The $a$-parameters and scalar soft mass can, in general, be Hermitian matrices in family space. Again, however, this could lead to unobserved flavor and CP violation. Therefore, we will assume they all are diagonal and real. Furthermore, we assume that only the (3,3) components of the up, down quark and charged lepton $a$-parameters are significant and that the neutrino $a$ parameters are negligible for the RG running and all other purposes. For more explanation of these assumptions, see \cite{Ovrut:2015uea}.

As discussed in \cite{Ovrut:2015uea}, without loss of generality one can assume that the third generation right-handed sneutrino, since it carries the appropriate $T_{3R}$ and $B-L$ charges, spontaneous breaks the $B-L$ symmetry by developing a non-vanishing VEV
\begin{equation}
\left< \tilde \nu^c_3 \right> \equiv \frac{1}{\sqrt 2} v_R \ .
\label{eq:7}
\end{equation}
This VEV spontaneously breaks $U(1)_{3R}\otimes U(1)_{B-L}$ down to the hypercharge gauge group $U(1)_Y$. We denote the associated gauge parameter by $g^{\prime}$. However, since sneutrinos are singlets under the $SU(3)_C\otimes SU(2)_L\otimes U(1)_{Y}$ gauge group, it does not break any of the SM symmetries.  
At a lower mass scale, electroweak symmetry is spontaneously broken by the neutral components of both the up and down Higgs multiplets acquiring non-zero VEV's.  In combination with the right-handed sneutrino VEV, this also induces a VEV in each of the three generations of left-handed sneutrinos. The notation for the relevant VEVs is
\beq
        \left<\tilde \nu_{i}\right> \equiv \frac{1}{\sqrt 2} {v_L}_i, \ \
	\left< H_u^0\right> \equiv \frac{1}{\sqrt 2}v_u, \ \ \left< H_d^0\right> \equiv \frac{1}{\sqrt 2}v_d,
	\label{eq:8}
\eeq
where $i=1,2,3$ is the generation index. The neutral gauge boson that becomes massive due to $B-L$ symmetry breaking, $Z_R$, has a mass at leading order, in the relevant limit that $v_R \gg v$, of
\beq
	M_{Z_R}^2 = \frac{1}{4}\left(g_R^2+g_{BL}^2 \right) v_R^2
				\left(1+\frac{g_R^4}{g_R^2+g_{BL}^2} \frac{v^2}{v_R^2}\right) \ ,
	\label{eq:9}
\eeq
where
\beq
	v^2 \equiv v_d^2 + v_u^2 \ .
\eeq
The second term in the parenthesis is a small effect due to mixing in the neutral gauge boson sector. 

A discussion of the neutrino masses is presented in the next section, where they are shown to be roughly proportional to the ${Y_\nu}_{ij}$ and ${v_L}_i$ parameters. It follows that ${Y_\nu}_{ij} \ll 1$ and ${v_L}_i\ll v_{u,d}, v_R$. In this phenomenologically relevant limit, the minimization conditions of the potential are simple, leading to the VEV's
\begin{align}
	\label{eq:10}
	v_R^2=&\frac{-8m^2_{\tilde \nu_{3}^c}  + g_R^2\left(v_u^2 - v_d^2 \right)}{g_R^2+g_{BL}^2} \ ,
	\\
	{v_L}_i=&\frac{\frac{v_R}{\sqrt 2}(Y_{\nu_{i3}}^* \mu v_d-a_{\nu_{i3}}^* v_u)}
			{m_{\tilde L_{i}}^2-\frac{g_2^2}{8}(v_u^2-v_d^2)-\frac{g_{BL}^2}{8}v_R^2} \ , \label{eq:11}
	\\
	\label{eq:12}
	\frac{1}{2} M_{Z^0}^2 =&-\mu^2+\frac{m_{H_u}^2\tan^2\beta-m_{H_d}^2}{1-\tan^2\beta} \ ,
	\\
	\label{eq:13}
	\frac{2b}{\sin2\beta}=&2\mu^2+m_{H_u}^2+m_{H_d}^2 
\end{align}
where 
\begin{equation}
{\rm tan} \beta= \frac{v_{u}}{v_{d}} \ .
\label{eq:14}
\end{equation}
Here, the first two equations correspond to the sneutrino VEVs. The third and fourth equations are of the same form as in the MSSM, but new $B-L$ scale contributions to $m_{H_u}$ and $m_{H_d}$ shift their values significantly compared to the MSSM. Eq.~(\ref{eq:10}) can be used to re-express the $Z_R$ mass as
\beq
	\label{eq:15}
	M_{Z_R}^2 = -2 m_{\tilde \nu^c_3}^2
				\left(1+\frac{g_R^4}{g_R^2+g_{BL}^2} \frac{v^2}{v_R^2}\right) \ .
\eeq
This makes it clear that, to leading order, the $Z_R$ mass is determined by the soft SUSY breaking mass of the third family right-handed sneutrino. The term proportional to $v^2/v_R^2$ is insignificant in comparison and, henceforth, neglected  in our calculations.

Recall that $R$-parity is defined as 
\begin{equation}
R=(-1)^{3(B-L)+2s}.
\label{16}
\end{equation}
It follows that a direct consequence of generating a VEV for the third family sneutrino is the spontaneous breaking of $B-L$ symmetry and, hence, $R$-parity. The $R$-parity violating operators induced in the superpotential are given by
\begin{equation}
	\label{eq:17}
	W \supset \epsilon_i \,  e_i \,  H^{+}_u - \frac{1}{\sqrt 2 }{Y_e}_i \, {v_L}_i \,  H_d^- \,   e^c_i \ ,
\end{equation}
where
\begin{equation}
	\epsilon_i  \equiv \frac{1}{\sqrt 2} {Y_\nu}_{i3} v_R \ 
	\label{eq:18}
\end{equation}
and $Y_{ei}$ is the $i$th component of the diagonal lepton Yukawa coupling.
This general pattern of $R$-parity violation is referred to as bilinear $R$-parity breaking and has been discussed in many different contexts \cite{Mukhopadhyaya:1998xj,Chun:1998ub,Chun:1999bq,Hirsch:2000ef}. In addition, the Lagrangian contains bilinear terms generated by ${v_L}_i$ and $v_R$ in the super-covariant derivatives. These are
\begin{align}
\begin{split}
	\mathcal{L} \supset &
	- \frac{1}{2}{v_L}_i^* \left[ g_2 \left(\sqrt 2 \, e_i \tilde W^+ 
	+  \nu_{L_i} \tilde W^0\right) - g_{BL} \nu_{L_i} \tilde B' \right] \label{eq:19}
	\\
	&
	-\frac{1}{2} v_R \left[-g_R \nu_3^c \tilde W_R + g_{BL} \nu_3^c \tilde B' \right]+ \text{h.c.}
\end{split}
\end{align}

The consequences of spontaneous $R$-parity violation are quite interesting, and have been discussed in a number of papers \cite{FileviezPerez:2008sx,Barger:2008wn,FileviezPerez:2009groo,Everett:2009vy,FileviezPerez:2012mj,Perez:2013kla,Ghosh:2010hy,Barger:2010iv,Mohapatra:1986aw}. 
In this section, we will present the decay channels for arbitrary mass charginos and neutralinos, and analytically determine their decay rates. In the following sections, we will explore the phenomenological consequences of the $R$-parity violating (RPV) decays of the lightest, and next-to-lightest, supersymmetric particles; referred to as the LSP and NLSP respectively. These decays are potentially observable at the ATLAS detector of the LHC. Hence, if detected, these explicit decays could verify the existence of low energy $N=1$ supersymmetry, shed light on the structure of the precise supersymmetric model-- such as the $B-L$ MSSM --and, as will become apparent, even constrain whether the neutrino mass hierarchy is ``normal'' or ``inverted''. 
However, as is clear from expressions \eqref{eq:17} and \eqref{eq:19}, these results will depend explicitly on the values of the parameters $\epsilon_{i}$, $i=1,2,3$ and $v_{L_{i}}$, $i=1,2,3$ defined in \eqref{eq:18} and \eqref{eq:11} respectively. In turn, these parameters are dependent on the present experimental values of the neutrino masses. These reduce the number of independent RPV parameters from six to one and potentially restrict the value of the remaining independent coefficient. For that reason, we will discuss the neutrino masses and their direct  relationship to the $\epsilon_{i}$ and $v_{L_{i}}$ parameters next.

\subsection{Neutrino Masses and the RPV Parameters}
\label{sec:3}
As discussed in \cite{Marshall:2014cwa,Marshall:2014kea,Ovrut:2014rba,Ovrut:2015uea,Barger:2010iv}, it follows from the above Lagrangian that the third family right-handed neutrino and the three left-chiral neutrinos $\nu_{i}$, $i=1,2,3$ mix with the fermionic superpartners of the neutral gauge bosons and with the up- and down- neutral Higgsinos. In other words, the neutralinos now mix with the neutral fermions of the standard model. The mixing with the third-family right-handed sneutrino, through terms proportional to $\epsilon_i = Y_{\nu_{i3}}v_R/\sqrt 2 $ and $v_{L_i}$, allows the third-family right-handed sneutrino to act as a seesaw field giving rise to Majorana neutrino masses. This is reviewed here.

First, we note that this analysis is focused on the consequences of RPV decays at the LHC. There is the possibility that the RPV parameters are so small that the LSP decay length is too long for it to decay within the detector. Then the LSP would be effectively stable within the detector. For certain cases, such effectively stable sparticles have been searched for in \cite{ATLAS:2014fka}. If the LSP decay length is small enough to decay within the detector, but greater than about 1 mm, this would lead to ``displaced'' vertices, such as those searched for in, for example, \cite{Aad:2015rba}. We will choose parameters so that the decay length of the LSP, whatever sparticle that may be, is less than about 1 mm. We refer to such decays as ``prompt'' decays. Therefore, even though the analysis in this work is valid for any mass chargino and neutralino, should we choose the initial conditions so that they are the LSP, then their RPV decays will be prompt.

As was shown in the case of stops and sbottoms in \cite{Marshall:2014cwa,Marshall:2014kea}, prompt decays require the RPV parameters to be large enough to allow for significant Majorana neutrino masses. We expect the same to hold true for a variety of LSPs.  Therefore, we focus on the case of significant Majorana neutrino masses.
Note that, in addition to these Majorana neutrino masses, there can be pure Dirac mass contributions coming from the neutrino Yukawa coupling. The components $Y_{\nu_{i3}}$, which couple the left-handed neutrinos to the third-family right-handed neutrino, allow the third-family right-handed neutrino to act like a seesaw field and give rise to Majorana neutrino masses. The other components, $Y_{\nu_{i1}}$ and $Y_{\nu_{i2}}$, couple the left-handed neutrinos to the first- and second-family right-handed neutrinos. Note that in this model, the heavy third-family right-handed neutrino acts as a seesaw field, while the first- and second-family right-handed neutrinos remain as light sterile neutrinos. This means that the Dirac mass terms related to $Y_{\nu_{i1}}$ and $Y_{\nu_{i2}}$ can give rise to active-sterile oscillations in the neutrino sector. There have been some experimental hints of such oscillations, see \cite{PDG} for review. However, it is not yet clear that these results are due to true active-sterile oscillations. Hence, we proceed under the assumption that no such oscillations exist and that the $Y_{\nu_i1}$ and $Y_{\nu_i2}$ components of the neutrino Yukawa coupling must, therefore, be negligible, so they do not appear in the neutralino mass matrix below. It may be interesting to revisit the question of active-sterile neutrino oscillations in the $B-L$ MSSM in the future, perhaps after there is more experimental data.

In the basis $\left(\tilde W_R, \ \tilde W^0, \ \tilde H_d^0, \ \tilde H_u^0, \ \tilde B', \ \nu_3^c, \ \nu_{i}\right)$ with $i=1,2,3$, the neutralino mass matrix is of the form
\begin{equation}
	\mathcal{M}_{\tilde\chi^0} =
	\begin{pmatrix}
		M_{\tilde\chi^0}
		&
		m_D
		\\
		m_D^T
		&
		0_{3 \times 3}
	\end{pmatrix},
	\label{eq:20}
\end{equation}
where $M_{\tilde\chi^0}$ is a six-by-six matrix of order a TeV given by
\begin{equation}
{\small
\label{eq:21}
	M_{\tilde\chi^0} =
	\begin{pmatrix}
			M_R
		&
			0
		&
			-\frac{1}{2} \, g_{R} \, v_d
		&
			\frac{1}{2} \, g_R \, v_u
		&
			0
		&
			-\frac{1}{2} g_R v_R
	\\
			0
		&
			M_2
		&
			\frac{1}{2} \, g_2 \, v_d
		&
			-\frac{1}{2} \, g_2 \, v_u
		&
			0
		&
			0
	\\
			-\frac{1}{2} \, g_{R} \, v_d
		&
			\frac{1}{2} \, g_{2} \, v_d
		&
			0
		&
			-\mu
		&
			0
		&
			0
	\\
			\frac{1}{2} \, g_R \, v_u
		&
			-\frac{1}{2} \, g_2 \, v_u
		&
			-\mu
		&
			0
		&
			0
		&
			0
	\\
			0
		&
			0
		&
			0
		&
			0
		&
			M_{BL}
		&
			\frac{1}{2} \, g_{BL} \, v_{R}
	\\
			-\frac{1}{2} g_R v_R
		&
			0
		&
			0
		&
			0
		&
			\frac{1}{2} \, g_{BL} \, v_{R}
		&
			0
	\\
	\end{pmatrix}, 
}
\end{equation}
and $m_D$ is a six-by-three matrix 
\begin{equation}
{\tiny
\label{eq:22}
        m_{D} =
	\begin{pmatrix}
	&
  	                0_{1 \times 3}
	\\
         &
			\frac{1}{2} \, g_2 \, {v_L}_i^*
	\\
        &
	         	0_{1 \times 3}
	\\
        &
			\epsilon_i
	\\
	&
			-\frac{1}{2} \, g_{BL} \, {v_L}_i^*
	\\
         &
			\frac{1}{\sqrt 2} \, {Y_\nu}_{i3} \, v_u
			
	\end{pmatrix}
}
\end{equation}
\noindent of order an MeV. This allows the mass matrix to be diagonalized perturbatively. Note that we have suppressed all terms of the form $v_{L_{i}}Y_{\nu ij}$ in $\mathcal{M}_{\tilde\chi^0}$ since both $v_{L_{i}}$ and the neutrino Yukawa parameters are small. In addition, we emphasize that since only the third family right-handed sneutrino gets a non-vanishing VEV, only $\nu_3^c$ couples to the gauginos/Higgsinos. It follows that only the Dirac mass of the third-family neutrino enters the above mass matrix, whereas the first and second family Dirac neutrino masses are excluded.

The entire mass matrix $\mathcal{M}_{\tilde\chi^0}$ in \eqref{eq:20} can be diagonalized to
\begin{equation}
	\label{eq:23}
	\mathcal{M}_{\tilde\chi^0}^D = \mathcal{N}^* \mathcal{M}_{\tilde\chi^0} \mathcal{N}^\dagger
	\end{equation}
with
\begin{equation}
	\mathcal{N} =
	\begin{pmatrix}
		N & 0_{3 \times3}
		\\
		 0_{3\times3} & V_\pmns^\dagger
	\end{pmatrix}
	\begin{pmatrix}
		1_{6\times6} & -\xi_0
		\\
		\xi_0^\dagger & 1_{3 \times 3}     \label{eq:24}
	\end{pmatrix},
	\end{equation}
where $N$ is the matrix that diagonalizes $M_{\tilde\chi^0}$ given in eq. \eqref{eq:21}. Requiring that $\mathcal{M}^D_{{\tilde \chi}^0}$ be diagonal yields
\begin{equation}
\xi_0 = M_{\tilde\chi^0}^{-1} m_D.
\label{eq:25} 
\end{equation}
The second matrix on the right-hand side of $\mathcal{N}$ rotates away the neutralino/left-handed neutrino mixing, whereas the first matrix diagonalizes the six neutralino/third family right-handed neutrino states as well as the three left-chiral neutrino states. In this section, we will consider the diagonal $3 \times 3$ left-handed neutrino Majorana mass matrix only, returning to the diagonal neutralino mass matrix later. 

The diagonal left-chiral neutrino Majorana mass matrix is found to be
\begin{align}
\begin{split}
	\label{eq:26}
	{m_\nu^D}_{ij} & = \left(V_\pmns^T \, m_\nu \, V_\pmns\right)_{ij} \ .
\end{split}
\end{align}
The $3 \times 3$ matrix $m_{\nu}$ is given by \cite{Marshall:2014cwa}
\begin{equation}
	\label{eq:27}
	{m_\nu}_{ij} = A {v_L}_i^* {v_L}_j^* + B \left({v_L}_i^* \epsilon_j + \epsilon_i {v_L}_j^* \right) + C \epsilon_i \epsilon_j \ ,
\end{equation}
where
\begin{align}
\label{eq:28}
	A & = \frac
		{
			\mu \, M_{\tilde \gamma}
		}
		{
			2 \, M_{\tilde \gamma} v_u v_d - 4 M_2 M_{\tilde Y} \mu
		}
	\\
\label{eq:29}	
	B & = \frac
		{
			M_{\tilde \gamma} v_d \left( 2 M_{Z_R}^2+ g_{Z_R}^2 v_u^2\right) - 2 g_{Z_R}^2 g_{BL}^2 M_2 M_R \, \mu \, v_u
		}
		{
			4 M_{Z_R}^2 (M_{\tilde \gamma} v_u v_d - 2 M_{\tilde Y} M_2 \mu)
		}
	\\
	\begin{split}
	\label{eq:30}
	C & = 
		\Big(
				 2 g_{Z_R}^4 M_2 M_{BL} M_R \, \mu^2 v_u^2\\
&\quad\quad
				- g_{Z_R}^2 M_{BL} \mu
				\left(
					g_2^2 \, g_{Z_R}^2 M_R v_u^2
					+ g_R^2 M_2 \left(4 M_{Z_R}^2 + g_{Z_R}^2 v_u^2\right)
				\right) v_d v_u
		\Big)\\
		&\quad\quad/\Big(
			4 M_{Z_R}^4 \mu \left(2 M_{\tilde Y} M_2 \, \mu - M_{\tilde \gamma} v_d v_u\right)
		\Big)
		\\
		& \quad \quad 
		-\frac
		{
			M_{\tilde \gamma}  v_d^2
		}
		{
			2 \mu \left(2 M_{\tilde Y} M_2 \, \mu - M_{\tilde \gamma} v_d v_u\right)
		} 
	\end{split}
\end{align}
and 
\begin{equation}
	g_{Z_R}^2 \equiv g_{BL}^2+g_{R}^2 \ . \label{eq:31}
\end{equation}
As will be discussed in detail below, the soft mass parameters are all initialized statistically at the scale $M_{I}$, whereas the measured values of the gauge couplings are introduced at the electroweak scale. All of these parameters are then run to the appropriate energy scale using the RGEs discussed in detail in \cite{Ovrut:2015uea}. Additionally, the value of tan$\beta$ will be chosen statistically within a physically relevant interval and, for a given value of tan$\beta$, the parameters $v_{u}$ and $v_{d}$ are the measured Higgs VEVs. Finally, for any given set of statistical initial data, we fine-tune the value of the parameter $\mu$ using equation \eqref{eq:12}, so as to obtain the experimental value of the electroweak gauge boson $Z^{0}$ and, hence, the measured values for $W^{\pm}$ as well. 
 The $3 \times 3$ Pontecorvo-Maki-Nakagawa-Sakata matrix is
\begin{eqnarray}
	V_\pmns &=& 
	\begin{pmatrix}
		c_{12} c_{13}
		&
		s_{12} c_{13}
		&
		s_{13} e^{-i \delta}
		\\
		-s_{12} c_{23} - c_{12} s_{23} s_{13} e^{i \delta}
		&
		c_{12} c_{23} - s_{12} s_{23} s_{13} e^{i \delta}
		&
		c_{13} s_{23}
		\\
		s_{12} s_{23} - c_{12}  c_{23} s_{13} e^{i \delta}
		&
		-c_{12} s_{23} - s_{12}  c_{23} s_{13} e^{i \delta}
		&
		c_{13} c_{23}
	\end{pmatrix}\nonumber\\ &&\times \text{diag}(1, e^{i \mathcal{A}/2}, 1) \ , \label{eq:32}
\end{eqnarray}
with $c_{ab} (s_{ab}) = \cos \theta_{ab} (\sin \theta_{ab})$. The mixing angles and phases are determined by neutrino experiments. For the mixing angles, we use the values and uncertainties from \cite{Capozzi:2018ubv}. They are
\begin{equation}
\sin^2\theta_{12}=0.307\pm0.013~, \qquad  \sin^2\theta_{13}=(2.12\pm0.08)\times 10^{-2}
\label{eq:33}
\end{equation}
for both the normal and inverted neutrino mass hierarchies. For $\theta_{23}$, however, the best-fit values depend on the hierarchy, and the data admits multiple best-fit values. In the normal hierarchy, one finds
\begin{equation}
\sin^2\theta_{23}=0.417^{~+0.025}_{~-0.028}\qquad\mbox{or}\qquad0.597^{~+0.024}_{~-0.030}\ ,
\label{eq:normalHierarchy}
\end{equation}
while in the inverted hierarchy
\begin{equation}
\sin^2\theta_{23}=0.421^{~+0.033}_{~-0.025}\qquad\mbox{or}\qquad0.529^{~+0.023}_{~-0.030}\ .
\label{eq:invertedHierarchy}
\end{equation}
In this analysis, we will do a complete study of  all four of the cases in equations \eqref{eq:normalHierarchy} and \eqref{eq:invertedHierarchy}.
Regarding the CP-violating phase, $\delta$, we use the recent results in \cite{Capozzi:2018ubv} that in the normal hierarchy
\begin{equation}
\delta = {231.6^{\circ}}^{~+41.4^{\circ}}_{~-30.6^{\circ}},
\end{equation}
while in the inverted hierarchy
\begin{equation}
\delta = {273.6^{\circ}}^{~+18.7^{\circ}}_{~-27.0^{\circ}}.
\end{equation}
In addition, note that there is only one ``Majorana'' phase, that is, the parameter $\mathcal{A}$, since in both the normal and the inverted hierarchy one of the neutrinos is massless and, therefore, does not have a Majorana mass. The value of $\mathcal{A}$ is unknown and, hence, we will simply throw it statistically in the interval $[0^{\circ}, 360^{\circ}]$.

The mathematical expressions for the mass eigenvalues of the Majorana neutrino mass matrix $m^{D}_{\nu ij}$ can be constructed from the A,B,C components of $m_{\nu ij}$ given in \eqref{eq:28}, \eqref{eq:29} and \eqref{eq:30} respectively, as well as from the PMNS matrix given in \eqref{eq:32}. This has been done in detail in \cite{Marshall:2014cwa}, to which we refer the reader for details. Given values for all the relevant parameters discussed above, and the measured values for the neutrino mass eigenvalues for the normal and inverted hierarchies, this allows one to solve for the RPV parameters $\epsilon_{i}$, $v_{L_{i}}$ $i=1,2,3$. Respectively, the experimental values of the mass eigenvalues of the normal and inverted hierarchies are \cite{PDG}

\begin{itemize}
\item Normal Hierarchy:
\begin{equation}
	m_1 = 0, \quad m_2 = (8.68 \pm 0.10) \times 10^{-3} ~{\rm eV},\quad m_3 = (50.84 \pm 0.50) \times 10^{-3}~{\rm eV}
\label{b2}
\end{equation}
\item Inverted Hierarchy:
\begin{equation}
	m_1 = (49.84 \pm 0.40) \times 10^{-3} ~{\rm eV}, \quad m_2 = (50.01 \pm 0.40) \times 10^{-3} ~{\rm eV},\quad m_3 = 0.
\label{b3}
\end{equation}
\end{itemize}

In each case, all three $v_{L}$ parameters, as well as two of the $\epsilon$ parameters, can be determined in terms of a third $\epsilon$ parameter. The explicit expressions, of course, differ in the normal and inverted hierarchy cases, and are presented in detail in \cite{Marshall:2014cwa}. These are encoded into the computer program by which we determine all decay rates and branching ratios and won't be presented here. Suffice it to say that, in each case, which parameter $\epsilon_i$ is inputted is undetermined. Thus, we will statistically decide which of the three $\epsilon_{i}$ parameters is selected. Furthermore, we choose its value by randomly throwing it to be in the interval $[10^{-4}, 1.0]$ GeV with a log-uniform distribution. We limit the upper bound to $1.0$ GeV to avoid excessive fine-tuning in the neutrino masses. Furthermore, we cut off the lower bound at $10^{-4}$ GeV-- although this could be taken to $0$ --to enhance the readability of our branching ratio plots. It is important to note that, having statistically chosen one of the $\epsilon$ parameters in the above range, the other two $\epsilon$ parameters are determined by the computer code and are not necessarily bounded by this interval. For example, at least one of the remaining two $\epsilon$ parameters could, in principle, be considerably larger than $1.0$ GeV. 
If so, this could have important consequences for the suppression of lepton number violating interactions. The reason is the following.

It was shown in \cite{Barbier:2004ez}, and discussed in \cite{Marshall:2014cwa}, that the experimental bound on the decay of $\mu \longrightarrow e\gamma$ leads to the constraint on $\epsilon_{1}$ and $\epsilon_{2}$ that
\begin{equation}
\big|\frac{\epsilon_{1}\epsilon_{2}}{\mu^{2}} \big| \lesssim 2.5 \times 10^{-3}\big(\frac{m_{{\tilde{\nu}}^{c}_{3}}}{100~{\rm GeV}} \big)^{-2} \ .
\label{burta}
\end{equation}
Scanning the initial parameters in the $B-L$ MSSM, using \eqref{eq:9}, \eqref{eq:15} and the values for the gauge parameters discussed in \cite{Ovrut:2012wg}, we find that this becomes
\begin{equation}
\epsilon_{1}\epsilon_{2} \lesssim 68~ {\rm GeV}^{2} \ .
\label{burtb}
\end{equation}
Therefore, to adequately suppress the lepton number violating decays, it is essential to show that this bound is satisfied for any physically interesting set of initial data in this analysis. Later, we will demonstrate that for the LSPs of interest this constraint is easily satisfied.

\subsection{Physically Acceptable Vacua}
\label{sec:4}

Having stated the structure of the $B-L$ MSSM, and defined all of the associated parameters, we now use the computer code specified in detail in \cite{Ovrut:2015uea} to find the initial values of the parameters leading to completely acceptable physical vacua. The choice of initial parameters will be subject to all experimental observations presented above. For example, as discussed in Section \ref{sec:2}, we will assume that the ratio of the Wilson line masses $M_{\chi_{B-L}}$ and $M_{\chi_{3R}}$ is such that all gauge parameters unify at $M_{U}$, that all quark and charged lepton Yukawa couplings can be taken to be nearly diagonal and real and, as discussed in Section \ref{sec:3}, only the $Y_{\nu i3}$ components of the neutrino Yukawa couplings are non-negligible. In addition, we solve for the physically acceptable initial conditions subject to several new, and important, constraints. These are the following:

\begin{itemize}

\item Scattering Range of the Dimensionful Soft SUSY Breaking Parameters:\\
\begin{equation}
\big[~\frac{M}{f},Mf~\big] \quad {\rm where}~~~M=1.5~{\rm TeV}~, f=6.7  \ .
\label{eq:36}
\end{equation}
\end{itemize}
The median supersymmetry breaking mass $M$ and the parameter $f$ are chosen so that the range of dimensionful soft masses can have random values from just above the electroweak scale to a scale approaching the upper bound of what will be observable at the LHC. That is, range (\ref{eq:36}) is approximately $[200$ GeV$\>, 10$ TeV$]$.

\begin{itemize}
\item Random Sign of the Soft SUSY Breaking Parameters $\mu$, $M$ and $a$:
\begin{equation}
\big[-,+\big ] \ .
\label{eq:37}
\end{equation}
\end{itemize}
The sign of $\mu$ and the various soft parameters of the form $M$ and $a$ are chosen randomly to have either a + or - sign. \\

\begin{itemize}
\item Randomly Scattered Choice of ${\rm tan}\beta$:
\begin{equation}
{\rm tan}\beta \in [1.2~,~65] \ .
\label{eq:38}
\end{equation}
\end{itemize}
The upper and lower bounds for ${\rm tan} \beta$ are taken from \cite{Martin:1997ns} and are consistent with present bounds that ensure perturbative Yukawa couplings.\\

In addition to being subject to the above constraints, physically acceptable initial conditions are those which lead to the following phenomenological results. First, $B-L$ gauge symmetry must be spontaneously broken at a sufficiently high scale. Presently, the measured lower bound on the $Z_{R}$ mass is given by \cite{Aaboud:2017buh}\\
\begin{equation}
M_{Z_{R}} = 4.1 ~{\rm TeV} \ .  \label{eq:39}
\end{equation}

\noindent Secondly, electroweak (EW) symmetry must be spontaneously broken so that the $Z^{0}$ and $W^{\pm}$ masses have the measured values of \cite{PDG}

\begin{equation}
\quad M_{{Z^{0}}}= 91.1876 \pm 0.0021~{\rm GeV} , \quad M_{W^{\pm}}= 80.379 \pm0.012~{\rm GeV} \ .  \label{eq:40}
\end{equation}
\\
\noindent Third, the remaining sparticles must be above their measured lower bounds \cite{Ovrut:2015uea} given in Table \ref{tab:lower_bounds}.\\

\begin{table}[t]

\begin{center}

\begin{tabular}{ |c|c| }

\hline

SUSY Particle & Lower Bound \\

\hline

Left-handed sneutrinos & 45.6 GeV\\

Charginos, sleptons& 100 GeV \\

Squarks, except stop or bottom LSP& 1000 GeV \\

Stop LSP (admixture)& 550 GeV \\

Stop LSP (right-handed)& 400 GeV\\

Sbottom LSP& 500 GeV\\

Gluino& 1300 GeV\\

\hline

\end{tabular}

\caption{Current lower bounds on the SUSY particle masses.}

\label{tab:lower_bounds}

\end{center}

\end{table}
\noindent Finally, the Higgs mass must be within the $3\sigma$ allowed range from ATLAS combined run 1 and run 2 results \cite{Aaboud:2018wps}. This is found to be
\begin{equation}
M_{h^{0}}=124.97 \pm 0.72~ {\rm GeV} \ .
\label{eq:41}
\end{equation}
\\
\indent We now want to search for physically acceptable initial data, subject to all of the constraints and phenomenological conditions introduced above. Before applying any of these constraints, the number of parameters appearing in the $B-L$ MSSM greatly exceeds 100. However, subject to the constraints discussed above, this number is significantly reduced-- down to only 24 soft SUSY breaking parameters, as well as ${\rm tan} \beta$ and $\mu$. The RG code \cite{Ovrut:2015uea} that we use in this analysis involves all 24 SUSY breaking parameters.  It is, 
however, helpful to point out that many of the RGE's are dominated by two specific sums of these parameters given by
\begin{eqnarray}
	\label{eq:S.BL}
	&S_{BL'}=\Tr(2m_{\tilde Q}^2-m_{\tilde u^c}^2-m_{\tilde d^c}^2-2m_{\tilde L}^2+m_{\tilde \nu^c}^2+m_{\tilde e^c}^2) \label{eq:42} \\
	\label{eq:S.R}
	&\qquad \quad S_{3R}=m_{H_u}^2-m_{H_d}^2+\Tr\left(-\frac{3}{2}m_{\tilde u^c}^2+\frac{3}{2}m_{\tilde d^c}^2-\frac{1}{2} m_{\tilde \nu^c}^2+\frac{1}{2} m_{\tilde e^c}^2\right) \label{eq:43}\ ,
\end{eqnarray}
where the traces are over generational indices. This is helpful in that one can now reasonably plot initial data points in two-dimensional $S_{BL^{\prime}}- S_{3R}$ space, rather than in the full 24-dimensional space of all parameters. At the electroweak scale, we randomly set the value of ${\rm tan}\beta$. Furthermore, and importantly, we do not run the parameter $\mu$. Rather, after running all other parameters down to the electroweak scale, we fine-tune $\mu$ to give the measured values for the electroweak gauge bosons, as discussed above.

\begin{figure}[t]

\centering

\begin{subfigure}[b]{0.7\textwidth}

\includegraphics[width=1.\textwidth]{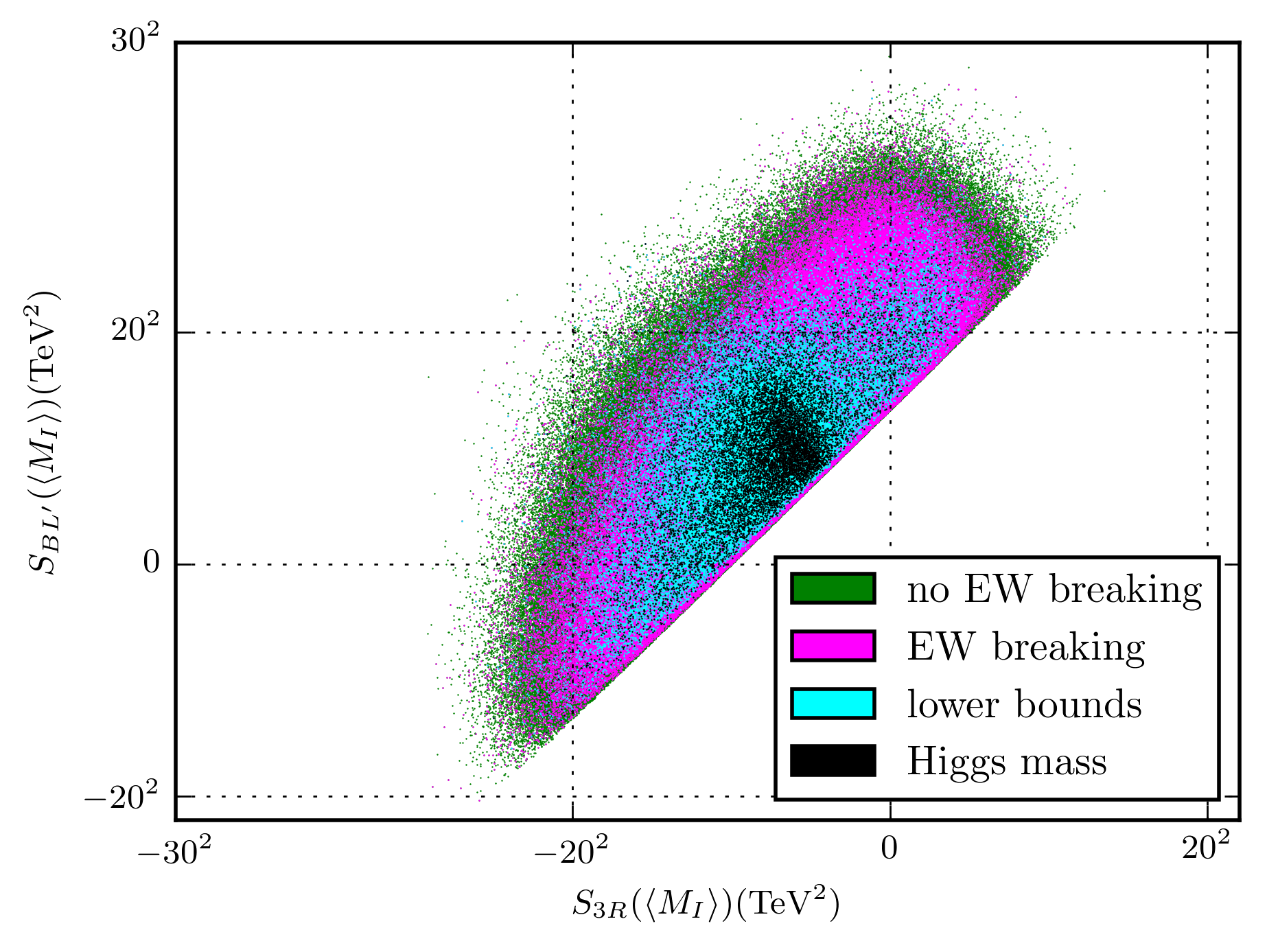}

\end{subfigure}

\caption{Plot of the 100 million initial data points for the RG analysis evaluated at $M_{I}$. The 4,351,809 green points lead to the appropriate breaking of the $B-L$ symmetry. Of these, the 3,142,657 purple points also break the EW symmetry with the correct vector boson masses. The cyan points correspond to 342,236 initial points that, in addition to appropriate $B-L$ and EW breaking, also satisfy all lower bounds on the sparticle masses. Finally, as a subset of these 342,236 initial points, there are 67,576 valid black points that lead to the experimentally measured value of the Higgs boson mass.}
\label{fig:eye}
\end{figure}

Searching for physically acceptable initial data, subject to all of the constraints and phenomenological conditions above, we find the following. For 100 million sets of randomly scattered initial conditions, it is found that 4,351,809 break $B-L$ symmetry with the $Z_{R}$ mass above the lower bound in equation \eqref{eq:39}. These are plotted as the green points in Figure \ref{fig:eye}. Running the RG down to the EW scale, one finds that of these 4,351,809 appropriate $B-L$ initial points, only 3,142,657 break electroweak symmetry with the experimentally measured values for $M_{Z^{0}}$ and $M_{W^{\pm}}$ given in equation \eqref{eq:40}. These are shown as the purple points in the Figure. Now applying the constraints that all sparticle masses be at or above their currently measured lower bounds presented in Table \ref{tab:lower_bounds}, we find that of these 3,142,657 initial points, only 342,236 are acceptable. These are indicated by cyan-colored points in the Figure. Finally, it turns out that of these 342,236 points, only 67,576 also lead to the currently measured Higgs mass given in equation \eqref{eq:41}. That is, of the 100 million sets of randomly scattered initial conditions, 67,576 satisfy all present phenomenological requirements. In Figure \ref{fig:eye}, we represent these ``valid'' points in black. 
That is, of the 100 million randomly scattered initial points, approximately .067\% satisfy all present experimental conditions. Although this might-- at first sight --appear to be a small percentage, it is worth noting that these initial points not only break $B-L$ symmetry appropriately and have all sparticle masses above their present experimental lower bounds, but also give the measured experimental values for the mass of the EW gauge bosons and, remarkably, the Higgs boson mass as well! From this point of view, this percentage of valid black points seems remarkably high. The electroweak gauge boson masses were obtained, as discussed above, by fine-tuning the parameter $\mu$. For example, a typical value of the fine-tuning of $\mu$ is of the order of 1 in 1000 \cite{Ovrut:2014rba}. However, one might also be concerned that getting the Higgs mass correct might require some other fine-tuning of the 24 initial parameters that may not be apparent. However, in previous work \cite{Ovrut:2015uea} it was shown that the 24 parameters associated with any given black point are generically widely disparate with no apparent other fine-tuning.
%

We conclude that the $B-L$ MSSM, in addition to arising as a vacuum of heterotic M-theory and having exactly the mass spectrum of the MSSM, satisfies all present experimental low-energy physical bounds for a remarkably large number of disparate initial data points. Given this, it becomes of real interest to determine whether the RPV decays of the $B-L$ MSSM can be directly observed at the LHC at CERN. These decays are most easily observed in the lightest sparticles in the mass spectrum; that is, the LSP has the best prospects for RPV detection in general. There are, however, cases in which the next lightest supersymmetric particle (NLSP) is highly degenerate in mass with the LSP-- see examples presented in \cite{Ovrut:2015uea} --and, hence, their RPV decay channels become relevant as well. 
Hence, in the sections to follow, we compute the decays of charginos and neutralinos without making any assumptions regarding their masses.
 The particle spectrum of each of the 67,576 valid black points is exactly determined by the computer code \cite{Ovrut:2015uea}.  It follows that we can compute the LSP associated with each valid black point. It turns out that there are many possible different LSPs. Before enumerating these, however, we must be more specific about the definition and structure of any LSP. Although the original fields entering the $B-L$ MSSM Lagrangian are ``gauge'' eigenstates, the LSP associated with a given valid black point  is, by definition, a ``mass'' eigenstate-- generically a linear combination of the original fields. For example, the lightest mass eigenstate chargino of either charge, which we denote by $\tilde \chi_{1}^{\pm}$, is found to be an $R$-parity conserving linear combination of the charged Wino, ${\tilde{W}}^{\pm}$, and the charged Higgsino, ${\tilde{H}}^{\pm}$, added to  RPV terms proportional to the left and right chiral charged leptons. The RPV coefficients are very small and, hence, can be ignored in the discussion of the masses of the charginos. Therefore, in this section, since we are analyzing the possible LSPs, we will consider the $R$-parity conserving part of the chargino states only. It then follows that when  $M_2<|\mu|$ the lightest chargino is given by
\begin{equation}
\tilde \chi^{\pm}_{1}=\cos \phi_{\pm}{\tilde{W}}^{\pm} + \sin \phi_{\pm}{\tilde{H}}^{\pm} \ ,
\label{cham2}
\end{equation}
whereas for $|\mu|<|M_2|$
\begin{equation}
\tilde \chi^{\pm}_{1}=-\sin \phi_{\pm}{\tilde{W}}^{\pm} + \cos \phi_{\pm}{\tilde{H}}^{\pm} \ .
\label{chamonix}
\end{equation}
The angles $\phi_{\pm}$ are exactly determined for any given black point. It follows that for some black points the mass eigenstate $\tilde \chi^{\pm}_{1}$ is predominantly a charged Wino, whereas for other black points it is mainly a charged Higgsino. We will, henceforth, denote the first and second type of mass eigenstates by $\tilde \chi^{\pm}_{W}$ and $\tilde \chi^{\pm}_{H}$, and refer to them as ``Wino charginos'' and ``Higgsino charginos'' respectively. That is, instead of labeling a chargino LSP simply as $\chi^{\pm}_{1}$, and counting the number of valid black points associated with it, we can be more specific-- breaking the chargino LSP into two different types of states, $\tilde \chi^{\pm}_{W}$ and $\tilde \chi^{\pm}_{H}$ respectively, and counting the number of black points associated with each type individually. This gives additional information about the structure of the LSPs.

\begin{figure}[t]

\centering

\begin{subfigure}[b]{0.7\textwidth}
\includegraphics[width=1.\textwidth]{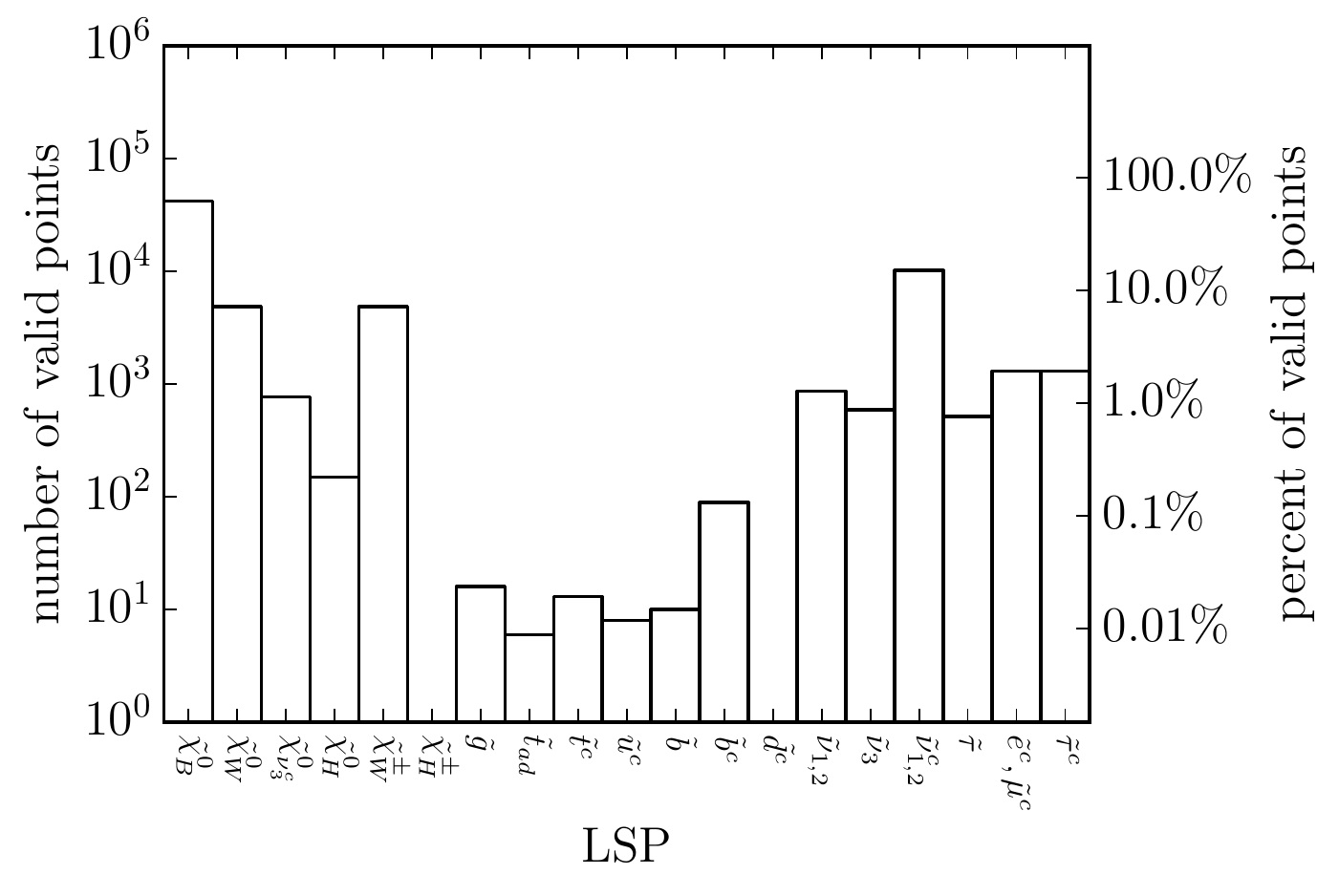}
\end{subfigure}
\caption{A histogram of the LSPs associated with a random scan of 100 million initial data points, showing the percentage of valid black points with a given LSP. Sparticles which did not appear as LSPs are omitted. The y-axis has a log scale. The notation and discussion of the sparticle symbols on the x-axis is presented in Table \ref{tab:notation}}
\label{fig:lspHistogram}
\end{figure}

With this in mind, we have calculated the LSP associated with each of the 67,576 valid black points and plotted the results as a histogram in Figure \ref{fig:lspHistogram}. The notation for the various possible LSPs is specified in Table \ref{tab:notation}. 
For example, out of the 67,576 valid black points, there are 4,858 that have a ${\tilde{\chi}}_{W}^{\pm}$ Wino chargino as their LSP. Similarly, out of all the valid black point initial conditions, 4,869 have a  ${\tilde{\chi}}_{W}^{0}$ Wino neutralino as their LSP. And so on.
Notice that the cases in which the chargino LSP is dominantly a charged Higgsino-- that is, $\tilde \chi^\pm_H$ --are rare. In fact, in Figure \ref{fig:lspHistogram} there is precisely one such black point. As discussed above and shown in Section \ref{sec:5}, the lighter chargino state is dominantly Wino if $|M_2|<|\mu|$, and dominantly Higgsino if $|\mu|<|M_2|$. The little hierarchy problem tells us that $\mu$ is generally large, of the order of a few TeV. However, the $M_2$ parameter generally takes smaller values in our simulation. For this reason, the instances in which $|\mu|<|M_2|$-- required for the Higgsino chargino to be the LSP --are scarce. 

%

\begin{figure}[!ht]
\centering
\begin{subfigure}[b]{0.45\textwidth}
\centering
\includegraphics[width=1.0\textwidth]{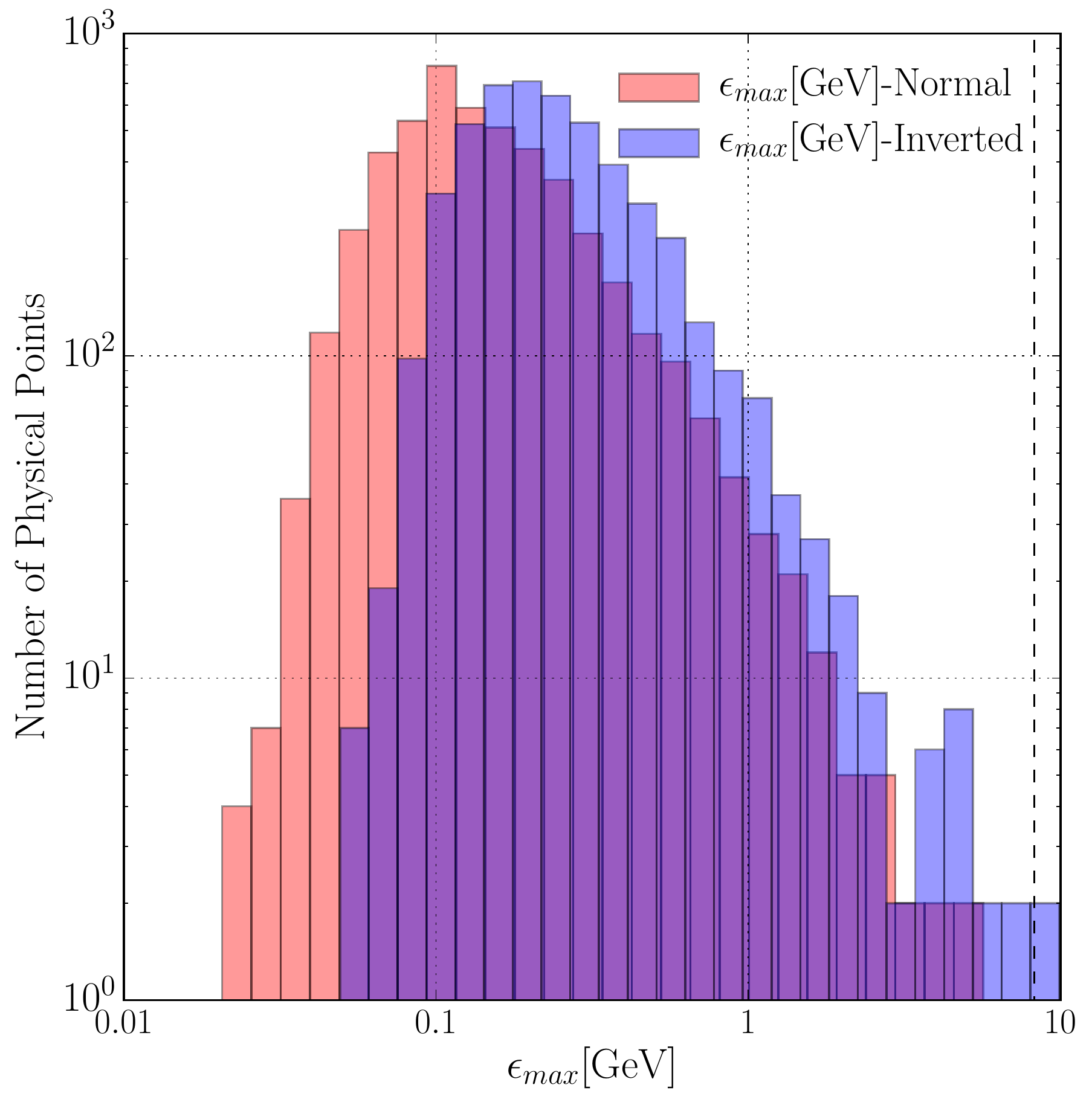}
\end{subfigure}
\caption{For each of the 4,858 black points with a Wino chargino LSP, we statistically sample one of the three $\epsilon_i$ parmeters in the range $[10^{-4}, 1.0]$ GeV and, for this graph, choose 1) a normal neutrino hierarchy and $\sin \theta_{23}=0.597$-- indicated in red --and 2) an inverted hierarchy with $\sin \theta_{23}=0.529$-- indicated in blue. We then solve for the other two $\epsilon$ values numerically. For each case, we plot a histogram of the number of valid black points associated with a Wino chargino LSP against the value of $\epsilon_{\text{max}}$, the largest of the three epsilon parameters. We find that most values of $\epsilon_{\text{max}}$ are smaller than 1 GeV. For larger values, the viable points become much less numerous, since such values would require more fine-tuning to match the existent neutrino data. The bound $\sqrt{68}$ GeV, beyond which unphysical lepton number violation is possible, is indicated by the dashed line. We find only 1  and 4 points beyond this line for the normal and inverted hierarchy cases respectively. Hence, lepton number violation, if it occurs at all, is statistically insignificant in our simulation. }
\label{fig:hope}
\end{figure}

Associated with a given choice of LSP, there are a fixed number of valid initial points. For example, as mentioned above, a Wino chargino LSP arises from 4,858 black points. For each such black point, we 1) statistically throw one of the parameters $\epsilon_{i}, i=1,2,3$ in the interval $[10^{-4},1.0]$ GeV, 2) choose the neutrino mass hierarchy to be either normal or inverted and, having done so, choose the associated value of $\theta_{23}$, 3) then, using \eqref{eq:27}, determine the remaining two epsilon parameters and the three $v_{L}$ parameters using the computer code. Let us denote the maximum one of the three $\epsilon$ parameters by $\epsilon_{{\rm max}}$. By running over all 4,858 black points subject to a fixed choice of the neutrino hierarchy and $\theta_{23}$, one can create a histogram of the 
number of valid points associated with a given value for $\epsilon_{{\rm max}}$. For example, the results of, first, choosing a normal neutrino hierarchy and $\theta_{23}$ such that $\sin \theta_{23}=0.597$ and, second, choosing an inverted neutrino hierarchy and $\theta_{23}$ with $\sin \theta_{23}=0.529$ are graphically depicted in Figure \ref{fig:hope}.
\noindent We find only a statistically insignificant number of points, in the case of the normal hierarchy 1 point and in the case of the inverted hierarchy 4 points, that exceed $\sqrt{68}$ GeV. If $\epsilon_{{\rm max}}=\epsilon_{3}$, then  constraint \eqref{burtb} is immediately satisfied. Even if this parameter is, say, $\epsilon_{2}$, it remains statistically extremely likely that constraint \eqref{burtb} remains satisfied. It follows that, for the choice of a normal neutrino hierarchy and  $\sin \theta_{23}=0.597$ and an inverted hierarchy with $\sin \theta_{23}=0.529$, lepton number violation via $\mu \rightarrow e\gamma$ is statistically highly  suppressed in our theory. We find  similar results for each of the other two choices of $\theta_{23}$. The identical conclusion can be drawn for  the Wino neutralino. We conclude that lepton number violation is highly suppressed in the $B-L$ MSSM-- at least when the LSP is a Wino chargino or a Wino neutralino.

In series of papers \cite{Marshall:2014kea,Marshall:2014cwa} analyzed the RPV decays of the ``admixture'' stop. Stops have a very high production cross section from proton-proton collisions. Furthermore, their decay products are relatively easy to observe at the LHC detectors. For these reasons, the ATLAS group at the LHC did a detailed study of the RPV decays of the admixture stop LSPs \cite{Aaboud:2018wps,ATLAS:2015jla,Jackson:2015lmj,ATLAS:2017hbw}. However, it is clear from Figure \ref{fig:lspHistogram} that neutralinos and charginos are much more prevalent as LSPs of the $B-L$ MSSM. Therefore, we begin a study of the RPV decays of neutralinos and charginos. 

\begin{table}[t]
\begin{center}
\begin{tabular}{|c|c|}
\hline
\ Symbol \ & Description
\\
\hline
\hline
$\tilde \chi^0_{B}$ & Mostly a neutral Bino.
\\
$\tilde \chi^0_{W}$ & Mostly a neutral Wino.
\\
${\tilde{\chi}}^{0}_{\nu^{c}_{3a,b}}$ & Mostly third generation right-handed neutrino.
\\
$\tilde \chi^0_{H}$ & Mostly a neutral Higgsino.
\\
\hline
$\tilde \chi^{\pm}_{W}$ & Mostly a charged Wino.
\\
$\tilde \chi^{\pm}_{H}$ & Mostly a charged Higgsino. 
\\
\hline
$\tilde g$ & Gluino. 
\\
\hline
 $\tilde t_{ad}$ & Left- and right-handed stop admixture. 
\\
\hline
$\tilde t_{r}$ & Mostly right-handed stop (over 99\%). 
\\
\hline
$\tilde q^{c}$ & Right-handed 1st and 2nd generation squarks. 
\\
\hline
$\tilde b$ & Mostly left-handed sbottom. 
\\
\hline
$\tilde b^{c}$ & Mostly right-handed sbottom. 
\\
\hline
$\tilde \nu_{{1,2}}$ & 1st and 2nd generation left-handed sneutrinos.				
\\
&LSPs evenly split among two generations.
\\
\hline
$\tilde \nu_{3}$ & Third generation left-handed sneutrino.
\\
\hline
$\tilde \nu^c_{{1,2}}$ & 1st and 2nd generation right-handed sneutrinos.
\\
\hline
{$\tilde \tau$} &  {Third generation left-handed stau.}
\\
\hline
{$\tilde e^{c}, \tilde \mu^{c}$} &  1st and 2nd generation right-handed sleptons. 
\\
	& LSPs evenly split among two generations.	
\\
\hline
$\tilde \tau^{c}$ &  Third generation right-handed stau. 
\\
\hline
\end{tabular}
\end{center}
\caption{The notation used for the LSP states on the x-axis of Figure \ref{fig:lspHistogram}.}
\label{tab:notation}
\end{table}

\section{Chargino and Neutralino states}\label{sec:5}
\subsection{Chargino mass eigenstates}\label{Chargino_masses}

After EW breaking, the Higgs fields acquire a VEV which induces off-diagonal couplings 
between the charged gauginos of the theory. The terms that enter the chargino mass matrix, {\it in the
absence of RPV effects}, are
\begin{equation}
\mathcal{L}\supset \frac{-g_2}{\sqrt{2}}[v_u\tilde W^-\tilde H_u^+ +v_d\tilde W^+H_d^-]-
M_2|\tilde W|^2-\mu \tilde H_u^+\tilde H_d^-+h.c.
\end{equation}
The first terms come from the supercovariant derivative of the Higgs chiral fields, the Wino mass term originates in the soft SUSY breaking Lagrangian, while the last term is introduced in the superpotential $W$.
Combining the charged Higgsinos and the charged Winos into $\psi^+=(\tilde W^+,\> \tilde H_u^+)$ and 
$\psi^-=(\tilde W^-,\> H_d^-)$, we can write the previous terms in the form
\begin{equation}
\mathcal{L}\supset -\frac{1}{2}\left(\psi^+ \> \psi^-\right)
\left(\begin{matrix}
0&{M_{\tilde \chi^\pm}}^T\\
{M_{\tilde \chi^\pm}}&0
\end{matrix}
\right)
\left(\begin{matrix}\psi^+\\ \psi^-\end{matrix}\right)
+h.c.
\label{eq:generalCharginoMatrix}
\end{equation}
where $M_{\tilde \chi^\pm}$ is the $2 \times 2$ matrix given by
\begin{equation}
{M_{\tilde \chi^\pm}}=\left(\begin{matrix}
M_2&\frac{1}{\sqrt{2}}g_2 v_u\\
\frac{1}{\sqrt{2}}g_2v_d&\mu 
\end{matrix}
\right).
\end{equation}
The mass eigenstates ${\tilde \chi}^+=V\psi^+$ and ${\tilde \chi}^-=U\psi^-$ diagonalize $M_{\tilde \chi^\pm}$ to
\begin{equation}
U^*M_{\tilde \chi^\pm}V^{-1}=M^D=
\left(\begin{matrix}M_{{\tilde \chi}^\pm_1}&0\\0&M_{{\tilde \chi}^\pm_2}\end{matrix}\right)
\end{equation}
with $M_{{\tilde \chi}^\pm_1}$ and $M_{{\tilde \chi}^\pm_2}$ positive.
One can solve analytically for the eigenvalues and obtain
\begin{multline}
M^2_{{\tilde \chi}^\pm_1}, M^2_{{\tilde \chi}^\pm_2}=
\frac{1}{2} \Big[|M_2|^2+|\mu|^2+2M^2_{W^\pm} \\
\mp \sqrt{ \left(|M_2|^2+|\mu|^2+2M_{W^\pm}^2 \right)^2-4|\mu M_2-M^2_{W^\pm} \sin{2 \beta}|^2} \>\Big] \ ,
\end{multline}
where $M^2_{{\tilde \chi}^\pm_1}$ and $M^2_{{\tilde \chi}^\pm_2}$ correspond to the $-$ and $+$ sign in front of the square root respectively.
We will always choose the square root to be positive, so that the "minus" sign-- and, hence, ${\tilde{\chi}}^{\pm}_1$ --corresponds to the lighter mass eigenstate. 
That is, with this convention $M_{{\tilde \chi}^\pm_1}<M_{{\tilde \chi}^\pm_2}$. The expressions for the mass eigenvalues can be simplified by noting that the lower bounds on sparticle masses are well above $M_{W^\pm}$. It follows that $M^2_{W^\pm}\ll M^2_2, \mu^2$. Therefore, the mass eigenvalues depend primarily on the parameters $M_{2}$ and $\mu$.
When $|M_{2}| \lesssim |\mu|$, we find that
%
\begin{equation}\label{eq:Wino_mass}
M_{{\tilde \chi}^\pm_1} \simeq|M_2|-\frac{M^2_{W^\pm}(M_2+\mu \sin 2\beta)}{\mu^2-M_2^2},
\end{equation}
\begin{equation}
M_{{\tilde \chi}^\pm_2} \simeq|\mu|+\frac{\text{sgn}(\mu)M^2_{W^\pm}(\mu+M_2 \sin 2\beta)}{\mu^2-M_2^2}
\label{eq:Higgsino_mass}
\end{equation}
whereas for $|\mu| \lesssim |M_{2}|$, the expressions for the mass eigenvalues are simply exchanged; that is, 
\begin{equation}
M_{{\tilde \chi}^\pm_1} \simeq |\mu|+\frac{\text{sgn}(\mu)M^2_{W^\pm}(\mu+M_2 \sin 2\beta)}{\mu^2-M_2^2} \ ,
\label{again1}
\end{equation}
\begin{equation}
M_{{\tilde \chi}^\pm_2} \simeq |M_2|-\frac{M^2_{W^\pm}(M_2+\mu \sin 2\beta)}{\mu^2-M_2^2}.
\label{again2}
\end{equation}
%

The mixing matrices $U$ and $V$, defined by 
\begin{equation}
\left( \begin{matrix}{\tilde \chi}^-_1\\{\tilde \chi}^-_2\end{matrix}  \right)=
U\left( \begin{matrix}\tilde W^-\\\tilde H_d^-\end{matrix}  \right),\quad
\left( \begin{matrix}{\tilde \chi}^+_1\\{\tilde \chi}^+_2\end{matrix}  \right)=
V\left( \begin{matrix}\tilde W^+\\ \tilde H_u^+\end{matrix}  \right) \ ,
\end{equation}
also are dependent on the relative sizes of $M_{2}$ and $\mu$. For $|M_{2}| \lesssim |\mu|$ they are found to be
\begin{equation}\label{eq:U_matrix}
U=O_-~~ , ~~ V=\begin{cases}
    O_+, & \det M_{\tilde \chi^\pm}>0\\
    \sigma_3 O_+, & \det M_{\tilde \chi^\pm}<0 \ ,
  \end{cases}
  \end{equation}
 where
\begin{equation}
O_\pm=\left(\begin{matrix}\cos \phi_\pm&\sin \phi_\pm\\-\sin \phi_\pm&\cos \phi_\pm\end{matrix}\right) \ .
\label{pass1}
\end{equation}
The Pauli matrix $\sigma_3$ is inserted so that the diagonal entries of $M^D$ are always positive. The angles $\phi_\pm$ are given by
\begin{equation}
\tan 2\phi_-=2\sqrt{2}M_{W^\pm}\frac{\mu \cos \beta +M_2 \sin \beta}{\mu^2-M_2^2-2M_{W^\pm}^2
\cos 2\beta}
\label{bernard1}
\end{equation}
\begin{equation}
\tan 2\phi_+=2\sqrt{2}M_{W^\pm}\frac{\mu \sin \beta +M_2 \cos \beta}{\mu^2-M_2^2+2M_{W^\pm}^2
\cos 2\beta}
\label{bernard2}
\end{equation}
respectively. On the other hand, when $|\mu| \lesssim |M_2|$, we find that \eqref{eq:U_matrix} remains the same, as do the expressions \eqref{bernard1} and \eqref{bernard2} for the angles $\phi_{\pm}$. However, the matrix $O_{\pm}$ now becomes
\begin{equation}
O_\pm=\left(\begin{matrix}-\sin \phi_\pm&\cos \phi_\pm\\-\cos \phi_\pm&-\sin \phi_\pm\end{matrix}\right) \ .
\label{talk1}
\end{equation}
It is important to note that the $|\mu| \lesssim |M_2|$ results for the $U$, $V$ matrices can be obtained from the $|M_{2}| \lesssim |\mu|$ expressions \eqref{eq:U_matrix}, \eqref{pass1},
\eqref{bernard1} and \eqref{bernard2} simply by replacing
\begin{equation}
\phi_{\pm} \longrightarrow \phi_{\pm} + \frac{\pi}{2}
\label{pass2}
\end{equation}
in all expressions. We will use this replacement, when required, in the main numerical analysis to follow.

It is useful to note that when  $|M_{2}| \lesssim |\mu|$, it follows from \eqref{pass1} that the lightest chargino eigenstate is
\begin{equation}
{\tilde{\chi}}_{1}^{\pm}= \cos \phi_{\pm} {\tilde{W}}^{\pm} + \sin \phi_{\pm} {\tilde{H}}^{\pm} \ .
\label{ok1}
\end{equation}
Since $M^2_{W^\pm}\ll M^2_2, \mu^2$, we see from \eqref{bernard1} and \eqref{bernard2} that ${\rm tan} ~2\phi_{\pm} \ll 1$ and, hence, $|\cos \phi_{\pm}| > |\sin \phi_{\pm}|$. It follows that 
\begin{equation}
{\tilde{\chi}}_{1}^{\pm} \simeq  {\tilde{W}}^{\pm}  \ .
\label{ok2}
\end{equation}
We say that ${\tilde{\chi}}_{1}^{\pm}$ is ``predominantly'' a charged Wino and, regardless of the exact value of $\phi_{\pm}$, denote it by  ${\tilde{\chi}}^{\pm}_{{{W}}}$. On the other hand, when $|\mu| \lesssim |M_2|$ it follows from 
\eqref{talk1} that 
\begin{equation}
{\tilde{\chi}}_{1}^{\pm}= -\sin \phi_{\pm} {\tilde{W}}^{\pm} + \cos \phi_{\pm} {\tilde{H}}^{\pm} \ .
\label{ok3}
\end{equation}
Expressions \eqref{bernard1} and \eqref{bernard2} again tell us that $|\sin \phi_{\pm}| < |\cos \phi_{\pm}|$ and, hence
\begin{equation}
{\tilde{\chi}}_{1}^{\pm} \simeq  {\tilde{H}}^{\pm}  \ .
\label{ok4}
\end{equation}
That is,  ${\tilde{\chi}}_{1}^{\pm}$ is ``predominantly'' a charged Higgsino and, regardless of the exact value of $\phi_{\pm}$, we denote it by  ${\tilde{\chi}}^{\pm}_{{{H}}}$. Having made this analysis, we note that the results could have been read off directly from the leading term in the expressions for the mass eigenvalues given in \eqref{eq:Wino_mass} and \eqref{again1}. Specifically, for $|M_{2}| \lesssim |\mu|$, the leading term in \eqref{eq:Wino_mass} is $|M_{2}|$, the soft mass associated with the charged Wino-- indicating that ${\tilde{\chi}}_{1}^{\pm} \simeq  {\tilde{W}}^{\pm}$. Similarly, for $|\mu| \lesssim |M_2|$, the leading term in \eqref{again1} is $|\mu|$, the parameter associated with the charged Higgsino-- indicating that
${\tilde{\chi}}_{1}^{\pm} \simeq  {\tilde{H}}^{\pm}$, as above. As stated previously, we will refer to the mass eigenstates ${\tilde{\chi}}_{W}^{\pm}$ and
${\tilde{\chi}}^{\pm}_{{{H}}}$ simply as a Wino chargino and Higgsino chargino respectively, even though they are only ``predominantly'' the pure states of $W^{\pm}$ and $H^{\pm}$ respectively. 

%
%
%
%

Let us now {\it include the $R$-parity violation terms} in the Lagrangian. These will not significantly affect the chargino masses, but they do introduce mixing between the charginos and the standard model charged leptons. These mixings are central to our study because they allow RPV chargino decays.
In the extended bases $\psi^+=(\tilde{W}^+, \tilde{H}_u^+, {e^c}_i)$ and $\psi^-=(\tilde{W}^-, \tilde{H}^-_d, e_i)$, the mixing matrix can again be written in the form of eq. \eqref{eq:generalCharginoMatrix},
\begin{equation}
\mathcal{L}\supset -\frac{1}{2}\left(\psi^+ \> \psi^-\right)
\left(\begin{matrix}
0&{{\cal{M}}_{\tilde \chi^\pm}}^T\\
{{\cal{M}}_{\tilde \chi^\pm}}&0
\end{matrix}
\right)
\left(\begin{matrix}\psi^+\\ \psi^-\end{matrix}\right)
+h.c.
\end{equation}
now, however, where
\begin{equation}
{\cal{M}}_{\tilde \chi^\pm}=\left(
\begin{matrix}
	M_2&\frac{1}{\sqrt{2}}g_2v_u&0&0&0\\
	\frac{1}{\sqrt{2}}g_2v_d&\mu&-\frac{{v_{L1}}}{v_d}m_e&-\frac{{v_{L2}}}{v_d}m_{\mu}&-\frac{v_{L3}}{v_d}m_{\tau}\\
\frac{1}{\sqrt{2}}g_2{v_{L1}}^*&{-\epsilon_1}&m_e&0&0\\
\frac{1}{\sqrt{2}}g_2{v_{L2}}^*&{-\epsilon_2}&0&m_{\mu}&0\\
\frac{1}{\sqrt{2}}g_2{v_{L3}}^*&{-\epsilon_3}&0&0&m_{\tau}
\end{matrix}
\right)
\end{equation}
and $m_e$, $m_\mu$, and $m_\tau$ denote the Dirac masses of the standard model charged leptons.
This matrix can be expressed in a schematic form that will be useful in diagonalizing it. Let us write
\begin{equation}
{\cal{M}}_{\tilde \chi^\pm}=\left(
\begin{matrix}
{M_{\tilde \chi^\pm}}&\Gamma\\
G^T&m_{e_i}
\end{matrix}
\right)
\end{equation}
where
\begin{equation}
\begin{split}
{M_{\tilde \chi^\pm}}=\left(
\begin{matrix}
M_2&\frac{1}{\sqrt{2}}g_2v_u \\
\frac{1}{\sqrt{2}}g_2v_d&\mu
\end{matrix}
\right),
\end{split}
\end{equation}
and
\begin{equation}
\begin{split}
\Gamma=\left(\begin{matrix}0&0&0\\-\frac{{v_{L1}}}{v_d}m_e&-\frac{{v_{L2}}}{v_d}m_{\mu}&-\frac{v_{L3}}{v_d}m_{\tau} \end{matrix}\right),\quad
G^{T}=\left(\begin{matrix} \frac{1}{\sqrt{2}}g_2{v_{L1}}^*&{-\epsilon_1}\\ \frac{1}{\sqrt{2}}g_2{v_{L2}}^*&{-\epsilon_2}\\ \frac{1}{\sqrt{2}}g_2{v_{L3}}^*&{-\epsilon_3} \end{matrix}\right)
\end{split}
\end{equation}
and $m_{e_i}$ is the $3 \times 3$ matrix with diagonal entries $(m_e,m_\mu,m_\tau)$.
The $G$, $\Gamma$, and $m_{e_i}$ matrices have entries that are much smaller than the entries of the ${M_{\tilde \chi^\pm}}$ matrix and, therefore, can be used to perturbatively diagonalize the ${\cal{M}}_{\tilde \chi^\pm}$ matrix. The mass eigenstates are related to the gauge eigenstates by unitary matrices $\mathcal{V}$ and $\mathcal{U}$ defined by
\begin{equation}
\left(
\begin{matrix}
\tilde{{\chi}}_1^-\\
\tilde{{\chi}}_2^-\\
\tilde{{\chi}}_3^-\\
\tilde{{\chi}}_4^-\\
\tilde{{\chi}}_5^-\\
\end{matrix}
\right)
=\mathcal{U}\left(
\begin{matrix}
\tilde{W}^-\\
\tilde{H}^-_d\\
e_1\\
e_2\\
e_3\\
\end{matrix}
\right),
\quad \quad
\left(
\begin{matrix}
\tilde{{\chi}}_1^+\\
\tilde{{\chi}}_2^+\\
\tilde{{\chi}}_3^+\\
\tilde{{\chi}}_4^+\\
\tilde{{\chi}}_5^+\\
\end{matrix}
\right)
=\mathcal{V}\left(
\begin{matrix}
\tilde{W}^+\\
\tilde{H}^+_u\\
{e^c_1}\\
{e^c_2}\\
{e^c_3}\\
\end{matrix}
\right).
\end{equation}
%
They are chosen so that
\begin{equation}
\mathcal{U}^*\mathcal{M}_{\tilde \chi^\pm}\mathcal{V}^{-1}=\mathcal{M}_{\tilde \chi^\pm}^D=
\text{diag}\left(M_{\tilde{{\chi}}^\pm_1},M_{\tilde{{\chi}}^\pm_2},M_{\tilde{{\chi}}^\pm_3},M_{\tilde{{\chi}}^\pm_4}, M_{\tilde{{\chi}}^\pm_5}
\right) \ ,
\end{equation}
where all eigenvalues are positive.
The first two eigenstates, $M_{\chi^\pm_{1,2}}$, are mostly charged Wino and charged Higgsino. The RPV couplings are small enough that their masses are still given by eqs. \eqref{eq:Wino_mass} and \eqref{eq:Higgsino_mass} or \eqref{again1} and \eqref{again2}, depending on the values of the parameters $M_{2}$ and $\mu$. Similarly, the masses of the leptons are 
basically unchanged; that is, $M_{\chi^\pm_{2+i}}$ are the standard model charged lepton masses, $m_{e_i}$, where $e_{1,2,3}\equiv e,\mu,\tau$. The phenomenologically important effect of the RPV mixing is the RPV decays of the charginos.
The matrices $\mathcal{U}$ and $\mathcal{V}$ can be written schematically as
\begin{equation}
\label{eq:matrixUV_PLM}
\mathcal{U}=
\left(
\begin{matrix}
U&0_{2\times3}\\
0_{3\times2}&1_{3\times3}\\
\end{matrix}
\right)
\left(
\begin{matrix}
1_{2\times2}&-\xi_-\\
\xi^{\dag}_-&1_{3\times3}\\
\end{matrix}
\right)~\ ,
\quad \>
\mathcal{V}=
\left(
\begin{matrix}
V&0_{2\times3}\\
0_{3\times2}&1_{3\times3}\\
\end{matrix}
\right)
\left(
\begin{matrix}
1_{2\times2}&-\xi_+\\
\xi^{\dag}_+&1_{3\times3}\\
\end{matrix}
\right).
\end{equation}
Next, requiring that $\mathcal{U^*}\mathcal{M}_{\tilde \chi^\pm}\mathcal{V}^{-1}$ is diagonal allows one to compute the 
$\xi_+$ and $\xi_-$ matrices. To first order, they are given by
\begin{equation}
\xi_-=-\left( {M_{\tilde \chi^\pm}}^T \right)^{-1}G~, \quad \xi_+=-({M_{\tilde \chi^\pm}})^{-1}\Gamma \ .
\end{equation}
The values of all the $\mathcal{U}$ and $\mathcal{V}$ matrix elements are presented in the Appendix \ref{appendix:A1}. These matrices will be crucial in calculating the decay rates of the charginos via RPV processes. In general, the exact expressions for the five charged mass eigenstates are complicated, and will be dealt with numerically in our calculations. However, it of interest to present the complete analytic expressions, including the RPV terms, for the mass eigenstates ${\tilde{\chi}}^{\pm}_{1}$. We find that the positive eigenvector ${\tilde{\chi}}^{+}_{1}$ is
\begin{equation}
{\tilde{\chi}}^{+}_{1}=\mathcal{V}_{1\>1} {\tilde{W}}^{+}+\mathcal{V}_{1\>2} {\tilde{H}}_{u}^{+}+ \mathcal{V}_{1\>2+i} e^{c}_{i} \ ,
\label{finish1}
\end{equation}
where one sums over $i=1,2,3$. When $|M_{2}|<|\mu|$, the $\mathcal{V}$ coefficients are given by
\begin{equation}
\mathcal{V}_{1\>1}=\cos \phi_{+}~,~\mathcal{V}_{1\>2} =\sin \phi_{+}
\label{finish2}
\end{equation}
and
\begin{equation}
\mathcal{V}_{1\>2+i}=-\cos \phi_+ \frac{g_2 \tan \beta m_{e_i}}{\sqrt{2}M_2\mu}v_{L_i}+\sin \phi_+\frac{m_{e_i}}{\mu v_d}v_{L_i} \ .
\label{finish3}
\end{equation}
The result for $|\mu|<|M_{2}|$ is found by replacing $\phi_{+} \rightarrow \phi_{+} + \frac{\pi}{2}$ in these expressions, as discussed above. Similarly, the negative eigenvector ${\tilde{\chi}}^{-}_{1}$ is found to be
\begin{equation}
{\tilde{\chi}}^{-}_{1}=\mathcal{U}_{1\>1} {\tilde{W}}^{-}+\mathcal{U}_{1\>2} {\tilde{H}}_{d}^{-}+ \mathcal{U}_{1\>2+i} e_{i} \ ,
\label{finish1}
\end{equation}
where one sums over $i=1,2,3$. When $|M_{2}|<|\mu|$, the $\mathcal{U}$ coefficients are given by
\begin{equation}
\mathcal{U}_{1\>1}=\cos \phi_{-}~,~\mathcal{U}_{1\>2} =\sin \phi_{-}
\label{finish2}
\end{equation}
and
\begin{equation}
\mathcal{U}_{1\>2+i}=-\cos \phi_- \frac{g_2 v_d}{\sqrt{2}M_2\mu}\epsilon_i^*+\sin \phi_-\frac{\epsilon_i^*}{\mu} \ .
\label{finish3}
\end{equation}
The result for $|\mu|<|M_{2}|$ is found by replacing $\phi_{-} \rightarrow \phi_{-} + \frac{\pi}{2}$ in these expressions. Note that the first two terms of ${\tilde{\chi}}^{\pm}_{1}$ are independent of RPV and are identical to the expressions given in \eqref{ok1} and \eqref{ok3} for $|M_{2}|<|\mu|$ and $|\mu|<|M_{2}|$  respectively. Also, {\it note that these terms dominate over the RPV terms and, hence, are the main contributors to the mass eigenvalues. However, although they are numerically smaller, the RPV terms in ${\tilde{\chi}}^{\pm}_{1}$ play a crucial role in the $R$-parity violating decays of chargino LSPs and, hence, cannot be ignored}.

\subsection{Neutralino mass eigenstates}\label{Neutralino_masses}

\indent In the {\it absence of the RPV violating terms proportional to $\epsilon_i$ and $v_{L_i}$}, the neutral Higgsinos and gauginos of the theory mix with
the third generation right handed neutrino. In the gauge eigenstate basis $\psi^0=\left( \tilde{W}_R, \tilde{W}_0, \tilde{H}_d^0,
\tilde{H}_u^0, \tilde{B}^\prime,{\nu}^c_3 \right)$, 
\begin{equation}
\mathcal{L}\supset-\frac{1}{2}\left(\psi^0\right)^T{M}_{\tilde{{ \chi}}^0}\psi^0+c.c
\end{equation}
where 
{\small
\begin{equation}
M_{{\tilde \chi}^0}=
\left(
\begin{matrix}
M_R&0&-\frac{1}{2}g_Rv_d&\frac{1}{2}g_Rv_u&0&-\frac{1}{2}g_Rv_R\\
0&M_2&\frac{1}{2}g_2 v_d&-\frac{1}{2}g_2 v_u&0&0\\
-\frac{1}{2}g_Rv_d&\frac{1}{2}g_2 v_d&0&-\mu&0&0\\
\frac{1}{2}g_Rv_u&-\frac{1}{2}g_2 v_u&-\mu&0&0&0\\
0&0&0&0&M_{BL}&\frac{1}{{2}}g_{BL}v_R\\
-\frac{1}{{2}}g_Rv_R&0&0&0&\frac{1}{{2}}g_{BL}v_R&0\\
\end{matrix}
\right) \ .
\label{eq:neutralinoMassMatrixWithoutEpsilonx}
\end{equation}
}
The mass eigenstates are related to the gauge states by the unitary matrix $N$ where ${\tilde{\chi }}^{0}=N \psi^{0}$. $N$ is chosen so that

\begin{equation}
N^*M_{\tilde{{ \chi}}^0}N^{\dag}= M_{\tilde{{ \chi}}^0}^{D} =\text{diag}
\left(
M_{{\tilde \chi}^0_1},M_{{\tilde \chi}^0_2},M_{{\tilde \chi}^0_3},M_{{\tilde \chi}^0_4},M_{{\tilde \chi}^0_5},M_{{\tilde \chi}^0_6}
\right) \ ,
\end{equation}
where all eigenvalues are positive.
The $B-L$ MSSM does not explicitly contain a Bino, associated with the hypercharge group $U(1)_Y$.  Instead, it contains a Blino and a Rino, the gauginos associated with $U(1)_{B-L}$ and $U(1)_{3R}$ respectively. Nevertheless, the theory does effectively contain a Bino. To see this, consider the limit $M_{W^\pm}^2,\> M_{Z^0}^2\ll M_{R}^{2}, \> M_{2}^{2},\> M_{BL}^{2}$ --- that is, when the 
EW scale is much lower than the 
soft breaking scale so that the Higgs VEV's are negligible. In this limit, the mass matrix in eq. \eqref{eq:neutralinoMassMatrixWithoutEpsilonx} becomes
{\small
\begin{equation}
M_{{\tilde \chi}^0}=
\left(
\begin{matrix}
M_R&0&0&0&0&-\frac{1}{ 2}g_Rv_R\\
0&M_2&0&0&0&0\\
0&0&0&-\mu&0&0\\
0&0&-\mu&0&0&0\\
0&0&0&0&M_{BL}&\frac{1}{{2}}g_{BL}v_R\\
-\frac{1}{{2}}g_Rv_R&0&0&0&\frac{1}{{2}}g_{BL}v_R&0\\
\end{matrix}
\right)
\end{equation}
}
The first, fifth, and sixth columns, corresponding to the Blino, the Rino and the third generation right-handed neutrino, are now decoupled from the others and mix only with each other. In the reduced basis $\left({\nu}_3^c, \tilde W_R, \tilde B^\prime \right)$, the mixing matrix is
\begin{equation}
\left(
\begin{matrix}
0&-\cos \theta_R M_{Z_R}&\sin \theta_R M_{Z_R}\\
-\cos \theta_R M_{Z_R}&M_R&0\\
\sin \theta_R M_{Z_R}&0&M_{BL}\\
\end{matrix}
\right)\ , \quad \cos \theta_R = \frac{g_R}{\sqrt{g_R^2+g_{BL}^2}} \ .
\end{equation}
The limit in which the gaugino masses are much smaller than $M_{Z_R}$ is phenomenologically relevant due to the lower bound on $M_{Z_R}$ being much higher than typical gaugino mass lower bounds. This limit is also motivated theoretically because RG running makes the gauginos masses lighter.
In this limit, the mass eigenstates can be found as an expansion in the gaugino masses. At zeroth order, they are
\begin{equation}
\tilde B=\tilde W_R \sin \theta_R+\tilde B^\prime \cos \theta_R \ ,
\end{equation}
\begin{equation}
{\nu}_{3a}^c=\frac{1}{\sqrt 2}({\nu^c}_3-\tilde W_R \cos \theta_R+\tilde B^\prime 
\sin \theta_R) \ ,
\end{equation}
\begin{equation}
{\nu}_{3b}^c=\frac{1}{\sqrt{2}}({\nu^c}_3+\tilde W_R \cos \theta_R-
\tilde B^\prime \sin \theta_R)
\end{equation}
with masses, calculated to first order, given by
\begin{equation}
M_1=\sin^2 \theta_R M_R+\cos^2 \theta_R M_{BL},\quad m_{{\nu^c}_{3a}}=M_{Z_R},
\quad m_{{\nu^c}_{3b}}=M_{Z_R} \ .
\end{equation}
respectively.
The state $\tilde B$ with mass $M_1$ is effectively a Bino.
We can rotate from the old basis, $\left( \tilde{W}_R, \tilde{W}_0, \tilde{H}_d^0,
\tilde{H}_u^0, \tilde{B}^\prime ,{\nu^c}_3 \right)$, to the new one,
$\left( \tilde{B}, \tilde{W}_0, \tilde{H}_d^0,
\tilde{H}_u^0, {\nu}_{3a}^c ,{\nu}_{3b}^c \right)$, using a rotation matrix, which at zeroth order has the form:

{\small
\begin{equation}
\left(
\begin{matrix}
\sin \theta_R&0&0&0&\cos \theta_R&0\\
0&1&0&0&0&0\\
0&0&1&0&0&0\\
0&0&0&1&0&0\\
-\frac{1}{\sqrt  2}\cos \theta_R&0&0&0&\frac{1}{\sqrt  2} \sin \theta_R&\frac{1}{\sqrt 2}\\
\frac{1}{\sqrt  2}\cos \theta_R&0&0&0&-\frac{1}{\sqrt 2}\sin \theta_R&\frac{1}{\sqrt 2}\\
\end{matrix}
\right).
\end{equation}
}
We then get a new neutralino mass matrix, which is in agreement with the MSSM model after B-L breaking. It is given by
{\small
\begin{equation}
M_{{\tilde \chi}^0}=
\left(
\begin{matrix}
M_1&0&-\frac{1}{\sqrt 2}g^\prime v_d&\frac{1}{\sqrt 2}g^\prime v_u&0&0\\
0&M_2&\frac{1}{\sqrt2}g_2 v_d&-\frac{1}{\sqrt 2}g_2 v_u&0&0\\
-\frac{1}{\sqrt 2}g^\prime v_d&\frac{1}{\sqrt 2}g_2 v_d&0&-\mu&0&0\\
\frac{1}{\sqrt 2}g^\prime v_u&-\frac{1}{\sqrt 2}g_2 v_u&-\mu&0&0&0\\
0&0&0&0&m_{{\nu^c}_{3a}}&0\\
0&0&0&0&0&m_{{\nu^c}_{3b}}\\
\end{matrix}
\right)
\end{equation}
}
Since the EW scale is generally much lower than the gaugino mass scale, the off-diagonal
terms are small,

\begin{equation}\label{eq:Bino_mass}
M_{{\tilde \chi}^0_1} \simeq |M_1|-\frac{M_{Z^0}^2\sin^2 \theta_W(M_1+\mu \sin 2\beta)}{\mu^2-M_1^2} \ ,
\end{equation}
\begin{equation}\label{eq:Wino_Neutralino_mass}
M_{{\tilde \chi}^0_2} \simeq |M_2|-\frac{M_{W^\pm}^2(M_2+\mu \sin 2\beta)}{\mu^2-M_2^2} \ ,
\end{equation}
\begin{equation}
M_{{\tilde \chi}^0_3} \simeq |\mu|+\frac{M_{Z^0}(\text{sgn}(\mu)\times 1-\sin 2\beta)(\mu+M_1\cos^2 \theta_W+M_2\sin^2 \theta_W)}
{2(\mu+M_1)(\mu+M_2)} \ ,
\end{equation}
\begin{equation}
M_{{\tilde \chi}^0_4} \simeq |\mu|+\frac{M_{Z^0}(\text{sgn}(\mu)\times 1+\sin 2\beta)(\mu-M_1\cos^2 \theta_W-M_2\sin^2 \theta_W)}
{2(\mu-M_1)(\mu-M_2)} \ ,
\end{equation}
\begin{equation}
M_{{\tilde \chi}^0_5} \simeq M_{Z_R} \ ,
\end{equation}
\begin{equation}\label{eq:neutrino_mass}
M_{{\tilde \chi}^0_6} \simeq M_{Z_R}
\end{equation}
where $\theta_W$ is the Weinberg angle.  Unlike for charginos discussed previously, our {\it labels do not imply any mass ordering}.
The exact eigenstates are more difficult to compute than in the chargino case. They are, generically, linear combinations of the six gauge neutralino states. However, as was discussed in detail for charginos, the dominant gauge neutralino in an eigenstate can be read off directly from the leading term in the associated mass eigenvalue. Using this, as well as explicit numerical computation, we find that 
\begin{equation}
{\tilde{\chi}}_{1}^{0} \simeq  {\tilde{B}}^{0}~~ ,~~{\tilde{\chi}}_{2}^{0} \simeq  {\tilde{W}}^{0}~~ ,~~{\tilde{\chi}}_{3}^{0} \simeq  {\tilde{H}}_{d}^{0} ~~,~~{\tilde{\chi}}_{4}^{0} \simeq  {\tilde{H}}_{u}^{0}~ ~,~~{\tilde{\chi}}_{5}^{0} \simeq  {{\nu}}_{3a}^{c}~~ ,~~{\tilde{\chi}}_{6}^{0} \simeq  {{\nu}}_{3b}^{c} \ .
\label{finally1}
\end{equation}
As with the charginos, we henceforth denote these mass eigenstates by ${\tilde \chi}^0_B$,~ ${\tilde \chi}^0_W$,~${\tilde \chi}^0_{H_d}$,~${\tilde \chi}^0_{H_u}$,~${\tilde \chi}^0_{\nu^{c}_{3a}}$,~${\tilde \chi}^0_{\nu^{c}_{3b}}$ respectively; and refer to them as a Bino neutralino, a Wino neutralino and so on, even though they are only ``predominantly'' the pure neutral state.


Let us now {\it add the RPV couplings $\epsilon_i$ and $v_{L_i}$}. This introduces mixing between the neutralinos and the neutral fermions of the standard model --- the neutrinos. As discussed at the beginning of Section \ref{sec:3}, mixing with the first- and second-family right-handed neutrino would lead to active-sterile neutrino oscillations. Unless and until there is more experimental evidence of such oscillations, we will continue to assume that they do not exist and, therefore, that the mixing with the first- and second-family right-handed neutrinos is negligible. Therefore, the neutrino mass matrix given below includes only mixing with the three families of left-handed neutrinos-- the seventh column --and the third-family right-handed neutrino-- the sixth column. As in the case of the charginos, the effect of adding these RPV couplings is important because it will allow RPV decays of the neutralinos. It's effect on the physical masses of the neutralinos, however, is negligible.
 The new basis is then extended to 
$\left( \tilde{W}_R, \tilde{W}_0, \tilde{H}_d^0,
\tilde{H}_u^0, \\ \tilde{B}^\prime ,{\nu}_3^c, \nu_1, \nu_2,\nu_3\right)$. The extended mass matrix is given by
{\footnotesize
\begin{equation}
\mathcal{M}_{{\tilde \chi}^0}=
\left(
\begin{matrix}
M_R&0&-\frac{1}{ 2}g_Rv_d&\frac{1}{2}g_Rv_u&0&-\frac{1}{ 2}g_Rv_R&0_{1\times 3}\\
0&M_2&\frac{1}{2}g_2 v_d&-\frac{1}{ 2}g_2 v_u&0&0&\frac{1}{ 2}g_2 v_{L_i}^*\\
-\frac{1}{ 2}g_Rv_d&\frac{1}{ 2}g_2 v_d&0&-\mu&0&0&0_{1\times 3}\\
\frac{1}{2}g_Rv_u&-\frac{1}{ 2}g_2 v_u&-\mu&0&0&0& \epsilon_i\\
0&0&0&0&M_{BL}&\frac{1}{{2}}g_{BL}v_R& -\frac{1}{ 2} g_{BL}v_{L_i}^*\\
-\frac{1}{{2}}g_Rv_R&0&0&0&\frac{1}{{2}}g_{BL}v_R&0&\frac{1}{\sqrt 2}Y_{\nu i3}v_u\\
0_{3\times 1}&\frac{1}{\sqrt 2} g_2 v_{L_j}^* &0_{3\times 1}& \epsilon_j&-\frac{1}{\sqrt 2}
g_{BL}v_{Lj}^*&\frac{1}{\sqrt 2} Y_{\nu j3}v_u&0_{3\times 3}\\
\end{matrix}
\right)
\end{equation}
}
This is the matrix that was introduced in eq. \eqref{eq:20}.
Just as for charginos, we can write the neutralino matrix in a schematic form that will help us diagonalize it perturbatively. As discussed in detail in Section \ref{sec:3},  $\mathcal{M}_{{\tilde \chi}^0}$ can be expressed as
\begin{equation}
\mathcal{M}_{{\tilde \chi}^0}=
\left(
\begin{matrix}
M_{{\tilde \chi}^0}&m_D\\
m_D^T&0_{3\times 3}
\end{matrix}
\right)
\end{equation}
where $M_{\chi_{0}}$ and $m^{D}$ are given in \eqref{eq:21} and \eqref{eq:22} respectively.
The mass eigenstates are related to the gauge eigenstates by the unitary matrix $\mathcal{N}$, which diagonalizes the neutralino mixing matrix
\begin{equation}
\mathcal{M}^D_{{\tilde \chi}^0}=\mathcal{N}^* \mathcal{M}_{{\tilde \chi}^0} \mathcal{N}^{\dag} \ .
\end{equation}
$\mathcal{N}$ can be written in the perturbative form
\begin{equation}
\label{eq:matrixN_PLM}
\mathcal{N}=\left(
\begin{matrix}
N&0_{3\times 3}\\
0_{3\times 3}&V^{\dag}_{PMNS}\\
\end{matrix}
\right)
\left(
\begin{matrix}
1_{6\times 6}& -\xi_0\\
\xi_0^{\dag}&1_{3\times 3}\\
\end{matrix}
\right),
\end{equation}
where $N$ is the unitary matrix introduced below eq. \eqref{eq:neutralinoMassMatrixWithoutEpsilon}. It is a $6\times 6$ matrix, analogous to the $2\times2$ matrices $U$ and $V$ in Section \ref{Chargino_masses}. However, while eq. \eqref{eq:U_matrix}, \eqref{pass1} and \eqref{talk1} provide simple analytic expressions for $U$ and $V$ in terms of the rotation angles $\phi_\pm$, it is much harder to solve for $N$ without approximations. We will compute $N$ numerically, using the the relevant soft mass terms and couplings as input. 

The equation $\xi^0=M_{\tilde \chi^0}^{-1}m_D$ is obtained by requiring that $\mathcal{M}^D_{{\tilde \chi}^0}$ be diagonal.  We denote the mass eigenstates as $\tilde \chi^0=\mathcal{N}\psi^0$. The entries of $\mathcal{N}$ are central in calculating the neutralino decay rates and are all presented in Appendix B. However, {\it exactly as in the case of the charginos discussed above, the physical masses of the ``proper'' neutralinos are not significantly changed by introducing the RPV couplings}, so eqs. \eqref{eq:Bino_mass} -- \eqref{eq:neutrino_mass} remain valid. The states $\chi^0_{6+i}$ for $i=1,2,3$ are the three left-handed neutrinos which now receive Majorana masses. This process has been discussed in detail in Section \ref{sec:3}. As in the case of charginos, {\it it is important to note that, although small compared to the $R$-parity preserving coefficients, the RPV terms make important contributions to the RPV decays of the neutralino LSPs and, hence, cannot be ignored}.

\subsection{Chargino and neutralino decay channels}
\label{sec:6}


The general $B-L$ MSSM Lagrangian, written in terms of chiral multiplet component fields, ($\phi_i, \>\psi_i$), and vector multiplet components, ($A_{\mu}^a, \>\lambda^a$), has the generic form 
{\small
\begin{multline}\label{eq:Lagrangian}
\mathcal{L}=-\partial_{\mu}\phi^{*i}\partial^{\mu}\phi_i+i\psi^{\dag i}\bar \sigma^{\mu}\partial_{\mu}\psi_i \qquad \text{ (kinetic terms)}\\
-igT^a A^a_{\mu} \phi^{*}\partial^{\mu} \phi_i+c.c+g^2(T^aA^a_{\mu})(T^bA^{b\mu})\phi_i\phi^{*i}+g\psi^{\dag}\bar \sigma^{\mu}T^aA^a_{\mu}\psi_i \qquad \text{( covariant derivative)}\\
-\sqrt{2}g(\phi^{*}T^a \psi)\lambda^a- \sqrt{2} g \lambda^{a\dag} (\psi^{\dag}T^a \phi)+g (\phi^{*} T^a \phi)D^a \qquad \text{(covariant derivative supersymmterization)}\\
+i\lambda^{\dag a}
\bar \sigma^{\mu}\partial_{\mu}\lambda^a+igf^{abc}\lambda^{\dag a}
\bar \sigma^{\mu}A_{\mu}^b\lambda^{ c}-\frac{1}{4} F^{a \mu \nu}{ F^a}_{\mu\nu}+\frac{1}{2}D^aD^a \qquad \text{(gauge field self-interactions)}\\
-\frac{1}{2}M^{ij}\psi_{i}\psi_{j}-\frac{1}{2}M^*_{ij}\psi^{\dag i}\psi^{\dag j}+M^*_{ik}M^{kj}\phi^{*i}\phi_j \qquad \text{(superpotential mass terms)} \\
-\frac{1}{2}Y^{ijk}\phi_i \psi_j \psi_k  -\frac{1}{2}Y^*_{ijk}\phi^{*i}\psi^{\dag j}\psi^{\dag k} \qquad \text{(superpotential scalar-fermion-fermion Yukawa coupling)}
\\+\frac{1}{2}M^{in}Y^*_{jkn}\phi_i\phi^{*j} \phi^{*k}+\frac{1}{2}M^{*}_{in}Y^{jkn}\phi^{*i}\phi_{j} \phi_{k}+\frac{1}{4}Y^{ijn}Y^*_{kln}\phi_i\phi_j\phi^{*k} \phi^{*l} \qquad \text{(3 and 4-scalar interactions)}.
\end{multline}
}
The interaction vertices that allow the charginos and neutralinos to decay into standard model particles are encoded in its complicated interactions. In order to read these RPV vertices from this Lagrangian, one needs to follow a series of steps:

 First, one writes this general Lagrangian in terms of the component fields of the theory. The $B-L$ MSSM matter content is give in \eqref{eq:3} and \eqref{eq:4}, whereas the gauge fields and gauginos are those associated with gauge group \eqref{eq:2}.

 In the first step, the component fields are pure gauge states. After $B-L$ and electroweak symmetry breaking, these states mix to form massive states. We have already discussed how massive chargino and neutralino states are constructed. The second step then, is to write all gauge eigenstates in the Lagrangian in terms of their mass eigenstate expansion. 

 After the second step is completed, one can identify the RPV vertices that couple a single chargino or a single neutralino to two standard model particles, typically a boson and a lepton. However, so far the theory has been written in terms of 2-component Weyl spinors, while the physical fermions are described by 4-component spinors. The final step, then, is to write the identified RPV vertices in  4-component spinor notation.

In the following sections, we will identify the RPV decay amplitudes for the charginos and neutralinos displayed in Table \ref{tab:decay_channels}, using the three steps described above. 
We find that such sparticle decays are due entirely to the RPV couplings proportional to $\epsilon_i$ and $v_{L_i}$, $i=1,2,3$ that mix the three
generations of leptons and the gauginos of the MSSM inside the chargino and neutralino mass matrices. That is, the mass eigenstate charginos and neutralinos  can decay into SM particles precisely
because they have lepton components. Only after we express the $B-L$ MSSM Lagrangian in terms of the mass eigenstates, will the decay processes in Table \ref{tab:decay_channels} become apparent. Henceforth, we use $\chi^{\pm, 0}$ when referring to chargino and neutralino 2-component Weyl fermions and $X^{\pm, 0}$ when referring to chargino and neutralino 4-component Dirac fermions. Furthermore, we use $e_i, \> i=1,2,3$ for the three families of charged leptons Weyl fermions, and $\ell_i, \>i=1,2,3$ for the three families expressed as Dirac fermions. The Dirac fermion states are defined in eq. \eqref{eq:DircaFermions}.

\begin{table}[t]
\begin{center}
\begin{tabular}{ |c|c| } 
 \hline
 Charginos & Neutralinos \\ 
 \hline
 ${\tilde X}^{\pm}\rightarrow W^{\pm}\nu_{i}$ & ${\tilde X}^0\rightarrow W^{\pm}\ell^\pm_i$\\ 
 ${\tilde X}^{\pm}\rightarrow Z^{0}\ell^\pm_i$& ${\tilde X}^{0}\rightarrow Z^0 \nu_{i}$ \\ 
  ${\tilde X}^{\pm}\rightarrow h^{0}\ell^\pm_i$& ${\tilde X}^{0}\rightarrow h^0 \nu_{i}$ \\  
 \hline
\end{tabular}
\end{center}
\caption{Chargino and Neutralino RPV decay channels, expressed in terms of 4-component spinors.}
\label{tab:decay_channels}
\end{table}

\subsubsection{Mass eigenstate expansion}

The chargino mass eigenstates ${\tilde \chi}^\pm=({\tilde \chi}_1^\pm,\> {\tilde \chi}_2^\pm,\> {\tilde \chi}_3^\pm,\> {\tilde \chi}_4^\pm,\> {\tilde \chi}_5^\pm)$ are related to the gauge eigenstates $\psi^+=(\tilde W^+, \> \tilde H_u^+,\> e_1^c, \>e_2^c,\> e_3^c)$ and  $\psi^-=(\tilde W^- \> \tilde H_d^-,\> e_1, \> e_2, \> e_3)$ via the unitary matrices $\mathcal{U}$ and $\mathcal{V}$ defined in Section \ref{sec:5} and given in Appendix \ref{appendix:A1} . That is,
\begin{equation}
{\tilde \chi}^-=\mathcal{U}\psi^-~, \quad \quad {\tilde \chi}^+=\mathcal{V}\psi^+ \ .
\end{equation}
There are two things worth pointing out here. First, only the mass eigenstates ${\tilde \chi}_1^\pm$ and ${\tilde \chi}_2^\pm$ are considered to be the actual charginos. They have dominant contributions from the MSSM gauginos and only small SM lepton components. Moreover, the mass eigenstates ${\tilde \chi}^\pm_3\simeq e_1, \> e_1^c$, ${\tilde \chi}^\pm_4\simeq e_2, \> e_2^c$, ${\tilde \chi}^\pm_5\simeq e_3, \> e_3^c$ are considered to be the three generations of charged leptons (to be more precise, the left-handed Weyl spinors of the negatively and positively charged leptons). Second, the $\mathcal{U}$ and $\mathcal{V}$ matrices are defined so that the chargino states ${\tilde \chi}_1^\pm$
are lighter than the chargino states ${\tilde \chi}_2^\pm$; that is, $M_{{\tilde \chi}_1^\pm} < M_{{\tilde \chi}_2^\pm}$. The state ${\tilde \chi}_1^\pm$ can be dominantly charged Wino or charged Higgsino, but it will be always be less massive than ${\tilde \chi}_2^\pm$. 

 In terms of the chargino mass eigenstates, the gauge eigenstate can be expressed as $\psi^-=\mathcal{U}^{\dag}{\tilde \chi}^-$ and  $\psi^+=\mathcal{V}^{\dag}{\tilde \chi}^+$. We then have the following mass eigenstate decomposition:
\begin{equation}\label{eq:mass_expanssion1}
e_i=\mathcal{U}^*_{1\>2+i}{\tilde \chi}^-_1+\mathcal{U}^*_{2\>2+i}{\tilde \chi}^-_2+\mathcal{U}^*_{3\>2+i}{\tilde \chi}^-_3+
\mathcal{U}^*_{4\>2+i}{\tilde \chi}^-_4+\mathcal{U}^*_{5\>2+i}{\tilde \chi}^-_5
\end{equation}
\begin{equation}\label{eq:mass_expanssion2}
 e_i^c=\mathcal{V}^*_{1\>2+i}{\tilde \chi}^+_1+\mathcal{V}^*_{2\>2+i}{\tilde \chi}^+_2+\mathcal{V}^*_{3\>2+i}{\tilde \chi}^+_3+
\mathcal{V}^*_{4\>2+i}{\tilde \chi}^+_4+\mathcal{V}^*_{5\>2+i}{\tilde \chi}^+_5
\end{equation}
Similarly, the Wino and Higgsino gauge eigenstate can be expressed as:
\begin{equation}\label{eq:mass_expanssion3}
\tilde W^-= \mathcal{U}^*_{1\>1}{\tilde \chi}^-_1+\mathcal{U}^*_{2\>1}{\tilde \chi}^-_2+\mathcal{U}^*_{3\>1}{\tilde \chi}^-_3+
\mathcal{U}^*_{4\>1}{\tilde \chi}^-_4+\mathcal{U}^*_{5\>1}{\tilde \chi}^-_5
\end{equation}
\begin{equation}\label{eq:mass_expanssion4}
\tilde W^+= \mathcal{V}^*_{1\>1}{\tilde \chi}^+_1+ \mathcal{V}^*_{2\>1}{\tilde \chi}^+_2+\mathcal{V}^*_{3\>1}{\tilde \chi}^+_3+
\mathcal{V}^*_{4\>1}{\tilde \chi}^+_4+\mathcal{V}^*_{5\>1}{\tilde \chi}^+_5
\end{equation}
\begin{equation}\label{eq:mass_expanssion5}
\tilde H^-_d=\mathcal{U}^*_{1\>2}{\tilde \chi}^-_1+ \mathcal{U}^*_{2\>2}{\tilde \chi}^-_2+\mathcal{U}^*_{3\>2}{\tilde \chi}^-_3+
\mathcal{U}^*_{4\>2}{\tilde \chi}^-_4+\mathcal{U}^*_{5\>2}{\tilde \chi}^-_5
\end{equation}
\begin{equation}\label{eq:mass_expanssion6}
\tilde H^+_u= \mathcal{V}^*_{1\>2}{\tilde \chi}^+_1+\mathcal{V}^*_{2\>2}{\tilde \chi}^+_2+\mathcal{V}^*_{3\>2}{\tilde \chi}^+_3+
\mathcal{V}^*_{4\>2}{\tilde \chi}^+_4+\mathcal{V}^*_{5\>2}{\tilde \chi}^+_5
\end{equation}

The neutralino mass eigenstates ${\tilde \chi}^0=({\tilde \chi}^0_1,\>{\tilde \chi}^0_2,\>{\tilde \chi}^0_3,\>{\tilde \chi}^0_4,\>{\tilde \chi}^0_5,\>{\tilde \chi}^0_6,\>{\tilde \chi}^0_7,\>{\tilde \chi}^0_8,\>{\tilde \chi}^0_9)$ are related to the gauge eigenstates 
$\psi^0=(\tilde W_R, \>\tilde W_0, \>\tilde H^0_d, \> \tilde H_u^0, \> \tilde B^', \> \nu^c_3, \>\nu_e,\>\nu_\mu,\>\nu_\tau )$ via the unitary matrix $\mathcal{N}$,  defined in Section \ref{sec:5} and given in  Appendix \ref{appendix:A2}. That is,
\begin{equation}
{\tilde \chi}^0=\mathcal{N}\psi^0 \ .
\end{equation}
Just as for the chargino states, it is important to remember that only the first six states ${\tilde \chi}_{1,2,3,4,5,6}$ are considered actual MSSM neutralinos, since their dominant contributions are from sparticles. The states ${\tilde \chi}_{7,8,9}$ are the three generations of left handed neutrinos, which obtain Majorana masses after the neutralino matrix diagonalization. However, the notation of the six neutralino mass eigenstates differs from that of the chargino mass eigenstates. In the case of charginos, the states are defined such that $M_{{\tilde \chi}^\pm_1}<M_{{\tilde \chi}^\pm_2}$, where both ${\tilde \chi}^\pm_1$ and ${\tilde \chi}^\pm_2$ could be dominantly charged Wino or charged Higgsino. As discussed above, usually $|M_2|<|\mu|$,  in which case ${\tilde \chi}^\pm_1$ would be dominantly charged Wino, while ${\tilde \chi}^\pm_2$ would be dominantly charged Higgsino.  In the rare case when $|\mu|<|M_2|$, $\tilde \chi_1^\pm$ would be dominantly charged Higgsino, while $\tilde \chi^\pm_2$ would be dominantly charged Wino. For the neutralino mass eigenstates, however, we always have ${\tilde \chi}_1^0$ mostly Bino,  ${\tilde \chi}_2^0$ mostly Wino,  ${\tilde \chi}_{3,4}^0$ mostly Higgsino, ${\tilde \chi}_{5,6}^0$ mostly right-handed third generation neutrino. We don't know, a priori, which state is the lightest, nor how to order them in terms of mass. Their masses are computed after we diagonalize the $6\times 6$ neutralino mixing matrix $M_{{\tilde \chi}^0}$-- neglecting RPV couplings --an operation significantly more complicated than the diagonalization of the $2\times2$ chargino mass matrix, $M_{{\tilde \chi}^\pm}$, in the absence of RPV couplings. 

In terms of the neutralino mass eigesntates, the gauge eigenstates are given by
\begin{equation}
\psi^0=\mathcal{N}^\dag {\tilde \chi}^0 \ .
\end{equation}
For the three neutrino gauge eigenstates $\nu_{e}, \nu_{\mu}, \nu_{\tau}=\nu_{{i}},~i=1,2,3$
\begin{multline}\label{eq:mass_expanssion7}
\nu_{i}=\mathcal{N}^*_{1\>6+i}
{\tilde \chi}^0_1+\mathcal{N}^*_{2\>6+i}{\tilde \chi}^0_2+
\mathcal{N}^*_{3\>6+i}{\tilde \chi}^0_3+\mathcal{N}^*_{4\>6+i}{\tilde \chi}^0_4\\+
\mathcal{N}^*_{5\>6+i}{\tilde \chi}^0_5+
\mathcal{N}^*_{6\>6+i}{\tilde \chi}^0_6+\mathcal{N}^*_{7\>6+i}{\tilde \chi}^0_7+
\mathcal{N}^*_{8\>6+i}{\tilde \chi}^0_8+\mathcal{N}^*_{9\>6+i}{\tilde \chi}^0_9 \ ,
\end{multline} 
while for the rest of the mass eigenstates, we have 
\begin{multline}\label{eq:mass_expanssion8}
\tilde W_R,\>\tilde W^0,\>\tilde H_d^0,\>\tilde H_u^0, \tilde B^', \> \tilde \nu_3^c=\mathcal{N}^*_{1\>1,2,3,4,5,6}
{\tilde \chi}^0_1+\mathcal{N}^*_{2\>1,2,3,4,5,6}{\tilde \chi}^0_2+
\mathcal{N}^*_{3\>1,2,3,4,5,6}{\tilde \chi}^0_3+\ \\+\mathcal{N}^*_{4\>1,2,3,4,5,6}{\tilde \chi}^0_4+
\mathcal{N}^*_{5\>1,2,3,4,5,6}{\tilde \chi}^0_5+
\mathcal{N}^*_{6\>1,2,3,4,5,6}{\tilde \chi}^0_6+\mathcal{N}^*_{7\>1,2,3,4,5,6}{\tilde \chi}^0_7+\\+
\mathcal{N}^*_{8\>1,2,3,4,5,6}{\tilde \chi}^0_8+\mathcal{N}^*_{9\>1,2,3,4,5,6}{\tilde \chi}^0_9 \ .
\end{multline} 

The Higgs scalar fields in the MSSM consist of two complex $SU(2)_L$ doublets; that is,
eight degrees of freedom. When electroweak symmetry is broken, 
three of them become the Goldstone bosons $G^0,\>G^{\pm}$, where  
$G^-=G^{+*}$. The rest will be Higgs scalar mass eigenstates; that is, CP-even neutral scalars $h^0$ 
and $H^0$, a CP-odd neutral scalar $\Gamma^0$ and a charged $H^+$ and a conjugate 
$H^-={H^+}^*$. They are defined by \cite{Martin:1997ns}.
\begin{equation}
\left(\begin{matrix}H_u^0\\H_d^0\end{matrix}\right)=
\left(\begin{matrix}v_u\\v_d\end{matrix}\right)+
\frac{1}{\sqrt{2}}R_{\alpha}\left(\begin{matrix}h^0\\H^0\end{matrix}\right)+
\frac{i}{\sqrt{2}}R_{\beta_0}\left(\begin{matrix}G^0\\\Gamma^0\end{matrix}\right) \ ,
\end{equation}
\begin{equation}
\left(
\begin{matrix}
H_u^+\\H_d^{-*}
\end{matrix}
\right)=
R_{\beta_\pm} \left(\begin{matrix}G^+\\H^+\end{matrix}\right) \ 
\end{equation}
where $R_{\alpha},\>R_{\beta_0}, \>R_{\beta_\pm}$ are rotation matrices.
Specifically, the matrix in front of the Standard Model Higgs Boson $h^0$ is
\begin{equation}
R_{\alpha}=\left( \begin{matrix}
\cos{\alpha}&\sin{\alpha}\\ -\sin{\alpha}&\cos{\alpha}
\end{matrix}\right),
\end{equation}
while, to lowest order, the other matrices are

\begin{equation}
R_{\beta_0}=
 R_{\beta_\pm}=\left( \begin{matrix}
 \sin \beta   & \cos \beta\\
 -\cos \beta & \sin \beta 
 \end{matrix}
 \right),
\end{equation}
where  $\tan \beta=v_u/v_d$. The mixing angle $\alpha$ is, at tree level: 
\begin{equation}\label{eq:alpha}
\frac{\tan 2 \alpha}{\tan 2 \beta}=\frac{M^2_{\Gamma^0}+M_{Z^0}^2}{M^2_{\Gamma^0}-M_{Z^0}^2}, \quad \quad
\frac{\sin 2 \alpha}{\sin 2 \beta}=-\frac{M^2_{H^0}+M^2_{h^0}}{M^2_{H^0}-M^2_{h^0}}
\end{equation}
where the masses of the Higgs eigenstates are
\begin{equation}
M^2_{\Gamma^0}=2b/\sin 2\beta=2|\mu|^2+m^2_{H_u}+m^2_{H_d} \ ,
\end{equation}
\begin{equation}
M^2_{h^0,\>H^0}=\frac{1}{2}\left(  M^2_{\Gamma^0}+M_{Z^0}^2\mp \sqrt{(M^2_{\Gamma^0}-M^2_{Z^0})^2+4M_{Z^0}^2M_{\Gamma^0}^2 
\sin^2(2\beta)} \right)
\end{equation}
and
\begin{equation}
M^2_{H^\pm}=M^2_{\Gamma^0}+M_{W^\pm}^2 \ .
\end{equation}
\subsubsection{Interaction vertices}

We now express Lagrangian \eqref{eq:Lagrangian}
in terms of all the matter and gauge fields in our $B-L$ MSSM theory, and then replace all gauge 
eigenstates with their mass eigenstate expansion. 
However, the full $B-L$ MSSM Lagrangian is complicated when expressed in its most general
form.  We proceed, therefore, by looking only for the terms that
can lead to chargino or neutralino decays into standard model particles. We identify the following tri-couplings:

\begin{itemize}
\item $g \psi^{i\dag} \bar \sigma^\mu T^a A_\mu^a \psi_i$, ~\quad\quad \qquad ~~from the covariant derivative of the fermionic matter fields

\item $-\sqrt{2}g(\phi^{i*}T^a \psi_i)\lambda^a$ and $-\sqrt{2}g\lambda^\dag(\psi^{i\dag}T^a\phi_i)$,~\qquad~~
from the supercovariant derivatives

\item $ig f^{abc}\lambda^{a\dag}\bar\sigma^\mu A_{\mu}^b\lambda^{c} $, ~\qquad \qquad \qquad \quad\quad \qquad \qquad \qquad  from the gauge self-interaction

\item $-\frac{1}{2}Y^{ijk}\phi_i\psi_j\psi_k$ and $-\frac{1}{2}Y^*_{ijk}\phi^{i*}\psi^{j\dag}\psi^{k\dag}$, \qquad
 from the superpotential Yukawa couplings

\end{itemize}
We now want to write these interactions in terms of the $B-L$ MSSM component fields.  The procedure is non-trivial. Hence, we split these interactions terms into two categories : 1) those responsible for the neutralino or chargino decays into a gauge boson ($Z^0$-boson or $W^\pm$-boson) and a lepton; that is 
\begin{equation}
\tilde \chi^{\pm, 0}\rightarrow Z^0, W^\pm-\text{lepton}
\end{equation}
and 2) those responsible for the decays into a Higgs boson and a lepton; that is
\begin{equation}
 {\tilde \chi}^{\pm,0}\rightarrow h^0 - \text{lepton}.
\end{equation}
 The terms with Yukawa couplings and those from the supercovariant derivatives are relevant only for 
the Higgs boson-lepton decay channel,
as we will show. 

\subsubsection{\boldmath${\tilde \chi}^{\pm,0}\rightarrow$$Z^0, W^\pm$-lepton } 

The part  of the Lagrangian responsible for the gauge boson-lepton decay channels is
\begin{equation}
\mathcal{L}_{{\tilde \chi}^{\pm,\>0}\rightarrow Z^0, W^\pm-{\rm lepton}} \supset g \psi^{i\dag} \bar \sigma^\mu T^a A_\mu^a \psi_i+ig f^{abc}\lambda^{a\dag}\bar\sigma^\mu A_{\mu}^b\lambda^{c},
\end{equation}
where this expression represents the sum over $SU(2)_L$ and $U(1)_Y$. The $i=1,2,3$ represents the three lepton families. Expressed in terms of the MSSM component fields this becomes
\begin{multline}
\mathcal{L}_{{\tilde \chi}^{\pm,\>0}\rightarrow Z^0, W^\pm-\text{lepton}} \supset g_2\left(L_i^{\dag}\bar{\sigma}^{\mu}\tau^a L_i +\tilde H_u^{\dag} \bar \sigma^{\mu
}\tau^a \tilde H_u + \tilde H_d^{\dag} \bar \sigma^{\mu
}\tau^a \tilde H_d\right)W^a_{\mu}\\
+g'\left(-\frac{1}{2}e_i^{\dag}\bar{\sigma}^{\mu}e_i+{e^c}_i^{\dag}\bar{\sigma}^
{\mu}{e^c}_i+\frac{1}{2}\tilde H_u^{\dag} \bar \sigma^{\mu
} \tilde H_u - \frac{1}{2}\tilde H_d^{\dag} \bar \sigma^{\mu
} \tilde H_d\right)B_{\mu}~~\\
+ig_2f^{abc}\tilde W^{a\dag}\bar \sigma^{\mu}\tilde{W}^{b}W_\mu^c, \quad \quad\quad ~~~
\end{multline}
where we sum over the $i$ index, $W^a_{\mu},\>a=1,2,3$ are the three vector bosons of the $SU(2)_L$ group and $B_\mu$ the vector boson of the hypercharge $U(1)_Y$ group. Here, $g_2$ and $g^{\prime}$ are the $SU(2)_L$ and $U_Y(1)$ couplings. In addition, $L_{i}$ represents the $i$-th $SU(2)_{L}$ left chiral lepton doublet defined in \eqref{eq:3}. We now make the replacements
\begin{equation}
\left(
\begin{matrix}
\gamma^0\\
Z^0
\end{matrix}
\right)=
\left(
\begin{matrix}
\cos \theta_W &  \sin \theta_W \\
-\sin \theta_W & \cos \theta_W
\end{matrix}
\right)
\left(
\begin{matrix}
B^0\\
W^0
\end{matrix}
\right) \ ,
\quad
W^\pm=\frac{1}{\sqrt 2}(W^1 \mp i W^2), 
\quad \tan \theta_W=\frac{g^{\prime}}{g_2},
\end{equation}
where $\theta_W$ is the Weinberg angle, and rearrange the previous expression to obtain
\small
\begin{multline}\label{eq:L_Z,A,W}
\mathcal{L}_{{\tilde \chi}^{\pm,\>0}
\rightarrow {Z^0, W^\pm}-{\rm lepton}}
\supset \frac{g_2}{\sqrt{2}}(J^{\mu}W^+_{\mu}+J^{\mu \dag}W_{\mu}^{-}
+J^{\mu}_HW_{\mu}^+ + J^{\mu \dag}_H W_{\mu}^{-})
+e(j^{\mu}_{{EM}}+j_{{EMH}}^{\mu})\gamma^0_{\mu}\\
+\frac{g_2}{2 \cos \theta_W}\Big(J^{\mu}_{n}
+J^{\mu}_{nH}\Big)Z^0_{\mu}
+g_2(\tilde W^{+\dag}\bar\sigma^{\mu}\tilde W^{+}-\tilde W^{-\dag}\bar \sigma^{\mu}\tilde W^{-})(\cos \theta_W Z^0_{\mu}+\sin \theta_W \gamma^0_{\mu})\\
+g_2(-\tilde W^{0\dag}\bar \sigma^{\mu}\tilde W^{+}+\tilde W^{-\dag}\bar\sigma^{\mu}\tilde W^{0})W^+_{\mu}
+g_2(\tilde W^{0\dag}\bar \sigma^{\mu}\tilde W^{-}-\tilde W^{+\dag}\bar\ \sigma^{\mu}\tilde W^{0})W^-_{\mu}.
\end{multline}
\normalsize
$J^{\mu}$, $J^\mu_{n}$ and $j^{\mu}_{{EM}}$ are the usual weak charged, neutral and electromagnetic currents from the standard model theory of EW breaking, while 
$J^{\mu}_H$, $J^{\mu}_{nH}$ and $j^{\mu}_{EMH}$ are the equivalent currents of the 
Higgsino fermionic fields. Also note that $e$ is the electromagnetic coupling $
e=\frac{g_2 g^{\prime}}{\sqrt{g_2^2+g^{\prime 2}}}. $
In 2-component Weyl notation these currents are:
\begin{itemize}

\item Weak charged currents, coupling to $W^\pm$ bosons

\begin{equation}
J^{\mu}=\nu_{i}^{\dag}\bar{\sigma}^{\mu}e_i \ , \quad \quad  
J^{\mu}_H=\tilde H_u^{+^{\dag}}\bar \sigma^{\mu} \tilde H_u^0
+\tilde H_d^{0^{\dag}}\bar \sigma^{\mu}\tilde H_d^-
\end{equation}

\item Electromagnetic currents, coupling to the photon $\gamma^0$

\begin{equation}
j^{\mu}_{{EM}}=+{e_i^c}^{\dag}\bar \sigma^{\mu}{e_i^c}-e_i^{\dag}\bar \sigma^{\mu}e_i  \ ,
\quad \quad
j^{\mu}_{EMH}=\tilde H_u^{+^{\dag}}\bar \sigma^{\mu}\tilde H_u^+
-\tilde H_d^{-^{\dag}}\bar \sigma^{\mu}\tilde H_d^-
\end{equation}

\item Neutral currents, coupling to the $Z^0$ boson

\begin{equation}
J_{n}^{\mu}=\nu_{i}^{\dag}\bar \sigma^{\mu}\nu_{i}-(1-2\sin^2\theta_W)e_i^{\dag}\bar{\sigma}^{\mu}e_i-2\sin^2\theta_W{e^c_i}^{\dag}\bar{\sigma}^{\mu}{e^c_i} \ ,
\end{equation}
\begin{multline}
J_{nH}^{\mu}=(1-2\sin^2\theta_W)\tilde H_u^{+^{\dag}}\bar \sigma^{\mu}\tilde H_u^+
-(1+2\sin^2\theta_W)\tilde H_u^{0^{\dag}}\bar \sigma^{\mu}\tilde H_u^0\\
+(1+2\sin^2\theta_W)\tilde H_d^{0^{\dag}}\bar \sigma^{\mu}\tilde H_d^0
-(1-2\sin^2\theta_W)\tilde H_d^{-^{\dag}}\bar \sigma^{\mu}\tilde H_d^-
\end{multline}

\end{itemize}
where in $J^{\mu}$, $j^{\mu}_{{EM}}$ and $J_{n}^{\mu}$ we sum over $i=1,2,3$.
Plugging these currents into eq. \eqref{eq:L_Z,A,W}, and arranging the couplings in terms of $W^\pm$, $Z^0$ and $\gamma^0$ respectively, we get
\footnotesize
\begin{multline}\label{eq:633}
\mathcal{L}_{{\tilde \chi}^{\pm,\>0}\rightarrow Z^0, W^\pm-\text{lepton}}\supset 
\frac{g_2}{\sqrt 2}W_\mu^- \Big[e_i^{\dag}\bar \sigma^\mu \nu_{i}+\tilde H_u^{0\dag}\bar
\sigma^\mu \tilde H^+_u+ \tilde H_d^{-\dag}\bar
\sigma^\mu \tilde H^0_d +\sqrt{2}(\tilde W^{0\dag}\bar \sigma^\mu \tilde W^{+}\\-
\tilde W^{-\dag}\bar\sigma^\mu \tilde W^{0}) \Big]
+\frac{g_2}{\sqrt 2}W_\mu^+\left[ \nu_{i}^{\dag} \bar \sigma^\mu e_i+\tilde H_u^{+\dag}\bar \sigma^\mu \tilde H^0_u+ \tilde H_d^{0\dag} \bar \sigma^\mu \tilde H^-_d
 +\sqrt 2(\tilde W^{+\dag}\bar \sigma^\mu \tilde W^{0}-
\tilde W^{0\dag}\bar \sigma^\mu \tilde W^{-}) \right]\\
+g_2c_W Z^0_{\mu}\Big[\tilde W^{+\dag}\bar \sigma^{\mu}\tilde W^{+}-\tilde W^{-\dag}\bar \sigma^{\mu}\tilde W^{-}  \Big]
+\frac{g_2}{2c_W}Z^0_\mu \Big[\nu_{i}^{\dag}\bar \sigma^{\mu}\nu_{i}-(1-2s_W^2)e_i^{\dag}\bar{\sigma}^{\mu}e_i-2 s_W^2{e_i^c}^{\dag}\bar{\sigma}^{\mu}{e_i^c}\\
+(1-2s_W^2)\tilde H_u^{+^{\dag}}\bar \sigma^{\mu}\tilde H_u^+
-(1+2s_W^2)\tilde H_u^{0^{\dag}}\bar \sigma^{\mu}\tilde H_u^0
+(1+2s_W^2)\tilde H_d^{0^{\dag}}\bar \sigma^{\mu}\tilde H_d^0
-(1-2s_W^2)\tilde H_d^{-^{\dag}}\bar \sigma^{\mu}\tilde H_d^-
\Big]\\
+g_2s_W \gamma^0_{\mu}\Big[\tilde W^{+\dag}\bar \sigma^{\mu}\tilde W^{+}-\tilde W^{-\dag} \bar \sigma^{\mu}\tilde W^{-}  \Big]
+e\gamma^0_\mu\Big[  e_i^{c\dag}\bar \sigma^\mu {e_i^c}-e_i^{\dag} \bar\sigma^\mu e_i
+\tilde H_u^{+^{\dag}}\bar \sigma^{\mu}\tilde H_u^+
-\tilde H_d^{-^{\dag}}\bar \sigma^{\mu}\tilde H_d^- \Big],
\end{multline}
\normalsize
where we have used the notation $s_W=\sin \theta_W$, $c_W=\cos \theta_W$ and summed over $i=1,2,3$.

Finally, we are in a position to expand all gauge eigenstates in terms of the mass eigenstates, as in equations \eqref{eq:mass_expanssion1}-\eqref{eq:mass_expanssion6} and \eqref{eq:mass_expanssion7}-\eqref{eq:mass_expanssion8}. After this procedure, the Lagrangian (\ref{eq:633}) is expressed in terms of the mass eigenstates $\tilde \chi_1^\pm$, $\tilde \chi_n^0$, $e_i$, $\nu_i$ for $i=1,2,3$, and their hermitian conjugates. Notice that we keep only $\tilde \chi^\pm_1$ to simplify our results, since $\tilde \chi^\pm_1$ are always lighter than $\tilde \chi^\pm_2$ and, hence, have better prospects to be detected. Their charged Wino and charged Higgsino content are determined from the rotation matrices $U$ and $V$ in eq. 
\eqref{eq:U_matrix}. At the same time, we study the vertices of general neutralino $\tilde \chi^0_n$ states for $n=1,2,3,4,5,6$, since we have no a priori mass ordering for these states. We remind the reader that $n=1$ means a mostly Bino neutralino ${\tilde \chi}^0_1={\tilde \chi}^0_B$,   $n=2$ means a mostly Wino neutralino ${\tilde \chi}^0_2={\tilde \chi}^0_W$,  $n=3,4$ means a mostly Higgsino neutralino ${\tilde \chi}^0_{3,4}={\tilde \chi}^0_H$ and $n=5,6$ means a mostly third generation right-handed neutrino neutralino ${\tilde \chi}^0_{5,6}={\tilde \chi}^0_{\nu_3^c}$. 

Until now, we have used 2-component Weyl spinor notation for all of our matter fields. Since we are interested in the decays of physical particles, we will henceforth introduce and use 4-component spinor notation for the initial and final states of the interacting particles. The 4-component spinors are defined in terms of the 2-component Weyl spinors as
\begin{equation}
\begin{split}\label{eq:DircaFermions}
&\ell_i^-=\left(\begin{matrix}e_i\\ {e_i^c}^\dag\end{matrix}\right),~
\ell_i^+=\left(\begin{matrix}{e_i^c}\\ e_i^\dag\end{matrix}\right),~
\nu_i=\left(\begin{matrix}{\nu_i}\\ \nu_i^\dag\end{matrix}\right),~
{\tilde X}_1^-=\left(\begin{matrix}{\tilde \chi}^-_1\\{\tilde \chi}^{+\dag}_1\end{matrix}\right),~
{\tilde X}_1^+=\left(\begin{matrix}{\tilde \chi}^+_1\\{\tilde \chi}^{-\dag}_1\end{matrix}\right),~
{\tilde X}_n^0=\left(\begin{matrix}{\tilde \chi}_n^0\\{\tilde \chi}^{0\dag}_n\end{matrix}\right).
\end{split}
\end{equation}
In our model, $\ell^\pm_i,\>X^\pm_1$ are Dirac fermions, while $\nu_i,\>X^0_n$ are Majorana fermions. Note that, for simplicity, we use the same symbol, $\nu_{i}$, for both a Weyl and Majorana neutrino. The Lagrangian (\ref{eq:633}) then becomes 
{\small
\begin{align}\label{eq:gauge_amplitudes}
&\mathcal{L}_{{\tilde \chi}^{\pm,\>0}\rightarrow Z^0, W^\pm-\text{lepton}}\supset \nonumber\\
&g_2Z^0_{\mu}\bar {{\tilde X}}^-_1\gamma^\mu\Bigg[\Bigg(
-\frac{1}{c_W}
\left(\frac{1}{2}-s_W^2\right)\mathcal{U}_{2+j\>1}\mathcal{U}_{2+j\>2+i}^*
-\frac{1}{c_W}\left(\frac{1}{2}-s_W^2 \right)\mathcal{U}_{1\>2}\mathcal{U}^*_{2+i\>2}-c_W\mathcal{U}^*_{2+i\>1}\mathcal{U}_{1\>1}
\Bigg)P_L  \nonumber \\
&+
\Bigg(\frac{1}{c_W} s_W^2\mathcal{V}_{2+i\>2+j}\mathcal{V}_{1\>2+j}^*-\frac{1}{c_W}\left(\frac{1}{2}-s_W^2\right)\mathcal{V}_{2+i\>2}\mathcal{V}_{1\>2}^*
-c_W\mathcal{V}^*_{1\>1}\mathcal{V}_{2+i\>1}
\Bigg)P_R
 \Bigg]\ell_i^- \nonumber \\
-&g_2Z^0_{\mu}\bar {{\tilde X}}^+_1\gamma^\mu\Bigg[
\left(-\frac{1}{c_W}\left(\frac{1}{2}-s_W^2\right)\mathcal{U}^*_{1\>2+j}\mathcal{U}_{2+i\>2+j}
-\frac{1}{c_W}\left(\frac{1}{2}-s_W^2 \right)\mathcal{U}^*_{1\>2}\mathcal{U}_{2+i\>2}-c_W\mathcal{U}_{2+i\>1}\mathcal{U}^*_{1\>1}
\right)P_R  \nonumber\\
&+
\left(\frac{1}{c_W} s_W^2\mathcal{V}^*_{2+i\>2+j}\mathcal{V}_{1\>2+j}-\frac{1}{c_W}\left(\frac{1}{2}-s_W^2\right)\mathcal{V}^*_{2+i\>2}\mathcal{V}_{1\>2}
-c_W\mathcal{V}_{1\>1}\mathcal{V}^*_{2+i\>1}
\right)P_L
 \Bigg]\ell_i^+ \nonumber \\
&+ \frac{g_2}{\sqrt 2}W_\mu^- \bar {{\tilde X}}^-_1 \gamma^\mu \Big[(\mathcal{U}_{1\>2}\mathcal{N}^*_{6+i\>3}+\mathcal{U}_{1\>2+j}\mathcal{N}^*_{6+j\>6+i}-\sqrt{2}\mathcal{N}^*_{6+i\>2}\mathcal{U}_{1\>1})P_L\nonumber \\
&\hspace{6cm}-(\mathcal{N}_{6+i\>4}\mathcal{V}^*_{1\>2}+\sqrt{2}\mathcal{V}^*_{1\>1}\mathcal{N}_{6+i\>2})P_R \Big]\nu_i\\
&-\frac{g_2}{\sqrt 2}W_\mu^+\bar {{\tilde X}}^+_1 \gamma^\mu\Big[
(\mathcal{U}^*_{1\>2}\mathcal{N}_{6+i\>3}+\mathcal{U}^*_{1\>2+j}\mathcal{N}_{6+j\>6+i}-\sqrt{2}\mathcal{N}_{6+i\>2}\mathcal{U}^*_{1\>1})P_R \nonumber \\
&\hspace{6cm}-(\mathcal{N}^*_{6+i\>4}\mathcal{V}_{1\>2}+\sqrt{2}\mathcal{V}_{1\>1}\mathcal{N}^*_{6+i\>2})P_L
 \Big] { \nu}_{i} \nonumber \\
& +\frac{g_2}{\sqrt 2}W_\mu^-\bar {{\tilde X}}^0_n \gamma^\mu
\Big[
(\mathcal{N}_{n\>4}\mathcal{V}^*_{2+i\>2}+\sqrt{2}\mathcal{V}^*_{2+i\>1}\mathcal{N}_{n\>2})P_L \nonumber \\
&\hspace{4cm}+(-\mathcal{U}_{2+i\>2+j}\mathcal{N}^*_{n\>6+j}-\mathcal{U}_{2+i\>2}\mathcal{N}^*_{n\>3}+\sqrt{2}\mathcal{N}^*_{n\>2}\mathcal{U}_{2+i\>1})P_R
\Big]\ell_i^+ \nonumber \\
&-\frac{g_2}{\sqrt 2}W_\mu^+\bar {{\tilde X}}^0_n \gamma^\mu
\Big[
(\mathcal{N}^*_{n\>4}\mathcal{V}_{2+i\>2}+\sqrt{2}\mathcal{V}_{2+i\>1}\mathcal{N}^*_{n\>2})P_R \nonumber \\
&\hspace{4cm}+(-\mathcal{U}^*_{2+i\>2+j}\mathcal{N}_{n\>6+j}-\mathcal{U}^*_{2+i\>2}\mathcal{N}_{n\>3}+\sqrt{2}\mathcal{N}_{n\>2}\mathcal{U}^*_{2+i\>1})P_L
\Big]\ell_i^- \nonumber \\
&+{g_2}Z^0_\mu \bar {{\tilde X}}^0_n \gamma^{\mu}\Big[
\Big(\frac{1}{2c_W}\mathcal{N}_{n\>6+j}\mathcal{N}^*_{6+j\>6+i}-\frac{1}{c_W}\left(\frac{1}{2}+s_W^2\right)\mathcal{N}_{n\>4}\mathcal{N}^*_{6+i\>4} \Big)P_L \nonumber \\
&\hspace{4cm}
- \frac{1}{c_W}\left(\frac{1}{2}+s_W^2\right) \mathcal{N}_{n\>3}\mathcal{N}^*_{6+i\>3}P_R
\Big]\nu_i \nonumber \\
&-{g_2}Z^0_\mu \bar {{\tilde X}}^0_n \gamma^{\mu}\Big[
\Big(\frac{1}{2c_W}\mathcal{N}^*_{n\>6+j}\mathcal{N}_{6+j\>6+i}-\frac{1}{c_W}\left(\frac{1}{2}+s_W^2\right)\mathcal{N}^*_{n\>4}\mathcal{N}_{6+i\>4} \Big)P_R \nonumber \\
&\hspace{4cm}
- \frac{1}{c_W}\left(\frac{1}{2}+s_W^2\right) \mathcal{N}^*_{n\>3}\mathcal{N}_{6+i\>3}P_L
\Big]\nu_i \nonumber \\
&-e\gamma^0_{\mu} \bar {{\tilde X}}^-_1\gamma^\mu\Bigg[\big(
\mathcal{U}_{2+j\>1}\mathcal{U}_{2+i\>2+j}^*
+\mathcal{U}_{1\>2}\mathcal{U}^*_{2+i\>2}+\mathcal{U}^*_{2+i\>1}\mathcal{U}_{1\>1}
\big)P_L \nonumber \\
&\hspace{4cm}
\big(\mathcal{V}_{2+i\>2+j}\mathcal{V}_{1\>2+j}^*+\mathcal{V}_{2+i\>2}\mathcal{V}_{1\>2}^*+\mathcal{V}^*_{1\>1}\mathcal{V}_{2+i\>1}
\big)P_R
\Bigg]\ell_i^- \nonumber \\
&+e\gamma^0_{\mu}\bar {{\tilde X}}^+_1\gamma^\mu\Bigg[\big(
\mathcal{U}^*_{2+j\>1}\mathcal{U}_{2+i\>2+j}
+\mathcal{U}^*_{1\>2}\mathcal{U}_{2+i\>2}+\mathcal{U}_{2+i\>1}\mathcal{U}^*_{1\>1}
\big)P_R
 \nonumber \\ &\hspace{4cm}+
\big(\mathcal{V}^*_{2+i\>2+j}\mathcal{V}_{1\>2+j}+\mathcal{V}^*_{2+i\>2}\mathcal{V}_{1\>2}
+\mathcal{V}_{1\>1}\mathcal{V}^*_{2+i\>1}
\big)P_L
\Bigg]\ell_i^+\nonumber
\end{align}
}
\normalsize
where we sum over all neutralino states $n=1,2,3,4,5,6$, all lepton families $i=1,2,3$ and $j=1,2,3$. $P_L$ and $P_R$ are the projection operators $\frac{1-\gamma^5}{2}$ and $\frac{1+\gamma^5}{2}$ respectively. 

It is important to note, however, that the last terms in this expression--that is, those proportional to the photon $\gamma_{0}$--exactly {\it cance}l. This occurs 
because the unitary matrices $\mathcal{U}$ and $\mathcal{V}$ satisfy the identities
\begin{equation}
\mathcal{U}^*_{2+j\>1}\mathcal{U}_{2+i\>2+j}
+\mathcal{U}^*_{1\>2}\mathcal{U}_{2+i\>2}+\mathcal{U}_{2+i\>1}\mathcal{U}^*_{1\>1}=0
\end{equation}
and
\begin{equation}
\mathcal{V}^*_{2+i\>2+j}\mathcal{V}_{1\>2+j}+\mathcal{V}^*_{2+i\>2}\mathcal{V}_{1\>2}
+\mathcal{V}_{1\>1}\mathcal{V}^*_{2+i\>1}=0 \ .
\end{equation}
It follows that the amplitude for the decay channel $\tilde X^\pm_1\rightarrow \gamma_0\ell^\pm$ vanishes. Therefore,  charginos cannot decay into a photon and a lepton

\subsubsection*{\boldmath${\tilde \chi}^{\pm, 0}\rightarrow$$h^0$-lepton }

The part  of the Lagrangian responsible for the Higgs boson-lepton decay channels is

\begin{equation}\label{eq:636}
\begin{split}
\mathcal{L}_{{\tilde \chi}^{\pm,\>0}\rightarrow h^0- \text{lepton}} \supset -\sqrt{2}g(\phi^{i*}T^a \psi_i)\lambda^a
&-\sqrt{2}g\lambda^{a\dag}(\psi^{i\dag}T^a\phi_i)\\&-\frac{1}{2}Y^{ijk}\phi_i\psi_j\psi_k-\frac{1}{2}Y^*_{ijk}\phi^{i*}\psi^{j\dag}\psi^{k\dag},
\end{split}
\end{equation}
where this expression represents the sum over $SU(2)_L$ and $U(1)_Y$. The index $i=1,2,3$ sums over the three lepton families.
The terms with Yukawa couplings arising from the $B-L$ MSSM superpotential
%
%
enter the Lagrangian as 
\begin{equation}\label{eq:Higgs1}
 -Y_{e_i}(H_d^0 e_i {e_i^c}+{H_d^0}^*{e_i^c}^{\dag}e_i^{\dag})
+Y_{\nu_i}(H_u^0 \nu_{i}{\nu^c}_{i}+{H_u^0}^*{\nu^c}_{i}^{\dag}\nu_{i}^{\dag}).
\end{equation}
Notice that we have kept only the terms with neutral Higgs scalar components, since only those have a Higgs boson mass eigenstate component $h^0$. Other terms responsible for this decay channel arise from the supercovariant derivatives of the Higgs fields of the type
\begin{equation}
-\sqrt{2}g(\phi_i^*T^a\psi_i)\lambda^a+h.c.
\end{equation}
in Lagrangian (\ref{eq:636}).
For the $B-L$ MSSM, these produce the terms
{\small
\begin{multline}\label{eq:Higgs2}
-\frac{1}{\sqrt{2}}g_2(H_u^{+^*}\tilde H_u^0)\tilde W^+-\frac{1}{\sqrt{2}}
g_2(H_u^{0^*}\tilde H_u^+)\tilde W^-
 -\frac{1}{\sqrt{2}}g_2(H_d^{0^*}\tilde H_d^-)\tilde W^+ -\frac{1}{\sqrt{2}}g_2(H_d^{-^*}
\tilde H_d^0)\tilde W^-\\
-\frac{1}{\sqrt{2}}g_2(H_u^{+^*}\tilde H_u^+)\tilde W^0+\frac{1}{\sqrt{2}}
g_2(H_u^{0^*}\tilde H_u^0)\tilde W^0
 +\frac{1}{\sqrt{2}}g_2(H_d^{-^*}\tilde H_d^-)\tilde W^0 -\frac{1}{\sqrt{2}}g_2(H_d^{0^*}
\tilde H_d^0)\tilde W^0\\
-\frac{1}{\sqrt{2}}g'(H_u^{+^*}\tilde H_u^+)\tilde B-\frac{1}{\sqrt{2}}
g'(H_u^{0^*}\tilde H_u^0)\tilde B
 +\frac{1}{\sqrt{2}}g'(H_d^{-^*}\tilde H_d^-)\tilde B +\frac{1}{\sqrt{2}}g'(H_d^{0^*}
\tilde H_d^0)\tilde B+h.c.
\end{multline}
}
 It follows that the part of the Lagrangian responsible for the ${\tilde \chi}^{\pm,\>0}\rightarrow h^0- \text{lepton}$ decays is the sum of the equations \eqref{eq:Higgs1} and \eqref{eq:Higgs2}. Note that these expressions are written in terms of the gauge eigenstates. As in the previous section, we expand the gauge eigenstates in terms of the mass eigenstates $\tilde \chi_1^\pm$, $\tilde \chi_n^0$, $e_i$, $\nu_i$ for $n=1,2,3,4,5,6$, $i=1,2,3$ and their hermitian conjugates. Once this step is completed, we group the terms into 4-component spinors to get
\small
\begin{align}\label{eq:Higgs_amplitudes}
\mathcal{L}_{{\tilde \chi}^{\pm,\>0}\rightarrow h^0-\text{lepton}}
\nonumber&\supset \\
-\frac{1}{\sqrt{2}}Y_{e_i}\sin \alpha h^0
\nonumber&\bar{ \tilde X}_1^-\Big[\mathcal{V}^*_{1\>2+j}\mathcal{U}^*_{2+i\>2+j}P_L+\mathcal{V}_{2+i\>2+j}\mathcal{U}_{1\>2+j}P_R\Big]\ell^-_j\\
 +\frac{1}{\sqrt{2}}Y_{e_i}\sin \alpha h^0\nonumber&\bar{ \tilde X}_1^+\Big[  \mathcal{V}^*_{2+i\>2+j}\mathcal{U}^*_{1\>2+j}P_L 
+\mathcal{V}_{1\>2+j}\mathcal{U}_{2+i\>2+j}P_R\Big]
\ell^+_j\\
+\frac{g_2}{{2}}
h^0 \bar {\tilde{{X}}}_1^- \nonumber\Big[(&-\cos \alpha \mathcal{V}^*_{1\>2}\mathcal{U}^*_{2+i\>1}-\sin \alpha
\mathcal{U}^*_{2+i\>2}\mathcal{V}_{1\>1}^*)P_L\\
\nonumber+(&-\cos\alpha \mathcal{V}_{2+i\>2}\mathcal{U}_{1\>1} -\sin \alpha  \mathcal{U}_{1\>2}\mathcal{V}_{2+i\>1})P_R
\Big] \ell_i^-\\
+\frac{g_2}{{2}}
h^0 \bar {\tilde{{ X}}}_1^+\nonumber \Big[& (\cos\alpha \mathcal{V}^*_{2+i\>2}\mathcal{U}^*_{1\>1} +\sin \alpha  \mathcal{U}^*_{1\>2}\mathcal{V}^*_{2+i\>1})P_L\\
\nonumber&+(\cos \alpha \mathcal{V}_{1\>2}\mathcal{U}_{2+i\>1}+\sin \alpha
\mathcal{U}_{2+i\>2}\mathcal{V}_{1\>1})P_R
\Big] \ell_i^+\\
+\frac{g_2}{{2}}\bar {\tilde{{ X}}}^0_n h^0\nonumber\Big[\Big(&
\cos \alpha (\mathcal{N}^*_{n\>4}\mathcal{N}^*_{6+i\>2}+\mathcal{N}^*_{6+i\>4}\mathcal{N}_{n\>2}^*)+\sin \alpha (\mathcal{N}^*_{n\>3}\mathcal{N}^*_{6+i\>2}+\mathcal{N}^*_{6+i\>3}\mathcal{N}_{n\>2}^*)\Big)P_L\\
\nonumber-&
\Big(
\cos \alpha (\mathcal{N}_{n\>4}\mathcal{N}_{6+i\>2}+\mathcal{N}_{6+i\>4}\mathcal{N}_{n\>2})+\sin \alpha (\mathcal{N}_{n\>3}\mathcal{N}_{6+i\>2}+\mathcal{N}_{6+i\>3}\mathcal{N}_{n\>2})\Big)P_R
\Big]\nu_{i}\\
-\frac{g'}{{2}}\bar {\tilde{{ X}}}^0_n h^0\nonumber\Bigg[\Big[&
\cos\alpha \Big(\sin \theta_R(\mathcal{N}^*_{n\>4}\mathcal{N}^*_{6+i\>1}+\mathcal{N}^*_{6+i\>4}\mathcal{N}^*_{n\>1})+\cos \theta_R(\mathcal{N}^*_{n\>4}\mathcal{N}^*_{6+i\>5}+\mathcal{N}^*_{6+i\>4}\mathcal{N}^*_{n\>5})   \Big)\\
\nonumber+&\sin \alpha \Big( \sin \theta_R(\mathcal{N}^*_{n\>3}\mathcal{N}^*_{6+i\>1}+\mathcal{N}^*_{6+i\>3}\mathcal{N}^*_{n\>1})+\cos \theta_R(\mathcal{N}^*_{n\>3}\mathcal{N}^*_{6+i\>5}+\mathcal{N}^*_{6+i\>3}\mathcal{N}^*_{n\>5}) \Big)\Big]P_L\\
\nonumber-&\Big[\cos\alpha \Big(\sin \theta_R(\mathcal{N}_{4\>n}\mathcal{N}_{1\>6+i}+\mathcal{N}_{4\>6+i}\mathcal{N}_{n\>1})
+\cos \theta_R(\mathcal{N}_{n\>4}\mathcal{N}_{6+i\>5}+\mathcal{N}_{6+i\>4}\mathcal{N}_{n\>5})   \Big)\\
\nonumber+&\sin \alpha \Big( \sin \theta_R(\mathcal{N}_{n\>3}\mathcal{N}_{6+i\>1}+\mathcal{N}_{6+i\>3}\mathcal{N}_{n\>1})+\cos \theta_R(\mathcal{N}_{n\>3}\mathcal{N}_{6+i\>5}+\mathcal{N}_{6+i\>3}\mathcal{N}_{n\>5}) \Big)\Big]P_R
 \Bigg]\nu_{i}\\
+\frac{1}{\sqrt 2}Y_{\nu3i}\nonumber&\cos\alpha {\tilde{{X}}}^0_n h^0
\Big[\Big(-\mathcal{N}^*_{n\>6+i}\mathcal{N}^*_{6+j\>6}
+\mathcal{N}^*_{6+j\>6+i}
\mathcal{N}^*_{n\>6}\Big)P_L\\&
+\Big(-\mathcal{N}_{n\>6+i}\mathcal{N}_{6+j\>6}
+\mathcal{N}_{6+j\>6+i}
\mathcal{N}_{n\>6}\Big)P_R
\Big]\nu_{j}
\end{align}
\normalsize
where the angle $\alpha$ is defined in equation (\ref{eq:alpha}) and we sum over all lepton families $i,j=1,2,3$.

One now has the information required to compute the amplitude for each of the processes listed in Table \ref{tab:decay_channels}. To do this, we need the exact expression for the vertex coefficient associated with each such process. These can be read off from the Lagrangians in Eqs. \eqref{eq:gauge_amplitudes} and \eqref{eq:Higgs_amplitudes}.
For example, consider the the decay channel $\tilde X_1^{-}\rightarrow Z^0\ell^{-}_i$. Then it follows from \eqref{eq:gauge_amplitudes} that the vertex coupling is
\begin{equation}
g_{\tilde X^-_1\rightarrow Z^0\ell^-_i}
={G_L}_{\tilde X^-_1\rightarrow Z^0\ell^-_i}P_L+ {G_R}_{\tilde X^-_1\rightarrow Z^0\ell^-_i} P_R \ ,
\end{equation}
where 
\begin{multline}
{G_L}_{\tilde X^-_1\rightarrow Z^0\ell^-_i}=g_{2}\gamma^{\mu}\Bigg(\frac{1}{c_W}
\left(\frac{1}{2}-s_W^2\right)\mathcal{U}_{2+j\>1}\mathcal{U}_{2+j\>2+i}^*
\\-\frac{1}{c_W}\left(\frac{1}{2}-s_W^2 \right)\mathcal{U}_{1\>2}\mathcal{U}^*_{2+i\>2}+2c_W\mathcal{U}^*_{2+i\>1}\mathcal{U}_{1\>1}
\Bigg) 
\label{hanger1}
\end{multline}
and
\begin{multline}
{G_R}_{\tilde X^-_1\rightarrow Z^0\ell^-_i}=g_{2} \gamma^{\mu}\Bigg(\frac{1}{c_W} s_W^2\mathcal{V}_{2+i\>2+j}\mathcal{V}_{1\>2+j}^*\\-\frac{1}{c_W}\left(\frac{1}{2}-s_W^2\right)\mathcal{V}_{2+i\>2}\mathcal{V}_{1\>2}^*
+2c_W\mathcal{V}^*_{1\>1}\mathcal{V}_{2+i\>1}
\Bigg) \ .
\label{hanger2}
\end{multline}
%


\subsection{Chargino decay diagrams}

The form and derivations of the interactions shown in the previous section are somewhat cumbersome. With this in mind, we provide a series of diagrams to pictorially express the origin of these interaction terms in the Lagrangian.
The following figures show the processes which contain the positively charged chargino state ${\tilde X}_1$ only. We assume that ${\tilde X}_1$ is the lightest chargino and is either dominantly charged Higgsino or charged Wino .The vertices associated with the negatively charged chargino decays are the hermitian conjugates of those. 

In each of the following figures, the diagrams on the left side are expressed in terms of gauge eigenstates, written as \emph{2-component Weyl spinors} and correspond to
interaction terms shown in eq. \eqref{eq:633} and \eqref{eq:Higgs2}. They represent the origin of the processes shown on the right.

The diagrams on the right side represent the Feynman diagrams of the same vertices, but expressed in terms of the mass eigenstates, obtained via the decompositions shown in eq. \eqref{eq:mass_expanssion1}-\eqref{eq:mass_expanssion6} and \eqref{eq:mass_expanssion7}-\eqref{eq:mass_expanssion8}. Note that we choose to represent these interactions in terms of 4-component Dirac mass eigenstates, shown in eq. \eqref{eq:DircaFermions}, thus reproducing the interaction terms shown in eq. \eqref{eq:gauge_amplitudes} and \eqref{eq:Higgs_amplitudes}.

The associated amplitudes of each process are also shown, expressed in terms of the gauge and Yukawa couplings of the $B-L$ MSSM, as well as
the elements of the rotation matrices $\mathcal{U}$, $\mathcal{V}$ and $\mathcal{N}$ given in eq. \eqref{eq:matrixUV_PLM} and \eqref{eq:matrixN_PLM}, respectively. We use "red" to represent the matrix elements which are proportional to the RPV coupling $\epsilon_i/M_{soft}$, and "blue" for the matrix elements which are of order unity, with small RPV corrections $1-\epsilon_i/M_{soft}$. At first order, the decay amplitudes are proportional to  $(1-\epsilon_i/M_{soft})\times \epsilon_i/M_{\text{soft}} \simeq \epsilon_i/M_{\text{soft}}$.

\subsubsection{\boldmath${\tilde X}_1^{+}\rightarrow W^{+}\nu_{{i}}$}

{
\centering
\begin{figure}[H]
 \begin{minipage}{0.48\textwidth}
     \centering
   \begin{subfigure}[b]{0.49\linewidth}
   \centering
       \includegraphics[width=0.83\textwidth]{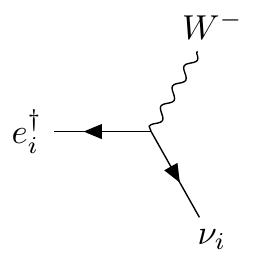}
\caption*{(a1)\\ $ \frac{g_2}{\sqrt{2}} \bar \sigma^\mu$}
       \label{fig:table2}
   \end{subfigure} \hspace{0.01\linewidth}\\
   \centering
   \begin{subfigure}[b]{0.49\linewidth}
   \centering
       \includegraphics[width=0.83\textwidth]{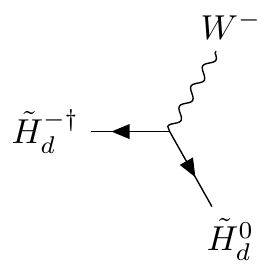}
\caption*{(a2)\\ $\frac{g_2}{\sqrt{2}} \bar \sigma^{\mu}$}
       \label{fig:table2}
\end{subfigure}
   \centering
     \begin{subfigure}[b]{0.49\textwidth}
   \centering
\includegraphics[width=0.83\textwidth]{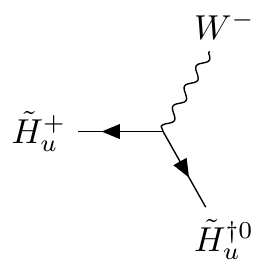}
\caption*{(a3) \\$\frac{g_2}{\sqrt{2}} \bar \sigma^{\mu}$}
\end{subfigure}\\
   \centering
   \begin{subfigure}[b]{0.49\textwidth}
   \centering
\includegraphics[width=0.8\textwidth]{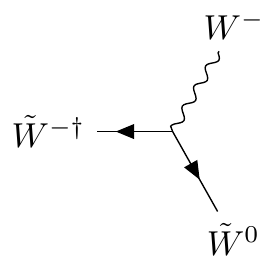}
\caption*{(a4) \\$ -g_2 \bar \sigma^{\mu}$}
\end{subfigure}
   \centering
   \begin{subfigure}[b]{0.49\textwidth}
   \centering
   \includegraphics[width=0.83\textwidth]{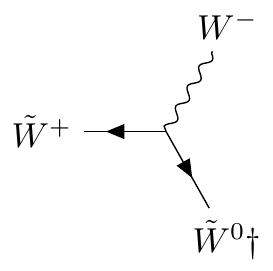}    
\caption*{(a5) \\$ g_2 \bar \sigma^{\mu}$}
\end{subfigure}\subcaption{ }\label{fig:X11}
\end{minipage}\hfill
\rulesep
 \begin{minipage}{0.48\textwidth}
     \centering
   \begin{subfigure}[b]{0.49\linewidth}
   \centering
    \includegraphics[width=0.83\textwidth]{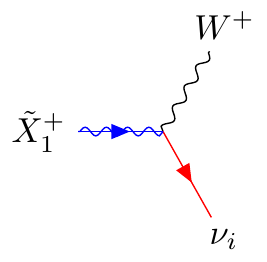} 
       \caption*{\centerline{(b1)}\\$+\frac{g_2}{\sqrt{2}}\textcolor{red}{\mathcal{U}_{1\>2+j}}
\textcolor{blue}{\mathcal{N}^*_{6+j\>6+i}} \gamma^{\mu}P_L$}
       \label{fig:table2}
   \end{subfigure}\\
   \begin{subfigure}[b]{0.49\linewidth}
   \centering
      \includegraphics[width=0.83\textwidth]{651b.pdf} 
\caption*{\centerline{(b2)}\\$\frac{g_2}{\sqrt{2}}\textcolor{blue}{\mathcal{U}_{1\>2}}
\textcolor{red}{\mathcal{N}^*_{6+i\>3}} \gamma^{\mu}P_L$}
       \label{fig:table2}
\end{subfigure}
   \begin{subfigure}[b]{0.49\textwidth}
   \centering
       \includegraphics[width=0.83\textwidth]{651b.pdf} 
\caption*{\centerline{(b3)}\\$-\frac{g_2}{\sqrt{2}}\textcolor{blue}{\mathcal{V}^*_{1\>2}}
\textcolor{red}{\mathcal{N}_{6+i\>4}} \gamma^{\mu}P_R$}
       \label{fig:X1}
\end{subfigure}\\
   \begin{subfigure}[b]{0.49\textwidth}
   \centering
        \includegraphics[width=0.83\textwidth]{651b.pdf} 
\caption*{\centerline{(b4)}\\$-g_2\textcolor{blue}{\mathcal{U}_{1\>1}}
\textcolor{red}{\mathcal{N}^*_{6+i\>2}} \gamma^{\mu}P_L$}
       \label{fig:X1}
\end{subfigure}
   \begin{subfigure}[b]{0.49\textwidth}
   \centering
       \includegraphics[width=0.83\textwidth]{651b.pdf} 
\caption*{\centerline{(b5)}\\ $-g_2\textcolor{blue}{\mathcal{V}^*_{1\>1}}
\textcolor{red}{\mathcal{N}_{6+i\>2}} \gamma^{\mu}P_R$}
       \label{fig:X1}
\end{subfigure}\\
\subcaption{  }\label{fig:X12}
\end{minipage}
\caption{a) Interaction vertices as they 
appear in the MSSM Lagrangian, in terms of the gauge eigenstates, expressed as 2-component Weyl fermions. Vertices (a1), (a2)  and (a3)
arise from the covariant derivatives of the lepton and Higgsino matter fields, respectively.
Vertices (a4) and (a5) come from the covariant derivative of the non-Abelian gaugino fields. b) We 
express the interactions in terms of the mass eigenstates relevant for the chargino decay into SM 
particles, expressed as 4-component Dirac fermions.}\label{fig:X1}
\end{figure}
}

\subsubsection{\boldmath${\tilde X}_1^{+} \rightarrow Z^0 \ell_i^+$}

\vspace{1.4cm}
\begin{figure}[H]
\begin{minipage}{0.48\textwidth}
     \centering
   \begin{subfigure}[b]{0.49\textwidth}
   \centering
     \includegraphics[width=0.87\textwidth]{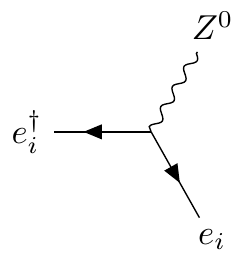} 
\caption*{\centerline{(a1)}\\ \centerline{$ -\frac{g_2}{c_W}\left(\frac{1}{2}-s_W^2\right) \bar \sigma^\mu$}}
       \label{fig:table2}
   \end{subfigure}
   \begin{subfigure}[b]{0.49\textwidth}
   \centering
        \includegraphics[width=0.87\textwidth]{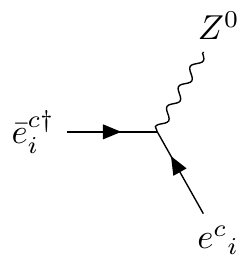} 
\caption*{\centerline{(a2)}\\ \centerline{$-\frac{g_2}{c_W}s_W^2 \bar \sigma^\mu$}}
       \label{fig:table2}
   \end{subfigure}\ \\
   \begin{subfigure}[b]{0.49\textwidth}
   \centering
       \includegraphics[width=0.87\textwidth]{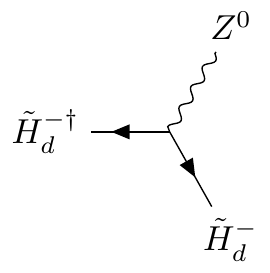} 
\caption*{\centerline{(a3)}\\ \centerline{$-\frac{g_2}{c_W}\left(\frac{1}{2}-s_W^2\right) \bar \sigma^{\mu}$}}
       \label{fig:table2}
       \bigskip
\end{subfigure}
   \begin{subfigure}[b]{0.49\textwidth}
   \centering
      \includegraphics[width=0.87\textwidth]{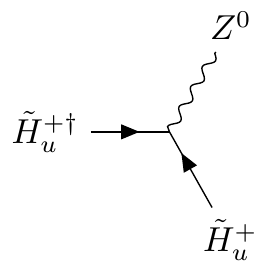} 
\caption*{\centerline{(a4)}\\ \centerline{$\frac{g_2}{c_W}\left(\frac{1}{2}-s_W^2\right)\bar \sigma^{\mu}$}}
       \label{fig:table2}
\bigskip
 \end{subfigure}\\
   \begin{subfigure}[b]{0.49\textwidth}
   \centering
      \includegraphics[width=0.87\textwidth]{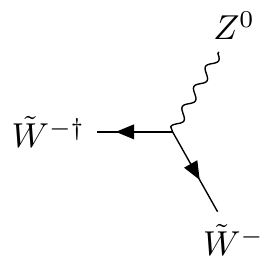}
\caption*{\centerline{(a5)}\\ \centerline{$-g_2c_W\bar \sigma^{\mu}$}}
       \label{fig:table2} 
\end{subfigure}
   \begin{subfigure}[b]{0.49\textwidth}
   \centering
      \includegraphics[width=0.87\textwidth]{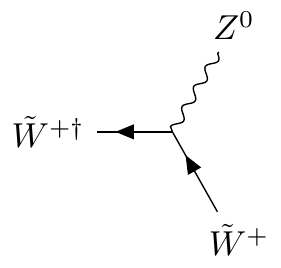}
 \caption*{\centerline{(a6)}\\ \centerline{$g_2c_W\bar  \sigma^{\mu}$}}
       \label{fig:table2}
\end{subfigure}
    \subcaption{ }\label{fig:X21}
\end{minipage}\hfill
\rulesep
\begin{minipage}{0.48\textwidth}
     \centering
   \begin{subfigure}[b]{0.49\textwidth}
   \centering
 \includegraphics[width=0.87\textwidth]{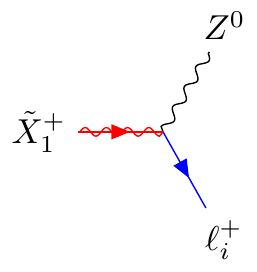}
      \caption*{\centerline{(b1)}\\ \centerline{$ -\frac{g_2}{c_W}\left(\frac{1}{2}-s_W^2\right) \textcolor{red}{\mathcal{U}_{1\>2+i}}\gamma^{\mu}P_L$}}
       \label{fig:table2}
   \end{subfigure}\
   \begin{subfigure}[b]{0.49\textwidth}
   \centering
 \includegraphics[width=0.87\textwidth]{652b.pdf}
       \caption*{\centerline{(b2)}\\ \centerline{$ \frac{g_2}{c_W}s_W^2 \textcolor{red}{\mathcal{V}^*_{1\>2+i}} \gamma^{\mu}P_R$}}
       \label{fig:table2}
   \end{subfigure}\\
   \begin{subfigure}[b]{0.49\textwidth}
   \centering
  \includegraphics[width=0.87\textwidth]{652b.pdf}
\caption*{\centerline{(b3)}\\ $-\frac{g_2}{c_W}\left(\frac{1}{2}-s_W^2\right) \times \\ \times \textcolor{blue}
{\mathcal{U}_{ 1\>2}}
\textcolor{red}{\mathcal{U}^*_{2+i\>2}} \gamma^{\mu}P_L$}
       \label{fig:table2}
\end{subfigure}
   \centering
   \begin{subfigure}[b]{0.49\textwidth}
   \centering
       \includegraphics[width=0.87\textwidth]{652b.pdf}
\caption*{\centerline{(b4)}\\ $-\frac{g_2}{c_W}\left(\frac{1}{2}-s_W^2\right) \times \\ \times \textcolor{blue}
{\mathcal{V}^*_{ 1\>2}}
\textcolor{red}{\mathcal{V}_{2+i\>2}} \gamma^{\mu}P_R$}
       \label{fig:table2}
 \end{subfigure}\\
   \begin{subfigure}[b]{0.49\textwidth}
   \centering
        \includegraphics[width=0.87\textwidth]{652b.pdf}
\caption*{\centerline{(b5)} \\\centerline{$-g_2c_W\textcolor{blue}{\mathcal{U}_{ 1\>1}}
\textcolor{red}{\mathcal{U}^*_{2+i\>1}} \gamma^{\mu}P_L$}}
       \label{fig:table2}
\end{subfigure}
   \begin{subfigure}[b]{0.49\textwidth}
   \centering
      \includegraphics[width=0.87\textwidth]{652b.pdf}
       \caption*{\centerline{(b6)}\\ \centerline{ $-g_2c_W\textcolor{blue}{\mathcal{V}^*_{ 1\>1}} \textcolor{red}{\mathcal{V}_{2+i\>1}} \gamma^{\mu}P_R$}}
       \label{fig:table2}
\end{subfigure}\\  
    \subcaption{}\label{fig:X22}
\end{minipage}
 \caption{a) Interaction vertices as they 
appear in the MSSM Lagrangian, in terms of the gauge eigenstates, expressed as 2-component Weyl fermions. Vertices (a1), (a2)  and (a3)
arise from the covariant derivatives of the lepton and Higgsino matter fields, respectively.
Vertices (a4) and (a5) come from the covariant derivative of the non-Abelian gaugino fields. b) We 
express the interactions in terms of the mass eigenstates relevant for the  chargino decay into SM 
particles, expressed as 4-component Dirac fermions.}\label{fig:X2}
\end{figure}

\subsubsection{\boldmath${\tilde X}_1^{+}\rightarrow h^0 \ell_i^+$}


\begin{figure}[H]
   \centering
   \begin{minipage}{0.48\textwidth}
   \begin{subfigure}[b]{0.49\textwidth}
   \centering
     \includegraphics[width=0.87\textwidth]{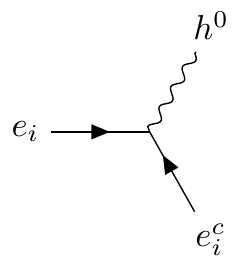}
\caption*{\centerline{(a1)}\\\centerline{ $ -\frac{1}{\sqrt 2}Y_{e_i} \sin \alpha$}}
       \label{fig:table2}
   \end{subfigure}
   \begin{subfigure}[b]{0.49\textwidth}
   \centering
      \includegraphics[width=0.87\textwidth]{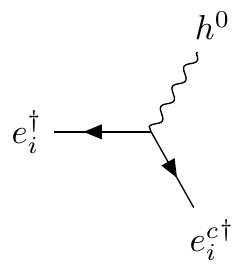}
\caption*{\centerline{(a2)}\\ \centerline{ $ -\frac{1}{\sqrt 2}Y_{e_i}\sin \alpha $}}
       \label{fig:table2}
   \end{subfigure}\\
   \begin{subfigure}[b]{0.49\textwidth}
   \centering
      \includegraphics[width=0.87\textwidth]{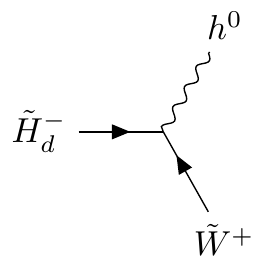}
\caption*{\centerline{(a3)}\\  \centerline{ $\frac{1}{ 2}g_2\sin \alpha$}}
       \label{fig:table2}
\end{subfigure}
   \begin{subfigure}[b]{0.49\textwidth}
   \centering
       \includegraphics[width=0.87\textwidth]{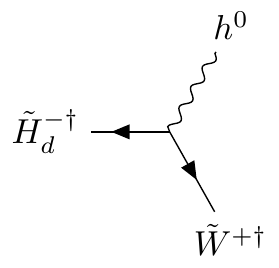}
\caption*{\centerline{(a4)}\\ \centerline{$\frac{1}{ 2}g_2\sin \alpha$}}
       \label{fig:table2}
 \end{subfigure}\\
   \begin{subfigure}[b]{0.49\textwidth}
   \centering
        \includegraphics[width=0.87\textwidth]{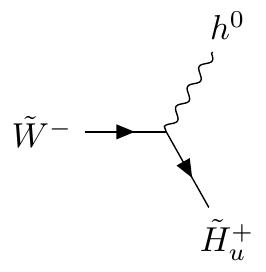}
\caption*{\centerline{(a5)}\\ \centerline{ $-\frac{1}{2}g_2\cos \alpha$}}
\bigskip
       \label{fig:table2}
\end{subfigure}
   \begin{subfigure}[b]{0.49\textwidth}
   \centering
      \includegraphics[width=0.87\textwidth]{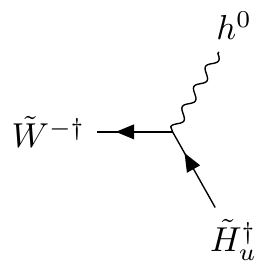}
\caption*{\centerline{(a6)}\\ \centerline{ $-\frac{1}{2}g_2\cos \alpha$}}
\bigskip
       \label{fig:table2}
 \end{subfigure}
    \subcaption{}\label{fig:X41}
\end{minipage}\hfill
\rulesep\begin{minipage}{0.48\textwidth}
\centering
   \begin{subfigure}[b]{0.49\textwidth}
   \centering
 \includegraphics[width=0.87\textwidth]{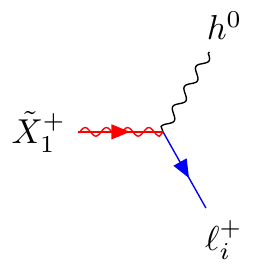}
       \caption*{\centerline{(b1)}\\ $ -\frac{1}{\sqrt 2}Y_{e_i}\sin \alpha 
\textcolor{red}{\mathcal{V}^*_{1\> 2+i}}P_L$}
       \label{fig:table2}
   \end{subfigure}
   \begin{subfigure}[b]{0.49\textwidth}
   \centering
       \includegraphics[width=0.87\textwidth]{654b.pdf}
       \caption*{\centerline{(b2)}\\ $-\frac{1}{\sqrt 2}Y_e\sin \alpha  \textcolor{red}{\mathcal{U}_{1\>2+i}}P_R$}
       \label{fig:table2}
   \end{subfigure}\\
   \begin{subfigure}[b]{0.49\textwidth}
   \centering
       \includegraphics[width=0.87\textwidth]{654b.pdf}
\caption*{\centerline{(b3)}\\ $-\frac{1}{ 2}g_2\sin \alpha \textcolor{blue}{\mathcal{V}^*_{ 1\>1}}
\textcolor{red}{\mathcal{U}^*_{2+i\>2}} P_L$}
       \label{fig:table2}
\end{subfigure}
   \begin{subfigure}[b]{0.49\textwidth}
   \centering
   \includegraphics[width=0.87\textwidth]{654b.pdf}
\caption*{\centerline{(b4) }\\ $-\frac{1}{ 2}g_2 \sin \alpha \textcolor{blue}{\mathcal{U}_{ 1\>2}}
\textcolor{red}{\mathcal{V}_{2+i\>1}} P_R$}
       \label{fig:table2}
 \end{subfigure}\\
   \begin{subfigure}[b]{0.49\textwidth}
   \centering
      \includegraphics[width=0.87\textwidth]{654b.pdf}
\caption*{\centerline{(b5)}\\ $-\frac{1}{2}g_2\cos \alpha \times \\ \times \textcolor{blue}{\mathcal{V}^*_{ 1\>2}}
\textcolor{red}{\mathcal{U}^*_{2+i\>1}} P_L$}
       \label{fig:table2}
\end{subfigure}
   \begin{subfigure}[b]{0.49\textwidth}
   \centering
       \includegraphics[width=0.87\textwidth]{654b.pdf}
\caption*{\centerline{(b6)} $-\frac{1}{ 2}g_2 \cos \alpha \times \\ \times \textcolor{blue}{\mathcal{U}_{ 1\>1}}
\textcolor{red}{\mathcal{V}_{2+i\>2}} P_R$}
       \label{fig:table2}
 \end{subfigure}\\
    \subcaption{}\label{fig:X42}
\end{minipage}
\caption{a)  Interaction vertices as they 
appear in the MSSM Lagrangian, in terms of the gauge eigenstates, expressed as 2-component Weyl fermions. Vertices (a1), (a2)  and (a3)
arise from the covariant derivatives of the lepton and Higgsino matter fields, respectively.
Vertices (a4) and (a5) come from the covariant derivative of the non-Abelian gaugino fields. b) We 
express the interactions in terms of the mass eigenstates relevant for the  chargino decay into SM 
particles, expressed as 4-component Dirac fermions.}
\label{fig:X4}
\end{figure}


\subsection{Neutralino decay diagrams}

We will now outline the RPV decay processes of the Neutralino into SM particles
Unlike for chargino, the lightest neutralino can be any state ${\tilde \chi}_n^0$ with $n=1$ Bino dominant,
$n=2$ for neutral Wino dominant, $n=3,4$ for neutral Higgsino dominant, and $n=5,6$ for right handed 
sterile neutrino dominant.

In each of the following figures, the diagrams on the left side correspond to
interaction terms shown in eq. \eqref{eq:633} and \eqref{eq:Higgs2}. They represent the origin of the processes shown on the right. The diagrams on the right side represent the Feynman diagrams of the same vertices, in terms of 4-component Dirac mass eigenstates, thus reproducing the interaction terms shown in eq. \eqref{eq:gauge_amplitudes} and \eqref{eq:Higgs_amplitudes}.

\subsubsection{\boldmath${\tilde X}^0_n\rightarrow Z^0\nu_{i}$}

\begin{figure}[H]
\begin{minipage}{0.48\textwidth}
\centering
   \begin{subfigure}[b]{0.49\textwidth}
       \includegraphics[width=0.87\textwidth]{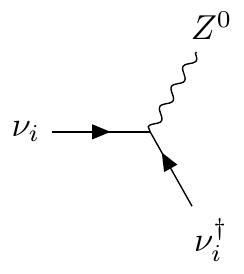} 
\caption*{\centerline{(a1)}\\ \centerline{ $ \frac{g_2}{2c_W}$}}
\bigskip
       \label{fig:table2}
   \end{subfigure}\\
   \begin{subfigure}[b]{0.49\textwidth}
       \includegraphics[width=0.87\textwidth]{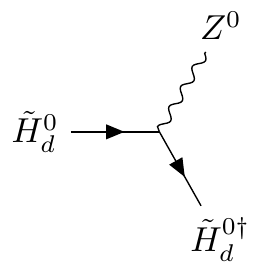} 
\caption*{\centerline{(a2)}\\ \centerline{ $\frac{g_2}{c_W}\left(\frac{1}{2}+s_W^2\right)  \sigma^{\mu}$}}
\bigskip
\bigskip
       \label{fig:table2}
\end{subfigure}
   \begin{subfigure}[b]{0.49\textwidth}
       \includegraphics[width=0.87\textwidth]{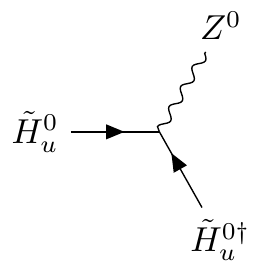} 
\caption*{\centerline{(a3)} \\ \centerline{ $-\frac{g_2}{c_W}\left(\frac{1}{2}+s_W^2\right) \sigma^{\mu}$}}
\bigskip
\bigskip
       \label{fig:table2}
 \end{subfigure}\\
    \subcaption{}\label{fig:X51}
\end{minipage}\hfill
\rulesep
\begin{minipage}{0.48\textwidth}
\centering
   \begin{subfigure}[b]{0.49\textwidth}
        \includegraphics[width=0.87\textwidth]{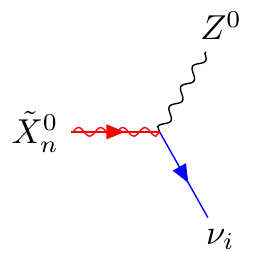} 
       \caption*{\centerline{(b1)} $\frac{g_2}{2c_W}[\textcolor{red}{\mathcal{N}_{n\>6+j}}\textcolor{blue}{\mathcal{N}^*_{6+j\>6+i}}P_L-\textcolor{red}{\mathcal{N}^*_{n\>6+j}}\textcolor{blue}{\mathcal{N}_{6+i\>6+j}}P_R]\gamma^{\mu}$}
       \label{fig:table2}
   \end{subfigure}\\
   \begin{subfigure}[b]{0.49\textwidth}
       \includegraphics[width=0.87\textwidth]{655b.pdf} 
\caption*{\centerline{(b2)}\\ $\frac{g_2}{c_W}\left(\frac{1}{2}+s_W^2\right) \times \\ \times 
[\textcolor{blue}{\mathcal{N}_{n\>3}^*}
\textcolor{red}{\mathcal{N}_{6+i\>3}}P_L\\-\textcolor{blue}{\mathcal{N}_{n\>3}}
\textcolor{red}{\mathcal{N}_{6+i\>3}^*}P_R ] \gamma^{\mu}$}
       \label{fig:table2}
\end{subfigure}
   \begin{subfigure}[b]{0.49\textwidth}
   \centering
        \includegraphics[width=0.87\textwidth]{655b.pdf} 
\caption*{\centerline{(b3) } \\$-\frac{g_2}{c_W}\left(\frac{1}{2}+s_W^2\right) \times \\ \times 
[\textcolor{blue}{\mathcal{N}_{n\>4}}
\textcolor{red}{\mathcal{N}^*_{6+i\>4}}P_L\\
-\textcolor{blue}{\mathcal{N}^*_{n\>4}}
\textcolor{red}{\mathcal{N}_{6+i\>4}}^*P_R] \gamma^{\mu}$}
       \label{fig:table2}
 \end{subfigure}\\
    \subcaption{}\label{fig:X52}
\end{minipage}
 \caption{a) We show the MSSM vertices, in terms of the gauge eigenstates, expressed as 2-component Weyl fermions.  (a1), (a2) and (a3) 
come from the covariant derivatives of the lepton and Higgsino matter fields, respectively. b) We 
express these interactions in terms of the mass eigenstates relevant for the neutralino decay into SM 
particles, expressed as 4-component fermions.
}\label{fig:X5}
\end{figure}

\subsubsection{\boldmath${\tilde X}_n^0 \rightarrow W^\pm \ell_i^\mp$}


\vspace{1.4cm}

\begin{figure}[H]
\centering
\begin{minipage}{0.48\textwidth}
\centering
   \begin{subfigure}[b]{0.49\textwidth}
        \includegraphics[width=0.87\textwidth]{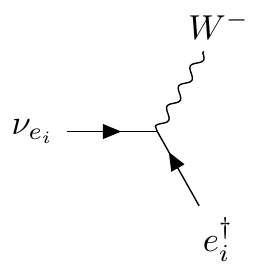} 
\caption*{\centerline{(a1)}\\ \centerline{ $ \frac{g_2}{\sqrt{2}} \bar \sigma^{\mu}$}}
       \label{fig:table2}
   \end{subfigure}\\
   \begin{subfigure}[b]{0.49\textwidth}
        \includegraphics[width=0.87\textwidth]{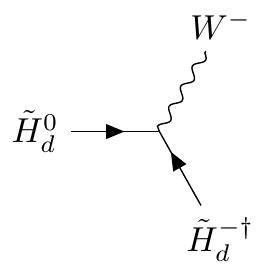} 
\caption*{\centerline{(a2)}\\ \centerline{ $\frac{g_2}{\sqrt{2}} \bar \sigma^{\mu}$}}
       \label{fig:table2}
\end{subfigure}
   \begin{subfigure}[b]{0.49\textwidth}
 \includegraphics[width=0.87\textwidth]{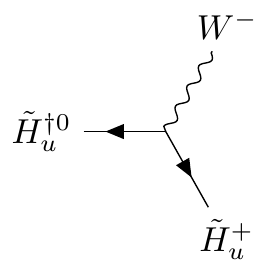} 
\caption*{\centerline{(a3)} \\ \centerline{ $\frac{g_2}{\sqrt{2}} \bar \sigma^{\mu}$}}
       \label{fig:table2}
 \end{subfigure}\\
   \begin{subfigure}[b]{0.49\textwidth}
   \centering
       \includegraphics[width=0.87\textwidth]{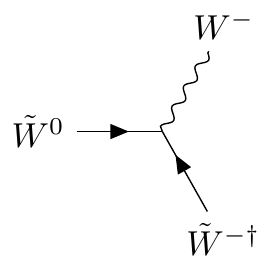} 
\caption*{\centerline{(a4)}\\ \centerline{ $-g_2 \bar \sigma^{\mu}$}}
       \label{fig:table2} 
\end{subfigure}
   \begin{subfigure}[b]{0.49\textwidth}
   \centering
    \includegraphics[width=0.87\textwidth]{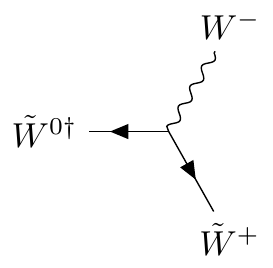}
      \caption*{\centerline{(a5)} \\\centerline{ $g_2 \bar \sigma^{\mu}$}}
       \label{fig:table2}
\end{subfigure}
    \subcaption{}\label{fig:X61}
\end{minipage}\hfill
\rulesep
\begin{minipage}{0.48\textwidth}
\centering
   \begin{subfigure}[b]{0.49\textwidth}
      \includegraphics[width=0.87\textwidth]{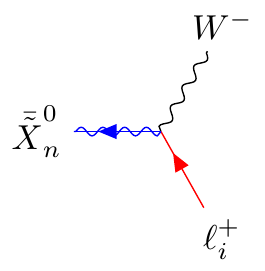}
       \caption*{\centerline{(b1)}\\ \centerline{ $-\frac{g_2}{\sqrt{2}}\textcolor{red}{\mathcal{N}^*_{n\>6+i}}
\textcolor{blue} \gamma^{\mu}P_R$}}
       \label{fig:table2}
   \end{subfigure}\\
   \begin{subfigure}[b]{0.49\textwidth}
           \includegraphics[width=0.87\textwidth]{656b.pdf}
\caption*{\centerline{(b2)}\\ $-\frac{g_2}{\sqrt{2}}\textcolor{blue}{\mathcal{N}^*_{n\>3}}
\textcolor{red}{\mathcal{U}_{2+i\>2}} \gamma^{\mu}P_R$}
       \label{fig:table2}
\end{subfigure}
   \begin{subfigure}[b]{0.49\textwidth}
   \centering
          \includegraphics[width=0.87\textwidth]{656b.pdf}
\caption*{\centerline{(b3)}\\ $\frac{g_2}{\sqrt{2}}\textcolor{blue}{\mathcal{N}_{n\>4}}
\textcolor{red}{\mathcal{V}^*_{2+i\>2}} \gamma^{\mu}P_L$}
       \label{fig:table2}
 \end{subfigure}\\
   \begin{subfigure}[b]{0.49\textwidth}
   \centering
           \includegraphics[width=0.87\textwidth]{656b.pdf}
\caption*{\centerline{(b4)} \\ $+g_2\textcolor{blue}{\mathcal{N}^*_{n\>2}}
                     \textcolor{red}{\mathcal{U}_{2+i\>1}} \gamma^{\mu}P_R$}
       \label{fig:table2}
\end{subfigure}
   \begin{subfigure}[b]{0.49\textwidth}
   \centering
            \includegraphics[width=0.87\textwidth]{656b.pdf} \caption*{\centerline{(b5)} \\ $g_2\textcolor{blue}{\mathcal{N}_{n\>2}}
                     \textcolor{red}{\mathcal{V}^*_{2+i\>1}} \gamma^{\mu}P_L$}
       \label{fig:table2}
\end{subfigure}
    \subcaption{}\label{fig:X62}
\end{minipage}
\caption{ a) We show the MSSM vertices, in terms of the gauge eigenstates, expressed as 2-component Weyl fermions.  (a1), (a2) and (a3) 
come from the covariant derivatives of the lepton and Higgsino matter fields, respectively. (a4) and (a5) come from the covariant derivatives of the non-Abelian gauge fileds. b) We 
express these interactions in terms of the mass eigenstates relevant for the neutralino decay into SM 
particles, expressed as 4-component fermions.
}\label{fig:X6}
\end{figure}

\subsubsection{\boldmath${\tilde X}_n^0\rightarrow h^0\nu_{L_i}$}

\vspace{1cm}


\begin{figure}[H]
   \centering
\begin{minipage}{0.48\textwidth}
\centering
   \begin{subfigure}[b]{0.49\textwidth}
      \includegraphics[width=0.76\textwidth]{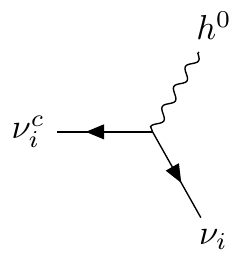}

\caption*{\centerline{(a1)}\\ \centerline{$ \frac{1}{\sqrt 2}Y_{\nu_{j}}\cos \alpha $}}
\bigskip
       \label{fig:table2}
   \end{subfigure}\\
   \begin{subfigure}[b]{0.49\textwidth}
       \includegraphics[width=0.76\textwidth]{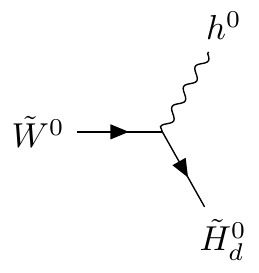}
\caption*{\centerline{(a2)}\\ \centerline{ $\frac{1}{2}g_2\sin \alpha$}}
\bigskip
\bigskip
       \label{fig:table2}
\end{subfigure}
   \begin{subfigure}[b]{0.49\textwidth}
       \includegraphics[width=0.76\textwidth]{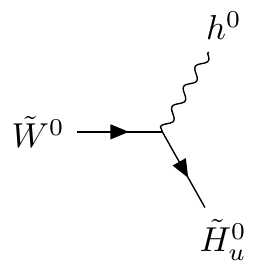}
\caption*{\centerline{(a3)} \\ \centerline{ $\frac{1}{2}g_2\cos \alpha$}}
\bigskip
\bigskip
       \label{fig:table2}
 \end{subfigure}\\
   \begin{subfigure}[b]{0.49\textwidth}
      \includegraphics[width=0.76\textwidth]{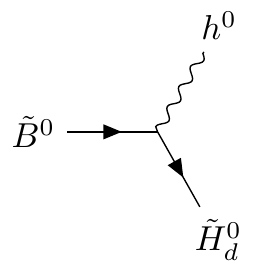}
\caption*{\centerline{(a4)} \\ \centerline{ $-\frac{1}{2}g^{\prime}\sin \alpha$}}
       \label{fig:table2}
\end{subfigure}
   \begin{subfigure}[b]{0.49\textwidth}
      \includegraphics[width=0.76\textwidth]{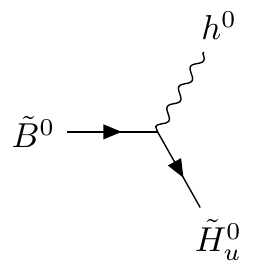}
\caption*{\centerline{(a5)} \\ \centerline{ $-\frac{1}{2}g^{\prime}\cos \alpha$}}
       \label{fig:table2}
 \end{subfigure}
\bigskip
\bigskip
\bigskip
    \subcaption{}\label{fig:X71}
\end{minipage}\hfill
\rulesep
\begin{minipage}{0.48\textwidth}
\centering
   \begin{subfigure}[b]{0.49\textwidth}
\centering
        \includegraphics[width=0.76\textwidth]{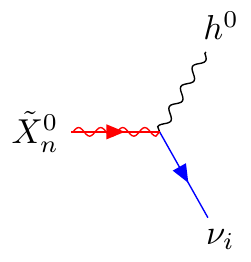}
       \caption*{\centerline{(b1)}\\ $\frac{1}{\sqrt 2}Y_{\nu_j}\cos \alpha \Big[ \textcolor{red}
{\mathcal{N}^*_{n\>6}} \textcolor{blue}{\mathcal{N}^*_{6+j\>6+i}}
- \textcolor{red}
{\mathcal{N}^*_{n\>6+i}} \textcolor{blue}{\mathcal{N}^*_{6+j\>6}}\Big] P_L$}
       \label{fig:table2}
   \end{subfigure}\\
   \begin{subfigure}[b]{0.49\textwidth}
         \includegraphics[width=0.76\textwidth]{657b.pdf}
\caption*{\centerline{(b2)} \\ $+\frac{1}{ 2}g_2\sin \alpha \times \\ \times (\textcolor{blue}{\mathcal{N}^*_{n\>2}}
\textcolor{red}{\mathcal{N}^*_{6+i\>3}}\\
+\textcolor{blue}{\mathcal{N}^*_{n\>3}}
\textcolor{red}{\mathcal{N}^*_{6+i\>2}})P_L$}
       \label{fig:table2}
\end{subfigure}
   \begin{subfigure}[b]{0.49\textwidth}
   \centering
         \includegraphics[width=0.76\textwidth]{657b.pdf}
\caption*{\centerline{(b3)} \\ $\frac{1}{ 2}g_2 \cos \alpha   \times \\ \times (-\textcolor{blue}{\mathcal{N}^*_{n\>2}}
\textcolor{red}{\mathcal{N}^*_{6+i\>4}  }\\
+\textcolor{blue}{\mathcal{N}^*_{n\>4}}
\textcolor{red}{\mathcal{N}^*_{6+i\>2}})P_L
$}
       \label{fig:table2}
 \end{subfigure}
   \begin{subfigure}[b]{0.49\textwidth}
         \includegraphics[width=0.76\textwidth]{657b.pdf}
\caption*{\centerline{(b4)} $-\frac{1}{ 2}g^{\prime}\sin \alpha \times \\ \times [\sin \theta_R(\textcolor{blue}{\mathcal{N}^*_{n\>1}}
\textcolor{red}{\mathcal{N}^*_{6+i\>3}}\\
+\textcolor{blue}{\mathcal{N}^*_{n\>3}}
\textcolor{red}{\mathcal{N}^*_{6+i\>1}}
)\\+\cos \theta_R (\textcolor{blue}{\mathcal{N}^*_{n\>5}}
\textcolor{red}{\mathcal{N}^*_{6+i\>3}}\\
+\textcolor{blue}{\mathcal{N}^*_{n\>3}}
\textcolor{red}{\mathcal{N}^*_{6+i\>5}})]P_L$}
       \label{fig:table2}
\end{subfigure}
   \begin{subfigure}[b]{0.49\textwidth}
   \centering
         \includegraphics[width=0.76\textwidth]{657b.pdf}
\caption*{\centerline{(b5)} $-\frac{1}{ 2}g^{\prime}\cos \alpha   \times \\ \times [\sin \theta_R(\textcolor{blue}{\mathcal{N}^*_{n\>1}}
\textcolor{red}{\mathcal{N}^*_{6+i\>4}}\\
+\textcolor{blue}{\mathcal{N}^*_{n\>4}}
\textcolor{red}{\mathcal{N}^*_{6+i\>1}}
)\\+\cos \theta_R (\textcolor{blue}{\mathcal{N}^*_{n\>5}}
\textcolor{red}{\mathcal{N}^*_{6+i\>4}}\\
+\textcolor{blue}{\mathcal{N}^*_{n\>4}}
\textcolor{red}{\mathcal{N}^*_{6+i\>5}})]P_L$}
       \label{fig:table2}
 \end{subfigure}\\
    \subcaption{}\label{fig:X72}
\end{minipage}
\caption{ a) We show the MSSM vertices in terms of the gauge eigenstates, expressed as 2-component Weyl fermions. (a1)
comes from the Yukawa couplings of the leptons to the Higgs
field in the superpotential, whereas
(a2), (a3), (a4), (a5) come from the supercovariant derivative terms. b) We 
express these interactions in terms of the mass eigenstates relevant for the neutralino decay into SM 
particles, expressed as 4-component fermions. }\label{fig:X7}
\end{figure}

\subsection{Decay rates}
\label{sec:7}

In the previous discussion, we presented all relevant RPV decay channels of charginos $\tilde \chi^\pm_1$ and neutralinos $\tilde \chi^0_n,\>n=1,2,3,4,5,6$ into standard model particles and gave the associated Lagrangian interactions. In this section, we will use these results to calculate the decay rates associated with each such process. The calculations are carried out using the dominant linear terms in the RPV couplings $\epsilon_i$ and $v_{L_i}$ only, since higher order terms are highly suppressed. The analysis is completely general, regardless of whether or not the charginos or the neutralinos are the LSPs. However, only the lightest sparticles decay exclusively via RPV processes into SM particles. Furthermore, they have best prospects for detection at the LHC. Therefore, in subsequent publications, we will use these results to compute the RPV branching ratios of chargino and neutralinos LSPs and NLSPs.

\subsection*{Wino/Higgsino Chargino}

We are ultimately interested in LSP and NSLP decays via RPV channels. Thus, we calculate the decay rates of the $\tilde \chi^\pm_1$ charginos only, which are defined to be lighter than the $\tilde \chi^\pm_2$ charginos. A chargino mass eigenstate is a superposition of a charged Wino, a charged Higgsino and an RPV sum over left-chiral and right-chiral charged leptons gauge eigenstates. 
In the present analysis, we will consider the decays of a {\it generic} chargino with arbitrary mixed charged Wino, charged Higgsino and RPV charged lepton content. More specialized decays involving predominantly Wino chargino or Higgsino chargino mass eigenstates, will be considered in the next sections.\\

\begin{itemize}

\item ${\tilde X}_1^\pm\rightarrow W^\pm \nu_{i}$

The terms in the Lagrangian associated with these decay channels have been calculated in eq. \eqref{eq:gauge_amplitudes} and illustrated in Figure \ref{fig:X1}. Summing all the vertices, one finds
\begin{multline}
g_{{\tilde X}^+_1\rightarrow W^-\nu_i}
=\gamma^\mu {G_L}_{{\tilde X}^+_1\rightarrow W^+ \nu_i} P_L+\gamma^\mu {G_R}_{{\tilde X}^+_1\rightarrow W^+ \nu_i} P_R  \label{F1} \\
~~=\frac{g_2}{\sqrt 2} \gamma^\mu \Big[(\mathcal{U}_{1\>2+j}\mathcal{N}^*_{6+j\>6+i}+\mathcal{U}_{1\>2}\mathcal{N}^*_{6+i\>3}-\sqrt{2}\mathcal{N}^*_{6+i\>2}\mathcal{U}_{1\>1})P_L\\
\hfill+(-\mathcal{N}_{6+i\>4}\mathcal{V}^*_{1\>2}-\sqrt{2}\mathcal{V}^*_{1\>1}\mathcal{N}_{6+i\>2})P_R \Big] \qquad
\end{multline}
and
\begin{multline}
g_{{\tilde X}^-_1\rightarrow W^+\nu_i}  =\gamma^\mu {G_L}_{{\tilde X}^-_1\rightarrow W^- \nu_i} P_L+\gamma^\mu {G_R}_{{\tilde X}^-_1\rightarrow W^- \nu_i} P_R \\
~~~~~=-\frac{g_2}{\sqrt 2} \gamma^\mu\Big[
(\mathcal{U}^*_{1\>2+i}\mathcal{N}_{6+j\>6+i}+\mathcal{U}^*_{1\>2}\mathcal{N}_{6+i\>3}-\sqrt{2}\mathcal{N}_{6+i\>2}\mathcal{U}^*_{1\>1})P_R\\
\hfill+(-\mathcal{N}^*_{6+i\>4}\mathcal{V}_{1\>2}-\sqrt{2}\mathcal{V}_{1\>1}\mathcal{N}^*_{6+i\>2})P_L
 \Big] \qquad 
\end{multline}
Next, using the expressions for the matrix elements of $\mathcal{U}$, $\mathcal{V}$ and $\mathcal{N}$ given in Appendices B.1 and B.2, we get
\begin{multline}\label{eq:72}
 {G_L}_{{\tilde X}^+_1\rightarrow W^+ \nu_i}= -{G^*_R}_{{\tilde X}^-_1\rightarrow W^- \nu_i} =\frac{g_2}{\sqrt 2}\Bigg[
\left( -\cos \phi_- \frac{g_2 v_d}{\sqrt{2}M_2\mu}\epsilon_j^*+\sin \phi_-\frac{\epsilon_j^*}{\mu}\right)\\
-\sin \phi_- \frac{1}{16d_{\tilde \chi^0}}\left( M_{\tilde \gamma}v_R^2v_u(v_d\epsilon_j-\mu v_{L_j}^*)-4M_2\mu(M_{\tilde Y}v_R^2+g_R^2M_{BL}v_u^2)\epsilon_j \right)\\
+\sqrt{2}\cos \phi_- \frac{g_2\mu}{8d_{\tilde \chi_0}}\left( 2g_R^2M_{BL}v_dv_u^2\epsilon_j+M_{\tilde Y}v_R^2(v_d\epsilon_j+\mu v_{L_j}^*) \right)\Bigg]\left[V_{\text{PMNS}}\right]_{ji}
\end{multline}
and
\begin{multline}\label{eq:73}
 {G_R}_{{\tilde X}^+_1\rightarrow W^+ \nu_i}= -{G^*_L}_{{\tilde X}^-_1\rightarrow W^- \nu_i} = \\
\frac{g_2}{\sqrt 2}\Bigg[\sin \phi_+ \frac{1}{16d_{{\tilde \chi}^0}}
[ M_{\tilde \gamma} v_R^2v_u (v_d\epsilon_j^*+\mu v_{L_j})
-4g_R^2\mu M_2M_{BL}v_d v_u\epsilon_j^*]  
 \\-\sqrt{2} \cos \phi_+ 
\frac{g_2\mu}{8d_{{\tilde \chi}^0}}
[2g_R^2M_{BL}v_dv_u^2\epsilon_j+M_{\tilde Y}v_R^2(v_d\epsilon_j^*+\mu v_{L_j})]
\Bigg] \left[V_{\text{PMNS}}^\dag \right]_{ij} \ .
 \end{multline}
 
 The decay width $\Gamma$ is proportional to the square of the amplitude of this process. Note that we account for the longitudinal degrees of freedom of the resultant $W^\pm$ gauge bosons (Goldostone equivalence theorem) in calculating this decay width. This results in an amplification of this channel, which becomes more significant as the mass of the decaying chargino increases. The result is
\begin{equation}
\begin{split}\label{eq:Chargino_Decay1}
\Gamma_{{\tilde X}_1^\pm\rightarrow W^\pm \nu_{i}}&=\frac{\left(|{G_L}|^2_{{\tilde X}^\pm_1\rightarrow W^\pm \nu_i}+|{G_R}|^2_{{\tilde X}^\pm_1\rightarrow W^\pm \nu_i} 
\right)}{64\pi}\times \\ &\times
\frac{M_{{\tilde \chi}_1^\pm}^3}{M_{W^\pm}^2}\left(1-\frac{M_{W^\pm}^2}{M_{{\tilde \chi}_1^\pm}^2}\right)^2\left(1+2\frac{M_{W^\pm}^2}{M_{{\tilde \chi}_1^\pm}^2}\right).
\end{split}
\end{equation}
Note that $|{G_L}|^2_{{\tilde X}^\pm_1\rightarrow W^\pm \nu_i} $ and $|{G_R}|^2_{{\tilde X}^\pm_1\rightarrow W^\pm \nu_i} $ are proportional to the RPV couplings $\epsilon_i$ and $v_{L_i}$ at first order. Therefore, the decay of the chargino into the SM particles $W^\pm$ boson and neutrino would vanish in the absence of RPV.\\

\item ${\tilde X_1}^\pm\rightarrow Z^0 \ell_{i}^\pm $

The terms in the Lagrangian associated with these decay channels have been calculated in eq. \eqref{eq:gauge_amplitudes} and illustrated in Figure \ref{fig:X2}. Summing all the vertices, one finds
\begin{equation}
\begin{split}
g_{{\tilde X}^+_1\rightarrow Z^0 \ell^+_i}&=\gamma^\mu {G_L}_{{\tilde X}^+_1\rightarrow Z^0 \ell^+_i} P_L+\gamma^\mu {G_R}_{{\tilde X}^+_1\rightarrow Z^0 \ell^+_i} P_R\\
&=g_2\gamma^\mu\Bigg[
\Bigg(\frac{1}{c_W}
\left(-\frac{1}{2}-s_W^2\right)\mathcal{U}_{2+j\>1}\mathcal{U}_{2+i\>2+j}^*\\
&\hspace{2cm}-\frac{1}{c_W}\left(\frac{1}{2}-s_W^2 \right)\mathcal{U}_{1\>2}\mathcal{U}^*_{2+i\>2}-c_W\mathcal{U}^*_{2+i\>1}\mathcal{U}_{1\>1}
\Bigg)P_L\\
&\>+
\Bigg(\frac{1}{c_W} s_W^2\mathcal{V}_{2+j\>2+i}\mathcal{V}_{1\>2+i}^*-\frac{1}{c_W}\left(\frac{1}{2}-s_W^2\right)\mathcal{V}_{2+i\>2}\mathcal{V}_{1\>2}^*\\
&\hspace{6cm}-c_W\mathcal{V}^*_{1\>1}\mathcal{V}_{2+i\>1}
\Bigg)P_R\Bigg]
\end{split}
\end{equation}
and
\begin{align}
g_{{\tilde X}^+_1\rightarrow Z^0 \ell^+_i}\nonumber&=\gamma^\mu {G_L}_{{\tilde X}^+_1\rightarrow Z^0 \ell^+_i} P_L+\gamma^\mu {G_R}_{{\tilde X}^+_1\rightarrow Z^0 \ell^+_i} P_R\\
\nonumber&=g_2\gamma^\mu\Bigg[
\Bigg(\frac{1}{c_W}
\left(-\frac{1}{2}-s_W^2\right)\mathcal{U}^*_{2+j\>1}\mathcal{U}_{2+i\>2+j}\\
&\hspace{2cm}-\frac{1}{c_W}\left(\frac{1}{2}-s_W^2 \right)\mathcal{U}^*_{1\>2}\mathcal{U}_{2+i\>2}-c_W\mathcal{U}_{2+i\>1}\mathcal{U}^*_{1\>1}
\Bigg)P_R\\
\nonumber&\>+
\Bigg(\frac{1}{c_W} s_W^2\mathcal{V}^*_{2+j\>2+i}\mathcal{V}_{1\>2+i}-\frac{1}{c_W}\left(\frac{1}{2}-s_W^2\right)\mathcal{V}^*_{2+i\>2}\mathcal{V}_{1\>2}\\
\nonumber&\hspace{6cm}-c_W\mathcal{V}_{1\>1}\mathcal{V}^*_{2+i\>1}
\Bigg)P_L\Bigg]
\end{align}

Using the expressions for the matrix elements of $\mathcal{U}$ and $\mathcal{V}$ from the Appendix B.1, we get
\begin{equation}
\begin{split}
{G_L}_{{\tilde X}^+_1\rightarrow Z^0 \ell^+_i}&=-{G_R}^*_{{\tilde X}^-_1\rightarrow Z^0 \ell^-_i}\\
&=-g_2c_W\Big(\frac{g_2}{\sqrt{2}M_2\mu}(v_d\epsilon_i+\mu v_{L_i}^*)\Big)
\cos \phi_- +\\
& +\frac{g_2}{c_W}
\left(\frac{1}{2}-s_W^2\right)
\Big(-\cos \phi_- \frac{g_2 v_d}{\sqrt{2}M_2\mu}\epsilon_i+\sin \phi_-\frac{\epsilon_i}{\mu}\Big)\\
&\hspace{3cm}-\frac{g_2}{c_W}\left(\frac{1}{2}-s_W^2\right)
\Big(\frac{\epsilon_i}{\mu} \Big)\sin \phi_-
\end{split}
\end{equation}
and
\begin{equation}
\begin{split}
{G_R}_{{\tilde X}^+_1\rightarrow Z^0 \ell^+_i}&=-{G_L}^*_{{\tilde X}^-_1\rightarrow Z^0 \ell^-_i}\\
&=-g_2c_W\cos \phi_+ \Big(-\frac{1}{\sqrt{2}M_2\mu}g_2\tan \beta m_{e_i}v_{L_i} \Big)-\\
&+
\frac{g_2}{c_W}s_W^2\Big( -\cos \phi_+ \frac{g_2 \tan \beta m_{e_i}}{\sqrt{2}M_2\mu}v_{L_i}+\sin \phi_+\frac{m_{e_i}}{\mu v_d}v_{L_i}\Big)\\
&\hspace{3cm}-\frac{g_2}{c_W}\left(\frac{1}{2}-s_W^2\right)\sin \phi_+\Big(\frac{m_{e_i}}{ v_d \mu}v_{L_i} \Big),
\end{split}
\end{equation}
where there is no sum over the $i$ in $v_{L_i}m_{e_i}$.

The decay width $\Gamma$ is proportional to the square of the amplitude of this process. We note that we have accounted for the longitudinal degrees of freedom of the resultant $Z^0$ gauge bosons (Goldstone equivalence theorem) in calculating this decay width. This results in an amplification of this channel which becomes more significant as the mass of the decaying chargino increases. We find that
\begin{equation}
\begin{split}\label{eq:Chargino_Decay2}
\Gamma_{{\tilde X}_1^\pm\rightarrow Z^0 \ell_i^\pm}=&\frac{\Big( |{G_L}|_{{\tilde X}^\pm_1\rightarrow Z^0 \ell^\pm_i}^2+|{G_R}|_{{\tilde X}^\pm_1\rightarrow Z^0 \ell^\pm_i}^2\Big)}{64\pi}\times \\\times &
\frac{M_{{\tilde \chi}_1^\pm}^3}{M_{Z^0}^2}\left(1-\frac{M_{Z^0}^2}{M_{{\tilde \chi}_1}^2}\right)^2
\left(1+2\frac{M_{Z^0}^2}{M_{{\tilde \chi}_1^\pm}^2}\right).
\end{split}
\end{equation}
Note that both $G_{\tilde X^\pm\rightarrow Z^0 \ell_i^\pm}$ coefficients are proportional to the RPV couplings $\epsilon_i$ and $v_{L_i}$ at first order. Therefore, there would be no decay of the chargino into the SM $Z^0$ boson and charged leptons if the RPV effects were non-existent.\\

\item ${\tilde X}^\pm_1\rightarrow h^0 \ell_i^\pm$\\
The terms in the Lagrangian associated with these decay channels have been calculated in eq. \eqref{eq:Higgs_amplitudes} and illustrated in Figure \ref{fig:X4}. Summing all the vertices, one finds
\begin{equation}
\begin{split}
g_{{\tilde X}^+_1\rightarrow h^0 \ell_i^+}&={G_L}_{{\tilde X}^-_1\rightarrow h^0 \ell_i^-} P_L+ {G_R}_{{\tilde X}^-_1\rightarrow h^0 \ell_i^-} P_R\\
&=-\frac{1}{\sqrt{2}}Y_{e_i}\sin \alpha \Big[\mathcal{V}^*_{1\>2+j}\mathcal{U}^*_{2+i\>2+j}P_L+\mathcal{V}_{2+i\>2+j}\mathcal{U}_{1\>2+j}P_R\Big] \qquad ~~\\
&\qquad+\frac{g_2}{{2}}\Big[(-\cos \alpha \mathcal{V}^*_{1\>2}\mathcal{U}^*_{2+i\>1}-\sin \alpha
\mathcal{U}^*_{2+i\>2}\mathcal{V}_{1\>1}^*)P_L\\
&\quad \quad\qquad\qquad+(-\cos\alpha \mathcal{V}_{2+i\>2}\mathcal{U}_{1\>1} -\sin \alpha  \mathcal{U}_{1\>2}\mathcal{V}_{2+i\>1})P_R
\Big]\\
\end{split}
\end{equation}
and
\begin{equation}
\begin{split}
g_{{\tilde X}^-_1\rightarrow h^0 \ell_i^-}&={G_L}_{{\tilde X}^+_1\rightarrow h^0 \ell_i^+} P_L+ {G_R}_{{\tilde X}^+_1\rightarrow h^0 \ell_i^+} P_R\\
&=\frac{1}{\sqrt{2}}Y_{e_i}\sin \alpha \Big[  \mathcal{V}^*_{2+j\>2+i}\mathcal{U}^*_{1\>2+j}P_L 
+\mathcal{V}_{1\>2+i}\mathcal{U}_{2+j\>2+i}P_R\Big] \qquad ~~~~~~\\
&\qquad+ \frac{g_2}{{2}}\Big[ (\cos\alpha \mathcal{V}^*_{2+i\>2}\mathcal{U}^*_{1\>1} +\sin \alpha  \mathcal{U}^*_{1\>2}\mathcal{V}^*_{2+i\>1})P_L\\
&
\qquad\qquad\qquad+(\cos \alpha \mathcal{V}_{1\>2}\mathcal{U}_{2+i\>1}+\sin \alpha
\mathcal{U}_{2+i\>2}\mathcal{V}_{1\>1})P_R 
\Big] \ .
\end{split}
\end{equation}
Next, using the expressions for the matrix elements of $\mathcal{U}$ and $\mathcal{V}$ from the Appendix \ref{appendix:A1}, we get

\begin{equation}
\begin{split}
{G_L}&_{{\tilde X}^+_1\rightarrow h^0 \ell_i^-}=-{G_R}^*_{{\tilde X}^+_1\rightarrow h^0 \ell_i^+} =\\
&=-\frac{1}{\sqrt 2}Y_{e_i}\sin \alpha 
\Big(-\cos \phi_+ \frac{g_2 \tan \beta m_{e_i}}{\sqrt{2}M_2\mu}v_{L_i}^*+\sin \phi_+\frac{m_{e_i}}{\mu v_d}v_{L_i}^* \Big)\\
&-
\frac{1}{ 2}g_2 \sin \alpha \cos \phi_+ \Big( \frac{\epsilon_i}{\mu} \Big)-
\frac{1}{2}g_2\cos \alpha \sin \phi_+\Big( \frac{g_2}{\sqrt{2}M_2\mu}(v_d\epsilon_i+\mu v_{L_i}^*) \Big)
\end{split}
\end{equation}
and
\begin{equation}
\begin{split}
{G_R}&_{{\tilde X}^-_1\rightarrow h^0 \ell_i^-}=-{G_L}^*_{{\tilde X}^+_1\rightarrow h^0 \ell_i^+}=\\
&=-\frac{1}{\sqrt 2}Y_{e_i}\sin \alpha 
\Big(-\cos \phi_- \frac{g_2v_d}{\sqrt{2}M_2\mu}\epsilon_i^*+\sin \phi_-\frac{\epsilon_i^*}{\mu}\Big)\\
&+\frac{1}{2}g_2 \sin \alpha \sin \phi_-\Big(-\cos \phi_+\frac{1}{\sqrt{2}M_2\mu}g_2\tan \beta m_{e_i}v_{L_i} -\sin \phi_+ \frac{m_{e_i}}{\mu v_d}v_{L_i}\Big)\\
&-\frac{1}{2}g_2\cos \alpha \cos \phi_-\Big( \frac{m_{e_i}}{ v_d \mu}v_{L_i} \Big),
\end{split}
\end{equation}

where we do not sum over the $i$ in either of these expressions. The decay width $\Gamma$ is proportional to the square of the amplitude of this process, and is found to be

\begin{equation}\label{eq:Chargino_Decay4}
\Gamma_{{\tilde X}_1^\pm\rightarrow h^0 \ell_i^\pm}=\frac{\Big(|{G_L}|_{{\tilde X}^\pm_1\rightarrow h^0 \ell_i^\pm}^2+|{G_R}|_{{\tilde X}^\pm_1\rightarrow h^0 \ell_i^\pm}^2\Big)}{64\pi}
M_{{\tilde X}_1^\pm}\left(1-\frac{M_{h^0}^2}{M_{{\tilde X}_1^\pm}^2}\right)^2.
\end{equation}

Again, note that $G_{\tilde X^\pm\rightarrow h^0 \ell_i^\pm}$ are proportional to the RPV couplings $\epsilon_i$ and $v_{L_i}$ at first order. Hence, there would be no decay of the chargino into the SM Higgs boson and charged leptons if there would be no RPV effects.

\end{itemize}

\subsection*{Neutralinos}\label{neut:sec}

Recall that the index $n$ indicates the neutralino species as follows:
\begin{equation}
{\tilde X}_1^0={\tilde X}_B^0,\quad  {\tilde X}_2^0={\tilde X}_W^0, \quad {\tilde X}_3^0={\tilde X}_{H_d}^0,
\quad {\tilde X}_4^0={\tilde X}_{H_u}^0, \quad {\tilde X}_5^0={\tilde X}_{\nu_{3a}}^0, \quad {\tilde X}_6^0={\tilde X}_{\nu_{3b}}^0.
\end{equation}
For a general neutralino state $n$, we found expressions for the parameters of the decay to two standard model particles. 

\begin{itemize}

\item{${\tilde X}^0_n\rightarrow Z^0 \nu_{i}$}

To begin with, it follows from
eq. \eqref{eq:gauge_amplitudes} and the vertices in Figure \ref{fig:X5} that
\begin{multline}
g_{{\tilde X}^0_n\rightarrow Z^0 \nu_{i}}=\gamma^\mu {G_L}_{{\tilde X}^0_n\rightarrow Z^0 \nu_{i}}+\gamma^{\mu}{G_R}_{{\tilde X}^0_n\rightarrow Z^0 \nu_{i}} \\
={g_2}\gamma^{\mu}\Big[
\Big(\frac{1}{2c_W}\mathcal{N}_{n\>6+j}\mathcal{N}^*_{6+j\>6+i}-\frac{1}{c_W}\left(\frac{1}{2}+s_W^2\right)\mathcal{N}_{n\>4}\mathcal{N}^*_{6+i\>4} \Big)P_L\\
-\Big( \frac{1}{c_W}\left(\frac{1}{2}+s_W^2\right) \mathcal{N}_{n\>3}\mathcal{N}^*_{6+i\>3}\Big)P_R
\Big]\\
-{g_2}\gamma^{\mu}\Big[
\Big(\frac{1}{2c_W}\mathcal{N}^*_{n\>6+j}\mathcal{N}_{6+j\>6+i}-\frac{1}{c_W}\left(\frac{1}{2}+s_W^2\right)\mathcal{N}^*_{n\>4}\mathcal{N}_{6+i\>4} \Big)P_R\\
-\Big( \frac{1}{c_W}\left(\frac{1}{2}+s_W^2\right) \mathcal{N}^*_{n\>3}\mathcal{N}_{6+i\>3}\Big)P_L
\Big] \,
\end{multline}
where we can read 
\begin{multline}
 {G_L}_{{\tilde X}^0_n\rightarrow Z^0 \nu_{i}}=
g_2\Big(\frac{1}{2c_W}\mathcal{N}_{n\>6+j}\mathcal{N}^*_{6+j\>6+i}-\frac{1}{c_W}\left(\frac{1}{2}+s_W^2\right)\mathcal{N}_{n\>4}\mathcal{N}^*_{6+i\>4} \Big)\\
+g_2\Big( \frac{1}{c_W}\left(\frac{1}{2}+s_W^2\right) \mathcal{N}^*_{n\>3}\mathcal{N}_{6+i\>3}\Big)
\end{multline}
and
\begin{multline}
{G_R}_{{\tilde X}^0_n\rightarrow Z^0 \nu_{i}}=g_2\Big(- \frac{1}{c_W}\left(\frac{1}{2}+s_W^2\right) \mathcal{N}_{n\>3}\mathcal{N}^*_{6+i\>3}\Big)\\
-{g_2}
\Big(\frac{1}{2c_W}\mathcal{N}^*_{n\>6+j}\mathcal{N}_{6+j\>6+i}-\frac{1}{c_W}\left(\frac{1}{2}+s_W^2\right)\mathcal{N}^*_{n\>4}\mathcal{N}_{6+i\>4} \Big).
\end{multline}

Using these results, one can compute the associated decay rate. It is found to be
\begin{equation}
\begin{split}
\Gamma_{{\tilde X}^0_n\rightarrow Z^0\nu_{i}}=&
\frac{\Big(|{G_L}|_{{\tilde X}^0_n\rightarrow Z^0\nu_{i}}^2
+|{G_R}|_{{\tilde X}^0_n\rightarrow Z^0\nu_{i}}^2 \Big)
}{64\pi}\\
&\frac{M_{{\tilde \chi}_n^0}^3}{M_{Z^0}^2}\left(1-\frac{M_{Z^0}^2}{M_{{\tilde \chi}_n^0}^2}\right)^2
\left(1+2\frac{M_{Z^0}^2}{M_{{\tilde \chi}^0_n}^2}\right),
\end{split}
\end{equation}

\item{${\tilde X}^0_n\rightarrow W^\pm \ell_i^\mp$}

Similarly, it follows from eq. \eqref{eq:gauge_amplitudes} and the vertices in Figure \ref{fig:X6} that
\begin{multline}
g_{{\tilde X}^0_n\rightarrow W^- \ell_i^+}=\gamma^\mu {G_L}_{{\tilde X}^0_n\rightarrow W^- \ell_i^+} P_L +\gamma^\mu {G_R}_{{\tilde X}^0_n\rightarrow W^- \ell_i^+} P_R \\
=\frac{g_2}{\sqrt 2} \gamma^\mu
\Big[
(\mathcal{N}_{n\>4}\mathcal{V}^*_{2+i\>2}+\sqrt{2}\mathcal{V}^*_{2+i\>1}\mathcal{N}_{n\>2})P_L\\+(-\mathcal{U}_{2+i\>2+j}\mathcal{N}^*_{n\>6+j}-\mathcal{U}_{2+i\>2}\mathcal{N}^*_{n\>3}+\sqrt{2}\mathcal{N}^*_{n\>2}\mathcal{U}_{2+i\>1})P_R
\Big]
\end{multline}
and its conjugate
\begin{multline}
g_{{\tilde X}^0_n\rightarrow W^+ \ell_i^-}=\gamma^\mu {G_L}_{{\tilde X}^0_n\rightarrow W^+ \ell_i^-} P_L +\gamma^\mu {G_R}_{{\tilde X}^0_n\rightarrow W^+ \ell_i^-} P_R\\
=-\frac{g_2}{\sqrt 2} \gamma^\mu
\Big[
(\mathcal{N}^*_{n\>4}\mathcal{V}_{2+i\>2}+\sqrt{2}\mathcal{V}_{2+i\>1}\mathcal{N}^*_{n\>2})P_R\\+(-\mathcal{U}^*_{2+i\>2+j}\mathcal{N}_{n\>6+j}-\mathcal{U}^*_{2+i\>2}\mathcal{N}_{n\>3}+\sqrt{2}\mathcal{N}_{n\>2}\mathcal{U}^*_{2+i\>1})P_L 
\Big] \ ,
\end{multline}
where we can read 
\begin{equation}
 {G_L}_{{\tilde X}^0_n\rightarrow W^- \ell_i^+}=-{G_R}_{{\tilde X}^0_n\rightarrow W^+ \ell_i^-}=\frac{g_2}{\sqrt{2}}\Big[\mathcal{N}_{n\>4}\mathcal{V}^*_{2+i\>2}+\sqrt{2}\mathcal{V}^*_{2+i\>1}\mathcal{N}_{n\>2}\Big]
\end{equation}
and
\begin{equation}
\begin{split}
{G_R}_{{\tilde X}^0_n\rightarrow W^- \ell_i^+}&=-{G_L}_{{\tilde X}^0_n\rightarrow W^+ \ell_i^-}=\\
&=\frac{g_2}{\sqrt{2}}\Big[-\mathcal{U}_{2+i\>2+j}\mathcal{N}^*_{n\>6+j}-\mathcal{U}_{2+i\>2}\mathcal{N}^*_{n\>3}+\sqrt{2}\mathcal{N}^*_{n\>2}\mathcal{U}_{2+i\>1}\Big].
\end{split}
\end{equation}
Using these results, one can compute the decay rate
\begin{equation}
\begin{split}
\Gamma_{{\tilde X}^0_n\rightarrow W^\mp \ell_i^\pm}&=\frac{\Big(|{G_L}|_{{\tilde X}^0_n\rightarrow W^\pm \ell_i^\mp}^2+|{G_R}|_{{\tilde X}^0_n\rightarrow W^\pm \ell_i^\mp}^2\Big)}{64\pi}\times \\
&\times\frac{M_{{\tilde \chi}_1^\pm}^3}{M_{W^\pm}^2}\left(1-\frac{M_{W^\pm}^2}{M_{{\tilde \chi}_n^0}^2}\right)^2
\left(1+2\frac{M_{W^\pm}^2}{M_{{\tilde \chi}_n^0}^2}\right),
\end{split}
\end{equation}

\item{${\tilde X}^0_n\rightarrow h^0 \nu_{i}$}

Finally, from eq. \eqref{eq:gauge_amplitudes}  and the vertices in Figure \ref{fig:X6} we find
{\small
\begin{multline}
g_{{\tilde X}^0_n\rightarrow h^0 \nu_{i}} ={G_L}_{{\tilde X}^0_n\rightarrow h^0 \nu_{i} } P_L +{G_R}_{{\tilde X}^0_n\rightarrow h^0 \nu_{i} } P_R \\
\qquad =+\frac{g_2}{{2}}\Big[\Big(
\cos \alpha (\mathcal{N}^*_{n\>4}\mathcal{N}^*_{6+i\>2}+\mathcal{N}^*_{6+i\>4}\mathcal{N}_{n\>2}^*)+\sin \alpha (\mathcal{N}^*_{n\>3}\mathcal{N}^*_{6+i\>2}+\mathcal{N}^*_{6+i\>3}\mathcal{N}_{n\>2}^*)\Big)P_L\\-
\Big(
\cos \alpha (\mathcal{N}_{n\>4}\mathcal{N}_{6+i\>2}+\mathcal{N}_{6+i\>4}\mathcal{N}_{n\>2})+\sin \alpha (\mathcal{N}_{n\>3}\mathcal{N}_{6+i\>2}+\mathcal{N}_{6+i\>3}\mathcal{N}_{n\>2})\Big)P_R
\Big]\\
-\frac{g'}{{2}}\Big[\Big(
\cos\alpha \left(\sin \theta_R(\mathcal{N}^*_{n\>4}\mathcal{N}^*_{6+i\>1}+\mathcal{N}^*_{6+i\>4}\mathcal{N}^*_{n\>1})+\cos \theta_R(\mathcal{N}^*_{n\>4}\mathcal{N}^*_{6+i\>5}+\mathcal{N}^*_{6+i\>4}\mathcal{N}^*_{n\>5})   \right)\\
+\sin \alpha \left( \sin \theta_R(\mathcal{N}^*_{n\>3}\mathcal{N}^*_{6+i\>1}+\mathcal{N}^*_{6+i\>3}\mathcal{N}^*_{n\>1})+\cos \theta_R(\mathcal{N}^*_{n\>3}\mathcal{N}^*_{6+i\>5}+\mathcal{N}^*_{6+i\>3}\mathcal{N}^*_{n\>5}) \right)\Big)P_L\\
-\Big(\cos\alpha \left(\sin \theta_R(\mathcal{N}_{n\>4}\mathcal{N}_{1\>6+i}+\mathcal{N}_{6+i\>4}\mathcal{N}_{n\>1})+\cos \theta_R(\mathcal{N}_{n\>4}\mathcal{N}_{6+i\>5}+\mathcal{N}_{6+i\>4}\mathcal{N}_{n\>5})   \right)\\
+\sin \alpha \left( \sin \theta_R(\mathcal{N}_{n\>3}\mathcal{N}_{6+i\>1}+\mathcal{N}_{6+i\>3}\mathcal{N}_{n\>1})+\cos \theta_R(\mathcal{N}_{n\>3}\mathcal{N}_{6+i\>5}+\mathcal{N}_{6+i\>3}\mathcal{N}_{n\>5}) \right)\Big)P_R
 \Big]\\
+\frac{1}{\sqrt 2}Y_{\nu i3}\cos\alpha 
\Big[\Big(\mathcal{N}^*_{n\>6+j}\mathcal{N}^*_{6+i\>6}
+\mathcal{N}^*_{6+i\>6+j}
\mathcal{N}^*_{n\>6}\Big)P_L
+\Big(\mathcal{N}_{n\>6+j}\mathcal{N}_{6+i\>6}
+\mathcal{N}_{6+i\>6+j}
\mathcal{N}_{n\>6}\Big)P_R
\Big],
\end{multline}
}
where
{\small
\begin{multline}
{G_L}_{{\tilde X}^0_n\rightarrow h^0 \nu_{i} }=\frac{g_2}{{2}}\Big(
\cos \alpha (\mathcal{N}^*_{n\>4}\mathcal{N}^*_{6+i\>2}+\mathcal{N}^*_{6+i\>4}\mathcal{N}_{n\>2}^*)+\sin \alpha (\mathcal{N}^*_{n\>3}\mathcal{N}^*_{6+i\>2}+\mathcal{N}^*_{6+i\>3}\mathcal{N}_{n\>2}^*)\Big)\\
-\frac{g'}{{2}}\Big(
\cos\alpha \left(\sin \theta_R(\mathcal{N}^*_{n\>4}\mathcal{N}^*_{6+i\>1}+\mathcal{N}^*_{6+i\>4}\mathcal{N}^*_{n\>1})+\cos \theta_R(\mathcal{N}^*_{n\>4}\mathcal{N}^*_{6+i\>5}+\mathcal{N}^*_{6+i\>4}\mathcal{N}^*_{n\>5})   \right)\\
+\sin \alpha \left( \sin \theta_R(\mathcal{N}^*_{n\>3}\mathcal{N}^*_{6+i\>1}+\mathcal{N}^*_{6+i\>3}\mathcal{N}^*_{n\>1})+\cos \theta_R(\mathcal{N}^*_{n\>3}\mathcal{N}^*_{6+i\>5}+\mathcal{N}^*_{6+i\>3}\mathcal{N}^*_{n\>5}) \right)\Big)\\
+\frac{1}{\sqrt 2}Y_{\nu i3}\cos\alpha \Big(\mathcal{N}^*_{n\>6+j}\mathcal{N}^*_{6+i\>6}
+\mathcal{N}^*_{6+i\>6+j}
\mathcal{N}^*_{n\>6}\Big)
\end{multline}
}
{\small
and
\begin{multline}
{G_R}_{{\tilde X}^0_n\rightarrow h^0 \nu_{i} }=\frac{g_2}{{2}}\Big(
\cos \alpha (\mathcal{N}_{n\>4}\mathcal{N}_{6+i\>2}+\mathcal{N}_{6+i\>4}\mathcal{N}_{n\>2})+\sin \alpha (\mathcal{N}_{n\>3}\mathcal{N}_{6+i\>2}+\mathcal{N}_{6+i\>3}\mathcal{N}_{n\>2})\Big)\\
+\frac{g'}{{2}}\Big(\cos\alpha \left(\sin \theta_R(\mathcal{N}_{n\>4}\mathcal{N}_{1\>6+i}+\mathcal{N}_{6+i\>4}\mathcal{N}_{n\>1})+\cos \theta_R(\mathcal{N}_{n\>4}\mathcal{N}_{6+i\>5}+\mathcal{N}_{6+i\>4}\mathcal{N}_{n\>5})   \right)\\
+\sin \alpha \left( \sin \theta_R(\mathcal{N}_{n\>3}\mathcal{N}_{6+i\>1}+\mathcal{N}_{6+i\>3}\mathcal{N}_{n\>1})+\cos \theta_R(\mathcal{N}_{n\>3}\mathcal{N}_{6+i\>5}+\mathcal{N}_{6+i\>3}\mathcal{N}_{n\>5}) \right)\Big)\\
+\Big(\mathcal{N}_{n\>6+j}\mathcal{N}_{6+i\>6}
+\frac{1}{\sqrt 2}Y_{\nu i3}\cos\alpha \Big(\mathcal{N}_{6+i\>6+j}
\mathcal{N}_{n\>6}\Big)
\end{multline}
}
The decay rate is given by
\begin{equation}
\Gamma_{{\tilde X}^0_n\rightarrow h^0\nu_{i}}=\frac{\Big(|{G_L}|_{{\tilde X}^0_n\rightarrow h^0\nu_{i}}^2+|{G_R}|_{{\tilde X}^0_n\rightarrow h^0\nu_{i}}^2\Big)}{64\pi}
M_{{\tilde \chi}_n^0}\left(1-\frac{M_{h^0}^2}{M_{{\tilde \chi}_n^0}^2}\right)^2.
\end{equation}

\end{itemize}

Note that in the above neutralino expressions, we sum over $j=1,2,3$. Using the matrix elements of $\mathcal{U}$, $\mathcal{V}$ and $\mathcal{N}$ given in Appendices \ref{appendix:A1} and \ref{appendix:A2}, one can can calculate the values of the decay rates numerically. Just as for charginos, it can be shown that the decay amplitudes are proportional to the RPV couplings $\epsilon_i$ and $v_{L_i}$ at first order. Hence, RPV is directly responsible for the neutralino decays into SM particles.

\section{Wino Chargino and Wino Neutralino LSP decays}\label{sec:winosec}

The $B-L$ MSSM appears to be the simplest possible phenomenologically realistic theory of heterotic superstring/M-theory; being exactly the MSSM with right-handed neutrino chiral supermultiplets and spontaneously broken R-parity. We would like to point out that, although the $B-L$ MSSM was originally derived from the ``top-down'' point of view of heterotic M-theory, it was also constructed from a low energy, ``bottom-up'' approach in \cite{FileviezPerez:2008sx,Barger:2008wn,FileviezPerez:2009groo,Everett:2009vy,FileviezPerez:2012mj,Perez:2013kla}. For all of these reasons, it would seem to be 
to be a rich arena to study the phenomenological predictions of the $B-L$ MSSM at energies low enough to be observable by the ATLAS detector at the LHC at CERN. This requires taking the interval of soft supersymmetry breaking parameters to be in the range $0.2-10$ TeV, as discussed in the previous section. 

The generic supersymmetric interactions of the $B-L$ MSSM are extremely complicated for arbitrary mass sparticles, with the RP conserving processes being much larger than, and, hence, potentially making unobservable, the RPV decays.
However, there is one very clean and obvious window where experimental observation of supersymmetric interactions becomes vastly simplified. That window is for the so-called lightest supersymmetric particle--the LSP. By definition, in an $R$-parity conserving theory, the LSP cannot further decay, either to other sparticles or to standard model particles. However, in a theory in which $R$-parity is spontaneously broken, the LSP, while still unable to decay via RP conserving interactions, can now decay through RPV processes to standard model particles. In the $B-L$ MSSM, these decay channels, their decay rates and the associated branching fractions can be explicitly calculated. 
We propose, therefore, that the RPV decays of the LSPs of the $B-L$ MSSM be searched for experimentally, and the results compared to the theoretical predictions. Any positive result obtained in this regard could be a first indication of the existence of $N=1$ supersymmetry, as well as a potential confirmation of the $B-L$ MSSM theory.

This program has already been carried out for the lightest admixture stop, which was shown to be one of the LSPs of the $B-L$ MSSM. The branching ratios for the RPV decay of the lightest stop LSP to a bottom quark and a charged lepton, the dominant decay mode, along with the relationship of these decays to the neutrino mass hierarchy and the $\theta_{23}$ neutrino mixing angle, were presented in \cite{Marshall:2014cwa,Marshall:2014kea}. The admixture stop LSP was chosen for two reasons. First, it carries both electric and color charge and, therefore, is ``exotic''; in the sense that in RP conserving theories such an LSP would contribute to ``dark matter'' which must be gauge neutral. Second, it has a high production cross section at the LHC. Based on the results of these two papers, a search for stop LSP decays in the recent ATLAS LHC data was carried out in \cite{Aaboud:2017opj,ATLAS:2017hbw,Jackson:2015lmj,ATLAS:2015jla}. No direct detection was observed. However, the lower bounds on the stop LSP mass were significantly strengthened. Be that as it may, as was discussed in \cite{Ovrut:2014rba,Ovrut:2015uea} and will be described in the next section, the number of physically realistic initial conditions leading to a stop LSP are relatively small compared to other sparticles. Therefore, in a series of papers, we will pursue this program focussing, however, on other sparticles that occur more frequently as LSPs of the $B-L$ MSSM.

In this section, we will explore the decay channels, the decay rates and calculate the branching ratios to standard model particles for Wino chargino and Wino neutralino LSPs, using the explicit results for generic chargino and neutralino sparticles presented in the previous section. As seen in Figure \ref{fig:lspHistogram}, the Wino charginos/neutralinos occur with much more frequency as LSPs of realistic initial conditions of the $B-L$ MSSM compared to the stops. 

We will discuss the relationship of their decays to standard model particles to the neutrino mass hierarchy and the $\theta_{23}$ neutrino mixing angle. Finally, we find that for a Wino chargino LSP, the NLSP is the Wino neutralino, and vice versa. Furthermore, the mass splitting between them is very small, on the order of several hundred MeV. It follows that a) the RPC decays of the Wino NLSP are highly suppressed relative to its RPV decays and b) that, in addition to the RPV decays of the Wino LSP,  the RPV decays of the Wino NLSP should be observable in the detector at the LHC as well.

Finally, we want to make three important statements concerning the computations in, and the context of, this section. These are:

\begin{enumerate}

\item All calculations in this section, as well as those in previous analyses of the $B-L$ MSSM, are carried out using the {\it one-loop 
corrected} $\beta$ and $\gamma$ renormalization group functions associated with the dimensionless and dimensionful parameters of the theory. However, we systematically {\it ignore all higher-loop corrections to the RGEs as well as any finite one-loop and higher-loop corrections} to the effective Lagrangian. For the purposes of this analysis this is sufficient, since our goal is to present the allowed RPV decay channels of Wino chargino and Wino neutralino LSPs in the $B-L$ MSSM theory and to give their {\it leading order} decay rates, branching ratios and the relationship of these to the neutrino mass hierarchy. However, we are well aware that some of these processes can be substantially effected by higher-loop corrections, both in the RG running of the parameters  and in finite quantities, such as particle masses. For example, in the Higgs mass calculation two-loop RGEs and higher-loop finite corrections could indeed be very important. We conclude that the calculations presented here could, and depending on specific experimental searches being performed to verify them should, be carried out to higher precision than the results presented in this analysis. This would put the $B-L$ MSSM computations on the same footing as the the more commonly studied MSSM. Indeed, the computational tools required to extend our work to finite one-loop and higher-loop RG and finite corrections already exist in the literature, such as in ISAJET \cite{Paige:2003mg}, FlexibleSUSY \cite{Athron:2014yba},
NMSPEC  \cite{Ellwanger:2006rn}, SUSPECT \cite{Djouadi:2002ze}, SARAH \cite{Staub:2008uz}, SPHENO \cite{Porod:2003um}, SUSEFLAV \cite{Chowdhury:2011zr} and the latest version of SOFTSUSY \cite {Allanach:2016rxd}. 
\item The initial soft supersymmetry breaking parameters are selected statistically using a {\it ``log-uniform''} distribution over a mass range compatible with LHC energies. This is the {\it standard} distribution used in analyzing  such initial conditions. We are well aware, however, that one could choose other statistical distributions for the initial parameters--such as a uniform distribution. However, the justification for which distribution to use depends on the choice of the explicit mechanism for spontaneous supersymmetry breaking. The analysis is restricted to the low energy phenomenology of the observable sector only, and does not specify the mechanism of supersymmetry breaking. This could be due, for example, to various non-vanishing F-terms, D-terms or gaugino condensation in the hidden sector of the theory, and is far from unique. Therefore we simply add to the effective Lagrangian the most general allowed soft supersymmetry breaking terms and choose the values of their parameters statistically. Since the log-uniform distribution is the {\it standard} distribution as justified above, we will employ it uniquely. 
Furthermore, as we now discuss, choosing a log-uniform distribution is sufficient for the purposes of this analyis.

To begin, we are simply seeking a set of ``viable'' initial parameters that, when scaled using the RG to lower energy, are completely consistent with all present phenomenological requirements. Using this log-uniform distribution, we show that there are indeed a very large number of such viable initial conditions. It is then demonstrated that, within this context, there is a subset of such parameters that lead to Wino chargino and Wino neutralino LSPs.
Were one to use a different initial distribution of parameters, one would find a potentially different set of viable points, some presumably already contained within the log-uniform distribution and, perhaps, some new viable points. However, any such new viable points will not greatly effect the calculations and conclusions. The first important example of this is the following. It is of some interest to ask, within the context of a log-uniform distribution, what the set of all allowed LSPs is and, furthermore, what percentage of  the viable initial points correspond to Wino charginos and Wino neutralinos. This information is presented in the histogram in Figure \ref{fig:lspHistogram}. We find, within the log-uniform context, that the percentage of valid initial points with Wino chargino and Wino neutralino LSPs is relatively large. Now, it is indeed possible that choosing a different initial distribution would change the percentages for the individual LSPs in this histogram. 
That being said, the actual content of this study is {\it independent} of whether or not these specific LSPs are statistically prominent. Rather, as was the case for the stop LSPs discussed in previous work \cite{Marshall:2014cwa,Marshall:2014kea}--which are statistically minimal in the histogram in Figure \ref{fig:lspHistogram}--we focus on Wino chargino and Wino neutralino LSPs because their RPV decays are {\it readily observable at the ATLAS detector at the LHC}. 

Secondly, it is obvious that the explicit decay channels, the {\it analytic expressions} for the the decay rates and, hence, the LSP lifetimes, as well as the {\it analytic expressions} for the associated branching ratios--both summing and not summing over the families of final states--are completely {\it independent} of the choice of initial parameters and, hence, the choice of the initial distribution. That being said, the statistical plots for the branching ratios and decay lifetimes of Wino charginos and Wino neutralinos, as well as the statistical decay rates of the RPC versus RPV processes for the NLSP presented in the analysis which follows in this work, are all calculated using the same {\it log-uniform} initial distribution. The choice of some different initial distribution of parameters, while reproducing much of these plots, can be expected to alter them somewhat. However, exactly as with the LSP histogram \ref{fig:lspHistogram}, these statistical plots are presented to give a concrete representation of what decay channels of the Wino charginos and Wino neutralinos should be observed at the ATLAS detector at the LHC, whether or not an individual such decay can be ``prompt'', has a ``displaced vertex'' or occurs outside the ATLAS detector, and what the relative probability is of observing a given decay channel as opposed to another channel. If one simply adds the new viable points of a different distribution to the log-uniform points, all of the LHC observational conclusions drawn from the log-uniform priors will remain, essentially, unchanged. However, were one to repeat the entire analysis using a completely different initial distribution--{\it not including the log-uniform priors}--then, although the explicit decay channels will remain the same, their decay lifetimes and relative branching ratios could be altered. However, there is no physical reason to expect the viable points of the log-uniform distribution to be excluded. Furthermore, the choice of a {\it ``non-standard''} initial distribution, not including the viable log-uniform priors, would require a physical and mathematical analysis of the exact mechanism of spontaneous supersymmetry  breaking which, as discussed above, is beyond the present scope. To conclude, the present analysis uses the standard log-prior distribution of initial points. The observational LHC conclusions will only be minimally altered if the viable initial points of additional distributions are added.

\item Finally, there is a long literature discussing RPV decays within a vast variety of contexts. Reference \cite{Barbier:2004ez} reviews the theoretical aspects of RPV violation with both bilinear and trilinear RPV couplings added in the superpotential. Relevant to the content of this analyis, this review discussed both explicit and, more briefly, spontaneous RPV due to both left- and right-chiral sneutrinos developing VEVs. More recently, the subject was reviewed in 2015 \cite{Mohapatra:2015fua}. This discussed explicit RPV in the MSSM but, in particular, focused on spontaneous breaking of $R$-parity in theories where the standard model symmetry is extended by a gauged $U(1)_{B-L}$.  More recently, there was a comprehensive paper \cite{Dercks:2017lfq} 
investigating the phenomenology of the MSSM extended by a single trilinear RPV coupling
at the unification scale. It goes on to discuss the RPV decay of some of the LSPs; specifically the Bino neutralino and the stau sparticle, within the context of the RPV-CMSSM. The mechanism of generating Majorana neutrino masses through RPV bilinear terms is treated in \cite {Hirsch:2008ur, Mitsou:2015eka, Mitsou:2015kpa}. This set of papers also studies the decay modes of some LSPs, with emphasis on the decay modes of the lightest neutralino. There are papers such as \cite{Bomark:2014rra, Csaki:2015uza, Dercks:2017lfq}, which study the RPV decay signatures of chargino, stop, gluinos and charged and neutral Higgsinos, using parameter scans in agreement with the existent experimental bounds. However, they work in different, more general theoretical contexts than our own.

The RPV decays of the Wino chargino and Wino neutralino LSPs share many of the concepts and techniques contained in these papers, such as RG evolution, the associated LSP calculations and their RPV decays, relationship to neutrino masses and so on. However, the purpose of our present work is to discuss the RPV decays of Wino charginos and Wino neutralinos precisely within the context of the $B-L$ MSSM; a minimal and specific extension of the MSSM with spontaneously broken $R$-parity. Furthermore, the initial conditions of this theory are chosen so as to be {\it completely consistent with all phenomenological requirements}, a property not shared by much of the previous literature. Our analysis is performed so as to predict RPV LSP decays amenable to observation at the LHC and arising from a minimal, realistic, $N=1$ supersymmetric theory. The calculation of the leading order RPV decays of the Wino chargino and Wino neutralino LSPs in this specific context have not previously appeared in the literature.

\end{enumerate}

\subsection{Mock LSP Data}

We use the data points shown in Figure \ref{fig:eye}. The RGE computer simulation used to obtain these data points was presented in the same section.

Here, we want to emphasize that the interval over which all 24 dimensionful soft supersymmetry breaking parameters are statistically scattered was chosen to be
\begin{equation}
\big[~\frac{M}{f},Mf~\big] \quad {\rm where}~~~M=1.5~{\rm TeV}~, ~f=6.7 \ .
\label{burt1}
\end{equation}
This guarantees, as mentioned above, that all mass parameters in the theory lie approximately in the range 
\begin{equation}
\big[200~{\rm GeV},10~{\rm TeV}\big]\ .
\label{burt2}
\end{equation}
The values of $M$ and $f$ were chosen to maximize the number of points that are of phenomenological interest --- that is, compatible with current LHC bounds while also being potentially amenable to observation at the LHC. Since this mass interval is ultimately used our analysis, it is important that our results do not significantly depend on these values. Varying $M$ and $f$ changes the range that the massive soft supersymmetry breaking parameters can take and, hence, effects how spread out the low energy mass spectrum can be.  However, the primary focus is the branching ratios and decay lengths of Wino LSPs and NLSPs. We have tested our code with substantially different values of $M$ and $f$ and found that our primary results, that is, the branching ratios and decay lengths of Wino LSPs and NLSPs, indeed do not significantly depend on the choice of $M$ and $f$.

The soft supersymmetry breaking parameters are statistically scattered in the range \eqref{burt1} with a log-uniform distribution. This is a {\it standard choice} of prior distribution. For examples and discussion see \cite{Athron:2017fxj,Fichet:2012sn,Fundira:2017vip,Bomark:2014rra}. There are at least three reasons for choosing a log-uniform distribution. First, it has the intuitive property of scattering masses evenly around the value $M$. That is, 50\% of the scattered masses will be above $M$ and 50\% will be below. A uniform distribution does not have this property. See \cite{Ovrut:2015uea} for further discussion. Second, using a log-uniform distribution addresses ambiguities in how the soft supersymmetry breaking parameters are scattered. For example, should we scatter the mass or the mass squared? The log-uniform distribution addresses this because it is actually invariant with respect to such choices. This ensures that our results are independent of these choices. See \cite{Fichet:2012sn,Fundira:2017vip} for discussion of this. Third, the statistical inference literature identifies the log-uniform distribution as a more objective one to use because it is non-informative in a formal sense \cite{Kass}.

 Out of 100 million initial statistical data points, we found that 65,576 satisfied all phenomenological requirements when scaled to low energy using the RGEs. These were called ``valid black points''. This result is explained in the previous section and displayed in Figure \ref{fig:eye}.

Although each such black point satisfies all physical requirements, they can have different LSPs (see Figure \ref{fig:lspHistogram}). 
A generic chargino is an $R$-parity conserving mixture of a charged Wino, ${\tilde{W}}^{\pm}$, and a charged Higgsino, ${\tilde{H}}^{\pm}$, along with a small RPV  component of charged leptons, $e^{c}_{i}, e_{i}$,~ $i=1,2,3$.  A Wino chargino is a chargino which is predominantly the charged Wino. A generic neutralino is an $R$-parity conserving linear combination of six neutralino sparticles, including  the neutral Wino, ${\tilde{W}}^{0}$, along with a small RPV  component of neutrinos, $\nu_{i}$,~ $i=1,2,3$. A Wino neutralino is a neutralino that is predominantly the neutral Wino. First, notice that the``Wino chargino'', ${\tilde{\chi}}^{\pm}_{W}$, and the associated ``Wino neutralino'', ${\tilde{\chi}}^{0}_{W}$ are statistically favourable LSP candidates. Our statistical analysis shows that the Wino chargino, ${\tilde{\chi}}^{\pm}_{W}$, and the Wino neutralino, ${\tilde{\chi}}^{0}_{W}$, arise from 4,858 and 4,869 valid black points respectively; that is, each occurring approximately 7.40\% of the time.

It will be helpful in our analysis to be more explicit about the properties of Wino chargino and Wino neutralino mass eigenstates.  We refer the reader to Section \ref{sec:5} for a detailed definition of these states respectively. Here, we simply point out that the RPV terms in the definition of both of these sparticles are always very small compared to the $R$-parity conserving terms. It follows that, although essential in the discussion of RPV decays to standard model particles in the following sections, these RPV terms give negligible contributions to the mass eigenvalues. Hence, in this section, where we are discussing their LSP properties, we will consider only the $R$-parity conserving components of the Wino chargino and Wino neutralino eigenstates.

\begin{figure}[t]
   \centering
   \begin{subfigure}[b]{0.6\textwidth}
\includegraphics[width=1.0\textwidth]{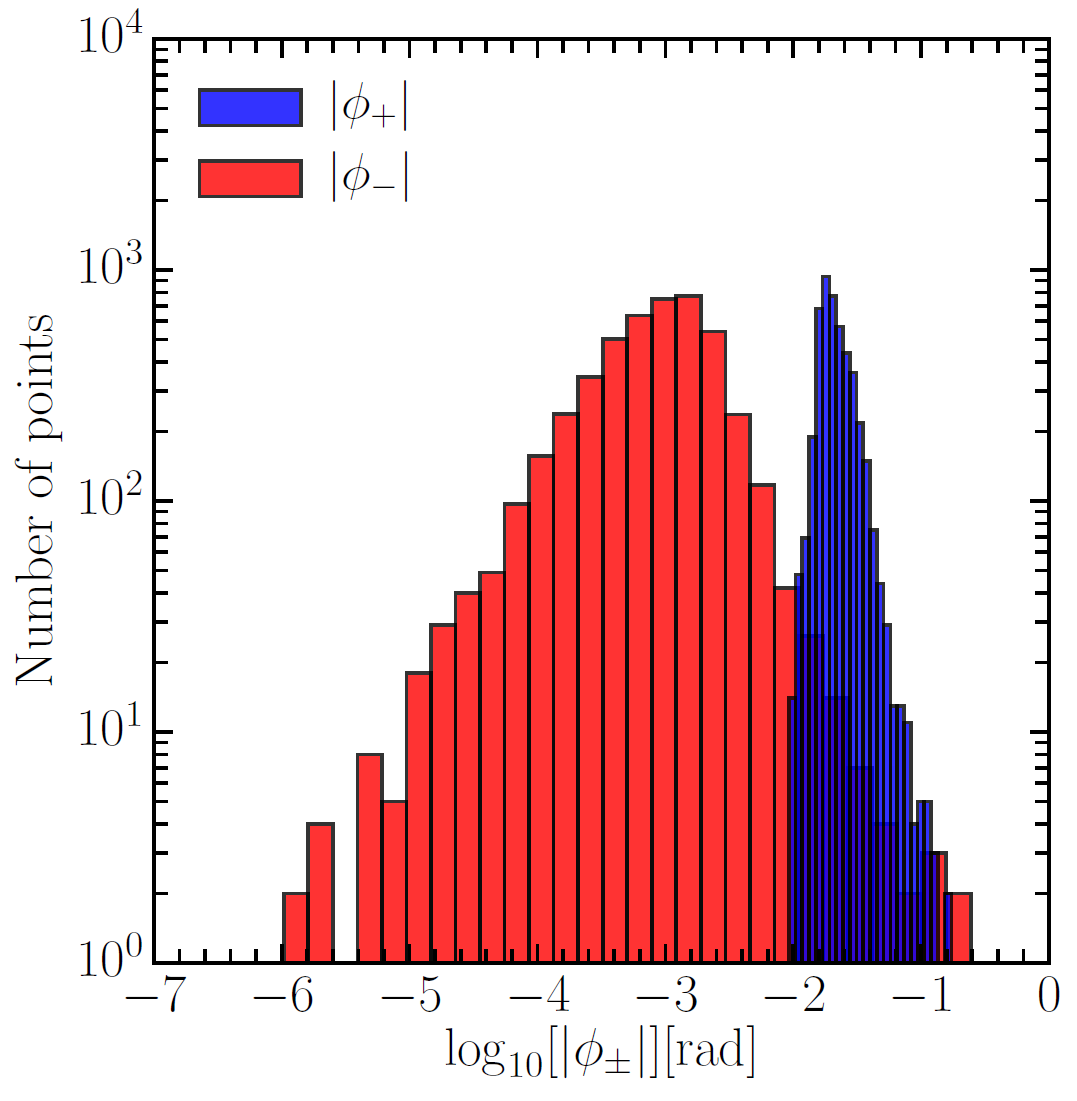}
\end{subfigure}
\caption{ The mixing angles $|\phi_{+}|$ and $|\phi_{-}|$ plotted against the 4,858 valid black points leading to a Wino chargino LSP's. It is clear that both angles are extremely small and, hence $\cos \phi_{\pm} \approx 1$ and $\sin \phi_{\pm} \approx 0$ for any such LSP.}
\label{fig:phi_scatter}
\end{figure}

First consider the Wino chargino. After diagonalizing the chargino mass mixing matrix, one obtains two chargino mass eigenstates, $\tilde \chi_1^\pm$ and $\tilde \chi_2^\pm$, labeled such that $\tilde \chi_1^\pm$ is lighter than $\tilde \chi_2^\pm$.
If the dimensionful parameters $M_{2}$ and $\mu$ in the $B-L$ MSSM satisfy $|M_{2}|<|\mu|$, then (ignoring the $R$-parity violating components) the lighter state $\tilde \chi^\pm_1$ is given by 
\begin{equation}
\tilde \chi^\pm_1=\cos \phi_\pm \tilde W^\pm+\sin \phi_\pm \tilde H^\pm \ ,
\label{car1}
\end{equation}
%
where $W^\pm$ and $\tilde H^\pm$ are the pure charged Wino and charged Higgsino respectively. The angles $\phi_{\pm}$ are defined in equations \eqref{bernard1} and \eqref{bernard2}. The 4,858 viable initial points leading to an observable chargino LSP will naturally tend to require $|M_{2}|<|\mu|$ and, hence, satisfy \eqref{car1}.
Generically, one finds $|M_{2}|$ to be of order of several hundred GeV to ensure that the associated LSP is observable at the LHC, whereas $\mu$ is much larger, of order a few TeV, to solve the ``little hierarchy problem''.
Importantly, these mass scales, along with one other important input, allow one to estimate the sizes of the $\phi_{\pm}$ angles. These angles are defined by
\begin{equation}
\tan 2\phi_-=2\sqrt{2}M_{W^\pm}\frac{\mu \cos \beta +M_2 \sin \beta}{\mu^2-M_2^2-2M_{W^\pm}^2
\cos 2\beta}
\label{bernard1}
\end{equation}
\begin{equation}
\tan 2\phi_+=2\sqrt{2}M_{W^\pm}\frac{\mu \sin \beta +M_2 \cos \beta}{\mu^2-M_2^2+2M_{W^\pm}^2
\cos 2\beta} \ .
\label{bernard2}
\end{equation}
\noindent Clearly, the final required input is an estimate of the size of ${\rm tan}~ \beta$. 
Although we sample $\tan \beta$ with a uniform prior between 1 and 65, we find that the 4,858 valid black points, subject to all low energy phenomenological constraints, tend to prefer larger values of $\tan \beta$ over the smaller ones. That is, for most of the black points with chargino LSPs, $\sin \beta \approx 1$ and $\cos \beta \ll 1$.
With these insights, we expect 
\begin{equation}
\phi_+ \approx \frac{M_{W^\pm}}{\mu} \sim \mathcal{O} \left( 10^{-2}, 10^{-1} \right)
\label{late2}
\end{equation}
 and 
 \begin{equation}
 \phi_-\approx \frac{M_{W^\pm}}{\mu}\left( \cos \beta+\frac{M_2}{\mu} \right) \sim \mathcal{O} \left(10^{-4}, 10^{-2}  \right),
 \label{late3}
 \end{equation}
where $M_{W^\pm}=80.379$ GeV is the measured mass of the $W^{\pm}$ weak gauge bosons. 
These angles $|\phi_\pm|$ can be evaluated numerically for each of the 4,858 associated black points. The results are shown as a histogram in Figure \ref{fig:phi_scatter}. It is clear that both angles are extremely small for any such black points,
%
%
which is in agreement with our estimates in \eqref{late2} and \eqref{late3}.
It then follows from \eqref{car1} that
\begin{equation}
\tilde \chi_{1}^{\pm} \simeq \tilde W^\pm 
\label{book1}
\end{equation}
Therefore, we will denote
\begin{equation}
\tilde{\chi}_{1}^{\pm} \equiv \tilde \chi^\pm_W 
\label{late1}
\end{equation}
and refer to $\tilde \chi^\pm_W$ as the Wino chargino.

Let us now consider the Wino neutralino. Ignoring the very small RPV corrections, there are 6 neutralino mass eigenstates, each a complicated linear combination of the neutral gauge eigenstates. Here, we will only consider one of them, the Wino neutralino LSP , ${\tilde {\chi}}^{0}_W$, and the 4,869 valid black points associated with it. The numerical calculation described in Section \ref{sec:4} allows us to compute, for each valid black point, the coefficients of the linear combination of neutral gauge eigenstates comprising the Wino neutralino. 
Here, we will simply state the result that the coefficient of the neutral Wino, $W^{0}$, component is largely predominant, whereas all other coefficients are very small. 
Hence, to a high degree of approximation,
\begin{equation}
{\tilde {\chi}}^{0}_W \simeq W^{0} \ .
\label{book2}
\end{equation}
That is, the Wino neutralino LSP mass eigenstate is almost the neutral Wino.

\begin{figure}[!ht]
   \centering
   \begin{subfigure}[b]{0.49\textwidth}
\includegraphics[width=1.0\textwidth]{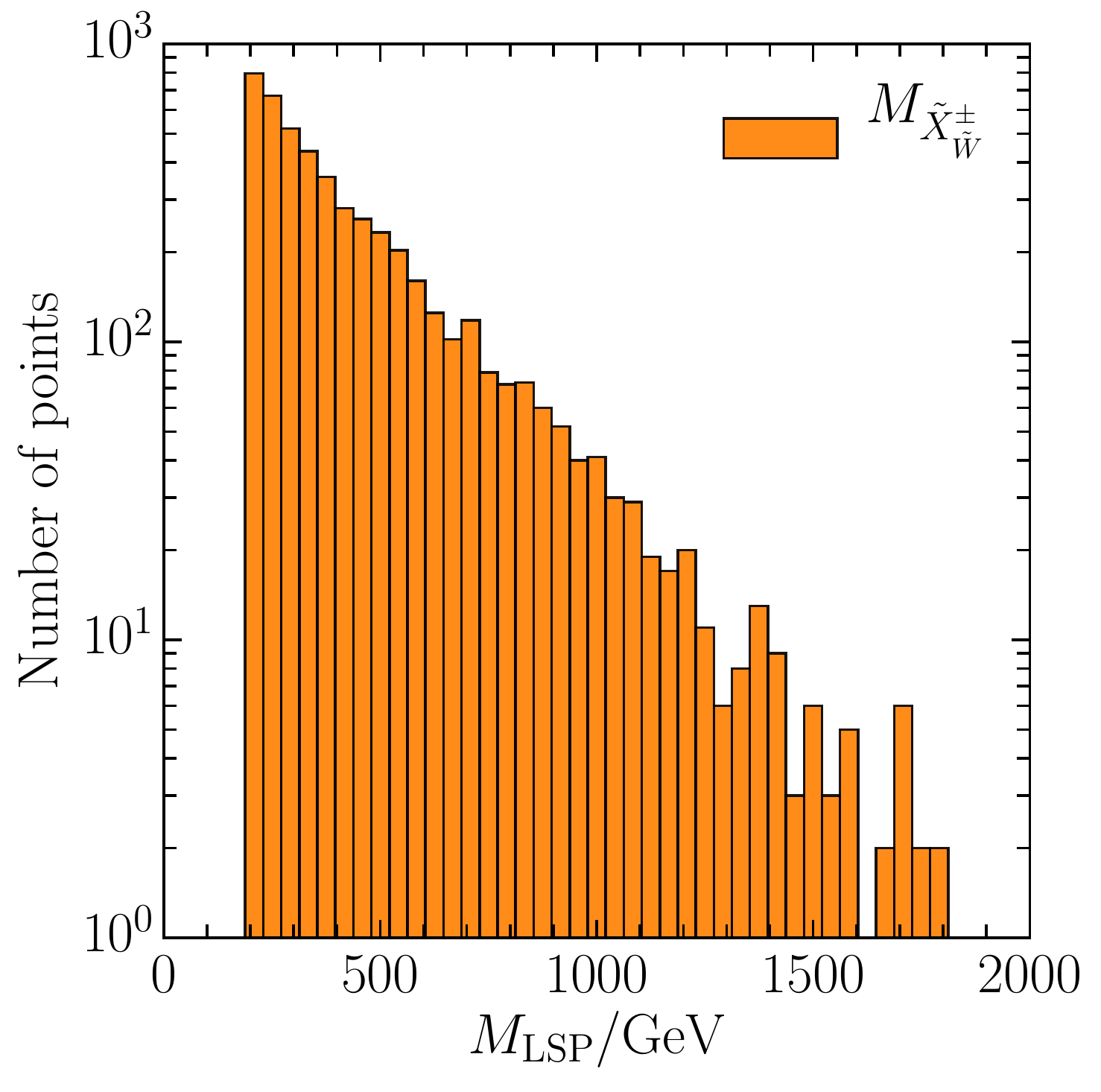}
\label{fig:mass_hist2}
\end{subfigure}
   \begin{subfigure}[b]{0.49\textwidth}
\includegraphics[width=1.0\textwidth]{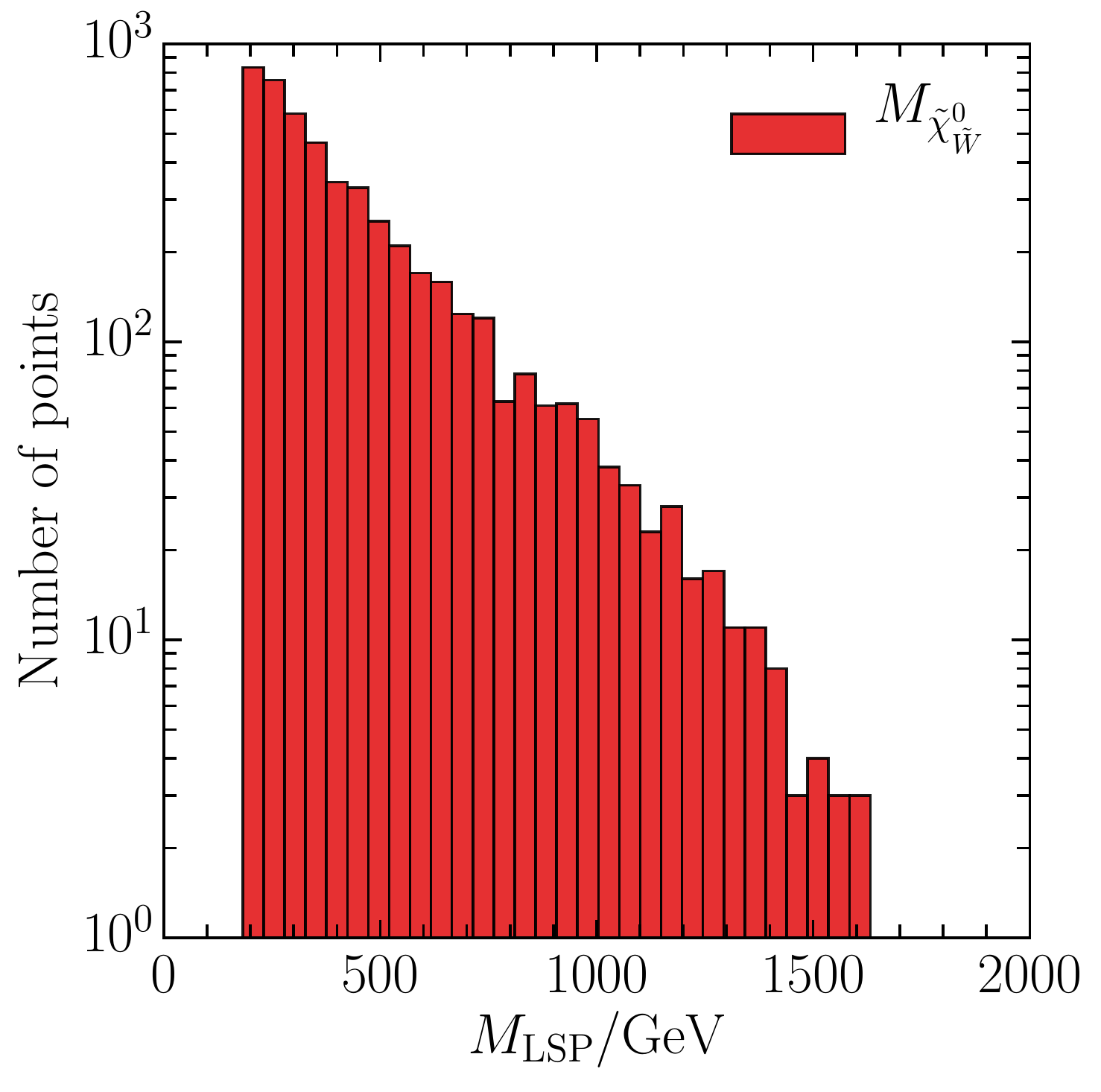}
\label{fig:mass_hist1}
\end{subfigure}
\caption{ a) Mass distribution of the Wino chargino LSP's for the 4,858 valid black points. The masses range from 200 GeV to 1820 GeV, peaking towards the low mass end. b) Mass distribution of the Wino neutralino LSP's for the 4,869 valid black points. The masses range from 200 GeV to 1734 GeV, peaking towards the low mass end.}
\label{fig:mass_hist}
\end{figure}

For any given choice of LSP, we can plot the number of such points as a function of their masses in GeV. As an example, Figure \ref{fig:mass_hist2} and \ref{fig:mass_hist1} presents such a mass distribution for Wino chargino and Wino neutralino LSPs respectively. We obtain viable supersymmetric spectra with Wino chargino and Wino neutralino LSP masses ranging from about 200 GeV to 1700 GeV. A striking feature of the Wino chargino and Wino neutralino LSP mass distributions in Figure \ref{fig:mass_hist} is the peak towards the low mass values. Higher LSP masses are exponentially less probable. The reason is that we sample all soft mass terms log-uniformly in the interval $[200$ GeV,  $10$ TeV]. This includes the $M_2$ gaugino mass term, which gives the dominant contribution for both the Wino chargino and Wino neutralino masses, see \eqref{eq:Wino_mass} and \eqref{eq:Wino_Neutralino_mass} respectively. If we would plot all the Wino chargino or Wino neutralino masses for all the viable points in our simulation, we would obtain an almost uniform mass distribution. However, for the Wino charginos or Wino neutralinos to be the LSPs, their masses must be lower than all the other random soft masses in our simulation. Conversely, it demands that all the other random soft mass terms be larger than a Wino chargino or Wino neutralino mass value for each viable point. This is exponentially less likely as this mass value increases, following a Boltzmann distribution. We point out that this discussion is a simplification of what actually happens, since it omits the running of the soft mass terms, as well as their mixing in the final mass eigenstates. These details, however, do not effect the essence of the above argument, since the mass runnings and the mass mixing couplings are generically very small.

\subsection{Wino Chargino LSP Decay Analysis}

The minimal B-L extension of the MSSM, that is, the $B-L$ MSSM, introduces RPV vertices that allow LSPs to decay directly into SM particles. In this section, we  will investigate the RPV decays of a Wino chargino LSP. A generic chargino mass eigenstate is a superposition of a charged Wino, a charged Higgsino and charged lepton gauge eigenstates. The $R$-parity conserving component of the Wino chargino is given by the linear combination of a charged Wino and charged Higgsino presented in \eqref{car1}, where the charged Wino component dominates.  For the case at hand, where $|M_{2}|<|\mu|$, the smaller RPV contribution to the Wino chargino was found to be
\begin{equation}
\mathcal{V}_{1\>2+i} e^{c}_{i} \quad {\rm where} \quad \mathcal{V}_{1\>2+i}=-\cos \phi_+ \frac{g_2 \tan \beta m_{e_i}}{\sqrt{2}M_2\mu}v_{L_i}+\sin \phi_+\frac{m_{e_i}}{\mu v_d}v_{L_i}
\label{wait1}
\end{equation}
for ${\tilde{\chi}}^{+}_{W}$ and 
\begin{equation}
\mathcal{U}_{1\>2+i} e_{i}  \quad {\rm where} \quad \mathcal{U}_{1\>2+i}=-\cos \phi_- \frac{g_2 v_d}{\sqrt{2}M_2\mu}\epsilon_i^*+\sin \phi_-\frac{\epsilon_i^*}{\mu}
\label{wait2}
\end{equation}
for ${\tilde{\chi}}^{-}_{W}$. We sum \eqref{wait1}
 and \eqref{wait2} over $ i=1,2,3$.

\begin{figure}[t]
 \begin{minipage}{1.0\textwidth}
     \centering
   \begin{subfigure}[b]{0.24\linewidth}
   \centering
       \includegraphics[width=1.0\textwidth]{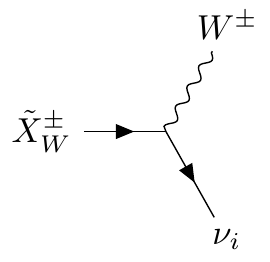} 
\caption*{(a) ~${\tilde X}^\pm_W\rightarrow W^\pm \nu_{i}$}
       \label{fig:table2}
   \end{subfigure} 
   \hfill
   \centering
   \begin{subfigure}[b]{0.24\linewidth}
   \centering
         \includegraphics[width=1.0\textwidth]{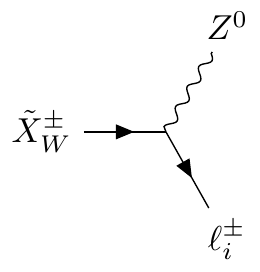} 
\caption*{(b) ~${\tilde X}^\pm_W\rightarrow Z^0 \ell_{i}^\pm$}
       \label{fig:table2}
\end{subfigure}\hfill
   \centering
   \begin{subfigure}[b]{0.24\textwidth}
   \centering
         \includegraphics[width=1.0\textwidth]{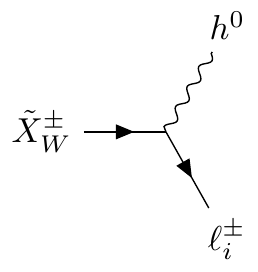} 
\caption*{(d) ~${\tilde X}^\pm_W \rightarrow h^0 \ell_{i}^\pm$}
\end{subfigure}
\end{minipage}
\caption{RPV decays of a general massive chargino state $X_W^\pm$. There are three possible channels, each with $i=1,2,3$, that allow for Wino chargino LSP decays. The decay rates into each individual channel were calculated in Section \ref{sec:7}, for generic charginos}\label{Figure4}
\end{figure}

One of the goals of of this analysis is to predict the possible signals produced by the RPV decays  of Wino chargino LSPs, were such particles to exist and be light enough to be detected at the LHC. In Section \ref{sec:7}, we analyzed RPV decay channels using 4-component spinor notation for the mass eigenstates. For example, the Dirac spinor associated with the  Weyl fermions ${\tilde{\chi}}^{\pm}_{W}$  is defined to be 

\begin{equation}
\tilde X^\pm_W=
\left(
\begin{matrix}
\tilde \chi_W^\pm\\
\tilde \chi_W^{\mp \dag}
\end{matrix}
\right) \ .
\end{equation}
We found that ${\tilde{ X}}^\pm_W$ can decay into standard model particles via three RPV channels. These are shown in Figure \ref{Figure4}.
\noindent  Each of these three channels have different properties concerning their potential experimental detection.
The  ${\tilde X}^\pm_W\rightarrow Z^0 \ell_i^\pm$
process is the Wino chargino decays most easily observed at the LHC. On the other hand, the left handed neutrinos produced during ${\tilde X}^\pm_W\rightarrow W^\pm \nu_i$ can only be detected as missing energy, while the Higgs boson $h^0$ resulting from the decay ${\tilde X}^\pm_W\rightarrow h^0 \ell_i^\pm$ couples to both quarks and leptons, leading to traces in the detector that are harder to interpret. 
In the following, we will explicitly compute the decay rates and branching ratios for all three channels. Sufficiently large probabilities for the process  ${\tilde X}^\pm_W\rightarrow Z^0 \ell_i^\pm$  may indicate favorable prospects for detecting Wino chargino LSPs at the LHC, whereas finding that this channel is subdominant would then make these experimental efforts more difficult. 

\subsection*{Branching ratios of the decay channels}

The decay rates into each individual channel were calculated analytically in Section \ref{sec:7}, for generic charginos. {\it For a fixed lepton family $i$, these decay rates depend explicitly on the choice of the neutrino hierarchy and the value of $\theta_{23}$}\footnote{Each of the measured values of $\theta_{23}$ in both the normal and inverted hierarchies have small uncertainties around a central value. These uncertainties are incorporated into our computer code in all calculations. However, for simplicity of notation, when we refer to the value of $\theta_{23}$, we will suppress these error intervals and indicate the central values only.}. We will discuss this in detail at the end of this section. However, for the present, we will confine ourselves to a calculation of the overall branching ratio for a given type of decay process, which explicitly involves a sum $\sum^{3}_{i=1}$ over the lepton families. The relative prevalence of each channel type is determined by its associated branching ratio. {\it A statistical analysis shows that, for any given decay channel, the sum over the three lepton families makes the branching ratio approximately independent of both the choice of the neutrino hierarchy and the value of $\theta_{23}$}. We will now evaluate these branching ratios for each of the 4,858 valid black points associated with a Wino chargino LSP, separating the data into statistically relevant bins of both $\tan \beta$ and the Wino chargino mass. To begin with, these calculations will be carried out assuming a normal hierarchy with $\theta_{23}=0.597$. Later on in this section, we will  discuss the small statistical differences that would occur had we chosen one of the other possible neutrino data sets. To make this process transparent, we now present our explicit calculational procedure.

For specificity, let us first discuss ${\tilde X}^\pm_W\rightarrow Z^0 \ell_i^\pm$ for any $i=1,2,3$. For this decay channel, the branching ratio is defined by
\begin{equation}\label{eq:Branching1}
\text{Br}_{{\tilde X}^\pm_W\rightarrow Z^0 \ell^\pm}=\frac{\sum_{i=1}^{3} \Gamma_{{\tilde X}^\pm_W\rightarrow Z^0 \ell_i^\pm}}{\sum_{i=1}^{3} \Big( \Gamma_{{\tilde X}^\pm_W\rightarrow W^\pm \nu_{i}}+\Gamma_{{\tilde X}^\pm_W\rightarrow Z^0 \ell_i^\pm} +\Gamma_{{\tilde X}^\pm_W\rightarrow h^0 \ell_i^\pm}\Big)} \ .
\end{equation}

\begin{figure}[!ht]
   \centering
\includegraphics[width=0.93\textwidth]{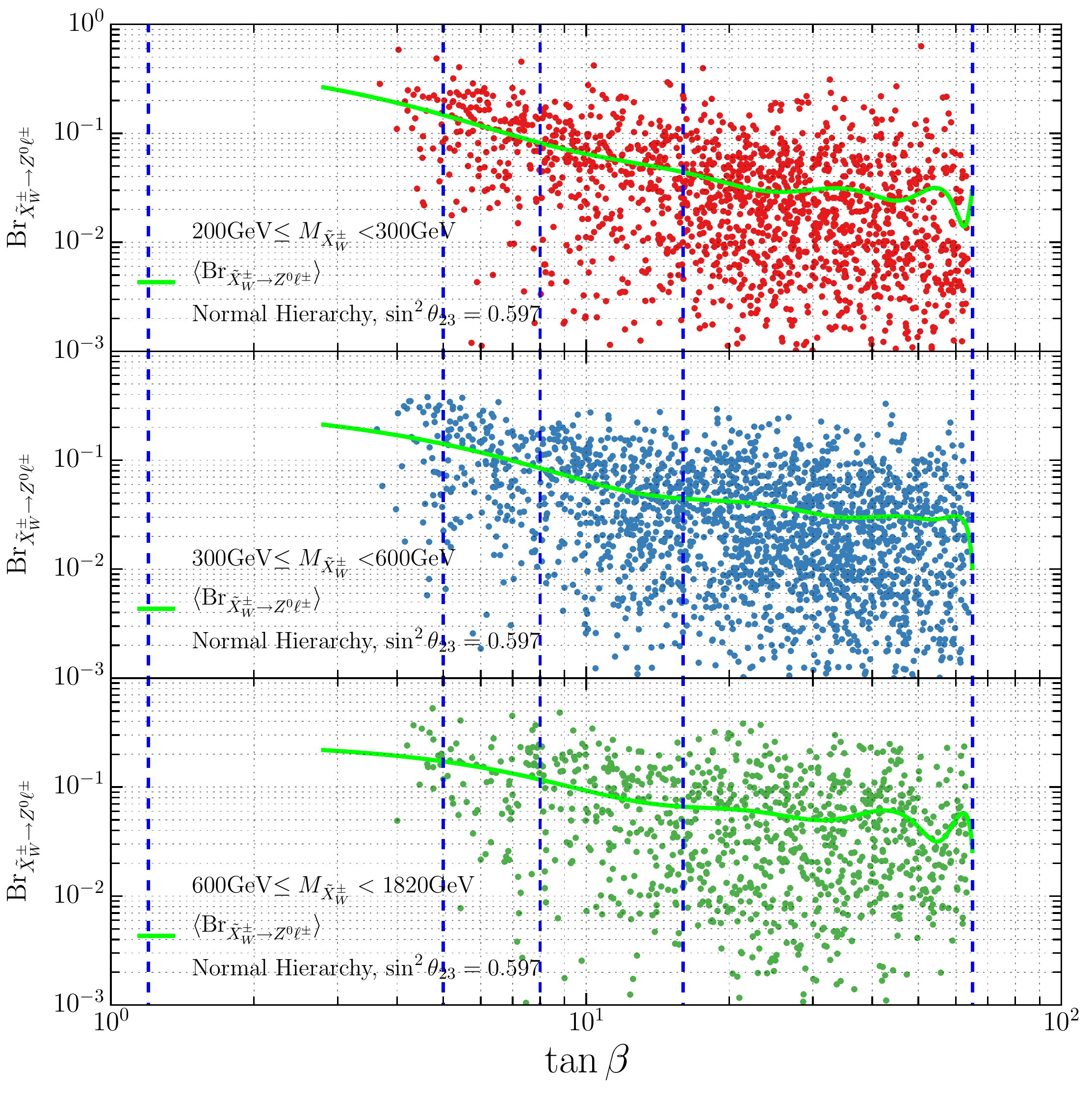}
\caption{A scatter plot of all 4,858 branching ratios, $\text{Br}_{{\tilde X}^\pm_W\rightarrow Z^0 \ell^\pm}$, associated with a Wino chargino LSP versus $\tan \beta$. The plot is broken up into the three  $M_{{\tilde X}_W^\pm}$ mass bins given in \eqref{bin1}. In each plot, the values of the branching fractions are highly scattered around the a green curve which represents the ``best fit'' to the data. The vertical blue lines mark the boundaries of the four regions where the behavior of the best fit lines are approximately identical.}\label{fig:ScatterGammaRatios}
\end{figure}

\noindent We now proceed to evaluate \eqref{eq:Branching1} for each of the valid black points associated with a Wino chargino LSP. Since there will be 4,858 different values of $\text{Br}_{{\tilde X}^\pm_W\rightarrow Z^0 \ell^\pm}$, we find it convenient to divide up this data into separate bins. Specifically, we will do the following. First, recall from Figure \ref{fig:mass_hist} that the physical mass of a Wino chargino is much more likely to be small, on the order of 200 GeV, and approximately $10^{-2}$ times less likely to be on the order of 1 TeV. This leads us to divide the Wino chargino LSP mass range into three bins given by
 \begin{equation}
 M_{{\tilde X}_W^\pm} \in [200, 300],\>[300,600],\> [600,1820]~ \text{GeV} \ .
 \label{bin1}
 \end{equation}

The range of each bin is chosen so that each contains approximately a third of the 4,858 valid black points. Second, as we will see below, the value of $\tan \beta$ plays a significant role in the relative sizes of the branching ratios of the three decay channels. With this in mind, we plot the values of $\text{Br}_{{\tilde X}^\pm_W\rightarrow Z^0 \ell^\pm}$ against $\tan \beta$ for each of the three mass bins in \eqref{bin1}. In each case, we present the ``best fit'' to the data as a green curve. We further partition each of these plots into bins-- represented by the vertical, dashed blue lines --where the best fit curves in each plot behave similarly. The results are presented in Figure \ref{fig:ScatterGammaRatios}. \noindent We see from these plots that the range of $\tan \beta$ is naturally broken into four regions approximately given by
\begin{equation}
 \tan \beta \in [1.2,5],\> [5,8],\>[8,16], \>[16,65] \ .
 \end{equation}

\begin{figure}[!ht]
\centering
\includegraphics[width=0.95\textwidth]{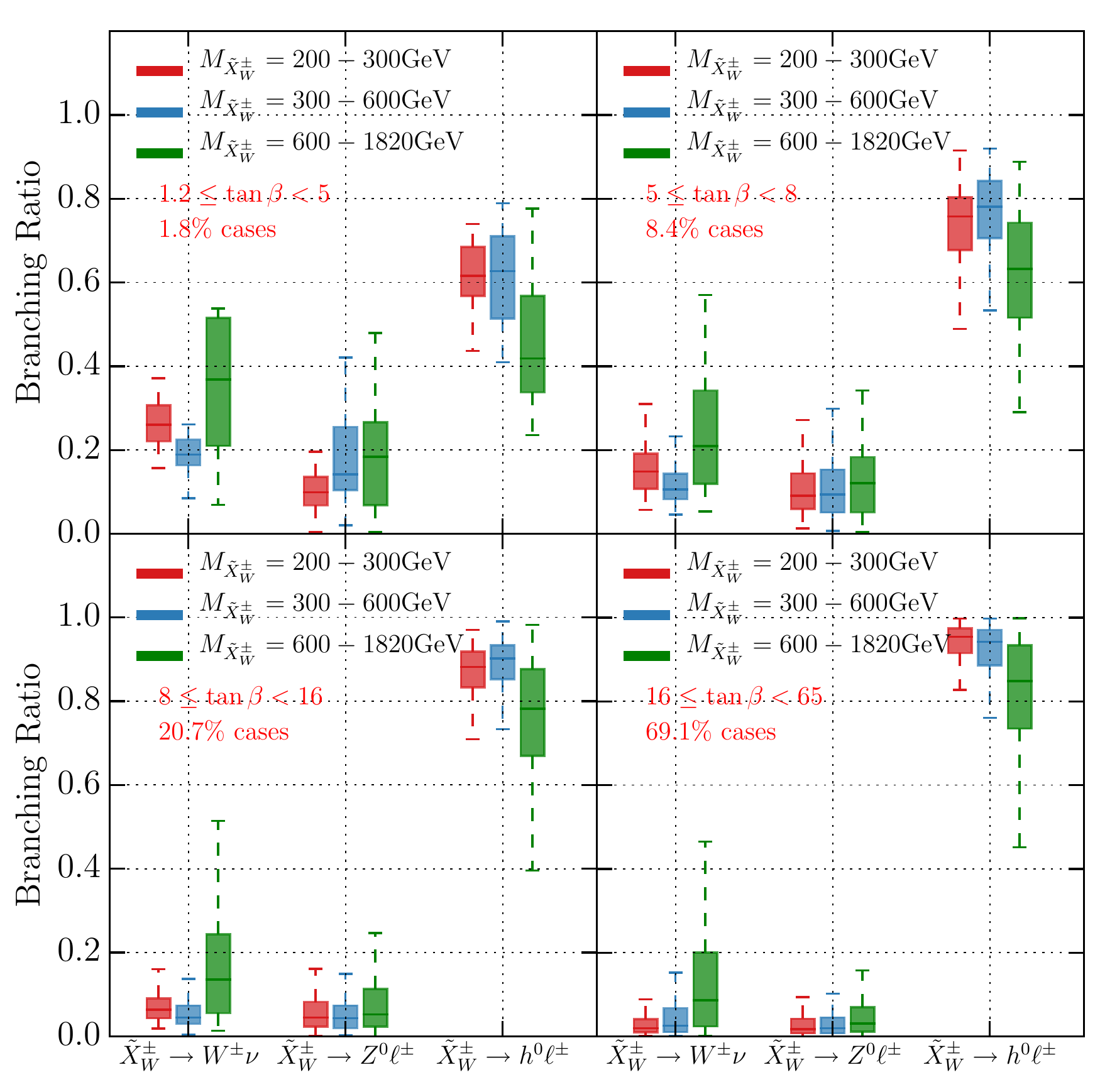}
    \caption{Branching ratios for the four possible decay channels of the Wino chargino LSP, presented for the three $M_{{\tilde X}_W^\pm}$ mass bins and four $\tan \beta$ regions. The colored horizontal line inside each box indicate the median value of the branching fraction in that bin, the colored box indicates the interquartile range in that bin, while the dashed error bars show the range between the maximum and the minimum values of the branching ratio for that bin. The case percentage indicate what percentage of the valid initial points have $\tan \beta$ values within the range indicated. For each channel, we sum over all three families of possible leptons.
Note that ${\tilde X}_W^\pm\rightarrow h^0 \ell^\pm$ is strongly favored-- except perhaps in the $1.2 < \tan\beta < 5$ bin. The calculations were performed assuming a normal neutrino hierarchy, with $\theta_{23}=0.597$.}
\label{BarPlotChargino.pdf}
\end{figure}

Having broken up the ranges of $M_{{\tilde X}_W^\pm}$ and $\tan \beta$ into 3 and 4 bins respectively, we now calculate the median, the interquartile range and the maximum and the minimum values\footnote{To make these terms explicit-- a) the ``median'' is the value of a quantity for which 50\% of that quantity have larger values and 50\% are smaller and b) the ``interquartile'' range is the interval of that quantity which contains 25\% of all values that lie above the median and 25\% that lie below it. 3) The meaning of the ``maximum'' and ``minimum'' values is self-evident} of the branching ratio $\text{Br}_{{\tilde X}^\pm_W\rightarrow Z^0 \ell^\pm} $ for the decay channel ${\tilde X}^\pm_W\rightarrow Z^0 \ell_i^\pm$ in each of the $3\times 4$ data bins. Using an identical procedure, one can compute the same quantities for the remaining two branching ratios ${ \rm Br}_{{\tilde X}^\pm_W\rightarrow W^\pm \nu} $ and $ {\rm Br}_{{\tilde X}^\pm_W\rightarrow h^0 \ell^\pm}$ as well. The results are displayed in Figure \ref{BarPlotChargino.pdf}. 

For example, for the valid black points with Wino chargino LSP mass between $200$ and $300$ GeV and with $\tan \beta$ between 8 and 16, Figure \ref{BarPlotChargino.pdf} can be interpreted in the following way: 
\begin{itemize}
\item{The branching ratios have median values of $0.064$ for the $\tilde X_W^\pm \rightarrow W^\pm \nu $ channel, 
$0.0445$ for the $\tilde X_W^\pm \rightarrow Z^0 \ell^\pm$ channel, and $0.882$  for the $\tilde X_W^\pm \rightarrow h^0 \ell^\pm $ channel. We therefore expect $h^{0}$ Higgs boson production via chargino RPV decays to dominate for these ranges of mass and $\tan \beta$.}

\item{After solving for the RPV couplings and the decay rates, we generally obtain different branching ratio values for different viable initial conditions in our simulation. The data is scattered around the ``best fit'' values, as shown in Figure \ref{fig:ScatterGammaRatios} . However, the branching ratios take values only within certain ranges, allowing for theoretical predictions for the decay patterns of the Wino Chargino LSPs.  The dashed error bars in Figure \ref{BarPlotChargino.pdf} indicate the full range of values that the branching ratios take. For example, in our chosen bin, the branching ratios for the $\tilde X_W^\pm \rightarrow Z^0 \ell^\pm $ channel  are not higher than approximately $0.18$ while they can be very close to $0$.  At the same time, the branching ratios for the $\tilde X_W^\pm \rightarrow h^0 \ell^\pm $ channel  are approximately between  $ 0.72$ and $0.95$. }
\item{The boxes show the interquartile ranges, within which $50\%$ of the points lie around the median value. In our chosen bin we learn, for example, that while the branching ratios for the $\tilde X_W^\pm \rightarrow Z^0 \ell^\pm $ channel can take any value between approximately $0.18$ and $0$, they tend to accumulate in the more restricted interval $0.03-0.08$.}

\end{itemize}

Generically, for all three mass ranges and all four $\tan\beta$ bins in Figure \ref{BarPlotChargino.pdf}, one can conclude the following. It is clear that the ${\tilde X}^\pm_W\rightarrow h^0 \ell^\pm$ channel is the most abundant and becomes increasingly so for higher values of $\tan \beta$.  
. For $\tan \beta<5$, the medians of the branching fractions for the most experimentally visible channel,
${\tilde X}^\pm_W\rightarrow Z^0 \ell^\pm$, lie between 0.20-0.65, depending on the mass bin. However, there are very few such cases in our simulation-only $1.8\%$. Much more likely is a scenario in which $\tan \beta$ is large. We find that $69.1\%$ of the total number of points have $\tan \beta>16$. For this parameter region, however, the branching fraction of ${\tilde X}^\pm_W\rightarrow Z^0 \ell^\pm$ drops between $0.05-0.18$, and the prospects of detecting it become slimmer.

The results in Figure \ref{BarPlotChargino.pdf} were calculated using numerical inputs into the complicated expressions for the decay rates given in Section \ref{sec:7}. Hence, the origin of the physical trends displayed in that Figure is obscure.
However, the formulas for the decay rates can, under certain assumptions, be simplified-- allowing for a physical interpretation for the observed relationships between the three decay channels. To do this, we note the following:

\begin{itemize}

\item{ As shown in Figure 2 above, the values for the angles $|\phi_{\pm}|$ are very small for each of the 4,858 black points associated with the Wino chargino; with $|\phi_{-}|$ being generically smaller than $|\phi_{+}|$. Hence, to a high degree of approximation, one can set $\cos \phi_{-}=1$ and $\sin \phi_{- }= 0$. However, due to the fact that the values for $|\phi_{+}|$, although very small, tend to be somewhat larger than $|\phi_{-}|$, we can only take $\cos \phi_{+}\approx 1$ and $\sin \phi_{+}\approx 0$.}

\item{ The lepton masses, $m_{\ell_i}$, are insignificant compared to the other masses in the expressions for the decay rates and, hence, the terms containing them can be neglected. Note that all occurences of the angle $\phi_{+}$ are contained in these terms. This facilitates our simplification even further, since the slightly larger values of $|\phi_{+}|$ no longer enter the approximation of the decay rates.}

\end{itemize}

Using these two approximations, we obtain simplified expressions for the decay rates of each of the three decay channels. They are

\begin{multline}\label{eq:decay_1}
\Gamma_{{\tilde X}_W^\pm\rightarrow W^\pm \nu_{i}} \approx \frac{g_2^4}{64\pi}
\Big(  \frac{ v_d}{\sqrt{2}M_2\mu}\epsilon_i-\frac{M_{BL} v_u}{M_1 v_R^2} \epsilon_j
 \left[V_{\text{PMNS}}\right]_{ji}  \Big)^2 \frac{M_{{\tilde X^\pm}_W}^3}{M_{W^\pm}^2}\left(1-\frac{M_{W^\pm}^2}{M_{{\tilde X}_W^\pm}^2}\right)^2
\left(1+2\frac{M_{W^\pm}^2}{M_{{\tilde X}_W^\pm}^2}\right) \ ,
\end{multline}

\begin{multline}\label{eq:decay_2}
\Gamma_{{\tilde X}_W^\pm\rightarrow Z^0 l_i^\pm}\approx\frac{g_{2}^4}{64\pi}
\Bigg( \frac{ \frac{1}{\sqrt 2} c_W(v_d\epsilon_i+\mu v_{L_i}^*)+
\frac{1}{c_W}\left( \frac{1}{2}-s_W^2 \right)v_d\epsilon_i}{M_2\mu} \Bigg)^2\times \\ 
\frac{M_{{\tilde X}_W^\pm}^3}{M_{Z^0}^2}\left(1-\frac{M_{Z^0}^2}{M_{{\tilde X}_W^\pm}^2}\right)^2
\left(1+2\frac{M_{Z^0}^2}{M_{{\tilde X}_W^\pm}^2}\right) \ ,
\end{multline}
\begin{equation}\label{eq:decay_4}
\Gamma_{{\tilde X}_W^\pm\rightarrow h^0 l_i^\pm}\approx\frac{g_2^2}{64\pi}\sin^2 \alpha
\Bigg(\frac{\epsilon_i}{2\mu} \Bigg)^2 M_{{\tilde X}_W^\pm}
\left(1-\frac{{M_{h^0}^2}}{M_{{\tilde X}_W^\pm}^2}\right)^2 \ .
\end{equation}

By examining \eqref{eq:decay_1}-\eqref{eq:decay_4}, we understand why the ${\tilde X}^\pm_W\rightarrow h^0 \ell^\pm$ channel dominates, being directly proportional to $\epsilon/\mu$, without the suppression $v_d/M_2$ that is present in the other decay channels for similar terms. However, the $v_d/M_2$ suppression becomes less pronounced for small $\tan \beta$ values, since $v_d=174~ \text{GeV}/(1+\tan \beta)$ increases.

\noindent Therefore, channels of interest such as ${\tilde X}^\pm_W\rightarrow Z^0 \ell^\pm$ become increasingly more significant towards smaller $\tan \beta$ values.
The Goldstone equivalence theorem tells us that the first two channels are amplified by the longitudinal degrees of freedom of the massive $Z_\mu^0$ and $W_\mu^\pm$ bosons, so the traces of these two decays become more apparent in scenarios with more massive LSP's.

\subsection*{Choice of neutrino data}

The neutrino mass hierarchy can be normal or inverted. Furthermore, for each of those possible hierarchies, two different values of the neutrino mixing angle $\theta_{23}$\footnote{As discussed above, each of the measured values of $\theta_{23}$ in both the normal and inverted hierarchies have small uncertainties around a central value-- which are incorporated into our computer code. However, for notational simplicity, we will ignore these uncertainties and write the central values only.} fit the existing data. See \cite{PDG,Capozzi:2018ubv}. For the normal hierarchy, the angle $\theta_{23}$ can be 0.597 or 0.417, while for the inverted one, $\theta_{23}$ can be 0.529 or 0.421. So far, out of the four possibilities, we have chosen a normal neutrino hierarchy with  $\sin \theta_{23}=0.597$ to compute the branching ratios-- each summed over all three families of leptons --and their relative properties for each decay channel.
The results were shown in Figures \ref{fig:ScatterGammaRatios} and \ref{BarPlotChargino.pdf}. Can choosing the other neutrino hierarchy and/or different values of $\theta_{23}$ 
modify those predictions? To explore this question, we begin by repeating the calculations for the inverted hierarchy with $\theta_{23}=0.529$. We find that the new median values of the branching ratios change, but are never outside the interquartile ranges displayed in Figure \ref{BarPlotChargino.pdf}. 
Furthermore, we find that switching between the two possible values of the angle $\theta_{23}$ while keeping the hierarchy the same has no impact on the results-- for either the normal or the inverted hierarchy. 

These results are best illustrated by plotting the branching ratios (summed over all mass and $\tan \beta$ bins) for each decay channel against the other two channels-- and doing this for each of the four choices of neutrino input data. Each such plot is simplified by using the fact 
\begin{equation}
\text{Br}_{{\tilde X}^\pm_W\rightarrow W^\pm \nu}+\text{Br}_{{\tilde X}^\pm_W\rightarrow Z^0 \ell^\pm}+\text{Br}_{{\tilde X}^\pm_W\rightarrow h^0 \ell^\pm}=1 \ .
\label{red1}
\end{equation}

\begin{figure}[!ht]
   \centering

   \begin{subfigure}[c]{0.495\textwidth}
\includegraphics[width=1.0\textwidth]{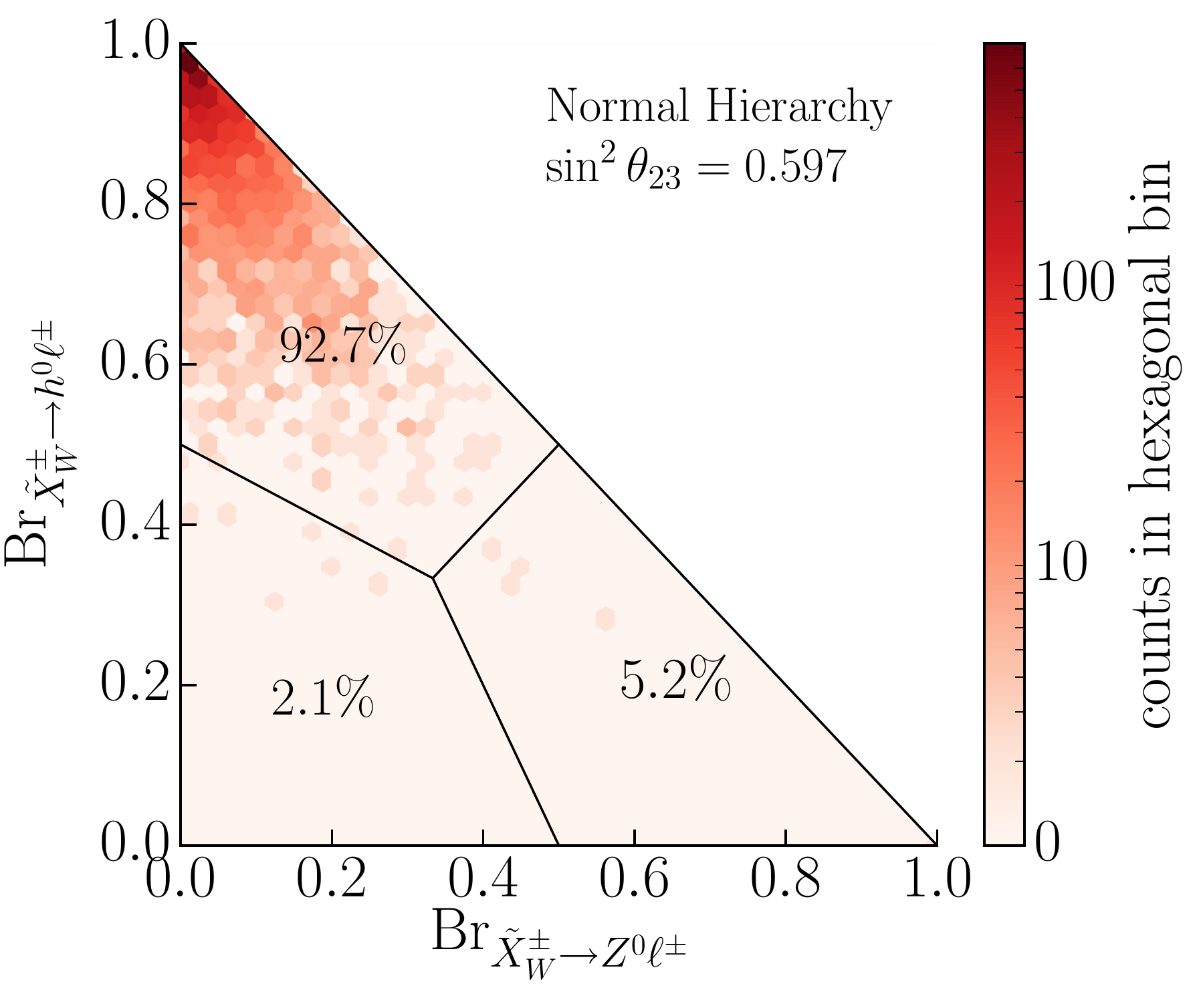}
\label{fig:ScatterHZ_normal}
\end{subfigure}
   \begin{subfigure}[c]{0.495\textwidth}
\includegraphics[width=1.0\textwidth]{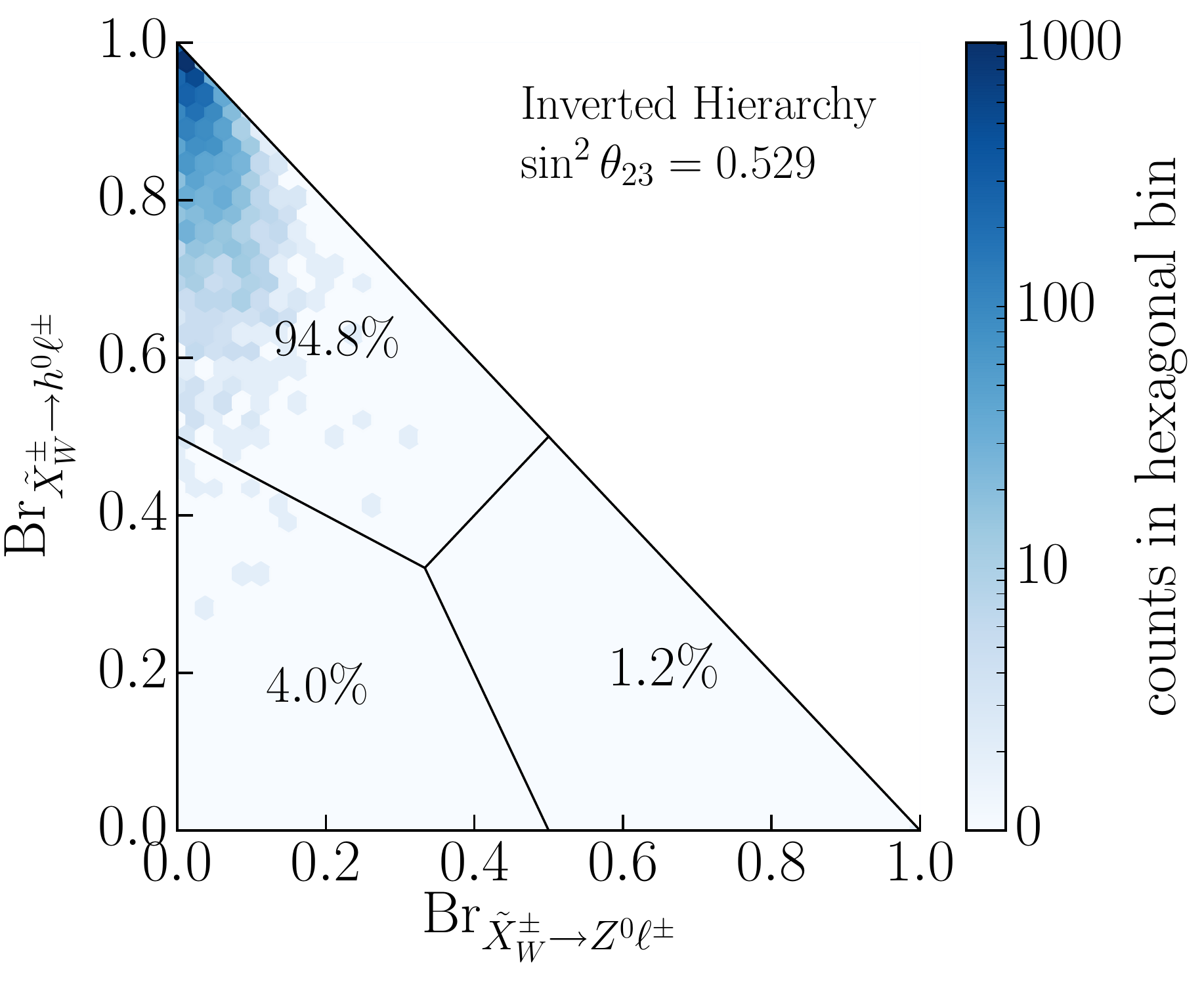}
\label{fig:ScatterHZ_normal}
\end{subfigure}
   \begin{subfigure}[c]{0.495\textwidth}
\includegraphics[width=1.0\textwidth]{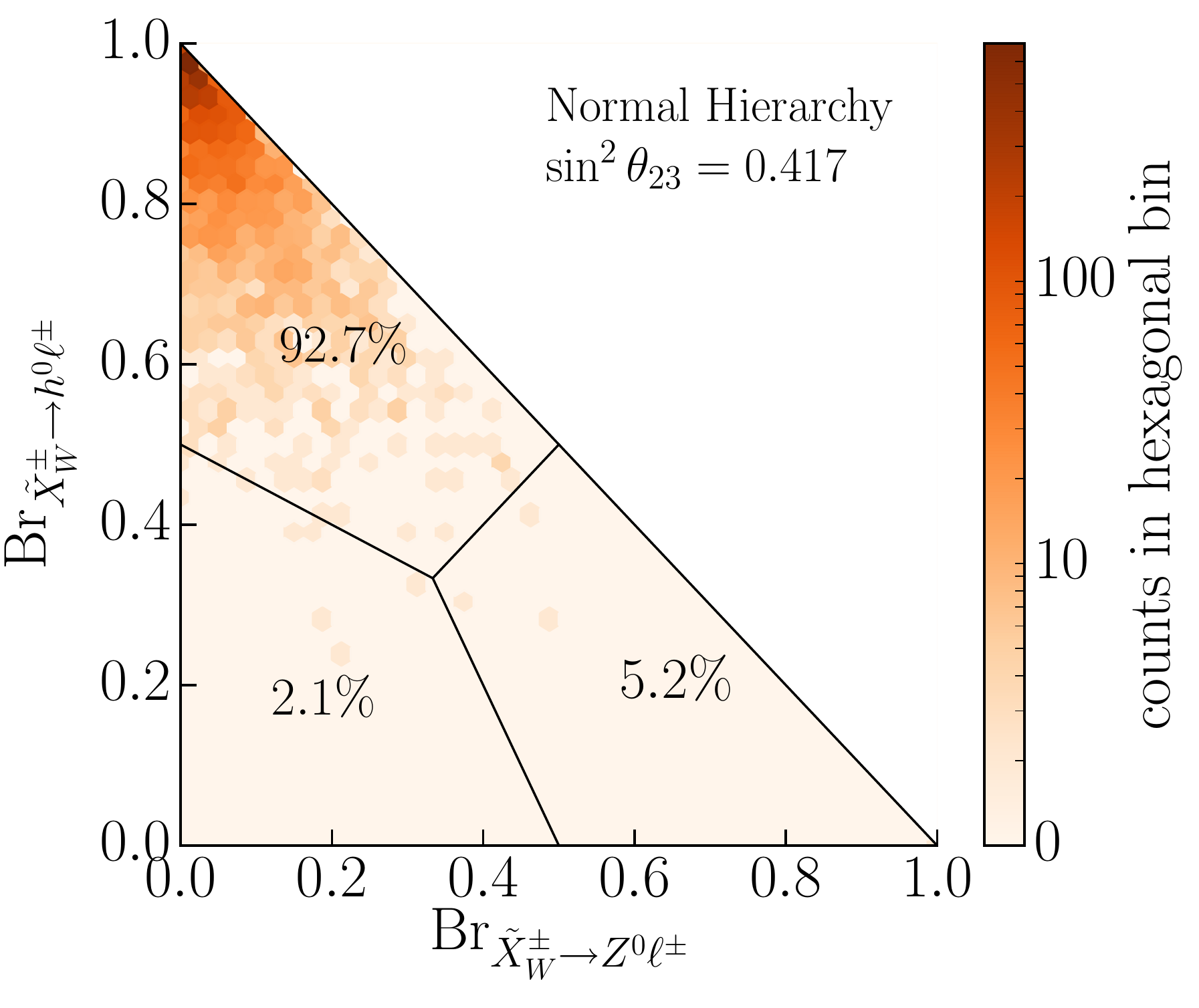}
\label{fig:ScatterHZ_normal}
\end{subfigure}
   \begin{subfigure}[c]{0.495\textwidth}
\includegraphics[width=1.0\textwidth]{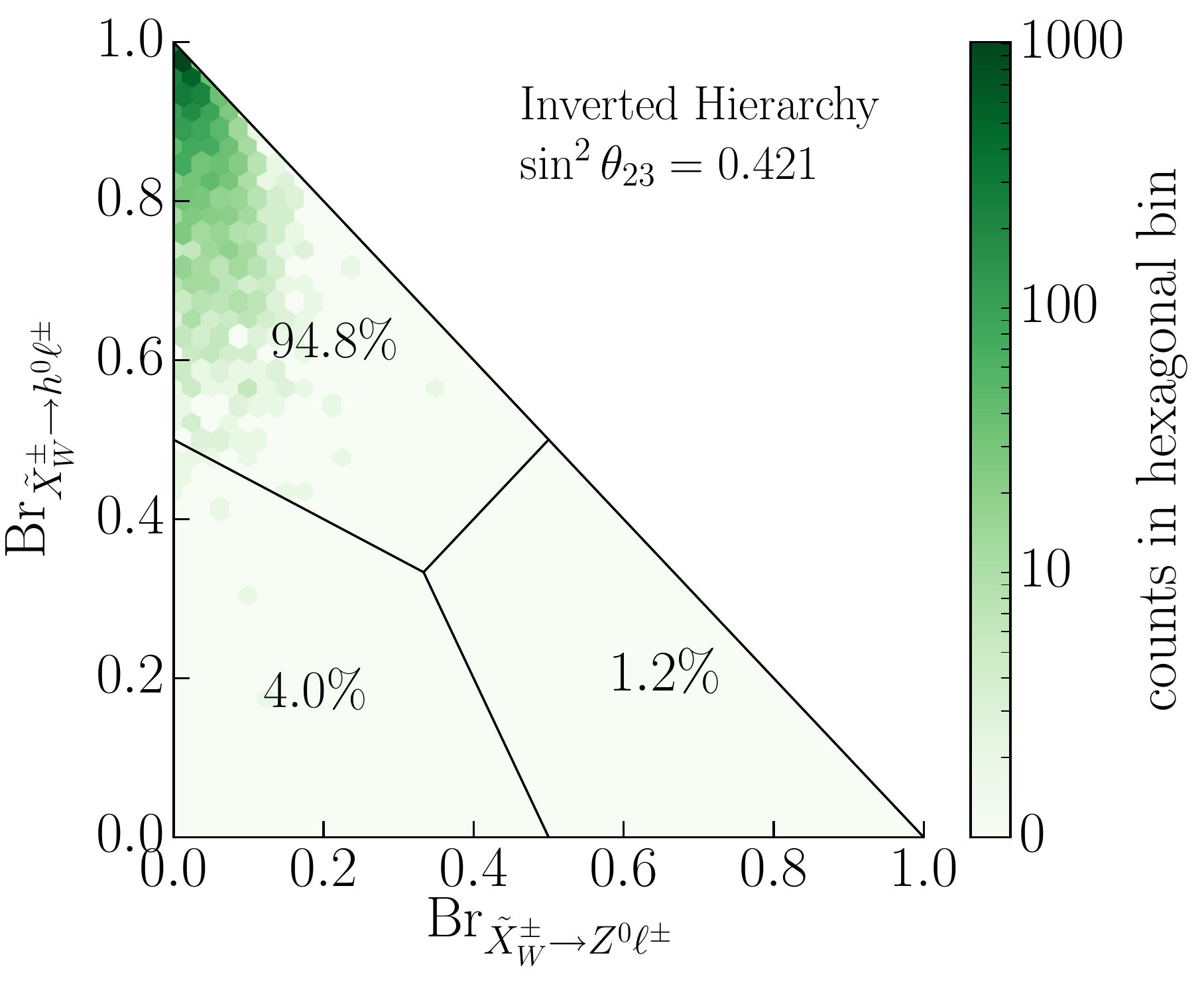}
\label{fig:ScatterHZ_normal}
\end{subfigure}

\caption{ Branching ratio to $h^0\ell^\pm$ versus branching ratio to $Z^0\ell^\pm$ for Wino chargino LSP decays, for both normal and inverted hierarchy. Wino chargino LSP decays via the  ${\tilde X}^\pm_W\rightarrow Z^0 \ell^\pm$ channel tend to be more abundant for a normal hierarchy. The choice of the angle $\theta_{23}$ has no impact on the statistics of these decays, for any of the two possible hierarchies. The percentages indicate what proportion of the points is contained within each third of the four plots.}
\label{fig:ScatterHZ}

\end{figure}

We have demonstrated \eqref{red1} explicitly for the normal hierarchy with $\theta_{23}=0.597$, and have numerically shown that it remains true for the other three neutrino input possibilities.

 It follows that $\text{Br}_{{\tilde X}^\pm_W\rightarrow W^\pm \nu}$ can be determined, using \eqref{red1}, from the remaining two decay channels. Hence, one can plot 2D histograms associated with all 4,858 valid black points associated with a Wino chargino LSP for each of the four possible neutrino input scenarios. These are presented in Figure \ref{fig:ScatterHZ}.

The most obvious fact that one learns by comparing the top and bottom plots for each individual neutrino hierarchy in Figure \ref{fig:ScatterHZ} is that the $\theta_{23}$ angles play no role in determining the branching ratios-- as stated above. The reason for this is the following. First, note that the simplified expressions \eqref{eq:decay_1} - \eqref{eq:decay_4}, although originally presented for the normal hierarchy with $\theta_{23}=0.597$, remain valid for the other three sets of neutrino data as well. When we sum over the three lepton families in these expressions, the decay rates for each individual channel are proportional to the squared amplitudes of the RPV couplings. Changing the value of the $\theta_{23}$ angle results in a different unitary $V_{\text{PMNS}}$ matrix, which rotates the $\epsilon_i$ and $v_{L_i}, \>i=1,2,3$ components differently, but does not change the squared amplitudes of these couplings to produce a statistically observable effect. For this reason, switching between different $\theta_{23}$ values inside any hierarchy doesn't result in different data patterns, as clearly shown in Figure \ref{fig:ScatterHZ}. This is why using only one value of the angle ( for example $\theta_{23}=0.597$ for the normal hierarchy and $\theta_{23}=0.529$ for the inverted hierarchy) is sufficient to make experimental predictions. Note, however, that if one does {\it not} sum over the three lepton families, this argument is no longer valid, and the value of $\theta_{23}$ can play a substantial role.

The second fact that one learns from comparing the left-hand and right-hand plots of Figure \ref{fig:ScatterHZ} is that there {\it is} a difference in the distribution of branching ratios between the normal and the inverted neutrino hierarchies. This is because, in our theory, the three generations of left handed neutrinos have Majorana masses, directly proportional to the squared amplitudes of these RPV couplings. In the normal hierarchy
\begin{equation}
	m_1 = 0, \quad m_2 = (8.68 \pm 0.10) \times 10^{-3} ~{\rm eV},\quad m_3 = (50.84 \pm 0.50) \times 10^{-3}~{\rm eV}
\label{b2}
\end{equation}
while in the inverted one
\begin{equation}
	m_1 = (49.84 \pm 0.40) \times 10^{-3} ~{\rm eV}, \quad m_2 = (50.01 \pm 0.40) \times 10^{-3} ~{\rm eV},\quad m_3 = 0.
\label{b3}
\end{equation} 
We expect, therefore, that the amplitudes of the couplings will change with the choice of neutrino hierarchy--  leading to the differences in the branching ratios that we observe in Figure \ref{fig:ScatterHZ}. Note, however, from the distribution of points-- plotted as percentages --in the subsections of each plot, that the difference in branching ratios between the normal and inverted hierarchies is relatively small, on the order of a few percent. This is consistent with our statement above that the ``new median values of the branching ratios (for the inverted hierarchy) change, but are never outside the interquartile ranges displayed in Figure 6 (the normal neutrino hierarchy).''
Moreover, in the next section we show that the chargino decay lengths are generally smaller when we assume the inverted hierarchy, compared to when we assume a normal one.

Finally, from Figure \ref{fig:ScatterHZ} we learn that the Wino chargino decays via the  ${\tilde X}^\pm_W\rightarrow h^0 \ell^\pm$ channel tend to be slightly more abundant for a normal hierarchy. However, the incremental difference is relatively small, since the bulk of the points lie in the top left corner, where the decay to $h^0\ell^\pm$ dominates. Although the effect is too small to be statistically distinguishable, it is of interest to note how the choice of neutrino hierarchy can have small influence over the decay rates.

\subsection*{Decay Length}

There is one more issue to be discussed; that is, are the decays of the Wino chargino ``prompt''-- defined to be decays where the {\it overall} decay length $L$, defined in \eqref{alan1},  satisfies  $L<1$mm? The key to this problem lies in the magnitudes of the RPV parameters, $\epsilon_i$ and $v_{L_i}$. We find that for prompt decays, at least one of the couplings $\epsilon_i$ needs to be larger than $10^{-4}$ GeV. The overall scale of neutrino masses guarantees that this is well satisfied. Putting the lower limit of this interval any lower would not change our results significantly. The upper limit of this interval eliminates the problem of unphysical finely tuned cancellations in the neutrino mass matrices. 
\begin{figure}[t]
\begin{subfigure}[t]{0.495\textwidth}
\includegraphics[width=1.\textwidth]{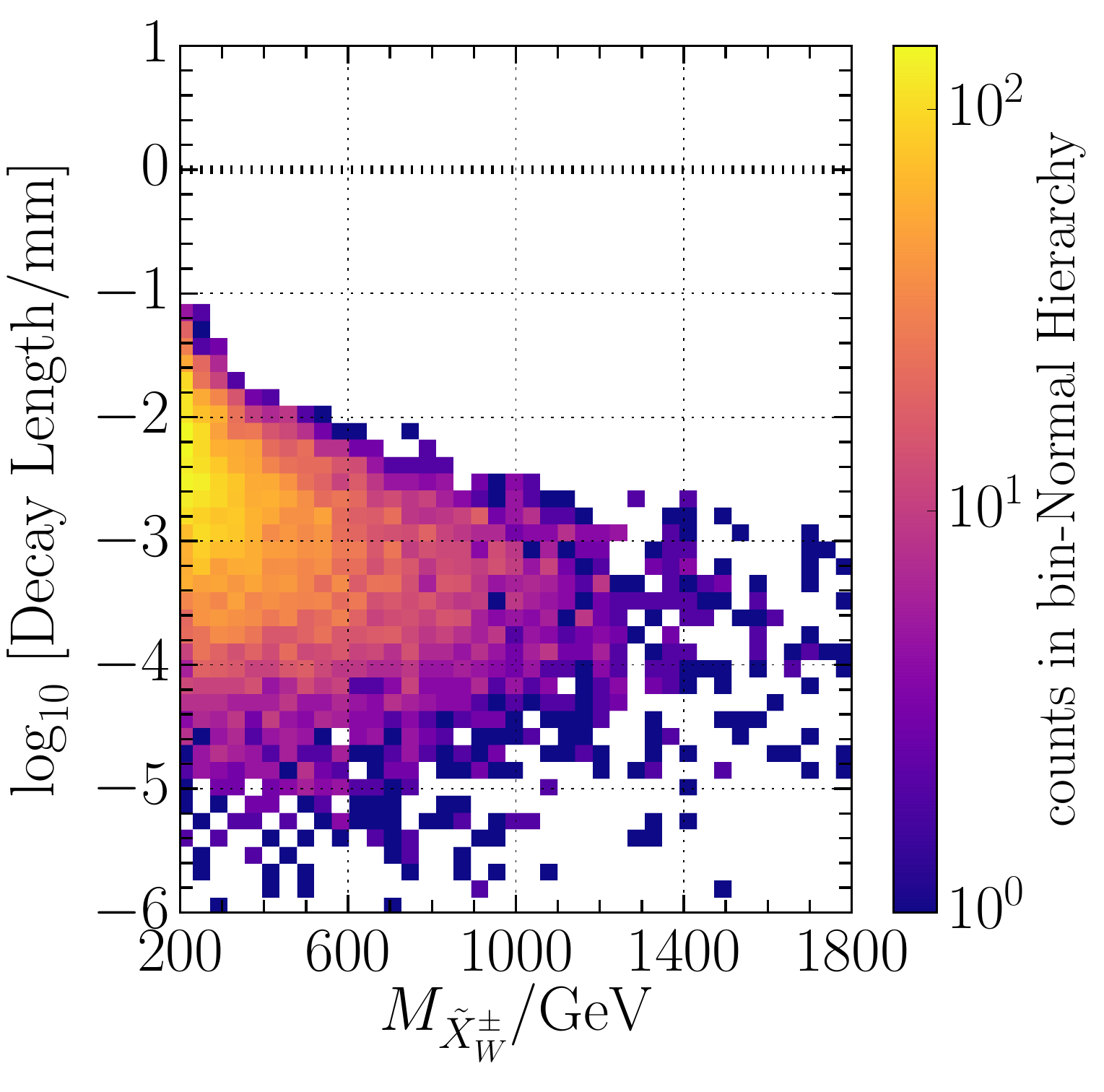}
\end{subfigure}
\begin{subfigure}[b]{0.495\textwidth}
\includegraphics[width=1.\textwidth]{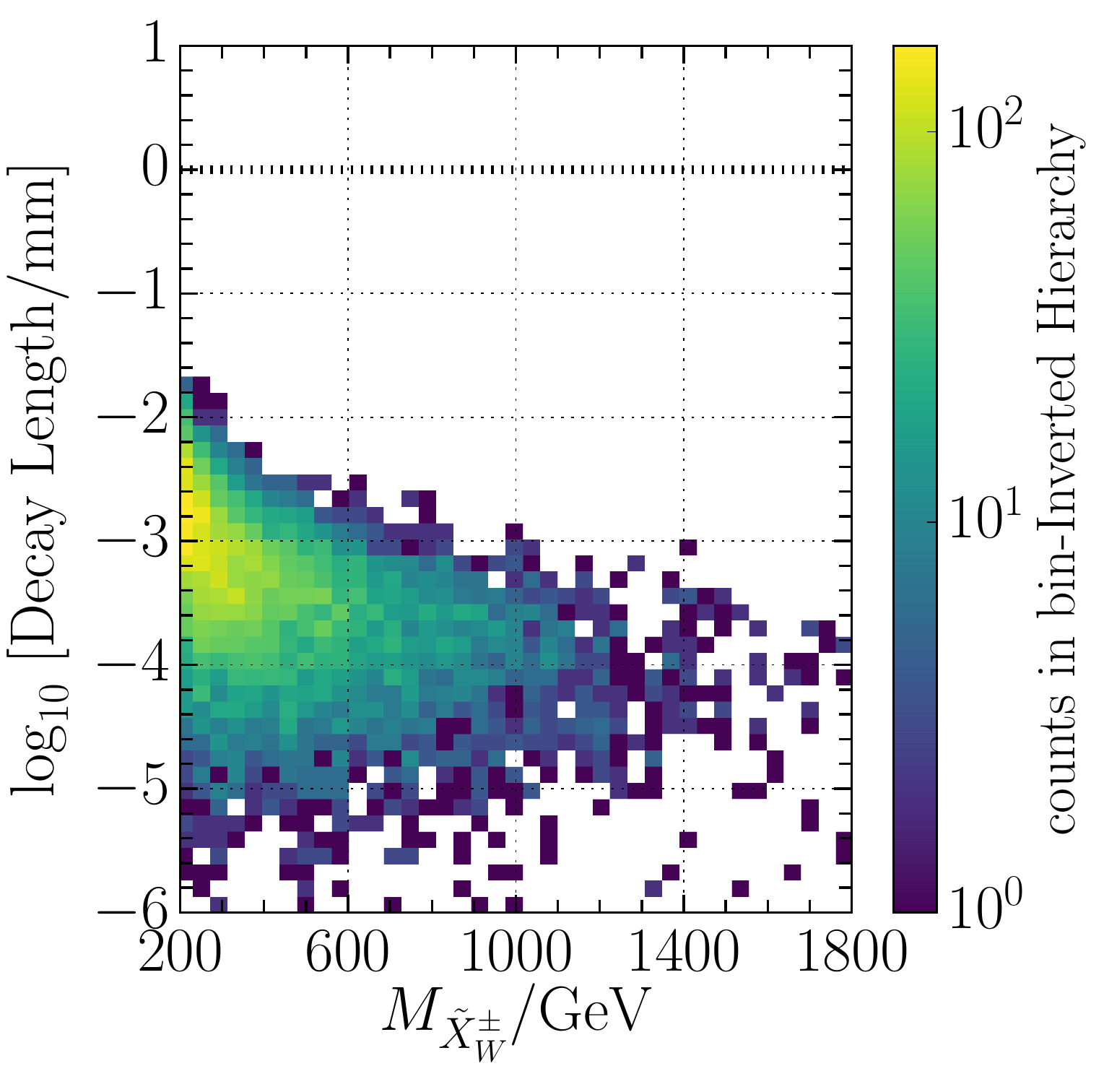}
\end{subfigure}
\caption{Wino Chargino LSP decay length in millimeters, for the normal and inverted hierarchies, summing over all three decay channels. The average decay length $L=c\times \frac{1}{\Gamma}$ decreases for larger values of $M_{\tilde X^\pm_W}$, since the decay rates are amplified because of the longitudinal degrees of freedom of the massive bosons produced. We have chosen $\theta_{23}=0.597$ for the normal neutrino hierarchy and $\theta_{23}=0.529$ for the inverted hierarchy. However, the choice of $\theta_{23}$ has no impact on the decay length.}
\label{fig:LSPprompt1}
\end{figure}
In Figure \ref{fig:LSPprompt1}, we present two scatter plots-- one for the normal and one for the inverted neutrino hierarchy --of the decay length 
\begin{equation}
L=c\times \frac{1}{\Gamma} \ , \quad  \Gamma= \sum_{i=1}^{3} \Big( \Gamma_{{\tilde {X}}_W^\pm\rightarrow W^\pm \nu_{i}} + \Gamma_{{\tilde {X}}^\pm_W \rightarrow Z^0 \ell^\pm_{i}}  + \Gamma_{{\tilde X}^\pm_W\rightarrow h^0 \ell^\pm_{i}} \Big) 
\label{alan1}
\end{equation}
against the Wino chargino LSP mass for all of the 4,858 valid black points with a Wino chargino LSP. The parameter $c$ is the speed of light. Since the {\it overall} decay rate involves a sum over the three lepton families, it follows from the results of the previous section that the value of $\theta_{23}$ plays no role for either hierarchy. We find that the viable Wino chargino LSPs in our simulation decay promptly and produce prompt vertices in the detector for both neutrino hierarchies. 
\noindent However, we note that the decay lengths tend to be slightly smaller in the case of the inverted hierarchy.
\noindent This follows from the fact that the masses of the neutrinos are, overall, slightly larger in the inverted case. Hence, the RPV couplings will be somewhat larger as well-- resulting in a tiny increase in the decay rates and, therefore, smaller decay lengths in the inverted hierarchy. 

  \begin{figure}[t]
\centering
\begin{subfigure}[b]{0.93\textwidth}
\includegraphics[width=1.0\textwidth]{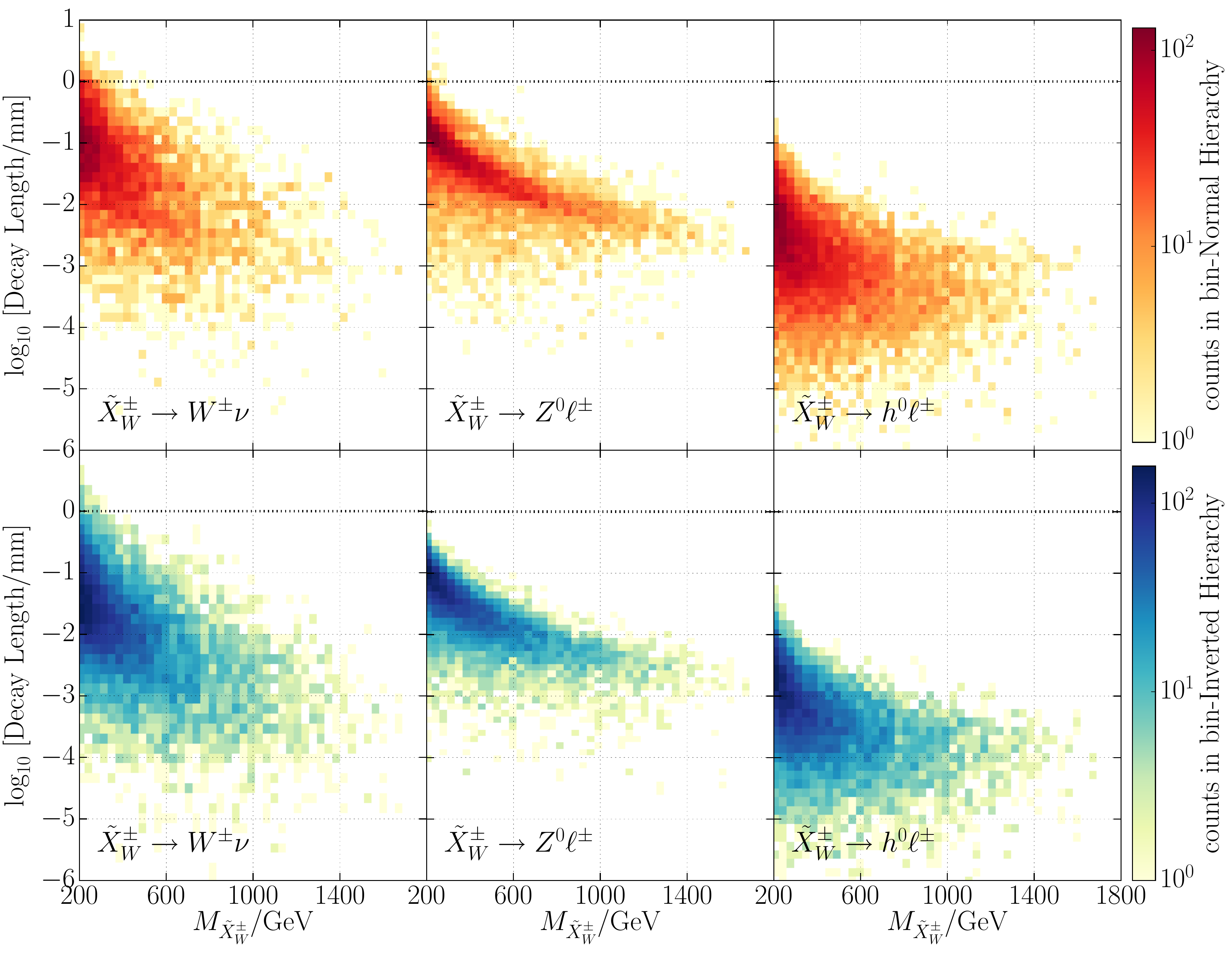}
\end{subfigure}
\caption{Wino Chargino LSP decay length in milimeters, for individual decay channels, for both normal and inverted hierarchies.  We have chosen $\theta_{23}=0.597$ for the normal neutrino hierarchy and $\theta_{23}=0.529$ for the inverted hierarchy. The choice of $\theta_{23}$ has no impact on the decay lengths. All individual channels have decay lengths $<1$mm }
\label{fig:LSPprompt3}
\end{figure}

Although Figure \ref{fig:LSPprompt1} shows that Wino chargino LSPs decay promptly for all viable initial points, their decay rates are strongly dominated by the ${\tilde X}^\pm_W\rightarrow h^0 \ell^\pm$ channel in general. Recall that the notion of ``prompt'' used above involved a sum over all three separate channels. This stimulates us to study the ``promptness'' of each individual decay channel independently-- although we continue to sum over the three lepton families. 

For example, the decay length of ${\tilde X}^\pm_W\rightarrow Z^0 \ell^\pm$  is given by
\begin{equation}
L_{{\tilde X}^\pm_W\rightarrow Z^0 \ell^\pm}=c\times \frac{1}{\sum_{i=1}^{3}\Gamma_{{\tilde X}^\pm_W\rightarrow Z^0 \ell^\pm_{i}}}.
\end{equation}

 In Figure \ref{fig:LSPprompt3} we show that the Wino chargino LSP has decay lengths smaller than 1mm when decaying via any of the channels ${\tilde X}^\pm_W\rightarrow Z^0 \ell^\pm$, ${\tilde X}^\pm_W\rightarrow h^0 \ell^\pm$ and ${\tilde X}^\pm_W\rightarrow W^\pm\nu$.

\subsection*{Lepton family production}

As discussed above, for any one of the three generic decay channels, the branching ratio for the decay into an single lepton family can, in principal, depend on the choice of the neutrino hierarchy and the value of $\theta_{23}$ used in determining the values of the $\epsilon_{i}$ and $v_{L_{i}}$ parameters. Using the available neutrino data with $3\sigma$ errors for the neutrino masses, along with the $V_{\text{PMNS}}$ rotation matrix angles and CP violating phases for one can calculate, for any valid black point associated with a Wino chargino LSP, the decay rate into each individual lepton family for a given  decay channel. Clearly, the value of the decay rate will depend explicitly on the choice of neutrino hierarchy-- either normal or inverted --and, for a given hierarchy, on the choice of the two allowed values of $\theta_{23}$. 
For example, to quantify the probability to observe an electron $e^\pm$ in the generic decay process ${\tilde X}^\pm_W\rightarrow Z^0 \ell^\pm$, we compute

 \begin{figure}[!ht]
   \centering

   \begin{subfigure}[b]{0.49\textwidth}
\includegraphics[width=1.0\textwidth]{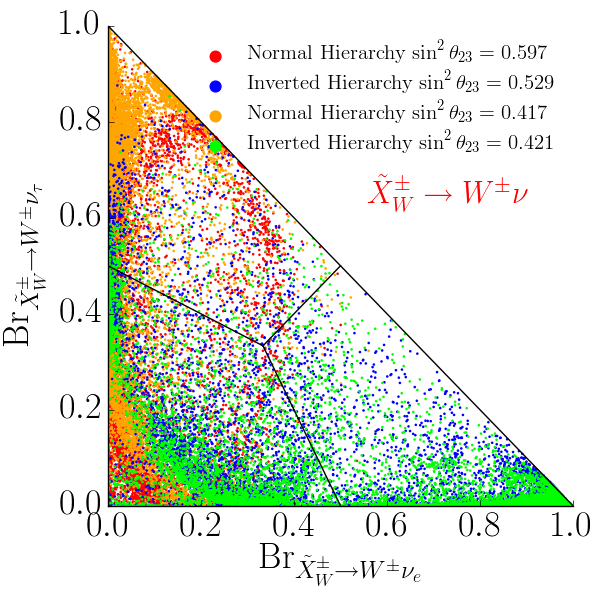}
\end{subfigure}
   \begin{subfigure}[b]{0.49\textwidth}
\includegraphics[width=1.0\textwidth]{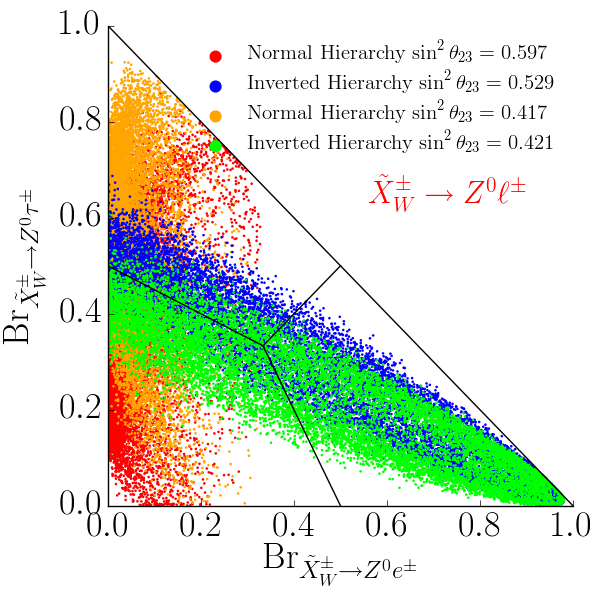}
\end{subfigure}\\
   \begin{subfigure}[b]{0.49\textwidth}
\includegraphics[width=1.0\textwidth]{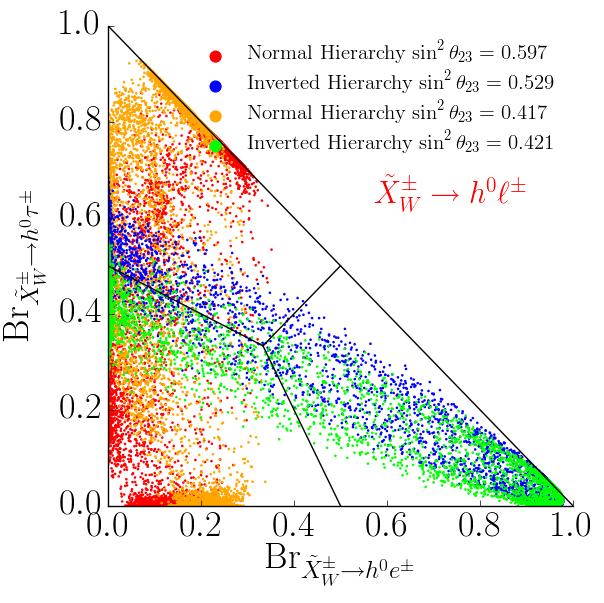}
\end{subfigure}
    \caption{Branching ratios into the three lepton families,
for each of the three main decay channels of a Wino chargino LSP. The associated neutrino hierarchy and the value of $\theta_{23}$ is specified by the color of the associated data point.}\label{fig:lepton_branching}
\end{figure}

\begin{equation}
\text{Br}_{{\tilde X}^\pm_W\rightarrow Z^0 e^\pm}=\frac{\Gamma_{{\tilde X}^\pm_W\rightarrow Z^0 e^\pm}}{\Gamma_{{\tilde X}^\pm_W\rightarrow Z^0 e^\pm}+\Gamma_{{\tilde X}^\pm_W\rightarrow Z^0 \mu^\pm} + \Gamma_{{\tilde X}^\pm_W\rightarrow Z^0 \tau^\pm} } \ ,
\end{equation}
and similarly for a muon, $\mu^{\pm}$, and a tauon, $\tau^{\pm}$, final state. Using this result, we proceed to quantify the branching ratios for each of the 3 decay processes ${\tilde X}^\pm_W\rightarrow W^\pm \nu_i$, ${\tilde X}^\pm_W\rightarrow Z^0 \ell_i^\pm$ and ${\tilde X}^\pm_W\rightarrow h^0 \ell_i^\pm$ into their individual lepton families. 

The results are shown in Figure \ref{fig:lepton_branching}.
Each subgraph in Figure \ref{fig:lepton_branching} has the following characteristics.
For a point near the top left corner of each subgraph, the branching ratio into a third family lepton is the largest, whereas for a point near 
the bottom right corner, the branching ratio into a first family lepton is the largest. Finally, using the fact that  
\begin{equation}
{\rm Br}_{{\tilde X}^\pm_W\rightarrow Z^0 e^\pm}+{\rm Br}_{{\tilde X}^\pm_W\rightarrow Z^0 \mu^\pm} + {\rm Br}_{{\tilde X}^\pm_W\rightarrow Z^0 \tau^\pm} = 1 \ ,
\label{cup1}
\end{equation}
it follows that for a point near the the bottom left corner, the branching ratio into a second family lepton is the largest.

Perhaps the most striking feature of each such graph is the connection between the Wino chargino decays, the neutrino hierarchy and the $\theta_{23}$ angle. Should experimental observation measure these branching ratios with sufficient precision, that could help shed light on the neutrino hierarchy and the value of $\theta_{23}$. For each neutrino hierarchy, there are two sets of points of different color, since the present experimental data allows for two values of $\theta_{23}$.

For example, let us consider the subgraph associated with the ${\tilde X}^\pm_W\rightarrow Z^0 \ell^\pm$ decay channels. If experimental observation finds that electrons are predominant after the Wino chargino LSP decays, then the hierarchy is inverted. Depending on whether the experimental result is a green or a blue point, implies that $\theta_{23}$ will be $0.421$ or $0.529$ respectively. However, if the branching ratios to either the second or third family leptons are highly dominant, then the hierarchy will be normal, with $\theta_{23}$ given, most likely, by $0.597$ and $0.417$ respectively. That is, with sufficiently precise measured branching ratios one could determine the type of neutrino hierarchy and the value of the $\theta_{23}$ mixing angle from the color of the associated data point.

\subsection{Wino Neutralino LSP Decay Analysis}

 In this section, we analyze the RPV decay signatures of the Wino neutralino LSPs. Written in 4-component spinor notation, the Wino neutralino Weyl spinor, $\tilde \chi_W^0$, becomes
\begin{equation}
\tilde X^0_W=
\left(
\begin{matrix}
\tilde \chi_W^0\\
\tilde \chi_W^{0 \dag}
\end{matrix}
\right) \ ,
\end{equation}
which is a Majorana spinor. In Section \ref{sec:7}, we computed the RPV decay rates  for all neutralino mass eigenstates. The Wino neutralino corresponds to the case where $n=2$.
Unlike the Wino chargino, the Wino neutralino has only three possible decay channels, reproduced here in Figure \ref{fig:NeutralinoDecays}.\\

\begin{figure}[t]
 \begin{minipage}{1.0\textwidth}
     \centering
   \begin{subfigure}[b]{0.245\linewidth}
   \centering
       \includegraphics[width=1.0\textwidth]{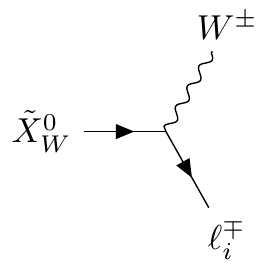} 
\caption*{${\tilde X}^0_W\rightarrow W^\pm \ell^\mp_{i}$}
       \label{fig:table2}
   \end{subfigure} 
   \hfill
   \centering
   \begin{subfigure}[b]{0.245\linewidth}
   \centering
      \includegraphics[width=1.0\textwidth]{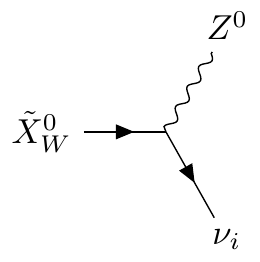} 
\caption*{ ${\tilde X}^0_W\rightarrow Z^0 \nu_{i}$}
       \label{fig:table2}
\end{subfigure}
\hfill
   \centering
     \begin{subfigure}[b]{0.245\textwidth}
   \centering
       \includegraphics[width=1.0\textwidth]{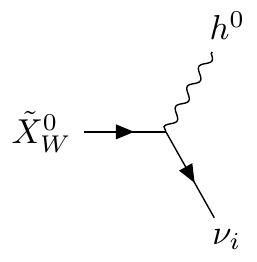} 
\caption*{ ${\tilde X}^0_W\rightarrow h^0 \nu_{i}$}
\end{subfigure}
\end{minipage}
\caption{RPV decays of a general massive Wino neutralino $X_W^0$. There are three possible channels, each with $i=1,2,3$, that allow for Wino neutralino LSP decays. The decay rates into each individual channel were calculated analytically in Section \ref{sec:7}.} \label{fig:NeutralinoDecays}
\end{figure}

\subsection*{Branching ratios of the decay channels}

The  ${\tilde X}^0_W\rightarrow W^\pm \ell^\mp_{i}$
processes is the most favored for detection at the LHC. Similarly to the Wino chargino decay products, the left handed neutrinos produced during ${\tilde X}^0_W\rightarrow Z^0 \nu_i$ decays can only be detected as missing energy, while the Higgs boson $h^0$ arising from ${\tilde X}^0_W\rightarrow h^0 \nu_i$ couples to both quarks and leptons, leading to decay remnants in the detector that are harder to interpret. Hence, the most interesting decay experimentally appears to be the Wino neutralino decay into a $W^\pm$ massive boson and a charged lepton. The abundance of each channel is proportional to its branching ratio. For example, for the process ${\tilde X}^0_W\rightarrow W^\pm \ell^\mp$ the branching ratio is defined to be
\begin{equation}
\text{Br}_{{\tilde X}^0_W\rightarrow W^\pm \ell^\mp}=\frac{\sum_{i=1}^{3}  \Gamma_{ {\tilde X}^0_W\rightarrow W^\pm \ell^\mp_{i}}}{\sum_{i=1}^3 \Big( \Gamma_{{\tilde X}^0_W\rightarrow Z^0 \nu_i}+  \Gamma_{{\tilde X}^0_W\rightarrow W^\pm \ell^\mp_i}+\Gamma_{{\tilde X}^0_W\rightarrow h^0 \nu_i}\Big)}  \ .
\end{equation}

We now study the decay patterns and branching ratios for each for the 3 decay channels of the Wino neutralino. There are 4,869 valid black points associated with Wino neutralino LSPs. For each of these, we compute the decay rates via RPV processes, using the expressions from Section \ref{sec:7} with $n=2$. The branching ratios of the main channels take different values for different valid points in our simulation. These values are scattered around the median values of these quantities. We compute the median values, interquartile ranges and the minimum and maximum values of the branching fractions in the same ``bins'' of the parameter space as we used in the study of the Wino chargino LSP decay channels.

 That is,
we sample the average branching fractions in the three bins for the LSP mass $M_{{\tilde X}_W^0} \in [200, 300],\>[300,600],\>[600,1734]\footnote{Note that the highest mass for a Wino neutralino is somewhat smaller than that for a Wino chargino.}$ GeV and in the four intervals for $\tan \beta \in [1.2,5],\> [5,8],\>[8,16], \>[16,65]$. The results are presented in Figure \ref{fig:bar_plot2}. To carry out the explicit calculations, we have chosen a normal neutrino hierarchy with $\theta_{23}=0.597$. We again find that assuming an inverted neutrino hierarchy changes these results only slightly, while the exact value of $\theta_{23}$ is statistically irrelevant.

\begin{figure}[!ht]
   \centering
   \begin{subfigure}[b]{1.\textwidth}
\includegraphics[width=1.0\textwidth]{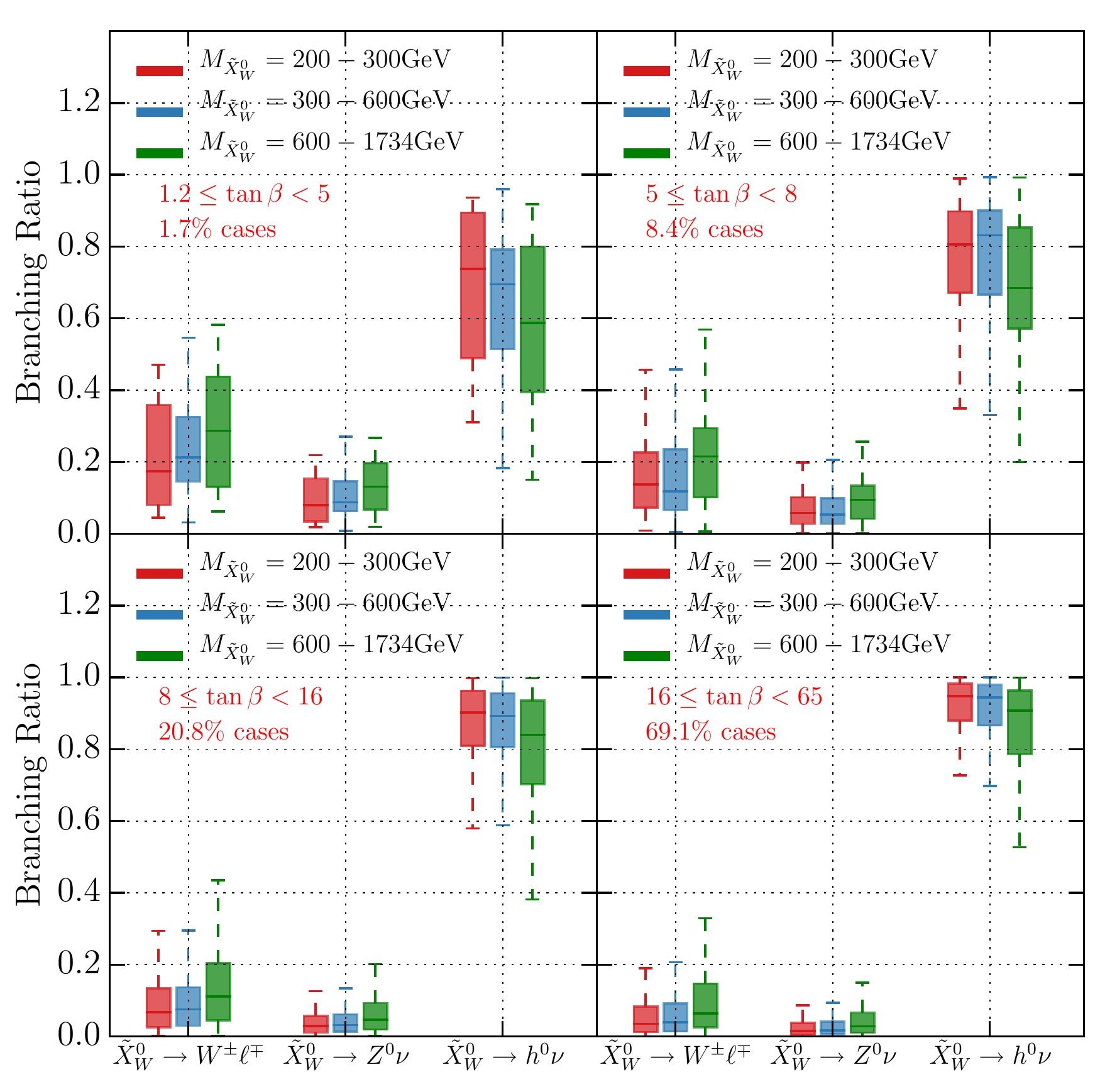}
\end{subfigure}
 
    \caption{Branching ratios for the three possible decay channels of a Wino neutralino LSP divided over three mass bins and four $\tan \beta$ regions. The colored horizontal lines inside the boxes indicate the median values of the branching fraction in each bin, the boxes indicate the interquartile range, while the dashed error bars show the range between the maximum and the minimum values of the branching fractions. The case percentage indicate what percentage of the physical mass spectra have $\tan \beta$ values within the range indicated.  We assumed a normal neutrino hierarchy, with $\theta_{23}=0.597$.}
\label{fig:bar_plot2}
\end{figure}

 Note that the ${\tilde X}^0_W\rightarrow h^0 \nu$ is dominant in all regions of the parameter space. The ${\tilde X}^0_W\rightarrow W^\pm \ell^\mp$ process has relatively high occurrence, especially for spectra characterized by small $\tan \beta$ values. Just as for charginos, the equations for the decay rates are complicated  and do not allow a simple explanation of the relative results. Furthermore, unlike for charginos, the rotation matrices involved are much more complicated since there are six neutralino species, while only two chargino species. Nevertheless, simplifying assumptions can be made. One such assumption is that the soft breaking terms have much larger magnitudes than the electroweak scale. This renders the Wino neutralino to be almost purely neutral Wino. Furthermore, using the fact that the charged lepton masses are much smaller than the soft breaking parameters further simplifies the equations. Using these approximations, one obtains the following simplified formulas for the decay rates. They are given by
\begin{multline}\label{eq:decay_neut1}
\Gamma_{{\tilde X}^0_W\rightarrow Z^0\nu_{i}} \approx
\frac{1}{64\pi}\Big({ \frac{g_2^2}{{2c_W}M_2\mu}(v_d\epsilon_i+\mu v_{L_i}^*)\left[V_{\text{PMNS}}\right]_{ij}^\dag  }\Big)^2\times\\
\times \frac{M_{{\tilde X}_W^0}^3}{M_{Z^0}^2}\left(1-\frac{M_{Z^0}^2}{M_{{\tilde X}_W^0}^2}\right)^2
\left(1+2\frac{M_{Z^0}^2}{M_{{\tilde X}^0_W}^2}\right) \ ,
\end{multline}
\begin{multline}
\Gamma_{{\tilde X}^0_W\rightarrow W^\mp \ell_i^\pm} \approx\frac{1}{64\pi}\Big(
\frac{g_2^2}{2M_2\mu}(v_d\epsilon^*_i+\mu v_{L_i}))
\Big)^2\times\\ \times
\frac{M_{{\tilde X}_W^0}^3}{M_{W^\pm}^2}\left(1-\frac{M_{W^\pm}^2}{M_{{\tilde X}_W^0}^2}\right)^2
\left(1+2\frac{M_{W^\pm}^2}{M_{{\tilde X}_W^0}^2}\right) \ ,
\end{multline}
\begin{equation}\label{eq:decay_neut4}
\Gamma_{{\tilde X}^0_W\rightarrow h^0\nu_{i}} \approx\frac{1}{64\pi}\Big({\frac{g_2}{{2}}\left[V_{\text{PMNS}}\right]_{ij}^\dag\Big(\sin \alpha \frac{\epsilon_j^*}{\mu}\Big)     }\Big)^2
M_{{\tilde X}_W^0}\left(1-\frac{M_{h^0}^2}{M_{{\tilde X}_W^0}^2}\right)^2 \ .
\end{equation}

 Unlike the approximate expressions for the decay rates of Wino charginos in eqs. \eqref{eq:decay_1}-\eqref{eq:decay_4}, the above expressions are less exact. The neutralino mass matrix contains a significantly larger number of soft mass parameters which can take values of a few GeV, close to the electroweak breaking scale, where the approximation breaks down. Nevertheless, the above expressions still provide valuable insights into which decay channel is expected to dominate in the chosen regions of parameter space. Analyzing \eqref{eq:decay_neut1}-\eqref{eq:decay_neut4}, we expect the decay channels to have comparable contributions. Interestingly, the channels ${\tilde X}^0_W\rightarrow W^\pm \ell_i^\mp$ and 
 ${\tilde X}^0_W\rightarrow Z^0 \nu_{i}$ receive a suppression proportional to $\frac{v_d}{M_2}=\frac{174 \text{GeV}}{M_2\sqrt{1+\tan^2 \beta}}$. Therefore, for large values of $\tan \beta$, the channel involving the Higgs boson, $h^{0}$, dominates for Wino neutralino decays, just as the Higgs channel dominated the Wino chargino LSP decays for this range of $\tan \beta$.

 \begin{figure}[t]
\begin{subfigure}[t]{0.49\textwidth}
\includegraphics[width=1.\textwidth]{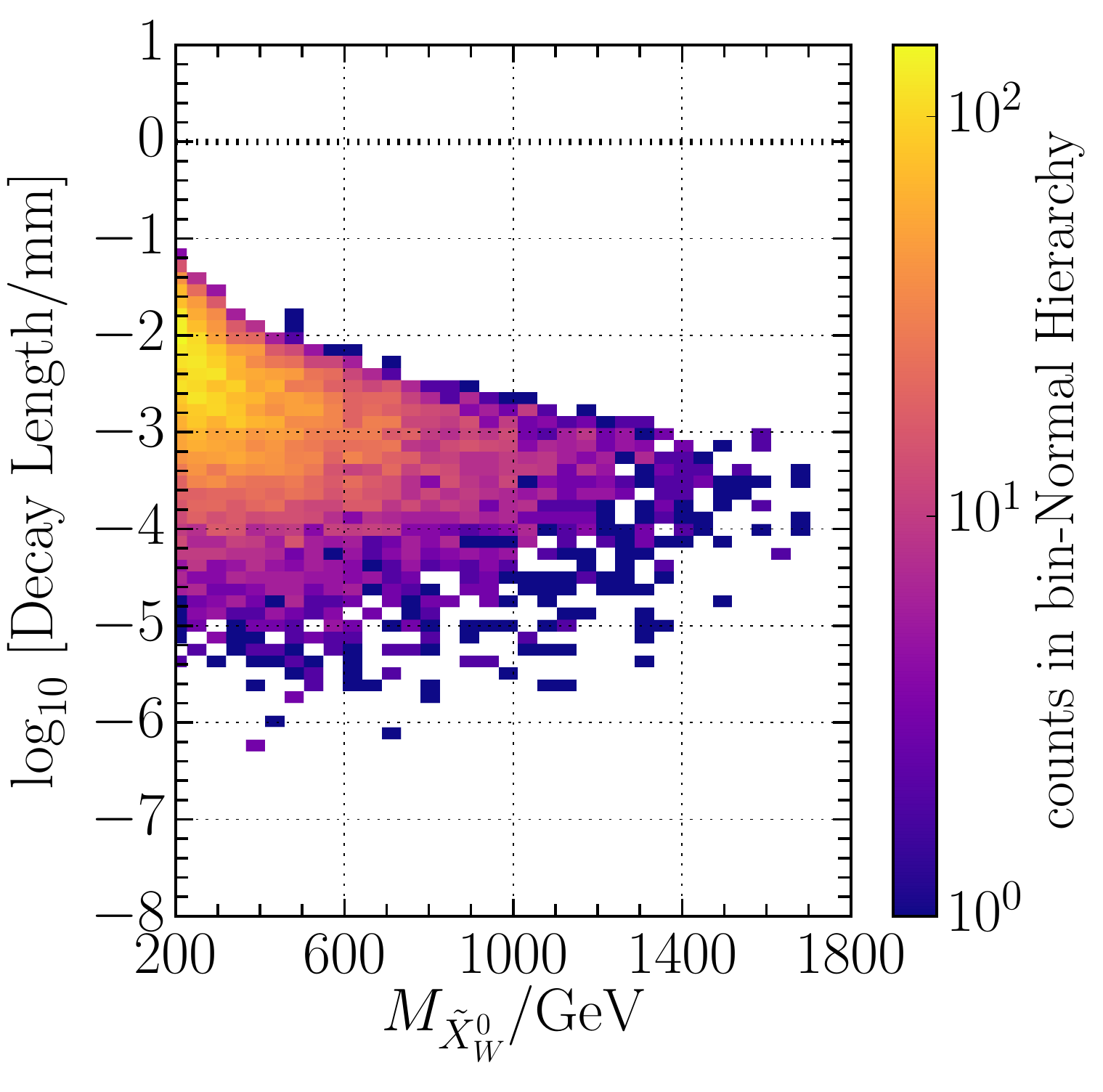}
\end{subfigure}
\begin{subfigure}[b]{0.49\textwidth}
\includegraphics[width=1.\textwidth]{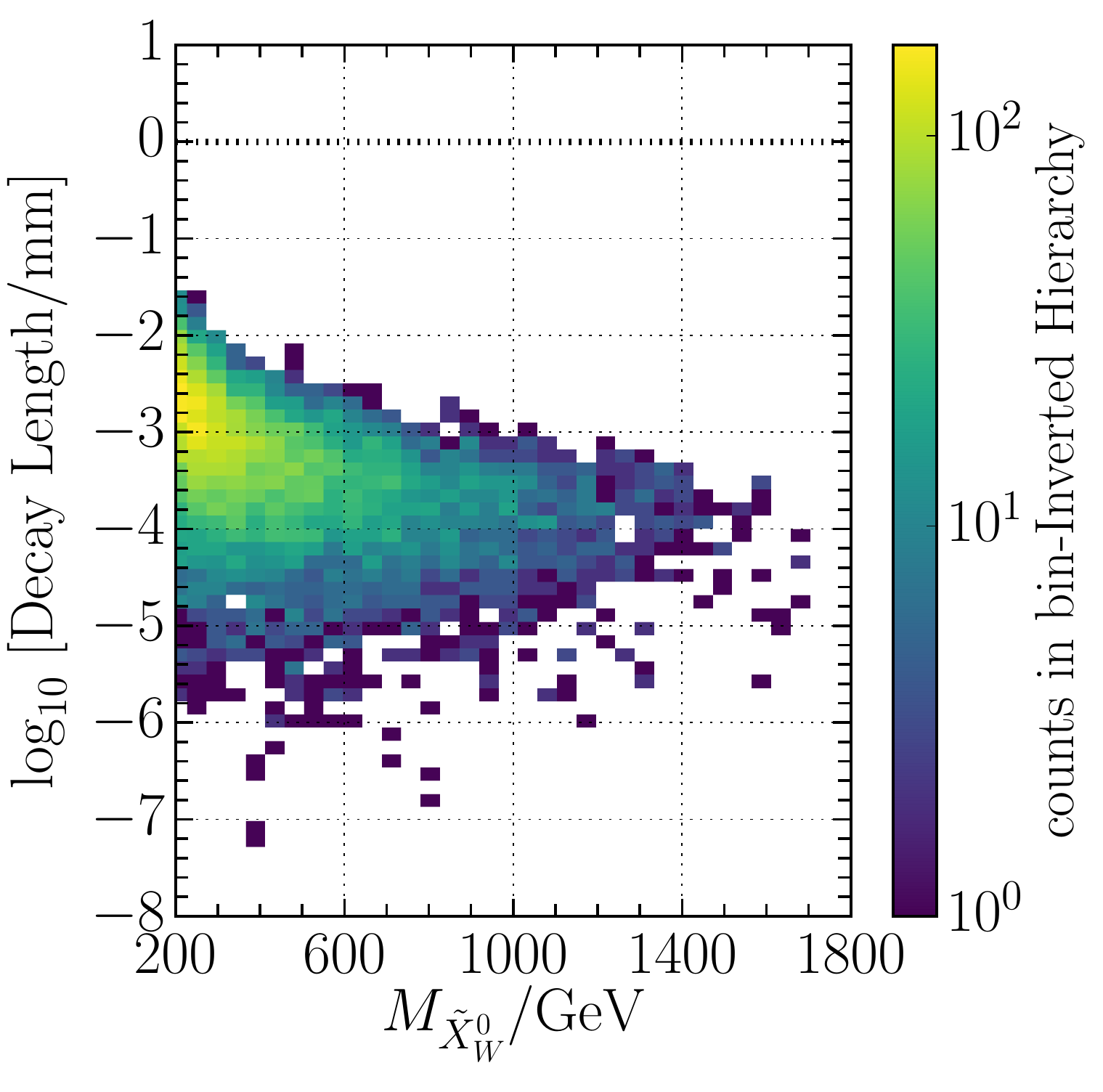}
\end{subfigure}
\caption{Wino neutralino LSP decay length in millimeters, for the normal and inverted hierarchies summed over all three channels. The average decay length $L=c\times \frac{1}{\Gamma}$ decreases for larger values of $M_{\tilde X^0_W}$, since the decay rates are amplified because of the longitudinal degrees of freedom of the massive bosons produced. We have chosen $\theta_{23}=0.597$ for the normal neutrino hierarchy and $\theta_{23}=0.529$ for the inverted hierarchy. However, the choice of $\theta_{23}$ has no impact on the decay length. The dotted line represents the 1mm line, below which all decays are considered prompt.}\label{fig:LSPprompt_neutx}
\end{figure}

\subsection*{Decay length}

Figure \ref{fig:LSPprompt_neutx} shows that Wino neutralino LSP decays are prompt-- that is, the {\it overall} decay length $L$ is less than 1mm --just as it is for Wino chargino LSP decays. Therefore, signals of both Wino chargino and Wino neutralino LSP decays produce point-like vertices. This insight is particularly useful when considering that the NLSPs of these two sparticle species (Wino neutralino NLSP for Wino chargino  LSP and Wino chargino NSLP for Wino neutralino LSP) are almost degenerate in mass with the LSPs. 
We observe that in the case of the inverted hierarchy, the decay lengths are generally a little smaller, since the values of the RPV couplings are somewhat larger, as we explained in the previous section.

\begin{figure}[t]
\centering
\begin{subfigure}[b]{1.\textwidth}
\includegraphics[width=1.0\textwidth]{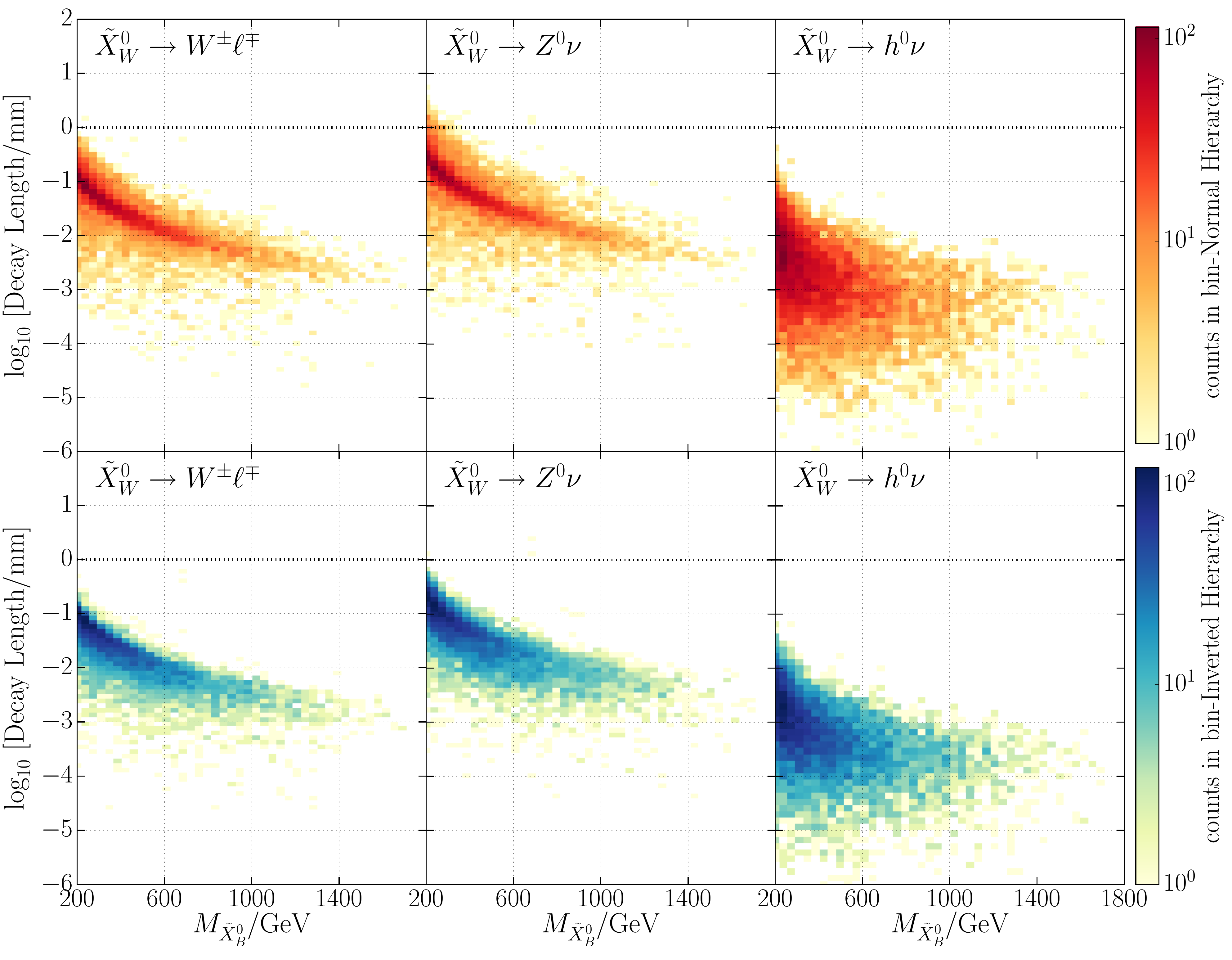}
\end{subfigure}
\caption{Wino neutralino LSP decay length in milimeters, for individual decay channels, for both normal and inverted hierarchies. We have chosen $\theta_{23}=0.597$ for the normal neutrino hierarchy and $\theta_{23}=0.529$ for the inverted hierarchy. The choice of $\theta_{23}$ has no impact on the decay length. The dotted line represents the 1mm line, below which all decays are considered prompt.}\label{fig:LSPpromptx}
\end{figure}

In Figure \ref{fig:LSPpromptx}, we study the decay lengths of the three decay channels separately.  We find that all three processes occur promptly in the detector. 

\begin{figure}[!ht]
   \centering

   \begin{subfigure}[b]{0.47\textwidth}
\includegraphics[width=1.0\textwidth]{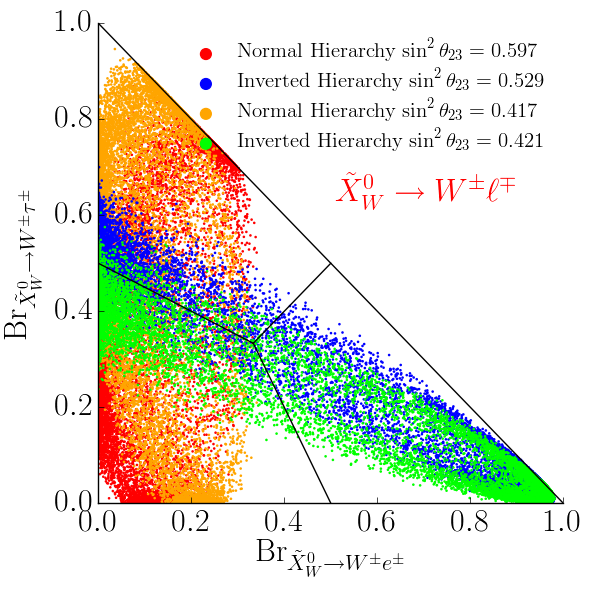}
\end{subfigure}
   \begin{subfigure}[b]{0.47\textwidth}
\includegraphics[width=1.0\textwidth]{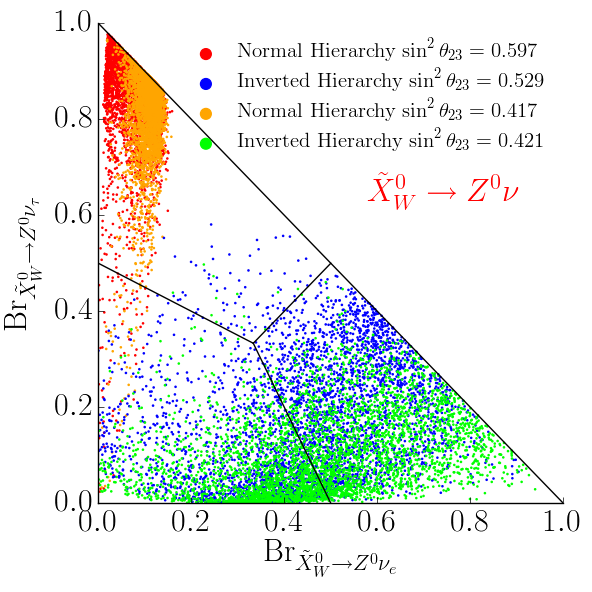}
\end{subfigure}\\
   \begin{subfigure}[b]{0.47\textwidth}
\includegraphics[width=1.0\textwidth]{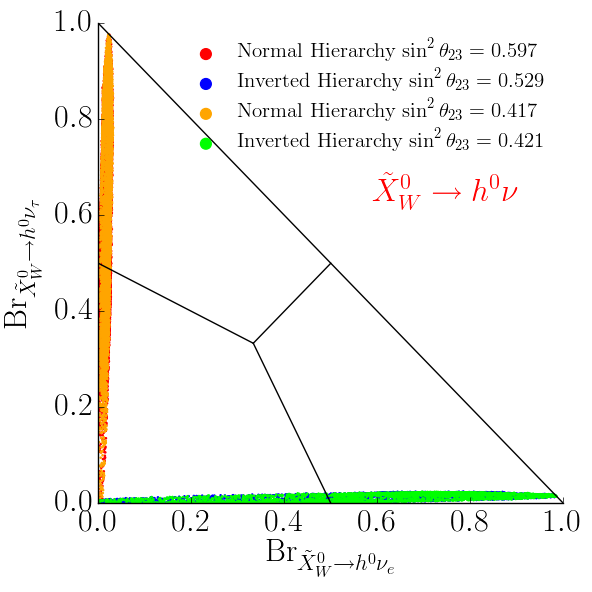} \ .
\end{subfigure}
    \caption{Branching ratios into the three lepton families,
for each of the three main decay channels of a Wino neutralino LSP. The associated neutrino hierarchy and the value of $\theta_{23}$ is specified by the color of the associated data point.}\label{fig:neutralino_lepton_familyx}
\end{figure}

\subsection*{Lepton family production}
 
We again study which of the three lepton families, if any, is favored within each of the three decay channels. For example, to quantify the probability to observe an electron $e^\mp$ in the ${\tilde X}^0_W\rightarrow W^\pm \ell^\mp$ process, over a muon $\mu^\mp$ or a tauon  $\tau^\mp$, we compute
\begin{equation}
\text{Br}_{{\tilde X}^0_W\rightarrow W^\pm e^\mp}=\frac{\Gamma_{{\tilde X}^0_W\rightarrow W^\pm e^\mp}}{\Gamma_{{\tilde X}^0_W\rightarrow W^\pm e^\mp}+\Gamma_{{\tilde X}^0_W\rightarrow W^\pm \mu^\mp} + \Gamma_{{\tilde X}^0_W\rightarrow W^\pm \tau^\mp} }
\end{equation}

\noindent Using this formalism, we proceed to quantify the branching ratios for each of the three decay processes ${\tilde X}^0_W\rightarrow W^{\pm} \ell^{\mp}$, ${\tilde X}^0_W\rightarrow Z^0 \nu_i$ and ${\tilde X}^0_W\rightarrow h^0 \nu_i$ into their individual lepton families. The results are shown in Figure \ref{fig:neutralino_lepton_familyx}.
\noindent In Figure \ref{fig:neutralino_lepton_familyx} we see that the ${\tilde X}^0_W\rightarrow W^\pm \ell^\mp$ process has an almost identical statistical distribution for lepton family production as does the chargino decay channel ${\tilde X}^\pm_W\rightarrow Z^0 \ell^\pm$. Additionaly, note that in a Wino neutralino decay via ${\tilde X}^0_W\rightarrow h^0 \nu_i$, the decay rate has a dominant term proportional to the square of $[V_{\text{PMNS}}^\dag]_{ij}\epsilon_j$. The combination leads to a branching ratio distribution as that observed in Figure \ref{fig:neutralino_lepton_familyx}--no $\nu_\tau$ neutrino is produced in the case of an inverted hierarchy and no $\nu_e$ is produced in the case of a normal hierarchy.

\subsection{Wino Neutralino NLSPs and Wino Chargino NLSPs }

\begin{figure}[t]
   \centering

   \begin{subfigure}[b]{0.49\textwidth}
 \centering
\includegraphics[width=1.0\textwidth]{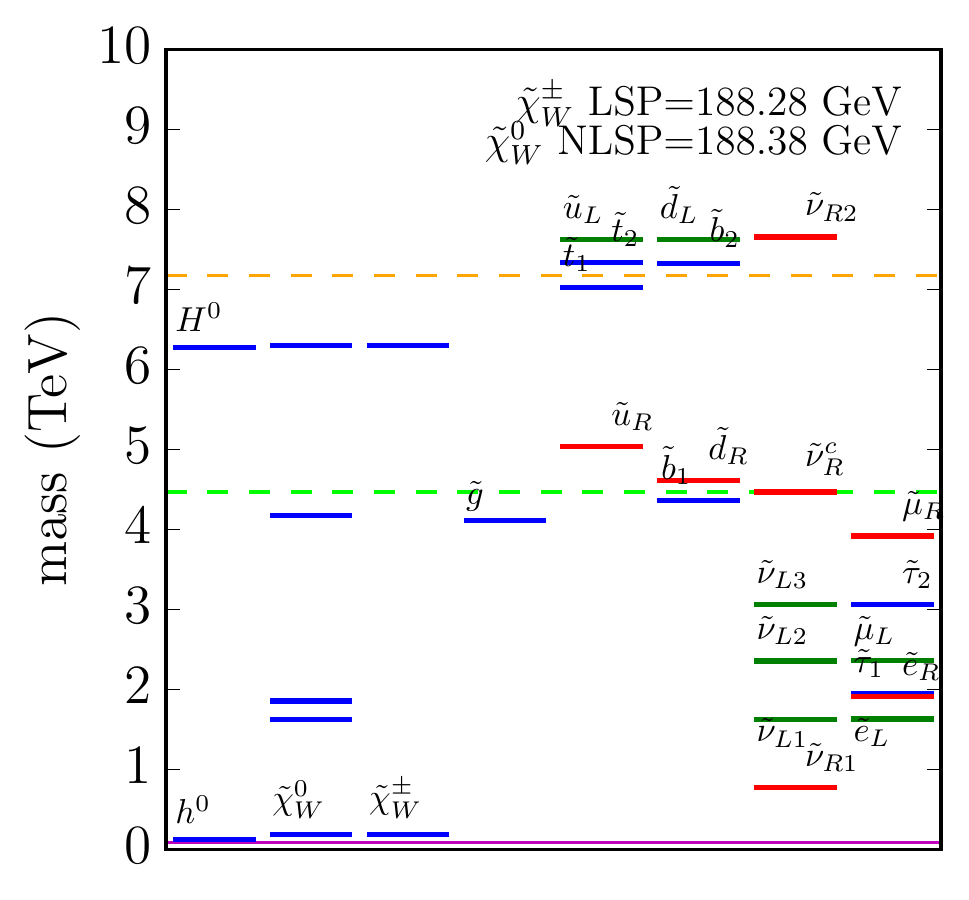}
\caption{}\label{fig:mass_spec1}
\end{subfigure}
   \begin{subfigure}[b]{0.49\textwidth}
 \centering
\includegraphics[width=1.0\textwidth]{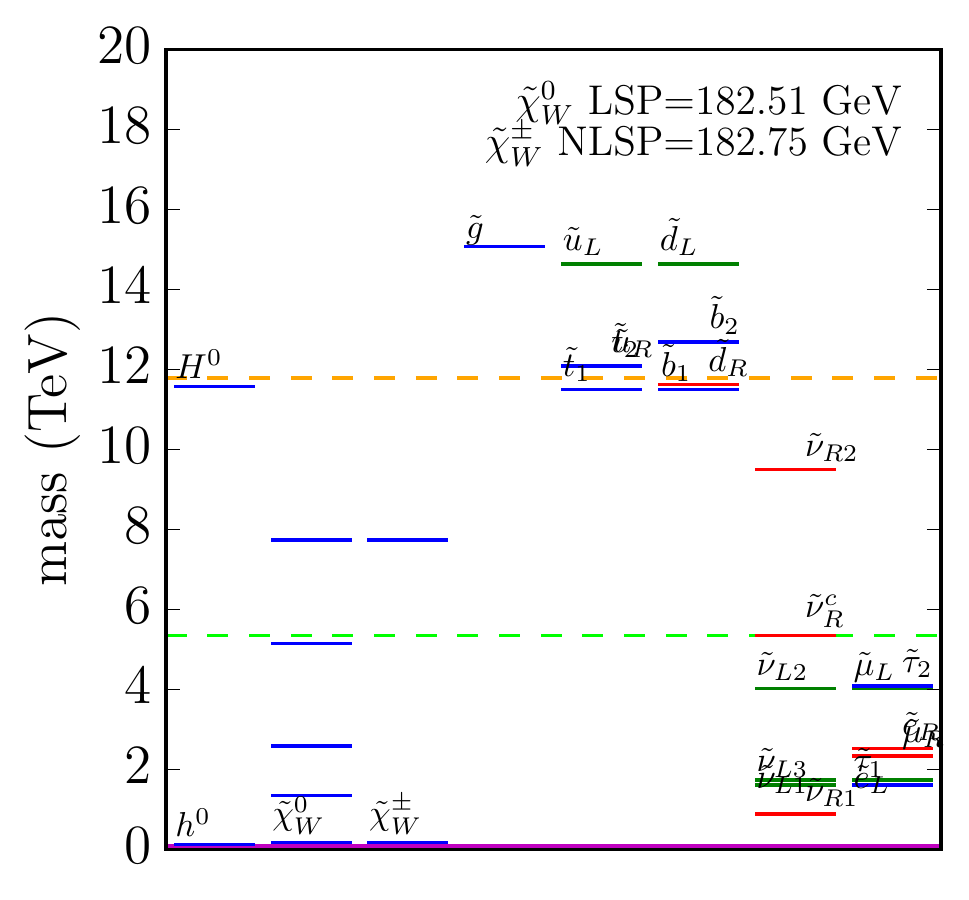}
\caption{}\label{fig:mass_spec2}
\end{subfigure}
\caption{a) A plot of the sparticle spectrum for a choice of one of the 4,858 valid black points associated with Wino chargino LSPs. The Wino neutralino NLSP is almost degenerate in mass with the LSP Wino chargino mass. b) A plot of the sparticle spectrum for a choice of one of the 4,869 valid black points associated with Wino neutralino LSPs. The Wino chargino NLSP is almost degenerate in mass with the LSP Wino neutralino mass.}
\end{figure}

Having analyzed the RPV decays of both Wino chargino LSPs and Wino neutralino LSPs, we now discuss the RPV decays of the NLSPs associated with each case. The reason this is important is the following. Let us begin with the Wino chargino LSPs associated with 4,858 valid black points. Now choose one of these black points. In Figure \ref{fig:mass_spec1}, we plot the entire sparticle spectrum of the theory for this fixed point. 

\begin{figure}[t]
   \centering

   \begin{subfigure}[b]{0.49\textwidth}
 \centering
\includegraphics[width=1.0\textwidth]{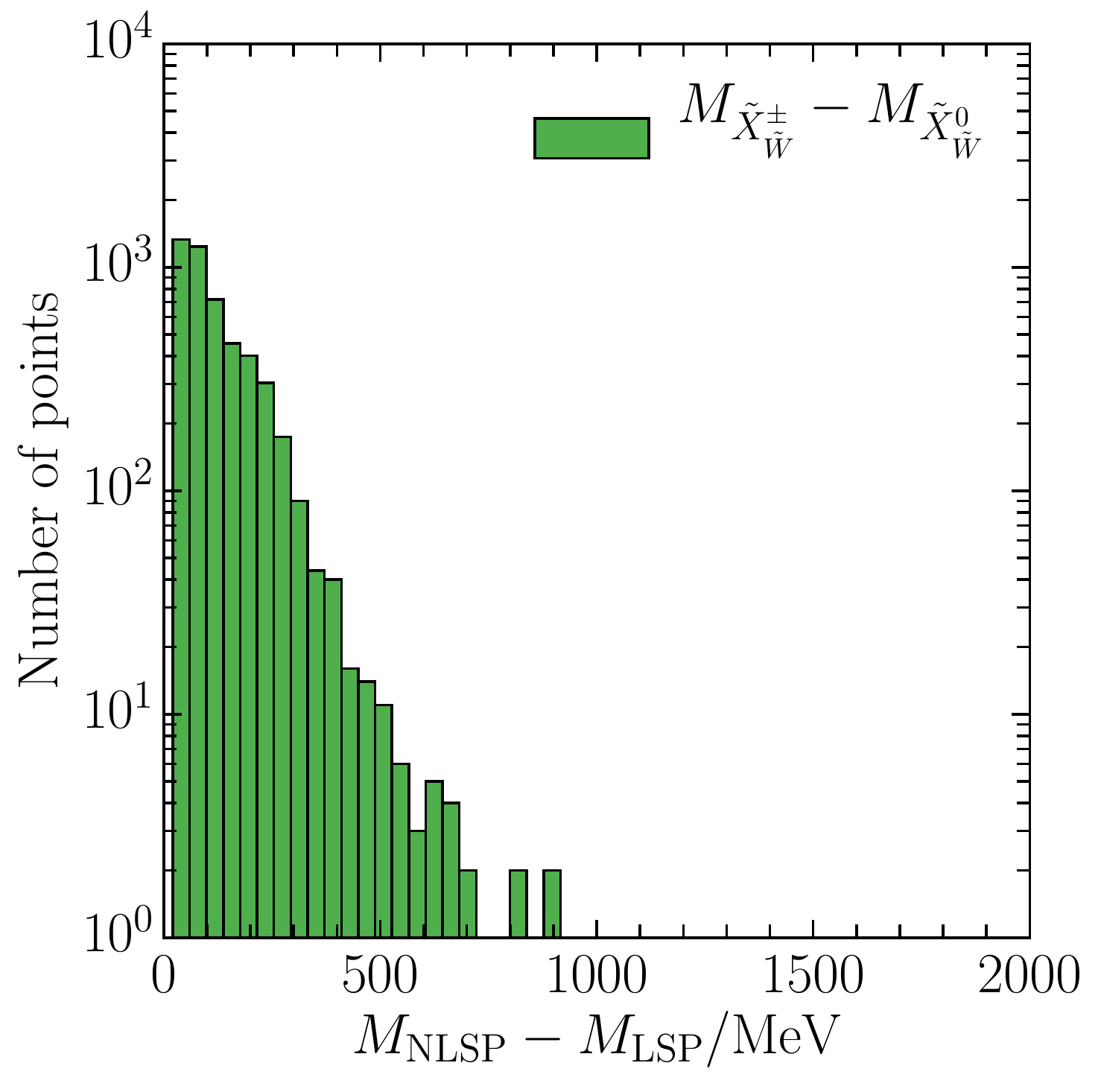}
\caption{}\label{fig:mass_diff1}
\end{subfigure}
   \begin{subfigure}[b]{0.49\textwidth}
 \centering
\includegraphics[width=1.0\textwidth]{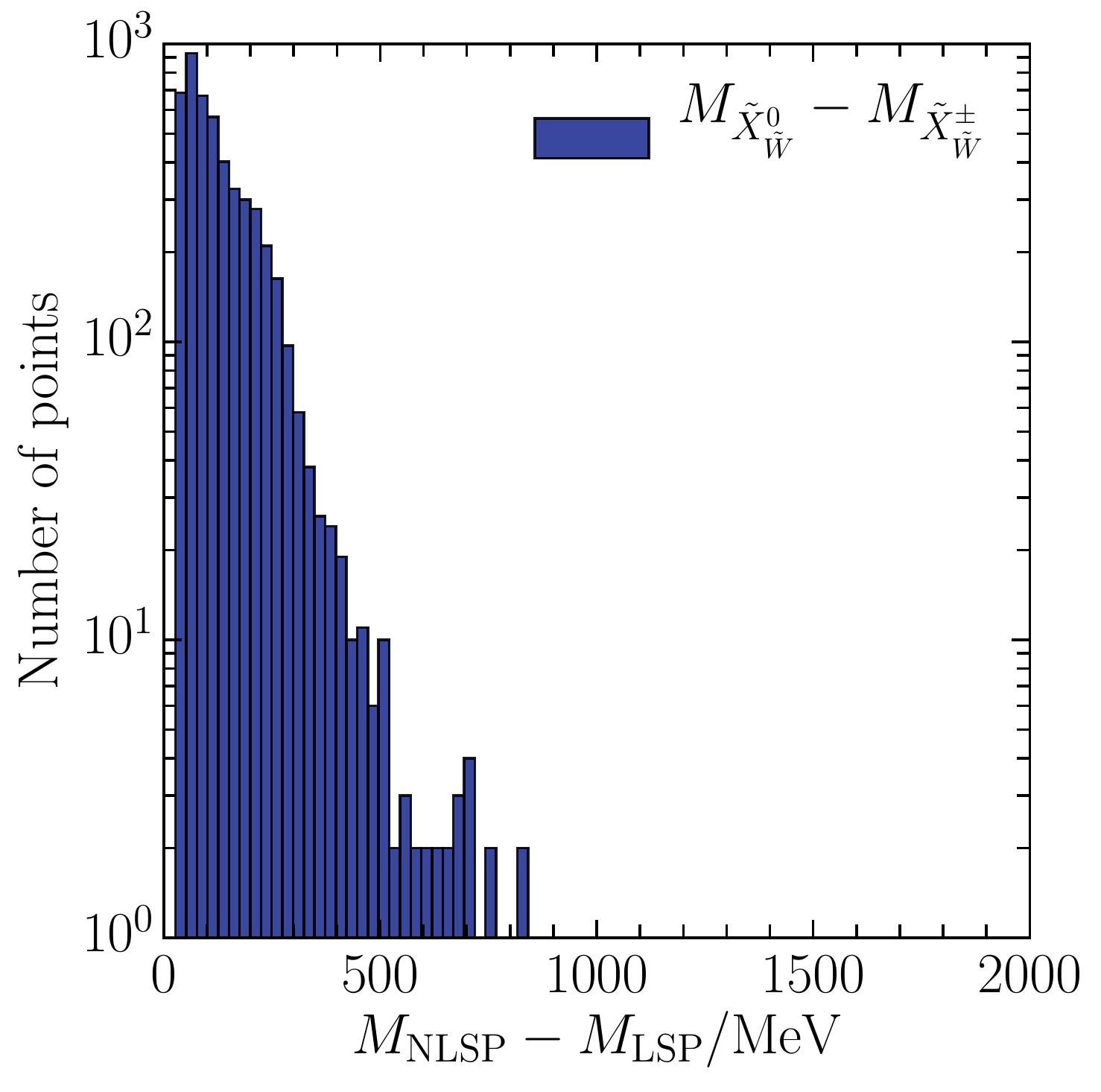}
\caption{}\label{fig:mass_diff2}
\end{subfigure}
\caption{a) The Wino neutralino NLSPs are all almost degenerate in mass with the LSPs, the Wino charginos. The mass difference is smaller than 200 MeV for most of the valid black points, as can be seen in the mass difference histogram. b) The Wino chargino NLSPs are all almost degenerate in mass with the LSPs, the Wino neutralinos. The mass difference is smaller than 200 MeV for most of the viable cases, as can be seen in the mass difference histogram}
\end{figure}

\noindent Of course, a Wino chargino is the LSP by construction.  
Interestingly, however, we see that the associated NLSP is, in fact, a Wino neutralino. This is not simply an accident of our specific choice of black point. In Figure \ref{fig:mass_diff1}, we plot the mass difference in MeV between the Wino neutralino NLSP and the Wino chargino LSP for all 4,858 black points. It is clear that for every Wino chargino LSP, the NLSP is a Wino neutralino whose mass is larger than, but very close to, the mass of the LSP-- as shown in Figure \ref{fig:mass_spec1} for a single such point. This is, perhaps, not surprising since the dominant contribution to the mass of both sparticles is given by the soft supersymmetry breaking  parameter $M_{2}$. 
Not surprisingly, we find that a similar, but reversed, situation occurs when the LSP is a Wino neutralino. Choosing one of the 4,869 associated valid black points, we find that the complete sparticle spectrum is given in Figure \ref{fig:mass_spec2}. \noindent Of course, a Wino neutralino is the LSP by construction.  
However, we now we find that the situation is reversed and that the associated NLSP is now a Wino chargino. Again, this is not simply an accident of our specific choice of black point. In Figure \ref{fig:mass_diff2}, we plot the mass difference in MeV between the Wino chargino and the Wino neutralino for all 4,869 Wino neutralino black points. It is clear that for every Wino neutralino LSP, the NLSP is a Wino chargino whose mass is larger than, but very close to, the mass of the LSP-- as in Figure \ref{fig:mass_spec2}. Once again, this is hardly surprising since the dominant contribution to the mass of both sparticles is given by the soft supersymetry breaking  parameter $M_{2}$.

Because the mass difference between the two species is so small, both the Wino chargino and the Wino neutralino will be produced at the LHC; assuming that one of them is the LSP and sufficiently light. We have already analyzed the decays of the LSP, both for the case in which the LSP is a Wino chargino and when the LSP is the Wino neutralino. These particles can decay into SM particles due to the RPV couplings in the B-L MSSM model we are studying. The NLSPs, however, as with any other sparticle in the mass spectrum that is not the LSP, can decay via channels that either violate R-parity or channels which conserve it. In general, the RPC couplings are much stronger than the RPV couplings introduced in our theory, since the latter need to be small enough to be consistent with the observed neutrino masses and not lead to unobserved effects such as proton decays. Therefore, the RPC decays of sparticles that are not the LSP are, in general, expected to have much higher branching ratio than the RPV decays. However, in the cases that we focus on, the NLSP is almost degenerate in mass with the LSP. The mass difference is so small that an RPC decay of a Wino neutralino NLSP into a Wino chargino LSP (or vice versa) might prove highly suppressed. Therefore, the NLSP would behave as though it was an LSP which decays via RPV decays. In the remainder of this section, we analyze both the RPV and the RPC decays of the Wino chargino and the Wino neutralino NSLPs and provide a quantitative comparison of the decay rates of those channels.

\subsection*{RPV decays of the NLSPs}

\begin{figure}[t]
   \centering

   \begin{subfigure}[b]{0.8\textwidth}
 \centering
\includegraphics[width=1.0\textwidth]{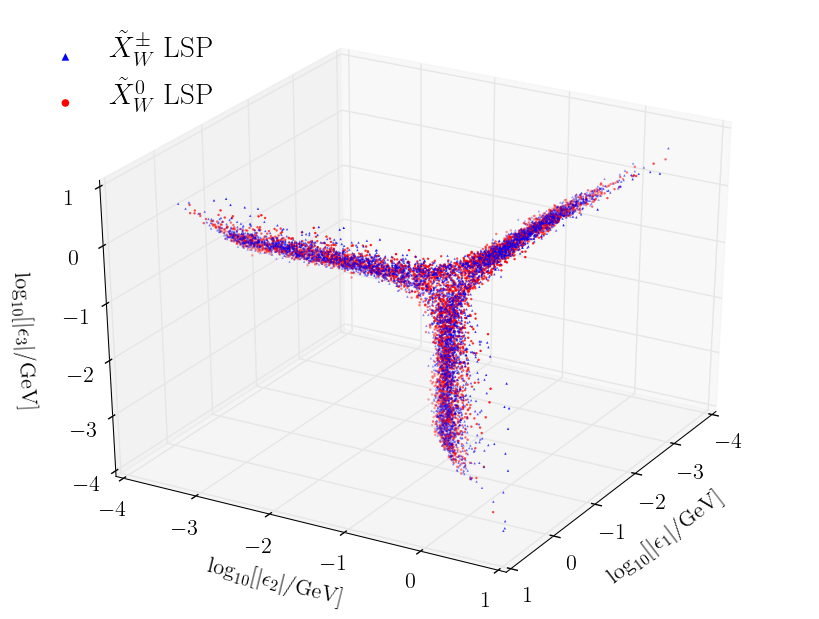}
\caption{}
\end{subfigure}

    \caption{Absolute values of the $\epsilon_1$, $\epsilon_2$ and $\epsilon_3$ parameters associated with the 4,858 black points with Wino chargino LSP (red) and with the 4,869 black points with Wino neutralino LSP (blue). We assume a normal hierarchy with $\theta_{23}$=0.597. We find that these RPV parameters lie in the same statistical regions, regardless of LSP species. }
    \label{fig:scatter_eps}
\end{figure}

We begin our discussion by analyzing the RPV decay channels of both the Wino chargino NLSP and the Wino neutralino NLSP. Wino chargino and Wino neutralino LSPs can only decay via RPV channels as presented above. However, are the NLSP RPV decay rates and branching ratios the same as though they were actual LSPs? Does a Wino chargino NLSP, associated with the initial conditions for a Wino neutralino LSP, decay in the same way as an actual Wino chargino LSP? The same question arises for the Wino neutralino NLSP. Even though, in these cases, the LSP and NLSP masses are very close, the answer is not immediately obvious, since the decay rates and branching ratios do not depend only on these masses. The decay rates for charginos and neutralinos given in Section \ref{sec:7} are completely general, and apply for any chargino and neutralino species, regardless of if they are the LSP or just another particle in the spectrum. Those equations depend on a large number of parameters of the theory, such as $M_{BL}$, $v_R$, $\tan \beta$, $M_R$, as well as on the RPV couplings $\epsilon_{i}, v_{L_{i}}$, $i=1,2,3$. A Wino chargino LSP and a Wino chargino NLSP (associated with a Wino neutralino LSP) have the same RPV decay patterns only if all these parameters are contained within similar statistical intervals.

For example, let us consider the $\epsilon$ parameters. In Figure \ref{fig:scatter_eps} we plot the absolute values of the $\epsilon_1$, $\epsilon_2$  and $\epsilon_3$ couplings, associated with the 4,858 black points with Wino chargino LSP and with the 4,869 black points with Wino neutralino LSP, respectively. We assumed a normal hierarchy, with $\theta_{23}$=0.597. We find that these RPV parameters are statistically similar, whether associated with a Wino chargino LSP, or with a Wino neutralino LSP. This is clear from  Figure \ref{fig:scatter_eps}, where the points lie substantially on top of each other. Plotting $\epsilon_1$ against $\epsilon_2$ and $\epsilon_3$ for both initial conditions with Wino chargino and Wino neutralino LSP is only one of the tests one can make, since other parameters could be relevant. However, this scatter plot is a particularly pertinent one, since the RPV couplings depend on the neutrino masses and neutrino mixing angles, as well as on numerous mass terms from the B-L MSSM Lagrangian of our theory. Indeed, further analysis quickly concludes that other initial parameters have negligable effect. We conclude that, when analyzing Wino chargino LSP decays, one should simultaneously look for the RPV decays of the Wino neutralino-- as though it were the LSP -- and, vice versa.

\subsection*{RPC decays of NLSPs}

\begin{figure}[t]
\includegraphics[width=1.0\linewidth]{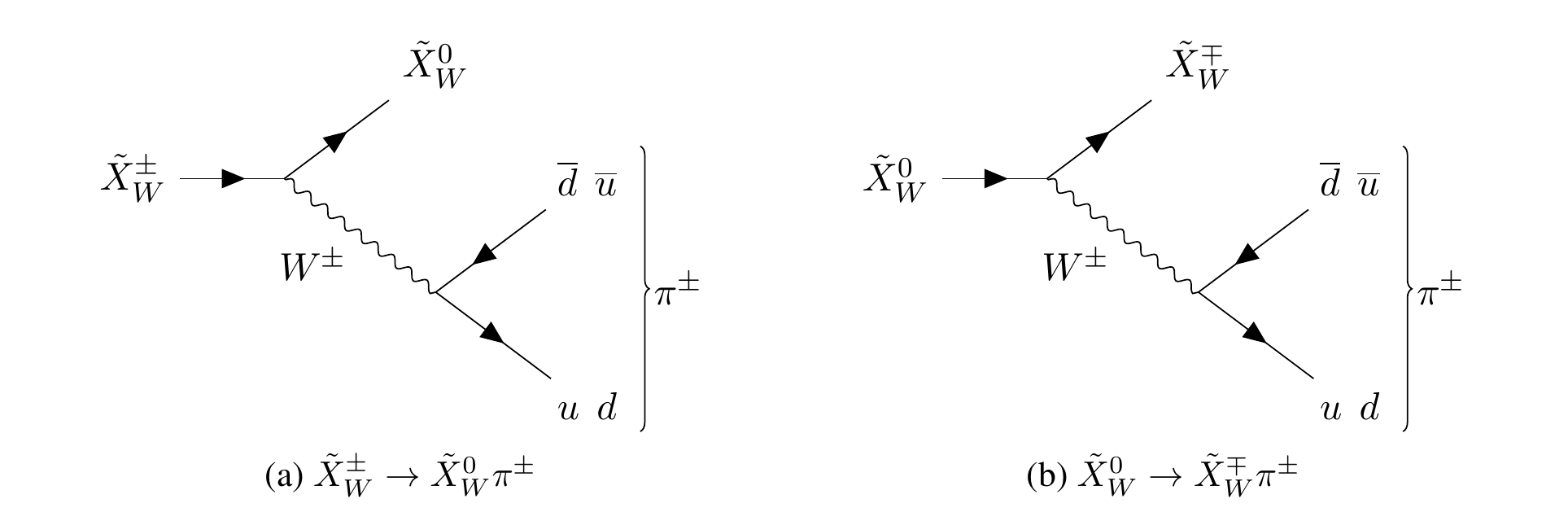}
\caption{Dominant RPC decay modes of (a) a Wino chargino NLSP and (b) a Wino neutralino NLSP. This decay mode dominates for NSLP-LSP mass difference $\delta M$ larger than the mass of the charged pions $\pi^\pm$; that is, $\delta M>m_{\pi^\pm}$}\label{fig:NLSPdecay}
\end{figure}

Figures \ref{fig:mass_diff1} and \ref{fig:mass_diff2} show that the mass differences between the Wino chargino LSP and the Wino neutralino NLSP, or between the Wino neutralino LSP and the Wino chargino NLSP, are generally smaller than 400 MeV. 
For this small mass splitting, there are only a limited number of possible RPC decay channels for the NLSP. If the mass splitting is larger than the charged pion mass, 
$m_{\pi}^{\pm} \sim140$~MeV, 
then the dominant RPC decay of the NLSP is into the LSP and a charged pion $\pi^\pm$. These processes, which involve the on-shell bosons $W^\pm$, are shown in Figure \ref{fig:NLSPdecay}.

\begin{figure}[t]
 \begin{minipage}{1.0\textwidth}
     \centering
        \begin{subfigure}[b]{0.48\linewidth}
   \centering
\includegraphics[width=0.8\textwidth]{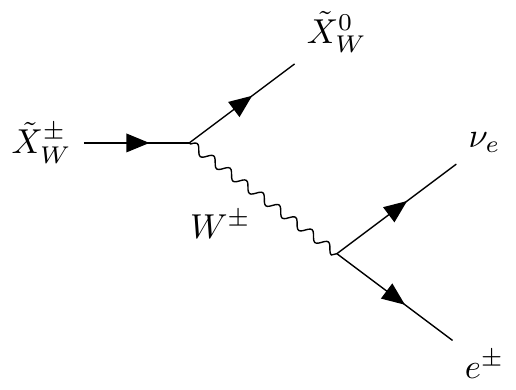}
\caption{${\tilde X}^\pm_W\rightarrow \tilde X_W^0 e^\pm \nu_e$}
       \label{fig:table2}
   \end{subfigure} 
   \begin{subfigure}[b]{0.48\linewidth}
   \centering
\includegraphics[width=0.8\textwidth]{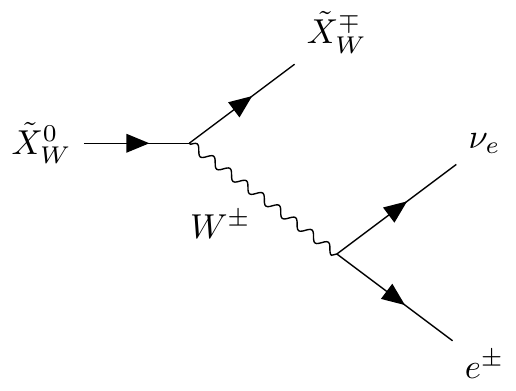}
\caption{${\tilde X}^0_W\rightarrow \tilde X_W^\mp e^\pm\nu_e$}
       \label{fig:table2}
   \end{subfigure} \\
\end{minipage}
\caption{Second most dominant RPC decay modes of (a) a Wino chargino NLSP and (b) a Wino neutralino NLSP. This decay mode dominates for NSLP-LSP mass difference $\delta M$ smaller than the mass of the charged pions $\pi^\pm$; that is, $\delta M<m_{\pi^\pm}$}\label{fig:NLSPdecay2}
\end{figure}
At leading order, the decay rate of the Wino chargino NLSP into a charged pion and the Wino neutralino LSP can be expressed in terms 
of the decay rate of the charged pion,
\begin{equation}\label{RPCrate}
\Gamma(\tilde X_W^\pm \rightarrow \tilde X_W^0 \pi^\pm)=\Gamma(\pi^\pm \rightarrow
\mu^\pm \nu_\mu)\times \frac{16\delta M^3}{m_\pi m_\mu^2}
\left(  1-\frac{m_\pi^2}{\delta M^2}   \right)^{1/2}\left( 1-\frac{m_\mu^2}{m_\pi^2}  \right)^{-2},
\end{equation}

\noindent where $\delta M=M_{\tilde X_W^\pm}-M_{\tilde X_W^0}$ is the mass difference between the NLSP and the LSP, and $m_{\pi}$ and $m_{\mu}$ denote the masses of the charged pion and the muon respectively. Conversely, in the case in which the Wino chargino is the LSP, the main RPC channel of the Wino neutralino NLSP is into a Wino chargino LSP and a charged pion. The decay rate is given by eq. \eqref{RPCrate}, but  now with $ \delta M=M_{\tilde X_W^0}-M_{\tilde X_W^\pm}$. 

Note that the processes shown in Figure \ref{fig:NLSPdecay} can only happen if the mass splitting between the LSP and the NLSP is larger than the charged pion mass $m_{\pi^\pm}$. For smaller mass differences, the decay modes shown in Figure \ref{fig:NLSPdecay2} then dominate. The decay rate of a Wino chargino NLSP decaying into a Wino neutralino LSP, an electron and a neutrino is given by

\begin{equation}\label{RPCrate2}
\Gamma(\tilde X_W^\pm \rightarrow \tilde X_W^0 e^\pm \nu_e)=\frac{2G_F^2}{15\pi^2}\delta M^5,
\end{equation}
where $\delta M=M_{\tilde X_W^\pm}-M_{\tilde X_W^0}$  and $G_F$ is the Fermi constant. Conversely,  the decay rate of a Wino neutralino NLSP into a Wino chargino LSP, an electon and a neutrino  is given by eq. \eqref{RPCrate2}, with $ \delta M=M_{\tilde X_W^0}-M_{\tilde X_W^\pm}$. Finally, we note that there is a similar RPC decay channel involving the muon. However, since the mass of the muon is much larger than that of the electron, this decay rate is greatly suppressed relative to \eqref{RPCrate2} and, hence, is irrelevant.

\subsection*{RPV vs RPC}

So far, we have shown that the RPV decays of the Wino chargino and the Wino neutralino NLSPs occur as if they were the both LSPs. We expect these RPV processes to allow for prompt decays of the NLSPs in the detector. In this section, we analyze whether the RPC processes of the NLSP can produce observable traces in the detector as well.

\begin{figure}[t]
   \centering

   \begin{subfigure}[b]{0.49\textwidth}
 \centering
\includegraphics[width=1.0\textwidth]{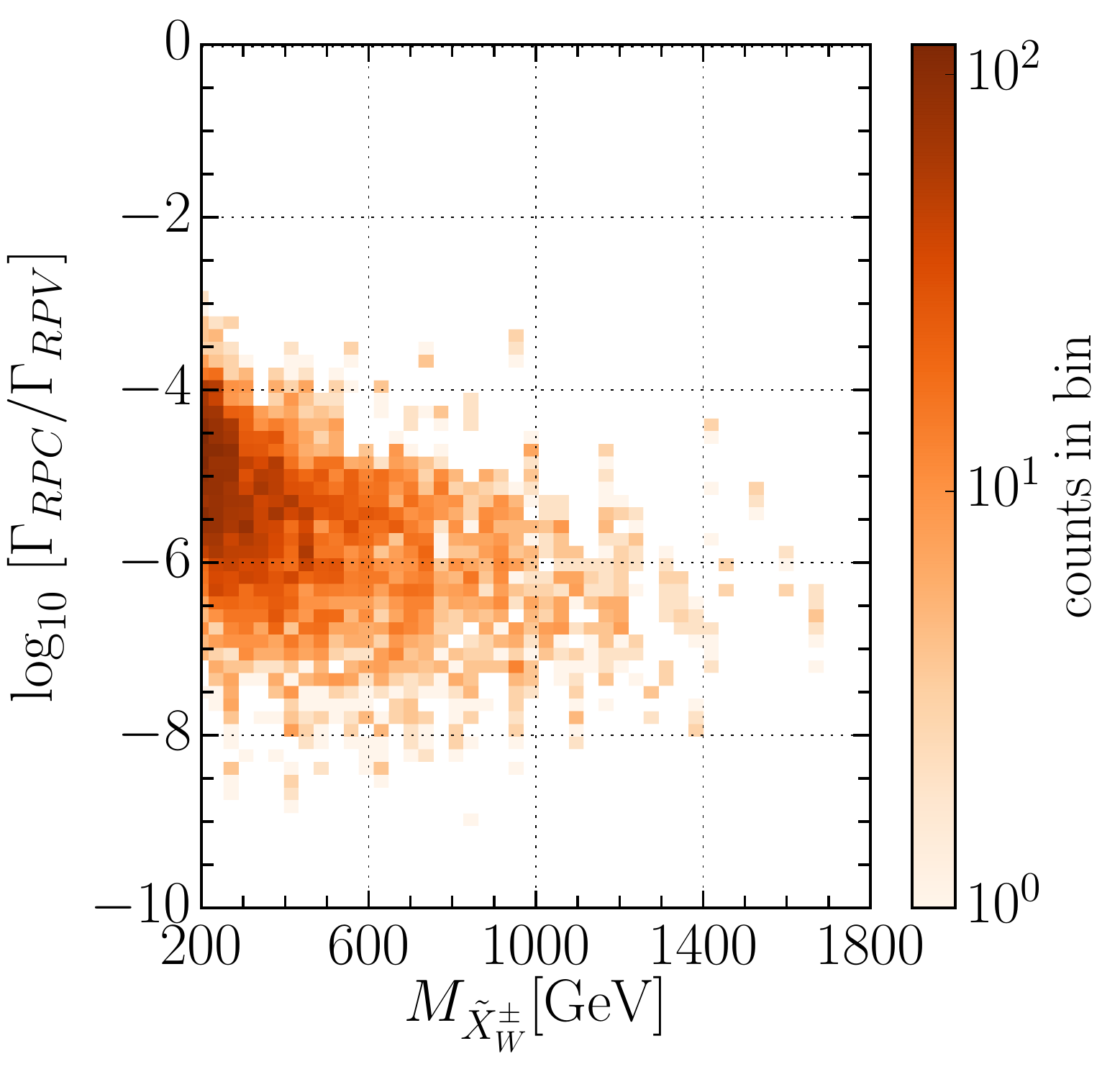}
\caption{}\label{fig:ratio1}
\end{subfigure}
   \begin{subfigure}[b]{0.49\textwidth}
 \centering
\includegraphics[width=1.0\textwidth]{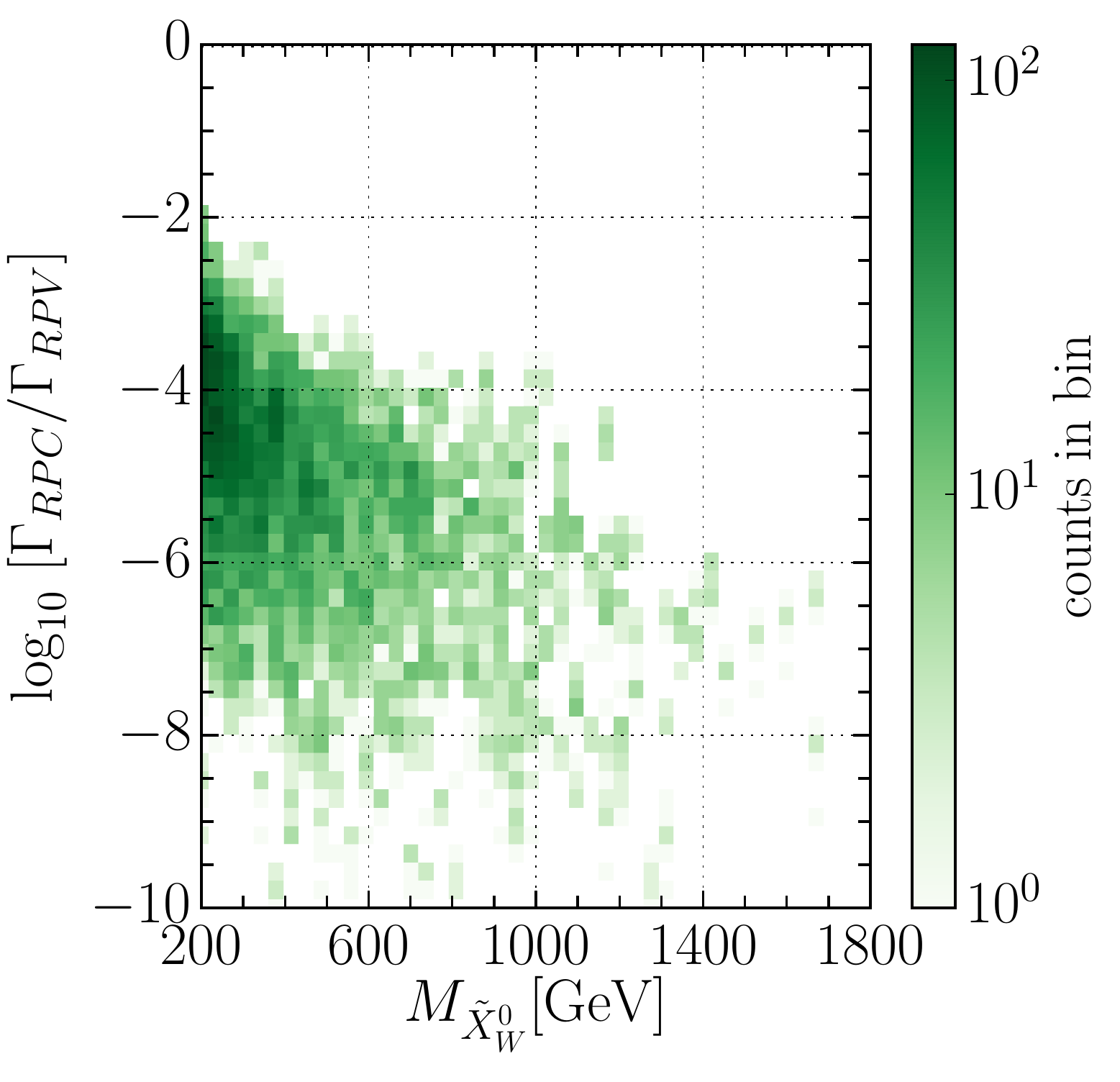}
\caption{}\label{fig:ratio2}
\end{subfigure}
\caption{Ratios between the decay rates of the RPC and the RPV channels. On the horizontal axis, we show the NLSP mass. We study both scenarios with (a) Wino chargino NLSPs and (b) Wino neutralino NLSPs. In all cases, the RPV processes are strongly dominant.}
\label{fig:ratios}
\end{figure}

 In Figure \ref{fig:ratios}, we present the ratios $\Gamma_{\text{RPC}}/\Gamma_{\text{RPV}}$ for all simulated NLSPs, where $\Gamma_{\rm RPC}$ is computed by summing over all RPC channels discussed above. We find that, {\it in all cases}, the RPC processes are strongly suppressed compared with the RPV ones. This suppression of the RPC decays is due to the near degeneracy in mass between the Wino chargino LSP and the Wino neutralino NLSP shown in Figure \ref{fig:mass_diff1}, and the similar mass degeneracy between a Wino neutralino LSP and its Wino chargino NLSP displayed in Figure \ref{fig:mass_diff2}-- with mass splittings ranging between 20 MeV and 500 MeV. Note that most mass splittings are $\lesssim 200$ MeV.
 Therefore, the RPC decays of Winos charginos and Wino neutralinos NLSPs are not expected to produce any visible traces in the detector. Hence, in both cases, the only decay mode of the LSP and the dominant decay mode of the NLSP are precisely the RPV decays.

Since this final conclusion rests on having a reliable computation of the mass splitting between the Wino NLSP and the Wino LSP, we end this Section by discussing the role that higher-loop contributions might have. As stated in the Introduction, all calculations performed are done using one-loop RG $\beta$ and $\gamma$ functions ignoring finite loop corrections. For the values of the parameters required to produce realistic low-energy phenomenology, such higher loop effects are not expected to have a large effect as regards the Wino NLSP and Wino LSP mass splitting--on the order of several hundred MeV. Indeed, such loop contributions to NLSP and LSP mass splitting have been explicitly computed to the one-loop and two-loop level in \cite{Gherghetta:1999sw, Cheng:1998hc, Feng1999xx}  and \cite{Ibe:2012sx} respectively--albeit in different theoretical contexts than our own, without the B-L extension, and without the RPV couplings which mix charginos and neutralinos with leptons. These papers found that, indeed, the higher loop corrections are small--on the order of 100-200 MeV. We conclude, therefore, that our expectation that the higher loop contributions to the mass splitting of the Wino NLSP and the Wino LSP in our theory would only be, at most, on the order of several hundred MeV, is indeed correct. Note that adding several hundred MeV might raise our maximum possible mass splitting to $\sim$ 700 MeV--with most other mass splittings being much smaller. Even at 700 MeV, the decay ratio $\Gamma_{\rm RPC}/\Gamma_{\rm RPV}\ll 1$ and, hence, our conclusions in the previous paragraph remain unaltered.

\subsection{Conclusion}

In this section, we have systematically derived the RPV decay channels, decay rates, branching ratios, and the relationship of these results to the neutrino mass hierarchy and the $\theta_{23}$ neutrino mixing angle, for both a Wino Chargino LSP and a Wino neutralino LSP-- all within the context of the explicit $B-L$ MSSM theory. It is shown that the Wino neutralino is the NLSP for a Wino chargino LSP and vice versa--with the mass splitting between them being small. Hence, while the Wino NLSP RPC decays are suppressed, its RPV decays should be observable at the LHC in addition to those of the Wino LSP. Since the $B-L$ MSSM is completely compatible with all low energy phenomenological data, the results of this study are explicit physical predictions that are of interest to the CERN SUSY ATLAS experimentalists. Run 1 and early run 2 data have already been used in the search for these processes--see \cite{Aaboud:2017opj,ATLAS:2017hbw,Jackson:2015lmj,ATLAS:2015jla}--and newer run 2 data is presently being analyzed.

\section{Bino Neutralino LSP decays}

The Wino chargino and Wino neutralino LSPs were chosen because their decay products are readily observable by ATLAS at the LHC. However, it was noticed that one particular LSP, the neutral fermionic superpartner ${\tilde{B}}$ of the hypercharge gauge field $Y$--referred to as the ``Bino''-- was a much more prevalent LSP associated with viable $B-L$ MSSM models. A statistical analysis showed that the Bino was approximately a factor of 10 more likely to occur that either a Wino chargino or a Wino neutralino. In fact, it is the most likely LSP to occur for any viable initial conditions. It seems well-motivated, therefore, to apply the results of \cite{Dumitru:2018jyb} to the Bino and to determine its decay modes to standard model particles, and the associated decay rates and branching ratios--as well as the relationship of these decays to the neutrino mass hierarchy. However, an {\it important new phenomenon} occurs for the Bino which is unique among all chargino and neutralino LSPs. That is, a careful analysis of the mass eigenstates of these LSPs shows that the Bino mass, although generically near, or above, the electroweak breaking scale, can be fine-tuned to be smaller than this scale. In fact, for sufficient fine-tuning it can be made to be arbitrarily small and even to vanish. On the other hand, this analysis reveals that the masses of all other charginos and neutralinos can never be smaller than the electroweak scale. Since the RPV decay products of the Bino must include either a $W^{\pm}$, a $Z^{0}$ or a neutral Higgs boson, the ``usual'' decays to standard model particles can no longer occur when the Bino mass drops below the electroweak scale. However, it can still decay via more complicated processes involving ``off-shell'' weak  interaction vector bosons or the Higgs. The decay products of these processes are also readily observable at the LHC. It is of interest, therefore, to also analyze these more complicated decays of a light Bino, and to compute their decay rates and branching ratios. Therefore, to summarize: the analysis of LSP Bino RPV decays carried out in this section breaks naturally into two parts--1) the RPV decays of a Bino with mass $M_{W^{\pm}} < M_{B}$ via ``on-shell'' $W^{\pm}$, $Z^{0}$ or $h^{0}$ bosons. Such decays are similar in form to those of both Wino chargino and Wino neutralino decays studied in Section \ref{sec:winosec} and 2) the RPV decays of a Bino with mass $M_{B} < M_{W^{\pm}}$ via ``off-shell'' $W^{\pm}$, $Z^{0}$ or $h^{0}$ bosons. These decays are more complicated and the decay rates and branching ratios are considerably suppressed relative to the on-shell case. Be that as it may, they lead to interesting signatures, which can offer unique signals for an ATLAS search. It is the purpose of this section to analyze the RPV decays of a Bino LSP--for the Bino mass both above and below the electroweak scale. 

\subsection{The Bino Neutralino}

\indent In the {\it absence of the RPV violating terms} proportional to $\epsilon_{i}$ and $v_{L_{i}}$,  the neutral Higgsinos and gauginos of the theory mix with
the third generation right handed neutrino. In the gauge eigenstate basis $\psi^0=\left( \tilde{W}_R, \tilde{W}_0, \tilde{H}_d^0,
\tilde{H}_u^0, \tilde{B}^\prime,{\nu}^c_3 \right)$, 
\begin{equation}
\mathcal{L}\supset-\frac{1}{2}\left(\psi^0\right)^T{M}_{\tilde{{ \chi}}^0}\psi^0+c.c
\end{equation}
where 
\begin{equation}
M_{{\tilde \chi}^0}=
\left(
\begin{matrix}
M_R&0&-\frac{1}{2}g_Rv_d&\frac{1}{2}g_Rv_u&0&-\frac{1}{2}g_Rv_R\\
0&M_2&\frac{1}{2}g_2 v_d&-\frac{1}{2}g_2 v_u&0&0\\
-\frac{1}{2}g_Rv_d&\frac{1}{2}g_2 v_d&0&-\mu&0&0\\
\frac{1}{2}g_Rv_u&-\frac{1}{2}g_2 v_u&-\mu&0&0&0\\
0&0&0&0&M_{BL}&\frac{1}{{2}}g_{BL}v_R\\
-\frac{1}{{2}}g_Rv_R&0&0&0&\frac{1}{{2}}g_{BL}v_R&0\\
\end{matrix}
\right) \ .
\label{eq:neutralinoMassMatrixWithoutEpsilon}
\end{equation}
In the neutralino mass mixing matrix shown in \eqref{eq:neutralinoMassMatrixWithoutEpsilon}, $M_2$, $M_R$ and $M_{BL}$ are the gaugino mass terms 
introduced in the soft SUSY breaking Lagrangian. They correspond to the symmetry groups $SU(2)_{L}$, $U(1)_{3R}$ and $U(1)_{B-L}$ respectively, The associated gauge couplings are  $g_2$, $g_R$ and $g_{B-L}$. In our simulation, we sample the absolute values of the gaugino masses between $\big[200~{\rm GeV},10~ {\rm TeV}\big]$, and further allow them to have either positive or negative signs, which are determined statistically.
The $\mu$ parameter is the Higgsino mass term. Its value is chosen so as to produce the correct $Z^0$ and $W^{\pm}$ boson masses, a process called the ``little hierarchy problem''  \cite{Ovrut:2015uea}. The dimensionful parameters $v_u$ and $v_d$ are the Higgs up and Higgs down VEVs that break electroweak symmetry, while $v_R$ is the third generation sneutrino VEV, which breaks $B-L$ symmetry at a much higher scale.

The $B-L$ MSSM does not explicitly contain a Bino, associated with the hypercharge group $U(1)_Y$.  Instead, it contains a Blino $\tilde B^'$ and a Rino $W_R$, the gauginos associated with the symmetry groups $U(1)_{B-L}$ and $U(1)_{3R}$, respectively. Nevertheless, the theory does effectively contain a Bino. This is most easily seen using the following approximation. Let us consider the limit $M_{W^\pm}^2,\> M_{Z^0}^2\ll M_{R}^{2}, \> M_{2}^{2},\> M_{BL}^{2},\mu^{2}$ --- that is, when the 
EW scale is much lower than the 
soft SUSY breaking scale so that the Higgs VEV's are negligible. Note that $\mu^{2}$ appears in these inequalities since, as discussed in \cite{Ovrut:2015uea}, it must be on the order of the soft SUSY breaking Higgs parameters $m_{H_{u}}^{2},m_{H_{d}}^{2}$ to solve the ``little hierarchy problem''. In this limit, the mass matrix in eq. \eqref{eq:neutralinoMassMatrixWithoutEpsilon} becomes
\begin{equation}
M_{{\tilde \chi}^0}=
\left(
\begin{matrix}
M_R&0&0&0&0&-\frac{1}{ 2}g_Rv_R\\
0&M_2&0&0&0&0\\
0&0&0&-\mu&0&0\\
0&0&-\mu&0&0&0\\
0&0&0&0&M_{BL}&\frac{1}{{2}}g_{BL}v_R\\
-\frac{1}{{2}}g_Rv_R&0&0&0&\frac{1}{{2}}g_{BL}v_R&0\\
\end{matrix}
\right)
\label{wow1}
\end{equation}
The first, fifth, and sixth columns, corresponding to the Blino, the Rino and the third generation right-handed neutrino, are now decoupled from the other three states and mix only with each other. In the reduced basis $\left({\nu}_3^c, \tilde W_R, \tilde B^\prime \right)$, the off-diagonal mass matrix is
\begin{equation}
\left(
\begin{matrix}
0&-\cos \theta_R M_{Z_R}&\sin \theta_R M_{Z_R}\\
-\cos \theta_R M_{Z_R}&M_R&0\\
\sin \theta_R M_{Z_R}&0&M_{BL}\\
\end{matrix}
\right)
\end{equation}
with
\begin{equation}
M_{Z_{R}}=\frac{1}{2}\sqrt{g_{R}^{2}+g_{BL}^{2}}~ v_{R}~~,~~\cos \theta_R = \frac{g_R}{\sqrt{g_R^2+g_{BL}^2}} \ .
\end{equation}
Note that the experimental lower bound on $M_{Z_R}$ is much higher than the typical physical gaugino mass lower bounds. This mass hierarchy is also motivated theoretically because RG running makes the gauginos masses lighter than $M_{Z_R}$; that is, $M^2_{R},M^2_{BL} \ll M^2_{Z_R}$. See  \cite{Ovrut:2015uea} for details. Taking this limit, the mass eigenstates and eigenvalues can be found as an expansion in the gaugino masses. To {\it zeroth order}, the mass eigenstates are
\begin{equation}
{\tilde B}=\tilde W_R \sin \theta_R+\tilde B^\prime \cos \theta_R \ ,
\label{train1}
\end{equation}
\begin{equation}
{\nu}_{3a}^c=\frac{1}{\sqrt 2}({\nu^c}_3-\tilde W_R \cos \theta_R+\tilde B^\prime 
\sin \theta_R) \ ,
\end{equation}
\begin{equation}
{\nu}_{3b}^c=\frac{1}{\sqrt{2}}({\nu^c}_3+\tilde W_R \cos \theta_R-
\tilde B^\prime \sin \theta_R) \ .
\end{equation}
Note that, to leading order, \eqref{train1} defines the ``Bino'' in terms of $W_R $ and $\tilde B^\prime$.
The associated mass eigenvalues, calculated to leading order, are given by
\begin{equation}\label{m1eqn}
M_1=\sin^2 \theta_R M_R+\cos^2 \theta_R M_{BL},
\end{equation}
for the Bino, and
\begin{equation}
 m_{{\nu^c}_{3a}}=M_{Z_R},
\quad m_{{\nu^c}_{3b}}=M_{Z_R} \ .
\end{equation}
for two species of massive right handed neutrinos.

\begin{figure}[t]
   \centering

   \begin{subfigure}[c]{0.49\textwidth}
\includegraphics[width=1.0\textwidth]{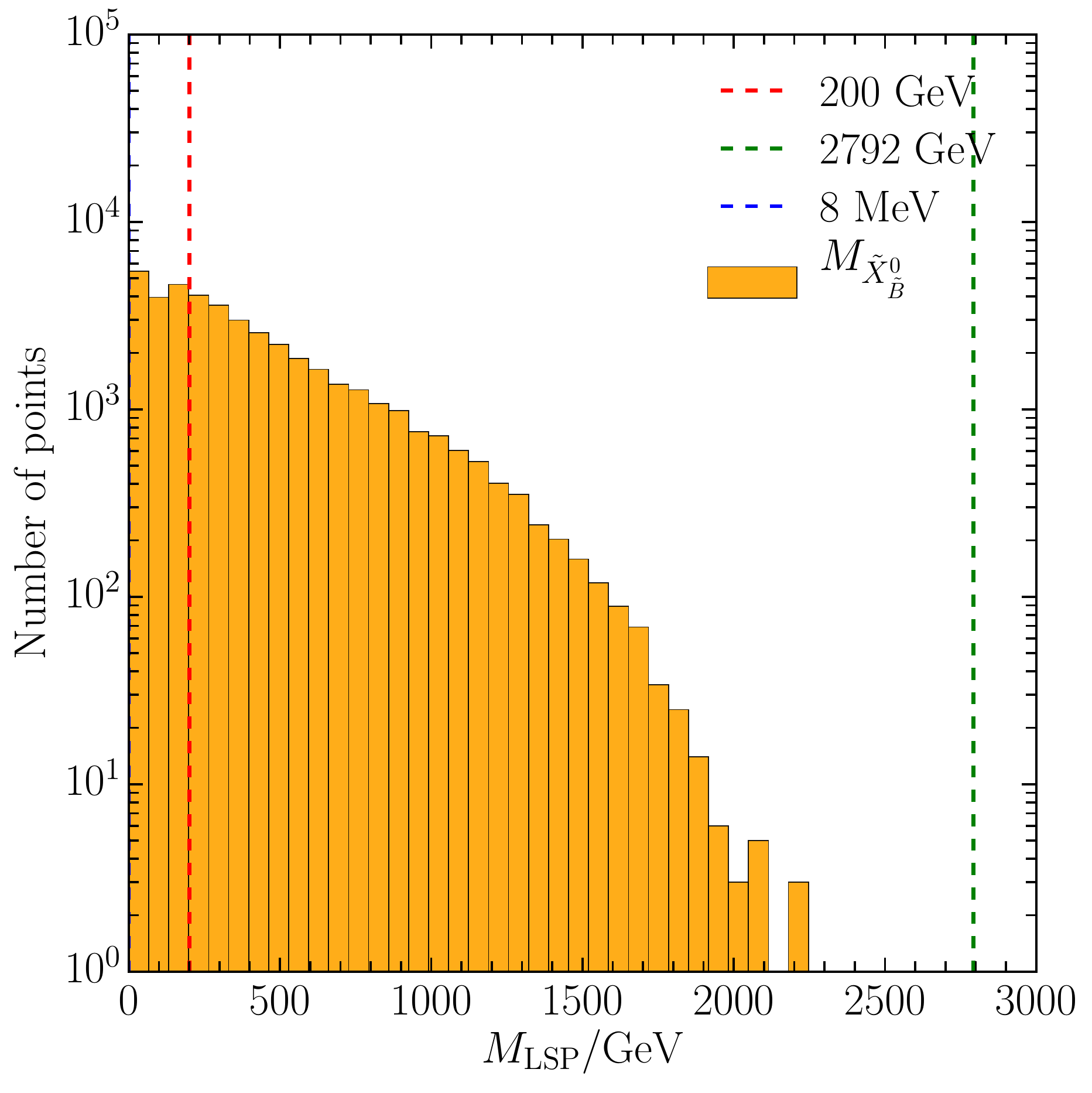}
\caption{}
\label{fig:mass_hist2}
\end{subfigure}
   \begin{subfigure}[c]{0.49\textwidth}
\includegraphics[width=1.0\textwidth]{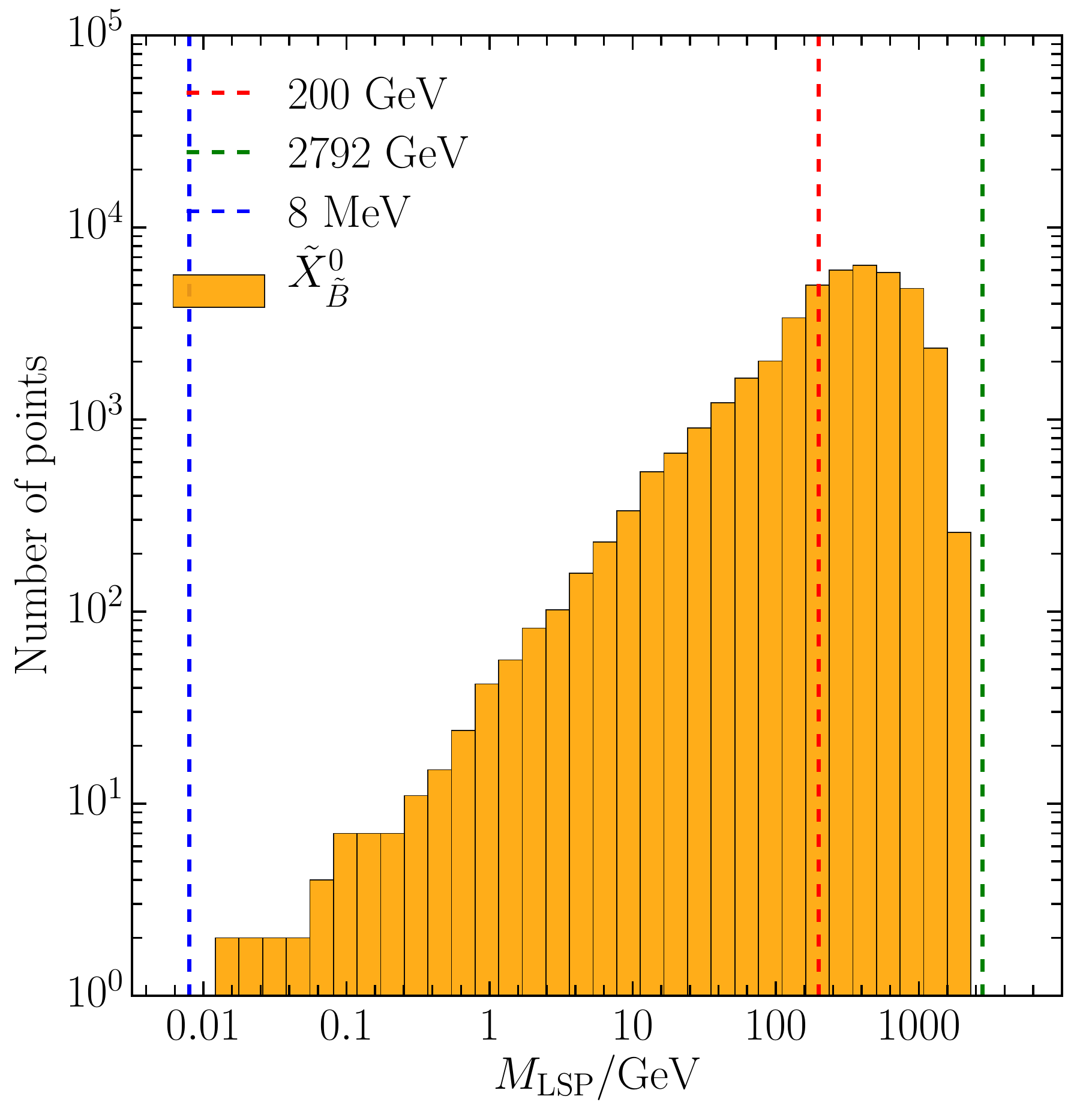}
\caption{}
\label{fig:mass_hist1}
\end{subfigure}
\caption{ The distribution of the Bino neutralino LSP masses for the 42,039 valid black points, shown with linear (a) and logarithmic (b) mass scales. The masses range from 8 MeV to 2792 GeV. Each of the boundary masses occurs only once out of the 42,039 valid points and, hence, they cannot be seen in the histogram.}
\end{figure}

Having defined the Bino, ${\tilde B}$, as well as the ${\nu}_{3a}^c$ , ${\nu}_{3b}^c$ states using the above approximations, we now return to the full gauge basis 
$\psi^0=\left( \tilde{W}_R, \tilde{W}_0, \tilde{H}_d^0,\tilde{H}_u^0, \tilde{B}^\prime,{\nu}^c_3 \right)$ and numerically diagonalize the complete mass matrix $M_{{\tilde \chi}^0}$ given in \eqref{eq:neutralinoMassMatrixWithoutEpsilon}-- without any approximations.
The mass eigenstates are related to the gauge states by the unitary matrix $N$ where ${\tilde{\chi }}^{0}=N \psi^{0}$. $N$ is chosen so that

\begin{equation}\label{eq:neutralino_states}
N^*M_{\tilde{{ \chi}}^0}N^{\dag}= M_{\tilde{{ \chi}}^0}^{D} =
\left(
\begin{matrix}
M_{{\tilde \chi}^0_1}&0&0&0&0&0\\
0&M_{{\tilde \chi}^0_2}&0&0&0&0\\
0&0&M_{{\tilde \chi}^0_3}&0&0&0\\
0&0&0&M_{{\tilde \chi}^0_4}&0&0\\
0&0&0&0&M_{{\tilde \chi}^0_5}&0\\
0&0&0&0&0&M_{{\tilde \chi}^0_6}\\
\end{matrix}
\right) \ ,
\end{equation}
where all eigenvalues are positive.  After diagonalizing the neutralino mass matrix, one obtains six neutralino mass eigenstates, $\tilde \chi_n^0$ with $n=1,2,3,4,5,6$. Unlike for charginos, the label $n$ does not automatically imply any mass ordering; for example, the $\tilde \chi_1^0$ neutralino is not necessarily the lightest. Each of the six neutralinos $\tilde \chi_n^0$ is a superposition of a Rino $\tilde W_R$, a Wino $\tilde W_2$, two neutral Higgsinos $\tilde H_d^0$, $\tilde H_u^0$, a Blino $\tilde B^'$ and a third generation right handed neutrino $\nu_3^c$.  In the theoretical context we work in,  the off-diagonal terms are much smaller than the diagonal ones. This allows one to determine which component dominates in each of the neutralino states $\tilde \chi_n^0$. We find that $\tilde \chi_1^0$ has a dominant Bino $\tilde B$ component, $\tilde \chi_2^0$ has a dominant Wino $\tilde W$ component, $\tilde \chi_{3,4}^0$ have dominant Higgsino $\tilde H_u^0, \> \tilde H_d^0$ components and $\tilde \chi_{5,6}^0$ have a dominant right-handed neutrino $\nu_3^c$ component. Therefore, we use the notation
\begin{equation}
{\tilde \chi}_1^0={\tilde \chi}_B^0,\quad  {\tilde \chi}_2^0={\tilde \chi}_W^0, \quad {\tilde \chi}_3^0={\tilde \chi}_{H_d}^0,
\quad {\tilde \chi}_4^0={\tilde \chi}_{H_u}^0, \quad {\tilde \chi}_5^0={\tilde \chi}_{\nu_{3a}}^0, \quad {\tilde \chi}_6^0={\tilde \chi}_{\nu_{3b}}^0
\end{equation}
to express which component dominates in each neutralino state. Although it is helpful to display the dominant component in each neutralino, in our calculations we use the {\it exact} content of each neutralino state. This is computed numerically, after diagonalizing the neutralino mass mixing matrix.

Our discussion thus far was carried out in the absence of the RPV couplings, which are central to our theory. These couple the gaugino, Higgsino and the third generation right-handed neutrino states to the three generations of left-handed neutrinos $\nu_1$, $\nu_2$, $\nu_3$. Therefore, with the RPV extension,  the gauge eigenstate basis is enlarged from  $\psi^0=\left( \tilde{W}_R, \tilde{W}_0, \tilde{H}_d^0
\tilde{H}_u^0, \tilde{B}^\prime,{\nu}^c_3 \right)$  to  $\Psi^0=\left( \tilde{W}_R, \tilde{W}_0, \tilde{H}_d^0,
\tilde{H}_u^0, \tilde{B}^\prime,{\nu}^c_3, \nu_1, \nu_2, \nu_3\right)$ and the neutralino mass matrix becomes $9 \times 9$. After diagonalization, the three generations of left-handed neutrinos receive non-zero Majorana masses, a process carefully outlined in \cite{Dumitru:2018jyb,Dumitru:2018nct}. The original six neutralino eigenstates, on the other hand, each receive a negligibly small RPV contribution containing the three left-handed neutrinos. For example, 
the RPV-extended Bino neutralino mass eigenstate $\tilde \chi_B^0$ is now a linear combination of nine gauge eigenstates; three gaugino states, two Higgsino states, a third generation right-handed neutrino and three left-handed neutrinos. The left-handed neutrino contributions to the Bino neutralino eigenstate and mass are negligible, since they have been introduced via small RPV couplings. Therefore, the Bino neutralino continues to generically have a strongly dominant Bino component $\tilde \chi_B^0\simeq {\tilde{B}}$. Expanding, as discussed above, in the limit that $M_{W^\pm}^2,\> M_{Z^0}^2\ll M_{R}^{2}, \> M_{2}^{2},\> M_{BL}^{2},\>\mu^{2}$-- that is, when the 
EW scale is much lower than the 
soft SUSY breaking scale--but now to  {\it first order}, we find that the Bino mass $M_{{\tilde \chi}^0_B}$ is given by
\begin{equation}\label{eq:Bino_mass}
M_{{\tilde \chi}^0_B} \simeq |M_1|-\frac{M_{Z^0}^2\sin^2 \theta_W(M_1+\mu \sin 2\beta)}{\mu^2-M_1^2} \ .
\end{equation}

In the second term in eq. \eqref{eq:Bino_mass}, the mass $\mu$ is of the order of the soft SUSY breaking mass parameters $m_{H_{u}}^{2},m_{H_{d}}^{2}$ to solve the ``little hierarchy problem''. Statistically, this is always much larger than the mass of the $Z^0$ boson in the numerator. 
Therefore, the mass of the Bino neutralino $\chi_B^0$ is approximately equal to $|M_1|$.
As discussed in \cite{Ovrut:2015uea}, the viable black points must satisfy the inequalities $M_{R}^2,M_{BL}^2 \ll M_{Z_R}^2$ to be consistent with low energy data. To lowest order, therefore, expression \eqref{m1eqn} is  a good approximation to the mass $M_{1}$. Generically, therefore, Bino LSP masses are expected to lie in the same the interval as $|M_{R}|$ and $|M_{BL}|$; that is  $\big[200~{\rm GeV}, 10~{\rm TeV}\big]$. That this is {\it generically} the case can be seen in Figure \ref{fig:mass_hist2}. However, there is a {\it very important caveat} to this statement. Note that 
$M_{R}$ and $M_{BL}$ do not enter expression \eqref{m1eqn} for $M_{1}$ as absolute values; that is, the only constraint on these mass terms in \eqref{m1eqn}  is that they be real--however, they can be either positive or negative. This leaves open the possibility, albeit more unlikely, that the terms in eq. \eqref{m1eqn} can almost, or even exactly, cancel. In such cases, one would obtain small Bino mass terms where $|M_1|<M_{W^\pm}$,  and lead to Bino LSP masses smaller than the EW scale. That such cancellations can indeed occur is shown in Figure \ref{fig:mass_hist1}. Note that such light Bino LSPs can only decay via suppressed off-shell processes, leading to interesting experimental signatures at the LHC.
 However, these light Bino LSPs are statistically much less probable. In Figure \ref{fig:mass_hist1} we see that Bino masses much smaller than 200 GeV are less and less likely for the range of scanned parameters. The analysis presented involves $10^{8}$ initial statistical samples, leading to 42,039 black points with Bino LSPs. For this sample, the smallest and largest Bino masses we find are 8 MeV and 2792 GeV respectively. However, a larger statistical sample can lead to much smaller, and larger, Bino LSP masses.
Note that the existence of very light Bino LSPs is exciting from a cosmological point of view, since it offers a possible dark matter candidate. 

In contrast, the masses of other neutralino species cannot become arbitrarily small. This is the case of the Wino neutralino and the Wino chargino, for example. Wino neutralinos and Wino charginos have a dominant Wino component. To first order, the masses of these sparticles are equal, given by
\begin{equation}\label{eq:Wino_Neutralino_mass}
M_{{\tilde \chi}^0_W}=M_{{\tilde \chi}^\pm_W} \simeq |M_2|-\frac{M_{W^\pm}^2(M_2+\mu \sin 2\beta)}{\mu^2-M_2^2} \ .
\end{equation}
Including the higher-order terms, the masses split and form almost degenerate pairs. Similarly as in eq. \eqref{eq:Bino_mass}, the second term in eq. \eqref{eq:Wino_Neutralino_mass} is very small compared to the leading term $|M_2|$, because the mass $\mu$ in the denominator must be much larger than the mass $M_{W^\pm}$ in the numerator. Therefore, the masses of the Wino chargino and the Wino neutralino are both approximately equal to $|M_2|$.
The mass of the Wino gaugino $M_2$ is introduced in the soft SUSY-breaking Lagrangian and, hence, we sample its absolute value in the interval $\big[200~{\rm GeV}, 10~{\rm TeV}\big]$. It is, therefore, fixed to be of the order of the SUSY breaking scale. Unlike the Bino gaugino mass term $M_1$,  soft mass parameter $M_2$ cannot get arbitrarily small. Hence, the masses of the Wino neutralinos and Wino charginos cannot get lower than the EW scale.

Finally, we note that even though the RPV left-handed neutrino components have a negligible contribution to the Bino neutralino eigenstate and mass, they remain central to our study of the RPV decays of the Bino neutralino regardless of its mass. From now on, we will use the 4-component spinor notation for the Bino neutralino state, which in terms of the Bino neutralino Weyl spinor, $\tilde \chi_B^0$ is given by
\begin{equation}
\tilde X^0_B=
\left(
\begin{matrix}
\tilde \chi_B^0\\
\tilde \chi_B^{0 \dag}
\end{matrix}
\right).
\end{equation}

\subsection{Bino Neutralino LSP RPV Decays with On-Shell  $W^\pm, Z^0, h^0$ Bosons }

We begin by studying the RPV decays of a Bino LSP with {\it mass greater than the electroweak scale} to standard model particles. In the $B-L$ MSSM model, such Bino neutralino LSPs can only have RPV decays into an on-shell massive boson and a lepton. The three possible decay channels, $\tilde X^0_B\rightarrow W^\pm \ell_i^\mp$,  $\tilde X^0_B\rightarrow Z^0 \nu_i$,  $\tilde X^0_B\rightarrow h^0 \nu_i$ for $i=1,2,3$, are shown in Figure \ref{fig:NeutralinoDecays}. Note, however, that all of these decay channels become forbidden at tree level if the mass of the Bino LSP is smaller than the mass of the lightest of the three boson species; that is, the $W^\pm$. 

\begin{figure}[t]
 \begin{minipage}{1.0\textwidth}
     \centering
   \begin{subfigure}[b]{0.31\linewidth}
   \centering
\includegraphics[width=0.8\textwidth]{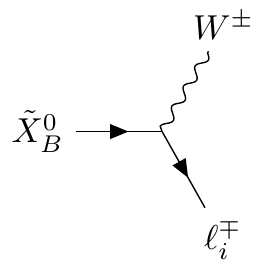}

\caption*{${\tilde X}^0_B\rightarrow W^\pm \ell^\mp_{i}$}
       \label{fig:table2}
   \end{subfigure} 
   \centering
   \begin{subfigure}[b]{0.31\linewidth}
   \centering
\includegraphics[width=0.8\textwidth]{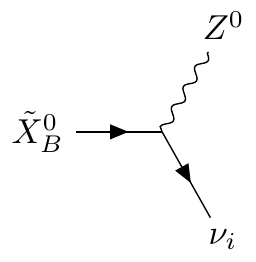}
 
\caption*{ ${\tilde X}^0_B\rightarrow Z^0 \nu_{i}$}
       \label{fig:table2}
\end{subfigure}
   \centering
     \begin{subfigure}[b]{0.31\textwidth}
   \centering
\includegraphics[width=0.8\textwidth]{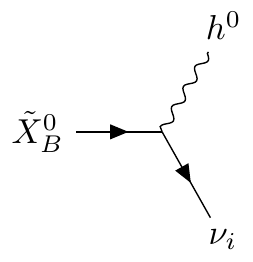}

\caption*{ ${\tilde X}^0_B\rightarrow h^0 \nu_{i}$}
\end{subfigure}
\end{minipage}
\caption{RPV decays of a general massive Bino neutralino $\tilde X_B^0$. There are three possible channels, each with $i=1,2,3$, that allow for Bino neutralino LSP decays. The decay rates into each individual channel were calculated analytically Section \ref{sec:7}.}\label{fig:NeutralinoDecays}
\end{figure}

\subsection{Branching ratios of the decay channels}

In this section, we analyze the RPV decay signatures of Bino neutralino LSPs with masses larger that of the $W^\pm$ boson. We will follow the methods used in our previous  study of Wino chargino and Wino neutralino RPV LSP decays~\cite{Dumitru:2018nct}. Note than in that study, the LSP masses were all found to be at least 200 GeV, so that only decays to on-shell bosons were considered. Of the three Bino decay channels, the  ${\tilde X}^0_B\rightarrow W^\pm \ell^\mp_{i}$
process provides an excellent target for LHC searches, since the final state can be fully reconstructed within the ATLAS detector. In the other two Bino decay processes, the left-handed neutrinos produced via ${\tilde X}^0_B\rightarrow Z^0 \nu_i$ and ${\tilde X}^0_B\rightarrow h^0 \nu_i$ can only be inferred through the presence of missing energy. Hence, the most experimentally clean signature appears to be the Bino neutralino decay into a $W^\pm$ massive boson and a charged lepton.

The relative abundance of each channel is presented in terms of the associated Bino decay branching ratio. For example, for the process ${\tilde X}^0_B\rightarrow W^\pm \ell^\mp$, the branching ratio is defined to be
\begin{equation}
\text{Br}_{{\tilde X}^0_B\rightarrow W^\pm \ell^\mp}=\frac{\sum_{i=1}^{3}  \Gamma_{ {\tilde X}^0_B\rightarrow W^\pm \ell^\mp_{i}}}{\sum_{i=1}^3 \Big( \Gamma_{{\tilde X}^0_B\rightarrow Z^0 \nu_i}+  \Gamma_{{\tilde X}^0_B\rightarrow W^\pm \ell^\mp_i}+\Gamma_{{\tilde X}^0_B\rightarrow h^0 \nu_i}\Big)}  \ ,
\label{fun1}
\end{equation}
In this section, we study the decay patterns and branching ratios for each for the 3 decay channels of the Bino neutralino. As discussed above, there are 42,039 valid black points associated with Bino neutralino LSPs. 
In the present analysis, we retain only the black points with LSPs whose masses are larger than that of the $W^\pm$ bosons.
For each of these, we compute the decay rates via RPV processes, using the expressions we calculated in Section \ref{sec:7} for the neutralino decay rates, using $n=1$. 
 \begin{figure}[!ht]
   \centering
   \begin{subfigure}[b]{0.99\textwidth}
\includegraphics[width=1.0\textwidth]{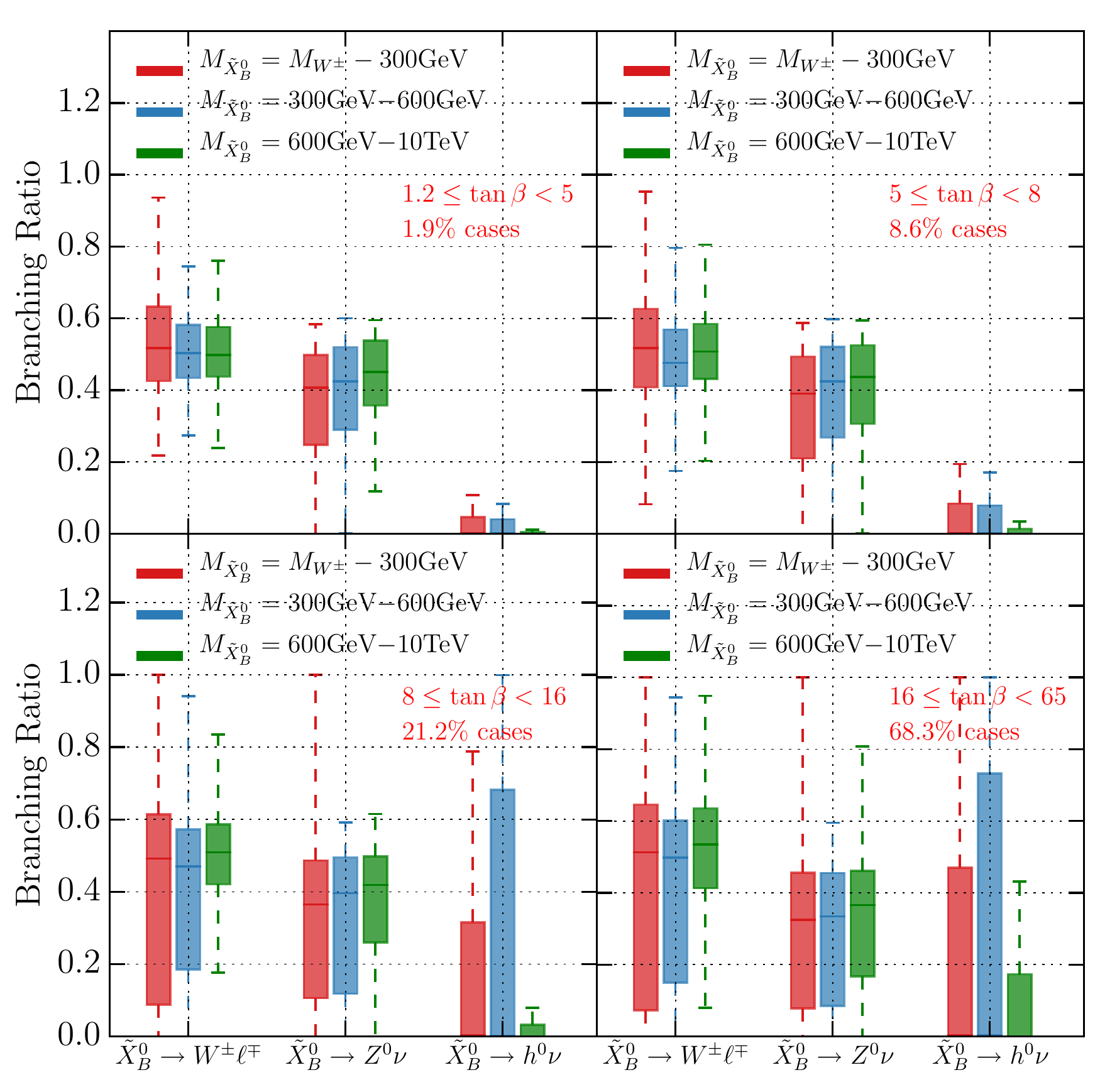}
\end{subfigure}
    \caption{Branching ratios for the three possible decay channels of a Bino neutralino LSP with mass $M_{{\tilde{X}}_{B}^{0}} \geq M_{W^{\pm}}$ divided over three mass bins and four $\tan \beta$ regions. The colored horizontal lines inside the boxes indicate the median values of the branching fraction in each bin, the boxes indicate the interquartile range, while the dashed error bars show the range between the maximum and the minimum values of the branching fractions. The case percentage indicate what percentage of the physical mass spectra have $\tan \beta$ values within the range indicated.  We assumed a normal neutrino hierarchy, with $\theta_{23}=0.597$. Note that the median values of the ${\tilde X}^0_B\rightarrow h^0 \nu$ decay channel approaches zero for all mass ranges and all values of $\tan \beta$.}
\label{fig:bar_plot2}
\end{figure}
The branching ratios to each channel take different values for every valid point in our simulation. We compute the median values, interquartile ranges and the minimum and maximum values of the branching fractions 
using the same categories of events as employed in our previous section for Wino charginos and Wino neutralinos \cite{Dumitru:2018nct}.
Specifically, we sample the average branching fractions in the three bins for the LSP mass $M_{{\tilde X}_B^0} \in [M_{W^\pm}, 300],\>[300,600],\>[600,10^{4}]$~GeV, and in the four intervals for $\tan \beta \in [1.2,5],\> [5,8],\>[8,16], \>[16,65]$. The results are presented in Figure \ref{fig:bar_plot2}. To carry out the explicit calculations, we have chosen a normal neutrino hierarchy with $\theta_{23}=0.597$. We find that assuming an inverted neutrino hierarchy instead produces only minimal changes to these results, while the exact value of $\theta_{23}$ is statistically irrelevant.

It was found--see Figure \ref{fig:bar_plot2}--that the median value of the ${\tilde X}^0_B\rightarrow h^0\nu$ decay channel approaches zero for every mass range and bin for $\tan \beta$. Although the distributions of the branching fractions are fairly broad, we find that they peak very strongly around the median values. It follows that the decay channel is generally  subdominant in all regions of the parameter space. Unlike for the case of Wino charginos and Wino neutralinos, however, we find that $\tan \beta$ has only minimal impact on the experimental predictions.
  While the full expressions for the decay rates are complicated, simplifying assumptions can allow for a better understanding of the relative results.

  One such assumption is that the soft breaking terms have much larger magnitudes than the electroweak scale. This renders the Bino neutralino to be almost purely neutral Bino. Furthermore, the fact that the charged lepton masses are much smaller than the soft breaking parameters further simplifies the equations. Using these approximations, one obtains the following simplified formulas for the decay rates. They are given by

\begin{multline}\label{eq:decay_neut1}
\Gamma_{{\tilde X}^0_B\rightarrow Z^0\nu_{i}} \approx
\frac{g_2^2}{16\pi c_W^2}\Big(  
\sin \theta_R 
\Big[ \frac{g_RM_{BL}v_u}{M_1 v_R^2}\epsilon_i +\frac{g_Rg_{BL}^2}{4M_1 \mu}(v_d \epsilon_i+\mu v_{L_i}^*)\Big]\left[ V_{\text{PMNS}}  \right]^\dag_{ij}\\
-\cos \theta_R\Big[\frac{g_R^2g_{BL}}{4M_1 \mu}(v_d \epsilon_i+\mu v_{L_i}^*) - \frac{g_{BL}v_uM_R}{M_1v_R^2}\epsilon_i\Big]\left[ V_{\text{PMNS}}  \right]^\dag_{ij}
\Big)^2
\frac{M_{{\tilde X}_B^0}^3}{M_{Z^0}^2}\left(1-\frac{M_{Z^0}^2}{M_{{\tilde X}_B^0}^2}\right)^2
\left(1+2\frac{M_{Z^0}^2}{M_{{\tilde X}^0_B}^2}\right) \ ,
\end{multline}
\begin{multline}
\Gamma_{{\tilde X}^0_B\rightarrow W^\mp \ell_i^\pm} \approx\frac{g_2^2}{32\pi}\Big(\sin \theta_R 
\Big[ \frac{2g_RM_{BL}v_u}{M_1 v_R^2}\epsilon_i +\frac{g_Rg_{BL}^2}{2M_1 \mu}(v_d \epsilon_i+\mu v_{L_i}^*)\Big]\\
-\cos \theta_R\Big[\frac{g_R^2g_{BL}}{2M_1 \mu}(v_d \epsilon_i+\mu v_{L_i}^*) - \frac{2g_{BL}v_uM_R}{M_1v_R^2}\epsilon_i\Big]
\Big)^2 \times
\frac{M_{{\tilde X}_B^0}^3}{M_{W^\pm}^2}\left(1-\frac{M_{W^\pm}^2}{M_{{\tilde X}_B^0}^2}\right)^2
\left(1+2\frac{M_{W^\pm}^2}{M_{{\tilde X}_B^0}^2}\right) \ ,
\end{multline}
\begin{equation}\label{eq:decay_neut4}
\Gamma_{{\tilde X}^0_B\rightarrow h^0\nu_{i}} \approx\frac{{g_2}^2}{64\pi}\Big( \sin \alpha (\cos^2 \theta_R-\sin^2 \theta_R)\left[ V_{\text{PMNS}}  \right]^\dag_{ij} \frac{\epsilon^*_j }{\mu} \Big)^2
M_{{\tilde X}_B^0}\left(1-\frac{M_{h^0}^2}{M_{{\tilde X}_B^0}^2}\right)^2 \ .
\end{equation}
The notation and derivation of these decay rates is outlined in more detail in \cite{Dumitru:2018jyb}. We learn that the approximate decay rate for the ${\tilde X}^0_B\rightarrow h^0\nu_{i}$ has an effective coupling proportional to $\frac{{g_2}^2}{64\pi}\Big( \sin \alpha (\cos^2 \theta_R-\sin^2 \theta_R)\Big)^2$. In our theory, $\tan \theta_R=g_{BL}/g_R$ is approximately equal to one. Therefore, $\sin \theta_R \approx \cos \theta_R$, which explains why this channel is subdominant in Bino decays. Furthermore, the expressions for the decay rates of the ${\tilde X}^0_B\rightarrow W^\mp \ell_i^\pm$ and the ${\tilde X}^0_B\rightarrow Z^0\nu_{i}$ channels contain terms that do not depend on $v_d=174~ \text{GeV}/(1+\tan \beta)$. Therefore, we do not observe the suppression of these channels for high $\tan \beta$ values, as is the case for the Wino neutralinos decays presented in \cite{Dumitru:2018nct}.

\subsection*{Decay length}

Knowing the branching ratios of the Bino LSP RPV decay channels does not offer a complete picture of the signals that such particle decays can produce in the detector. We further need to analyze the 
{\it overall} decay length $L$ of the Bino LSP in the frame of the detector, which we define to be
\begin{equation}
L=c~\times~\frac{1}{\Gamma}~,\quad \Gamma=\sum_{i=1}^{3} \Big( \Gamma_{{\tilde {X}}_B^0\rightarrow Z^0 \nu_{i}} + \Gamma_{{\tilde {X}}_B^0 \rightarrow W^\pm \ell^\mp_{i}}  + \Gamma_{{\tilde X}_{B}^0\rightarrow h^0 \nu_{i}} \Big) \ .
\label{Mary1}
\end{equation}
 \begin{figure}[t]
\begin{subfigure}[t]{0.49\textwidth}
\includegraphics[width=1.\textwidth]{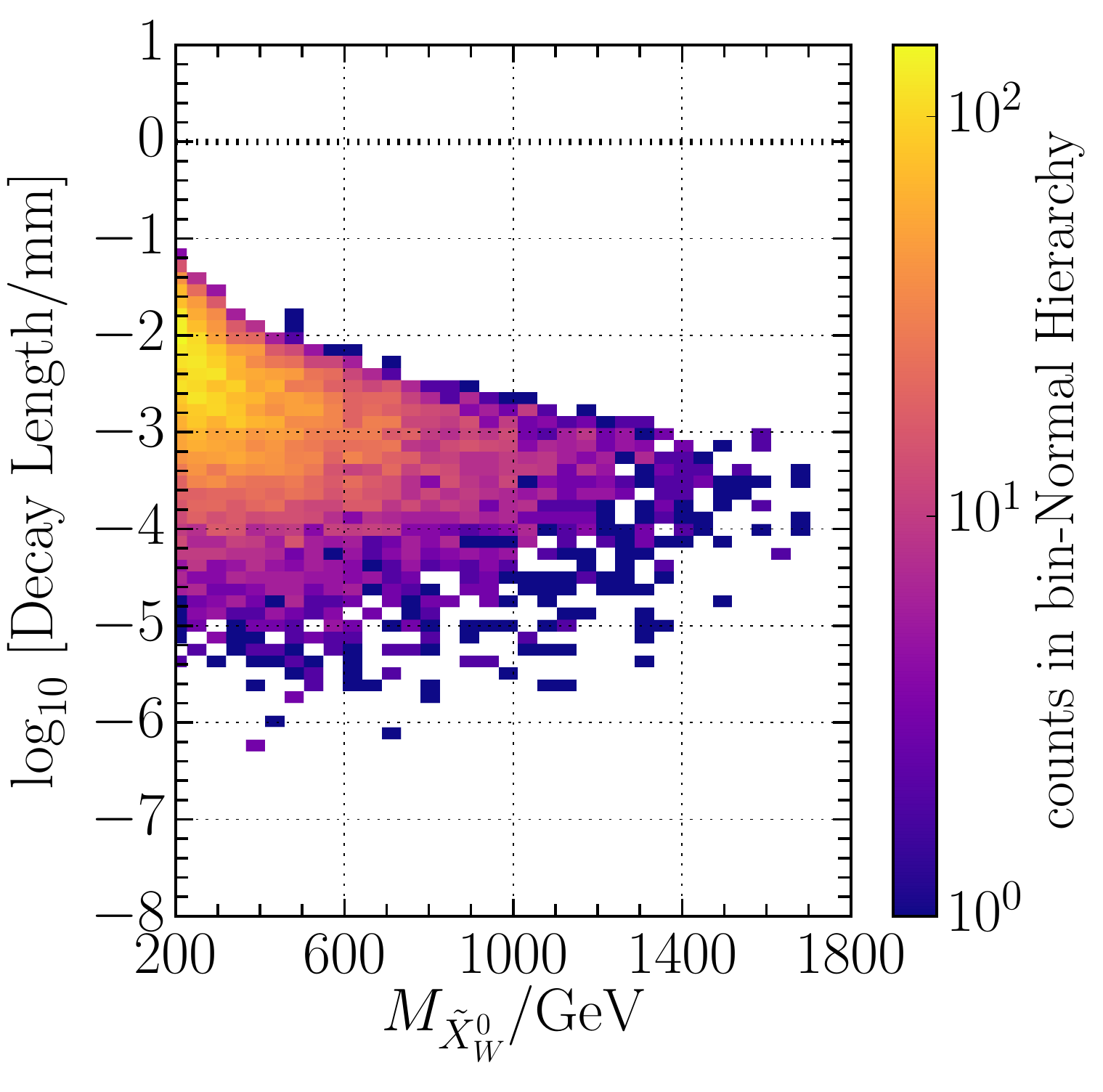}
\end{subfigure}
\begin{subfigure}[b]{0.49\textwidth}
\includegraphics[width=1.\textwidth]{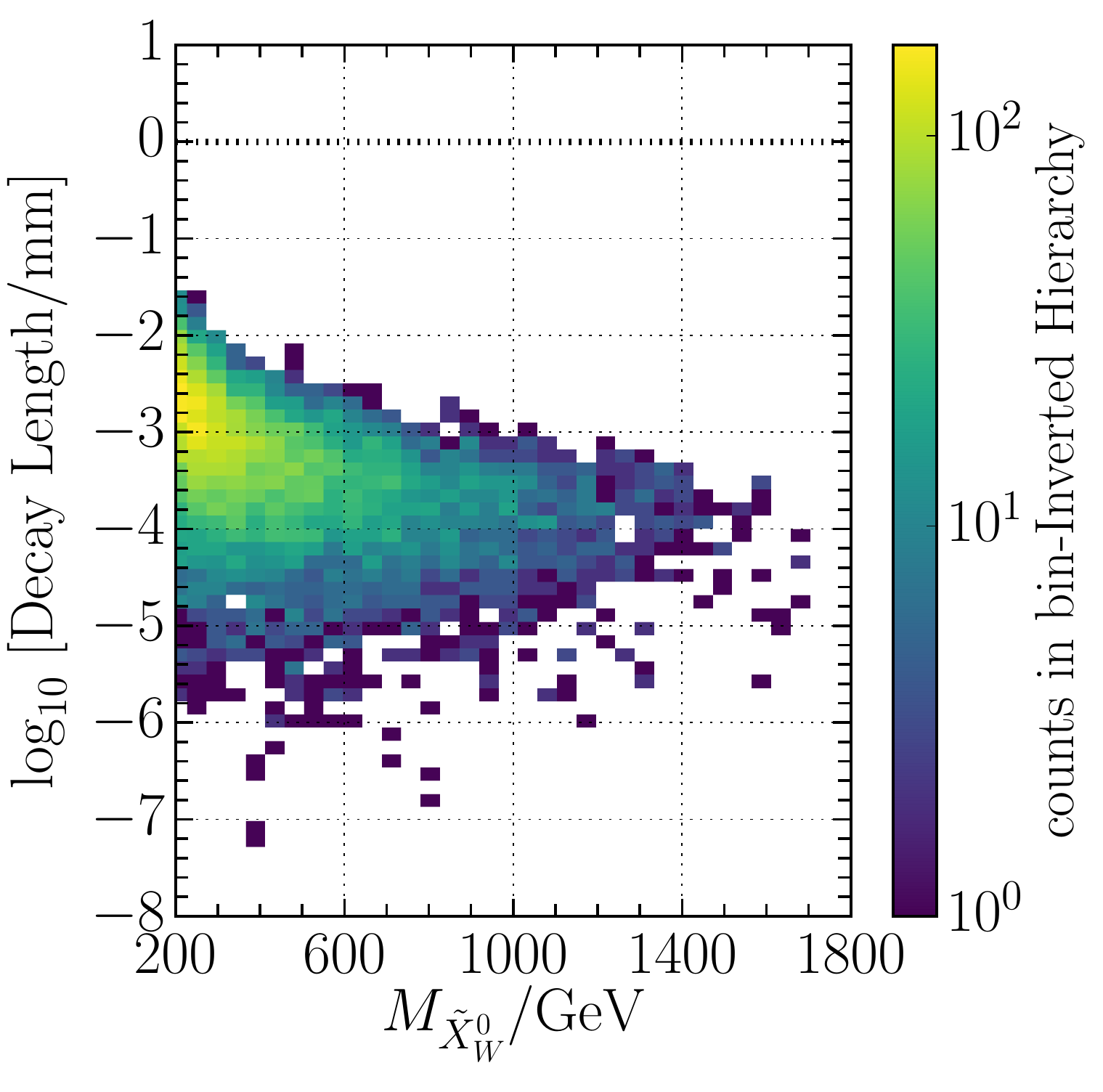}
\end{subfigure}
\caption{Bino neutralino LSP decay length in millimeters, for the normal and inverted hierarchies, summed over the three decay channels. The average decay length $L=c\times \frac{1}{\Gamma}$ decreases for larger values of $M_{\tilde X^0_B}$. We have chosen $\theta_{23}=0.597$ for the normal neutrino hierarchy and $\theta_{23}=0.529$ for the inverted hierarchy. However, the choice of $\theta_{23}$ has no impact on the decay length. The dashed blue line represents the 1~mm decay length, at and below which the decays are ``prompt''. Note that 1) both Figures begin at $M_{W^{\pm}}$ on the left-hand side and 2) there are no points above approximately 2300~GeV. This follows from the fact that, even though the maximum value we obtained for the Bino mass is 2792~GeV, points higher than 2300~Gev are statistically insignificant. }\label{fig:LSPprompt_neut}
\end{figure}

$L=\frac{c}{\Gamma}$ is the typical decay length of the Bino.
It is conventional at the LSP to divide sparticle decay lengths into four catagories. Defining $\tau=\frac{1}{\Gamma}$, these are:

\begin{itemize}
\item{\textit{Prompt decays:} where finite-lifetime effects are experimentally negligible--that is, they do not impact the efficiency of charged lepton reconstruction used by standard analyses. Prompt decays satisfy:~ $c\tau<1$~mm.}
\item{\textit{Displaced vertex} decays: where a \textit{secondary} Bino decay vertex may be identified via charged particle tracking, separate from that of the initial $pp$ interaction. Displaced vertex decays satisfy:~ $1~\text{mm}<c\tau<30$~cm.}

\item{\textit{Decays within the detector:} but outside the tracking apparatus, where measurements made in the muon system may allow observation of the decay. Decays in the detector, but outside the tracking apparatus satisfy:~ $30~\text{cm}<c\tau<10$~m.}

\item{\textit{Detector-stable decays:} where the lifetime is long enough that the only detector signature of the particle is momentum imbalance, that is, ``missing energy''.  Detector-stable decays satisfy:~ $10~m<c\tau$.}

\end{itemize}

In reality, search strategies focusing on each of these four cases overlap in sensitivity, partly due to the probabilistic variation in lifetimes of the individual particles produced in $pp$ collisions.
Dedicated searches for long-lived particles have recently been conducted by the ATLAS and CMS Collaborations, searching for displaced charged-particle vertices~\cite{Aaboud:2017iio,Sirunyan:2018vlw}, displaced charged-lepton pairs~\cite{Aad:2015rba,CMS:2014hka}, and displaced jets decaying in the ATLAS muon spectrometer~\cite{Aaboud:2018aqj}.
ATLAS has also studied the complementarity of searches targeting the production of promptly decaying, long-lived, and stable BSM particles to models predicting a wide range of lifetimes~\cite{ATLAS-CONF-2018-003}.
However, the simplified categorization presented above is sufficient for our analysis, indicating the most promising approaches for searching for long-lived Bino LSPs in a given mass range.

\noindent Figure 6 shows that Bino neutralino LSP RPV decays are generally prompt,
with the vast majority of decay lengths found to be less than 1~mm.
Therefore, the Bino decay products may be identified in conventional collider searches without the need for specialized experimental techniques.
We observe that in the case of the inverted hierarchy, the decay lengths are generally smaller, since the values of the RPV couplings are somewhat larger, as explained above.

\begin{figure}[!ht]
\centering
\begin{subfigure}[b]{1.\textwidth}
\includegraphics[width=1.0\textwidth]{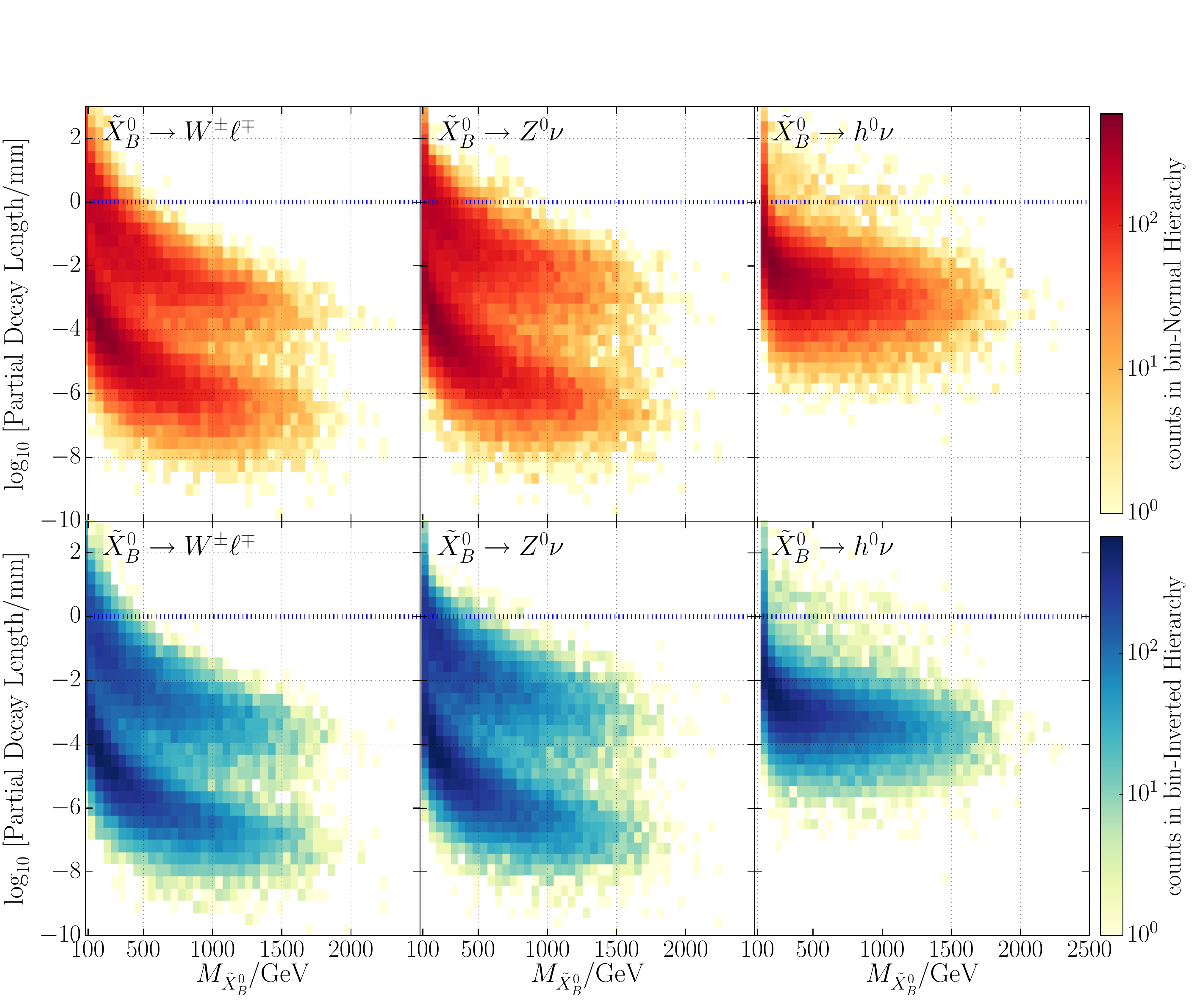}
\end{subfigure}
\caption{Bino neutralino LSP partial decay length	
	in millimeters, shown for the individual decay channels, for both normal and inverted hierarchies. We have chosen $\theta_{23}=0.597$ for the normal neutrino hierarchy and $\theta_{23}=0.529$ for the inverted hierarchy. The choice of $\theta_{23}$ has no impact on the decay length. The blue dashed line denotes a decay length of 1~mm, at and below which decays are ``prompt''. Note that on the x-axis 1) all Figures begin at $M_{W^{\pm}}$ on the left-hand side and 2) there are no points above approximately 2300~GeV. This follows from the fact that, even though the maximum value we obtained for the Bino mass is 2792~GeV, points higher than 2300~GeV are statistically insignificant. 
}
\label{fig:LSPprompt}
\end{figure}

Recall that the notion of decay length used above involved a {\it sum} over all three separate channels. It is, however, of interest to consider the decay length defined for each of the {\it individual} decay channels separately--although we continue to sum over the three lepton families. For example, the decay length for ${{\tilde {X}}_B^0\rightarrow Z^0 \nu}$ is given by
\begin{equation}
L_{{{\tilde {X}}_B^0\rightarrow Z^0 \nu}}=c~\times~\frac{1}{\sum_{i=1}^{3} \Gamma_{{{\tilde {X}}_B^0\rightarrow Z^0 \nu_{i}}}} \ .
\label{Mary2}
\end{equation}
In Figure \ref{fig:LSPprompt}, the decay length
for each of the three decay channels is shown separately. Note that, for both the normal and inverted neutrino hierarchies, the ${{\tilde {X}}_B^0\rightarrow Z^0 \nu}$  and ${\tilde X}^0_B\rightarrow W^\pm \ell^\mp$ processes have many more points with decay lengths much shorter than those of the ${\tilde X}^0_B\rightarrow h^0 \nu$ decays.
Since the decay rates are inversely proportional to decay length, it follows that Bino LSP RPV decays are predominantly through the ${{\tilde {X}}_B^0\rightarrow Z^0 \nu}$  and ${\tilde X}^0_B\rightarrow W^\pm \ell^\mp$ channels, consistent with the results in the previous subsection.

\subsection*{Lepton family production}

In this subsection, we study the correlation between the electroweak boson and the lepton {\it family} emitted in each of the possible Bino decays.
For example, to quantify the probability to observe an electron $e^\mp$ in the ${\tilde X}^0_B\rightarrow W^\pm \ell^\mp$ process, over a muon $\mu^\mp$ or a tauon  $\tau^\mp$, we compute the relative branching fraction

\begin{figure}[t]
   \centering

   \begin{subfigure}[b]{0.44\textwidth}
\includegraphics[width=1.0\textwidth]{ScatterW_LSP.png}
\end{subfigure}
   \begin{subfigure}[b]{0.44\textwidth}
\includegraphics[width=1.0\textwidth]{ScatterZ_LSP.png}
\end{subfigure}\\
   \begin{subfigure}[b]{0.44\textwidth}
\includegraphics[width=1.0\textwidth]{ScatterH_LSP.png} 
\end{subfigure}
   \begin{subfigure}[b]{0.44\textwidth}
\includegraphics[width=1.0\textwidth]{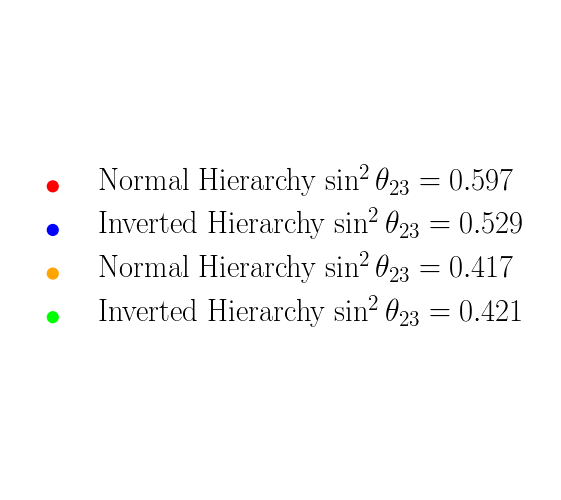}
\end{subfigure}
    \caption{Branching ratios into the three lepton families,
for each of the three main decay channels of a Bino neutralino LSP. The associated neutrino hierarchy and the value of $\theta_{23}$ is specified by the color of the associated data point.}\label{fig:neutralino_lepton_family}
\end{figure}

\begin{equation}
\text{Br}_{{\tilde X}^0_B\rightarrow W^\pm e^\mp}=\frac{\Gamma_{{\tilde X}^0_B\rightarrow W^\pm e^\mp}}{\Gamma_{{\tilde X}^0_B\rightarrow W^\pm e^\mp}+\Gamma_{{\tilde X}^0_B\rightarrow W^\pm \mu^\mp} + \Gamma_{{\tilde X}^0_B\rightarrow W^\pm \tau^\mp} } \ .
\end{equation}

\noindent Using this formalism, we proceed to quantify the branching ratios for each of the three decay processes ${\tilde X}^0_B\rightarrow W^{\pm} \ell^{\mp}$, ${\tilde X}^0_B\rightarrow Z^0 \nu_i$ and ${\tilde X}^0_B\rightarrow h^0 \nu_i$ into their individual lepton families. The results are shown in Figure \ref{fig:neutralino_lepton_family}.
We observe that the ${\tilde X}^0_B\rightarrow W^\pm \ell^\mp$ process has an almost identical statistical distribution for lepton family production as does the Wino chargino decay channel ${\tilde X}^\pm_W\rightarrow Z^0 \ell^\pm$ presented in \cite{Dumitru:2018nct}. Additionally, note that in a Bino neutralino decay via ${\tilde X}^0_B\rightarrow h^0 \nu_i$, the decay rate, given in \eqref{eq:decay_neut4}, has a dominant term proportional to the square of $[V_{\text{PMNS}}^\dag]_{ij}\epsilon_j$. This combination leads to a branching ratio distribution where  no $\nu_\tau$ neutrino is produced in the case of an inverted hierarchy and no $\nu_e$ is produced in the case of a normal hierarchy.

\subsection{Bino Neutralino LSP RPV Decays with Off-Shell $W^{\pm}, Z^0, h^0$ Bosons }

\begin{figure}[t]
 \begin{minipage}{1.0\textwidth}
     \centering
   \begin{subfigure}[b]{0.49\linewidth}
   \centering
\includegraphics[width=0.8\textwidth]{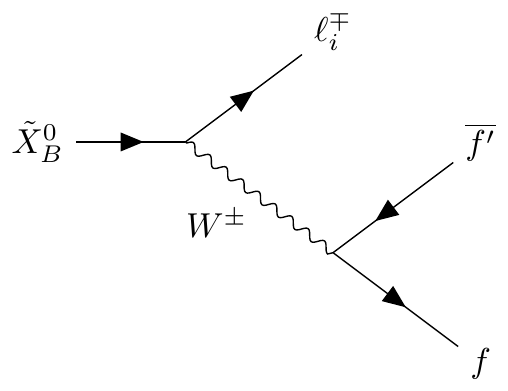}

\caption*{${\tilde X}^0_B\xrightarrow{W^\pm} \ell^\mp_{i}\overline {f^'} f$}
       \label{fig:table2}
   \end{subfigure} \\
   \centering
   \begin{subfigure}[b]{0.48\linewidth}
   \centering
\includegraphics[width=0.8\textwidth]{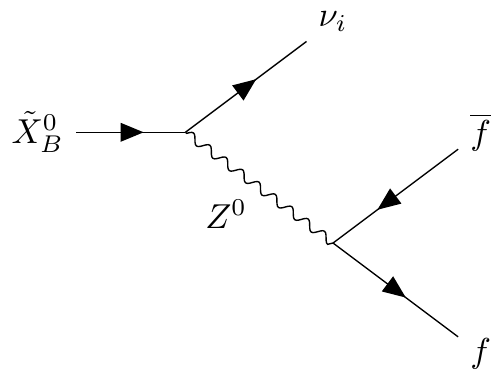}   

\caption*{ ${\tilde X}^0_B\xrightarrow{Z^0} \nu_{i}\overline f f$}
       \label{fig:table2}
\end{subfigure}
   \centering
     \begin{subfigure}[b]{0.48\textwidth}
   \centering
\includegraphics[width=0.8\textwidth]{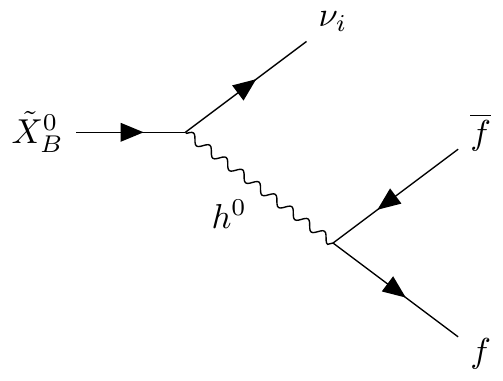}
\caption*{ ${\tilde X}^0_B\xrightarrow{h^0} \nu_{i}\overline f f$}
\end{subfigure}
\end{minipage}
\caption{RPV decays of an $M_{{\tilde{X}}_{B}^{0}} < M_{W^{\pm}}$ Bino neutralino $\tilde X_B^0$ via off-shell $W^\pm$, $Z^0$ and $h^{0}$ bosons, each with three possible channels $i=1,2,3$.
The $W^\pm$ and $Z^0$ bosons may decay to fermion-antifermion pairs, while bosonic decays are also possible in the case of the Higgs. In these Figures, ${\it f^{\prime}}$ represents a generic fermionic state, whereas {\it f} represents a possible fermion or boson decay product. }\label{fig:BinoDecays2}
\end{figure}

In Figures \ref{fig:mass_hist2} and \ref{fig:mass_hist1}, we found that the mass of the Bino neutralino LSP can be as low as a few MeV. For small enough masses, the Bino neutralino LSP can no longer decay via the emission of an on-shell boson as shown in Figure \ref{fig:NeutralinoDecays}. For example,  the process $\tilde X^0_B \rightarrow W^\pm \ell^\mp$ is forbidden if the mass of the Bino neutralino LSP is smaller than the total mass of the $W^\pm$ boson and the accompanying charged lepton. Similarly,  the processes $\tilde X^0_B \rightarrow Z^0 \nu$  and $\tilde X^0_B \rightarrow h^0 \nu$ cannot take place for Bino neutralino LSPs lighter than the $Z^0$ and $h^0$ bosons, respectively. However, in such cases the Bino neutralino LSP will still decay via the RPV processes illustrated in Figure \ref{fig:BinoDecays2}, with intermediate, {\it off-shell} $W^\pm, Z^0$ and $h^0$ bosons.

\subsection*{Calculation of off-shell decay widths}

For Binos lighter than $M_{W^{\pm}}$, the processes displayed in Figure \ref{fig:BinoDecays2} are similar to the familiar case of muon decay $\mu\to e \overline{\nu_e} \nu_\mu$. In muon decay, because the momentum transfer is much smaller than the $W^\pm$ mass, the process may be approximated as an effective 4-point interaction, so that the computation of the decay width becomes straightforward. One obtains
\begin{equation}
\Gamma_{\mu\to e \overline{\nu_e} \nu_\mu}=\left(\frac{g_2}{\sqrt{2}}\right)^2  \left(\frac{g_2}{\sqrt{2}}\right)^2 \frac{1}{2\times 192\pi^3}\frac{1}{{M_W^\pm}^4}m_{\mu}^5.
\label{cup1}
\end{equation}

If that were not the case however, and the mass of the incoming muon was close enough in magnitude to the mass of the off-shell $W^\pm$ bosons, then the low-momentum approximation would not be valid. In general, the decay rate $\Gamma$ is proportional to the coupling strength associated with each vertex, in addition to some dependence on the momentum transfer and the masses of the interacting particles. These contributions can be factorized in the form
\begin{equation}\label{eq:Gamma_form4}
\Gamma_{\mu\to e \overline{\nu_e} \nu_\mu}=\left(\frac{g_2}{\sqrt{2}}\right)^2  \left(\frac{g_2}{\sqrt{2}}\right)^2 F(m_\mu, M_{W^\pm}, m_e, m_{\nu_e}),
\end{equation}
where $g_2/\sqrt{2}$ are the couplings at the first and second vertex, and $F(m_\mu, M_{W^\pm}, m_e, m_{\nu_e})$ is a function that only depends on the masses of the particles involved and the width of the intermediate $W^\pm$ boson. The expression is obtained after integrating over the momenta of the final particle states. For example, when the mass of the incoming muon is much smaller than the width of the $W^\pm$ boson (4-point interaction approximation), the function $F$ in \eqref{eq:Gamma_form4} is given by
\begin{equation}
F=\frac{1}{2\times 192\pi^3}\frac{1}{{M_W^\pm}^4}m_{\mu}^5 \ ,
\end{equation}
thus explaining expression \eqref{cup1}.
Returning to our case, the incoming decaying Binos have large masses, close in magnitude to the width of the intermediate off-shell bosons, $W^\pm, \>Z^0$ or $h^0$, depending on the type of decay. We therefore expect the decay widths to factorize  into expressions similar to equation \eqref{eq:Gamma_form4}. For example, the decay width of the process ${\tilde X}^0_B  \xrightarrow{Z^0} \nu_{i}\overline f f$ is 
\begin{equation}\label{eq:Gamma_form}
\Gamma_{{\tilde X}^0_B  \xrightarrow{Z^0} \nu_{i}\overline f f}=G_{{\tilde X}^0_B\rightarrow Z^0 \nu_{i}}^2g_f^2 F_{{\tilde X}^0_B  \xrightarrow{Z^0} \nu_{i}\overline f f}(M_{\tilde X^0_B}, M_{Z^0}, m_f, m_{\nu_i}),
\end{equation}
where $G^2_{{\tilde X}^0_B\rightarrow Z^0 \nu_{i}}={G_L}_{{\tilde X}^0_B\rightarrow Z^0 \nu_{i}}^2+{G_R}_{{\tilde X}^0_B\rightarrow Z^0 \nu_{i}}^2$ is the squared RPV coupling from the Bino decay vertex,  $g_f^2={g}^2_L+{g}^2_R$ is the squared coupling of the $Z^{0}$  boson to $\bar{f}$,$f$ after EW breaking, and $F_{{\tilde X}^0_B  \xrightarrow{Z^0} \nu_{i}\overline f f}(M_{\tilde X^0_B}, M_{Z^0}, m_f, m_{\nu_i})$  is a function which only depends on the masses of the particles involved and the width of the intermediate $Z^0$ boson.   Note that $F_{{\tilde X}^0_B  \xrightarrow{Z^0} \nu_{i}\overline f f}(M_{\tilde X^0_B}, M_{Z^0}, m_f, m_{\nu_i})$ could, in principle, differ from the function $F$ in  \eqref{eq:Gamma_form4}, because the $W^-$ ( $W^+$) boson in the muon decay only couples to the left-handed (right-handed) fermionic states, while the $Z^0$ boson in the Bino decay ${\tilde X}^0_B  \xrightarrow{Z^0} \nu_{i}\overline f f$ couples to both left-handed and right-handed fermionic states. However, when the masses of the final particles are much smaller than the masses of the incoming muon or Bino and the width of the intermediate bosons, we can drop the dependence on the masses of the final particles from the $F$ functions. Hence, in this limit, the functional form for the $F$ function  for the ${\tilde X}^0_B  \xrightarrow{Z^0} \nu_{i}\overline f f$ decay becomes identical to the $F$ function of muon decay. 
Furthermore, we show that these $F$ functions are the same for all Bino processes of interest. That is, when $m_f, \>m_{f^'} \ll M_{B^0},\> M_{W^\pm},\> M_{Z^0}, M_{h^0}$, we have
\begin{equation}\label{equality_imp}
F_{{\tilde X}^0_B  \xrightarrow{W^\pm} \nu_{i}f^' f}=
F_{{\tilde X}^0_B  \xrightarrow{Z^0} \nu_{i}\overline f f}=
F_{{\tilde X}^0_B  \xrightarrow{h^0} \nu_{i}\overline f f}=F
\end{equation}
with $F$ given in \eqref{eq:Gamma_form4} where the small masses $m_{e}$ and $m_{\nu_e}$ are dropped. 

\begin{figure}[t]
 \begin{minipage}{1.0\textwidth}
     \centering
   \begin{subfigure}[b]{0.48\linewidth}
   \centering

\includegraphics[width=0.8\textwidth]{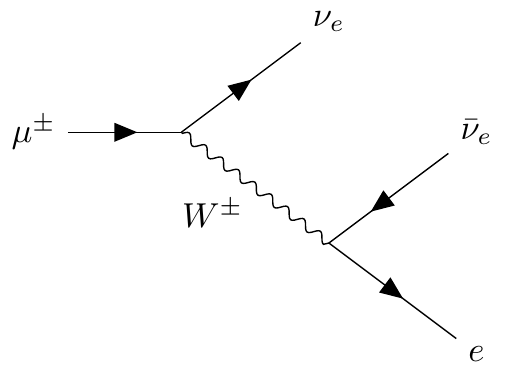}
\caption{ $\mu^\pm \xrightarrow{W^\pm} \nu_{e}\overline{\nu}_e e$}
\label{fig:mudecay}
\end{subfigure}
   \begin{subfigure}[b]{0.48\linewidth}
   \centering
\includegraphics[width=0.8\textwidth]{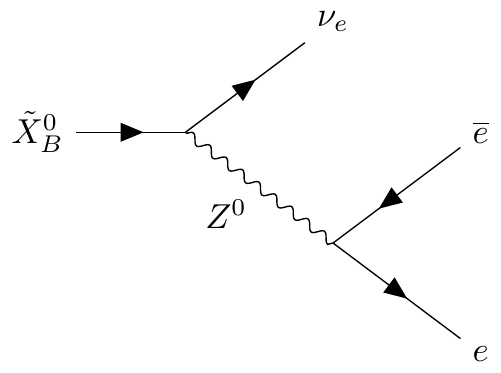}
\caption{${\tilde X}^0_B\xrightarrow{Z^0} \nu_e \overline {e} e$}
       \label{fig:BinoExample}
   \end{subfigure} \\
\end{minipage}
\caption{The Feynman diagram of a $\mu^\pm$ lepton decay (left) has a similar structure to that of a Bino LSP decay (right). We use this similarity to compute the decay rates of a Bino LSP via off-shell bosons, using the Madgraph software.}\label{fig:Madgraph}
\end{figure}

For the Bino decay channels, such as in \eqref{eq:Gamma_form}, analytical calculations of the $F$ functions are non-trivial. Therefore, in this analyis, we will compute these functions numerically. The mass dependence of the Bino decays are calculated using the \textsc{Madgraph5\_aMC@NLO 2.6.4} Monte Carlo event generation program (\madgraph)~\cite{Alwall:2014hca}, to leading order accuracy in the QCD coupling constant. This is where the equality in eq. \eqref{equality_imp} becomes useful.
Standard Model particle masses and widths are configurable parameters, allowing the kinematic dependence of Bino decay widths to be extracted from a modified calculation of the SM muon decay process $\mu\to e \overline{\nu}_{e} \nu_\mu$ shown in Figure \ref{fig:Madgraph}.
The particle test masses in the \madgraph\ calculation (denoted, henceforth, as $\hat m_{\mu}$,$\hat m_{\nu_\mu}$, and so on) can be set to match those of the desired process.
As an example, the dependence of the decay ${\tilde X}^0_B\rightarrow \nu_{i}Z^0$, $Z^0\to\overline e e$--shown in Figure \ref{fig:BinoExample}--on $M_{{\tilde X}^0_B}$ may be extracted by calculating the dependence of the decay 
$\mu\to e \overline{\nu}_{e} \nu_\mu$ on the $\mu$ mass.
Up to differences in couplings, this is accomplished by setting $\hat\Gamma_{W^\pm}=\Gamma_{Z^0}$, $\hat M_{W^\pm}= M_{Z^0}$ in the calculation.
This method is used to calculate the partial widths for each of the Bino decays via $W^\pm$, $Z^0$, and $h^0$ bosons, taking into account masses for all products of the three-body decays.

Returning to the decay rate \eqref{eq:Gamma_form}, we define
\begin{equation}\label{eq:def_f}
{\mathcal{ F}}(M_{\tilde X^0_B}, M_{Z^0})=\frac{1}{g_2^2}\sum_{f}g_f^2 F(M_{\tilde X^0_B}, M_{Z^0}),
\end{equation}

\begin{figure}[!ht]
\centering
\begin{subfigure}[b]{0.8\textwidth}
\includegraphics[width=1.0\textwidth]{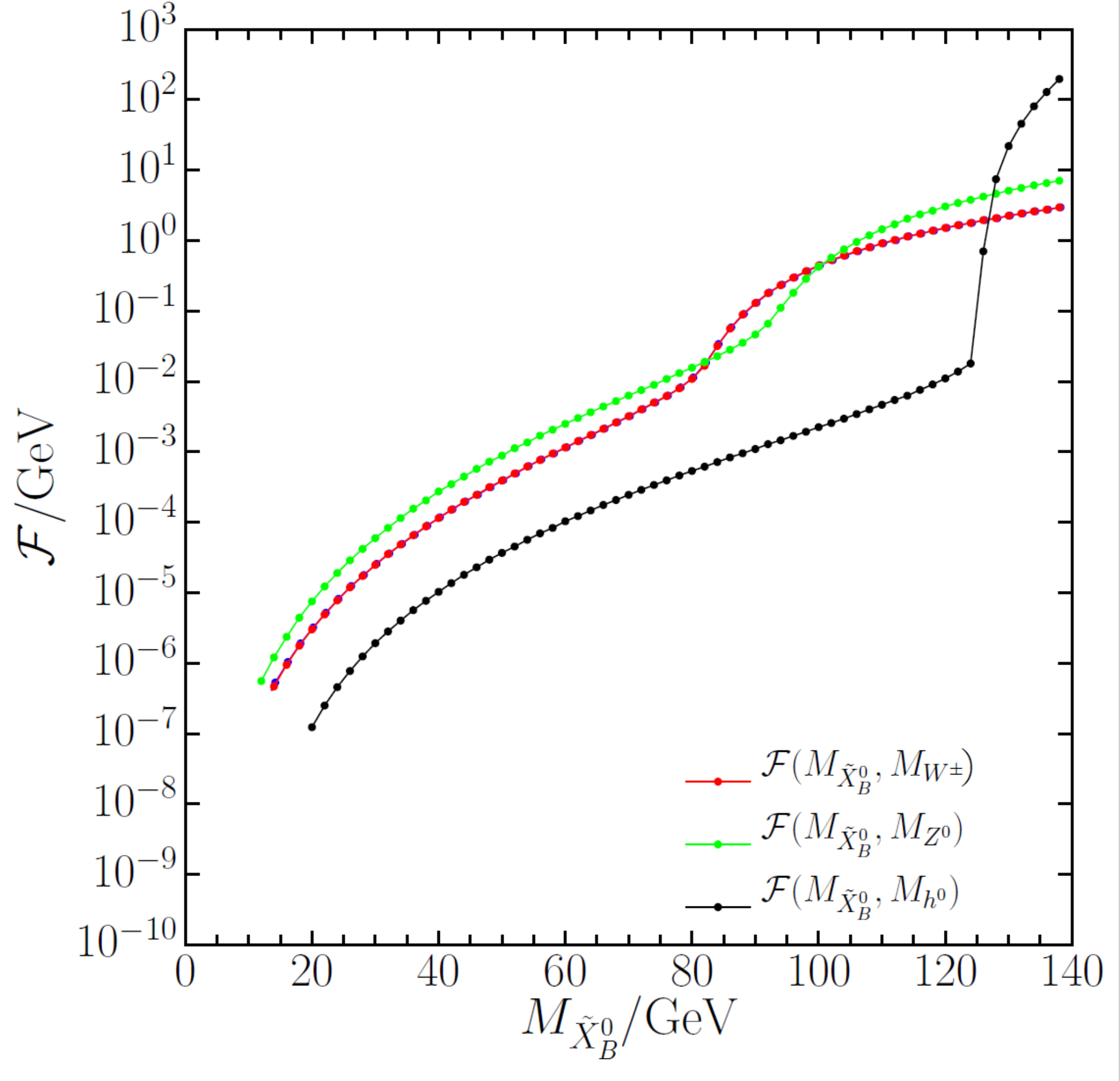}
\end{subfigure}
\caption{ Mass functions $\mathcal{F}(M_{\tilde X^0_{B}})$ defined analogously to \eqref{eq:def_f}, after summing over all final states from the electroweak boson decays.  Each function shows the dependence of the corresponding partial decay width on the mass of the decaying Bino LSP.  Note the rapid increase of each function as the LSP mass approaches, and then surpasses, the mass of the associated electroweak boson. For higher LSP mass, the decay can now proceed via an intermediate on-shell boson. For example, $d{\mathcal{F}}(M_{\tilde X^0_{B}}, M_{h^0})/dM_{\tilde X^0_{B}}$ increases rapidly near 125 GeV (the mass of the $h^0$ boson). Note that it is a more dramatic effect than in the case of the $W^\pm$ and $Z^0$ bosons, due to the fact that $\Gamma_{h^0} \ll \Gamma_{W^\pm}, \Gamma_{Z^0}$.}
\label{fig:fms}
\end{figure}

where the sum is over all possible decays of the $Z^0$ to fermion-antifermion pairs $\overline f f$. Note we dropped the dependence of $F$ on $m_f$, since we consider the masses of the final states negligible. We choose to normalize $\mathcal F$ by dividing by $g_2^2$, where $g_2$ is the $SU(2)_L$ gauge coupling. 
Defining the process ${\tilde X}^0_B\xrightarrow{Z^0} \nu_{i}$ to be ${\tilde X}^0_B\xrightarrow{Z^0} \nu_{i}\overline f f$ summed over the final states $\bar{f}$ and $f$, it follows that the associated decay rate is given by
\begin{equation}\label{eq:Gamma_form2}
\Gamma_{{\tilde X}^0_B  \xrightarrow{Z^0} \nu_{i}}=g_{{\tilde X}^0_B\rightarrow Z^0 \nu_{i}}^2 g_{2}^{2}\mathcal{F}(M_{\tilde X^0_B}, M_{Z^0}).
\end{equation}
Similar definitions apply to decays which involve off-shell $W^\pm$ and $h^0$ bosons. Figure~\ref{fig:fms} shows the mass functions for all Bino decays, which are calculated using \madgraph\ and were defined analagously to \eqref{eq:def_f}.

\subsection*{Lifetime of a light Bino LSP}

\begin{figure}[t]
   \centering

   \begin{subfigure}[c]{0.49\textwidth}
\includegraphics[width=1.0\textwidth]{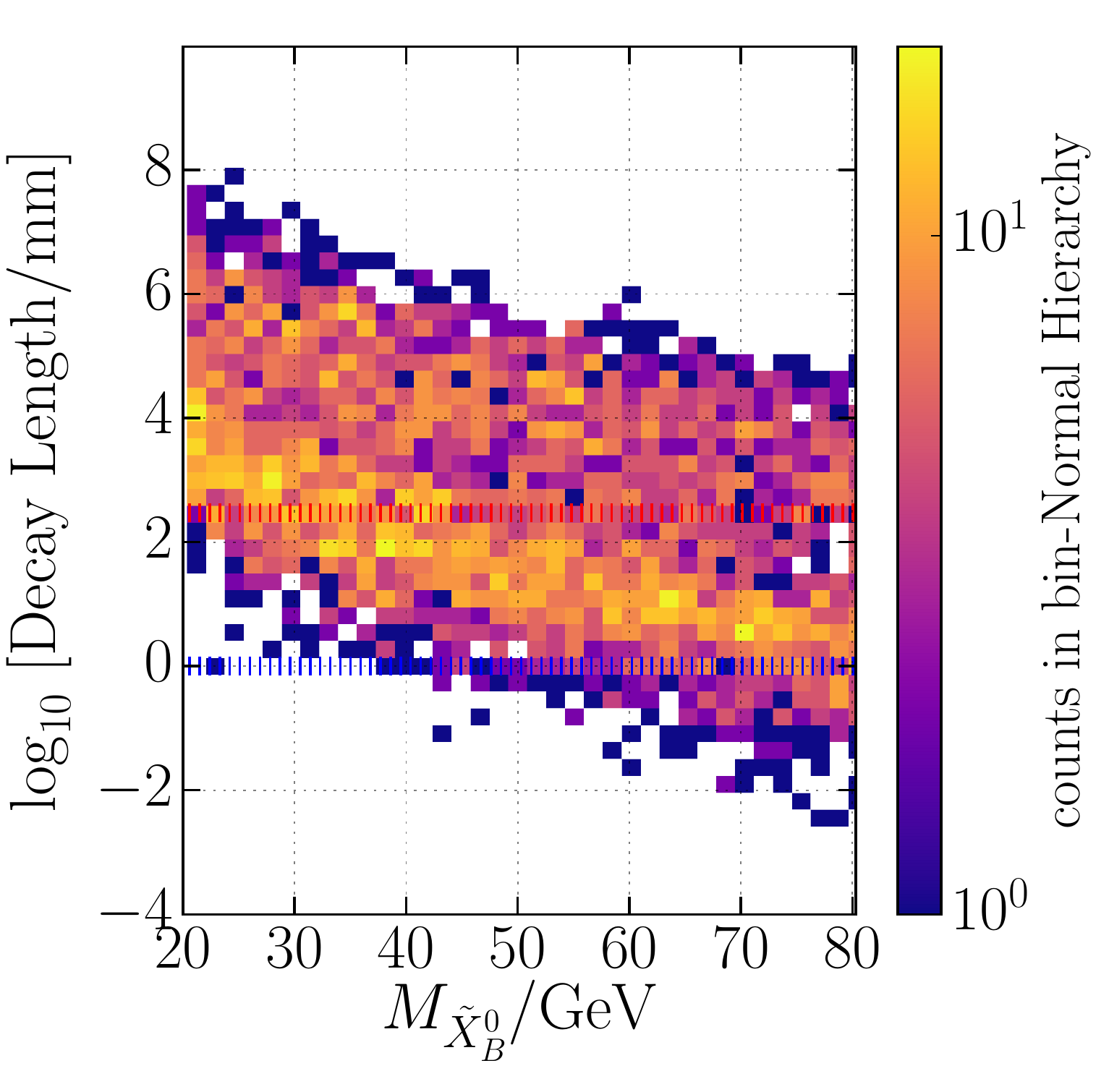}
\caption*{}
\label{fig:prompt_small_1}
\end{subfigure}
   \begin{subfigure}[c]{0.49\textwidth}
\includegraphics[width=1.0\textwidth]{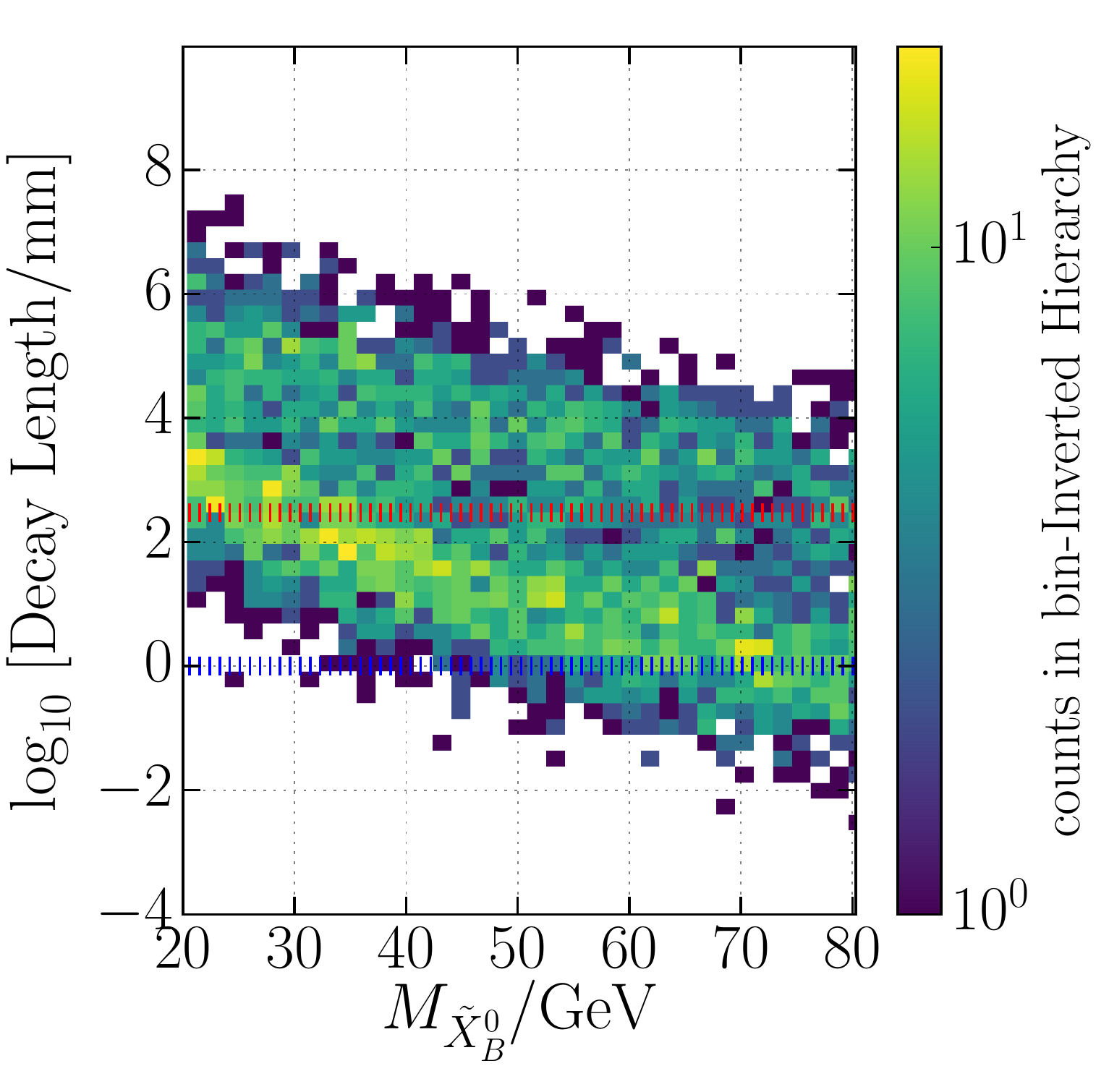}
\caption*{}
\label{fig:prompt_small_2}
\end{subfigure}
   \caption{Bino neutralino LSP RPV decay lengths, summed over all three channels, for Bino masses lighter than the $W^\pm$ and, hence, which can only decay through an off-shell boson. The results are in millimeters, for the normal and inverted hierarchies. The average decay length $L=c\times \frac{1}{\Gamma}$ increases for  smaller values of $M_{\tilde X^0_B}$. We have chosen $\theta_{23}=0.597$ for the normal neutrino hierarchy and $\theta_{23}=0.529$ for the inverted hierarchy to display the results. At and below the blue dashed line ($c\tau=1$~mm), the decays are considered prompt. 
   The red dashed line ($c\tau=30$~cm) denotes the largest decay lengths that may be measured via displaced vertices.}
   \label{fig:prompt_small_mass_total}
\end{figure}

In this section, we study whether light Bino neutralino LSPs can RPV decay promptly since their decays proceed only through {\it off-shell} bosons. We will compute these decay rates, summing over the partial widths of all possible final states produced. We use the mass functions $\mathcal{F}$ shown in Figure \ref{fig:fms}. The decay width of the Bino LSPs lighter than the EW scale is 
\begin{multline}
\Gamma_{{\tilde X}^0_B}=\sum_i \sum_{f} \Gamma_{{\tilde X}^0_B\xrightarrow{W^\pm} \ell^\mp_{i} \overline f^' f}+\sum_i \sum_{f} \Gamma_{{\tilde X}^0_B\xrightarrow{Z^0} \nu_{i} \overline f f}+\sum_i \sum_{f} \Gamma_{{\tilde X}^0_B\xrightarrow{h^0} \nu_{i} \overline f f}\\
=\sum_i g_{{\tilde X}^0_B\rightarrow W^\pm \ell^\mp_{i}}^2g_2^2 \mathcal{F}(M_{\tilde X^0_B}, M_{W^\pm})
+\sum_i g_{{\tilde X}^0_B\rightarrow Z^0 \nu_{i}}^2g_2^2 \mathcal{F}(M_{\tilde X^0_B}, M_{Z^0})+ \\ \sum_i g_{{\tilde X}^0_B\rightarrow h^0 \nu_{i}}^2g_2^2 \mathcal{F}(M_{\tilde X^0_B}, M_{h^0}).
\end{multline}
The decay lengths of the Bino LSPs,
\begin{equation}
L=\frac{c}{\Gamma_{{\tilde X}^0_B}},
\end{equation}
are calculated for both the normal and inverted hierarchy scenarios and shown in Figure \ref{fig:prompt_small_mass_total}. Prompt decays ($L<1$mm) are possible for Bino neutralino LSP masses as low as about 50~GeV, in both the normal and the inverted hierarchy scenarios. However, such Bino LSPs are most likely to decay with significant displacement from the production vertex, though still within the typical LHC detector volume. Bino LSPs with masses in the range from about 50~GeV to 20~GeV do not exhibit prompt decays, but can still decay with displaced vertices in the detector. 
Bino LSPs with very low masses (roughly $<20$~GeV) may be stable on the scale of LHC detectors.
Conventional ``missing-energy'' searches should have some sensitivity to these models, 
while ambitious next-generation experiments~\cite{Chou:2016lxi} may offer the possibility for direct detection of displaced decays. 

Note that Figure \ref{fig:prompt_small_mass_total} displays the decay lengths strictly for Bino LSPs lighter than $M_W^\pm$, which can only decay via off-shell processes. However, decays via on-shell $Z^0$ bosons become forbidden even earlier; that is, for Bino LSPs lighter than $M_{Z^0}$.  Similarly, Bino LSPs with masses smaller than $M_{h^0}$ cannot decay through an on-shell Higgs bosons. Therefore, Bino LSPs with masses in the interval between $M_{W^{\pm}}$ and $M_{h^0}$ could possibly, for example, decay via both on-shell $W^{\pm}$ bosons and off-shell $Z^0, \> h^0$ bosons. 
\begin{figure}[!ht]
\centering
\begin{subfigure}[b]{0.98\textwidth}
\includegraphics[width=1.0\textwidth]{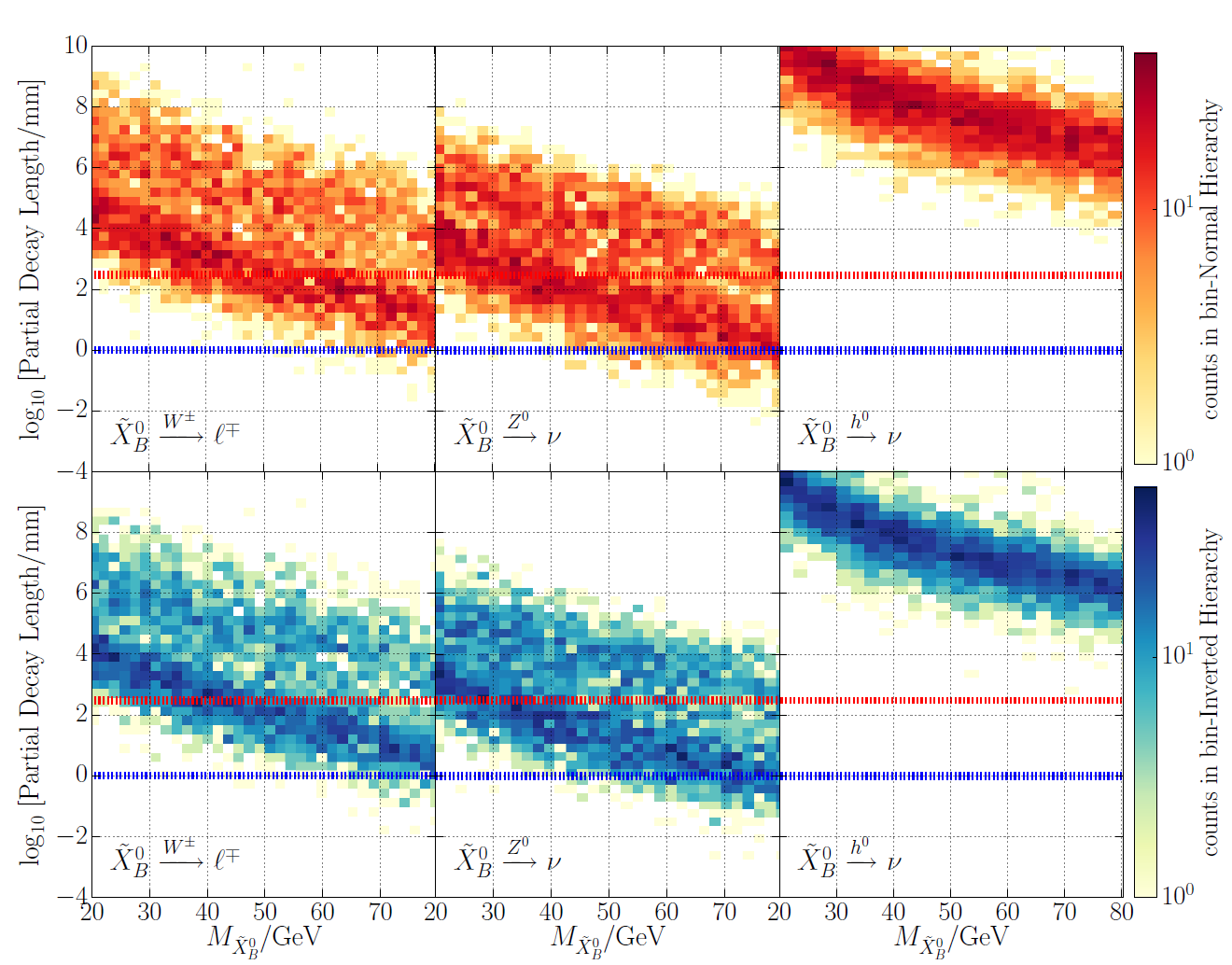}
\end{subfigure}
\caption{ Bino neutralino LSP partial decay lengths in millimeters, for individual decay channels, for both normal and inverted hierarchies.Widths are calculated for Bino masses when all decays must proceed through intermediate off-shell bosons. We have chosen $\theta_{23}=0.597$ for the normal neutrino hierarchy and $\theta_{23}=0.529$ for the inverted hierarchy. At and below the blue dashed line ($c\tau=1$~mm), the decays are considered prompt. The red dashed line ($c\tau=30$~cm) denotes the largest decay lengths that may be measured via displaced vertices.}
\label{fig:PromptnessSmall}
\end{figure}
\noindent In this region, decays via off-shell $Z^0, \> h^0$ bosons are strongly suppressed in general compared to the decays via the on-shell $W^{\pm}$ bosons. 
The effect of this suppression is seen in Figure \ref{fig:fms}. For decays via the $W^\pm$ boson, the red curve drops about two orders of magnitude when we move from the on-shell region, where the mass of the incoming Bino is larger than $M_{W^\pm}$, to the off-shell one,  where the mass of the incoming Bino is smaller than $M_{W^\pm}$.  A similar drop in magnitude occurs in the green line for decays via an on-shell versus an off-shell $Z^0$ boson. Even more pronounced is the drop from the on-shell to the off-shell region, approximately four orders of magnitude, for the decays via a Higgs boson-- the black curve in Figure \ref{fig:fms}.  It follows that for a Bino LSP mass above $M_{W^\pm}$, but below $M_{Z^{0}}$ and  $M_{h^0}$, the size of the $\mathcal{F}$ functions for $Z^{0}$, $h^{0}$ are significantly suppressed relative to $\mathcal{F}$ for the $W^{\pm}$. Hence, in this mass regime, the decay rate of the Bino LSP is dominated by decay via an on-shell $W^{\pm}$; the decay rates for the off-shell $Z^{0}$ and, particularly, the off-shell $h^{0}$ being suppressed.
Note that in all three cases the {\it transition interval} from on-shell to off-shell bosons is narrow, of order $\approx 10$ GeV.

Considering the relatively narrow transitions between the on-shell to the off-shell regions for all decay channels and the strong suppression of the off-shell processes, we neglect the off-shell decays via the $Z^0$ and $h^0$ bosons when decays via on-shell $W^\pm$ bosons are possible. Figure \ref{fig:LSPprompt_neut}, which takes into account only processes that occur via on-shell bosons, provides accurate estimates for the summed decay lengths of all Binos heavier than $M_{W^{\pm}}$. Figure \ref{fig:prompt_small_mass_total}, which presents the summed decay lengths for all Bino LSPs lighter than $M_{W^\pm}$,  completes the decay width analysis.

\subsection*{Branching ratios of the Bino LSP RPV decays}

For Bino LSPs masses smaller that the mass of $W^{\pm}$, there is a wide range that could lead to visible signatures in LHC detectors-- despite decaying via off-shell bosons.  Therefore, we separately analyze each of the decay channels, to determine the dominant decay modes. To mimic the analysis undertaken in Section 2, we classify the Bino neutralino decays into three categories, depending on which off-shell boson, $W^\pm, \> Z^0$ or $h^0$, the Bino neutralino LSP decays into. For each category, we compute the decay rates by summing over the three lepton families produced in the Bino decay and over all final-state particles associated with the electroweak boson decay.
\begin{figure}[!ht]
\centering
\begin{subfigure}[b]{1.\textwidth}
\includegraphics[width=1.0\textwidth]{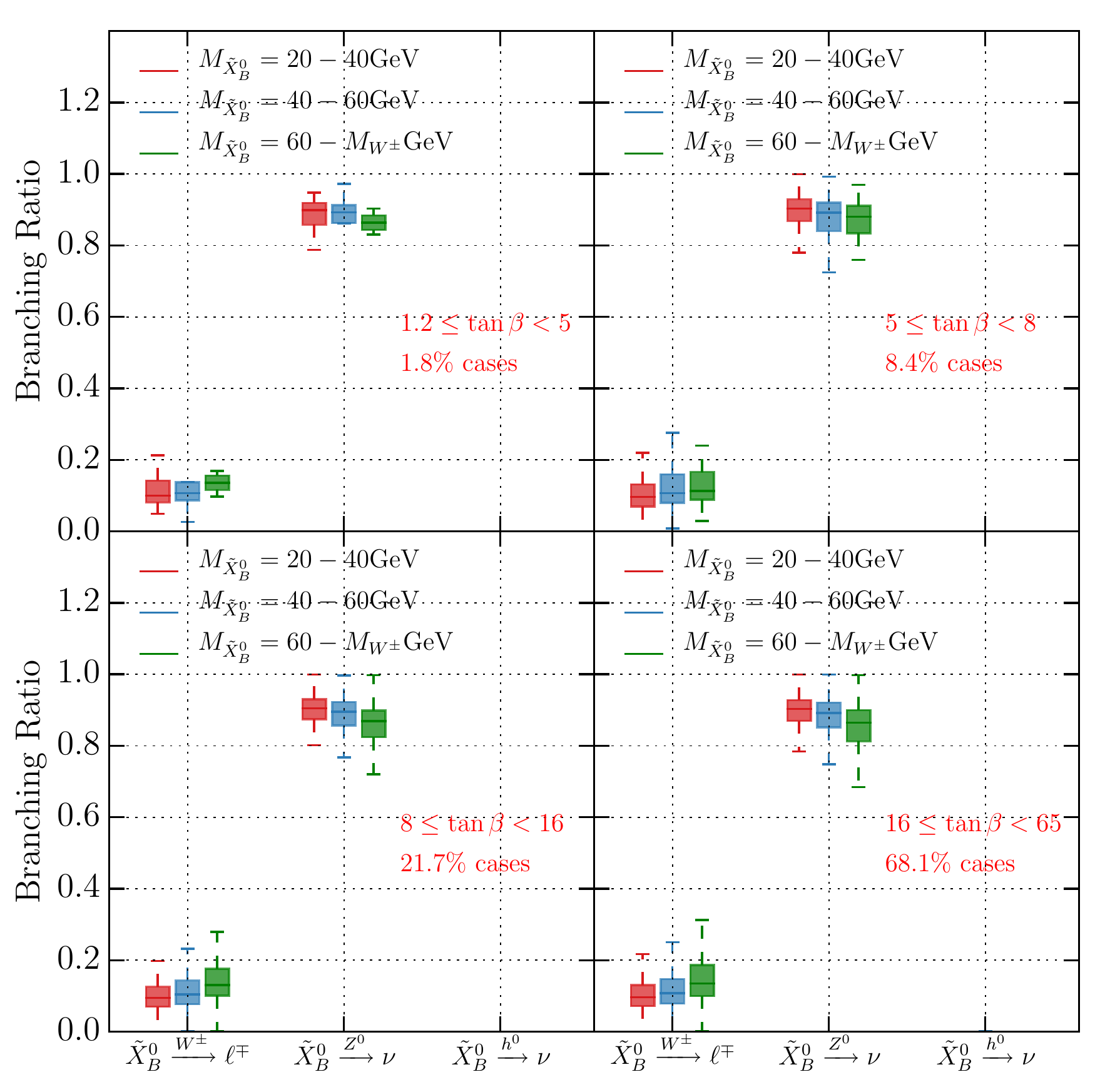}
\end{subfigure}
\caption{ Branching ratios for the three possible decay channels of a Bino neutralino LSP divided over three mass bins and four $\tan \beta$ regions. We studied the viable points for Bino LSPs with masses smaller than the mass of the $W^\pm$ bosons but larger than 20~GeV. The colored horizontal lines inside the boxes indicate the median values of the branching fraction in each bin, the boxes indicate the interquartile range, while the dashed error bars show the range between the maximum and the minimum values of the branching fractions. The case percentage indicate what percentage of the physical mass spectra have $\tan \beta$ values within the range indicated.  We assumed a normal neutrino hierarchy, with $\theta_{23}=0.597$. Note that the branching ratios via an off-shell $h^{0}$, while non-vanishing, are of order $10^{-3}$ and smaller and, hence, too small to appear in the Figure. }\label{fig:BarSmall}
\end{figure}
For example, the partial decay length $L$ of a Bino LSP associated with decays through an off-shell $Z^0$ boson is 
\begin{equation}
L_{\tilde X_B^0 \xrightarrow{Z^0} \nu}=\frac{c}{\Gamma_{\tilde X_B^0 \xrightarrow{Z^0} \nu}},\quad \text{where} \quad
\Gamma_{{\tilde X}^0_B\xrightarrow{Z^0} \nu}=\sum_{i} \sum_{f} \Gamma_{{\tilde X}^0_B\xrightarrow{Z^0} \nu_{i} \bar{f} f}
\end{equation}

The results are shown in Figure \ref{fig:PromptnessSmall}. We learn that Bino LSPs lighter than $M_{W^\pm}$ decay mainly via $W^\pm$ and $Z^0$ bosons, with decays proceeding via off-shell Higgs being completely negligible. It is also {\it important to note} that for Bino masses below approximately 20~GeV, all three channels have decay lengths larger than 30 cm--that is, are longer than displaced vertices--and, hence, are essentially stable within the ATLAS detector. This is consistent with our previous observation from Figure \ref{fig:prompt_small_mass_total}.

In Figure \ref{fig:BarSmall}, we analyze the branching ratios of the three main decay channels of these light Bino LSPs. We group the light Bino LSPs into three mass bins and four $\tan \beta$ bins, as we did for the heavier Bino LSPs in Figure \ref{fig:bar_plot2}. For consistency, we keep the same binning for the $\tan \beta$ parameter, and choose evenly spaced mass bins
\begin{equation}
[20\>\text{GeV}, 40\>\text{GeV}], \quad[40\>\text{GeV}, 60\>\text{GeV}], \quad[\text{60}\>\text{GeV}, M_{W^\pm}]. 
\end{equation}
Note that, as discussed above, the decay lengths for Bino LSP masses below 20 GeV are all generally very large and outside the detector. We therefore don't consider Bino masses smaller than 20~GeV in our analysis.
 We learn that Bino LSPs decay mainly via an off-shell $Z^0$ boson independent of mass. This type of decay is usually at least four times more probable than decays via off-shell $W^\pm$ bosons. Decays via off-shell Higgs, although non-vanishing, are significantly suppressed and thus not likely to be observed.
Variations in $\tan\beta$ do not significantly affect the Bino LSP branching fractions.

\subsection*{Experimental signatures of off-shell Bino LSP decays}
So far, we have seen that Bino LSPs as light as 20 GeV may be detected at the LHC if they decay via $W^\pm$ or $Z^0$ bosons. In this subsection, we analyze the experimental signature of such decays in the detector. That is, we compute two sets of branching ratios: one corresponding to the family of the lepton produced at the RPV vertex and another for the decay products of the electroweak boson. 

Note that our analysis of the Bino LSPs decays when they are lighter than the electroweak vector bosons differs slightly from our approach in Section 4, where we studied the RPV decays of the Bino LSPs heavier than the electroweak scale. In Section 4, we only considered the vertices shown in Figure \ref{fig:NeutralinoDecays}, in which the Bino decays into an on-shell boson and a lepton via RPV couplings. The physical boson that is produced would further decay into pairs of final states $\bar{f}$ and $f$. However, we did not discuss such decays, since these standard model processes are well-known. Hence, we limited ourselves to analyzing the statistical distributions for the families of the leptons produced at the RPV vertex only.

However, for decays via off-shell bosons, we need to consider the full diagrams shown in Figure \ref{fig:BinoDecays2}; that is, including the RPV vertex, the off-shell boson propagator and the vertex in which the final states $\bar{f}$ and $f$ are produced. 
Equation \eqref{eq:Gamma_form} shows that the decay rate for each such individual process is proportional to the RPV coupling, the $g_{f}$ coupling and a function $F$ which depends on the masses of the initial, intermediate and final states. The final states  $\bar{f}$ and $f$ at the second vertex and the lepton produced at the first vertex are interconnected through the function $F$, obtained after integrating over the momenta of all the final and initial states. An unified treatment for the branching fractions at the first and second vertices would be very complicated, as it would involve counting all possible combinations of decay products at these vertices. However, in the physical regime that we are working in, that is, for incoming Binos heavier than 20 GeV,  a simplifying assumption can be made. Because the final decay products are much lighter than the incoming Bino, the function $F$ has a weak dependence on the masses of these final states. Therefore, the two sets of branching ratios depend mainly on the value of the coupling at each vertex and become independent of each other. We will now explain how accurate this assumption is and what sets of branching fractions we expect at each of the two vertices.

First, we analyze the distributions of the branching ratios to different lepton families at the RPV vertex. In Figure \ref{fig:neutralino_lepton_family}, we have shown the results of a similar study, but in the case of heavy Bino LSPs  that could decay via on-shell bosons. When repeating the computation for light Bino LSPs with masses in the detectable interval from 20~GeV to $M_{W^\pm}$, we obtain statistically identical distributions to those shown in Figure \ref{fig:neutralino_lepton_family}. The similarity between the distributions in the off-shell and the on-shell regime is expected. The reason is that when calculating the branching fractions at the RPV vertex in the off-shell decay regime for a fixed pair of final states at the second vertex,  the $F$ functions and the couplings at the second vertex are divided out.   The cancellation is possible because the $F$ functions show no dependence on the family of the lepton produced at the first vertex. Therefore, the  generation of leptons produced at the first vertex depends on the RPV breaking parameters  $\epsilon_i$ and $v_{L_i}$ only,  just as in the on-shell decay regime. This result is independent of the nature of the particles $f$ produced at the second vertex.  Furthermore, the relative sizes of the RPV breaking parameters do not depend on the Bino mass and, therefore, the branching fraction distributions at the first vertex are identical to those in Figure \ref{fig:neutralino_lepton_family}.

Secondly, we compute the expected branching ratios to different pairs of final states  $\bar{f}$ and $f$ at the second vertex. This time, we fix the family of the lepton produced at the first vertex. The RPV couplings from the first vertex are divided out. The $F$ functions do not depend on the family of the lepton produced at the first vertex. Hence, the set of branching ratios to final states  $\bar{f}$ and $f$ at the second vertex is independent of the nature of the lepton at the first vertex. Note that when we computed the branching ratios at the first vertex for a fixed pair of final states  $\bar{f}$ and $f$, we divided out the $F$ functions as well, because they showed no dependence on the lepton family. However, the $F$ functions depend on the heavier pair of states  $\bar{f}$ and $f$, although weakly. If this was not the case, we would simply recover the same branching fractions at the second vertex as those calculated for physical $W^\pm$, $Z^0$ and $h^0$ bosons, existent in the standard model literature. Although the effect is weak, these branching fractions at the second vertex differ from their on-shell values, especially for the lightest Bino LSPs. As the mass $M_{\tilde X_B^0}$ of the incoming Bino is taken to be lighter, it becomes comparable to some of the masses of the $W^\pm$ and $Z^0$ decay products. Hence, the decays to these final states  $\bar{f}$ and $f$ get more and more suppressed.
To illustrate this effect, in Tables \ref{table:Zboson} and \ref{table:Hboson} we compare the relative fractions for each of the electroweak boson decays, for Bino LSP masses of 30 GeV and 60 GeV .  For comparison, in the columns labelled $M_{\tilde X_B^0}>M_{Z^0}$  and $M_{\tilde X_B^0}>M_{h^0}$ of Tables \ref{table:Zboson} and \ref{table:Hboson} respectively, we show the SM branching fractions calculated for on-shell $Z^0$ and $h^0$.

\begin{table}[t]

\begin{center}
  \begin{tabular}{ l |c|c|c }
    \hline
    {Process} &\multicolumn{3}{c}{$Z^0$ branching fractions for $Z^0 \rightarrow \overline{f} f$ decays (in \%)} \\
    \cline{2-4}
     & $M_{\tilde X_B^0}>M_{Z^0}$ & $M_{\tilde X_B^0}=60$~GeV & $M_{\tilde X_B^0}=30$~GeV \\ 
    \hline
    $Z^0 \rightarrow e^\pm e^\mp $       & \phantom{0}3.4 & \phantom{0}3.4 & \phantom{0}3.6  \\
    $Z^0 \rightarrow \mu^\pm \mu^\mp$    & \phantom{0}3.4 & \phantom{0}3.4 & \phantom{0}3.6  \\
    $Z^0 \rightarrow \tau^\pm \tau^\mp $ & \phantom{0}3.4 & \phantom{0}3.4 & \phantom{0}3.4  \\
    \hline %
    $Z^0 \rightarrow \nu \overline \nu $ & 20.0 &  20.3 &  21.3  \\
    \hline %
    $Z^0 \rightarrow u \overline u $     & 11.6 &  11.7 &  12.4  \\
    $Z^0 \rightarrow c \overline c $     & 12.0 &  12.0 &  12.2  \\
    \hline %
    $Z^0 \rightarrow d \overline d $     & 15.6 &  15.8 &  16.6  \\
    $Z^0 \rightarrow s \overline s $     & 15.6 &  15.8 &  16.6  \\
    $Z^0 \rightarrow b \overline b $     & 15.1 &  14.3 &  10.3  \\
    \hline
  \end{tabular}
\end{center}
\caption{Branching fractions for decays of the virtual $Z^0$ boson for the $Z^0 \rightarrow \overline{f} f$  process for several values of the $\tilde X_B^0$ mass. The reference values for on-shell $Z^0$ decays ($M_{\tilde X_B^0}>M_{Z^0}$) are taken from the Particle Data Group recommendations~\cite{PDG}.}
\label{table:Zboson}
\end{table}

\begin{table}[t]
\begin{center}
  \begin{tabular}{ l |c|c|c }
    \hline
    {Process} &\multicolumn{3}{c}{$h^0$ branching fractions for $h^0 \rightarrow \overline{f} f$  decays (in \%)} \\
    \cline{2-4}
     & $M_{\tilde X_B^0}>M_{h^0}$ & $M_{\tilde X_B^0}=60$~GeV & $M_{\tilde X_B^0}=30$~GeV \\ 
    \hline %
    $h^0 \rightarrow b\bar b $          & \phantom{}58.9   & \phantom{}84.2   & \phantom{}87.0   \\
    $h^0 \rightarrow c\bar c $          & \phantom{0}2.9   & \phantom{0}4.1   & \phantom{0}4.4   \\
    $h^0 \rightarrow \tau^\pm \tau^\mp$ & \phantom{0}6.3   & \phantom{0}8.0   & \phantom{0}6.9   \\
    $h^0 \rightarrow \mu^\pm \mu^\mp$   & \phantom{00}0.02 & \phantom{00}0.03 & \phantom{00}0.02 \\
    \hline %
    $h^0 \rightarrow gg $               &  \phantom{0}7.8   & \phantom{0}3.5    & \phantom{0}1.7    \\
    $h^0 \rightarrow W^\pm W^\mp$       &  \phantom{}21.0   & \phantom{00}0.02  & \phantom{}$<0.01$ \\
    $h^0 \rightarrow Z^0 Z^0$           &  \phantom{0}2.6   & \phantom{00}0.01  & \phantom{}$<0.01$ \\
    $h^0 \rightarrow \gamma\gamma$      &  \phantom{00}0.23 & \phantom{00}0.07  & \phantom{00}0.01  \\
    $h^0 \rightarrow Z^0\gamma$         &  \phantom{00}0.15 & \phantom{}$<0.01$ & \phantom{}$<0.01$ \\
   \hline
  \end{tabular}
\end{center}
\caption{Branching fractions for decays of the virtual Higgs boson for the $h^0 \rightarrow \overline{f} f$  process for several values of the $\tilde X_B^0$ mass. These values are adapted from the Higgs branching fractions presented as a function of mass, published by the CERN LHC Higgs Yellow Report~\cite{deFlorian:2016spz}. Decay modes which contribute $<0.01$\% for all values of the Bino LSP mass  are suppressed.}
\label{table:Hboson}

\end{table}

For Bino LSP decays via $Z^0$ bosons, shown in Table~\ref{table:Zboson}, the partial widths corresponding to quark-antiquark pairs are somewhat suppressed relative to the SM decays. This is most apparent in the decays to $b\bar b$, but this suppression also occurs in the widths corresponding to charm and tau decays, but to a lesser degree.  For decays via $W^\pm$ bosons, all final state particles are light enough that the impact of the Bino LSP mass on the relative branching fractions is negligible for the considered mass range, so a dedicated Table is not presented.
Bino LSP decays via the Higgs boson are very rare, as discussed above.
However, the relative branching fractions of these processes are given in Table~\ref{table:Hboson}, to compare the importance of each channel to the final result.  These figures are adapted from Higgs decay widths calculated in the CERN Yellow Report~\cite{deFlorian:2016spz} for various values of the Higgs boson mass.

\subsection*{Experimental outlook}

These findings demonstrate that the Bino LSP is a viable candidate for direct detection at the LHC across a wide range of masses.
For very low masses of the Bino LSP, the existing search program for $R$-parity conserving SUSY scenarios should be sensitive to final states with this new ``detector-stable'' particle.
Such searches may also be sensitive in the case of prompt Bino LSP decays to neutrinos (that is,  $\tilde X_B^0 \rightarrow Z^0 \nu$ and $\tilde X_B^0 \rightarrow h^0 \nu$).
Generally, a diverse set of searches for $R$-parity violating decays using prompt objects should also be pursued. 
In particular, maximal sensitivity could be obtained by taking advantage of the unconventional signatures produced in Bino LSP decays, such as $W^\pm$-lepton resonances.
Finally, the calculated distribution of possible lifetimes makes it abundantly clear that searches for displaced leptons and jets are an invaluable tool, particularly when the Bino LSP is lighter than the $W^\pm$ boson.

The Bino presents an attractive candidate to the experimentalist, as it is by far the most prevalent LSP in the space of models considered in the present analysis.
As has been shown, it may also be arbitrarily lighter than the soft SUSY breaking scale, due to cancelling contributions from unrelated soft mass terms.
On the other hand, pure Bino pairs cannot be produced directly from SM particle decays, so that experimental prospects will in general depend on the detailed spectrum of heavier SUSY particles.
However, this makes the prediction of a long-lived Bino LSP intriguing, as it is a process with no SM background.
This enables Bino LSP searches to be conducted without regard to the potentially complicated mechanism responsible for their production.
Hence, searches for displaced leptons and jets (independent of other activity in the detector) present a completely orthogonal method of probing otherwise challenging spectra of sparticle masses.

\subsection{Conclusion}

In this study, using the formalism developed in \cite{Ovrut:2015uea,Dumitru:2018jyb}, we have shown that the Bino neutralino is the most prevalent LSP of the $B-L$ MSSM. An accurate approximation to its mass formula is presented and compared to the mass formula for both Wino charginos and Wino neutralinos, that were discussed in detail in the previous section. It is shown that, whereas the Wino LSP masses must always exceed the $W^{\pm}$ electroweak boson mass, the mass of the Bino neutralino LSP, while generically also larger than $M_{W^{\pm}}$, can be smaller than this scale--although such ``light'' Binos  are less prevalent. The mass spectrum for the Bino neutralino LSP is displayed. We have shown, however, that for sufficient ``fine-tuning'' its mass can actually become vanishly small. 

We then proceed to analyze the decays channels, decay rates/lengths and branching ratios for the RPV decays of Bino neutralino LSPs in the $B-L$ MSSM. This analysis, following the above comments, naturally breaks into  two different parts: a) for the Bino neutralino mass $M_{{\tilde{X}}_{B}^{0}} > M_{W^{\pm}}$ and b) for  $M_{{\tilde{X}}_{B}^{0}} < M_{W^{\pm}}$. Since the Bino neutralino mass can be made arbitrarily small by fine-tuning, we put a lower bound of 20 GeV on its mass for two reasons--1) since below that value the degree of fine-tuning increases dramatically and 2) when $M_{{\tilde{X}}_{B}^{0}} < 20$~GeV its decay length becomes very large, outside the range of the ATLAS detector. The mass of the Bino neutralino LSP has an important impact on its RPV decays. For $M_{{\tilde{X}}_{B}^{0}} > M_{W^{\pm}}$, it can always directly decay to a lepton and at least one, and perhaps each, of the three {\it on-shell} $W^{\pm}$, $Z^{0}$and $h^{0}$ bosons. In this regime, we compute the branching ratios for each boson decay channel. The associated decay lengths are also presented, both summing over all three decay channels and for each channel independently. A discussion of whether the decays are ``prompt'', occur as ``displaced vertices'' or are longer is given. We also analyze the branching fractions for each boson channel into individual leptons. Finally, the relationship of the decay lengths and the individual branching fractions to the neutrino mass hierarchy--both normal and inverted, is discussed in detail.

For Bino neutralino LSPs with mass in the range $[20~{\rm GeV},M_{W^{\pm}}]$, the RPV decays must occur via one of three {\it off-shell} $W^{\pm}$, $Z^{0}$and $h^{0}$ bosons.
The analysis of decays channels, decay rates/lengths and branching ratios for these RPV off-shell processes is much more computationally involved. Our method of calculation is presented and used to compute the same quantities as in the on-shell case. The fact that the intermediate bosons are off-shell significantly lowers the decay rates--and, hence, there are fewer prompt decays in this category, most lengths being at least displaced vertices and much larger. However, the effect of the type of neutrino hierarchy does not greatly change from the previous analysis. The branching fractions to a specific lepton at the first RPV vertex is almost unchanged from the heavy Bino case. However, the analysis of the decay products arising from the decay of the off-shell boson does somewhat change. The branching fractions for these decays are analyzed separately.

We conclude that for an LSP Bino neutralino in the $B-L$ MSSM there is, regardless of its mass, a significant chance that its RPV decays through various specified channels can be observed in the run 2 data at the LHC. If discovered, the theoretical predictions presented here could be a first discovery of possible $N=1$ supersymmetry in nature and, secondly, partially validate the specific $B-L$ MSSM theory.

\appendix
\chapter*{Appendices}%
\addcontentsline{toc}{chapter}{Appendices}

\chapter{4D Anomaly Cancellation Mechanism}

At lowest order in string coupling, the bosonic part of the string frame Lagrangian takes the following form~\cite{Horava:1996ma,Green:1987mn}
\begin{equation}
\begin{split}
\label{eq:initial_act}
S_{\text{het}}&=\frac{1}{2\kappa^2_{10}}\int_{\mathcal{M}_{10}}e^{-2\phi}\left[R+4d\phi\wedge \star\phi-\frac{1}{2}H\wedge \star H
\right]\\
&-\frac{1}{2\kappa^2_{10}}\frac{\alpha^\prime}{4}\int_{\mathcal{M}_{10}} e^{-2\phi}\text{tr}(\mathcal{F}_1\wedge \star \mathcal{F}_1)+e^{-2\phi}\text{tr}(\mathcal{F}_2\wedge \star \mathcal{F}_2)\ .
\end{split}
\end{equation}
In the above action, $\phi$ is the 10D dilaton, $\mathcal{F}_1=d\mathcal{A}_1-i\mathcal{A}_1\wedge \mathcal{A}_1$ is the field strength on the observable sector, $\mathcal{F}_2=d\mathcal{A}_2-i\mathcal{A}_2\wedge \mathcal{A}_2$ is the field strength on the hidden sector and $H$ is the heterotic three-form
strength 
\begin{equation}
H=dB^{(2)}-\frac{\alpha^\prime}{4}(\omega_{YM}-\omega_L)\ .
\end{equation}
 $B^{(2)}$ is the Kalb-Ramond two-form. $\omega_{YM}$ and $\omega_L$ are the Chern-Simons three-forms defined in terms of 
the gauge potentials $\mathcal{A}_1$, $\mathcal{A}_2$ and the spin connection $\Omega$ by
\begin{equation}
\begin{split}
 d \omega_{YM}&=\text{tr}\mathcal{F}_1^2+\text{tr}\mathcal{F}_2^2\ ,\\
d \omega_{L}&=\text{tr}R^2\ .
\end{split}
\end{equation}
From the kinetic term of the three-form $H$ in \eqref{eq:initial_act}, that is
\begin{equation}
S_{\text{kin}}=-\frac{1}{4\kappa_{10}^2}\int_{\mathcal{M}_{10}}e^{-2\phi_{10}}H\wedge \star H\ ,
\end{equation}
we obtain
\begin{equation}
\begin{split}
\label{eq:kinetic_full}
S_{\text{kin}}&=-\frac{1}{4\kappa_{10}^2}\int_{\mathcal{M}_{10}}e^{-2\phi_{10}}dB^{(2)}\wedge \star dB^{(2)}
\\&+\frac{\alpha^\prime}{8\kappa_{10}^2}\int_{\mathcal{M}_{10}}(\text{tr}\mathcal{F}_1^2+\text{tr}\mathcal{F}_2^2-\text{tr}R^2)\wedge B^{(6)}+\mathcal{O}({\alpha^\prime}^2)\ .
\end{split}
\end{equation}
To obtain the result above we have used integration by parts and the duality
\begin{equation}
dB^{(6)}=e^{-2\phi}\star_{10}dB^{(2)}\ ,
\end{equation}
which relates the Kalb-Ramond two form $B^{(2)}$ to a six-form $B^{(6)}$.

Let us now assume that on the 6D compactification manifold of each sector we turn on a bundle with structure group $G^{(\alpha)}$, $\alpha=1,\>2$, such that the unbroken gauge group in 4D is $H^{(\alpha)}$. We can then write the ten-dimensional gauge field strengths $\mathcal{F}^\alpha$,  as $\mathcal{F}^\alpha\equiv F^\alpha+\bar F^\alpha$, where
$F^\alpha$ is the external four dimensional part taking values in the low energy gauge group $H^{(\alpha)}$ and 
$\bar F^\alpha$ denotes the internal six-dimensional part, which takes values in the structure group $G^{(\alpha)}$ of the
bundle. We do a similar decomposition for the gauge potentials $\mathcal{A}^\alpha=A^\alpha+\bar A^\alpha$. In the present work, we turn on a non-abelian bundle $G^{(1)}$ on the observable sector, leading to a low energy group $H^{(1)}$ in four dimensions. However, on the hidden sector, we turn on a $G^{(2)}$ bundle which contains a $U(1)$ sub-bundle. Note that there is a $U(1)$ factor in both the internal structure group of the hidden sector bundle, and in its effective theory. This type of $U(1)$ factor leads to an anomaly in the effective theory which is canceled via the four dimensional equivalent of the well-known ten dimensional Green-Schwarz mechanism~\cite{Green:1984sg}. Such a $U(1)$ is called an ``anomalous'' $U(1)$~\cite{Dine:1987xk,Dine:1987gj,Anastasopoulos:2006cz,Green:1987mn}. We denote by $f$ the $U(1)$ field strength associated to the $U(1)$ gauge connection from the low energy theory on the hidden sector, and by $\bar f$ the internal $U(1)$ field strength.

Now note that the second term in \eqref{eq:kinetic_full} can be expressed as
\begin{equation}
\begin{split}
&\frac{\alpha^\prime}{8\kappa_{10}^2}\int_{\mathcal{M}_{10}}(\text{tr}\mathcal{F}_1^2+\text{tr}\mathcal{F}_2^2-\text{tr}R^2)\wedge B^{(6)}\\
&=\frac{\alpha^\prime}{8\kappa_{10}^2}\int_{\mathcal{M}_{10}}\left(\text{tr}F_1^2+\text{tr}\bar F_1^2+2\text{tr}(F_1\bar F_1)+\text{tr}F_2^2+\text{tr}\bar F_2^2+2\text{tr}(F_2\bar F_2)-\text{tr}R^2\right)\wedge B^{(6)}\ .
\end{split}
\end{equation}
On the observable sector $\text{tr}(F_1\bar F_1)$ vanishes, because the non-abelian group $G^{(1)}$ and its commutant do not share any generators of $E_8$. On the other hand, on the hidden sector, because of the presence of the ``anomalous'' $U(1)$, we get
\begin{equation}
\text{tr}(F_2\bar F_2)=(\text{tr}\>Q^2)\>f\wedge \bar f \neq 0\equiv 4 a f\wedge \bar f\ ,
\end{equation} 
where $Q$ is the $E_8$ generator that the $U(1)$ bundle and the low energy $U(1)$ connection share.  Keeping this cross-term only, we then find
\begin{equation}
\begin{split}
\label{eq:terms_kin_reduce}
S_{\text{kin}}&=-\frac{1}{4\kappa_{10}^2}\int_{\mathcal{M}_{10}}e^{-2\phi_{10}}H\wedge \star H \\
&\supset-\frac{1}{4\kappa_{10}^2}\int_{\mathcal{M}_{10}}e^{-2\phi_{10}}dB^{(2)}\wedge \star dB^{(2)}+\frac{\alpha^\prime}{8\kappa_{10}^2}\int_{\mathcal{M}_{10}}8a\left(f\wedge \bar f\right)\wedge B^{(6)}\ .
\end{split}
\end{equation}

We will consider the compactification manifold is a Calabi-Yau (CY) threefold $X$. For the purpose of reducing from 10D to 4D, by integrating over the Calabi-Yau $X$, it is convenient to use a basis of K\"ahler $(1,1)$-forms $\omega_i$, $i=1,\dots, h^{1,1}$ and their 
Hodge duals $\hat\omega^i$ such that
\begin{equation}
\int_X\omega_i\wedge \hat \omega^j=\delta_i^j\ .
\end{equation}
Following \cite{Lukas:1999nh} (see also\cite{Brandle:2003uya,Blumenhagen:2006ux}) we reduce the first term in the sum shown in \eqref{eq:terms_kin_reduce} from 10D to 4D leads to the kinetic terms of the dilaton axion $\sigma$ and of the K\"ahler axions $\chi^i$, given by
\begin{equation}
\label{eqs:kinetic_terms}
S_{\text{kin}}\supset -\int_{M_4}\left(g_{S\bar S}\>d\sigma\wedge\star_4d\sigma+4g^T_{i\bar  j}d\chi^i\wedge\star_4 d\chi^{\bar j}\right)\ ,
\end{equation}
where 
\begin{equation}
\begin{split}
&\kappa^2_4g_{S\bar S}=\frac{1}{4V^2}\ ,\\
&\kappa_4^2g_{i\bar j}^T=-\frac{d_{ijk}t^k}{\tfrac{2}{3}d_{ijk}t^it^jt^k}+\frac{d_{ijk}t^kt^ld_{jmn}t^mt^n}{\left(\tfrac{2}{3}\right)^2(d_{ijk}t^it^jt^k)^2}\ .
\end{split}
\end{equation}
are the dilaton and K\"ahler moduli metrics, respectively.

Furthermore, dimensionally reducing the second term in the sum in \eqref{eq:terms_kin_reduce} leads to a coupling between the K\"ahler axions and the $U(1)$ gauge field $A^\mu$, given by
\begin{equation}
\begin{split}
\label{eq:1:10}
S_{\text{kin}}\subset \int_{M_4}{8a}\epsilon_S\epsilon_R^2\>g^T_{i\bar j}\chi^{\bar j}c^i_1(L)\wedge d\star_4A\ .
\end{split}
\end{equation}
Note that this coupling cannot exist unless the hidden sector contains an anomalous $U(1)$.
In the following, we will show 
that we can also find a coupling between the dilaton axion $\sigma$ and the $U(1)$ vector field in 4D, which, however, has a different origin inside the 10D theory.

It is well known that the heterotic 10D theory exhibits gravitational, gauge and mixed gauge-gravitational anomalies resulting from anomalous hexagon diagrams at one-loop in string perturbation theory. Non-factorisable anomalies vanish by themselves and the factorizable ones are canceled by adding a one-loop counter term. The Green-Schwarz anomaly canceling one-loop counter term is given by
\begin{equation}
S_{GS}=\frac{1}{48(2\pi)^5\alpha^\prime}\int_{\mathcal M_{10}}B^{(2)}\wedge X_8\ ,
\end{equation}
where the eight-form $X_8$ defined as
\begin{equation}
\begin{split}
X_8=&\frac{1}{4}(\text{tr}\mathcal{F}^2_1)^2+\frac{1}{4}(\text{tr}\mathcal{F}^2_2)^2-\frac{1}{4}(\text{tr}\mathcal{F}^2_1)(\text{tr}\mathcal{F}^2_2)\\
&-\frac{1}{8}(\text{tr}\mathcal{F}^2_1+\text{tr}\mathcal{F}^2_2)(\text{tr}R^2)+\frac{1}{8}\text{tr}R^4+\frac{1}{32}(\text{tr}R^2)^2\ .
\end{split}
\end{equation}
 As shown in ~\cite{Weigand:2006yj}, splitting further each field strength into its internal and low-energy parts, one can find that the Green-Schwarz term contains a term of the type
\begin{equation}
\label{eq:GS_term_axion}
S_{GS}\supset\sum_{\alpha=1}^2\frac{1}{4(2\pi)^3\alpha^\prime}\int_{\mathcal{M}_{10}}B^{(2)}\wedge{\text{tr}}(F_\alpha\bar F_\alpha)
\left[ \frac{1}{4(2\pi)^2}\left( \text{tr}\bar F_\alpha^2-\frac{1}{2}\text{tr}\bar R^2 \right)\right]\ ,
\end{equation}
where the sum runs over the observable and the hidden sectors. We have already explained that for our observable sector, $\text{tr}(F_1\bar F_1)$ vanishes, while on the hidden sector, which contains an anomalous $U(1)$ symmetry, $\text{tr}(F_2\bar F_2)=4a(f\wedge \bar f)$.
Then, \eqref{eq:GS_term_axion} becomes
\begin{equation}
\label{eq:GS_term_axion2}
S_{GS}\supset\frac{1}{4(2\pi)^3\alpha^\prime}\int_{\mathcal{M}_{10}}4aB^{(2)}\wedge f\wedge \bar f
\left[ \frac{1}{4(2\pi)^2}\left( \text{tr}\bar F_2^2-\frac{1}{2}\text{tr}\bar R^2 \right)\right]\ ,
\end{equation}
Reducing \eqref{eq:GS_term_axion} from 10D to 4D by integrating over the Calabi-Yau $X$, we find a term which couples the dilaton axion to the $U(1)$ gauge field in 4D:
\begin{equation}
\begin{split}
\label{eq:1.16}
S_{GS}\supset\int_{M_4}2g_{S\bar S}\>\pi a\epsilon_S^2\epsilon_R^2\beta_i \sigma c_1^i(L) \wedge d \star_4 A\ ,\quad i=1,\dots,h^{1,1}\ ,
\end{split}
\end{equation}
where $\beta^i$ are the integer charges on the hidden sector under consideration
\begin{equation}
\beta_i=-\frac{1}{v^{1/3}}\frac{1}{4(2\pi)^2}\int_X\left( \text{tr}\bar F_2^2-\frac{1}{2}\text{tr}\bar R^2 \right)\wedge \omega_i\ .
\end{equation}
The coupling between the dilaton axion and the $U(1)$ gauge field appears at one-loop in string theory, when the term responsible for the cancellation of the hexagonal diagrams in the $E_8\times E_8$ heterotic theory is included.

Combining \eqref{eqs:kinetic_terms} and \eqref{eq:1.16}, we find the following action for the dilaton axion 
\begin{equation}
\begin{split}
\label{eq:action_dilaton}
S_\sigma&=-\int_{M_4}g_{S\bar S}\>d\sigma\wedge\star_4d\sigma
+\int_{M_4}g_{S\bar S}\>2\pi a\epsilon_S^2\epsilon_R^2\beta_i \sigma c_1^i(L)\wedge d\star_4A\\
&=\int_{M_4}d^4x\sqrt{-g}\>g_{S\bar S}\>\left[-\partial_\mu\sigma\partial^\mu\sigma
+2\pi a\epsilon_S^2\epsilon_R^2 \beta_i l^i \sigma(\partial_\mu  A^\mu) \right]\ ,
\end{split}
\end{equation}
whereas for the K\"ahler axions, after we combine \eqref{eqs:kinetic_terms} and \eqref{eq:1:10}, we get
\begin{equation}
\begin{split}
\label{eq:action_axion}
S_{\chi}&=-\int_{M_4}4g^T_{i \bar j}\>d\chi^i\wedge\star_4 d\chi^{\bar j}+\int_{M_4}{8a}\epsilon_S\epsilon_R^2\>g^T_{i\bar j}\>\chi^ic_1^{\bar j}(L)\wedge d \star_4 A\\
&=\int_{M_4}d^4x\sqrt{-g}\>4g^T_{i\bar  j}\left[-\partial_\mu \chi^i\partial^\mu \chi^{\bar j}+\> 2a \epsilon_S\epsilon_R^2l^i\chi^{\bar j} (\partial_\mu A^\mu) \right]\ .
\end{split}
\end{equation}
The couplings of the axions $\sigma$, $\chi^i$ to the anomalous $U(1)$ gauge field in the effective theory induces transformation laws for these axions under the $U(1)$ symmetry.

Equations \eqref{eq:action_dilaton} and \eqref{eq:action_axion} have the generic form
\begin{equation}
\begin{split}
\label{eq:general_axion_action}
S_{\rho}&=\int_{M_4}d^4x\sqrt{-g}\left[- g_{ab}\partial_\mu \rho^a\partial^\mu\rho^b-2g_{ab}Q^a \rho^b (\partial_\mu A^\mu)\right]\\ 
&=\int_{M_4}d^4x\sqrt{-g}\left[- g_{ab}\partial_\mu \rho^a\partial^\mu\rho^b+2g_{ab}Q^a (\partial_\mu \rho^b)  A^\mu\right]\ .
\end{split}
\end{equation}
where the index $a$ runs over the dilaton and the $h^{1,1}$  K\"ahler moduli, that is $a=S, T^1, \dots, T^{h^{1,1}}$. We have used integration by parts to obtain the second line. The coefficients $Q^a$ can be read off from \eqref{eq:action_dilaton} and \eqref{eq:action_axion}. They depend on the coupling parameters $\epsilon_S$ and $\epsilon_R$ which, generically, depend on the values of the moduli. However, when evaluating these coefficients in our context, the values of the moduli are assumed fixed at a specified supersymmetric vacuum. Therefore, within our context, the coupling coefficients $\epsilon_S$, $\epsilon_R$ and, hence, $Q^a$ are non-dynamical fields and can be considered to be constants. Similarly, we assumed that the moduli space metric $g_{a\bar b}$ is fixed at the vacuum; that is, $g_{a\bar b}= \langle g_{a\bar b}\rangle$. Perturbations of the metric around the vacuum are neglected in this analysis. 
It can be also be shown that higher order contributions $(\mathcal{O}(\alpha^\prime))$ lead to terms of the type $-Q_a^2A^\mu A_\mu$ for  $a=S, T^1, \dots, T^{h^{1,1}}$; see discussions in ~\cite{Lukas:1999nh,Ibanez:2012zz}. These can be added to the generic action in \eqref{eq:general_axion_action} to obtain
\begin{equation}
\begin{split}
\label{eq:general_axion_action_full}
S_{\rho}&=\int_{M_4}d^4x\sqrt{-g}\left[- (\partial^\mu\rho_a)^2+2Q^a (\partial_\mu \rho_a)  A^\mu
-Q_a^2A^\mu A_\mu\right]\\
&=-\int_{M_4}d^4x\sqrt{-g}\left[\partial_\mu\rho_a-Q_a   A_\mu\right]^2\ .
\end{split}
\end{equation}
If the anomalous $U(1)$ symmetry of the low energy theory is gauged and the gauge vector field $A^\mu$ transforms as
\begin{equation}
A^\mu \rightarrow A^\mu+\partial^\mu \theta\ ,
\end{equation}
 then it can be shown that the action $S_{\rho}$ is gauge invariant only if the scalar fields $\rho^a$ transform as
\begin{equation}
\rho^a\rightarrow \rho^a+Q^a \theta\ .
\end{equation}
In the transformations above, $\theta=\theta(x)$ is an arbitrary gauge parameter. 
Comparing the generic case shown in equation \eqref{eq:general_axion_action}, to the actions 
\eqref{eq:action_dilaton} and \eqref{eq:action_axion}, involving the dilaton axion the and K\"ahler axions respectively, we learn that the axions in our low energy theory have the following transformations under the anomalous $U(1)$:
\begin{equation}
\begin{split}
\label{eq:axions_plm}
&\sigma \rightarrow \sigma+Q^0\theta\ ,\quad Q^0=-\pi a\epsilon_S^2\epsilon_R^2 \beta_i l^i\ ,\\
&\chi^i\rightarrow \chi^i+Q^i\theta\ ,\quad Q^i=-\pi a \epsilon_S\epsilon_R^2l^i\ , \quad i=1,\dots,h^{1,1}\ .
\end{split}
\end{equation}

These set of axions represent the imaginary components of the dilaton moduli $S$ and the K\"ahler moduli $T$. As shown in Section 2, the expressions for these fields in the presence of five-branes are
\begin{equation}
\begin{split}
\label{eq:def_scalar}
& S=V+\pi\epsilon_SW_it^i\left(\tfrac{1}{2}+\lambda\right)^2
+i\left[\sigma+2 \pi\epsilon_S W_i\chi^iz^2\right] \ ,\\
&T^i=t^i+2i\chi^i\ ,\quad i=1,\dots,h^{1,1}\ ,
\end{split}
\end{equation}
where $W_i$ and $\lambda_i$ specify properties of the internal five-brane.
Hence, the $S$ and $T^i$ moduli have the following transformations under $U(1)$:
\begin{equation}
\begin{split}
\label{eq:def_scalar_transform}
& \delta_\theta S=iQ^0\theta+2i\pi\epsilon_S Q^iW_iz^2  \theta \equiv k_S\theta,\\
&\delta_\theta T^i=2iQ^i \theta\equiv k_T^i\theta\ ,\quad i=1,\dots,h^{1,1}\ ,
\end{split}
\end{equation}
where we have defined
\begin{equation}
\begin{split}
\label{eq:def_scalar_killing}
 k_S&=iQ^0+2i\epsilon_S Q^iW_iz^2  \\
&=-2i\pi a\epsilon_S^2\epsilon_R^2\left( \tfrac{1}{2}\beta^{N+1}_i l^i + W_il^iz^2\right)\\
k_T^i&=2iQ^i=-2i a\epsilon_S\epsilon_R^2l^i ,\quad i=1,\dots,h^{1,1}\ .
\end{split}
\end{equation}
Note that the five-brane modulus,
\begin{equation}
Z=W_it^iz+2iW_i(-\eta^i\nu+\chi^iz)\ ,
\end{equation}
contains the K\"ahler axions in the definition of its imaginary component. Therefore, the five-brane modulus transforms inhomogenously under under $U(1)$ as well, 
\begin{equation}
\delta_\theta Z=2iW_iz Q^i\theta=k_Z\theta\ ,
\end{equation}
where we defined
\begin{equation}
k_Z=2iW_iz Q^i=-2ia\epsilon_S\epsilon_R^2W_il^iz\ .
\end{equation}
Hence, we have obtained the inhomogenous gauge transformations of the $S$, $T^i$ and $Z$ moduli fields, and defined them in terms of the Killing vectors $k_S$, $k_T^i$ and $k_Z$.

\chapter{ D-term Stabilization Mechanism - General Formalism}

The 4D effective theory of our heterotic $M$-theory vacua contains a set of moduli fields; specifically, the dilaton, $h^{1,1}$ K\"ahler moduli, as well as a number the complex structure moduli. However, it has been shown in~\cite{Anderson:2010mh, Anderson:2011cza} that the complex structure moduli can be stabilized at the compactication scale for certain CY types. We assume this to be the case, such that the complex structure moduli are absent from the low energy theory. In addition to the moduli, we assume that the hidden sector contributes a set of massless matter fields $C^L$, $L=1,\dots, \mathcal{N}$ to the effective theory. 
We assume that the gauge group of the hidden sector contains a $U(1)$ factor of the ``anomalous'' type. Both types of fields, namely moduli and matter multiplets, transform under this Abelian symmetry. The matter fields $C^L$ are charged under this $U(1)$, with charges $Q^L$, and transform homogeneously. . Furthermore, as shown in Appendix A, the K\"ahler moduli $T^i$ transform inhomogenously under this $U(1)$, through a tree-level string coupling, while the dilaton transforms inhomogenously as well, but though a coupling which originates at genus-one. We neglect the fields from the observable sector in this analysis, because they do not transform under the anomalous $U(1)$ symmetry.

The Lagrangian of the hidden sector is determined by a K\"ahler potential $K(S,\bar S, T^{i},\bar {T,}^{i}C^{L},\bar {C}^{\bar{L}})$,  holomorphic gauge kinetic function $f_{2}(S,T^{i})$, which depends non-trivially on the moduli fields $S$ and $T^i$,
and a holomorphic superpotential $\mathcal{W}$, which, at perturbative level, does not depend on the moduli fields. This Lagrangian is invariant under the inhomogenous transformations of the moduli fields 
$S$ and $T^i$, as well as the homogenous transformations of the matter fields $C^L$. 
The effective theories we are interested in are best analyzed using the formalism of the non-linear $N=1$ SUSY $\sigma-$model; see, for example, the analysis shown in~\cite{Freedman:2012zz}.

The role of this Appendix is to outline the generic characteristics of this formalism, for a set of chiral superfields which transform under an anamolous $U(1)$ symmetry. We will consider a set of $N$ complex chiral superfields $Z^A$, with scalar, fermionic and auxiliary field components $(z^A, \psi^A, F^A)$, coupled to the same $U(1)$ gauge supermultiplet. We will formulate the F-term and D-term stabilization conditions and, hence, show that, in general, the D-term stabilization mechanism can lead to non-zero VEVs for the scalar components and, consequently, to the formation of a massive $U(1)$ vector superfield.

We begin our general analysis by assuming a K\"ahler potential $K(z,\bar z)$ and a K\"ahler metric
\begin{equation}
g_{A\bar B}=\partial_A\partial_{\bar B}K(z,\bar z)\ ,
\end{equation}
which has a Lie group of symmetries generated by holomorphic Killing vectors. Under these symmetries, the components of the chiral multiplets have the following infinitesimal transformations~\cite{Wess:1992cp,Freedman:2012zz}:
\begin{equation}
\begin{split}
&\delta_\theta z^A=k^A(z)\theta\ ,\\
&\delta_\theta \psi^A= \frac{\partial k(z)^A}{\partial z^B}\psi^B\theta\,\\
&\delta_\theta F^A=\frac{\partial k^A(z)}{\partial z^B}F^B\theta-
\frac{1}{2}\frac{\partial^2k(z)^A}{\partial z^B \partial z^C}\psi^{B\dag}\psi^C\theta\ .
\end{split}
\end{equation}
The vectors $k^A(z),\>\bar k^{\bar A}(\bar z)$ are related to real scalar moment map $\mathcal{P}(z,\bar z)$, such as
\begin{equation}
k^A=-ig^{A\bar B}\partial_{\bar B}\mathcal{P}(z, \bar z)\ , \quad \bar k^{\bar A}=-ig^{A\bar B}\partial_{A}\mathcal{P}(z, \bar z)\ .
\end{equation}
Inverting the expressions above we get
\begin{equation}
\label{eq:Killing_vectors}
\mathcal{P}(Z,\bar Z)=ik^A\partial_AK(z,\bar z)=-i\bar k^{\bar A}\partial_{\bar A}K(z,\bar z)\ .
\end{equation}
In the following we will show that the D-term flatness condition is naturally expressed in terms of this moment map.

As we promote this Killing symmetry to a gauge symmetry, with parameter $\theta\rightarrow \theta(x)$, 
we introduce a $U(1)$ gauge supermultiplet $(A_\mu,\lambda,D)$ with gauge coupling constant $g$. The next step is to define covariant derivatives for the fields in the $N$ chiral multiplets $Z^A=(z^A, \psi^A, F^A)$,
\begin{equation}
\begin{split}
&\mathcal{D}_\mu z^A=\partial_\mu z^A-A_\mu k^A\ ,\\
&\mathcal{D}_\mu \bar z^{\bar A}=\partial_\mu \bar z^{\bar A}-A_\mu \bar k^{\bar A}\ ,\\
&\mathcal{D}_\mu \psi^A=\partial_\mu \psi^A-A_\mu \frac{\partial k^A}{\partial z^B}\psi^B\ .
\end{split}
\end{equation}
These covariant derivatives are used to build a Lagrangian with $N$ chiral superfields $Z^A$, which is invariant under the gauged anomalous $U(1)$ symmetry. The result is
\begin{equation}
\label{eq:initial_lagrangian}
\begin{split}
\mathcal{L}&\supset -g_{A\bar B}\mathcal{D}_\mu  z^A \mathcal{D}^\mu \bar z^{\bar B}  -ig_{A\bar B}\psi^A \slashed{\mathcal{D}} \psi^{\bar B\dag}- \frac{i}{g^2}\lambda \slashed{\partial} \lambda^\dag -\frac{1}{4g^2}F_{\mu\nu}F^{\mu \nu}
+\sqrt{2}g_{A\bar B}k^A\lambda^\dag \psi^{\bar B\dag}\\&+\sqrt{2} g_{A\bar B}\bar  k^{\bar B}\lambda \psi^A
-\frac{1}{2}g^2\mathcal{P}^2-g^{A\bar B}\frac{\partial \mathcal{W}}{\partial z^A}\frac{\partial\mathcal{\bar W}}{\partial \bar z^{\bar B}}\dots\ , \quad A,B=1,\dots, N\ .\\
\end{split}
\end{equation}
The coupling $g=g(z)$ is the $U(1)$ gauge coupling. It can be expressed in terms of the holomorphic gauge kinetic function $f(z)$ as
\begin{equation}
\frac{1}{g^2(z)}=\text{Re}f(z)\ .
\end{equation} 
The scalar potential contains a D-term potential $V_D$, defined in terms of the real moment maps and an F-term potential $V_F$, defined in terms of the superpotential $\mathcal{W}(Z)$.
\begin{equation}
 V_D+V_F=\frac{1}{2}g^2\mathcal{P}^2+g^{A\bar B}\frac{\partial \mathcal{W}}{\partial z^A}\frac{\partial\mathcal{\bar W}}{\partial \bar z^{\bar B}}\ .
\end{equation}

Unbroken supersymmetry requires both D-term and F-term potentials to vanish. These requirements are also called the D-flatness and F-flatness conditions. More explicitly, in a supersymmetric vacuum
\begin{equation}
\langle \mathcal{P}\rangle =0\ , \quad \langle  \frac{\partial \mathcal{W}}{\partial z^A} \rangle=0\ , \quad A=1,\dots, N\ .
\end{equation}
In general, these conditions alone do not fix all the VEVs of the scalar fields $z^A$ of the system. They do, however, restrict the possible range that the VEVs of the scalar fields $z^A$ can obtain. 
The D-flatness condition can be written
\begin{equation}
\label{eq:susy_condition_0}
\langle \mathcal{P}\rangle= \langle\> k^A(z)\partial_A K(z,\bar z)\>\rangle =0\ .
\end{equation}
The condition above does not lead to particularly interesting effects if the effective theory contains only chiral fields which transform homogeneously under the $U(1)$ symmetry. Indeed, when this is the case, $k^A\sim z^A$, and hence, the D-term potential always vanishes when the values of the scalar fields are equal to zero. Such scalars would not aquire non-zero VEVs. On the other hand, if the theory contains fields which transform inhomogenously under the $U(1)$ symmetry, the D-flatness condition can lead to non-trivial VEVs of the scalar fields. 

Let us assume that among the fields $z^A$ we find some which do transform inhomogenously under $U(1)$. In the following we will prove that in a D-flat, non-trivial vacuum, a massive vector supermultiplet is always produced. To display this process, we expand the scalar fields around
the vacuum $z^A=\langle z^A\rangle +\delta z^A$. Note that fixing the VEVs $\langle z^A \rangle$ results in fixing an 
expectation value for the $z$-dependent Killing vectors, as well as for the $z$-dependent K\"ahler metric $g_{A\bar B}$ and the gauge coupling $g$. We use the notation
\begin{equation}
\begin{split}
&\langle k^A(z)\rangle =\left.k^A(z)\right|_{ z=\langle z \rangle}\ , \quad \langle \bar k^{\bar A}(\bar z)\rangle =\left.\bar k^{\bar A}(\bar z)\right|_{ \bar z=\langle \bar z \rangle}\ ,\\
&\langle g_{A\bar B}(z,\bar z)\rangle =\left.g_{A\bar B}(z,\bar z)\right|_{ z,\bar z=\langle z\rangle, \langle \bar z\rangle}\ ,\quad \langle g(z)\rangle =\left.g(z)\right|_{ z=\langle z \rangle}\ .
\end{split}
\end{equation}
which means that the metric and the Killing vectors are evaluated in the vacuum delimited by the D-flatness condition.

In $N=1$ SUSY, the massive vector multiplet has one spin-$0$ component, two spin-$\frac{1}{2}$ components and one spin-$1$ component~\cite{Wess:1992cp,Shifman:2012zz}. More specifically, these components are: a real scalar, a massive vector boson and a Dirac fermion. Next, 
we will attempt to identify the origin of each of these elements.

\begin{itemize}

\item Massive real scalar $\phi$:

Expanding around the (assumed non-trivial) vacuum defined in \eqref{eq:susy_condition_0} we get:
\begin{equation}
\delta\mathcal{P}=\langle \frac{\partial \mathcal{P}}{\partial z^A} \rangle \delta z^A+\langle \frac{\partial \mathcal{P}}{\partial {\bar z}^{\bar B}}\rangle \delta z^{\bar B}\ .
\end{equation}
Now, inverting the equations from \eqref{eq:Killing_vectors} we find
\begin{equation}
\partial_{\bar B}\mathcal{P}=ik^Ag_{A\bar B}\ ,\quad \partial_{A}\mathcal{P}=-ik^{\bar B}g_{A\bar B}\ ,
\end{equation}
and therefore, we express the perturbation of the moment map as
\begin{equation}
\delta\mathcal{P}=i\langle k^{\bar B}g_{A\bar B}\rangle \delta z^A-i\langle k^Ag_{A\bar B}\rangle \delta \bar z^{\bar B}\equiv -2\sqrt{\langle g_{A\bar B}k^A\bar k^{\bar B}\rangle}\phi \ .
\end{equation}
In the expression above, we defined a scalar field $\phi $, as 
\begin{equation}
\label{eq:phi_def}
\phi=\frac{ i\langle k^Ag_{A\bar B}\rangle \tfrac{\delta \bar z^{\bar B}}{2}-i\langle k^{\bar B}g_{A\bar B}\rangle \tfrac{\delta z^A}{2}  }{\sqrt{ \langle g_{A\bar B}k^A\bar k^{\bar B}\rangle}}\\
\end{equation}
Hence, we found that as we expand around the vacuum, we obtain the following expression for the potential energy
\begin{equation}
\label{eq:potential_phi}
\mathcal{L}\supset- \frac{1}{2}\delta (g^2)\langle \mathcal{P}^2\rangle- \frac{1}{2}\langle g^2\rangle (\langle \mathcal{P}\rangle+\delta \mathcal{P})^2=-\frac{1}{2}\langle g^2\rangle \delta \mathcal{P}^2=-2\langle g^2g_{i\bar j}k^A\bar k^{\bar B}\rangle\phi^2\ ,
\end{equation}
which is a mass term for the real scalar $\phi$. 
The field $\phi$ has a canonically normalized kinetic term in the Lagrangian. To prove it, it is best to rotate the basis of scalar perturbations $\{\delta z^A\}$  into a new basis of scalars $\{ \xi^A\}$, which have canonically normalized kinetic energy. We interpret the fields $\xi^A$, $A=1,\dots, N$ to be the true mass eigenstates of the system. However, in the vacuum defined by the D-flatness condition, only one such eigenstate, which we denote by $\xi^1$, becomes massive. The rest of the scalars $\xi^A$, $A=2,\dots,N$ fields remain massless. 

The two sets of fields, $\{ \xi^A\}$ and $\{ \delta z^A\}$, $A=1,\dots, N$, are related by the rotation matrix $U$, such that
\begin{equation}
\begin{split}
&\xi^A=[U]^A_B\delta z^B\ , \quad \bar \xi^A=[U^*]^{\bar A}_{\bar B}\delta\bar z^B\ ,\\
&\delta z^A=[{U^{-1}}]^A_B\xi^B\ , \quad \delta\bar z^{\bar A}=[{U^{*-1}}]^{\bar A}_{\bar B}\bar\xi^{\bar B}\ .
\end{split}
\end{equation}
%
We demand that after the rotation, the fields $\xi^A$ have canonically normalized kinetic energy and therefore
\begin{equation}
\langle g_{A\bar B}\rangle \partial_\mu \delta z^A \partial^\mu\delta \bar z^{\bar B}=\langle g_{A\bar B}\rangle[{U^{-1}}]^A_C\bar [{U^{-1}}]^{\bar B}_{\bar D}
\partial_\mu  \xi^C \partial^\mu \bar \xi^{\bar D}=\delta_{C\bar D}\partial_\mu  \xi^C\partial^\mu\ \bar \xi^{\bar D}\ ,
\end{equation}
which is possible only if
\begin{equation}
\label{eq:matrix_condition_x}
\langle g_{A\bar B}\rangle[{U^{-1}}]^A_C [{U^{*-1}}]^{\bar B}_{\bar D}=\delta_{C\bar D}\ .
\end{equation}
The condition shown above is not enough by itself to determine all the elements of the rotation matrix. One possible ansatz for the rotation matrix $[U^{-1}]^A_B$, which solves \eqref{eq:matrix_condition_x}, is 
\begin{equation}
\label{eq:matrix_sol}
[U^{-1}]^A_B=
\left(
\begin{matrix}
\frac{\langle k^1\rangle}{\sqrt{ \langle g_{A\bar B}k^A\bar k^{\bar B}\rangle}}&\frac{u^1_{2}}{\sqrt{\langle g_{A\bar B}u_2^A\bar u_2^{\bar B}\rangle}}&\dots&\frac{u^1_{N}}{\sqrt{\langle g_{A\bar B}u_N^A\bar u_N^{\bar B}\rangle}}\\
\frac{\langle k^2 \rangle}{\sqrt{\langle g_{A\bar B}k^A\bar k^{\bar B}\rangle}}&\frac{u^2_{2}}{\sqrt{\langle g_{A\bar B}u_2^A\bar u_2^{\bar B}\rangle}}&\dots&\frac{u^2_{N}}{\sqrt{\langle g_{A\bar B}u_N^A\bar u_N^{\bar B}\rangle}}\\
\vdots& \vdots & \ddots &\vdots\\
\frac{\langle k^N \rangle}{\sqrt{\langle g_{A\bar B}k^A\bar k^{\bar B}\rangle}}&\frac{u^N_{2}}{\sqrt{\langle g_{A\bar B}u_2^A\bar u_N^{\bar B}\rangle}}&\dots&\frac{u^N_{N}}{\sqrt{\langle g_{A\bar B}u_N^A\bar u_N^{\bar B}\rangle}}
\end{matrix}
\right)\ \equiv\left(\underline k\>,\underline{u_2},\>\dots,\>\underline{u_N} \right)\ ,
\end{equation}
The matrix can be expressed as a set of column vectors $(\underline{k}, \underline{u_n})$, $n=2,\dots,N$, as shown above (note that we considered $u_1\equiv k$).
After solving the $D$-flatness condition shown in \eqref{eq:susy_condition_0}, we can fix the vector $\underline{k}$ of the matrix only. The reason for this ansatz, as well as the physical implications that follow, will become apparent later.
The rest of the column vectors in this matrix, $\underline{u_n}$, $n=2,\dots,N$ cannot be completely determined in the absence of yet another vacuum stabilization potential in the theory. They not are not entirely unconstrained, though. They are orthogonal to the vector $\underline k$. 
To prove this, we write condition \eqref{eq:matrix_condition_x} in terms of the elements of the matrix shown above and obtain
 \begin{equation}
 \label{orthog1}
 \langle g_{A\bar B}k^A\rangle \bar u^{\bar B}_n=\delta_{1n}=0 \quad \text{for}\quad n\in[2,\dots,N]\ .
 \end{equation}
 The index $n$ runs over the subscript of the column vectors $\left(\underline{u_2},\>\dots,\>\underline{u_N} \right)$. We introduce the tensor product $\langle g_{A\bar B}k^A\rangle =({g\cdot k})_{\bar B}$, which defines the contravariant vector $(\underline{g\cdot k})$, and write
 \begin{equation}
 \label{equ:orthogo_exp}
\langle  g_{A\bar B}k^A\rangle \bar u_n^{\bar B}=({g\cdot k})_{\bar B} \bar u_n^{\bar B}=(\underline{g\cdot k})\cdot \overline{\underline{ u_n}}=\delta_{1n}\ .
 \end{equation}
Note that we have introduced the dot product between the vectors $(\underline{g\cdot k})$ and $\overline{\underline{ u_n}}$. The expression in \eqref{equ:orthogo_exp} represents a set of $N-1$ equations that express the orthogonality between each of the $N-1$ column vectors $\underline{u_n}$, $n=2,\dots,N$ and the vector $\underline{k}$.
This set of orthogonality relations is not the only one we obtain; remember that we demand that all the $N$ fields $\xi^A$ have canonically normalized kinetic terms; that is, condition \eqref{eq:matrix_condition_x} constraints the set of vectors $\underline{u_n}$, $n=2,\dots, N$ to be orthogonal to each other and have unit length:
\begin{equation}
\begin{split}
\label{orthog2}
&\langle g_{A\bar B}\rangle u_n^A\bar u_m^{\bar B}=(\underline{g\cdot u_n}) \cdot \overline{\underline{ u_m}}=\delta_{nm}=1 \quad \text{if}\quad n= m\ , \quad \text{for}\quad n,\>m\in[2,\dots,N]\ .\\
&\langle g_{A\bar B}\rangle u_n^A\bar u_m^{\bar B}=(\underline{g\cdot u_n}) \cdot \overline{\underline{ u_m}}=\delta_{nm}=0 \quad \text{if}\quad n\neq m\ , \quad \text{for}\quad n,\>m\in[2,\dots,N]\ .
 \end{split}
\end{equation}
We have defined the tensor product $g_{A\bar B}u_n^A=(\underline{g\cdot u_n})_{\bar B}$, which defines the $N-1$ contravariant vectors $(\underline{g\cdot u_n})$.

The total number of undetermined matrix entries within the set of $N-1$ column vectors $\underline{u_n}$, $n=2,\dots,N$ is $N-1\times N$. We have shown that these parameters are not completely independent, however. The orthogonality constraints \label{eq:matrix_condition} significantly reduces the number of degrees of freedom of the system. $N-1$ constraints arise because these $N-1$ column vectors have unit length ($|u_n|=1$). The requirement that these $N-1$ column vectors are orthogonal to each other, as well as to the fixed direction $\underline k$ introduces another set of $\frac{1}{2}N(N-1)$ constraints. Therefore, the real number of independent free parameters, ``hidden'' in the matrix \eqref{eq:matrix_sol} is actually equal to
\begin{equation}
\begin{split}
\label{N_degrees}
N_{\text{dof}}&=(N-1) N-(N-1)-\frac{1}{2}(N-1)\\
&=\frac{1}{2}(N-1)(N-2)\ .
\end{split}
\end{equation}
The inverse matrix $U$ can be inferred from the condition $UU^{-1}=1_N$, together with the orthogonality requirement shown in \eqref{orthog1} and \eqref{orthog2}:
\begin{equation}
\label{eq:matrix_solada}
[U]^A_B=
\left(
\begin{matrix}
\frac{\langle g_{1\bar A}\bar k^{\bar A}\rangle}{\sqrt{\langle g_{A\bar B}k^A\bar k^{\bar B}\rangle}}& \frac{\langle g_{2\bar A}\bar k^{\bar A}\rangle}{\sqrt{\langle g_{A\bar B}k^A\bar k^{\bar B}\rangle}}&\dots&\frac{\langle g_{N\bar A}\bar k^{\bar A}\rangle}{\sqrt{\langle g_{A\bar B}k^A\bar k^{\bar B}\rangle}}\\
\frac{\langle g_{1\bar A}\bar u_2^{\bar A}\rangle}{\sqrt{\langle g_{A\bar B}u_2^A\bar u_2^{\bar B}\rangle}}&\frac{\langle g_{2\bar A}\bar u_2^{\bar A}\rangle}{\sqrt{\langle g_{A\bar B}u_2^A\bar u_2^{\bar B}\rangle}}&\dots&\frac{\langle g_{N\bar i}\bar u_2^{\bar A}\rangle}{\sqrt{\langle g_{A\bar B}u_2^A\bar u_2^{\bar B}\rangle}}\\
\vdots& \vdots& \ddots&\cdots\\
\frac{\langle g_{1\bar A}\bar u_N^{\bar A}\rangle}{\sqrt{\langle g_{A\bar B}u_2^A\bar u_2^{\bar B}\rangle}}&\frac{\langle g_{2\bar A}\bar u_N^{\bar A}\rangle}{\sqrt{\langle g_{A\bar B}u_2^A\bar u_2^{\bar B}\rangle}}&\dots&
\frac{\langle g_{N\bar A}\bar u_N^{\bar A}\rangle}{\sqrt{\langle g_{A\bar B}u_2^A\bar u_2^{\bar B}\rangle}}\\
\end{matrix}
\right)\ 
\equiv\left(
\begin{matrix}
\underline{g\cdot \bar k}\\ \underline{g\cdot \bar u_2}\\ \dots \\ \underline{g\cdot \bar u_N}
\end{matrix}
\right)\ .
\end{equation}
As shown above, we can write this matrix as a set of row vectors $\underline{g\cdot k}$, $\underline{g\cdot u_n}$, for $n=2,\dots,N$, which were first encountered in the orthogonality relations \eqref{equ:orthogo_exp} and \eqref{orthog2}. As mentioned previously, the $U$ matrix combines the set of $N$ scalar linear perturbations around the vacuum, $\delta z^A$, into a set of $N$ physical states, $\xi^A$. However, the $D$-flatness condition shown in \eqref{eq:susy_condition_0} determines only the first row of this matrix and hence, we can determine exactly one physical state:
\begin{equation}
\label{eq:define_xi}
\xi^1=\frac{\langle g_{A\bar B}\bar k^{\bar B}\rangle}{\sqrt{\langle g_{A\bar B}k^A\bar k^{\bar B}\rangle}}\delta z^A\equiv (\underline{g\cdot k})\cdot \underline{\delta z}\ .
\end{equation}

It is now time to reveal why we have chosen the ansatz shown in eq. \eqref{eq:matrix_solada} and \eqref{eq:matrix_sol} for the matrix $U$ and its inverse, respectively. Let us first split the field $\xi^1$ into its real and imaginary parts.
Using the definition of $\xi^1$, one can then check that we can find the following expression
\begin{equation}
\begin{split}
\label{eq:phi_eta_def}
&\text{Re}\xi^1=\frac{\langle  g_{A\bar B}k^A \rangle  \tfrac{\delta \bar z^{\bar B}}{2}+\langle  g_{A\bar B}\bar k^{\bar B}\rangle \tfrac{\delta  z^{A}}{2}}
{\sqrt{ \langle g_{A\bar B}k^A\bar k^{\bar B}\rangle}}\ ,\\
&\text{Im}\xi^1=\frac{ i\left( \langle g_{A\bar B}k^A\rangle\tfrac{\delta \bar z^{\bar B}}{2}-\langle g_{A\bar B}\bar k^{\bar B}\rangle \tfrac{\delta z^A}{2}\right)  }{\sqrt{\langle  g_{A\bar B}k^A\bar k^{\bar B}\rangle}}\ .
\end{split}
\end{equation}
We have recovered the field $\phi=\text{Im}\xi^1$, first defined in eq. \eqref{eq:phi_def}, which represents
the direction in which we develop the quadratic potential shown in \eqref{eq:potential_phi}. This quadratic potential is responsible for stabilizing the theory to a supersymmetric vacuum. 

Note that both fields $\phi$ and $\eta$ have canonically normalized kinetic terms, that is
\begin{equation}
\mathcal{L}\supset -\partial_\mu \xi^1\partial^\mu \bar \xi^1=-\partial_\mu \phi\partial^\mu \phi-\partial_\mu \eta 
\partial^\mu \eta\ .
\end{equation}
Combining the kinetic terms of the scalars with the quadratic potential shown in \eqref{eq:potential_phi} it follows that
\begin{equation}
\mathcal{L}\supset -\partial_\mu \phi \partial^\mu \phi -\partial_\mu \eta 
\partial^\mu \eta-\partial_\mu \xi_A\partial^\mu \bar \xi^A-2\langle g^2g_{A\bar B}k^A\bar k^{\bar B}\rangle\phi^2\ ,
\quad A=2,\dots, N\ .
\end{equation}
We see that the D-term potential produced a mass term for the scalar $\phi$,
\begin{equation}
m_\phi=\sqrt{ 2\langle g^2g_{A\bar B}k^A\bar k^{\bar B}\rangle}\ ,
\end{equation}
while the other $N-1$ scalars $\xi^A$ remained massless.
After it aquires a mass, $\phi$ becomes the real scalar component of a supersymmetric massive vector multiplet. 

The real part of $\xi^1$, which we denote $\eta=\text{Re}\xi^1$ also plays an important role in the theory. In the next part we will show that $\eta$ is the Goldstone scalar which becomes the longitudinal degree of freedom of the massive vector boson $A^\mu$. All other $N-1$ states $\xi^A$, remain undetermined. We know, however, that they must be orthogonal to the direction in which we develop a potential, and hence, they remain as flat directions in the theory. The role played by the matrix $U$ and its inverse, therefore, is to project the field perturbations around the supersymmetic vacuum into these $N$ physical eigenstates $\xi^A$.

\item Massive vector boson $A^\mu$:

From the covariant derivative of the scalar field we get a mass term for the vector boson
\begin{equation}
\begin{split}
\label{eq:covariant_der_term}
\mathcal{L}\subset &  -g_{A\bar B}\mathcal{D}_\mu  z^A  \mathcal{D}^\mu \bar z^{\bar B}=-g_{A\bar B}\mathcal{D}_\mu  (\langle z^A\rangle +\delta z^A) \mathcal{D}^\mu (\langle\bar z^{\bar B}\rangle +\delta \bar z^{\bar B})\\
&=-g_{A\bar B}\partial_\mu \delta z^A \partial^\mu\delta \bar z^{\bar B}+g_{A\bar B}(\langle k^A\rangle \partial_\mu \delta \bar z^{\bar B}+\langle \bar k^{\bar B}\rangle \partial_\mu\delta  z^{A})A_\mu-\langle g_{A\bar B} k^A \bar k^{\bar B}\rangle A^\mu A_\mu\\
&=-g_{A\bar B}\partial_\mu \delta z^A \partial^\mu\delta \bar z^{\bar B}+2\sqrt{\langle g_{A\bar B}k^A\bar k^{\bar B}\rangle}\partial_\mu \eta A^\mu- \langle g_{A\bar B} k^A \bar k^{\bar B}\rangle A^\mu A_\mu \ .
\end{split}
\end{equation}

We combined the terms linear in $\partial_\mu \delta z^{A}, \partial_\mu \delta \bar z^{\bar A}$ into the field
\begin{equation}
\eta=\frac{\langle g_{A\bar B}k^A\rangle \tfrac{\delta \bar z^{\bar B}}{2}+\langle g_{A\bar B} \bar k^{\bar B}\rangle \tfrac{\delta  z^{A}}{2}}
{\sqrt{\langle g_{A\bar B}k^A\bar k^{\bar B}\rangle}}\ ,
\end{equation}
which we first encountered in eq. \eqref{eq:phi_eta_def}. We can now write \eqref{eq:covariant_der_term} in the form

\begin{equation}
\begin{split}
\mathcal{L}\subset & - g_{A\bar B}\mathcal{D}_\mu  z^A  \mathcal{D}^\mu \bar z^{\bar B}=\\
&=\dots+- \partial_\mu \eta\partial^\mu \eta+2\sqrt{\langle g_{A\bar B}k^A\bar k^{\bar B}\rangle}\partial_\mu \eta A^\mu- \langle g_{A\bar B} k^A \bar k^{\bar B}\rangle A^\mu A_\mu\\
&=\dots -\langle g_{A\bar B} k^A \bar k^{\bar B}\rangle \left(A_\mu-\frac{\partial_\mu \eta}{\sqrt{\langle g_{A\bar B}k^A\bar k^{\bar B}\rangle}}\right)\left(A^\mu-\frac{\partial^\mu \eta}{\sqrt{\langle g_{A\bar B}k^A\bar k^{\bar B}\rangle}}\right)\\
&=-\langle g_{A\bar B} k^A \bar k^{\bar B}\rangle A^\prime_\mu A^{\prime\mu}\ .
\end{split}
\end{equation}
In the above expression, we have used the ``unitary gauge'' to define the field 
\begin{equation}
A_\mu^\prime = A_\mu-\frac{\partial_\mu \eta}{\sqrt{\langle g_{A\bar B}k^A\bar k^{\bar B}\rangle}}\ .
\end{equation}
We found that after the scalar fields $z^i$ aquire VEVs, $\partial_\mu \eta/\sqrt{\langle g_{A\bar B}k^A\bar k^{\bar B}\rangle}$ becomes the longitudinal component of a now massive vector field $A_\mu^\prime$. The field $\eta$ plays the role of a Goldstone boson ``eaten'' by the vector boson. 

Including the kinetic energy term of the vector boson we have
\begin{equation}
\begin{split}
\mathcal{L}&\supset -\frac{1}{4g^2}F_{\mu\nu}F^{\mu \nu}-\langle g_{A\bar B}k^A \bar k^{\bar B}\rangle A_\mu A^{\mu}\\
&=-\frac{1}{4\langle g^2\rangle }F_{\mu\nu}F^{\mu \nu}-\langle g_{A\bar B}k^A \bar k^{\bar B}\rangle A_\mu A^{\mu}+\dots
\Rightarrow m_A^2=2\langle g^2g_{A\bar B}k^A\bar k^{\bar B}\rangle\ .
\end{split}
\end{equation}
Note that we have dropped the prime on $A_{2\mu}$ for simplicity. The dots represent higher order interaction terms obtained after expanding the gauge kinetic function around its fixed value in the vacuum.

\item Massive Dirac fermions $\Psi$:

The Lagrangian shown in \eqref{eq:initial_lagrangian} contains cross couplings between the gaugino $\lambda$ and the fermions $\psi^A$. After the fields $z^A$ obtain VEVs, these couplings mix the gaugino with the chiral fermions into a new mass eigenstate. We have
\begin{equation}
\begin{split}
\mathcal{L}\subset \sqrt{2}\langle g_{A\bar B}\rangle \langle k^A\rangle \lambda^\dag \psi^{\bar B\dag}+\sqrt{2} \langle g_{A\bar B}\rangle \langle\bar k\rangle^{\bar B}\lambda \psi^A\equiv &\sqrt{2 \langle g_{A\bar B}k^A\bar k^{\bar B}\rangle}
\left(\psi_\xi^{1\dag} \lambda^\dag+\lambda \psi^1_\xi\right)\\
=&\sqrt{2 \langle g^2 g_{A\bar B}k^A\bar k^{\bar B}\rangle}\bar \Psi \Psi\ .
\end{split}
\end{equation}
To obtain the last expression, we had to define:
\begin{equation}
\begin{split}
\label{eq:psi_xi_def}
\Psi=\left(\begin{matrix}\lambda^\dag/\langle g\rangle\\ \psi^1_{\xi}\end{matrix}\right)\ , \quad \text{and}\quad \psi^1_{\xi}=\frac{ \langle g_{A\bar B}\bar k^{\bar B}\rangle \psi^A }{\sqrt{\langle g_{A\bar B}k^A\bar k^{\bar B}\rangle}}\ , \quad 
\psi_{\xi}^{1\dag}=\frac{ \langle g_{A\bar B} k^{A}\rangle \psi^{\bar B\dag} }{\sqrt{\langle g_{A\bar B}k^A\bar k^{\bar B}\rangle}}\ .
\end{split}
\end{equation}
We learned that when we evaluate the cross couplings in the newly defined vacuum, a single linear combination of fermions, called $\psi^1_\xi$, combines with the existing gaugino to form a massive Dirac spinor,
Examining eq. \eqref{eq:psi_xi_def}, we learn that the state $\psi^1_{\xi}$ is a linear combination of the $N$ fermions $\psi^A$. Note that we can use the same rotation matrix $U$, defined for the scalars, to express this linear combination.
\begin{equation}
\label{eq:both_rotations}
\xi^1=[U]^1_A\delta z^A\ , \quad \psi^1_{\xi}=[U]^1_A\psi^A\ .
\end{equation}
What this shows is that the fermion eigenstate $\psi^1_{\xi}$ corresponds to the scalar eigenstate $\xi^1$ They belong in the same chiral supermultiplet $(\xi^1,\psi^{1}_{\xi})$. This chiral supermultiplet becomes massive after the system is stabilized into a supersymmetric vacuum, but is absorbed into a massive vector multiplet. The imaginary component of $\xi^1$, $\phi$, becomes the real scalar component of this vector multiplet. We have also shown that $\eta$, the real component of $\xi^1$, becomes the longitudinal degree of freedom of the massive vector boson $A_\mu$. We are left to discuss the fermionic component of this vector multiplet, and show that it has the same mass as the field $\phi$, a direct consequence of unbroken supersymmetry.

We found a mass term for a Dirac fermion $\Psi$: $\sqrt{2 \langle g^2g_{A\bar B}k^A\bar k^{\bar B}\rangle}\bar \Psi \Psi$. To determine the mass of $\Psi$, we have to ensure its kinetic energy is canonically normalized. We have already seen that the fields $\xi^1$ and $\psi^1_{\xi}$ are defined with the same rotation matrix $U$. In fact, we form other $N-1$ chiral multiplets with the same rotation matrix. These eigenstates, however, pair into massless multiplets $(\xi_A,\psi_{A\xi})$, $i=2,\dots, N$. We write
\begin{equation}
\begin{split}
\label{eq:rest_states1}
&\xi^A=[U_B^A]\delta z^B\ , \quad A=2,\cdots,N\ ,\\
&\psi_\xi^A=[U]^A_B\psi^B\ ,\quad B=1,\cdots, N\ .
\end{split}
\end{equation}

 In terms of the new fermion eigenstates, the kinetic term of the chiral fermionic fields $\psi^i$ becomes
\begin{equation}
 -ig_{A\bar B}\psi^A  \slashed{\partial} \psi^{\dag \bar B}\rightarrow -i g_{A\bar B}[U^{-1}]^A_C[U^{*-1}]^{\bar B}_{\bar D}\psi^C_\xi
 \slashed{\partial}\psi_{\xi}^{\dag \bar D}=-i\delta_{C\bar D}\psi^C_\xi \slashed{\partial}\psi_{\xi}^{\dag \bar D}\ ,
\end{equation}
where we made use of the property of the matrix $U^{-1}$ of diagonalizing the metric, as shown in \eqref{eq:matrix_condition_x}. 
 We learn that the massive field $\psi^1_{\xi}$, as well as the massless fermions $\psi^A_{ \xi}$, for $A=2,\dots,N$ have canonically normalized kinetic terms, which was to be expected. We can therefore write the following Lagrangian for the fermions (neglecting interactions)
\begin{equation}
\begin{split}
\mathcal{L}\supset& -i\psi^1_\xi \slashed{\partial} \psi_\xi^{1\dag}  -i \psi_{A\xi} \slashed{\partial} \psi_\xi^{A\dag}-\tfrac{i}{g^2} \lambda \slashed{\partial} \lambda^\dag+\sqrt{2 \langle g^2 g_{A\bar B}k^A\bar k^{\bar B}\rangle}
\left(\psi^{1\dag}_\xi \frac{\lambda^{\dag}}{\langle g\rangle}+\psi^1_\xi\frac{\lambda}{\langle g\rangle} \right)\\
&=i\bar \Psi \slashed\partial \Psi  -i \psi_{A\xi} \slashed{\partial} \psi_\xi^{A\dag}+\sqrt{2 \langle g^2 g_{A\bar B}k^A\bar k^{\bar B}\rangle}\bar \Psi \Psi\ , \quad A=2,\dots, N
\end{split}
\end{equation}
This is the equation of motion for a Dirac fermion of the form $\Psi=\left( \begin{matrix} \lambda^\dag/\langle g\rangle\\\psi \end{matrix} \right)$, with mass $M_\Psi=\sqrt{2 \langle g^2 g_{A\bar B}k^A\bar k^{\bar B}\rangle}$, and $N-1$ massless Weyl spinors $\Psi^A_\xi$.

\end{itemize}
 
In conclusion, we found that after we fix the a linear combination of the VEVs of the scalar fields $z^A$ to obtain a supersymmetric vacuum, we obtain 
a massive vector multiplet, containing a real scalar, a massive vector boson and a Dirac fermion. That is,
\begin{equation}
\left( \phi, A_\mu, \Psi \right)\ .
\end{equation}
The real scalar and the fermions are linear combinations of fields which receive mass terms after we fix the VEVs of the scalars and expand around the vacuum:
\begin{equation}
\begin{split}
\phi=\frac{ i\langle k^Ag_{A\bar B}\rangle \tfrac{\delta \bar z^{\bar B}}{2}-i\langle k^{\bar B}g_{A\bar B}\rangle \tfrac{\delta z^A}{2}  }{\sqrt{ \langle g_{A\bar B}k^A\bar k^{\bar B}\rangle}}\\
\end{split}
\end{equation}

All components of the massive vector multiplet have the same mass,
\begin{equation}
\label{eq:masses_equal}
m_{A}=M_\Psi=m_\phi=\sqrt{ 2\langle  g^2 g_{A\bar B}k^A\bar k^{\bar B}\rangle}\ ,
\end{equation}
as expected, since we fixed the vacuum to be supersymmetric, with $V_D=0$.

Furthermore, we found that the effective theory contains another $N-1$ massless chiral multiplets  $(\xi^A, \psi^A)$, for $A=2,\cdots N$. These chiral components are given by the linear combinations given in \eqref{eq:rest_states1}.  The 
scalar components $\xi^A$, $A=2,\dots,N$ are linear combinations of the scalar perturbations around the vacuum, which, however, receive no mass terms. A D-term potential develops in the direction of the $\xi^A$ scalar only, while the rest of the states $\xi^A$, $A=2,\dots,N$, which are orthogonal to it, remain flat. 

Having computed all the mass eigenstates after the D-term stabilization, we are ready to express the complete Lagrangian of the low-energy theory. We find
\begin{equation}
\begin{split}
\mathcal{L}&\supset
-\partial_\mu \phi \partial^\mu \phi - m_\phi^2\phi^2+i\bar \Psi \slashed\partial \Psi+M_\Psi\bar \Psi \Psi\ 
 -\frac{1}{4g^2}F_{\mu\nu}F^{\mu \nu}-\frac{1}{2\langle g^2\rangle}m_A^2A_\mu A^{\mu} \\
 &-\sum_{A=2}^{N}\partial_\mu \xi_A\partial^\mu \bar \xi^A-\sum_{A=2}^{N}i\psi_{A\xi}\slashed{\partial}\psi^{A\dag}_\xi+\dots\ , \qquad A=2,\cdots, N\ .
 \end{split}
\end{equation}  
We have omitted the interaction terms. The masses $m_\phi$, $M_\Psi$ and $m_A$ are equal, as shown in eq. 
\eqref{eq:masses_equal}. The vector boson is defined in the "unitary" gauge.

\chapter{K\"ahler metrics}

The complete K\"ahler potential for the $S$, $T^i$, $Z$ and $C^L$ fields is given by
\begin{equation}
K=K_S+K_T+K_{\text{matter}}\ ,
\end{equation}
where
\begin{equation}
\begin{split}
\label{}
&K_S=-\kappa_4^{-2}\ln\left(S+\bar S-\frac{\pi}{2}\epsilon_S \frac{(Z+\bar Z)^2}{W_i(T^i+\bar T^i)}\right)\ , \\
&K_T=-\kappa_4^{-2}\ln\left(\frac{1}{48}d_{ijk}(T^i+\bar T^i)(T^j+\bar T^j)(T^k+\bar T^k)\right)\ ,\\
&K_{\text{matter}}=e^{ \kappa_4^2K_T/3}\mathcal{G}_{L\bar M}C^L\bar C^{\bar M}\ .
\end{split}
\end{equation}
The first derivatives of the K\"ahler potential with respect to $S$, $T^i$, $Z$ and $C^L$ are
\begingroup
\allowdisplaybreaks
\begin{align}
\frac{\partial K}{\partial S}=&\frac{\partial K_S}{\partial S}=-\frac{1}{\kappa_4^2\left(S+\bar S-\frac{\pi}{2}\epsilon_S \frac{(Z+\bar Z)^2}{W_i(T^i+\bar T^i)}\right)}=-\frac{1}{2\kappa^2_4V}\ ,\\
\frac{\partial K}{\partial T^i}
=&\frac{\partial K_T}{\partial T^i}    -\frac{\frac{\pi}{2}\epsilon_S \frac{W_i(Z+\bar Z)^2}{\left(W_j(T^j+\bar T^j)\right)^2}}{\kappa_4^2\left(S+\bar S-\frac{\pi}{2}\epsilon_S \frac{(Z+\bar Z)^2}{W_j(T^j+\bar T^j)}\right)}    +\frac{\kappa_4^2\partial K_T}{\partial T^i}\frac{e^{\kappa_4^2K_T/3}}{3}\mathcal{G}_{L\bar M}C^L\bar C^{\bar M}\\
=&-\frac{d_{ijk}a^ja^k}{4\kappa_4^2\hat RV^{2/3}}-\frac{1}{4\kappa_4^2V}\pi\epsilon_Sz^2W^i-\frac{d_{ijk}a^ja^k}{4\hat RV^{2/3}}\frac{e^{\kappa_4^2K_T/3}}{3}\mathcal{G}_{L\bar M}C^L\bar C^{\bar M}\ ,\\
\frac{\partial K}{\partial Z}=& \frac{\partial K_S}{\partial Z}= \frac{{\pi}\epsilon_S \frac{(Z+\bar Z)}{W_i(T^i+\bar T^i)}}{\kappa_4^2\left(S+\bar S-\frac{\pi}{2}\epsilon_S \frac{(Z+\bar Z)^2}{W_i(T^i+\bar T^i)}\right)}  =\frac{\pi\epsilon_Sz}{2\kappa^2_4V}\ ,\\
\frac{\partial K}{\partial C^L}=&\frac{\partial K_{\text{matter}}}{\partial C_L}=e^{\kappa_4^2K_T/3}\mathcal{G}_{L\bar M}\bar C^{\bar M}\ .
\end{align}
\endgroup

The second derivatives of the K\"ahler potential with respect to $S$, $T^i$, $Z$ and $C^L$ are

\begin{align}
\label{black1}
\nonumber
g_{S\bar S}=\frac{\partial^2 K}{\partial S\partial \bar S}&=\frac{1}{\kappa_4^2\left(S+\bar S-\frac{\pi}{2}\epsilon_S \frac{(Z+\bar Z)^2}{W_k(T^k+\bar T^k)}\right)^2}\\\nonumber
&=\frac{1}{4\kappa^2_4V^2}\ ,\\\nonumber
g_{T^i\bar S}=\frac{\partial^2K}{\partial T^i\partial \bar S}&=\frac{\frac{\pi}{2}\epsilon_SW_i \frac{(Z+\bar Z)^2}{\left(W_k(T^k+\bar T^k)\right)^2}}{\kappa_4^2\left(S+\bar S-\frac{\pi}{2}\epsilon_S \frac{(Z+\bar Z)^2}{W_k(T^k+\bar T^k)}\right)^2}  \\
&=\frac{1}{8\kappa_4^2V^2}\pi\epsilon_Sz^2W^i\ ,\\\nonumber
g_{Z\bar S}=\frac{\partial^2K}{\partial Z\partial \bar S}&=-\frac{{\pi}\epsilon_S \frac{(Z+\bar Z)}{W_k(T^k+\bar T^k)}}{\kappa_4^2\left(S+\bar S-\frac{\pi}{2}\epsilon_S \frac{(Z+\bar Z)^2}{W_k(T^k+\bar T^k)}\right)^2} \\\nonumber
& =-\frac{\pi\epsilon_Sz}{4\kappa^2_4V^2}\ ,\\\nonumber
g_{T^i\bar Z}=\frac{\partial^2K}{\partial T^i\partial \bar Z}&=
- \frac{{\pi}\epsilon_SW_i \frac{(Z+\bar Z)}{\left(W_k(T^k+\bar T^k)\right)^2}}{\kappa_4^2\left(S+\bar S-\frac{\pi}{2}\epsilon_S \frac{(Z+\bar Z)^2}{W_k(T^k+\bar T^k)}\right)}
 -\frac{{\pi}\epsilon_S \frac{(Z+\bar Z)}{W_k(T^k+\bar T^k)}\frac{\pi}{2}\epsilon_S W_i\frac{(Z+\bar Z)^2}{\left(W_k(T^k+\bar T^k)\right)^2}
 }{\kappa_4^2\left(S+\bar S-\frac{\pi}{2}\epsilon_S \frac{(Z+\bar Z)^2}{W_k(T^k+\bar T^k)}\right)^2}\\\nonumber
 &=-\frac{{\pi}\epsilon_SW_i{z}}{4\kappa_4^2V(W_kt^k)}-\frac{{\pi^2}\epsilon_S^2W_iz^3}{8\kappa_4^2V^2}\ ,\\ \nonumber
g_{T^i\bar T^j}&=\frac{\partial^2K}{\partial T^i\partial \bar T^j}\\
&=\frac{\partial^2 K_T}{\partial T^i\partial \bar T^j}
  +\frac{{\pi}\epsilon_SW_iW_j \frac{(Z+\bar Z)^2}{\left(W_k(T^k+\bar T^k)\right)^3}}{\kappa_4^2\left(S+\bar S-\frac{\pi}{2}\epsilon_S \frac{(Z+\bar Z)^2}{W_k(T^k+\bar T^k)}\right)}  
    +\frac{\frac{\pi^2}{4}\epsilon_S^2 W_iW_j\frac{(Z+\bar Z)^4}{\left(W_k(T^k+\bar T^k)\right)^4}}{\kappa_4^2\left(S+\bar S-\frac{\pi}{2}\epsilon_S \frac{(Z+\bar Z)^2}{W_k(T^k+\bar T^k)}\right)^2} \\\nonumber
    &+\frac{\kappa_4^2\partial^2 K_T}{\partial T^i\partial \bar T^j}\frac{e^{\kappa_4^2K_T/3}}{3}\mathcal{G}_{L\bar M}C^L\bar C^{\bar M}+\frac{\kappa_4^2\partial K_T}{\partial T^i}\frac{\kappa_4^2\partial K_T}{\partial \bar T^j}\frac{e^{\kappa_4^2K_T/3}}{9}\mathcal{G}_{L\bar M}C^L\bar C^{\bar M}\\\nonumber
&=g^T_{ij}+\frac{\pi \epsilon_SW_iW_jz^2}{4\kappa_4^2V(W_kt^k)}
+\frac{\pi^2\epsilon_S^2W_iW_jz^4}{16\kappa_4^2V^2} +\left(\kappa_4^2g_{ij}^T+\frac{1}{3}\left(\frac{d_{ijk}a^ja^k}{4\hat RV^{2/3}}\right)^2\right)\frac{e^{\kappa_4^2K_T/3}}{3}\mathcal{G}_{L\bar M}C^L\bar C^{\bar M}\ ,\\\nonumber
 g_{Z\bar Z}&=\frac{\partial K}{\partial Z\partial Z}=  \frac{{\pi}\epsilon_S \frac{1}{W_k(T^k+\bar T^k)}}{\kappa_4^2\left(S+\bar S-\frac{\pi}{2}\epsilon_S \frac{(Z+\bar Z)^2}{W_k(T^k+\bar T^k)}\right)} 
+\frac{{\pi}\epsilon_S \frac{(Z+\bar Z)}{W_k(T^k+\bar T^k)}{\pi}\epsilon_S \frac{(Z+\bar Z)}{W_k(T^k+\bar T^k)}}{\kappa_4^2\left(S+\bar S-\frac{\pi}{2}\epsilon_S \frac{(Z+\bar Z)^2}{W_k(T^k+\bar T^k)}\right)^2} 
,\\\nonumber
&=\frac{{\pi}\epsilon_S}{4\kappa_4^2V}\frac{1}{W_kt^k}+\frac{\pi^2\epsilon_S^2z^2}{4\kappa_4^2V^2}\\\nonumber
 \nonumber g_{T^i\bar C^{L}}&=\frac{\partial^2K}{\partial T^i\partial \bar C^{\bar L}}=\frac{1}{3}\frac{\kappa_4^2\partial K_T}{\partial T^i}e^{\kappa_4^2K_T/3}\mathcal{G}_{L\bar M}\bar C^{\bar M}\\\nonumber
&=\frac{d_{ijk}a^ja^k}{12\hat RV^{2/3}}e^{\kappa_4^2K_T/3}\mathcal{G}_{L\bar M}\bar C^{\bar M}\ ,\\\nonumber
  g_{C^L\bar C^{\bar M}}&=-\frac{\partial^2K}{\partial C^L\partial \bar C^{\bar M}}=e^{\kappa_4^2K_T/3}\mathcal{G}_{L\bar M}\ .
\end{align}
In the above, we have defined
\begin{equation}
g_{i\bar j}^T=\frac{\partial^2 K_T}{\partial T^i\partial {\bar T^{\bar j}}}=-\frac{d_{ijk}t^k}{\kappa_4^2(\tfrac{2}{3})d_{lmn}t^lt^mt^n}+\frac{d_{ikl}t^kt^ld_{jmn}t^mt^n}{\kappa_4^2\left(\tfrac{2}{3}\right)^2(d_{lmn}t^lt^mt^n)^2}\ .
\end{equation}
Of course, we also have
\begin{equation}
g_{S\bar T^i}=g_{T^i\bar S}\ ,\quad g_{S\bar Z}=g_{Z\bar S}\ ,\quad g_{Z\bar T^i}=g_{T^i\bar Z}\ ,\quad \text{and}\quad g_{C_L\bar T^i}=g^*_{T^i\bar C^{\bar L}}\ .
\end{equation}

\chapter{Root Diagrams}\label{sec:Conventions}

\begin{figure}[t]
	\centering
	\includegraphics[width=0.9\textwidth]{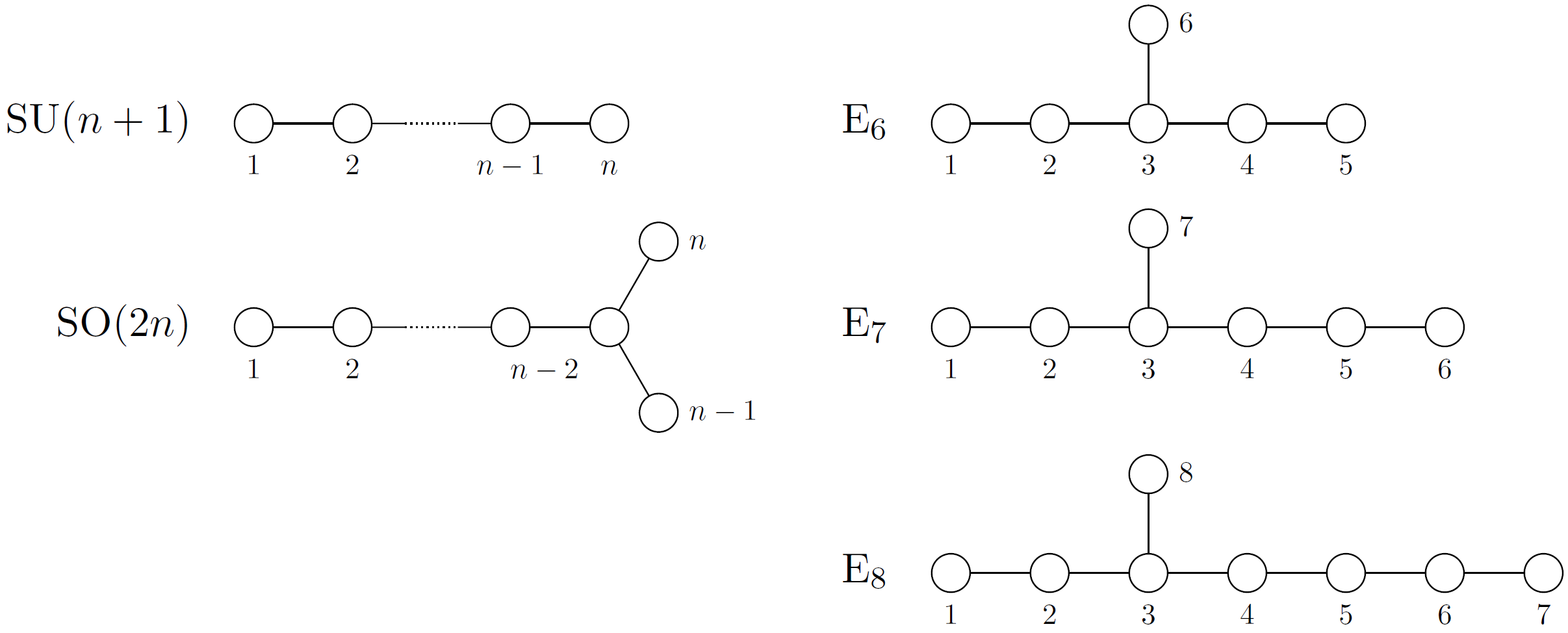}
	\caption{Dynkin diagrams of the classical Lie groups. We follow the Dynkin convention for labelling simple roots, as used in \cite{Feger:2012bs,Feger:2019tvk}.}
	\label{fig:e8}
\end{figure}

We follow the Dynkin convention for labelling the simple roots of $\ex 8$, in agreement with the \emph{Mathematica} package LieART~\cite{Feger:2012bs,Feger:2019tvk} which we used in many of the line bundle vector calculations. In particular, we number the nodes of the Dynkin diagram as in Figure \ref{fig:e8}. We mostly work in the ``orthogonal basis'' $\{e_{a}\}$ with $a=1,\ldots,8$ (see \cite{Feger:2012bs} for more details), where the components of the root vectors are given with respect to an orthogonal basis. In particular, the line bundle vectors $\boldsymbol{V}_{i}$ and roots $\boldsymbol{r}$ in the main text are expressed in the orthogonal basis. The eight simple roots $\alpha_{I}=\alpha_{I}^{a}e_{a}$ of $\ex 8$ are given in this basis by
\begin{equation}
	\begin{aligned}\alpha_{1}^{a} & =\tfrac{1}{2}(1,-1,-1,-1,-1,-1,-1,1)\ , & \alpha_{5}^{a} & =(0,0,0,-1,1,0,0,0)\ ,\\
		\alpha_{2}^{a} & =(-1,1,0,0,0,0,0,0)\ , & \alpha_{6}^{a} & =(0,0,0,0,-1,1,0,0)\ ,\\
		\alpha_{3}^{a} & =(0,-1,1,0,0,0,0,0)\ , & \alpha_{7}^{a} & =(0,0,0,0,0,-1,1,0)\ ,\\
		\alpha_{4}^{a} & =(0,0,-1,1,0,0,0,0)\ , & \alpha_{8}^{a} & =(1,1,0,0,0,0,0,0)\ .
	\end{aligned}
	\label{eq:roots}
\end{equation}

In addition to the orthogonal basis we also have the $\alpha$-basis and the $\omega$-basis. The $\alpha$-basis is the basis of simple roots. This has the advantage that it shows precisely how a given root is made from a sum of simple roots. In this basis, the components of the simple roots are given by
\begin{equation}
	\tilde{\alpha}_{1}^{a}=(1,0,0,0,0,0,0,0)\ ,\qquad\tilde{\alpha}_{2}^{a}=(0,1,0,0,0,0,0,0)\ ,
\end{equation}
and so on, so that $\alpha_{I}=\tilde{\alpha}_{I}^{a}\alpha_{a}=\delta_{I}^{a}\alpha_{a}$. Finally, the $\omega$-basis is the basis of fundamental weights, also known as the Dynkin basis. This basis is such that the simple roots correspond to the rows of the Cartan matrix $A_{ab}$ (for $\ex 8$ in this case), so that, for example, the first two simple roots can be written as
\begin{equation}
	\hat{\alpha}_{1}^{a}=(2,-1,0,0,0,0,0,0)\ ,\qquad\hat{\alpha}_{2}^{a}=(-1,2,-1,0,0,0,0,0)\ ,
\end{equation}
where we have written the simple roots in the $\omega$-basis as $\alpha_{I}=\hat{\alpha}_{I}^{a}\omega_{a}$. Note that for algebras (such as $\ex 8$) whose roots are all of length 2, the $\alpha$-basis and $\omega$-basis are dual
\begin{equation}
	(\alpha_{a},\omega_{b})=\delta_{ab}\ .
\end{equation}
In particular, this implies
\begin{equation}
	(\alpha_{I},\alpha_{J})=\tilde{\alpha}_{I}^{a}\hat{\alpha}_{J}^{b}(\alpha_{a},\omega_{b})=\delta_{I}^{a}\hat{\alpha}_{J}^{b}\delta_{ab}=A_{IJ}\ .
\end{equation}
The transformations between these bases are given by
\begin{equation}
	\alpha_{a}=\sum_{b}A_{ab}\omega_{b}\ ,\qquad\omega_{a}=\sum_{b}\hat{\Omega}_{ab}e_{b}\ ,
\end{equation}
where $\hat{\Omega}_{ab}$ is the matrix whose rows are the fundamental weights in the orthogonal basis and $A_{ab}$ is the Cartan matrix of $\ex 8$, given by
\begin{equation}
	A_{ab}=\left(\begin{array}{cccccccc}
		2 & -1 & 0 & 0 & 0 & 0 & 0 & 0\\
		-1 & 2 & -1 & 0 & 0 & 0 & 0 & 0\\
		0 & -1 & 2 & -1 & 0 & 0 & 0 & -1\\
		0 & 0 & -1 & 2 & -1 & 0 & 0 & 0\\
		0 & 0 & 0 & -1 & 2 & -1 & 0 & 0\\
		0 & 0 & 0 & 0 & -1 & 2 & -1 & 0\\
		0 & 0 & 0 & 0 & 0 & -1 & 2 & 0\\
		0 & 0 & -1 & 0 & 0 & 0 & 0 & 2
	\end{array}\right)\ .
\end{equation}

Given the transformations between the various bases, it is simple to write down line bundle vectors that break particular combinations of the simple roots. Consider the inner product of the line bundle vector $V$ with the $I^{\text{th}}$ simple root $\alpha_{I}$:
\begin{equation}
	V\cdot\alpha_{I}=\hat{V}^{a}\tilde{\alpha}_{I}^{b}(\omega_{a},\alpha_{b})=\hat{V}^{a}\delta_{I}^{b}\delta_{ab}=\hat{V}_{I}\ .
\end{equation}
This means that the inner product of a line bundle vector $V$ with a simple root $\alpha_{I}$ is given by the $I^{\text{th}}$ component of the line bundle vector written in the $\omega$-basis. If, for example, we want to pick a line bundle vector that breaks the first simple root of $\ex 8$ and is orthogonal to the others, we can take $\hat{V}^{a}=(1,0,0,0,0,0,0,0)$. Transforming back to the orthogonal basis (which is the basis used in the main text), we would then have
\begin{equation}
	\boldsymbol{V}=(0,0,0,0,0,0,0,2)\ .
\end{equation}
This would lead to an unbroken $SO(14)$ group.

\chapter{Subbundles of Isomorphic Extension Bundles}\label{app:subbundles}

In Table \ref{tab:branches}, we presented the six different extension  branches for deforming the Whitney sum $V_{\repd 3}=\mathcal{F}\oplus\mathcal{K}\oplus\mathcal{E}$ away 
from the decomposable locus. For each such branch, there are two different pairs of sequences which lead, however, to isomorphic $SU(3)$ bundles. Because of this isomorphism, in Table 3 we presented only one pair of these sequences.  Here, however, we need to discuss both of them. For simplicity, let us restrict the discussion to the {\it first} extension branch only. However, the conclusions will apply to the remaining five branches as well. Let us briefly review the two sets of extension sequences in the first extension branch. These are
\begin{equation}
\begin{split}
\label{eq:sequences_app1}
\op{Ext}^1(\mathcal{E},\mathcal{F})=H^1(X,\mathcal{F}\otimes \mathcal{E}^*)\neq 0\quad  \Rightarrow\quad  &0\rightarrow \mathcal{F}\rightarrow W\rightarrow \mathcal{E}\rightarrow  0\ ,\\\op{Ext}^1(W,\mathcal{K})=H^1(X,\mathcal{K}\otimes \mathcal{E}^*) \neq 0\quad \Rightarrow \quad &0\rightarrow \mathcal{K}\rightarrow V^{\prime}_{\repd 3} \rightarrow W \rightarrow 0\ ,
\end{split}
\end{equation}
or
\begin{equation}
\begin{split}
\label{eq:sequences_app2}
\op{Ext}^1(\mathcal{E},\mathcal{K})=H^1(X,\mathcal{K}\otimes \mathcal{E}^*)\neq 0\quad  \Rightarrow\quad  &0\rightarrow \mathcal{K}\rightarrow W^\prime\rightarrow \mathcal{E}\rightarrow  0\ ,\\
\op{Ext}^1(W^\prime,\mathcal{F})=H^1(X,\mathcal{F}\otimes \mathcal{E}^*) \neq 0\quad \Rightarrow \quad &0\rightarrow \mathcal{F}\rightarrow \mathcal{V}^{\prime}_{\repd 3} \rightarrow W^\prime \rightarrow 0\ .
\end{split}
\end{equation}
However, following the calculation in \cite{Anderson:2010ty}, it can be shown that the resulting $SU(3)$ bundles are actually isomorphic $V^\prime_{\repd 3}\simeq \mathcal{V}_{\repd 3}^\prime$ and so it does not matter which extension one uses.

From the first set definition of the extension in \eqref{eq:sequences_app1}, we learn there is an embedding
\begin{equation}
\mathcal{K}\hookrightarrow V^\prime_{\repd 3}\ ,
\end{equation}
and, hence, $\mathcal{K}$ is a rank-one subbundle of $V_{\repd 3}^\prime$. This means $V_{\repd 3}^\prime$ is stable only if the slope of $\mathcal{K}$ is less than the slope of $V_{\repd 3}^\prime$ (which vanishes):
\begin{equation}
\mu(\mathcal{K})<0\ .
\end{equation}
From the second definition of the extension in \ref{eq:sequences_app2}, we learn that $\mathcal{F}$ is a subbundle of $\mathcal{V}_3^\prime$, which itself is isomorphic to $V_{\repd 3}^\prime$. An obvious question is whether $\mathcal{F}$ is thus a subbundle of $V_{\repd 3}^\prime$ as well, which would then constrain the slope of $\mathcal{F}$ to be negative.

Recall that a sheaf $\mathcal{F}$ is a sub-sheaf of $V$ if it has smaller rank and 
and there exists an embedding $\mathcal{F} \hookrightarrow V$~\cite{Anderson:2009nt}. The space of homomorphisms from $\mathcal{F}$
to $V$, denoted $\op{Hom}_X(\mathcal{F},V)$, is then isomorphic to the space of global sections $H^0(X, \mathcal{F}^*\otimes V)$. If $V$ is an $SU(N)$ bundle, it is stable if all its sub-sheaves $\mathcal{F}$ have negative slope. Hence, we have that
\begin{equation}\label{eq:stability}
	\begin{gathered}
		V \text{ is stable}\\
		\Updownarrow\\
\mu(\mathcal{F})<0\quad\forall\,\mathcal{F} \text{ with } 
0<\rank \mathcal{F}<\rank V\text{ and } H^0(X, \mathcal{F}^*\otimes V)\neq0\ .
\end{gathered}
\end{equation}
Applying this statement to our case, the bundle $\mathcal{F}$ is a subbundle of $V_{\repd 3}^\prime$ if we can find a homomorphism 
$\mathcal{F}\hookrightarrow V^\prime_{\repd 3}$ or, equivalently, if
\begin{equation}
\label{eq:condition}
\op{Hom}_X(\mathcal{F},V^\prime_{\repd 3})=H^0(X,\mathcal{F}^*\otimes V^\prime_{\repd 3} )\neq 0\ .
\end{equation}
In the following, we will show that such a homomorphism does indeed exist. Let us start by tensoring with $\mathcal{F}^*$ the sequences
\begin{equation}
\label{eq:sequences_app3}
 0\rightarrow \mathcal{K}\rightarrow V^{\prime}_{\repd 3} \rightarrow W \rightarrow 0\ 
\end{equation}
and
\begin{equation}
\label{eq:sequences_app4}
 0\rightarrow \mathcal{F}\rightarrow W \rightarrow \mathcal{E} \rightarrow 0\ 
\end{equation}
to obtain
\begin{equation}
\begin{split}
\label{eq:sequences_app5}
& 0\rightarrow \mathcal{F}^*\otimes \mathcal{K}\rightarrow \mathcal{F}^*\otimes V^{\prime}_{\repd 3} \rightarrow \mathcal{F}^*\otimes W \rightarrow 0\ ,\\
& 0\rightarrow \mathcal{F}^*\otimes \mathcal{F}\rightarrow \mathcal{F}^*\otimes W \rightarrow \mathcal{F}^*\otimes \mathcal{E} \rightarrow 0\ .
\end{split}
\end{equation}
Taking long exact sequences in cohomology of these, gives
\begin{equation}
\begin{split}
\label{eq:12341}
 0&\rightarrow H^0(X,\mathcal{F}^*\otimes \mathcal{K})\rightarrow H^0(X,\mathcal{F}^*\otimes V^{\prime}_{\repd 3}) \rightarrow H^0(X,\mathcal{F}^*\otimes W )
 \\ 
 &\xrightarrow{\delta_1} H^1(X,\mathcal{F}^*\otimes \mathcal{K})\rightarrow \dots\ ,\\
 \end{split}
\end{equation}
and
\begin{equation}
\label{eq:12342}
\begin{split}
 0&\rightarrow H^0(X,\mathcal{F}^*\otimes \mathcal{F})\rightarrow H^0(X,\mathcal{F}^*\otimes W) \rightarrow H^0(X,\mathcal{F}^*\otimes \mathcal{E} )
 \\ 
 &\xrightarrow{\delta_2} H^1(X,\mathcal{F}^*\otimes \mathcal{F})\rightarrow \dots\ .\\
 \end{split}
\end{equation}

For a line bundle and its dual we have
\begin{equation}
\begin{split}
H^0(X,\mathcal{F}^*\otimes \mathcal{F})=H^0(X,\mathcal{O}_X)=\mathbb{C}\ ,\\
H^1(X,\mathcal{F}^*\otimes \mathcal{F})=H^1(X,\mathcal{O}_X)=0\ .
\end{split}
\end{equation}
Furthermore, if the line bundles $\mathcal{F}^*\otimes \mathcal{K}$ and $\mathcal{F}^*\otimes \mathcal{E}$ have negative
slopes somewhere in the Kähler cone, the zeroth cohomology classes
\begin{equation}
	H^0(X,\mathcal{F}^*\otimes \mathcal{K})=0\ ,\qquad H^0(X,\mathcal{F}^*\otimes \mathcal{E})=0\ ,
\end{equation}
vanish, as explained in Footnote 4 of \cite{Anderson:2012yf}. It can be shown that for our particular Schoen manifold, this condition is always satisfied if the line bundles $\mathcal{F}^*\otimes \mathcal{K}=L_1L_2^3$ and $\mathcal{F}^*\otimes \mathcal{E}=L_1^{-1}L_2^3$ have both positive and negative entries $m^i+3n^i$ and $-m^i+3n^i$, when written as $\mathcal{F}^*\otimes \mathcal{K}=\mathcal{O}_X(m^1+3n^1,m^2+3n^2,m^3+3n^3)$ and $\mathcal{F}^*\otimes \mathcal{E}=\mathcal{O}_X(-m^1+3n^1,-m^2+3n^2,-m^3+3n^3)$. This is generally the case for the line bundles we sample. 

Hence equations \eqref{eq:12341} and \eqref{eq:12342} become
\begin{equation}
\label{eq:delta_1_seq}
 0\rightarrow H^0(X,\mathcal{F}^*\otimes V^{\prime}_{\repd 3}) \rightarrow H^0(X,\mathcal{F}^*\otimes W )\xrightarrow{\delta_1} H^1(X,\mathcal{F}^*\otimes \mathcal{K})\rightarrow \dots\ ,\\
\end{equation}
and
\begin{equation}
\label{eq:C_seq}
 0\rightarrow \mathbb{C}\rightarrow H^0(X,\mathcal{F}^*\otimes W) \rightarrow 0
\end{equation}
respectively. 

From the sequence \eqref{eq:delta_1_seq} we learn that $H^0(X,\mathcal{F}^*\otimes V^{\prime}_{\repd 3})=\kernel \delta_1$. Therefore, to evaluate $H^0(X,\mathcal{F}^*\otimes V^{\prime}_{\repd 3})$, we must first analyze the coboundary map $\delta_1$. First note that from eq.~\eqref{eq:C_seq} we learn that $H^0(X,\mathcal{F}^*\otimes W)=\mathbb C$. Furthermore, we have that the fields $\tilde C_1$ are counted by $H^1(X,\mathcal{F}^*\otimes \mathcal{K})$. In the chosen vacuum branch all VEVs for the $\tilde C_1$ fields vanish. Since $\delta_1$ is determined by the vacuum state configuration, it follows that it maps only to the zero element of $H^1(X,\mathcal{F}^*\otimes \mathcal{K})$. Hence, $\kernel \delta_1=H^0(X,\mathcal{F}^*\otimes W)=\mathbb C$. 

 Putting this together, we conclude that
\begin{equation}
 H^0(X,\mathcal{F}^*\otimes  V^{\prime}_{\repd 3})=\mathbb C\ ,
\end{equation}
which is indeed non-zero. Therefore, according to \eqref{eq:condition}, there exists a homomorphism $\mathcal{F}\hookrightarrow V^{\prime}_{\repd 3}$ such that $\mathcal{F}$ is a subbundle of $V^{\prime}_{\repd 3}$.
Sequences \eqref{eq:sequences_app1} and \eqref{eq:sequences_app2} then tell us that both $\mathcal{F}$ and $\mathcal{K}$ are subbundles of $V_{\repd 3}^\prime$. Therefore, according to \eqref{eq:stability}, $V_{\repd 3}^\prime$ is stable only if
\begin{equation}
\mu({\mathcal{F}})<0 \quad\text{and}\quad \mu({\mathcal{K}})<0\ .
\end{equation}

\chapter {$B-L$ MSSM field content }

In this Appendix, we present for clarity all the $B-L$ MSSM field content.

\section{Gauge Eigenstates}

\begin{itemize}
\item{Bosons}

\underline{{\it vector gauge bosons}}

~~~ $SU(2)_L-\quad\> W^1_\mu\>, W^2_\mu\>, W^3_\mu$, $\quad $   
coupling parameter $g_2$

~~~ $U(1)_{B-L}-\quad \> B^'_\mu \> $, $\quad $   
coupling parameter $g_{BL}$

~~~ $U(1)_{3R}-\quad \> {W_R}_\mu \> $, $\quad $   
coupling parameter $g_R$

~~~ $U(1)_{Y}-\quad \> {B}_\mu \> $, $\quad $   
coupling parameter $g'$

~~~ $U(1)_{EM}-\quad \> {\gamma}^0_\mu \> $, $\quad $   
coupling parameter $e$

~~~B-L Breaking:    $U(1)_{3R}\otimes U(1)_{B-L}\rightarrow U(1)_Y,\quad$    massive boson ${Z_R}_\mu$,$\>\>$ coupling $g_{Z_R}$

~~~EW Breaking:    $SU(2)_L\otimes U(1)_Y \rightarrow U(1)_{EM},\quad$    massive bosons $Z^0_\mu
,\>W^\pm_\mu\quad$

\underline{{\it Higgs scalars}}

~~~$H_u^0\>, H_u^+\>, H_d^0\>, H_d^-\quad $

\item{Weyl Spinors}

\underline{ {\it gauginos}}

~~~ $SU(2)_L-\> \tilde W^0\>, \tilde W^\pm$,\quad
$U(1)_{B-L}- \>\tilde B^' , \quad$ $U(1)_{3R}-\> {\tilde W_R} \> $,\quad
$U(1)_{Y}-\> \tilde {B},\quad$ $U(1)_{EM}- \> \tilde {\gamma}^0 \> $

\underline{{\it Higgsinos}}

~~~$\tilde H_u^0\>, \tilde H_u^+\>, \tilde H_d^0\>, \tilde H_d^-$

\underline{{\it leptons}}

 ~~~left chiral\quad  $e_i,\>\nu_i, \>\> i=1,2,3 \quad \text{where} \quad e_1=e,\>e_2=\mu,\>e_3=\tau$
          
 ~~~right chiral\quad  $e^c_i,\>\nu^c_i, \>\> i=1,2,3 \quad \text{where} \quad e^c_1=e^c,\>e_2^c=\mu^c,\>e_3^c=\tau^c$

\underline{{\it sleptons}}

 ~~~left chiral\quad  $\tilde e_i,\>\tilde \nu_i, \>\> i=1,2,3 \quad \text{where} \quad \tilde e_1=\tilde e,\>\tilde e_2=\tilde \mu,\>\tilde e_3=\tilde \tau$
          
 ~~~right chiral\quad  $\tilde e^c_i,\>\tilde \nu^c_i, \>\> i=1,2,3 \quad \text{where} \quad \tilde e^c_1=\tilde e^c,\>\tilde e_2^c=\tilde \mu^c,\>\tilde e_3^c=\tilde \tau^c$

\end{itemize}

\section{Mass Eigenstates}

\begin{itemize}

\item{Weyl Spinors}

\underline{{\it leptons}}\\
 \quad  $e_i,\>\nu_i, \>\> i=1,2,3 \quad \text{where} \quad e_1=e,\>e_2=\mu,\>e_3=\tau$

\underline{{\it charginos and neutralinos}}\\
  \quad $\tilde \chi^\pm_1, \quad \tilde \chi^\pm_2, \quad \tilde \chi_n^0,\quad n=1,2,3,4,5,6$

\item{4-component Spinors}

\underline{{\it  leptons}}\\
~~~$\ell_i^-=\left(\begin{matrix}e_i\\ {e_i^c}^\dag\end{matrix}\right),\quad
\ell_i^+=\left(\begin{matrix}{e_i^c}\\ e_i^\dag\end{matrix}\right), \quad
\nu_i=\left(\begin{matrix}\nu_i\\ {\nu_i}^\dag\end{matrix}\right)\quad i=1,2,3$

\underline{{\it charginos and neutralinos}}

$\tilde X^-_1=\left(\begin{matrix}\tilde \chi^-_1\\ \tilde {\chi}^{+\dag}_1\end{matrix}\right),\quad
\tilde X^+_1=\left(\begin{matrix}\tilde \chi^+_1\\ \tilde {\chi}^{-\dag}_1\end{matrix}\right), \quad
\tilde X^0_n=\left(\begin{matrix}\tilde \chi^0_n\\ \tilde {\chi}^{0\dag}_n\end{matrix}\right)$

\end{itemize}

\subsection{VEV's}

\begin{itemize}
\item {sneutrino VEV's \\
\quad $\left<\tilde \nu^c_{3}\right> \equiv \frac{1}{\sqrt 2} {v_R} \quad 
\epsilon_i=\frac{1}{2}Y_{\nu i3}v_R \quad \left<\tilde \nu_{i}\right> \equiv \frac{1}{\sqrt 2} {v_L}_i, \quad i=1,2,3$}

\item {Higgs VEV's \\
$\left< H_u^0\right> \equiv \frac{1}{\sqrt 2}v_u, \ \ \left< H_d^0\right> \equiv \frac{1}{\sqrt 2}v_d, \quad \tan \beta=v_u/v_d$}

\end{itemize}

\chapter{ Mass Matrix elements}

\section{Chargino mass matrix}\label{appendix:A1}

The matrices $\mathcal{U}$ and $\mathcal{V}$ can be written schematically as

\begin{equation}
\mathcal{U}=
\left(
\begin{matrix}
U&0_{2\times3}\\
0_{3\times2}&1_{3\times3}\\
\end{matrix}
\right)
\left(
\begin{matrix}
1_{2\times2}&-\xi_-\\
\xi^{\dag}_-&1_{3\times3}\\
\end{matrix}
\right) \ ,
\quad \> 
\mathcal{V}=
\left(
\begin{matrix}
V&0_{2\times3}\\
0_{3\times2}&1_{3\times3}\\
\end{matrix}
\right)
\left(
\begin{matrix}
1_{2\times2}&-\xi_+\\
\xi^{\dag}_+&1_{3\times3}\\
\end{matrix}
\right)
\end{equation}
Assuming that the lighter chargino is ${\tilde \chi}_1^\pm$, and since we are interested in its decays, it follows that we will need the elements $\mathcal{U}_{1\>2+i}$ and $\mathcal{V}_{1\>2+i}$ and their conjugates when replacing 
a lepton state with the lightest chargino mass eigenstate. It follows from the above that

\begin{equation}
\mathcal{U}_{1\>2+i}=-\cos \phi_- \frac{g_2 v_d}{\sqrt{2}M_2\mu}\epsilon_i^*+\sin \phi_-\frac{\epsilon_i^*}{\mu} \ ,
\end{equation}
\begin{equation}
\mathcal{V}_{1\>2+i}=-\cos \phi_+ \frac{g_2 \tan \beta m_{e_i}}{\sqrt{2}M_2\mu}v_{L_i}+\sin \phi_+\frac{m_{e_i}}{\mu v_d}v_{L_i} \ .
\end{equation}

When replacing a charged Wino gaugino with the lightest chargino mass eigenstate,
 we need the elements $\mathcal{U}_{1\>1}$ and $\mathcal{V}_{1\>1}$ given by  
\begin{equation}
\mathcal{U}_{1\>1}=\cos \phi_-  \ ,\quad \quad
\mathcal{V}_{1\>1}=\cos \phi_+
\end{equation}
and their conjugates. Similarly for replacing a charged Higgsino, one needs $\mathcal{U}_{1\>2}$ and $\mathcal{V}_{1\>2}$, which we find to be
\begin{equation}
\mathcal{U}_{1\>2}=\sin \phi_-  \ , \quad \quad
\mathcal{V}_{1\>2}=\sin \phi_+
\end{equation}
and their conjugates.

We also need the elements $\mathcal{U}_{2+i\>1}$ and $\mathcal{V}_{2+i\>1}$ and their complex conjugates when 
replacing a  charged Wino state with a charged lepton, where
\begin{equation}
\mathcal{U}_{2+i\>1}=\frac{g_2}{\sqrt{2}M_2\mu}(v_d\epsilon^*_i+\mu v_{L_i}) \ ,
\quad \quad
\mathcal{V}_{2+i\>1}=-\frac{1}{\sqrt{2}M_2\mu}g_2\tan \beta m_{e_i}v_{L_i} \ .
\end{equation}
The elements $\mathcal{U}_{2+i\>2}$ and $\mathcal{V}_{2+i\>2}$ and their complex conjugates when required when
replacing a charged Higgsino state with a charged lepton,
\begin{equation}
\mathcal{U}_{2+i\>2}=\frac{\epsilon_i^*}{\mu} \ , \quad \quad
\mathcal{V}_{2+i\>2}=\frac{m_{e_i}}{ v_d \mu}v_{L_i} \ .
\end{equation}

The angles $\phi_\pm$ are defined in Section 5.1. They express the charged Wino and charged Higgsino content of the chargino mass eigenstates, in the absence of the RPV couplings $\epsilon_i$ and $v_{L_i}$
\begin{equation}
\tilde \chi^\pm_1=\cos \phi_\pm {\tilde{W}}^\pm+\sin \phi_\pm {\tilde{H}}^\pm
\end{equation}
and
\begin{equation}
\tilde \chi^\pm_2=-\sin \phi_\pm {\tilde{W}}^\pm+\cos \phi_\pm {\tilde{H}}^\pm.
\end{equation}

Hence, for $\phi^\pm=0$, we have purely Wino chargino states $\tilde \chi^\pm_1$ and purely Higgsino chargino
states $\tilde \chi^\pm_2$. Conversely, for $\phi^\pm=\pi/2$, we have purely Higgsino chargino states $\tilde \chi^\pm_1$ and purely Wino chargino
states $\tilde \chi^\pm_2$.

\section{Neutralino mass matrix}\label{appendix:A2}

The $\mathcal{N}$ matrix can be written schematically as
\begin{equation}
\mathcal{N}=\left(
\begin{matrix}
N&0_{3\times 3}\\
0_{3\times 3}&V^{\dag}_{PMNS}\\
\end{matrix}
\right)
\left(
\begin{matrix}
1_{6\times 6}& -\xi_0\\
\xi_0^{\dag}&1_{3\times 3}\\
\end{matrix}
\right) \ .
\end{equation}
The rows of $\xi_0$ are the gaugino gauge eigenstates, whereas the columns correspond to the
neutrino gauge eigenstates. These are explicitly labeled and presented below. They are
\begin{equation}
\xi_{0_{\tilde W_R\nu_{L_i}}}=\frac{g_R\mu}{8d_{{\tilde \chi}^0}}
[2M_{BL}v_u(g_2^2v_dv_u-2M_2\mu)\epsilon_i-g_{BL}^2M_2v_R^2(v_d\epsilon_i+\mu v_{L_i}^*)] \ ,
\end{equation}
\begin{equation}
\xi_{0_{\tilde W_2\nu_{L_i}}}=\frac{g_2\mu}{8d_{{\tilde \chi}^0}}
[2g_R^2M_{BL}v_dv_u^2\epsilon_i+M_{\tilde Y}v_R^2(v_d\epsilon_i+\mu v_{L_i}^*)] \ ,
\end{equation}
\begin{equation}
\xi_{0_{\tilde H^0_d\nu_{L_i}}}=\frac{1}{16d_{{\tilde \chi}^0}}
[ M_{\tilde \gamma} v_R^2v_u (v_d\epsilon_i-\mu v_{L_i}^*)
-4M_2\mu(M_{\tilde Y}v_R^2+g^2_RM_{BL}v_u^2)\epsilon_i] \ ,
\end{equation}
\begin{equation}
~~\qquad \xi_{0_{\tilde H^0_u\nu_{L_i}}}=\frac{1}{16d_{{\tilde \chi}^0}}
[ M_{\tilde \gamma} v_R^2v_u (v_d\epsilon_i+\mu v_{L_i}^*)
-4g_R^2\mu M_2M_{BL}v_d v_u\epsilon_i] \ ,
\end{equation}
\begin{multline}
\qquad\qquad \qquad \xi_{0_{\tilde B^'\nu_{L_i}}}=-\frac{1}{8d_{{\tilde \chi}^0}}
[ g_{BL}g_R^2 M_2\mu v_R^2 (v_d\epsilon_i+\mu v_{L_i}^*)\\
+2g_{BL}\mu v_u((  g_R^2M_2+g_2^2 M_R  )v_dv_u-2M_RM_2\mu)\epsilon_i] \ , \qquad\qquad \qquad
\end{multline}
\begin{multline}
\qquad\qquad \qquad\xi_{0_{\tilde \nu_3^c\nu_{L_i}}}=\frac{\mu}{8v_Rd_{{\tilde \chi}^0}}[(
M_{\tilde \gamma}v_R^2v_dv_u-2g_{BL}^2M_RM_2\mu v_R^2)v_{L_i}^*\\
+2M_{BL}(M_2(g_R^2v_R^2v_d-4M_R\mu v_u)+2(g_R^2M_2+g_2^2M_R)v_dv_u^2)\epsilon_i] \qquad
\end{multline}
where
\begin{equation}
d_{{\tilde \chi}^0}=\frac{1}{4}M_2M_1\mu^2v_R^2-\frac{1}{8}M_{\tilde \gamma}\mu v_R^2v_dv_u \ ,
\end{equation}
\begin{equation}
M_Y=g_R^2M_{BL}+g^2_{BL}M_R \ ,
\end{equation}
\begin{equation}
M_{\gamma} =g^2_{BL}g^2_RM+g_2^2g^2_RM_{BL}+g_2^2g_{BL}^2M_R \ .
\end{equation}
We can now express the matrix elements of 
\begin{equation}\label{eq:A19}
\mathcal{N}=\left(
\begin{matrix}
N&0_{3\times 3}\\
0_{3\times 3}&V^{\dag}_{PMNS}\\
\end{matrix}
\right)
\left(
\begin{matrix}
1_{6\times 6}& -\xi_0\\
\xi_0^{\dag}&1_{3\times 3}\\
\end{matrix}
\right)=\left(
\begin{matrix}
N&-N\xi_0\\
V_{PMNS}^\dag\xi_0^\dag&V_{PMNS}^\dag
\end{matrix}
\right)
\end{equation}
$N$ is the matrix that diagonalizes the neutralino mass matrix in the absence of RPV couplings to the three families of left-handed neutrinos. If the soft masses in the neutralino mass matrix, eq. \eqref{eq:A19}, are much larger than the Higgs VEV's $v_u$ and $v_d$, then, at zeroth order, we have
\begin{equation}
N=\left(
\begin{matrix}
\sin \theta_R&0&0&0&\cos \theta_R&0\\
0&1&0&0&0&0\\
0&0&0&1&0&0\\
0&0&1&0&0&0\\
-\frac{1}{\sqrt 2}\cos \theta_R&0&0&0&\frac{1}{\sqrt 2}\sin \theta_R&\frac{1}{\sqrt 2}\\
\frac{1}{\sqrt 2}\cos \theta_R&0&0&0&-\frac{1}{\sqrt 2}\sin \theta_R&\frac{1}{\sqrt 2}\\
\end{matrix}
\right) \ .
\end{equation}
However, in the regimes with small chargino and neutralino masses that we analyze, this approximation is no longer valid. The elements of $N$ will, in general,  have complicated expressions and we choose to evaluate them numerically. We use as input the numerical values of the neutralino mass matrix. We expect, however, based on the zeroth order form of $N$, that $N_{11}$, $N_{22}$, $N_{34}$, $N_{43}$, $N_{15}$, $N_{51}$, $N_{61}$,
$N_{55}$, $N_{65}$, $N_{56}$ and $N_{66}$ are of order $ \mathcal{O}(1)$, while the remaining matrix elements are of order $\mathcal{O}({M_{EW}/M_{soft}})<<1$.

Elements form the top-right block $N\xi_0$ have the form 
\begin{multline}
\mathcal{N}_{n\>6+i}=-N_{n\>1}\xi_{0_{\tilde W_R\nu_{L_i}}}-N_{n\>2}\xi_{0_{\tilde W_2\nu_{L_i}}}-N_{n\>3}\xi_{0_{\tilde H_d^0\nu_{L_i}}}-N_{n\>4}\xi_{0_{\tilde H_u^0\nu_{L_i}}}-N_{n\>5}\xi_{0_{\tilde B^'\nu_{L_i}}}-N_{n\>6}\xi_{0_{\tilde \nu_3^c\nu_{L_i}}}\\
\simeq N_{n\>1}\left[\frac{2g_RM_{BL}v_u}{M_1v_R^2}\epsilon_i+\frac{g_Rg_{BL}^2}{2M_1}v_{L_i}^*\right]
-N_{n\>2}\left[\frac{g_2v_d}{2M_2\mu}\epsilon_i+\frac{g_2}{2M_2}v_{L_i}^*\right]\\
+N_{n\>3}\left[\frac{\epsilon_i}{16\mu}\right]-N_{n\>4}\left[\frac{\tilde M_\gamma v_u}{4M_2M_1\mu^2}(v_d\epsilon_i+\mu v_{L_i})-\frac{g_R^2M_{BL}v_uv_d}{M_1v_R^2\mu}\epsilon_i\right]\\
-N_{n\>5}\left[ \frac{g_{BL}g_R^2}{2M_1\mu}(v_d\epsilon_i+\mu v_{L_i}^*) -\frac{2g_{BL}M_Rv_u}{M_1v_R^2}\epsilon_i \right] \qquad \\
-N_{n\>6}\left[ \frac{-g_{BL}^2M_R}{v_RM_1}v_{L_i}^*+\frac{ M_{BL}}{v_R^3M_2M_1}\mu(g_R^2M_2v_R^2v_d-4M_R\mu v_u)\epsilon_i \right] \qquad \qquad
\end{multline}
for $n=1,2,3,4,5,6$ and $i=1,2,3$. Elements of the bottom left block $V_{PMNS}^\dag\xi_0^\dag$ are computed in a similar fashion. One can then determine $\mathcal{N}_{6+i\>n}$ as
\begin{equation}
\mathcal{N}_{6+i\>n}={[V_{PMNS}^\dag]}_{6+i\>6+j}{[\xi_0^\dag]}_{6+j\>n}
\end{equation}
%


 
\bibliographystyle{utphys}
\bibliography{citationsbib2}

\end{document}